%% file: main_report.tex
\documentclass{elsart}

\usepackage{hyperref}
\hypersetup{a4paper=true}

\usepackage{cite}
\usepackage{stmaryrd}
\usepackage{amssymb}
\usepackage[tbtags]{amsmath}
\usepackage{graphicx}
\usepackage{wrapfig}
\usepackage{psfrag}
\usepackage{epsfig}
\usepackage{axodraw}
\usepackage{capt-of}
\usepackage{floatflt}
\usepackage{dsfont}

\newcommand{\bc}{\begin{center}}
\newcommand{\ec}{\end{center}}
\newcommand{\be}{\begin{equation}}
\newcommand{\ee}{\end{equation}}
\newcommand{\bea}{\vspace{-0mm}\begin{eqnarray}}
\newcommand{\eea}{\end{eqnarray}}
\newcommand{\ms}{\mskip 1.5mu}
\newcommand{\qbar}{\overline q}
\newcommand{\sq}[1]{\text{$ #1 $}}
\newcommand{\mbf}{\mathbf}
\def\t0{t\text{{\ttfamily =}}0}
\def\ix0{\xi\text{{\ttfamily =}}0}
\def\eql{\text{{\ttfamily =}}}
\newcommand{\Dlr}{{D^{\hspace{-0.8em}%
      \raisebox{0.8ex}{$\scriptstyle\leftrightarrow$}}}{}}
\newcommand{\Dl}{{D^{\hspace{-0.8em}%
      \raisebox{0.8ex}{$\scriptstyle\leftarrow$}}}{}}
\newcommand{\Dr}{{D^{\hspace{-0.8em}%
      \raisebox{0.8ex}{$\scriptstyle\rightarrow$}}}{}}

\def\eps{\epsilon}
\def\MSbar{\overline{\text{MS}}}
\def\myfiguresFFs{figures/FFs}
\def\myfiguresPDFsGPDs{figures/PDFsGPDs}
\def\myfiguresDAs{figures/DAs}
\def\myfiguresInProg{figures/inprogress}
\def\lat{\text{lat}}
\def\phys{\text{phys}}
\def\GeV{\text{ GeV}}
\def\MeV{\text{ MeV}}
\def\fm{\text{ fm}}
\def\twist{\text{pTBC}}
\def\snk{\text{snk}}
\def\src{\text{src}}
\def\mytr{\text{Tr}}
\def\Dslash{\hspace{-0.2ex}\not \hspace{-0.8ex}D}
\def\mybox{\!\boxempty\!}
\def\myDelta{\text{{\small $\Delta$}}}

\def\min{\text{min}}
\def\max{\text{max}}
\def\EM{\text{EM}}
\def\pheno{\text{pheno}}
\def\myIm{\text{Im}}

\begin{document}

\begin{frontmatter}

\title{ \hfill {\rm \small{TUM/T39-09-12}} \\ \vspace{1cm} Hadron structure from lattice quantum chromodynamics}

\author{Ph.~H\"agler}

\address{Institut f\"ur Theoretische Physik T39,
Physik-Department der TU M\"unchen,
James-Franck-Strasse,
D-85747 Garching,
Germany}

\begin{abstract}
This is a review of hadron structure physics from lattice QCD.
Throughout this report, we place emphasis on the contribution of lattice results to
our understanding of a number of fundamental physics questions related 
to, e.g., the origin and distribution of the charge, magnetization, momentum and spin of hadrons. 
Following an introduction to some of the most important hadron structure observables, 
we summarize the methods and techniques employed for their calculation in lattice QCD.
We briefly discuss the status of relevant chiral perturbation theory
calculations needed for controlled extrapolations of the lattice results to the physical point.
In the main part of this report, we give an overview of lattice calculations 
on hadron form factors, moments of (generalized) parton distributions, moments of
hadron distribution amplitudes, and other important hadron structure observables.
Whenever applicable, we compare with results from experiment and phenomenology, taking
into account systematic uncertainties in the lattice computations.
Finally, we discuss promising results based on new approaches, ideas and techniques,
and close with remarks on future perspectives of the field.
\end{abstract}

\end{frontmatter}
\newpage
\tableofcontents
\newpage

\include{Introduction}

\include{Concepts}
\include{FFs}

\include{PDFs}

\include{DAs}
\include{InProgress}
\include{Summary}
\bibliographystyle{halpha}
\bibliography{report_bib}

\end{document}

%% file: Introduction.tex
\section{Introduction}
\label{sec:Introduction}
\subsubsection*{Preface and disclaimer}
A rarely appreciated fact is that most of the visible matter in our universe, 
composed of protons and neutrons, is dynamically generated by the strong interactions 
between the quarks and the gluons inside the nucleons, as described by QCD. 
Indeed, while the gluons are exactly massless, the light quark current masses
are negligible compared to the observed nucleon mass of $m_N=938\MeV$.
This fascinating and fundamental observation already gives a first indication that 
the inner structure of QCD bound states, the hadrons, must be immensely rich.

Lattice QCD calculations of fundamental hadron properties, in particular the hadronic masses and decay constants, 
go back to the early 1980's  \cite{Hamber:1981zn,Weingarten:1982qe,Fucito:1982ip}.
First lattice QCD studies of hadron structure in terms of the pion distribution amplitude and the pion form factor
followed in the mid to late 1980's \cite{Martinelli:1987si,Martinelli:1987bh,Draper:1988bp}. 
Since then, remarkable progress has been made with respect to the theoretical foundations of gauge theories
on the lattice as well as the methods and algorithms required
for their numerical implementation and large scale simulations on supercomputers.
As a consequence, 
\emph{ab initio} computations of the hadron spectrum can nowadays 
be performed in full lattice QCD
very close to the physical pion mass \cite{Aoki:2008sm,Durr:2008zz},
and hadron structure calculations
have been pushed down to pion masses of around two times $m_\pi^{\phys}\approx139\MeV$. 
The aim of this report is to give a compact 
overview and review of the achievements in lattice QCD in the field of
hadron structure, excluding mere hadronic spectrum studies.
To be specific, the focus of this work is on the light up- and down-quark 
and gluon structure of the lowest lying spin-$0$, $1/2$, $1$ and  $3/2$
(the pion, nucleon, $\rho$-meson and $\Delta$-baryon, respectively) bound states of QCD. 
We will mostly concentrate on the leading twist vector-, axial vector- and tensor-operators,
which offer in general a probability interpretation of the corresponding observables.

In many cases, the amount of quenched and unquenched lattice
QCD data is so large that a detailed discussion of all the available results
is not possible. The inevitable selection of particular results is based roughly on the following criteria:
\begin{itemize}
\item dynamical fermion calculations vs simulations in the quenched approximation
\item published in peer-reviewed articles vs published in proceedings
\item lattice technology and ensembles; 
lowest accessible pion masses, operator renormalization, etc.,
\end{itemize}
where the order of the above items does not imply a strict hierarchy.
This is of course to some extent based on personal experience and therefore
not necessarily objective. 

Before presenting results from lattice QCD, we give brief and basic introductions to 
hadron structure observables in theory and in experiment, lattice methods and techniques, 
and chiral perturbation theory (ChPT) calculations. Readers who are interested in the details
of these topics are referred to the very useful books, reviews, overview articles and progress reports
on the structure of the nucleon \cite{Thomas:2001kw} and its spin structure \cite{Burkardt:2008jw},
form factors \cite{HydeWright:2004gh,Perdrisat:2006hj,Arrington:2006zm}, 
PDFs \cite{Stirling:2008sj}, GPDs \cite{Ji:1998pc,Goeke:2001tz,Diehl:2003ny,Belitsky:2005qn,Boffi:2007yc}, 
lattice hadron structure calculations \cite{Orginos:2006zz,Hagler:2007hu,Zanotti2008}, 
and chiral effective field theory and ChPT \cite{Holstein:1995mw,Scherer:2002tk,Scherer:2005ri},
as well as to the references in the corresponding sections below.

Substantial effort went into a consistent presentation of the relevant
concepts, methods, techniques and results in this field. However,
due to the large number of different sources 
that have been considered, it cannot be guaranteed that all notations, symbols 
and definitions in different parts of the report are perfectly consistent. 

The following section gives a first introduction to and motivation for
the particular hadron structure observables that will be discussed in the remainder of this review.

\subsubsection*{Hadron structure observables}
\label{sec:PhysicsCase}
It is well known that many fundamental properties hadrons, e.g.,
the distribution of their charge, the origin and strength of their magnetization,
and their (possibly deformed) shape can be studied on the basis
of \emph{hadron form factors} $F(Q^2)$, only depending on the squared momentum transfer
$Q^2=-(P'-P)^2$, with initial and final hadron momenta $P$ and $P'$, respectively.
Specifically,  many of the classical hadron structure
observables can be directly defined from form factors, including 
the \emph{charges} or coupling constants, particularly the 
axial\footnote{More precisely denoted as axial vector coupling constant.},
pseudoscalar, and tensor charge of the nucleon, the \emph{magnetic moments}
of the nucleon, $\rho$-meson and $\Delta$-baryon, their quadrupole moments (present
for particles of spin $\geq1$), 
as well as (root) mean square \emph{charge radii} of the pion, nucleon and other hadrons.

A second class of important observables\footnote{Throughout this report, we use the phrase ``observable''
for quantities that are at least in principle measurable, 
even if they are renormalization scale and scheme dependent.} is given by the \emph{parton distribution functions} (PDFs) $f(x,\mu^2)$.
They encode essential information about the distribution of momentum and spin of quarks and gluons
inside hadrons and have in general an interpretation as probability densities 
in the longitudinal momentum fraction $x$ carried by the constituents.
Their dependence on the renormalization (``resolution'') scale $\mu^2$ is described by 
the well-known DGLAP evolution equations. 

It turns out that both the form factors and the PDFs are fully encoded within 
the so-called \emph{generalized parton distributions} (GPDs) \cite{Mueller:1998fv,Ji:1996nm,Radyushkin:1997ki}. 
Generalized parton distributions, which we may generically denote by $H(x,\xi,t,\mu^2)$,
provide a comprehensive description of the partonic content of a given hadron,
simultaneously as a function of the (longitudinal) momentum fractions $x$
and $\xi$
(representing to the longitudinal momentum transferred to the hadron) and
the total momentum transfer squared, $t=\Delta^2=(P'-P)^2$. Similarly to the PDFs,
their dependence on the renormalization scale $\mu$ is described by corresponding evolution equations.
Apart from precisely reproducing the form factors and PDFs in certain limiting cases,
GPDs contain vital information about the decomposition of the total hadron spin in terms
of (orbital) angular momenta carried by quarks and gluons, 
and correlations between the momentum, coordinate and spin degrees of freedom.
Specifically, GPDs describe how partons carrying a certain fraction of the longitudinal momentum
of the parent hadron moving in, e.g., $z$-direction
are spatially distributed in the transverse $(x,y)$-plane.
In summary, GPDs represent a modern and encompassing approach to the partonic structure of hadrons.

Another class of important hadron structure observables closely related to the PDFs are the hadronic 
distribution amplitudes (DAs), $\phi(x,\mu^2)$.
In contrast to the PDFs, they have the interpretation of probability \emph{amplitudes}
for finding a parton with momentum fraction $x$ 
in a hadron at small transverse parton separations.
Distribution amplitudes play a central role in the description 
of exclusive processes at very large momentum transfers $\mu^2=Q^2$ and in 
many factorization theorems related to, e.g., $B$-meson decays. 

Finally, in addition to the charges and magnetic moments,
important information about hadron structure at low
energies is provided by the electric and magnetic polarizabilities $\alpha_E$ and $\beta_M$.
They describe the response of hadrons to external electric and magnetic fields
in form of \emph{induced} electric and magnetic dipole moments.

The quantities introduced above, i.e. the form factors,
PDFs, GPDs, DAs and polarizabilities, clearly represent only a subset of
all observables that provide important information about the structure of hadrons.
Among the other relevant observables that have been studied quite extensively in lattice QCD
are in particular decay constants, most prominently the pion decay constant $f_\pi$, and  
transition form factors, for example related to $N$-to-$\Delta$ transitions.
We will not discuss them in this review. Recent lattice results for, e.g., 
$f_\pi$ from different collaborations can be found in
\cite{Allton:2008pn,Dimopoulos:2008sy,Baron:2008xa,Aoki:2008sm,Noaki:2008iy,Aubin:2008ie,Bernard:2007ps,Hashimoto:2007vv,Gockeler:2006ns}. 
Lattice calculations of nucleon-to-$\Delta$ vector and axial-vector
transition form factors have been performed mainly by the Athens-Cyprus-MIT group,
and some of the more recent results can be found in \cite{Alexandrou:2006mc,Alexandrou:2007dt}.

\subsubsection*{Why lattice QCD?}
All the observables described above are universal (process independent), inherently non-\-perturbative objects.
As we will see in the following sections, 
they also share exact definitions
in terms of (forward or off-forward) hadron matrix elements of QCD quark and gluon
operators. These matrix elements can, in turn, be written in the form of QCD path integrals,
which makes them directly amenable to the methods of lattice gauge theory.
Specifically, with QCD properly discretized on a finite Euclidean space-time lattice,  
the path integrals can be numerically and fully non-perturbatively computed
in a systematic and controlled fashion.
Most importantly, the statistical and systematic uncertainties
of lattice QCD simulations can, at least in principle 
and increasingly also in practice, be systematically reduced.
To this date, lattice QCD represents the only known and working approach
to quantitatively study the non-perturbative aspects of QCD and QCD bound states from first principles.

To illustrate the strengths of the lattice approach, we first briefly recapitulate
how hadron structure observables are accessed in experiment and phenomenology.
Nature, as described by the standard model,
provides only a very limited number of currents
that can be used to study the quark content of hadrons directly: 
The spin-$1$ vector and axial-vector currents related to
exchanges of the electroweak gauge bosons.
Probably the most prominent example of an application of these couplings 
is elastic electron-nucleon scattering, representing
the classic way to measure the nucleon vector form factors $F(Q^2)$ over a large 
range of photon virtualities (squared momentum transfers) $Q^2$.
It is well known, however, that there exist other types of fundamental local couplings,
for example the tensor coupling, related to a parton helicity flip and so-called transversity observables, 
spin-$2$ couplings giving access to the energy-momentum and angular-momentum structure
of hadrons (described by the ``spin-$2$'' energy momentum tensor $T^{\mu\nu}$) as well as higher-spin couplings.
Remarkably, all these can only be investigated \emph{indirectly}
on the basis of QCD-factorization theorems for, e.g., 
deep inelastic scattering (DIS), Drell-Yan (DY) production,
deeply virtual Compton scattering (DVCS) and related processes\footnote{
A coupling to the spin-$2$ graviton under controlled experimental conditions 
is obviously not feasible because of the smallness of the gravitational coupling constant.}. 
At large scales $\mu^2=Q^2\gg\lambda^2_{\text{QCD}}$, DIS and DY-production processes give access
in particular to the quark PDFs $f(x,\mu^2)$ of the nucleon over a wide range of momentum fractions, $x$.
Deeply virtual Compton scattering
at large photon virtualities $Q^2\gg\lambda_{\text{QCD}}^2$ and 
squared momentum transfers $-t\ll Q^2$ is sensitive
to the nucleon GPDs $H(x,\xi,t,\mu^2)$ and their correlated 
dependence on $x$, 
$\xi$ and $t$.
So it turns out that experimentally, a rather large number of 
different processes must be studied in order to access the structure of hadrons 
in great detail in terms of form factors, PDFs and GPDs.
Further challenges arise in studies of polarized distribution functions, which
in general demand a sophisticated preparation of polarized beams and targets.
It is also important to recall that the electromagnetic (EM) current, 
which plays a central role in most of the
processes discussed above, only provides access to the
charge weighted combination of different quark flavors. A full mapping of the flavor structure 
therefore requires the use of different targets, e.g. protons and neutrons 
(for example in form of hydrogen and deuterium/deuteron targets), and/or a study of different
final states in semi-inclusive or exclusive scattering processes.
Finally, in contrast to the nucleon, other hadrons of interest
like the pion, $\rho$-meson and $\Delta$-baryon are unstable and 
therefore do not easily form a target or beam in an experimental setup.
Taken together, all these issues require enormous efforts on the 
experimental and theoretical sides. However during many decades 
of dedicated work remarkable progress has been made and invaluable insight 
has been gained into the structure of hadrons.

Coming back to the lattice approach, one of its main characteristics is that 
many of the above mentioned difficulties are absent. Most importantly, the lattice simulations
are set up in a way that allows one to study a large number of local currents of interest
(vector, axial-vector, tensor, spin-$2$ gravitational, higher spin,... -currents) 
directly and almost simultaneously for a given hadron. 
In addition, there is no need to use charge weighted currents, or to multiply them with tiny coupling constants.
In other words, just `bare' currents that couple to a single quark flavor with unit weight may be used. 
With respect to the hadron matrix elements, it is noteworthy that
different polarizations of the initial and final hadron states can be easily included and accessed
in a lattice calculation. 
A non-zero momentum transfer $q=\Delta=P'-P$ to the hadron can also be taken into account in
a straightforward manner, giving direct access to the $Q^2=-q^2$ and $t=\Delta^2$ 
dependencies of form factors and GPDs in the space-like region, $(q^2,t) <0$, without
increasing the cost of the calculation dramatically compared to the overall simulation costs. 
Although it is comparatively straightforward to put
different meson and baryon ground states on the lattice,
it is important to note that unstable hadrons require special care
on finite Euclidean space-time lattices used in practical calculations,
see, e.g., Ref.~\cite{Luscher:1991cf} for a discussion of the $\rho$-resonance in finite periodic box.
An exception is the pion, which is stable in QCD, so that 
its structure in terms of the pion form factors, PDFs, and GPDs can be directly studied. 
Issues related to unstable hadrons on the lattice may be provisionally evaded by investigating them
in a range of unphysically large quark masses where the hadron mass is below the decay threshold.

Following this praise of the lattice approach to hadron structure, we now briefly comment
on some of its shortcomings.
One of the most serious limitations 
is that the full $x$-dependence, which is related to bi-local operators on the light-cone,
of parton distributions and GPDs cannot be studied directly on the lattice.
Only the lowest $x$-moments of the distributions given by integrals of the form
$\int dx x^{n-1}$, corresponding to matrix elements of \emph{local} operators, 
can so far be reliably computed. In practice, calculations have been performed for
$n=1,\ldots,4$ in selected cases.
This is clearly insufficient to allow for a model-independent reconstruction of the $x$-dependence,
and it also implies that contributions from quarks and anti-quarks cannot in principle be separated.
It turns out that higher moments suffer from increasingly bad statistics.
Further complications arise from the loss of continuum space symmetries on the lattice. 
As a consequence, even the local lattice vector current is not conserved and has to be renormalized.
Lattice operators corresponding to higher moments (higher $n$) also require renormalization,
and special care has to be taken to properly account for possible operator mixing,
particularly with operators of lower dimensions.
To this date, observables in the singlet channel, in particular moments of gluon
PDFs and GPDs, have received little attention because of very low signal-to-noise ratios.
This may change in the not-too-far future, however for the time being we will mostly have to
concentrate on 
light quark operators in the isovector channel.

Among the more practical limitations due to limited computational resources
are the rather small lattice sizes and coarse lattice spacings\footnote{
What is `large' and `coarse' clearly has to be judged with respect to the object under investigation.} 
of present lattice simulations.
However the arguably most important
one is that the lowest up- and down-quark masses that can be reached 
in up-to-date hadron structure calculations
are still unphysically large, $m_q^\lat\sim4\ldots 10\;m_q^\phys$, 
corresponding to $m_\pi^\lat\sim 2\ldots 3\;m_\pi^\phys$ in terms of pion masses. 
As we will see and discuss throughout this report, this has dramatic consequences for many observables.
Results from chiral perturbation theory provide in many cases a qualitative,
and in some cases already a quantitative explanation for the characteristics of lattice
results at these pion masses.
Remarkable efforts are under way to perform lattice simulations with significant statistics
directly at the physical pion mass, $m_\pi\approx139\MeV$.
The situation might therefore improve substantially in the near future at least for the pion structure.
In the case of nucleon correlators, however, it follows from quite general arguments 
that the signal-to-noise ratio decreases exponentially towards smaller quark masses. 
Although it has to be seen in practice how large this effect really is,
it may turn out that a substantial increase in statistics is required
to retain a meaningful precision for nucleon structure observables 
much below the presently accessible pion masses.
In this case, results from chiral perturbation theory will be of crucial importance
to extrapolate the lattice data to the physical point.

%% file: Concepts.tex
\section{Concepts}
\label{sec:Concepts}
%
\subsection{Operators and observables}
\label{sec:observables}
In this section, we shall define and discuss some observables that are
important for the investigation of the structure of hadrons. We will focus on the form factors, PDFs and GPDs
that can be directly related to probability distributions of quarks and gluons in hadrons. 
Their phenomenological importance will be discussed in section \ref{sec:exppheno}.
Hadronic distribution amplitudes and polarizabilities will be introduced below in sections 
\ref{secDAs} and \ref{sec:polarizabilities}, respectively.
The gluonic structure of hadrons has so far only been investigated in lattice QCD in a few rare cases, which will be
briefly discussed in section \ref{sec:MomentumFractions}. 
In the following, we will concentrate on the quark structure and begin 
with a definition of bilocal quark operators,
\bea
\label{QuarkOp}
  O^q_\Gamma(x)=
\int_{-\infty}^{\infty} \frac{d \eta}{2 \pi} e^{i \eta x}
  \overline{q} \big(\!-\!\frac{\eta}{2}n\big)\!
  \,\Gamma\, \mathcal{U}_{[-\frac{\eta}{2}n,\frac{\eta}{2}n]}
  q\big(\frac{\eta}{2} n\big)  \,,
\eea
where $q=u,d,s$ are the quark fields, the variable $x$ is directly related to the quark momentum fraction, 
and $n$ is a light cone vector to be specified below.
In the following, we will often drop the label $q$ of $O^q$ denoting the 
quark flavor of the operator for notational simplicity. 
We will focus on the Dirac structures $\Gamma=\gamma^\mu,\gamma^\mu\gamma_5,i\sigma^{\mu\nu}$, 
and refer to the corresponding operators as vector, $O^\mu_V(x)$, axial-vector, $O^\mu_A(x)$
and tensor operators, $O^{\mu\nu}_T(x)$, respectively. 
The Wilson-line $\mathcal{U}$ in Eq.~(\ref{QuarkOp}) ensures gauge invariance of the operators
and is given by a path-ordered exponential,
\bea
\label{WilsonLine}
  \mathcal{U}_{[-\frac{\eta}{2}n,\frac{\eta}{2}n]}={\cal P} e^{ig \int_{\lambda / 2}^{-\lambda / 2} d \lambda \, n \cdot A(\lambda n)}\,.
\eea

In many cases, in particular in the framework of lattice QCD, calculations are based on towers of \emph{local} operators
rather than the bilocal operators in Eq.~(\ref{QuarkOp}).
The relevant local leading twist operators are given by
\bea
\nonumber
  \mathcal{O}_{V[A]}^{\mu\mu_1\cdots\mu_{n-1}}(z)
&=& \mathcal{S}_{\mu\mu_1\cdots\mu_{n-1}}\; \overline{q}(z) \, \gamma^\mu[\gamma_5]
      i\Dlr^{\mu_1} \cdots\ms i\Dlr^{\mu_{n-1}}\ms q(z) -\text{traces}\\
  \mathcal{O}_{T}^{\mu\nu\mu_1\cdots\mu_{n-1}}(z)
&=& \mathcal{A}_{\mu\nu}\mathcal{S}_{\nu\mu_1\cdots\mu_{n-1}}\; \overline{q}(z) \, i\sigma^{\mu\nu}
      i\Dlr^{\mu_1} \cdots\ms i\Dlr^{\mu_{n-1}}\ms q(z) -\text{traces}\,,
\label{localOps}
\eea
with left- and right-acting covariant derivatives $\Dlr=(\Dr-\Dl)/2$, and where
the symmetrization and anti-symmetrization of indices is denoted by 
$\mathcal{S}_{\mu\mu_1\cdots\mu_{n-1}}$ and $\mathcal{A}_{\mu\nu}$, respectively.
Below, symmetrization will also be denoted by $\{\mu_1\mu_2\}$, and 
the subtraction of traces will be implicit.

The bilocal, Eq.~(\ref{QuarkOp}), and local, Eq.~(\ref{localOps}), operators 
as well as the hadron structure observables that 
are based on them are related through moments in the momentum fraction $x$, given by the integral
\bea
\label{moments1}
  f^{n}=\int_{-1}^1 dx x^{n-1}f(x)\,.
\eea

Operators as defined in Eq.~(\ref{QuarkOp}) and Eq.~(\ref{localOps}) have to be renormalized in QCD and therefore 
lead in general to renormalization scale, $\mu$, and scheme dependent quantities. 
We will suppress the explicit scale- and scheme-dependence of the
observables most of the time for better readability.
To be specific, we will consider in the following 
the quark structure of the pion (spin-$0$), nucleon (spin-$1/2$), $\rho$-meson (spin-$1$),
and $\Delta$-baryon (spin-$3/2$),
employing corresponding hadron matrix elements of the operators $\mathcal{O}_\Gamma(x)$. 
In the most general case, these matrix elements are off-diagonal (off-forward) 
in the incoming and outgoing hadron momenta, $P$ and $P'$, and spins, $S$ and $S'$. 
We set $\overline P=(P'+P)/2$,
choose the light-cone vector $n$ such that $n\cdot\overline P=1$, and 
denote the momentum transfer by $\Delta=P'-P$ and the longitudinal momentum transfer 
(or skewness parameter) by $\xi=-n\cdot\Delta/2$.
To avoid confusion with the additional scale (`$Q^2$'-)dependence of many observables,
we will mostly denote the squared momentum transfer by $t=\Delta^2$ instead of $Q^2(=-q^2=-t)$, which is 
commonly used for form factors.
Following their transformation properties
under Lorentz, parity and time-reversal transformations, 
one can parametrize the
hadron matrix elements of the bilocal operators $O_\Gamma(x)$ 
in terms of real-valued generalized parton distributions (GPDs), which depend,
apart from the renormalization scale $\mu$, on the three kinematic
variables $x$, $\xi$ and $t$.
\subsubsection{Form factors, PDFs and GPDs}
\label{sec:FFsPDFsGPDs}
\subsubsection{Spin-0}
\label{sec:spin0}
For the pion, we have
\bea
\label{PionVec1}
  \langle \pi(P')|O_V^\mu(x)|\pi(P)\rangle &=&
   2\overline P^\mu H^{\pi}(x,\xi,t)+\mbox{higher twist}\,, \\
   \label{PionAxial1}
    \langle \pi(P')|O_A^\mu(x)|\pi(P)\rangle &=&0\,, \\
    \label{PionTensor1}
   \langle \pi(P')|O_T^{\mu\nu}(x)|\pi(P)\rangle &=&
    \frac{ \overline P^{[\mu}\Delta^{\nu]} } {m_\pi} E_{T}^{\pi}(x,\xi,t)+\mbox{higher twist}\,,
\eea
with the leading twist vector and tensor GPDs $H^{\pi}(x,\xi,t)$ and $E_{T}^{\pi}(x,\xi,t)$, respectively,
and where $A^{[\mu}B^{\nu]}=A^{\mu}B^{\nu}-A^{\nu}B^{\mu}$.
The vanishing of the axial vector matrix element in Eq.~(\ref{PionAxial1}) 
follows directly from parity transformation properties.
The only structure that contributes in the forward limit, $\Delta=0$, is the pion parton distribution,
$q^\pi(x)=H^{\pi}(x,0,0)$.
The off-forward pion matrix element of the local vector operator, corresponding
to $n=1$ in Eq.~(\ref{localOps}), can be in parametrized by a single
form factor $A^\pi_{(n=1)0}(t)$,
\be
\label{PionFF}
\langle \pi(P')|\overline q(0)\gamma^\mu q(0)|\pi(P)\rangle =
   2\overline P^\mu A^\pi_{10}(t)\,,
\ee
which is for, e.g., up-quarks in the $\pi^+$
identical to the well-known pion form factor, i.e. $A^{\pi^{+}}_{u,10}(t)=F_\pi(t=-Q^2)$.
Comparison with Eq.~(\ref{PionVec1}) shows that $\int_{-1}^{1}dxH^{\pi^+}_u(x,\xi,t)=F_\pi(t)$. 
Similarly, the matrix element of the local tensor operator is parametrized by the pion 
tensor form factor $B_{T10}^{\pi}(t)$,
\be
\label{PionTensorFF}
\langle \pi(P')|\overline q(0)i\sigma^{\mu\nu} q(0)|\pi(P)\rangle =
   \frac{ \overline P^{[\mu}\Delta^{\nu]} } {m_\pi} B_{T10}^{\pi}(t)\,,
\ee
with $\int_{-1}^{1}dxE_{T}^{\pi}(x,\xi,t)=B_{T10}^{\pi}(t)$.
We will see later that the forward limit of $B_{T10}^{\pi}(t)$ may be identified
with the tensor anomalous magnetic moment, $\kappa^\pi_T$, of the pion.
Corresponding matrix elements of the local operators for $n=2$ are given by 
\be
\label{PionVectorGFFn2}
\langle \pi(P')|\mathcal{O}_V^{\mu\mu_1}(0)|\pi(P)\rangle =
   2\overline P^\mu \overline P^{\mu_1} A^{\pi}_{20}(t)+2\Delta^\mu \Delta^{\mu_1}C^{\pi}_{20}(t)\,,
\ee
for the vector, and 
\be
\label{PionTensorGFFn2}
\langle \pi(P')|\mathcal{O}_T^{\mu\nu\mu_1}(0)|\pi(P)\rangle =
  \mathcal{A}_{\mu\nu}\mathcal{S}_{\nu\mu_1} \frac{ \overline P^{\mu}\Delta^{\nu} } {m_\pi} 
  \overline P^{\mu_1}  B_{T20}^{\pi}(t)\,,
\ee
for the tensor case, where we have introduced the pion generalized form factors
$A^{\pi}_{20}(t)$, $C^{\pi}_{20}(t)$ and $B_{T20}^{\pi}(t)$. They are related to the
$x$-moments of the pion GPDs by
\bea
\int_{-1}^{1}dxxH_{}^{\pi}(x,\xi,t)&=&A_{20}^{\pi}(t)+(-2\xi)^2 C_{20}^{\pi}(t)\,, \label{PionGPDsn2}\\
\int_{-1}^{1}dxxE_{T}^{\pi}(x,\xi,t)&=&B_{T20}^{\pi}(t)\,,
\label{PionGPDsn2b}
\eea
In the forward limit, we find that $A_{20}^{\pi}$ is equal to the momentum fraction carried by quarks
in the pion, $\langle x\rangle^\pi=\int_{-1}^{1}dxxq^{\pi}(x)=A_{20}^{\pi}(\t0)$.

\subsubsection{Spin-1/2}
\label{sec:spin0p5}
In the case of the nucleon, the matrix elements can be parametrized by two vector GPDs $H(x,\xi,t)$ and $E(x,\xi,t)$,
two axial vector GPDs $\widetilde H(x,\xi,t)$ and $\widetilde E(x,\xi,t)$, and four tensor GPDs
$H_T(x,\xi,t)$, $E_T(x,\xi,t)$, $\widetilde H_T(x,\xi,t)$ and $\widetilde E_T(x,\xi,t)$ as follows
\cite{Ji:1996nm,Diehl:2001pm}
\bea
\label{NuclVec1}
  \langle N(P')|O_V^\mu(x)|N(P)\rangle &=&
  \overline U(P')\left\{\gamma^\mu H(x,\xi,t)+\frac{ i\sigma^{\mu\nu}\Delta_\nu} {2m_N}     
   E(x,\xi,t)\right\}U(P)+\mbox{ht}\,, \\
   \label{NuclAxial1}
     \langle N(P')|O_A^\mu(x)|N(P)\rangle &=&
     \overline U(P')\left\{\gamma^\mu \gamma_5 \widetilde H(x,\xi,t)+\frac{\gamma_5\Delta^\mu} {2m_N} 
      \widetilde E(x,\xi,t)\right\}U(P)+\mbox{ht}\,, \\
    \label{NuclTensor1}
   \langle N(P')|O_T^{\mu\nu}(x)|N(P)\rangle &=&
     \overline U(P')\left\{
    i\sigma^{\mu\nu} H_T(x, \xi, t) + \frac{\gamma^{[\mu}\Delta^{\nu]}} {2 m_N} E_T(x, \xi, t) \right. \nonumber\\
    &+&  \left. \frac{\overline P^{[\mu} \Delta^{\nu]}} {m_N^2 } \widetilde H_T(x, \xi, t)
 + \frac{\gamma^{[\mu}\overline P^{\nu]}} {m_N} \widetilde E_T(x, \xi, t)
   \right\}U(P)+\mbox{ht}\,,
\eea
where ``ht'' denotes higher twist contributions, and where we have suppressed the dependence on the spins $S$ and $S'$.
It turns out that in many practical applications it is useful to work with the linear combination
$\overline E_T=E_T+2\widetilde H_T$ instead of the GPD $E_T$.
For vanishing momentum transfer, $\Delta=0$, the matrix elements in Eqs.~(\ref{NuclVec1}),
(\ref{NuclAxial1}) and (\ref{NuclTensor1}) can be parametrized by the well-known unpolarized, $q(x)$,
polarized, $\Delta q(x)$, and transversity, $\delta q(x)$, parton distribution functions 
(PDFs)\footnote{Another common notation for these twist-2 PDFs is $f_1=q$, $g_1=\Delta q$, and $h_1=\delta q$.},
which are directly related to the corresponding GPDs by 
\bea
\label{NuclVec2}
 q(x)&=&H(x,0,0)\,, \\
   \label{NuclAxial2}
    \Delta q(x)&=&\widetilde H(x,0,0)\,, \\
    \label{NuclTensor2}
    \delta q(x)&=& H_T(x,0, 0)\,.
\eea
Matrix elements of the local vector and axial-vector currents,
\bea
\label{NuclVec3}
  \langle N(P')|\overline q(0)\gamma^\mu q(0)|N(P)\rangle &=&
  \overline U(P')\left\{\gamma^\mu F_1(t)+\frac{ i\sigma^{\mu\nu}\Delta_\nu} {2m_N}  F_2(t)\right\}U(P)\,, \\
   \label{NuclAxial3}
     \langle N(P')|\overline q(0)\gamma^\mu\gamma_5 q(0)|N(P)\rangle &=&
     \overline U(P')\left\{\gamma^\mu \gamma_5 G_A(t)+\frac{\gamma_5\Delta^\mu} {2m_N} 
      G_P(t)\right\}U(P)\,, 
\eea
are parametrized by the familiar Dirac, $F_1(t)$, Pauli, $F_2(t)$, axial-vector, $G_A(t)$, and induced pseudo-scalar
, $G_P(t)$, nucleon form factors. 
The corresponding equation for the tensor current reads
\bea
    \label{NuclTensor3}
   \langle N(P')| \overline q(0)i\sigma^{\mu\nu} q(0)|N(P)\rangle &=&
     \overline U(P')\left\{
    i\sigma^{\mu\nu} A_{T10}(t) + \frac{\gamma^{[\mu}\Delta^{\nu]}} {2 m_N} B_{T10}(t) \right. \nonumber\\
    &+&  \left. \frac{\overline P^{[\mu} \Delta^{\nu]}} {m_N^2 } \widetilde A_{T10}(t)
   \right\}U(P)\,,
\eea
where we have introduced the tensor (quark helicity flip)
form factors $A_{T10}(t)$, $B_{T10}(t)$ and $\widetilde A_{T10}(t)$. A comparison of
$x$-integrals of Eqs.~(\ref{NuclVec1},\ref{NuclAxial1},\ref{NuclTensor1}) with
Eqs.~(\ref{NuclVec3},\ref{NuclAxial3},\ref{NuclTensor3}) immediately reveals the relation between the form factors and the lowest moment of the GPDs,
\bea
\label{NuclVec4}
 F_1(t)&=&\int_{-1}^{1}dx H(x,\xi,t)\,,\quad
 F_2(t)=\int_{-1}^{1}dx E(x,\xi,t)\,, \\
   \label{NuclAxial4}
 G_A(t)&=&\int_{-1}^{1}dx \widetilde H(x,\xi,t)\,,\quad
 G_P(t)=\int_{-1}^{1}dx \widetilde E(x,\xi,t)\,, \\
 A_{T10}(t)&=&\int_{-1}^{1}dx H_T(x,\xi,t)\,,\quad
 B_{T10}(t)=\int_{-1}^{1}dx E_T(x,\xi,t)\,,\quad
 \widetilde A_{T10}(t)=\int_{-1}^{1}dx \widetilde H_T(x,\xi,t)\,,\nonumber\\
 \label{NuclTensor4}
\eea
while time reversal transformation properties lead to
\bea
\int_{-1}^{1}dx \widetilde E_T(x,\xi,t)=0\,.
 \label{NuclTensor4b}
\eea
In the framework of lattice studies, the nucleon vector and axial-vector form factors are 
also denoted by $A_{(n=1)0}(t)=F_1(t)$, $B_{10}(t)=F_2(t)$, 
$\widetilde A_{10}(t)=G_A(t)$ and $\widetilde B_{10}(t)=G_P(t)$, emphasizing their 
relation (Eqs.~\ref{NuclVec4},\ref{NuclAxial4}) to the $(n\eql1)$-moments of the corresponding GPDs.
The Dirac and Pauli form factors are related to Sachs nucleon electric and magnetic form factors by
$G_E(t)=F_1(t)+t/(4m_N^2)F_2(t)$ and $G_M(t)=F_1(t)+F_2(t)$, respectively. The forward values, $t=0$, of the
form factors $F_1(t)$, $G_A(t)$ and $A_{T10}(t)$ can be identified as coupling constants or ``charges", where
$F_1(\t0)$ just counts the number of valence quarks (carrying electrical charge). The ``axial charge" or, to be more precise, 
the axial vector coupling constant is given by $g_A=G_A(\t0)$, and 
$g_T=A_{T10}(\t0)$  will be called the tensor charge. In the case of the induced pseudoscalar 
form factor $G_P(t)$, it is
common practice to define the pseudoscalar coupling constant not at $t=0$ but by $g_P=(m_\mu/2m_N)G_P(-0.88m_\mu^2)$,
i.e. for a momentum transfer squared of $t=-0.88m_\mu^2$, where $m_\mu$ is the mass of the muon.
The Pauli form factor at $t=0$ gives the anomalous magnetic moment $\kappa=F_2(\t0)$, 
which is directly related to the magnetic moment $\mu=G_M(\t0)$ by $\kappa=\mu-F_1(\t0)$.
It turns out that the combination of tensor form factors 
$\overline B_{T10}=B_{T10}+2\widetilde A_{T10}$ in the forward limit plays a role very similar to
that of $\kappa$ and therefore may be identified with a tensor magnetic moment, $\kappa_T=\overline B_{T10}(\t0)$ 
\cite{Burkardt:2005hp}.

The matrix elements of the local operators for $n=2$ can be decomposed as follows
\bea
\label{NuclVec5}
  \langle N(P')|\mathcal{O}_V^{\mu\mu_1}(0)|N(P)\rangle &=&
  \mathcal{S}_{\mu\mu_1}  \overline U(P')\bigg\{\gamma^\mu \overline P^{\mu_1}
    A_{20}(t)+\frac{ i\sigma^{\mu\nu}\Delta_\nu} {2m_N} \overline  P^{\mu_1} B_{20}(t)\nonumber\\
    &+&\frac{\Delta^{\mu}\Delta^{\mu_1}}{m_N} C_{20}(t) \bigg\}U(P)\,, 
\eea
\bea
     \langle N(P')|\mathcal{O}_A^{\mu\mu_1}(0)|N(P)\rangle &=&
    \mathcal{S}_{\mu\mu_1}  \overline U(P')\bigg\{\gamma^\mu \gamma_5 \overline P^{\mu_1} 
    \widetilde A_{20}(t)+\frac{\gamma_5\Delta^\mu} {2m_N} \overline P^{\mu_1} \widetilde B_{20}(t)\bigg\}U(P)\,,
     \nonumber\\     \label{NuclAxial5}
\eea
\bea   
    \label{NuclTensor5}
   \langle N(P')| \mathcal{O}_T^{\mu\nu\mu_1}(0)|N(P)\rangle &=&
    \mathcal{A}_{\mu\nu}\mathcal{S}_{\nu\mu_1}  \overline U(P')\bigg\{
    i\sigma^{\mu\nu} \overline P^{\mu_1} A_{T20}(t) \nonumber\\
    &+& \frac{\gamma^{[\mu}\Delta^{\nu]}} {2 m_N} \overline P^{\mu_1} B_{T20}(t) 
    +   \frac{\overline P^{[\mu} \Delta^{\nu]}} {m_N^2 } \overline P^{\mu_1} \widetilde A_{T20}(t)\nonumber\\
    &+& \frac{\gamma^{[\mu}\overline P^{\nu]}} {m_N} \Delta^{\mu_1} \widetilde B_{T21}(t) \bigg\}U(P)\,,
\eea
in terms of vector, $A_{20}(t)$, $B_{20}(t)$, $C_{20}(t)$, axial-vector, $\widetilde A_{20}(t)$, 
$\widetilde B_{20}(t)$ and tensor, $A_{T20}(t)$, $B_{T20}(t)$, $\widetilde A_{T20}(t)$,
$\widetilde B_{T21}(t)$ generalized form factors (GFFs).
Comparing with the $x$-moments of Eqs.(\ref{NuclVec3}), (\ref{NuclAxial3}) and (\ref{NuclTensor3}), 
one finds the relations 
\bea
\label{NuclVec6}
 \int_{-1}^{1}dx x H(x,\xi,t)&=&A_{20}(t)+(2\xi)^2C_{20}(t)\,,\quad
 \int_{-1}^{1}dx x E(x,\xi,t)=B_{20}(t)-(2\xi)^2C_{20}(t)\,, 
\eea
\bea
   \label{NuclAxial6}
\int_{-1}^{1}dx x \widetilde H(x,\xi,t)&=&\widetilde A_{20}(t) \,,\quad
\int_{-1}^{1}dx x \widetilde E(x,\xi,t)=\widetilde B_{20}(t) \,,
\eea
\bea
 \int_{-1}^{1}dx x H_T(x,\xi,t)&=&A_{T20}(t)\,,\quad
 \int_{-1}^{1}dx x E_T(x,\xi,t)=B_{T20}(t)\,,\quad\nonumber\\
 \int_{-1}^{1}dx x \widetilde H_T(x,\xi,t)&=& \widetilde A_{T20}(t)\,,\quad
 \int_{-1}^{1}dx x \widetilde E_T(x,\xi,t) = -2\xi\widetilde B_{T20}(t)\,.
 \label{NuclTensor6}
\eea

As an example, and for later convenience, we finally give the form factor decomposition for
the local vector operator in Eq.~(\ref{localOps}) 
for arbitrary $n$,
\begin{eqnarray}
\langle N(P')|\mathcal{O}_{V}^{\mu\mu_1\cdots\mu_{n-1}}|N(P)\rangle
 &=&\overline U(P')\bigg[
 \sum_{\substack{i=0\\\text{even}}}^{n-1}\bigg\{ \gamma
^{\{\mu }\Delta ^{\mu _{1}}\cdots \Delta ^{\mu _{i}}\overline{P}^{\mu
_{i+1}}\cdots \overline{P}^{\mu_{n-1}\}}A_{n,i}(t)
\nonumber \\
&& -i\frac{\Delta_{\alpha }\sigma ^{\alpha \{\mu }}{2m_N}\Delta ^{\mu
_{1}}\cdots \Delta^{\mu_{i}}\overline{P}^{\mu_{i+1}}\cdots \overline{P}
^{\mu_{n-1}\}}B_{n,i}(t)\bigg\}  \nonumber \\
&& + \frac{\Delta^{\mu }\Delta^{\mu_1}\cdots \Delta^{\mu _{n-1}}}{m_N}
C_{n,0}(\Delta ^{2})\big|_{n\text{ even}}\bigg] U(P).  \label{parav}
\end{eqnarray}
Comparing this expression with the $x^{n-1}$-moment of Eq.~(\ref{NuclVec1}), one finds the following
decomposition of the unpolarized GPDs 
$H$ and $E$ in terms of generalized form factors \cite{Ji:1998pc},
\begin{eqnarray}
\int_{-1}^{1}dxx^{n-1}H(x,\xi ,t)&=&
\sum_{\substack{i=0\\\text{even}}}^{n-1}( -2\xi )^{i} A_{n,i}(t)
+ (-2\xi)^{n}C_{n,0}(t)\big|_{n\text{ even}},  \label{Hn} \\
\int_{-1}^{1}dxx^{n-1}E(x,\xi ,t)&=&
\sum_{\substack{i=0\\\text{even}}}^{n-1}( -2\xi )^{i} B_{n,i}(t)
- (-2\xi)^{n}C_{n,0}(t)\big|_{n\text{ even}}.  \label{En}
\end{eqnarray}
Corresponding results for the nucleon axial-vector and tensor GPDs can be found in \cite{Hagler:2004yt}.
\subsubsection{Spin-1}
\label{sec:spin1}
For the spin-1 case (e.g. the $\rho$-meson), we concentrate on matrix elements of the vector and axial vector operators 
\cite{Berger:2001zb},
\bea
\label{RhoVec1}
  \langle \rho(P')|n_\mu O_V^\mu(x)|\rho(P)\rangle &=&
  \eps_\beta(P')^*\bigg\{ 
   - g^{\alpha\beta} n\cdot \overline P H_1(x,\xi,t)
   + n_\mu ( g^{\mu\alpha} \overline P_\beta + g^{\mu\beta} \overline P_\alpha)H_2(x,\xi,t)\nonumber\\
   &-& \frac{\overline P_\alpha \overline P_\beta}{2m_\rho^2} n\cdot \overline P H_3(x,\xi,t) 
   + n_\mu ( g^{\mu\alpha} \overline P_\beta - g^{\mu\beta} \overline P_\alpha)H_4(x,\xi,t) \nonumber\\
   &+&   \left( m_\rho^2 \frac{n_\alpha  n_\beta}{n\cdot\overline P} + \frac{1}{3} g^{\alpha\beta}n\cdot 
   \overline P\right)H_5(x,\xi,t)  \bigg\} \eps_\alpha(P)\,,
 \eea
parametrized by the five vector GPDs $H_{i=1\cdots5}$, and  
 \bea
 \label{RhoAxial1}
  \langle \rho(P')|n_\mu O_A^\mu(x)|\rho(P)\rangle &=&
  i\eps_\alpha(P')^*\bigg\{ 
  - n_\mu \eps^{\mu\alpha\beta\nu} \overline P_\nu \widetilde H_1(x,\xi,t)\nonumber\\
  &+&2 n_\mu \eps^{\mu\gamma\delta\nu} \frac{\Delta_\gamma \overline P_\delta}{m_\rho^2}
  \left(g_\nu^\beta \overline P^\alpha + g_\nu^\alpha \overline P^\beta\right)\widetilde H_2(x,\xi,t)\nonumber\\
  &+&2 n_\mu \eps^{\mu\gamma\delta\nu} \frac{\Delta_\gamma \overline P_\delta}{m_\rho^2}
  \left(g_\nu^\beta \overline P^\alpha - g_\nu^\alpha \overline P^\beta\right)\widetilde H_3(x,\xi,t)\nonumber\\
   &+&\frac{1}{2} n_\mu \eps^{\mu\gamma\delta\nu}\Delta_\gamma \overline P_\delta
  \frac{\left(g_\nu^\beta n^\alpha + g_\nu^\alpha n^\beta\right)}{n\cdot \overline P}
  \widetilde H_4(x,\xi,t) \bigg\} \eps_\beta(P)\,,
\eea
parametrized by the four axial vector GPDs  $\widetilde H_{i=1\cdots4}$.
The only structures that contribute in the forward limit are the two unpolarized distributions 
\bea
\label{RhoVec2}
 H_1(x)&=&H_1(x,0,0)\,, \\
  \label{RhoVec2b}
 H_5(x)&=&H_5(x,0,0)\,,
\eea
which are related to the unpolarized structure functions $F_1(x)$ and $b_1(x)$ 
parametrizing the hadronic tensor for spin-$1$ particles, and one polarized distribution
\bea
   \label{RhoAxial2}
 \widetilde H_1(x)&=&\widetilde H_1(x,0,0)\,,
\eea
which is related to the corresponding polarized structure function $g_1(x)$.

In the case of a spin-$1$ hadron, three vector, $G_{i=1\ldots3}$, and two axial-vector, $\widetilde G_{i=1\ldots2}$,
form factors are needed for the parametrization of the matrix elements of the corresponding local currents,
\bea
\label{RhoVec3}
  \langle \rho(P')|\overline q(0)\gamma^\mu q(0)|\rho(P)\rangle &=&
  \eps_\alpha(P')^*\bigg\{ 
   - g^{\alpha\beta}\overline P^\mu G_1(t)
   + ( g^{\mu\alpha} \overline P_\beta + g^{\mu\beta} \overline P_\alpha)G_2(t)\nonumber\\
   &-& \frac{\overline P_\alpha \overline P_\beta}{2m_\rho^2} \overline P^\mu G_3(t) \bigg\} \eps_\beta(P)\,,\\
 \label{RhoAxial3}
  \langle \rho(P')|\overline q(0)\gamma^\mu\gamma_5 q(0)|\rho(P)\rangle &=&
  i\eps_\alpha(P')^*\bigg\{ 
  - \eps^{\mu\alpha\beta\nu} \overline P_\nu \widetilde G_1(t)
 + 2  \eps^{\mu\gamma\delta\nu} \frac{\Delta_\gamma \overline P_\delta}{m_\rho^2}
  \left(g_\nu^\beta \overline P^\alpha + g_\nu^\alpha \overline P^\beta\right)\widetilde G_2(t)\nonumber\\
  &+& 2 \eps^{\mu\gamma\delta\nu} \frac{\Delta_\gamma \overline P_\delta}{m_\rho^2}
  \left(g_\nu^\beta \overline P^\alpha - g_\nu^\alpha \overline P^\beta\right)\widetilde G_3(t) \bigg\} \eps_\beta(P)\,,
\eea
where we dropped the contraction with $n_\mu$. 

Comparing with Eqs.~(\ref{RhoVec1},\ref{RhoAxial1}) integrated over $x$, we find
\bea
\label{RhoVec4}
 G_i(t)&=&\int_{-1}^{1}dx H_i(x,\xi,t)\,,\quad
 i=1,\ldots 3\,,\\
 \label{RhoAxial4}
 \widetilde G_i(t)&=&\int_{-1}^{1}dx \widetilde H_i(x,\xi,t)\,,\quad
 i=1,\ldots 2\,.
\eea
Since the GPDs $H_5$ and $\widetilde H_4$ parametrize structures proportional to $n\cdot\eps^*n\cdot\eps$ which 
cannot appear in the decomposition of the matrix elements in Eqs.~(\ref{RhoVec3},\ref{RhoAxial3}),
their integrals over $x$ must vanish. Furthermore, the integrals over $x$ of $H_4$ and $\widetilde H_3$ 
vanish due to time reversal transformation properties, so that
\bea
\label{RhoVecAxial5}
 \int_{-1}^{1}dx H_{4,5}(x,\xi,t)=\int_{-1}^{1}dx \widetilde H_{3,4}(x,\xi,t)=0\,.
\eea
The vector form factors $G_{i=1\ldots3}$ are related to the charge (or electric), $G_C(t)$, magnetic, $G_M(t)$, and quadrupole,
$G_Q(t)$, form factors by
\bea
\label{RhoVec6}
 G_C(t)&=&\left(1+\frac{2}{3}\eta\right)G_1(t)-\frac{2}{3}\eta G_2(t)+\frac{2}{3}\eta\left(1+\eta\right)G_3(t)\,,\\
\label{RhoVec7}
 G_M(t)&=&G_2(t)\,,\\
\label{RhoVec8}
 G_Q(t)&=&G_1(t)-G_2(t)+\left(1+\eta\right)G_3(t)\,,
\eea
where $\eta=-t/(4m_\rho^2)$. Analogously to the nucleon case, we define a $\rho$-magnetic moment by $\mu_\rho=G_M(\t0)$,
which may be understood as given in terms of a `$\rho$-magneton' $e/(2m_\rho)$ (i.e. natural units). 
The forward limit of the quadrupole form factor gives the $\rho$-quadrupole moment, $Q_\rho=G_Q(\t0)/m_\rho^2$.

\subsubsection{Spin-$3/2$}
\label{sec:spin32}
In the case of a spin-$3/2$ state, e.g. the $\Delta$-baryon, the matrix element of 
the local vector current can be parametrized by four form factors $a_{1,2}$ and $c_{1,2}$,
\bea
 \langle \Delta(P') | \overline q(0)\gamma^\mu q(0)| \Delta(P)\rangle
    &=& \overline U_\alpha(P')\Bigg\{
    -g^{\alpha \beta}\left( \gamma^\mu a_1(t) +\frac{\overline P^\mu}{m_\Delta} a_2(t)\right) 
    \nonumber \\
    &-&\frac{\Delta^\alpha \Delta^\beta}{4m_\Delta^2}\left(\gamma^\mu  c_1(t)
    + \frac{\overline P^\mu}{m_\Delta} c_2(t)\right)\Bigg\} U_\beta(P)\,,
    \label{spin32FFs}
\eea
where the $U_\alpha(P)$ are the well-known Rarita-Schwinger spinors, which can be constructed from
products of a polarization vector (spin-$1$) and a Dirac spinor (spin-$1/2$).
They obey the relations $P^\alpha U_\alpha(P)=0$ and $\gamma^\alpha U_\alpha(P)=0$ for an on-shell 
$\Delta$-baryon.
The relation of the form factors $a_{1,2}$ and $c_{1,2}$ in Eq.~(\ref{spin32FFs})
to Sachs' charge (or electric), $G_{E0}$, electric quadrupole, $G_{E2}$, 
magnetic dipole, $G_{M1}$, and magnetic octupole, $G_{M3}$, form factors
is given by \cite{Nozawa:1990gt}
\bea
\label{spin32GE0}
 G_{E0}(t)&=& \bigg(1+\frac{2}{3}\eta\bigg)\Big\{a_1(t) + (1+\eta)a_2(t) \Big\} 
 - \frac{1}{3}\eta( 1+\eta )\Big\{c_1(t) + (1+\eta)c_2(t) \Big\} \,,\\
\label{spin32GE2}
 G_{E2}(t)&=& \Big\{a_1(t) + (1+\eta)a_2(t) \Big\} 
 - \frac{1}{2}( 1+\eta )\Big\{c_1(t) + (1+\eta)c_2(t) \Big\} \,,\\
\label{spin32GM1}
 G_{M1}(t)&=& \bigg( 1 + \frac{4}{5}\eta \bigg)a_1(t) - \frac{2}{5}\eta( 1+\eta )c_1(t)\,,\\
\label{spin32GM3} 
 G_{M3}(t)&=& a_1(t) - \frac{1}{2}( 1+\eta )c_1(t)\,,
\eea
where $\eta=-t/(4m_\Delta^2)$. At zero momentum transfer, the magnetic dipole form factor
defines the magnetic moment, $\mu_\Delta=G_{M1}(0)=a_1(0)$, in units of natural magnetons $e/(2m_\Delta)$.
The only form factors that contribute in Eq.~(\ref{spin32FFs}) at $t=0$ and that are
therefore directly accessible in the forward limit are $a_1$ and $a_2$, but only in form
of the linear combination $G_{E0}(0)=a_1(0)+a_2(0)$, corresponding to the 
number of quarks minus anti-quarks in the $\Delta$-baryon, or equivalently its electric charge
in the case that a charge weighted current is used in Eq.~\ref{spin32FFs}.

\subsubsection{Physics content of form factors, PDFs and GPDs}
\label{sec:Interpretation}

First, we note that the moments of PDFs, which
can be obtained from forward matrix elements of the local quark operators in Eq.~(\ref{localOps}), 
always correspond to linear combinations of moments of quark and anti-quark distributions. 
Concentrating for definiteness on the twist-2 PDFs of the nucleon, one finds
\bea
\label{VecMoments1}
 \langle x^{n-1}\rangle_{q}&=&\int_{-1}^{1}dx x^{n-1}q(x)
     =\int_{0}^{1}dx x^{n-1}\big[q(x)-(-1)^{n-1}\overline q(x)\big]\,, \\ \label{AxVecMoments1}
 \langle x^{n-1}\rangle_{\Delta q}&=&\int_{-1}^{1}dx x^{n-1}\Delta q(x)
     =\int_{0}^{1}dx x^{n-1}\big[\Delta q(x)+(-1)^{n-1} \Delta \overline q(x)\big]\,, \\ \label{TensorMoments1}
 \langle x^{n-1}\rangle_{\delta q}&=&\int_{-1}^{1}dx x^{n-1}\delta q(x)
     =\int_{0}^{1}dx x^{n-1}\big[\delta q(x)-(-1)^{n-1} \delta \overline q(x)\big]\,,
\eea
where it has been used that quark PDFs for $x<0$ correspond to anti-quark distributions at $-x>0$, i.e.
$q(-x)=-\overline q(x)$, $\Delta q(-x)=\Delta \overline q(x)$ and $\delta q(-x)=-\delta \overline q(x)$.
Since the sign factors $\pm(-1)^{n-1}$ only depend on the charge conjugation properties of the operators, i.e. 
$-(-1)^{n-1}$ for the vector- and tensor-, and $+(-1)^{n-1}$ for the axial-vector operator, similar results hold of course for other hadrons. 
In particular, Eq.~(\ref{VecMoments1}) also holds for the pion PDF, $q(x)\rightarrow q_\pi(x)$,
as well as for the unpolarized $\rho$-meson PDFs in Eqs.~(\ref{RhoVec2}, \ref{RhoVec2b}),
i.e. for $q(x)\rightarrow H_1(x),H_5(x)$.
Similarly, for the polarized PDF of the $\rho$-meson in Eq.~(\ref{RhoAxial2}),
one obtains Eq.~(\ref{AxVecMoments1}) with $\Delta q(x)\rightarrow \widetilde H_1(x)$.
Such simple decompositions 
in terms of quark and anti-quark contributions are also possible for moments of GPDs at $\xi=0$ and non-zero $t$,
but not for $\xi\not=0$.

Observables like, e.g., the proton and neutron electromagnetic (EM)
form factors, i.e. $F^{p,n}_1(t)$ and $F^{p,n}_2(t)$, parametrize 
matrix elements of the EM current, which is a charge weighted linear combinations of quark currents, 
\bea
\label{WeightedFFs1}
  \langle N(P')|\sum_{q=u,d,\ldots}e_q\overline q(0)\gamma^\mu q(0)|N(P)\rangle &=&
  \overline U(P')\left\{\gamma^\mu F^N_1(t)+\frac{ i\sigma^{\mu\nu}\Delta_\nu} {2m_N}  F^N_2(t)\right\}U(P)\,,
\eea
where $N=p,n$.
Taking into account only up- and down-quark contributions, and using isospin symmetry,
these charge weighted FFs can be related to (unweighted) FFs in the 
$u-d$ (`isovector') and $u+d$ (`isosinglet') channels,
\bea
\label{WeightedFFs2}
  F^p_i(t)-F^n_i(t)&=&F_{i}^{p,u-d}(t)\nonumber\\
  3\big(F^p_i(t)+F^n_i(t)\big)&=&F^{p,u+d}_i(t)\,,
\eea
where $i=1,2$, and, e.g., $F_1^{p,u-d}=F_1^{p,u}-F_1^{p,d}$ stands for (unweighted) 
up- minus down-quark contributions to $F_1$ of the proton, which is often just written as $F_1^{u-d}$.
In lattice calculations, there is no need to use charge weighted currents, and one therefore usually
works with individual up- and down-quark currents or isovector and isosinglet combinations.
Isospin symmetry is also used to relate the neutron beta decay matrix element to the proton
matrix elements of the isovector axial vector current,
\bea
\label{BetaDecay1}
  \langle n|\overline d\gamma^\mu\gamma_5 u|p\rangle &=&
  \langle p|\overline u\gamma^\mu\gamma_5 u-\overline d\gamma^\mu\gamma_5 d|p\rangle 
   = \overline U(P)\gamma^\mu\gamma_5 U(P) g_A^{u-d}\,,
\eea
where $g_A^{u-d}=g_A^3$ is the isovector axial vector coupling constant.

The physical relevance of form factors at zero momentum transfer, 
i.e. the charges, magnetic moments and quadrupole moments is well established. 
We just recall that, e.g., a non-zero positive (negative) quadrupole moment for
the spin-$1$ $\rho$-meson indicates a deviation from a
spherically symmetric to a prolate (oblate) spheroidal charge distribution.
The slopes in $Q^2=-t$ of the form factors at $Q^2=0$ define
mean square (charge) radii (rms radii), i.e.
\bea
\label{radii}
\langle r_{i}^2\rangle
  =-\frac{6}{F_i(0)}\left.\frac{dF_i(Q^2)}{dQ^2}\right|_{Q^2=0}\,,
\eea
where $i=1,2,\ldots$, and similar definitions hold for Sachs' electric, magnetic etc. 
form factors $G_i$ with $i=E,M,\ldots$. 
The normalization factor $F_i(0)$ is usually replaced by unity
in the case that the corresponding form factor vanishes at $Q^2=0$, 
for example for the Dirac form factor of the neutron, $F^{n}_1$.
At non-zero momentum transfer $\Delta=(0,\mbf{\Delta})$ in the Breit frame $\mbf{P}=-\mbf{P'}$, 
the three-dimensional Fourier-transforms with respect to $\mbf{\Delta}$
of Sachs' electric and magnetic form factors have the classical interpretations
of charge and magnetization densities, respectively \cite{Sachs:1962zzc}. 
Strictly speaking, these interpretations hold only in the non-relativistic limit, 
$m_h\rightarrow \infty$, a restriction that does not apply to the density 
interpretation of FFs and GPDs in so-called impact parameter space 
to be discussed below.

A key feature of PDFs is their well-known interpretation as probability distributions in the momentum fraction $x$. 
The simplest example is the pion PDF $q^\pi(x)$, which represents the probability of finding an `unpolarized' quark
in the pion, $q^\pi(x)=q^\pi_+(x)+q^\pi_-(x)$, summed over the quark helicities $\lambda=+,-$. 
The twist-2 unpolarized and polarized PDFs of the nucleon in Eqs.(\ref{NuclVec2},\ref{NuclAxial2})
can be decomposed as $q(x)=q_+(x) + q_-(x)$ and $\Delta q(x)=q_+(x) - q_-(x)$, where 
$q_{+/-}(x)=q^{\Lambda=+1}_{+/-}=q^{\Lambda=-1}_{-/+}$ 
have the interpretation of number densities of quarks with, e.g., positive/negative 
helicity in a nucleon with positive helicity $\Lambda=+1$. 
The probability interpretation
of $\delta q(x)$ in Eqs.(\ref{NuclTensor2}) is concealed in the helicity basis but becomes clear in the 
transversity basis, 
\bea
|P,S_\perp=(1,0)=\uparrow\rangle &=&\frac{1}{\sqrt{2}}\big(|P,+\rangle + |P,-\rangle\big)\,,\nonumber\\
|P,S_\perp=(0,1)=\downarrow\rangle &=&\frac{1}{\sqrt{2}}\big(|P,+\rangle + i|P,-\rangle\big)\,,
\label{TransvBasis1}
\eea
where $\delta q(x)= q_\uparrow(x) - q_\downarrow(x)$, and $q_{\uparrow/\downarrow}(x)$
are densities of $\uparrow/\downarrow$-transversely polarized quarks in a $\uparrow$-transversely polarized hadron.
In the case of a spin-$1$ hadron, e.g. the $\rho$-meson, the two unpolarized distribution
functions in Eqs.~\ref{RhoVec2} and \ref{RhoVec2b} have a slightly different density interpretation
of the form \cite{Hoodbhoy:1988am}
\bea
H_1(x) &=& \frac{1}{3} \sum_{\substack{\lambda=\pm\\\Lambda=0,\pm1}} q^{\Lambda}_{\lambda}(x)\,,
\label{H1density}\\
H_5(x) &=& \frac{1}{2} \sum_{\lambda=\pm} \left( 2 q^{0}_{\lambda}(x) - q^{+1}_{\lambda}(x) - q^{-1}_{\lambda}(x)\right)\,,
\label{H5density}
\eea
while the polarized distribution $\widetilde H_1(x)$ in Eq.~\ref{RhoAxial2} can be decomposed in
the same way as for the nucleon, $\widetilde H_1(x)=q^{+1}_+(x) - q^{+1}_-(x)$.
Similar interpretations hold for anti-quark as well as gluon distribution functions. 
Note that a gluon transversity distribution does not exist for a spin-$1/2$ hadron.
It turns out that the interpretation of the PDFs as probability distributions (i.e. their positivity) 
is valid under evolution at leading order to higher renormalization scales $\mu$ \cite{LlewellynSmith:1978me,Bourrely:1997bx}, but is in general not guaranteed for higher order evolution.

Turning our attention to the GPDs, we note that at first sight a general probabilistic interpretation
seems to be lost: 1) The underlying hadron matrix elements are non-diagonal for non-zero momentum transfer,
i.e. $\xi\not=0$, $t\not=0$. 2) The physical interpretation of the GPDs in the
the so-called DGLAP region, $0\le\xi\le x\le1$ ($0\ge\xi\ge x\ge-1$), where the GPDs generically describe
the emission and reabsorption of quarks (anti-quarks) similarly to PDFs, differs strongly from the 
so-called ERBL region, $|x|\le|\xi|$, where the GPDs describe the emission of a quark anti-quark pair.

\subsubsection{Geometrical interpretation}
\begin{figure}[t]
  \begin{centering}
    \includegraphics[width=0.35\textwidth]{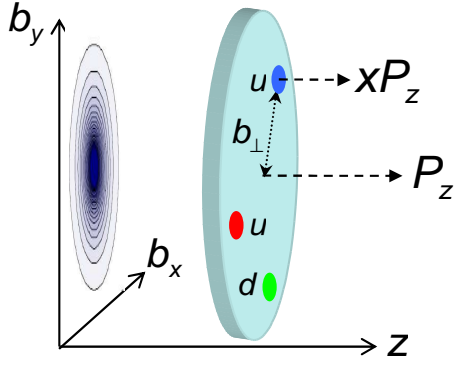}
  \caption{\label{ill_impact_1}Illustration of a quark distribution in impact parameter space.}
  \par\end{centering}
  \end{figure}

Importantly, it has been noted by Burkardt \cite{Burkardt:2000za} that a probability density
interpretation of the GPDs is possible for vanishing longitudinal momentum transfer, $\xi=0$.
In this case, the momentum transfer to the hadron is purely transverse, $t=-\Delta_\perp^2$, 
and the Fourier-transform of, e.g., the GPD $H(x,\xi=0,t)$ with respect to $\Delta_\perp$
to the so-called impact parameter space described by the variable $b_\perp$ has the interpretation
of a probability density in $x$ and $b_\perp$. 
In simple terms, this is because the Fourier-transformation
diagonalizes the underlying hadron matrix elements in terms of hadron states in a mixed representation
$|P^+,R_\perp\rangle$, where the center of momentum of the hadron may be set to zero, $R_\perp=0$. That is,
\bea
H_h(x,b_\perp)=\int_{-\infty}^{\infty}\frac{d^2\Delta_\perp}{(2\pi)^2}e^{-i\Delta_\perp\cdot b_\perp}H_h(x,\xi=0,t)\,,
\label{Impact1}
\eea
is the probability density of quarks carrying a momentum fraction $x$ at distance $b_\perp$ 
to the center of momentum of the parent hadron $h$, as illustrated in Fig.~\ref{ill_impact_1}. 
Probability density interpretations, as for the PDFs discussed above,
also hold for, e.g., the polarized and tensor/transversity nucleon GPDs, $\widetilde H(x,0,t)$ and $H_T(x,0,t)$,
respectively. 
An interpretation of the nucleon GPD $E(x,0,t)$ in the framework of impact parameter densities
has already been given in \cite{Burkardt:2002ks}, and a comprehensive physical interpretation of the GPDs in impact
parameter space can be given based on probability densities of (longitudinally or transversely) 
polarized quarks in a (longitudinally or transversely) polarized nucleon \cite{Diehl:2005jf}. 
To give an example, the corresponding density for transverse polarization is given by
\bea
\rho(x,b_\perp,s_\perp,S_\perp)&=&
\left\langle N_\perp\right|
\int\limits_{-\infty}^{\infty} \frac{d \eta}{2 \pi} e^{i \eta x}
  \overline{q} \big(\!-\!\frac{\eta}{2}n,b_\perp\big)\!
  \,\frac{1}{2}\big[\gamma^+ - s_\perp^j i \sigma^{+j}\gamma_5\big]\, 
  q\big(\frac{\eta}{2} n,b_\perp\big)
\left|N_\perp\right\rangle
\nonumber\\
&=& \frac{1}{2}\left\{H(x,b_\perp^2)
+ s_\perp^i S_\perp^i \left( H_{T}(x,b_\perp^2)
- \frac{1}{4m_N^2} \Delta_{b_\perp} \widetilde{H}_{T}(x,b_\perp^2) \right) \right.  \nonumber \\
&+& \frac{ b_\perp^j \eps ^{ji}}{m_N} \left( S_\perp^i E'(x,b_\perp^2)
+ s_\perp^i \overline{E}_{T}'(x,b_\perp^2) \right) 
+ \left. s_\perp^i \frac{( 2 b_\perp^i b_\perp^j
- b_\perp^2 \delta^{ij} )}{m_N^2} S_\perp^j \widetilde{H}_{T}''(x,b_\perp^2)
\right\}\,, \nonumber\\
\label{density1}
\eea
where the nucleon states are $\left|N_\perp\right\rangle=\left| P^+,R_\perp=0,S_\perp \right\rangle$,
and $f'(b_\perp^2)=\partial_{b_\perp^2}f(b_\perp^2)$. The interpretation of the different GPDs
becomes now very clear: While $H(x,b_\perp^2)$ is the spherically symmetric charge distribution,
the GPD $E(x,b_\perp^2)$ is responsible for dipole-like distortions $\propto \mbf{S}\times\mbf{b}$
of the charge density. Similarly, the tensor GPD $E_T$ accounts for dipole-distortions
of the form $\mbf{s}\times\mbf{b}$ for transversely polarized quarks.

Finally, the tensor GPDs $H_{T}$ and $\widetilde{H}_{T}$ contribute to the monopole
structure $\propto \mbf{S}\cdot\mbf{s}$, and to the quadrupole distortion given by the
last term in Eq.~(\ref{density1}).
Similar expressions hold for longitudinal polarizations \cite{Diehl:2005jf}, 
as well as for transversely polarized quarks in the pion \cite{Brommel:2007xd}.

In particular with respect to lattice QCD calculations it is interesting to study
$x$-moments of the density in Eq.~(\ref{density1}). The first moment, $n=1$, 
is then entirely given in terms of nucleon vector, $F_{1,2}$, and tensor form factors (Fourier
transformed to impact parameter space) and corresponds to the 
$x$-integrated density of quarks minus the density of anti-quarks, 
according to Eqs.~(\ref{VecMoments1}),(\ref{TensorMoments1}).
All $n$-even moments are given by the sum of quark and anti-quark densities and are therefore
strictly positive. We note that the probability density interpretation 
of GPDs, \emph{including} the standard form factors as their first moments, in impact parameter
space holds independently of a non-relativistic approximation or a special frame like the Breit-frame,
in contrast to the classical interpretation of the three-dimensional Fourier-transforms introduced by Sachs
\cite{Sachs:1962zzc}.

\subsubsection{Fundamental sum rules}
Based on Noethers theorem,
fundamental momentum and spin sum rules can be derived \cite{Jaffe:1989jz,Ji:1996ek}
from the energy-momentum and angular momentum density tensor of QCD.
Here, we will concentrate on the nucleon, but we note that it is straightforward
to derive similar results for the pion as well as hadrons of higher spin.
First we note that the off-forward nucleon matrix element of the gauge invariant, symmetric and traceless QCD 
energy-momentum tensor $T^{\mu\nu}$ can be parametrized in terms of three form factors $A(t)$, $B(t)$ and $C(t)$,
\bea
\label{EMT1}
  \langle N(P')|T^{\mu\nu}|N(P)\rangle &=&
  \mathcal{S}_{\mu\nu}
  \overline U(P')\left\{\gamma^\mu \overline P^\nu A(t)
  +\frac{ i\sigma^{\mu\rho}\Delta_\rho} {2m_N} \overline P^\nu B(t)
  +\frac{\Delta_\mu \Delta^\nu} {m_N}  C(t)\right\}U(P)\,,
\eea
where the energy-momentum tensor and the form factors in Eq.~(\ref{EMT1})
contain implicit sums over all fundamental fields, i.e. the quarks and gluons.
A comparison with the definition of the vector GPDs in Eq.~(\ref{NuclVec1}) reveals that
the form factors of the energy-momentum tensor are \emph{identical} to the GFFs that
parametrize the second, $(n\eql2)$-moments of the GPDs $H$ and $E$ (for quarks and gluons), 
see Eq.~(\ref{NuclVec6}), i.e.
\bea
 A(t)=A_{20}(t)\,,\quad
 B(t)=B_{20}(t)\,,\quad
 C(t)=C_{20}(t)\,.
 \label{EMTFFsGFFs1}
\eea
Following Noethers theorem, the nucleon momentum sum rule can then be written as
\bea
 1=A(0)=\sum_q A^q_{20}(0)+A^g_{20}(0)=\sum_q \langle x\rangle_q+\langle x\rangle_g\,,
 \label{MomSumrule1}
\eea
which holds in practically identical form of course also for all other hadrons.
For the spin-$1/2$ nucleon, the corresponding spin sum rule reads \cite{Ji:1996ek}
\bea
 \frac{1}{2}=\frac{1}{2}\big(A(0)+B(0)\big)=\frac{1}{2}\big(\sum_q \langle x\rangle_q+\langle x\rangle_g
 +\sum_q B^q_{20}(0)+B^g_{20}(0)\big)\equiv\sum_q J_{q}+J_{g}\,,
 \label{SpinSumrule1}
\eea
and is therefore, in addition to the momentum fractions carried by the quarks and gluons, completely determined by
the second moments of the GPDs $E^{q,g}(x,\xi,t)$ in the forward limit $\xi=0$, $t=0$, see Eq.~(\ref{NuclVec6}). 
Furthermore, since the GFF $C_{20}(t)$ contributes with different sign to the second moments of the GPDs $H$ and $E$ 
in Eq.~(\ref{NuclVec6}),
we can also write the total quark and gluon angular momenta $J_{q,g}$ in Eq.~(\ref{SpinSumrule1}) as 
\bea
 J_{q,g}&=&\frac{1}{2}\int_{-1}^1dxx\big\{H^{q,g}(x,\xi,0)+E^{q,g}(x,\xi,0)\big\}
 = \frac{1}{2}\big(A^{q,g}_{20}(0)+B^{q,g}_{20}(0)\big)\,,
 \label{J1}
\eea
which are independent of $\xi$. Clearly, the sum rules in Eq.~(\ref{MomSumrule1}) and (\ref{SpinSumrule1})
as well as the individual terms on their right hand sides are all gauge invariant.
We note that $B(0)$ is called the \emph{anomalous gravitomagnetic moment} \cite{Kobzarev:1962wt,Kobsarev:1970qm,Teryaev:1999su,Brodsky:2000ii}
and, according to Eq.~(\ref{MomSumrule1}) and (\ref{SpinSumrule1}), has to vanish identically when summed over quarks and gluons,
\bea
 B(0)=\sum_q B^q_{20}(0)+B^g_{20}(0)=0\,.
 \label{AGM1}
\eea

From a study of the underlying (local) QCD operators \cite{Ji:1996ek}, 
one finds that the total quark angular momentum $J_q$ can be naturally
decomposed,
\bea
 J_{q}&=&\frac{1}{2}\Delta\Sigma+L_{q}\,,
 \label{Jq}
\eea
in terms of the quark spin contribution, $\Delta\Sigma$, and the quark orbital angular
momentum $L_q$, which are separately gauge invariant. 
As repeatedly stated 
in the literature, such a gauge invariant decomposition in terms of the gluon spin and OAM
is \emph{not} possible for $J_g$, using just \emph{local} operators. 
However, one has to keep in mind that the gauge invariant and measurable
gluon spin contribution, $\Delta G$, is given by the $x$-integral of 
the polarized distribution $\Delta g(x)$,
\bea
 \Delta G=\int_{-1}^{1}dx\Delta g(x)\,,
 \label{DeltaG}
\eea
which is in turn defined through a \emph{non-local} gauge-invariant gluon operator. 
It is well known that the integration over $x$ of this non-local gluon spin
operator cannot be analytically performed to obtain a gauge-invariant, local gluon
spin operator \cite{Jaffe:1995an}. 
In view of this, one might just define a gauge invariant gluon orbital angular momentum as \cite{Ji:1996ek}
\bea
 L_{g}&\equiv&J_{g}-\Delta G\,.
 \label{Lg}
\eea
We note that such a definition has been critically discussed in \cite{Burkardt:2008jw}.
In summary, based on Eqs.~(\ref{SpinSumrule1}), (\ref{Jq}) and (\ref{Lg}), we can write down a
decomposition of the nucleon spin,
\bea
 \frac{1}{2}&=&\frac{1}{2}\Delta\Sigma + L_{q} + J_{g}\nonumber\\
 \bigg(&=&\frac{1}{2}\Delta\Sigma + L_{q} + \Delta G + L_{g}\bigg)_\text{using Eq.~(\ref{Lg})}\,,
 \label{SpinSumrule2}
\eea
where each term is separately gauge invariant and measurable. 
We would like to point out that the individual terms on the right hand side
of Eq.~(\ref{SpinSumrule2}) are scale, $\mu$, and scheme dependent in QCD.
At least the quark spin and total angular momentum, $J_{q}$, are defined through
nucleon matrix elements of local quark operators and are therefore directly 
calculable in lattice QCD.

\subsubsection{Hadronic distribution amplitudes}
\label{secDAs}
Hadronic distribution amplitudes (DAs) $\phi(x,\mu)$ are universal, non-perturbative functions
that play a fundamental role in hard exclusive processes and QCD-factorization theorems. 
Meson DAs in particular are 
essential ingredients in the description of 
heavy-to-light form factors in the framework of light cone sum rules, see, e.g., \cite{Ball:2004ye},
and in corrections to the ``naive'' factorization of 
non-leptonic decays of $B$-mesons to $\pi\pi$, $K\pi$ etc. \cite{Beneke:1999br,Beneke:2001ev}
as well as radiative and semi-leptonic decays like $B\rightarrow\rho\gamma,K^{*}\gamma, \pi l\nu,\ldots$
\cite{Beneke:2001at,Ali:2001ez,Bosch:2001gv} (for an overview we refer to Table 1 in \cite{Stewart:2003gt}).
They are therefore of direct phenomenological importance for 
the study of $CP$-violation and the determination of CKM-parameters.
Distribution amplitudes can be written as momentum integrals of
Bethe-Salpeter wave functions $\varphi(x,k_\perp)$, 
$\phi(x,\mu)\sim\int^{\mu^2}d^2k_\perp\varphi(x,k_\perp)$,
and have the interpretation of probability amplitudes for finding a parton
with momentum fraction $x$ in a hadron at small transverse separation
(related to the cut-off $\mu$) of the constituents.
At very large momentum transfer $Q^2$, hadron form factors $F(Q^2)$
can be written as convolutions of two corresponding distribution amplitudes
and a hard scattering kernel $T$ that can be calculated perturbatively,
$F(Q^2)\sim\int dx dy \phi(x,Q^2)T(x,y,Q^2)\phi(y,Q^2)$.

The distribution amplitude for, e.g., the $\pi^+$, $\phi_\pi(\xi)$,  parametrizes the pion-to-vacuum matrix
element of a bi-local light cone operator similar to the ones defined in Eq.~(\ref{QuarkOp}),
\bea
\label{PionDA1}
  \langle 0| \overline{d} (-z)\!
  \,\gamma^\mu\gamma_5\, \mathcal{U}_{[-z,z]} u(z) |\pi(P)\rangle 
  &=&if_\pi P^\mu\int_{-1}^{1} d\xi e^{-i \xi P\cdot z}\phi_\pi(\xi)\,.
\eea
The DA $\phi_K(\xi)$ for a $K^+$ is defined in the same way, with replacements $d\rightarrow s$ and $f_\pi\rightarrow f_K$.
Comparing with the standard definition of the pion and kaon decay constants, one finds
\bea
\label{PionDA2}
\int_{-1}^{1} d\xi \phi_{\pi,K}(\xi)=1\,.
\eea
The variables $x=(1+\xi)/2$ and $\overline x=1-x=(1-\xi)/2$ describe the longitudinal momentum fractions
of the quark and the anti-quark in the meson. 
Higher moments of the DAs are defined by
\bea
\label{PionDA3}
\int_{-1}^{1} d\xi \xi^{n-1}\phi_{\pi,K}(\xi)=\langle\xi^{n-1}\rangle_{\pi,K}\,.
\eea
They parametrize meson-to-vacuum matrix elements of local quark operators, e.g. for the pion
\bea
\label{PionDA4}
  \langle 0| \mathcal{S}_{\mu\mu_1\cdots\mu_{n-1}}\; \overline{d} \, \gamma_\mu\gamma_5\ms
      i\Dlr^{\mu_1} \cdots\ms i\Dlr^{\mu_{n-1}}\ms u -\text{traces} |\pi(P)\rangle 
  &=&if_\pi P_{\mu}P^{\mu_1}\cdots P^{\mu_{n-1}}\langle\xi^{n-1}\rangle_{\pi}\,.
\eea
In the isospin-symmetric case one has $\phi_{\pi}(\xi)=\phi_{\pi}(-\xi)$, so that $\langle\xi^{2n-1}\rangle_{\pi}=0$.
In the case of vector mesons,
for example the $\rho$ and the $K^*$, two types of distributions amplitudes exist at twist-2 level,
$\phi_\Vert(\xi)$ and $\phi_\perp(\xi)$. The DA $\phi_\perp(\xi)$ describes transversely polarized mesons and
parametrizes matrix elements of the form $\langle 0|\overline{d}(-z)\Gamma\mathcal{U}_{[-z,z]} u(z) |\rho(P)\rangle$
where $\Gamma =i\sigma_{\mu\nu}$, and $\phi_\Vert(\xi)$ describes longitudinally polarized mesons, parametrizing 
matrix elements of vector and axial vector operators, $\Gamma = \gamma_\mu$ and $\Gamma = \gamma_\mu\gamma_5$,
respectively. For details we refer to \cite{Ball:1996tb}.
Distribution amplitudes can be conveniently expanded in terms of Gegenbauer polynomials $C_n^{3/2}(\xi)$,
\begin{equation}
\phi_{\pi,K}(\xi) = \frac{3}{4} (1-\xi^2) \left\{1 + \sum\limits_{n=1}^\infty
  a^{\pi,K}_{n} C_{n}^{3/2}(\xi)\right\}\,
\label{Gegenbauer1}
\end{equation}
where the Gegenbauer moments carry the full information about the scale
dependence of the DA, $a^{\pi,K}_{n}=a^{\pi,K}_{n}(\mu^2)$.

For the proton, one studies the nucleon-to-vacuum matrix element of a tri-local
three-quark light-cone operator, given by \cite{Chernyak:1984bm}
\bea
\label{ProtonDA1}
  \eps^{abc}\langle 0| \left\{\mathcal{U}_{[z_3,z_1]}u_i(z_1)\right\}^a
  \left\{\mathcal{U}_{[z_3,z_2]}u_j(z_2)\right\}^b d_k(z_3)^c|P,S\rangle \,,
\eea
where $z_i^2=0$, and Dirac and color indices are denoted by $i,j,k$ and $a,b,c$, respectively.
The matrix element in Eq.~(\ref{ProtonDA1}) can be conveniently parametrized at leading twist in terms
of an overall non-perturbative factor $f_N$, the nucleon `decay constant', times
vector, axial-vector and tensor Dirac structures, including the respective
(Fourier-transformed) distribution amplitudes $V$, $A$ and $T$.
Using isospin symmetry, it can be shown that at twist-2 level, all three amplitudes
may be written in terms of just a single
nucleon distribution amplitude $\varphi(x_1,x_2,x_3)$, which is a function of the momentum fractions of the
three valence quarks $x_i$, with $\sum_{i=1\ldots3}x_i=1$ \cite{Dziembowski:1987es}.
The moments of the proton DA are defined by
\bea
\label{ProtonDA2}
  \varphi^{klm}=\int_0^1 dx_1dx_2dx_3 \delta\big(\sum_{i=1\ldots3}x_i-1\big) x_1^k x_2^l x_3^m\varphi(x_1,x_2,x_3)\,,
\eea
and parametrize proton-to-vacuum matrix elements of towers of 
\emph{local} three-quark operators including covariant derivatives.
Some more details will be given in section \ref{sec:DAsNucleon} below.

\subsubsection{Polarizabilities}
\label{sec:polarizabilities}
The electric polarizability describes the response of a hadron to an external electric
field $\mbf{E}$ in form of an \emph{induced}
electric dipole moment (EDM)\footnote{In distinction to a possible non-zero
\emph{permanent} EDM. For, e.g., the neutron,
the current upper limit is $|\mbf{d}_n^{\text{perm}}|<0.29 \;10^{-25}\text{ecm}$ \cite{PDG2008}.} 
$\mbf{d}^{\text{ind}}=\alpha_E \mbf{E}$.
Similarly, an external magnetic field $\mbf{B}$ induces a magnetic dipole moment
proportional to the magnetic polarizability, $\boldsymbol{\mu}^{\text{ind}}=\beta_M \mbf{B}$, 
in addition to the static magnetic moment related to the
magnetic form factor, $\mu=G_M(Q^2=0)$.

In a low energy expansion in the photon energy $\omega$, 
the nucleon Compton scattering amplitude $T_\text{Compt.}$ to order  $\mathcal{O}(\omega^2)$
can be parametrized by the nucleon electric and magnetic polarizabilities, $\alpha_E$ and $\beta_M$, 
\bea
\label{Compton1}
 T_\text{Compt.}=T_\text{Compt.}^\text{point}
 +\omega \omega'\boldsymbol{\epsilon}'^*\cdot\boldsymbol{\epsilon}\,\alpha_E
 +\big(\mathbf{q}'\times\boldsymbol{\epsilon}'^*\big)\cdot\big(\mathbf{q}\times\boldsymbol{\epsilon}\big)\,\beta_M
 +\mathcal{O}(\omega^3)\,,
\eea
with photon polarization $\epsilon$ ( $\epsilon'$) and 
four momentum $(\omega,\mathbf{q})$ ($(\omega',\mathbf{q}')$) of the incoming (outgoing) photon.
Clearly, the polarizabilities describe the response of the nucleon to the electric ($\propto \omega\boldsymbol{\epsilon}$)
and magnetic ($\propto\mathbf{q}\times\boldsymbol{\epsilon}$) components of the photon fields.

With respect to lattice calculations, it is important
to note that the effects of the external fields can be accounted for by using an 
effective non-relativistic Hamiltonian with interaction term
\bea
\label{Heff1}
H_\text{eff}^\text{int}=-\frac{1}{2}\big(\alpha_E \mathbf{E}^2+\beta_M \mathbf{B}^2  \big)\,,
\eea
from which in particular the amplitude in Eq.~(\ref{Compton1}) at low energies can be derived.
The polarizabilities thus lead to shifts of the hadron masses 
of $\mathcal{O}( \mathbf{E}^2, \mathbf{B}^2)$ in the presence of external EM fields. 
%
%

\subsection{Hadron structure in experiment and phenomenology}
\label{sec:exppheno}
Here we give a very brief overview of selected hadron structure observables in experiment
and phenomenology.
\subsubsection{Form factors and polarizabilities}
\label{sec:expphenoFFs}
The pion form factor $F_\pi(Q^2)$ at low $Q^2\simeq 0.01,\ldots,0.3$ GeV$^2$
has been measured in experiments where a pion beam is scattered off
the electrons of a liquid hydrogen target \cite{Amendolia:1986wj}.
Investigations of the pion form factor in pion electroproduction, $ep\rightarrow e\pi^+n$, 
\cite{Bebek:1977pe,Ackermann:1977rp,Brauel:1979zk,Volmer:2000ek}
at larger $Q^2\simeq 0.3,\ldots,10$ GeV$^2$ are based on the assumption of
pion exchange dominance and described by quasi-elastic scattering on a virtual pion in the proton.
These studies are in general subject to larger systematic uncertainties.

The magnetic moments of the proton and the neutron have been measured
to excellent accuracy in experiments based on nuclear resonance \cite{PDG2008},
$\mu_p=1+\kappa_p=2.7928$ and $\mu_n=-1.9130$.
The isovector axial vector coupling constant 
is known to very high precision from neutron beta decay \cite{PDG2008}, 
$g_A^{u-d}=g_A^{(3)}=a_3=1.2695(29)$\footnote{Strictly speaking, 
what is measured is the ratio $\lambda=g_A/g_V$. However, assuming isospin symmetry, 
and since the vector current ist strictly conserved,
$g_V=1$, equal to the net number of $u-d$ quarks in the proton.
We therefore set $\lambda=g_A$ throughout this review.}.
Using $SU(3)$ flavor symmetry, $g_A^{(3)}$ and the octet axial coupling constant, $g_A^{(8)}=a_8$,
can be related to the hyperon decay constants $F$ and $D$ by $g_A^{(3)}=F+D$ and $g_A^{(8)}=3F-D$ 
\cite{Anselmino:1994gn},
which are measured to fair accuracy in hyperon $\beta$-decay \cite{PDG2008}. 
The proton and neutron electromagnetic form factors, i.e. Sachs' form factors $G^{p,n}_{E,M}(Q^2=-t)$, 
at non zero momentum transfer are in general accessible in unpolarized elastic electron-nucleon 
scattering described by the standard Rosenbluth cross section. 
The results of these Rosenbluth-separation measurements  
have been improved and challenged in the last couple of years by experiments with polarized electron beams on
polarized and unpolarized targets (beam-target asymmetry and polarization transfer measurements, 
respectively). For reviews of recent experimental results on nucleon EM form 
factors we refer to \cite{HydeWright:2004gh,Perdrisat:2006hj,Arrington:2006zm}.
In summary, results are available for
\begin{itemize}
\item $G^p_E$ from Rosenbluth-separation in a range of $Q^2\leq0.1,\ldots,2$ GeV$^2$
\item $G^p_M$ from Rosenbluth-separation in a range of $Q^2\leq0.1,\ldots,30$ GeV$^2$
\item $G^n_M$ from quasi-elastic electron scattering from deuterium (inclusive and
with neutron tagging) using Rosenbluth-separation for $Q^2\leq10$ GeV$^2$, however with
increasing uncertainties for increasing values $Q^2$
\item $G^n_E$ from polarization transfer and beam-target asymmetry measurements in a range of $Q^2\leq0.1,\ldots,2$ GeV$^2$
\item $G^p_E/G^p_M$ from polarization transfer and beam-target asymmetry measurements in a range of 
$Q^2\leq0.2,\ldots,6$ GeV$^2$
\end{itemize}
Note that the Rosenbluth measurements of $G_E$ are limited to small $Q^2$ since
the cross section is dominated by $G_M$ at large momentum transfers.
Rosenbluth-separation analyses have in the past been based on the single-photon-exchange approximation. 
The difference that is seen for the ratio $G^p_E/G^p_M$ between 
Rosenbluth-separation and (beam and target) polarization measurements at large values of 
the momentum transfer squared might indicate that two-photon exchange contributions are non-negligible \cite{Guichon:2003qm}. 
Further uncertainties in FF measurements are related to, e.g.,
nuclear effects in the analysis of electron scattering from deuterium and normalization uncertainties
in the Rosenbluth-separation. 

Assuming isospin symmetry, the charge weighted form factor results for the proton and neutron can be used to perform
a flavor separation to obtain isovector and isosinglet or individual unweighted up- and down quark
contributions, see Eq.~\ref{WeightedFFs2}.

Strange quark contributions to the form factors have been obtained from parity violating electron proton scattering.
At present, experimental results for the strange quark contributions to the proton form factors are 
largely compatible with zero \cite{Acha:2006my,Baunack:2009gy}.

Proton electric and magnetic polarizabilities, $\alpha^p_E$ and $\beta^p_M$, 
are mainly known from real Compton scattering (RCS), e.g. at MAMI \cite{OlmosdeLeon:2001zn}, 
with reasonable accuracy. 
Recent average values by the particle data group are $\alpha^p_E=(12.0\pm0.5)10^{-4}\fm^{3}$ 
and $\beta^p_M=(1.9\pm0.5)10^{-4}\fm^{3}$ \cite{PDG2008}.
Experimental results for the neutron polarizabilities still suffer from
rather large statistical and systematic uncertainties and do not yet
provide a consistent picture.
For reviews on nucleon polarizabilities, see e.g. \cite{HydeWright:2004gh,Schumacher:2005an,Drechsel:2007sq}.

The pion polarizability, more precisely $\alpha^{\pi^+}_E-\beta^{\pi^+}_M$,
has been determined from radiative pion photoproduction off the proton, 
$\gamma p\rightarrow \gamma \pi^+ n$, at, e.g., MAMI \cite{Ahrens:2004mg},
and the scattering of pions on the Coulomb field of a large nucleus
using the Primakoff effect at, e.g., Serpukhov \cite{Antipov:1982kz,Antipov:1984ez}. 
Experiments using the Primakoff effect at COMPASS/CERN are ongoing.
The situation is however somewhat unclear since the available results, which are 
based on different measurements and analysis methods, are not quite consistent within errors.
For an overview of results from experiment and chiral perturbation theory calculations, see 
\cite{Gasser:2006qa}.
\subsubsection{PDFs and GPDs}
\label{sec:expphenoPDFsGPDs}
Parton distribution functions (PDFs) of the nucleon are mainly accessible in deep 
inelastic lepton-nucleon scattering (DIS), semi-inclusive DIS (SIDIS), 
Drell-Yan lepton pair production and inclusive jet production in proton-(anti-)proton collisions.
The basis of global PDF analyses are QCD factorization theorems, allowing to decompose
cross sections and the corresponding structure functions in terms of hard scattering kernels
and the PDFs, which parametrize the non-perturbative physics. 
The underlying perturbative QCD calculations of, e.g., the coefficient functions for unpolarized DIS and splitting
functions for the evolution have already been pushed to impressive three-loop order (NNLO),
see \cite{Vermaseren:2005qc} and references therein. 
Global phenomenological analyses of unpolarized PDFs have been carried out at NLO 
\cite{Martin:2002aw,Martin:2003sk,Nadolsky:2008zw,Ball:2008by} 
and NNLO \cite{Alekhin:2006zm,Martin:2007bv,DelDebbio:2007ee}.
Recently, a first global analysis of polarized (helicity) PDFs at NLO has been presented \cite{deFlorian:2008mr},
and previous global analyses are described in \cite{Gluck:2000dy,Bluemlein:2002be,Leader:2005ci}.
From the phenomenological studies, unpolarized valence quark PDFs, $(u,d)(x)$, are known to high accuracy. 
Further results are available for the unpolarized antiquark, $(\overline u,\overline d)(x)$,
and strange (sea) quark, $(s,\overline s)(x)$, distributions, 
the corresponding polarized quark PDFs, as well as the unpolarized gluon distribution, $g(x)$.
For a review of the spin structure of the proton and polarized PDFs we refer to \cite{Bass:2004xa}, and
a discussion of recent progress in unpolarized PDFs can be found in \cite{Stirling:2008sj}.
A precise measurement of the spin structure function $g_1(x)$ and corresponding results for
the quark spin contributions to the nucleon spin have been recently reported by the HERMES
collaboration \cite{Airapetian:2007aa}.
Despite enormous theoretical and experimental efforts over the last decade, the polarized
gluon distribution, $\Delta g(x)$ is still only roughly constrained 
\cite{Hirai:2008aj,Alekseev:2008cz,deFlorian:2008mr,PHENIX:2008px}.
Potential sources of uncertainties in global PDF analyses include 
contributions from higher twist, higher-order corrections, 
treatment of heavy flavors, and the need for phenomenological
ans\"atze for the $x$-dependence of the PDFs at the input scale.

The $x$-moments of PDFs, which may be compared to lattice QCD calculations,
can in principle be directly obtained by integrating the phenomenological PDFs 
(weighted with some power of $x$) over $x$.
In practice, structure functions are of course only known for a certain range, $x=x_{\min},\ldots,x_{\max}$,
due to a limited kinematical coverage in the experiments. The systematic uncertainty that is introduced by a
necessarily model dependent extrapolation of the experimental results to $x=0$ and $x=1$, or by 
phenomenological parametrizations of the PDFs that determine (at least to some extent) their behavior for 
$x\rightarrow 0$ and $x\rightarrow 1$, is in general difficult to quantify. 

Generalized parton distribution functions can be accessed in deeply virtual Compton scattering (DVCS) 
\cite{Mueller:1998fv,Ji:1996nm,Radyushkin:1997ki,Belitsky:2001ns},
wide-(large-) angle Compton scattering \cite{Diehl:1998kh,Radyushkin:1998rt},
and related exclusive meson production processes.
Of particular importance in the case of DVCS is the interference with the Bethe-Heitler (BH) process. 
The BH-amplitude
can be described by the comparatively well-known nucleon form factors, see the discussion above.
As usual, beam- and target-spin, as well as beam-charge asymmetries are constructed to reduce systematic
uncertainties and to facilitate the analysis in terms of the individual unpolarized and polarized GPDs. 
A separation of contributions from up- and down-quark GPDs is attempted by using proton (hydrogen) 
and quasi-free neutron (deuterium) targets. First experimental DVCS results in form of 
the beam spin asymmetry have been presented in 2001 by HERMES at DESY \cite{Airapetian:2001yk} 
and CLAS at JLab \cite{Stepanyan:2001sm}.
Since then, much more results became available from measurements performed by the H1, ZEUS and HERMES 
collaborations at DESY (see, e.g., \cite{Chekanov:2003ya,Aktas:2005ty,Airapetian:2006zr,H1:2007cz}) 
and the Hall A and Hall B/CLAS collaborations at JLab 
(see, e.g., \cite{MunozCamacho:2006hx,Chen:2006na,CLAS:2007jq}) .
Recently, transverse (proton) target spin asymmetries measured at HERMES \cite{HERMES:2008jga} and 
DVCS cross section measurements with a longitudinally polarized beam on a neutron target
at Hall A \cite{HallA:2007vj} have been published and used to put first model dependent constraints
on the total angular momentum of up- and down-quarks, $J_{u,d}$, in the proton.
Compared to PDFs it turns out that the experimental and phenomenological
analysis of GPDs is significantly more demanding.
Based on a QCD factorization theorem \cite{Collins:1996fb,Collins:1998be}, 
DVCS and deeply virtual meson production at leading order 
can be described by complex valued Compton form factors of the form
\bea
\mathcal H(\xi,t)=\int^1_{-1} dx\left\{\frac{1}{\xi-x-i\eps}-\frac{1}{\xi+x-i\eps}\right\}H(x,\xi,t)\,,
\label{DVCSamp1}
\eea
and similar for the expressions involving the GPDs $E$, $\widetilde H$ and $\widetilde E$.
The integration over $x$ clearly makes a direct extraction of GPDs as functions of the
three variables $x$, $\xi$ and $t$ impossible (this is part of the ``deconvolution problem''
discussed in, e.g., \cite{Diehl:2003ny}). 
Furthermore, the imaginary part of Eq.(\ref{DVCSamp1})
is given by $\text{Im}\mathcal H(\xi,t)=\pi(H(\xi,\xi,t)-H(-\xi,\xi,t))$
and therefore only sensitive to the GPD at the crossover trajectory $x=\pm\xi$, so that only the real 
part of Eq.(\ref{DVCSamp1}) may provide access to the full GPD $H(x,\xi,t)$ for $x\not=\xi$.
However, using dispersion relations for the DVCS amplitude, 
it has been shown that the real part of Eq.(\ref{DVCSamp1}) can be written as
\bea
\text{Re}\mathcal H(\xi,t)=\text{PV}\int^1_{-1} dx\left\{\frac{1}{\xi-x}-\frac{1}{\xi+x}\right\}H(x,x,t)
+2\sum_{m=1}^{\infty}2^{2m}C_{2m,0}(t)\,,
\label{DVCSamp2}
\eea
where $\text{PV}$ denotes the principle value of the integral, and the $C_{2m,0}(t)$ are
the generalized form factors, contributing to higher moments $n=m+1$ of the GPDs
$H$ and $E$ as in Eq.~\ref{NuclVec6} for $n=2$. For studies of the 
analytic properties of the DVCS amplitudes, we refer to \cite{Teryaev:2001qm,Teryaev:2005uj,Diehl:2007jb,Polyakov:2007rv,Kumericki:2007sa}.
In summary, one finds that DVCS and deeply virtual meson production at leading order only give access to GPDs 
at $x=\pm\xi$, apart from the contribution of the generalized form factors $C_{2m,0}(t)$ in Eq.(\ref{DVCSamp2}).
From the above discussion, it is clear that a description of these DVCS processes
in terms of GPDs requires at least a partial modeling of their combined $x$-, $\xi$- and $t$-dependence. 
Attempts in this direction are for example described in 
\cite{Vanderhaeghen:1999xj,Goeke:2001tz,Freund:2002qf,Diehl:2004cx,Guzey:2006xi,Ahmad:2006gn,Ahmad:2007vw}, 
and a more critical discussion of this topic can be found in \cite{Kumericki:2007sa}.
For the above reasons, a direct comparison of moments of GPDs from lattice QCD
and moments of the partially modeled GPDs has to be handled with some care.

Wide-angle Compton scattering, i.e. Compton scattering at large energy and momentum transfer squared, $s\sim |t|\sim |u|$,
gives access to $1/x$-moments of GPDs at $\xi=0$, the so-called Compton form factors
$R_V(t)=\int dx x^{-1}H(x,0,t)$, etc. \cite{Diehl:1998kh,Radyushkin:1998rt}. 
Measurements of wide-angle Compton scattering
and also wide-angle meson photoproduction cross sections and helicity correlations
could in particular help to understand the correlated $x$- and $t$-dependence of the GPDs.
For a short review we refer to \cite{Kroll:2007fu} and references therein. 

While PDFs of valence quarks in the pion, $q^\pi_V(x)$, can be well constrained from $\pi N$ DY 
lepton pair production \cite{Sutton:1991ay,Gluck:1999xe}, the corresponding sea quark
distribution can only be determined in a model dependent way \cite{Gluck:1999xe}.
The distribution of gluons in the pion, $g^\pi(x)$, can be accessed in prompt photon production
$\pi^{\pm}p\rightarrow \gamma X$ only at large $x$.
An overview of early attempts to measure $g^\pi(x)$ can be found in \cite{Bordner:1996wq}.
The systematic uncertainties in the moments of pion PDFs obtained from
such phenomenological analyses should be kept in mind when comparing to results
from lattice QCD.

%
%

\subsection{Hadrons in lattice quantum chromodynamics}
\label{sec:hadronslattice}

\subsubsection{Basics of lattice QCD}
\label{sec:basicsLQCD}
For an introduction to lattice gauge theories, we refer to the excellent monographs
\cite{Rothe:1997kp,Rothe:2005nw,Creutz:1984mg,Montvay:1994cy,Smit:2002ug,DeGrand:2006zz}.

Common to the observables discussed in the previous sections is that they share exact definitions
in terms of hadronic matrix elements of QCD quark and gluon operators, which in turn
can be written in the form of QCD path integrals.
For a typical QCD correlator or expectation value, one has
\bea
\label{PI1}
\langle \mathcal{O}\rangle = \frac{1}{Z}\int [dq][d\overline q][dA]\;e^{iS_{QCD}[q,\overline q,A]}\;\mathcal{O}
\,,
\eea
where $\mathcal{O}$ consists of a product of quark and gluon fields, and 
\bea
\label{SQCD}
S_{QCD}=S_q+S_g=\int \!d^4x\left\{\overline q\left(i\Dslash-m\right) - \frac{1}{2}\mytr F^{\mu\nu}F_{\mu\nu}\right\}\,,
\eea
denotes the standard continuum QCD action, where we have suppressed all flavor, 
Dirac and color indices for notational simplicity.

\begin{figure}[t]
     \centering
          \includegraphics[angle=0,width=0.6\textwidth,clip=true]{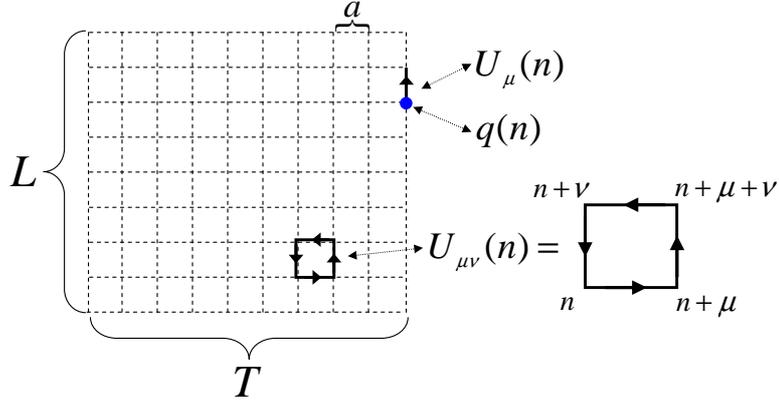}
  \caption{Illustration of basic lattice quantities.}
  \label{ill_latt_v1}
 \end{figure}

As a prerequisite for the numerical computation of the path integral, 
the theory is formulated in Euclidean space-time, as obtained from an analytic
continuation to imaginary times, $t=x_0\rightarrow -i t_E=-i x_4$, where
$t_E$ is the real Euclidean time. Since $iS^M_{QCD}\rightarrow -S^E_{QCD}$, 
the Euclidean formulation avoids the oscillating exponential
in Minkowski space-time that would prevent a numerical evaluation of the path integral.
An ultra-violet regularization of the theory is provided by the discretization of space-time in form of a 
hypercubic lattice with lattice spacing $a$, corresponding to a cut-off $\mu=1/a$. 
An exactly local gauge invariant discretized path integral can be formulated in
terms of discretized quark fields, $q(x=a n)$, living on the lattice sites $n$, and the 
so-called link variables $U_\mu(a n)=e^{iagA_\mu(a n)}\in SU(3)$, which replace the gluon fields and which are connecting the sites, as illustrated in Fig.~\ref{ill_latt_v1}.
%
%

We will set $a=1$ in the following for better readability and restore factors of $a$ later when necessary.
For the correlator in Eq.~(\ref{PI1}), the discretization corresponds to the replacement
\bea
\label{PI2}
\langle \mathcal{O}\rangle \rightarrow \frac{1}{Z}\int\bigg[\prod_i dq(i)\bigg]\bigg[\prod_j d\overline q(j)\bigg] \bigg[\prod_{k,\mu} dU_\mu(k)\bigg]\; e^{- \sum_{n,m} \overline q(n)M_{n,m}q(m) - S_{g}[U]}\;\mathcal{O}
\,,
\eea
where we have introduced the fermion matrix $M$. For, e.g., the standard Wilson gauge action we have
\bea
\label{WilsonG}
 S_{g}^W[U]=\beta\sum_P\bigg\{1-\frac{1}{2N_c}\text{Tr}\big(U_P+U^\dagger_P \big) \bigg\}
\,,
\eea
involving a sum over all distinct elementary plaquettes $U_P\hat= U_{\mu\nu}(n)$,
which are built up from a product of four link
variables as illustrated in Fig.~\ref{ill_latt_v1}. 
The coupling is given by $\beta=2N_c/g^2$. The Wilson gauge action
reproduces the continuum action at small lattice spacings up to terms $\propto a^2$.
The construction of a proper fermion action, i.e. $M_{n,m}$,
is non-trivial. According to the Nielsen-Ninomiya theorem \cite{Nielsen:1980rz,Nielsen:1981xu}, 
it requires an explicit breaking of 
continuum chiral symmetry of the massless theory to obtain a theory that is free of fermion doublers,
is local and leads to the correct equation of motion in the continuum limit.
An introduction to chiral symmetry and lattice QCD is given in \cite{Chandrasekharan:2004cn}.
Different lattice actions that are used in practice will be discussed in the following section \ref{sec:actions}.
We note that in general, improvement terms in the lattice action and the lattice operators are required to avoid
discretization errors starting at $\mathcal{O}(a)$. 

Most frequently, one is interested in correlators built up from quark fields $\mathcal{O}(n_1,n_2,\ldots)
=q(n_1)q(n_2)\ldots \overline q(m_1)\overline q(m_2)\ldots$, for which one finds after 
integration over the Grassmann valued quark fields
\bea
\label{PI3}
\langle\mathcal{O}(n_1,n_2,\ldots)\rangle =
 \frac{1}{Z}\int\bigg[\prod_{k,\mu} dU_\mu(k)\bigg] e^{-S_{g}[U]}\det M[U] \sum_{\text{contract.}} M^{-1}_{n_i,m_j}[U]
 M^{-1}_{n_k,m_l}[U]\ldots
\,,
\eea
where the sum runs over all contributing contractions of the $q(n_i)$ and $\overline q(m_j)$.
The numerical integration over the link variables in Eq.~(\ref{PI3}) requires a finite lattice
volume. Typical state-of-the-art lattice QCD simulations are performed for lattices of size, e.g.,  
$L^3\times L_t=32^3\times 64$. The number of integration variables for such lattices is of the order
of $32^3\times 64\times (N_c^2-1)\times 4\approx67000000$, clearly showing that statistical methods
must be employed. 
The basic idea is to generate a sequence or chain of so-called configurations ${U_i}$ 
where each $U_i$ represents link variables $U_\mu(n)$ for all $\mu$ and $n$,
which sample the probability distribution given by $e^{-S[U]}$ (importance sampling).
This can be achieved using a Markov-process (update-process) 
where the transition probability for going from one field configuration, $U$, to the next, $U'$, satisfies
detailed balance.
Detailed balance can,
for example, be implemented in form of the Metropolis algorithm (Metropolis accept-reject steps),
where the new configuration $U'$ is always accepted when the action decreases,
and only accepted with probability $e^{-S[U']}/e^{-S[U]}<1$ when the action increases.
However, the ordinary Metropolis algorithm will be extremely inefficient when
non-local updates are required in the generation of the configurations, which is 
in particular the case when the full fermion determinant, $\det M[U]$, is taken into account.
Non-local updates can be performed by evolving the system in a deterministic way
in ``simulation time'' with finite time steps
according to Hamiltonian equations of motion along a ``classical'' trajectory to a new 
field configuration, which is finally accepted or rejected based on the Metropolis acceptance test.
Since the Hamiltonian for the evolved configuration stays close to
its initial value, unacceptably high rejection rates as in the case of
a random global change of the field configuration can be avoided.
All this can be properly implemented in form of the widely used Hybrid Monte Carlo (HMC) algorithm,
which is free of systematic errors.
The effects of the fermion determinant can be included into the Hamiltonian and the HMC algorithm
employing pseudo-fermionic fields and a corresponding pseudo-fermonic action.
With an ensemble of $N_{\text{conf}}$ configurations $U_{i=1,\ldots,N_{\text{conf}}}$ generated 
by, e.g., the HMC algorithm, an estimate for the expectation value $\langle\mathcal{O}(n_1,n_2,\ldots)\rangle$ is given by
\bea
\label{PI4}
\langle\mathcal{O}(n_1,n_2,\ldots)\rangle \approx \overline{\mathcal{O}}(n_1,n_2,\ldots)=
 \frac{1}{N_{\text{conf}}} \sum_{i=1}^{N_{\text{conf}}} \sum_{\text{contract.}} M^{-1}_{n_1,m_1}[U_{i}]
 M^{-1}_{n_2,m_2}[U_{i}]\ldots
\,,
\eea
with a statistical error proportional to $N_{\text{conf}}^{-1/2}$. 
A lattice QCD calculation of a correlator as in Eq.~(\ref{PI4}) therefore requires
1) the generation of a set of configurations, which is particularly expensive due to the presence
of the fermion determinant $\det M[U]$, and
2) the computation of quark propagators $M^{-1}_{n,m}[U_{i}]$, 
i.e. the numerical inversion of the fermion matrix for a given ensemble of configurations.
In the past, many lattice simulations have been done in the quenched approximation
where $\det M[U]\rightarrow 1$, corresponding to neglecting sea quark loops, 
in order to reduce the computational expense.
The quenched approximation is not a controlled approximation and only becomes 
exact in the limit of infinitely heavy quarks, which is why dynamical (unquenched)
calculations including the full fermion determinant are indispensable.

A typical $N_f=2+1$ lattice QCD calculation with two light (up and down) and one
heavier (strange) quark involves three dimensionless input parameters
(assuming fixed spatial, $L$, and temporal, $L_t$, lattice extents): 
The coupling constant $\beta$ in the gauge action, Eq.~(\ref{WilsonG}), 
and the quark masses $m_u=m_d$ and $m_s$ in the fermion determinant.
The calculation of dimensionful quantities requires the determination of 
the lattice spacing $a$ in physical units, which can only be done a posteriori
by a matching of lattice results for e.g. the mass of the nucleon or the rho,
or the heavy quark quark potential at the ``Sommer scale'' \cite{Sommer:1993ce}, with results from
experiment and/or phenomenology. The determination of the lattice scale
is therefore subject to, and a source of, statistical and systematic uncertainties.

Costs of dynamical fermion simulations typically rise approximately with some power of the 
lattice extent $L$ and powers of the inverse lattice spacing and the inverse light quark mass.
Until a couple of years ago, unquenched simulations where found to be restricted
to the ``heavy quark'' regime, corresponding to pseudoscalar (pion) masses 
in a range of $m_\pi\approx600\MeV\ldots1000\MeV$. Due to the predicted rise of cost
with powers of $1/m_q$, $L$ and $1/a$, simulations below $m_\pi\approx300\MeV$,
let alone close to the physical pion mass, were regarded as unattainable.
However, due to ongoing progress in the development of machines
and the resulting increase in computer power, significantly improved algorithms and 
other conceptual advances, lattice hadron structure calculations
at pion masses of around $300\MeV$ actually became feasible during the recent years.
Based on these developments, current predictions are way more optimistic than before,
giving hope that in the not-too-far future lattice hadron structure calculations
of selected observables may be possible with reasonable statistics close to 
or even directly at the physical pion mass.
For a recent review of the status of lattice QCD simulations,
including cost estimates, we refer to \cite{Jansen:2008vs}.

Concerning the investigation of the structure of hadrons, the choice of 
the lattice parameters and the lattice size is roughly speaking a two-scales
problem. On the one hand, the quark masses and/or the lattice extent must be sufficiently large  
so that not only the hadron under investigation, but also other relevant
degrees of freedom, in particular the virtual pions (the ``pion cloud"), which are essential for
the hadron structure, fit into the 
lattice volume\footnote{We do not discuss in this work the interesting developments
in the so-called $\epsilon$-regime where $m_\pi L<<1$.}. 
At the same time, the lattice spacing should be small enough (the coupling large enough) so that the
internal structure of the hadron can be resolved in the first place and discretization effects
can be kept under control.
The spatial dimensions of the relevant states may be measured in terms of their de Broglie wavelengths and rms radii. 
As a rule of thumb it is often required that $m_\pi L\ge4$, but if this is sufficient to suppress
finite volume effects clearly must be studied case-by-case and will depend on the observable under consideration. 
Typical lattice spacings are $a\approx0.1\fm$ or smaller, which may be compared
to the rms radius of, e.g., the nucleon
at currently accessible quark masses on the lattice, which is $\langle r^2\rangle^{1/2}\approx0.5\fm$.
A further restriction in conventional lattice hadron structure calculations comes from the fact
that the lattice momenta accessible in a finite volume, $V=L^3$, 
are discrete and given by $p=2\pi n/(aL)$, $-L/2\le n\le L/2$, 
for periodic boundary conditions in spatial directions, 
i.e. $q(\vec x+L\vec e_j)=q(\vec x)$ for quark fields.
This leads to lowest non-vanishing momentum components ($n=1$) that
are rather large, $p\gtrapprox 0.3\GeV$.

In order to finally obtain predictions at the physical point, 
the continuum and infinite volume limits have to be taken, 
and the quark masses eventually have to be tuned to their physical values.
The continuum limit is in general achieved by tuning the coupling to the value where the
lattice correlation length, generically given by the inverse mass gap in lattice units, 
$\xi^\lat=(aE-aE_0)^{-1}=(am)^{-1}$, diverges, such that the physical (continuum) mass stays finite.
In practice, the expected discretization errors of the lattice results proportional to $a$ or $a^2$ 
may be studied directly by performing simulations at different couplings $\beta$, and the continuum limit
can be taken by performing extrapolations in the lattice spacing.
As will be briefly discussed in section \ref{sec:ChEFT} below,
extrapolations to $a=0$, $L=\infty$, and in particular $m^\lat_q=m^\phys_q$ may be 
systematically studied in the framework of the low energy effective field theory of QCD.
In summary, lattice QCD simulations are subject to a number of statistical and systematic uncertainties:
\begin{itemize}
\item statistical errors from the Monte Carlo evaluation of the path-integral
\item discretization effects due to finite lattice spacing
\item finite size (volume) effects due to finite lattice volume
\item large unphysical quark (pion) masses 
\item statistical and systematic errors in the setting of the lattice scale
\end{itemize}
Additional uncertainties in hadron structure observables arise for example in 
the calculation of lattice renormalization constants,
from contributions of excited states to hadron correlation functions
and from extrapolations required by large lowest non-vanishing lattice momenta.
Of central importance is that all these uncertainties can at least in principle 
be systematically and continuously reduced by using larger ensembles of configurations,
reducing the lattice spacing (increasing the coupling), 
increasing the lattice size, and lowering the quark masses.
This is different from model calculations, which certainly can and in many cases 
do provide a better understanding of how QCD works, but generically suffer from uncontrollable
systematic uncertainties.
%

\subsubsection{Lattice actions: Overview}
\label{sec:actions}

\subsubsection*{Gauge actions}

Starting from the original Wilson gluon action in Eq.~(\ref{WilsonG}), many 
$\mathcal{O}(a^n)$-improved gauge actions
have been introduced over the years. This is systematically described in the form of the
Symanzik improvement program \cite{Symanzik:1983dc,Symanzik:1983gh}. 
A large variety of gauge field actions have been used in large scale 
lattice simulations, mostly differing in the numerical values of certain improvement coefficients.
For example, the standard Wilson gauge action has been used 
in dynamical Wilson fermion simulations 
by QCDSF/UKQCD since 2000 \cite{Stuben:2000wm,Irving:2000hs} and by JLQCD \cite{Aoki:2002uc}.
A tree-level Symanzik improved action is employed by ETMC \cite{Urbach:2007rt} 
in dynamical Wilson twisted mass fermion simulations, and by QCDSF in recent dynamical Wilson fermion 
studies \cite{Cundy:2009yy}.
Simulations by MILC are based on a one-loop Symanzik improved action with
Asqtad staggered quarks, see \cite{Bernard:2001av,Aubin:2004wf} and references therein.
A tapole-improved Symanzik action was used in \cite{Zanotti:2001yb}.
Similar (renormalization-group improved) gauge actions in the form of the Iwasaki
action were used in, e.g., dynamical overlap fermion simulations
by JLQCD \cite{Aoki:2008tq}, Wilson fermion simulations by PACS-CS \cite{Aoki:2008sm}
and domain wall fermion calculations by RBC/UKQCD \cite{Allton:2008pn},
and in the form of the doubly-blocked Wilson (DBW2) action by RBC \cite{Aoki:2004ht}.

A further way of improving lattice simulations, for example concerning chiral properties of Wilson fermions and 
taste symmetry breaking of staggered fermions, is to replace the original ``thin'' link variables by so-called 
fat or smeared gauge links. Smeared links can be iteratively constructed by replacing each link
by a weighted average of the link and nearby gauge paths represented by products of three or more links
(e.g. staples). Typical examples are APE smearing
\cite{Albanese:1987ds} and HYP blocking/smearing \cite{Hasenfratz:2001hp}, where the latter
only includes staples that lie within the hypercube attached to the original gauge link.
The differentiable\footnote{Differentiability is crucial with respect to HMC simulations.} 
stout smearing \cite{Morningstar:2003gk} has been used in 
a number of recent studies \cite{Durr:2008zz,Lin:2008pr,Cundy:2009yy}, and a differentiable form of APE
smearing was proposed in \cite{Kamleh:2004xk} in the context of Fat Link Clover Improved (FLIC) fermions \cite{Zanotti:2001yb}.

Smearing generically suppresses short-range fluctuations of the gauge fields and is therefore
useful in many practical applications.

\subsubsection*{Lattice fermions}
Wilson fermions \cite{Wilson:1975id} are numerically simple to implement, well understood and tested, 
and cost-efficient for not too small quark masses. 
The Wilson fermion action action is obtained by adding the so-called Wilson term
of the form $-ar/2\sum_n\overline q(n)\mybox q(n)$ to the
naively discretized action. The Wilson term removes the fermion doubler modes
and vanishes in the continuum limit, but breaks chiral symmetry explicitly at finite lattice spacing.
The Wilson fermion action can be written as
\bea
\label{WilsonF}
S^{WF} 
       &=& \frac{1}{2\kappa}\sum_{n,m}\overline q(n) M_{n,m} q(m)\,,
\eea
with 
\be
\label{WilsonF2}
 M_{n,m}=\delta_{nm}-\kappa\sum_\mu\bigg\{(r-\gamma_\mu)U_\mu(n) \delta_{n+\hat\mu,m} 
        + (r+\gamma_\mu)U^\dagger_\mu(n-\hat\mu) \delta_{n-\hat\mu,m}  \bigg\}
\ee
where $\hat\mu$ denotes a unit vector in $\mu$-direction, 
$r$ is the Wilson parameter (in most applications $r=1$), 
and the hopping parameter $\kappa$ is defined by $\kappa=(8r+2m_q)^{-1}$.
Lattice spacing errors for Wilson fermions
are generically of $\mathcal{O}(a)$. The Wilson fermion action can be $\mathcal{O}(a)$-improved 
using the so-called clover term $ ia/4c_{SW} \sum_n\overline q(n)\sigma_{\mu\nu}F_{\mu\nu}(n)q(n)$
\cite{Sheikholeslami:1985ij}
with a properly discretized field strength tensor. 
The $\mathcal{O}(a)$-improvement can be achieved numerically by tuning 
the Sheikholeslami-Wohlert coefficient $c_{SW}$. 
Wilson fermions are used in large scale numerical simulations
by, e.g., QCDSF/UKQCD with $n_f=2$ flavors \cite{Stuben:2000wm,Irving:2000hs}, 
QCDSF with $n_f=2+1$ flavors\cite{Gockeler:2007rm}, 
and PACS-CS with $n_f=2+1$ flavors \cite{Aoki:2008sm}. Due to improved methods,
e.g. the rational hybrid Monte Carlo (RHMC) algorithm \cite{Clark:2003na},
mass preconditioning and Hasenbusch acceleration \cite{Hasenbusch:2001ne}, it is by now feasible
to employ Wilson fermions at low quark masses where $m_\pi<300$ MeV.	

In their original form, Wilson fermions suffer from additive quark mass renormalization, 
which leads in particular to numerical instabilities in simulations at small quark masses.
This can be avoided by working with twisted-mass Wilson Fermions, which can be obtained
by adding a twisted mass term of the form
$i\mu_q\overline q\gamma_5\tau_3q$ to the Wilson fermion action \cite{Frezzotti:2000nk},
with mass parameter $\mu_q$, and where $\tau_3$ acts in flavor space.
Another advantage of twisted mass fermions is that discretization errors
for many observables can be reduced to be $\mathcal O(a^2)$ by tuning the 
bare quark mass to its critical value, i.e. when working at 
so-called `maximal twist' \cite{Frezzotti:2003ni}. 
On the downside, the twisted mass term leads to a violation of isospin and parity symmetry at $\mathcal O(a^2)$.
Twisted mass Wilson fermions are cost-efficient and used in large scale numerical
simulations by the ETMC \cite{Urbach:2007rt}. A review of twisted mass lattice QCD
is given in \cite{Shindler:2007vp}.		

Staggered (Kogut-Susskind) fermions 
represent another highly cost-efficient fermion discretization that is free of the usual
doublers and possesses a remnant chiral symmetry for vanishing quark masses, which is sufficient to prevent
additive quark mass renormalization. However, staggered
fermion fields carry an additional (unphysical) quantum number called `taste'. 
The four different tastes of staggered fermions complicate the physical interpretation of the results,
and in many practical applications the fourth root of the staggered fermion determinant is
taken in order to reduce the four taste degrees of freedom to a single fermionic DOF.
The validity of this `fourth-root-trick' is currently under intense debate, see
\cite{Bernard:2007eh,Creutz:2008kb} and references therein.
Staggered fermions have been widely used in numerical simulations over the last decade
by the MILC collaboration in the form of the Asqtad (\emph{a-squared tadpole} improved)
staggered quark action \cite{Bernard:2001av,Aubin:2004wf}. 
The lowest pion masses that have been reached are $m_\pi\approx270$ MeV.

Recently, interest has shifted towards using chiral (or Ginsparg-Wilson \cite{Ginsparg:1981bj}) 
fermions in large scale lattice simulations. The Dirac operator $D$ of
chiral fermions is required to satisfy the Ginsparg-Wilson relation $\{D,\gamma_5\}=aD\gamma_5D$,
which has a non-zero RHS at finite lattice spacing, corresponding to broken continuum 
chiral symmetry\footnote{Chiral symmetry in the continuum can be expressed in the form $\{D,\gamma_5\}=0$}.
However, GW-fermion theories are invariant under a \emph{lattice chiral symmetry} transformation 
\cite{Luscher:1998pqa}, which is not only appealing from a theoretical point of view, but
also proves beneficial for a number of practical reasons: 
The fermion action is automatically $\mathcal O(a)$ improved, 
operator renormalization is simplified by reduced/eliminated 
mixing of operators (operator renormalization will be discussed in the following
sections), it is in particular guaranteed that the vector and axial-vector current
renormalization constants agree, $Z_A=Z_V$, 
and additive quark mass renormalization is absent. 
The two types of chiral fermions that are employed in 
numerical simulations are domain wall \cite{Kaplan:1992bt} and overlap fermions \cite{Neuberger:1997fp}.
At the moment, computational resources are not sufficient as to allow for
full overlap fermion simulations. Overlap fermion calculations by JLQCD have been 
performed in a fixed topological sector \cite{Aoki:2008tq}. 
Domain wall (DW) fermions are formulated in 5 dimensions, 
and provide the full lattice chiral symmetry only when the extent of the fifth dimension, $L_5$,
is approaching infinity. They have the advantage that $L_5$ can be tuned to achieve a compromise between
residual chiral symmetry breaking on the one hand and computational cost on the other.
In typical DW lattice simulations $L_5=16$ or $32$, with pion masses as low as $m_\pi\approx300$ MeV.
Domain wall fermions have been employed by RBC in simulations with $n_f=2$ \cite{Aoki:2004ht},
and by RBC/UKQCD with $n_f=2+1$ \cite{Allton:2008pn} flavors.

In the ideal case, the fermion action that is being used for the computation
of the lattice gauge configurations (i.e. including the fermion determinant)
is identical to the fermion action that is being used for the calculation of
quark propagators. In other words, one consistently uses the same type of lattice fermion for
the sea and valence quarks. In order to benefit from the numerical efficiency of, 
e.g., Wilson or staggered fermions, and at the same time from the desired chiral properties 
of overlap or DW fermions, so-called hybrid or mixed action schemes have been devised, where
cost-efficient fermion discretizations are used for the time-consuming and expensive calculation of the
gauge configurations, and where the quark propagators are based on chiral fermion formulations.
In this case, the valence bare quark masses may be tuned in order to match
the masses of mesons and baryons in the pure sea quark and hybrid formulations.
One has to keep in mind, however, that mixed action simulations
suffer from unitarity violation at finite lattice spacing.
Mixed action calculations have been performed by, e.g., the LHPC collaboration \cite{Hagler:2007xi,WalkerLoud:2008bp}, 
using domain wall valence fermions in combination with Asqtad staggered sea quarks of the
gauge configurations provided by MILC.

%

\subsubsection{Basic methods and techniques}
\label{sec:methods}
An excellent introduction to lattice hadron structure calculations can be found in \cite{Horsley:2000}.
Below, we only briefly introduce some of the basic methods and techniques employed in numerical
studies of hadrons in lattice QCD.
\subsubsection*{Two- and three-point functions}
Lattice studies of hadron matrix elements introduced in section \ref{sec:observables} are mostly based on
hadron two- and three-point functions.
Two-point functions have the following form:
\bea
\label{C2pt}
C_{\text{2pt}}(t,\mbf{P})&=&\langle h(t,\mbf{P})h^\dagger(0,\mbf{P})\rangle \,.
\eea
Here $h^\dagger(t,\mbf{P})$ and $h(t,\mbf{P})$ are interpolating fields for the source and sink, respectively, 
of a hadron with lattice momentum $\mbf{P}$ at the Euclidean time $t$. 
The interpolating fields are given by products of quark fields, which are combined in color, Dirac and flavor space
in order to provide the quantum numbers of the hadron under consideration. The exact form of the interpolating
field for a given hadron is not unique. With $h(t,\mbf{P})=\sum_{\mbf{x}}e^{-i\mbf{x}\cdot\mbf{P}}h(t,\mbf{x})$, 
standard choices for the $\pi^+$, the proton, $p$, and the $\rho^+$ are
\bea
\label{interpol1}
\pi^+\big(J^{P}=0^{-}\big)&:&\quad h(t,\mbf{x})=\overline d(t,\mbf{x})[\gamma_4]\gamma_5u(t,\mbf{x})\,,\\
\label{interpol2}
p\big(J^{P}=\sq{\frac{1}{2}}^{+}\big)&:&\quad h_i(t,\mbf{x})=\eps_{abc}  
             \left\{u^a(t,\mbf{x})^TC\gamma_5d^b(t,\mbf{x})\right\} u^c_i(t,\mbf{x}) \quad i=1\ldots4\,,\\
\label{interpol3}
\rho^+\big(J^{P}=1^{-}\big))&:&\quad h_j(t,\mbf{x})=\overline d(t,\mbf{x})\gamma_j u(t,\mbf{x})\quad j=1\ldots3\,
\eea
where $[\gamma_4]$ indicates the possibility to include $\gamma_4$, and $C$ denotes the charge conjugation matrix.
It should be noted that the interpolating field in Eq.(\ref{interpol2}) has no definite parity, 
and a projection operator may be used to suppress the negative parity contributions, see below.
A Rarita-Schwinger interpolating field for, e.g., the spin-$3/2$ $\Delta^+$-baryon is given by 
\bea
\Delta^+\big(J^{P}=\sq{\frac{3}{2}}^{+}\big):
\quad h_{i,\alpha}(t,\mbf{x}) &=& \eps_{abc}\bigg(
     2\left\{u^{a}(t,\mbf{x})^T C\gamma_\alpha d^b(t,\mbf{x})\right\} u^c_i(t,\mbf{x})\nonumber\\
     &+& \left\{u^{a}(t,\mbf{x})^T C\gamma_\alpha u^b(t,\mbf{x})\right\} d^c_i(t,\mbf{x}) \bigg)
\quad i=1\ldots4\,.
\label{interpol4}
\eea
Clearly, these interpolating fields do not directly
create the physical pion, nucleon, $\rho$-meson and $\Delta$-baryon ground states. 
However, they do have a non-zero overlap with the
physical states, which can be enhanced by using spatially smeared instead of 
point-like quark fields,
\bea
\label{smear1}
q^S(t,\mbf{x})=\sum_\mbf{y}H(\mbf{x},\mbf{y},t;[U])q(t,\mbf{y}) \,,
\eea
where the smearing kernel $H$ is hermitian and transform as
\bea
\label{smear2}
H(\mbf{x},\mbf{y},t;[U])\rightarrow G(\mbf{x},t)H(\mbf{x},\mbf{y},t;[U])G^{-1}(\mbf{y},t)\,
\eea
under gauge transformations. In practice, the source and sink smearing is often implemented
in form of the iterative Jacobi smearing \cite{Allton:1993wc}, which depends on
two smearing parameters. These parameters can be tuned to provide an optimal overlap with the physical 
hadron wave function.

Hadron three-point functions are given by
\bea
\label{C3pt}
C^{\mathcal{O}}_{\text{3pt}}(\tau,\mbf{P}',\mbf{P})&=&\langle h(\mbf{P}',t_{\text{snk}})\sum_{\mbf{x}}e^{i\mbf{x}\cdot\mbf{\Delta}} \mathcal{O}(\tau,\mbf{x})h^\dagger(\mbf{P},0)\rangle \,,
\eea
describing a hadron that is probed by an operator $\mathcal{O}(\tau,\mbf{x})$ at time $\tau$, 
which is Fourier transformed with respect to $\mbf{x}$ to momentum space with 
a momentum transfer $\mbf{\Delta}=\mbf{P}'-\mbf{P}$.  
The Euclidean time dependence of two- and three-point functions can be obtained by inserting time translation
operators, $\exp(-\hat P_4 t )$, to shift the interpolating fields to $t=0$, and by inserting
complete sets of energy-momentum eigenstates, $|E_m(\mbf{P})\rangle$, 
into Eqs.~(\ref{C2pt}) and (\ref{C3pt})\footnote{The time dependence of correlation functions
on finite Euclidean lattices is properly treated in the transfer matrix formalism \cite{Luscher:1976ms}}.
One obtains for the pion
\bea
C^{\pi}_{\text{2pt}}(t,\mbf{P})&=&Z(P)\big\{e^{-E(\mbf{P})t} + e^{-E(\mbf{P})(T-t)}\big\}+\cdots= 2Z(P)\cosh((T/2-t)E(\mbf{P}))+\cdots\nonumber\\
\label{C2ptPi}\\
C^{\pi,\mathcal{O}}_{\text{3pt}}(\tau,\mbf{P}',\mbf{P})&=&\big(Z(P')Z(P)\big)^{1/2}\langle \pi(\mbf{P'})|\mathcal{O}(0)|\pi(\mbf{P})\rangle\times
\nonumber\\
&&\left\{\begin{array}{c}
e^{-(t_\text{snk} - \tau) E(\mbf{P'}) - \tau E(\mbf{P})} +\cdots                \quad, t_\text{snk} \ge \tau\\
(-1)^{n_4+n_5} e^{-(\tau - t_\text{snk}) E(\mbf{P'}) - (T - \tau) E(\mbf{P})} +\cdots  \quad, T \ge \tau \ge t_\text{snk}
\end{array}\right.  \,,
\label{C2ptC3ptPi}
\eea
where $T$ denotes the time extent of the lattice, $T\ge t_{snk}$, while $n_4$ and $n_5$ are the number of
$\gamma_4$ and $\gamma_5$ matrices, respectively, in the operator. Contributions from excited states with energies 
$\ldots>E''>E'$
and from wrap-around effects 
are denoted by the dots in Eqs.(\ref{C2ptPi},\ref{C2ptC3ptPi}). 
For e.g. $t_\text{snk} \ge \tau$, they are suppressed if $T-t,t,\tau \gg(E'-E)^{-1}$
and $t_\text{snk}-\tau \gg(E'-E)^{-1}$.
The nucleon two-point function can be written as
\bea
C^{N}_{\text{2pt};i,j}(t,\mbf{P})&=&\big(Z(P)\overline Z(P)\big)^{1/2}
\sum_S\big\{\overline U_i(P,S)U_j(P,S)e^{-E(\mbf{P})t} 
\big\}
+\cdots\,.
\label{C2ptN}
\eea
where the dots represent contributions due to wrap-around effects (e.g. anti-nucleons moving in opposite
time direction) that are suppressed for $T\gg E^{-1}$, and from excited states, for example the first excited positive
parity state corresponding to the Roper resonance $N(1440)$, that are suppressed for $t\gg(E'-E)^{-1}$.
Contributions from negative parity partners, for example the $N^*(1535)$ resonance, and anti-nucleons 
are additionally suppressed by projection onto the upper Dirac components, 
\bea
C^{N}_{\text{2pt};\Gamma_\text{unpol}}(t,\mbf{P})=\mytr\big(\Gamma_\text{unpol}C^{N}_{\text{2pt}}(t,\mbf{P})\big) \,,
\label{C2ptN2}
\eea
where $\Gamma_\text{unpol}=(1+\gamma_4)/2$.
The nucleon three-point function can be written as
\bea
C^{N,\mathcal{O}}_{\text{3pt};\Gamma_\text{pol}}(\tau,\mbf{P}',\mbf{P})&=&
e^{-(t_\text{snk} - \tau) E(\mbf{P'}) - \tau E(\mbf{P})} 
\big(\overline Z(P')Z(P)\big)^{1/2}\times\nonumber\\
&&\sum_{S,S'}\mytr\big\{\Gamma_\text{pol}\overline U(P,S) U(P',S') \big\}
\langle P',S'|\mathcal{O}(0)|P,S\rangle
 +\cdots\,,
\label{C3ptN}
\eea
where we have introduced a polarized projector $\Gamma_\text{pol}$, e.g. $\Gamma_\text{pol}=\Gamma_\text{unpol}(1-\gamma_5\gamma_3)/2$,
which suppresses contributions from negative parity states and at the same time allows to access the spin structure of the nucleon.
The two- and three-point functions for spin-1 states like the $\rho$-meson can be written in a way similar to 
the pion correlators in Eq.(\ref{C2ptC3ptPi}), except for the appearance of polarization vectors
$\eps_j(\mbf{P'})^*$ and $\eps_i(\mbf{P})$ for the sink and source, respectively, see, for example, 
Ref.~\cite{Hedditch:2007ex}.
The corresponding formalism for spin-$3/2$ $\Delta$-baryon correlators can be found in \cite{Leinweber:1992hy}.

Lattice calculations of two-point functions clearly provide access to the spectrum of mesons and baryons, including
hybrid and exotic states via the exponential decay in the Euclidean time $t$. 
Particularly relevant for hadron structure calculations discussed in this work are of
course the masses (and dispersion relations) of the respective ground states, which may be directly obtained
from plateaus in the effective mass $m_\text{eff}(t)=\ln C_{\text{2pt}}(t-1,\mbf{0})/C_{\text{2pt}}(t,\mbf{0})$,
or from (multi-) exponential fits to the two-point functions.
The hadron three-point functions in Eqs.(\ref{C2ptC3ptPi}) and (\ref{C3ptN}) give direct access to the
hadron matrix elements of the local quark operators that have been discussed in section \ref{sec:observables}, 
and thereby the form factors and generalized form factors that parametrize the non-perturbative physics. 
One option to extract the lattice estimates of the hadronic matrix element
is to perform simultaneous fits to the time dependences of the two- and three-point functions,
based on Eqs.~(\ref{C2ptC3ptPi}), (or (\ref{C2ptN}), (\ref{C3ptN})), 
possibly also including excited state contributions,
with energies $E(\mbf{P}),E(\mbf{P}'),\ldots$, overlap factors $Z$, and the 
sought-after matrix elements $\langle h(P')|\mathcal{O}|h(P)\rangle$ as fitting parameters.
In the case that a spin-projection has been used in the hadron three-point function, for example 
as for the nucleon in Eq.~(\ref{C3ptN}), the extracted hadron matrix element will be understood as 
implicitly including this spin-projection.
Alternatively, ratios of three- and two-point functions can be constructed in order to
remove overlap and exponential factors in Eqs.~(\ref{C2ptC3ptPi}) and (\ref{C3ptN}).
An often used ratio is 
\begin{equation}
  \label{ratio1}
  R_{\mathcal O}(\tau,P',P) = \frac{C^{\mathcal{O}}_{\text{3pt}}(\tau,\mbf{P}',\mbf{P})}
  {C_{\text{2pt}}(t_{\text{snk}},\mbf{P}')}
\left[
    \frac{C_{\text{2pt}}(t_{\text{snk}}-\tau,\mbf{P})
      \;C_{\text{2pt}}(\tau,\mbf{P}')\;C_{\text{2pt} }(t_{\text{snk}},\mbf{P}')}
    {C_{\text{2pt}}(t_{\text{snk}}-\tau,\mbf{P}')
      \;C_{\text{2pt}}(\tau,\mbf{P})\;C_{\text{2pt}}(t_{\text{snk}},\mbf{P})}
  \right]^{1/2}\,,
\end{equation}
which is approximately time-independent, i.e. exhibits a plateau in $\tau$, for an operator insertion
sufficiently far away from source and sink, $t_{\text{snk}}\gg\tau\gg0$. Note that this holds
for arbitrary choices of momenta $P$ and $P'$ (in particular $P\not=P'$), 
projectors $\Gamma$, and operators $\mathcal{O}$. The situation is slightly different for the
pion, since pion masses in modern lattice calculations are approaching values for 
which wrap-around effects are in general non-negligible, i.e. where
the full $\cosh(T/2-t)$ time dependence of the pion two-point functions has to be taken into account.
In this case, a residual time dependence of $R_{\mathcal O}(\tau,P',P)$ remains, which is however
well understood and, up to small systematic errors, does not spoil the usefulness of the ratio. 
Different ratios
have been used in the literature, see, e.g.,
\cite{Wilcox:1991cq,Leinweber:1992hy,Hedditch:2007ex,Boyle:2007wg}.
A  salient feature of ratios as in Eq.(\ref{ratio1}) is that statistical fluctuations
of the involved quark propagators in the numerator and denominator cancel out to some extent. 
On the other hand, the square root in Eq.(\ref{ratio1}) poses a potential disadvantage as two-point functions
at or close to $t_\text{snk}$ (where they are most noisy) may take values that are very close to zero or even negative,
in which case it is not clear what to do with the square root of a negative number. 
This does happen in practice in particular at large momenta, and 
in such cases it is most safe to exclude the corresponding two-point functions/hadron momenta from the analysis.
An alternative way of dealing with this problem using shifted two-point functions has been suggested and used 
in \cite{Brommel:2006ww}.
Once a plateau in the ratio Eq.~\ref{ratio1} has been identified, an average may be taken to increase the statistics,
and a lattice estimate of the hadronic matrix element is then given by
\begin{equation}
  \label{ratio2}
  \langle h(P')|\mathcal{O}|h(P)\rangle_\lat=C(P',P)\frac{1}{\tau_\max-\tau_\min+1}
  \sum_{\tau=\tau_\min}^{\tau_\max} R_{\mathcal O}(\tau,P',P) \,,
\end{equation}
where $C(P',P)$ is a known kinematical coefficient.

The lattice estimate of $\langle h(P')|\mathcal{O}|h(P)\rangle$ 
(which may already include a spin projection) obtained from Eq.~\ref{ratio2},
or from a simultaneous fit to the two- and three-point functions as described above, 
can then be set equal to the corresponding continuum parametrization of the hadron matrix element
in terms of (generalized) form factors, as given in Eqs.~(\ref{PionFF},\ref{PionTensorFF},\ref{PionVectorGFFn2},\ref{PionTensorGFFn2},\ref{NuclVec3},\ref{NuclAxial3},\ref{NuclTensor3},...).
Since the lattice momenta $\mbf{P},\mbf{P}'$ are discrete, only specific values of the squared momentum
transfer $t=\Delta^2=(P'-P)^2$ can be reached. Typically, several different combinations 
of the accessible $\mbf{P},\mbf{P}'$, e.g. 
all $L/(2\pi)\mbf{P}=(\pm1,0,0),(0,\pm1,0),(0,0,\pm1)$ for fixed $\mbf{P'}=2\pi/L(0,0,0)$, 
give the same value of $t$. Furthermore, we note that the (generalized) form factors for a given hadron and 
type of operator (vector, axial-vector, tensor operator with $l$-derivatives) 
by definition only depend on $t$ (or $Q^2$), but are in particular 
independent of the hadron spins $S,S'$ (or the spin-projection)
and of the specific indices $\mu,\nu,\mu_1,\ldots$ of the local operators given in Eq.~(\ref{localOps}).
For a given value of $t$, it is therefore possible to construct a set of equations of the form
``$\langle h(P')|\mathcal{O}|h(P)\rangle_\lat=$ continuum parametrization in terms of (generalized) form factors'',
for all available and contributing $\mbf{P},\mbf{P}'$ with $t=(P'-P)^2$, 
spins or spin projections, and operator-indices.
This gives an in general overdetermined set of linear equations, which may finally be numerically solved
by $\chi^2$-minimization to obtain the (generalized) form factors.
For details, we refer to \cite{Hagler:2003jd}.
Alternatively to the construction of an overdetermined, potentially very large set of linear equations,
one might use in the first place only linear combinations of matrix elements of particular momenta, polarizations and
operator indices that are parametrized by just a single or two (generalized) form factors at a time. 
This approach has, for example, been successfully employed in studies of the
electromagnetic form factors of vector mesons in \cite{Hedditch:2007ex} and the $\Delta$-baryon 
in \cite{Alexandrou:2008bn}.

It is important to note that the observables extracted in this way not only carry statistical errors
due to the Monte-Carlo computation of the correlation functions, but are also subject to
systematic uncertainties. Discretization errors have to be expected
from equating a \emph{lattice} estimate of a hadron matrix element with the 
corresponding \emph{continuum} parametrization.
In particular, for the calculation of the kinematical coefficients in the system of equations, 
one might either employ the continuum or the lattice dispersion relation for $E(\mbf{P})$, or
even use the ``measured'' dispersion relation obtained from a numerical analysis of the
hadron two-point functions. 
Additionally, special attention has to be paid to the classification 
of local operators in discretized Euclidean space-time, as will be discussed in the following section.

\subsubsection*{Lattice operators and renormalization}
The lattice operator $\mathcal{O}(\tau,\mbf{x})$ in Eq.~(\ref{C3pt}) is the Euclidean, 
discretized version of the continuum operators
that have been discussed in section \ref{sec:observables}. 
Three-point functions have in general 
discretization errors of $\mathcal{O}(a)$\footnote{So called ``automatic $\mathcal{O}(a)$-improvement''
of operators may be achieved by using twisted-mass Wilson fermions at maximal twist.}, 
which may be removed by adding improvement terms to the lattice operators. 
The improvement terms have to respect the symmetry properties of the original operator and must 
vanish in the continuum limit. An $\mathcal{O}(a)$-improved variant
of, for example, the point-like lattice vector current, 
$\mathcal{J}^V_\mu(x)=\overline q(x)\gamma_\mu q(x)$, is given by
\bea
\label{improved1}
 \mathcal{J}^{V,\text{impr}}_\mu(x)=\big(1+ac_0m_q\big)\overline q(x)\gamma_\mu q(x)
      + a c_2 \hat\partial_\nu\big(\overline q(x)i\sigma_{\mu\nu} q(x)\big)   \,,
\eea
with $\hat\partial_\mu q(x)=(q(x+\hat\mu)-q(x-\hat\mu))/(2a)$, and where the $c_{0,2}$ are real-valued improvement coefficients,
which depend on the coupling $\beta$. Similar improvements are possible for the axial and tensor currents
and the corresponding operators including covariant derivatives like, e.g., the vector operator
with one covariant derivate
\bea
 \mathcal{O}_{\mu\nu}(x)=\overline q(x)\gamma_\mu \hat\Dlr_\nu q(x)
 &=& \frac{1}{4}\bigg\{\qbar(x)\gamma_\mu\big[ U_{\nu}(x) q(x+\hat\nu) - U_{-\nu}(x) q(x-\hat\nu)\big]\nonumber\\ 
      &-& \big[\qbar(x+\hat\nu) U^\dagger_{\nu}(x) - \qbar(x-\hat\nu) U^\dagger_{-\nu}(x) \big]\gamma_\mu q(x) \bigg\}   \,, 
 \label{op1}
\eea
which also suffers from discretization errors of $\mathcal{O}(a)$.
Presently, improvement coefficients are often only known to leading-order 
in lattice perturbation theory, and $\mathcal{O}(a)$-improved operators
are therefore in practice unfortunately of limited use. 

Although the lattice regularization of QCD makes an infinite renormalization unnecessary,
a finite renormalization is still required in order to be able to compare 
results obtained in the ``lattice regularization scheme'' with experiment and phenomenology, where
the (in general) scheme and scale dependent quantities are 
often quoted in the modified minimal subtraction, $\MSbar$, scheme
at a certain renormalization scale $\mu$. Also, different lattice actions correspond to 
different regularization schemes, so that lattice results have to be transformed
to a common scheme in order to be comparable.

Many complications arise from the loss of continuum space symmetries on the lattice.
To begin with, it is important to note that the \emph{local} vector current $\mathcal{J}^V_\mu(x)$,
is not conserved on the lattice.
To obtain the correct values for the charges or baryon numbers of hadrons,
the local lattice vector current operator has to be renormalized, 
$\mathcal{J}^{V,\text{ren}}_\mu=Z_V\mathcal{J}^V_\mu$.
The renormalization constant $Z_V$ can be fixed by, e.g., demanding that the number
of $u-\overline u$ minus $d-\overline d$ quarks, as measured by the isovector vector current, 
in the proton is equal to one. 
Alternatively, a conserved vector current can be constructed on the lattice based on 
the Noether procedure \cite{Karsten:1980wd},
which is for Wilson type fermions of the form
\bea
\label{conserved1}
 \mathcal{J}^{V,\text{cons}}_\mu(x)&=&
 \frac{1}{4}\big\{\overline q(x)\left( \gamma_\mu-r \right)U_\mu(x)\gamma_\mu q(x+\hat\mu)
 + \overline q(x+\hat\mu)\left( \gamma_\mu+r \right)U^\dagger_\mu(x)\gamma_\mu q(x) 
 \nonumber\\
 &+& \left( x\rightarrow x-\hat\mu\right)\big\}\,.
\eea
The operators in Eqs.~(\ref{QuarkOp},\ref{localOps}) in continuous Minkowski space-time
can be classified according to their Lorentz transformation
properties.
In Euclidean space-time, the relevant symmetry group is $O(4)$, which is for the case
of a discretized lattice theory reduced to the hypercubic group $H(4)$.
Corresponding traceless and symmetric 
discrete lattice operators, $\mathcal{O}_{j}$, in Euclidean space-time may therefore be identified
by studying the irreducible subspaces of $H(4)$ \cite{Gockeler:1996mu}. 
Multiplicatively renormalized lattice operators $\mathcal{O}^\text{ren}_{j}$ are given by
\bea
 \mathcal{O}^\text{ren}_{i}=Z_{ij}\mathcal{O}_{j}\,,
 \label{opRen}
\eea
where a possible operator mixing under renormalization is taken into account by the renormalization matrix $Z$.
Following the notation of \cite{Gockeler:1996mu}, the relevant index combinations for the 1-derivative vector operators, Eq.~(\ref{op1}), are 
\bea
 \mathcal{O}^b_{1234}&=&\frac{1}{2}\big(\mathcal{O}_{44}-\frac{1}{3}
    \left\{\mathcal{O}_{11}+\mathcal{O}_{22}+\mathcal{O}_{33}\right\}\big)\,,\;\;\nonumber\\
 \mathcal{O}^b_{34}&=&\frac{1}{\sqrt{2}}\big(\mathcal{O}_{33}-\mathcal{O}_{44}\big)\,, \;\;\nonumber\\
 \mathcal{O}^b_{12}&=&\frac{1}{\sqrt{2}}\big(\mathcal{O}_{11}-\mathcal{O}_{22}\big)\,,
 \label{op2}
\eea
belonging to the 3-dimensional representation $\tau^3_1$. Together with the ``non-diagonal'' operators of $\tau^6_3$,
\bea
 \mathcal{O}^a_{\mu\nu}=\frac{1}{\sqrt2}\big(\mathcal{O}_{\mu\nu}+\mathcal{O}_{\nu\mu}\big)\,,\quad \mu,\nu=1\ldots4\,\,,\mu\not=\nu
  \label{op3}
\eea
this gives a total of 9 linearly independent index combinations which can be used in a lattice analysis.
Since no operator mixing is present in this case, two renormalization constants, $Z_V^a$ and $Z_V^b$,
are required for the renormalization of the lattice operators in Eq.~(\ref{op2}) and Eq.~(\ref{op3}).
The full classification according to the irreducible representations of $H(4)$ for the
vector and axial vector operators with up to three covariant derivatives can be found in \cite{Gockeler:1996mu}.
In the case of the 1-derivative tensor operators $\mathcal{O}^T_{\mu\nu\mu_1}$, Eq.~(\ref{localOps}) 
with $\Gamma=\sigma_{\mu\nu}$ and $n=2$,
the relevant lattice operators transform according to the two 
8-dimensional representations $\tau^8_1$ and $\tau^8_2$. 
Examples for operators belonging to these multiplets
are given by $(\mathcal{O}^T_{122}-\mathcal{O}^T_{133})$ for $\tau^8_1$ and 
$(\mathcal{O}^T_{123}+\mathcal{O}^T_{132})$
for $\tau^8_2$ \cite{Gockeler:2005cj}.

Operator mixing, which is present for two derivative operators, has been studied in some detail
in \cite{Gockeler:2004xb}, where also mixing coefficients for the improved Wilson action
have been worked out in lattice perturbation theory at 1-loop level.
It has been noted in particular that vector operators with two covariant derivatives that transform
according to the representation $\tau_1^{(8)}$ of $H(4)$, i.e. operators of the type $\mathcal{O}^{DD}_{\{\mu\nu\nu\}}=\overline q\gamma_{\{\mu}\Dlr_{\nu}\Dlr_{\nu\}}q$
with $\mu\not=\nu$, mix with lower dimensional operators of the type 
$\mathcal{O}^{\partial}_{\mu\nu\omega}=\partial_\omega(\overline q\sigma_{\mu\nu}q)$.

Operator renormalization constants can be calculated in lattice perturbation theory 
or non-perturbatively. 
At one-loop order in lattice perturbation theory, they are of the form
\bea
 Z^{\MSbar}_{ij}=\delta_{ij}-\frac{g^2}{(4\pi)^2}\big(-\gamma_{ij}\log(a^2\mu^2)
       + \big[ B_{ij}^{\text{latt}} - B_{ij}^{\MSbar}\big]\big) + \mathcal{O}(g^4)\,,
  \label{ren1}
\eea
with the anomalous dimensions $\gamma_{ij}$, and where the non-trivial part is given by
the difference between the finite constants, $B_{ij}^{\text{latt}} - B_{ij}^{\MSbar}$.
Over the years, lattice perturbation theory calculations of operator renormalization constants
have been performed by a number of authors for different lattice actions. More recent
studies for Wilson, domain wall and overlap fermions can be found in, e.g., \cite{Aoki:2002iq,Capitani:2005vb,Ioannou:2006ds,Gockeler:2006nb}.
For an introduction to lattice perturbation theory and renormalization, we refer to \cite{Capitani:2002mp}.
In many cases, it turns out that even improved 1-loop lattice PT renormalization constants
deviate quite strongly from the result obtained from a ``non-perturbative''
calculation. The use of 1-loop perturbative renormalization constants
can therefore be a serious source of systematic uncertainties.
Non-perturbative renormalization constant can be calculated following, e.g., the so-called Rome-Southampton method 
\cite{Martinelli:1994ty}.
Similar to a momentum subtraction scheme, the operator renormalization constant $Z^\text{NP-MOM}_\mathcal{O}$ 
for a scale $\mu^2=p^2$ is non-perturbatively fixed by demanding that 
\bea
Z_q^{-1}(\mu) Z^{\text{NP-MOM}}_\mathcal{O}(\mu)
\frac{1}{4 N_c}\mytr\big\{\Gamma^{\text{latt}}_\mathcal{O}(p)(\Gamma^{\text{tree}}_\mathcal{O}(p))^{-1}\big\}=1 \,, 
  \label{ren3}
\eea
where the trace runs over color and Dirac indices, and where
$\Gamma^{\text{latt}}_\mathcal{O}(p)$ and $\Gamma^{\text{tree}}_\mathcal{O}(p)$
are lattice and tree-level off-shell vertex functions, respectively, given by
amputated quark correlation functions,
\bea
\Gamma^{\text{}}_\mathcal{O}(p)=
S_{\langle q\overline q\rangle}^{-1}(p)C_{\langle q\mathcal{O}\overline q\rangle}(p)
S_{\langle q\overline q\rangle}^{-1}(p) \,.
  \label{ren2}
\eea
The calculation of the non-gauge-invariant, gauge-field averaged quark propagators,
$S_{\langle q\overline q\rangle}^{-1}$, and quark correlation functions, $C_{\langle q\mathcal{O}\overline q\rangle}$,
has to be done in a fixed gauge, for example Landau gauge.
The lattice operators in the NP-MOM scheme, 
$\mathcal{O^\text{NP-MOM}}=Z^{\text{NP-MOM}}_\mathcal{O}\mathcal{O^\text{latt}}$,
can finally be transformed to the (perturbative) $\MSbar$ scheme with a 
multiplicative renormalization constant $Z^{\MSbar}_{\mathcal{O},\text{NP-MOM}}$, 
which can be calculated in continuum perturbation theory in, e.g., Landau gauge to a given order.
Care has to be taken that the scale $\mu^2=p^2$ is chosen in a range
so that a perturbative calculation of $Z^{\MSbar}_{\mathcal{O},\text{NP-MOM}}(\mu)$
is justified, and at the same time discretization effects due to the high resolution
$1/\mu$ in the numerical calculation of $Z^{\text{NP-MOM}}_\mathcal{O}(\mu)$ are under control.
May details concerning the non-perturbative renormalization of quark bilinears
can be found in \cite{Gockeler:1998ye}.

Non-perturbative operator renormalization has also been studied in the Schr\"odinger functional
scheme \cite{Luscher:1992an}, see, e.g., \cite{Guagnelli:1999wp,Guagnelli:2003hw} and references therein.

\subsubsection*{Lattice computation of hadron correlators; The sequential source technique}
\begin{figure}[t]
   \begin{minipage}{0.48\textwidth}
      \centering
          \includegraphics[angle=0,width=0.9\textwidth,clip=true]{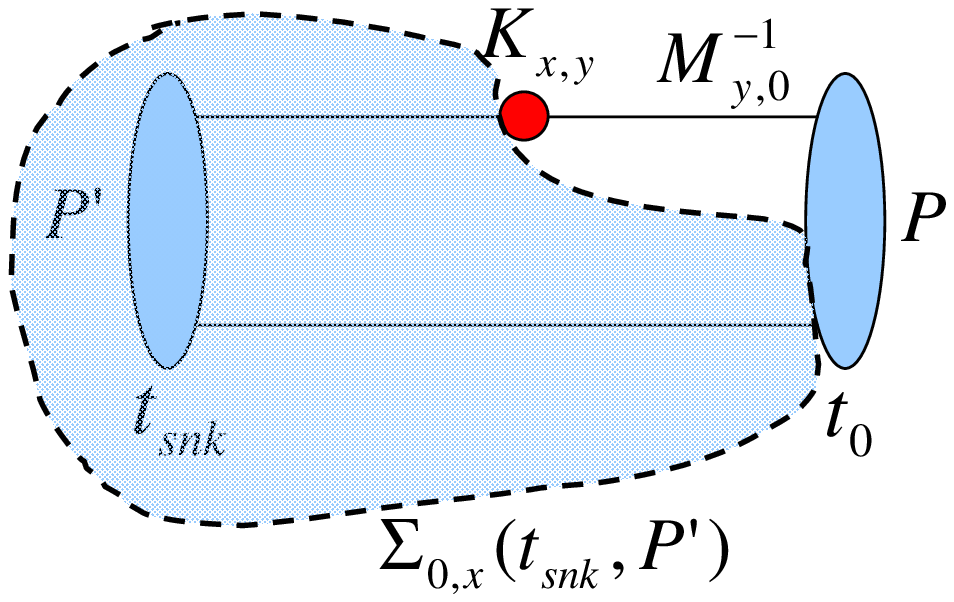}
  \caption{Illustration of the sequential source technique.}
  \label{ill_3pt_seq_v2}
     \end{minipage}
     \hspace{0.5cm}
    \begin{minipage}{0.48\textwidth}
      \centering
          \includegraphics[angle=0,width=0.9\textwidth,clip=true]{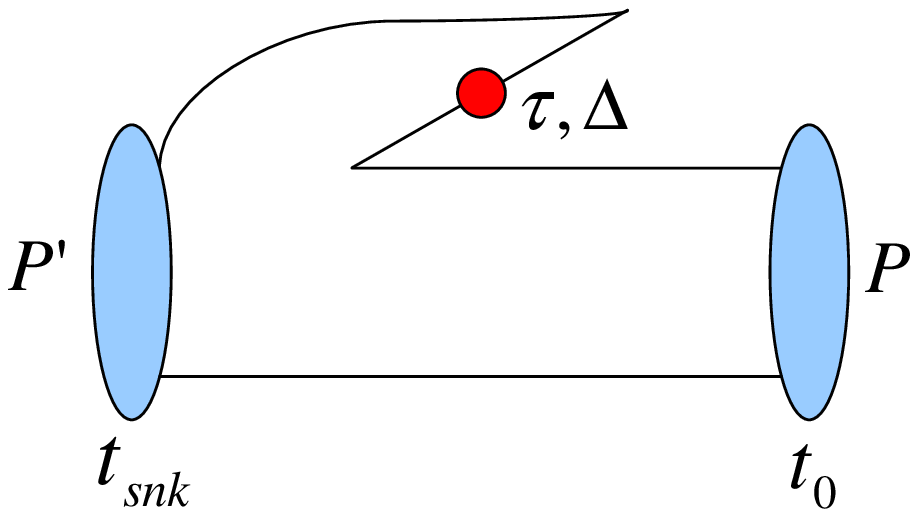}
  \caption{Illustration of anti-quark contributions to a meson three-point function.}
  \label{ill_3pt_Z_v1}
     \end{minipage}
 \end{figure}

Inserting Eqs.~(\ref{interpol1}) to (\ref{interpol3}) into Eq.~(\ref{C3pt}), and integrating out
the $q$ and $\overline q$ fields in the path integral, see Eq.~(\ref{PI3}), the three-point
correlation functions are given by the path integral of products of (gauge field dependent) quark
propagators $M^{-1}_{x,y}[U_{i}]$. 
To give a simple example, we consider a standard meson two-point function
$C^\Gamma_{\text{2pt}}(t,\mbf{P})$ as defined in Eq.~(\ref{C2pt}), where the Dirac matrix
$\Gamma$ specifies the type of meson, e.g. $\Gamma=\gamma_5, \gamma_4\gamma_5$ for a pseudoscalar (pion) 
and $\Gamma=\gamma_i$ for a vector meson ($\rho$), cf. Eqs.~(\ref{interpol1},\ref{interpol3}).
%
%
\begin{figure}[t]
   \begin{minipage}{0.48\textwidth}
      \centering
       \vspace{0.5cm}
          \includegraphics[angle=0,width=0.9\textwidth,clip=true]{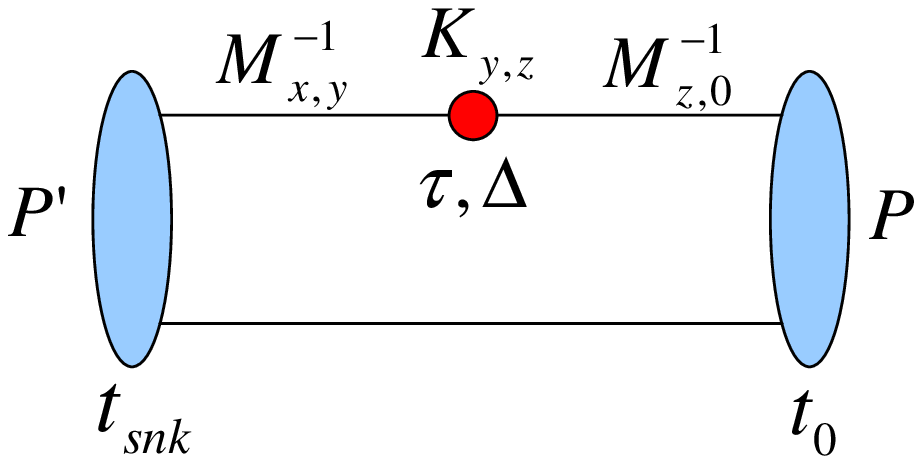}
  \caption{Illustration of a connected contribution to a meson three-point function.}
  \label{ill_3pt_v1}
     \end{minipage}
     \hspace{0.5cm}
    \begin{minipage}{0.48\textwidth}
      \centering
          \includegraphics[angle=0,width=0.9\textwidth,clip=true]{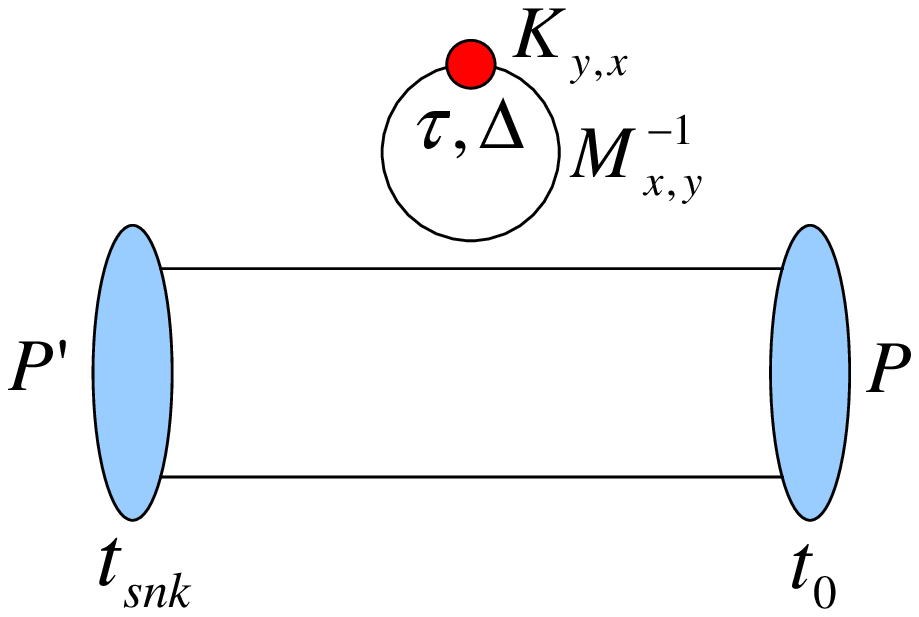}
  \caption{Illustration of a quark line disconnected contribution to a meson three-point function.}
  \label{ill_3pt_disc_v1}
     \end{minipage}
 \end{figure}
%
Using the $\gamma_5$-hermiticity of 
the fermion matrix, i.e. $\gamma_5M\gamma_5=(M)^\dagger$, translation invariance, $m_u=m_d$, and
setting $\Gamma=\gamma_5$ for definiteness, one finds for a pion with momentum $\mbf{P}$
\bea
C^\pi_{\text{2pt}}(t,\mbf{P})=\sum_{\mbf{x}} e^{-i\mbf{x}\mbf{P}}
  \langle \text{Tr}\{M^{-1}_{(\mbf{x},t);(0,0)}(M^{-1}_{(\mbf{x},t);(0,0)})^\dagger \}\rangle_U 
  = \langle G^\pi(t,\mbf{P})\rangle_U\,,
  \label{C2ptProps}
\eea
i.e. the two-point function (or pion propagator $G^\pi(t,\mbf{P})$) 
can be readily expressed in terms of the point-to-all quark propagator $M^{-1}_{x,0}$.
The corresponding three-point function, Eq.~(\ref{C3pt}), for a $\pi^+$ probed by an up-quark operator 
$\mathcal{O}=\overline u K^{\mathcal{O}} u$ can be written as
\bea
C^{\pi,\mathcal{O}}_{\text{3pt}}(\tau,\mbf{P}',\mbf{P})&=&
\sum_{\mbf{x},y,z} e^{-i\mbf{x}\mbf{P}'} 
\langle \text{Tr}\{M^{-1}_{(\mbf{x},t_\snk);y} \tilde K^{\mathcal{O}}_{y,z}(\tau,\mbf{\Delta})M^{-1}_{z,0} 
(M^{-1}_{(\mbf{x},t_\snk);0})^\dagger \}\rangle_U\nonumber\\
&-&\langle G^\pi(t_\snk,\mbf{P}') \sum_{y,z}\mytr\{ M^{-1}_{z,y} \tilde K^{\mathcal{O}}_{y,z}(\tau,\mbf{\Delta}) \} \rangle_U
  \,,
  \label{C3ptProps}
\eea
which receives contributions from quark-line connected and disconnected parts in the first and 
the second line, respectively, 
as illustrated in Figs.~\ref{ill_3pt_v1} and \ref{ill_3pt_disc_v1}.
Both quark-line disconnected and connected diagrams involve ``point-to-all'' and
``all-to-all'' quark propagators, $M^{-1}_{x,y=0}$ and $M^{-1}_{x,y}$, respectively.
While the point-to-all propagators are routinely and efficiently computed by solving
sets of linear equations of the type $M_{xz}(M^{-1})_{zy=0}=M_{xz}\phi_{z}=s_{x}$ 
with source $s$ using, for example,
the conjugate gradient method, the inversion of the full Dirac matrix to obtain the
all-to-all propagators is prohibitively expensive. Fortunately, for connected diagrams,
all-to-all propagators can be completely avoided by using the sequential source technique \cite{Maiani1986445}.
The sequential source technique requires a fixing of the sink 
interpolating field, in particular the sink time $t_\snk$, momentum $\mbf{P}'$, and spin (or the spin-projection).
The inconvenient all-to-all propagator in Fig.~\ref{ill_3pt_v1}
can then be incorporated into the propagator structure
denoted by $\Sigma$ in Fig.~\ref{ill_3pt_seq_v2}, called the sequential propagator, which is
obtained from $M\gamma_5\Sigma^\dagger=s'$, where the source $s'$ itself 
is written in terms of a point-to-all quark propagator. 
The advantage of fixing the sink momentum and quantum numbers
is that once a set of propagators $M^{-1}_{x,0}$ and sequential propagators $\Sigma$ has been calculated, 
one is free to study any quark operator $\mathcal{O}$ of interest with some
momentum transfer of $\Delta=P'-P$ by performing a simple contraction,
\bea
C^{\pi,\mathcal{O}}_{\text{3pt,con.}}(\tau,\mbf{P}',\mbf{P})=
\sum_{x,y} \big\langle \mytr\big(\Sigma_{0,x}(t_\snk,\mbf{P}')
\tilde K^{\mathcal{O}}_{x,y}(\tau,\mbf{\Delta})M^{-1}_{y,0}\big)\big \rangle_U \,, 
  \label{C3pt2_seq}
\eea
see the illustration in Fig.~\ref{ill_3pt_seq_v2}.
An alternative to fixing the sink is to construct sequential propagators based 
on a fixed quark operator $\mathcal{O}$ and momentum transfer $\Delta=P'-P$.

%
%
%

We note that the omission of disconnected diagrams, Fig.~\ref{ill_3pt_disc_v1}, 
does \emph{not} correspond to a calculation
in the pure \emph{valence} quark sector, not even in the quenched approximation. 
Contributions from anti-quarks can be thought of as already implicitly included in the connected diagrams 
in the form of so-called ``z-graphs'', as illustrated in Fig.~\ref{ill_3pt_Z_v1}. 
Lattice calculations that are based on a finite set of local quark operators
therefore do not allow to disentangle contributions from quarks and anti-quarks to observables.

\subsubsection{Disconnected diagrams and stochastic methods}
\label{sec:stochastic}
Contributions from disconnected diagrams, Fig.~\ref{ill_3pt_disc_v1}, 
which are computationally very expensive due to the unavoidable
presence of all-to-all propagators, are often neglected in practical lattice calculations.
In the case that the lattice action is isospin symmetric with $m_u=m_d$, disconnected diagrams
cancel out in the isovector channel, i.e. for $\mathcal{O}^{u-d}$. 
Neglecting disconnected contributions clearly leads to an additional systematic uncertainty 
for flavor singlet observables.
Attempts to compute 
them are based on stochastic methods. 
For an introduction to ``noise methods'', see, e.g., \cite{Wilcox:1999ab}. 
First, a set of random noise vectors (or random sources, stochastic sources),
for example based on complex $Z(2)$ noise \cite{Dong:1993pk}, 
$\eta^{(j)}=\pm1,\pm i$, $(j=1,\ldots,N_\eta)$, is generated, with the crucial property
$N_\eta^{-1}\sum_j\eta^{(j)}_x\eta^{(j)\dagger}_y=\delta_{xy}+\mathcal{O}(N_\eta^{-1/2})$
(for simplicity, we again suppress Dirac and color indices).
The random noise vectors may then be used as random sources in the inversion of
$M_{xz}(M^{-1})_{zy=0}=M_{xz}\phi^{}_{z}=\eta^{(j)}_{x}$, which is solved for all $j$ to obtain $\phi^{(j)}$,
as in the case of the point-to-all propagators described above.
An unbiased estimate of the all-to-all propagator is then given by
\bea
 \frac{1}{N_\eta}\sum_j \phi^{(j)}_x \eta^{(j)\dagger}_{y}&=&(M^{-1})_{xy} 
 + (M^{-1})_{xz}\left\{\frac{1}{N_\eta}\sum_j\eta^{(j)}_z\eta^{(j)\dagger}_y - \delta_{zy}\right\}
 =(M^{-1})_{xy} +\mathcal{O}(N_\eta^{-1/2}) \,, \nonumber\\
 \label{stochastic1} 
\eea
showing that the stochastic error of $\mathcal{O}(N_\eta^{-1/2})$ is directly related to the (near) off-diagonal
elements in $\sum_j\eta^{(j)}_z\eta^{(j)\dagger}_y$. In practice, depending on the number of 
gauge configurations and stochastic sources and the observable under consideration, 
the stochastic error can be larger than the usual gauge noise from the Monte Carlo integration. 
In such cases, a reduction of the stochastic noise is essential. 
Several noise reduction methods and improvement techniques
have been invented and used in practical computations over the years.

The method of ``partitioning'' \cite{Wilcox:1999ab} or ``dilution'' \cite{Foley:2005ac} is based on a replacement of
each noise vector $\eta^{(j)}$ by a set of new noise vectors $\eta^{(j,p)}$, $p=1,\ldots,N_p$,
where the $\eta^{(j,p)}$ have zero entries for many color, Dirac and space-time indices.
Hence $N_p$ times as many inversions 
are required to get the solution vectors $\phi^{(j,p)}$ for all $j,p$. 
Assuming that $\eta^{(j)}=\sum_p \eta^{(j,p)}$, an estimate of the propagator is then
given by $\sum_{j,p}\phi^{(j,p)}_x \eta^{(j,p)\dagger}_{y}$, with reduced statistical noise from
the off-diagonal contributions. Clearly, the computational overhead due to the larger number of
inversions must be taken into account in judging the efficiency of the partitioning.

The truncated eigenmode approach \cite{Neff:2001zr} is based on the orthonormal eigensystem
of the hermitian fermion matrix $Q\mbf{v}^{(j)}=\gamma_5M\mbf{v}^{(j)}=\lambda_{j}\mbf{v}^{(j)}$
and the corresponding spectral decomposition of its inverse,
\bea
 Q^{-1}_{xy}=\sum^{N_\lambda}_j \frac{1}{\lambda_j}\mbf{v}^{(j)}_{x}\mbf{v}^{(j)\dagger}_{y} \,.
  \label{stochastic2} 
\eea
In the region of small quark masses, the sum will be dominated by the lowest
eigenvalues. They can be calculated explicitly, providing a truncated version
of the propagator. The remaining subdominant part of $M^{-1}=Q^{-1}\gamma_5$ can then be 
estimated using random noise vectors as described above,
reducing the stochastic noise compared to a full stochastic calculation. 
This approach has been successfully used in, e.g., 
the calculation of string breaking \cite{Bali:2005fu}, and
meson two-point-functions \cite{Foley:2005ac}.

Another (perturbative) noise reduction technique, based on the hopping parameter expansion (HPE), 
is useful in the heavy quark regime and particularly for the (ultralocal) Wilson action \cite{Thron:1997iy}.
To briefly introduce this method, we note that the disconnected diagrams as in 
Fig.~\ref{ill_3pt_disc_v1} are given by correlators of the form
$\langle\text{Tr}\{\Gamma M^{-1} \}\times G_h \rangle_U$, where $G_h$ is the
hadron propagator, and $\Gamma$ is one of the 16 elements of the Dirac algebra.
An unbiased reduction of the noise in the stochastic
estimate of the trace can then be achieved by subtracting suitable \emph{traceless}
matrices from $\Gamma M^{-1}$. To this end, 
the Wilson fermion matrix is written in the form $M=1-\kappa D$ with the lattice
Dirac matrix $D$ and the hopping parameter $\kappa=(8r+2m_q)^{-1}$, 
see Eqs.~(\ref{WilsonF},\ref{WilsonF2}) and related discussion. 
For large $m_q$, the quark propagator can then be expanded in powers of the hopping parameter 
in a geometric series, $M^{-1}=1+\kappa D+ \kappa^2 D^2 +\ldots$, which suggest to use
$\kappa D, \kappa^2 D^2, \ldots$ as a basis to construct the traceless matrices required for the 
subtraction \cite{Thron:1997iy}.
Similar methods based on these ideas, for example the ``hopping parameter acceleration'', 
were already successfully employed in practice \cite{Bali:2005fu,Collins:2007mh,Bali:2008sx}.
\subsubsection{``One-end-trick''}
\label{sec:oneendtrick}
Stochastic sources, $\eta^{(j)}$, are not only an essential tool for the evaluation of disconnected diagrams, but 
can also be very useful in the calculation of connected contributions to hadron correlators,
using the so-called ``one-end-trick'' \cite{Foster:1998vw}.
For the two-point function Eq.~(\ref{C2ptProps}) for the pion, $\Gamma=\gamma_5$, a 
stochastic source $\eta^{(j)}_{(\mbf{x},t)}=\eta^{(j)}_{(\mbf{x})}\delta_{t,t_0}$,
with the corresponding solution vector (see section above)
\bea
\phi^{(j)}_{(\mbf{x},t);t_0}=\sum_{\mbf z}(M^{-1})_{(\mbf{x},t);(\mbf{z},t_0)}\eta^{(j)}_{(\mbf{z},t_0)} \,,   \label{stochastic3}
\eea
is sufficient to give an unbiased estimate of the form
\bea
\sum_{\mbf{x}}
  \langle \text{Tr}\{\sum_j^{N_\eta} \phi^{(j)}_{(\vec{x},t);t_0} 
  \phi^{(j)\dagger}_{(\vec{x},t);t_0}\}\rangle_U 
&\approx& 
\frac{1}{V}\sum_{\mbf{x},\mbf{y}} 
  \langle \text{Tr}\{ 
  M^{-1}_{(\mbf{x},t);(\mbf{y},t_0)}
  (M^{-1}_{(\mbf{x},t);(\mbf{y},t_0)})^\dagger 
  \}\rangle_U
 = C^\pi_{\text{2pt}}(t,\mbf{P})
  \,,
  \label{C2ptPropsOneEnd}
\eea
with a stochastic error of $\mathcal{O}(N_\eta^{-1/2})$. 
Compared to Eq.~\ref{C2ptProps}, one finds that the estimate with
stochastic sources leads to an additional volume average, $V^{-1}\sum_{\mbf{y}}$,
and therefore potentially a dramatic increase in statistics. 
The reason is that the summation over the source position $\mbf{y}$
is already implicit in the stochastic average. 
This has been numerically studied and also extended to other mesons and 
for the calculation of three- and four-point
functions in hadron structure calculations
\cite{Simula:2007fa,Boyle:2008rh,Alexandrou:2008ru,Boyle:2008yd},
where in many cases a substantial increase in statistics 
was found at fixed cost for the pion using stochastic sources 
compared to conventional point sources.
Some recent applications will be discussed in more detail in sections 
\ref{pionFFs} and \ref{sec:deformations} below.
\subsubsection{Partially twisted boundary conditions}
\label{sec:ptbcs}
Many of the fundamental hadron structure observables that were introduced and discussed
in section \ref{sec:observables} above are defined at vanishing momentum transfer squared, 
$t=0$ or $q^2=-Q^2=0$. 
Typical examples are mean square radii, Eq.~(\ref{radii}), the anomalous
magnetic moment of the nucleon, $\kappa=F_2(Q^2=0)$, the quadrupole moment of the $\rho$-meson, 
$Q_\rho=G_Q(Q^2=0)/m_\rho^2$, and angular momentum contributions to the nucleon spin, Eq.~(\ref{J1}),
$J=(A_{20}(\t0)+B_{20}(\t0))/2$.
The numerical analysis of these quantities requires, however, small \emph{non-zero} values of the momentum transfer 
$q=\Delta=P'-P$: Derivatives in the definition of charge radii have to be approximated by finite differences,
and the other relevant form factors and generalized form factors are multiplied
by factors of $\Delta$ in the parametrization of the respective hadron matrix
elements, cf. Eqs.~(\ref{PionTensorFF},\ref{NuclVec3},\ref{NuclAxial3},\ref{NuclTensor3},\ref{RhoVec3},\ref{RhoAxial3},\ref{EMT1},...),
and therefore \emph{cannot} be extracted in the forward limit, $P'=P$.
As already noted in section \ref{sec:basicsLQCD},
periodic boundary conditions are conventionally used in spatial directions for the
discretized fields, leading to discrete momenta of the form $\mbf{P}=(2\pi/(aL))\mbf{n}$.
The analysis of the observables mentioned above is therefore challenging in current 
lattice simulations, where lattice spacings $a$ and spatial extents $L$ are such that the 
lowest non-zero momentum components are $P\gtrapprox 0.3\GeV$, resulting in comparatively large
non-zero momentum transfers of $|t|^{\not=0}\gtrsim0.15-0.4\GeV^2$.
Furthermore, because of the discrete lattice momenta, neighboring values of $t$ are 
often separated by large gaps of up to $\approx0.5\GeV^2$.
The interpolation between accessible values of the momentum transfer, and the extrapolation
to the forward limit therefore requires in general a parametrization of the
($t$-) $Q^2$-dependence of the (generalized) form factors based on a certain ansatz, e.g.,
a monopole or dipole, which introduces additional systematic uncertainties.
These difficulties may be avoided to some extent by 
using so-called partially twisted boundary conditions (pTBCs) 
\cite{Sachrajda:2004mi,Bedaque:2004ax} in spatial directions. 
The advantage of pTBCs over (fully) twisted BCs
is that they are only applied to the valence quark fields
in the quark propagators, while the sea quarks obey
the conventional periodic BCs. Hence existing dynamical 
gauge configurations generated using periodic BCs may be employed
for the calculation of correlators. The price one has to pay is 
the appearance of additional finite volume effects, see below.
%
%
%
\begin{figure}[t]
      \centering
          \includegraphics[width=0.30\textwidth]{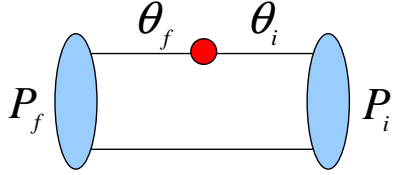}
  \caption{Illustration of pTBCs.}
  \label{ill_pTBCs_v3}
 \end{figure}
%
%
As illustrated in Fig.~\ref{ill_pTBCs_v3}, pTBCs are of the form 
\bea
q_{f,i}(\mbf{x}+a L\hat{\mbf{e}}_j)=e^{i(\boldsymbol{\theta}_{f,i})_j}q_{f,i}(\mbf{x})\,
\nonumber
\eea
with
twisting angles $(\boldsymbol{\theta}_{f,i})_j=0,\ldots,2\pi$, $j=1,2,3$, for the
`final' and and `initial' valence quark fields.
Denoting the conventional discrete (`Fourier') momenta 
by $\mbf{P}^{F}_{f,i}=(2\pi/(aL))\mbf{n}_{f,i}$, and 
setting $\mbf{P}^\twist_{f,i}=\mbf{P}^{F}_{f,i}+\boldsymbol{\theta}_{f,i}/(aL)$,  
the squared momentum transfer for twisted BCs is given by
\bea
t=\big(E(\mbf{P}^\twist_f)-E(\mbf{P}^\twist_i)\big)^2-\big(\mbf{P}^\twist_f-\mbf{P}^\twist_i\big)^2\,,
\nonumber
\eea
where as usual $E^{}(\mbf{P})^2=m^2+\mbf{P}^2$. 
Clearly, setting the conventional discrete (`Fourier') momenta 
equal to zero, $\mbf{n}_{f,i}=0$, and tuning the twisting angles 
$\boldsymbol{\theta}_{f,i}$, arbitrarily small values of the 
momenta and thereby the momentum transfer can be accessed.
Non-zero Fourier momenta and twisting angles can also be combined to reach larger fine-tuned
values of $t$ allowing to, e.g., fill the gaps between the $t$-values
that are accessible with periodic BCs.
Partially twisted BCs can be implemented in practice
by replacing the pTBC quark fields by new quark fields, $\tilde q(x)$,  which
obey periodic BCs, $q(x)=\exp(i\boldsymbol{\theta}\cdot\mbf{x}/L)\tilde q(x)$. Since this
can be seen as gauge transformation, in order to implement pTBCs, 
effectively only the link variables in the propagators have to be replaced, 
$U_{x,x+a\hat{\mbf{e}}_j}\rightarrow \exp(i\boldsymbol{\theta}_ja/L) U_{x,x+a\hat{\mbf{e}}_j}$ \cite{Flynn:2005in}.

Finite volume effects (FVEs) due to pTBCs were shown to be in general exponentially suppressed
in the volume \cite{Sachrajda:2004mi,Bedaque:2004ax}, i.e. $\propto\e^{-m_\pi L}$.
Detailed studies \cite{Jiang:2006gna,Jiang:2008te} in partially quenched chiral 
perturbation theory (pQChPT), see following section, showed that they can indeed expected to be small in
current lattice calculations of the pion form factor, and that 
isospin symmetry breaking effects in particular can be avoided by working in the Breit frame, 
i.e. choosing $\boldsymbol{\theta}_f=-\boldsymbol{\theta}_i$.
Similar  investigations in the nucleon sector showed that the
FVEs from pTBCs are small, of the order of a few percent,
for the isovector electric form factor \cite{Tiburzi:2006px}, 
but can be as large as $\approx20\%$ in the case of the magnetic
or Pauli form factor \cite{Tiburzi:2006px,Jiang:2008ja}, 
for small twisting angles and typical parameters of current lattice simulations.
This has to be kept in mind regarding the preliminary lattice results for
$F_2^{u-d}$ obtained with pTBCs that will be discussed below in section \ref{sec:NuclPTBCs}.
Some applications of pTBC in meson structure calculations
will be presented in sections \ref{pionFFs}.
%

%
%
%
%
%

%
\subsection{Chiral effective field theory and chiral perturbation theory}
\label{sec:ChEFT}

Due to the substantial progress in lattice QCD simulations over the last couple of years,
the lowest pion masses that can be reached in hadron structure investigations 
today are $\sim300$ MeV in physical volumes of $\sim(3 \mbox{fm})^3$, with lattice spacings $\sim0.1$ fm.  
Clearly, to exploit the lattice results as much as possible, extrapolations to the infinite volume limit,
the continuum limit, and to the physical pion mass are necessary.
In order not to spoil the ``ab-initio''-character of lattice QCD calculations, 
such extrapolations must be addressed in a systematic way. 
This is in general possible using the methods of low energy effective field theories (EFT),
based on a systematic expansion in powers $p/\Lambda$, where $p$ generically denotes a 
small mass, energy or momentum, and where $\Lambda$ is a ``large'' energy scale. In massless QCD,
the chiral $SU(3)_L\times SU(3)_R$ (where $SU(3)_{L/R}$ acts on the left and right
handed $u$, $d$, and $s$ quark fields, respectively) symmetry is spontaneously broken
at the fundamental chiral symmetry breaking scale $\Lambda_{\chi}$,
giving rise to eight Goldstone bosons. 
The Goldstone bosons can be identified with the pions, kaons and the eta, and
to leading order in $p/\Lambda$, the low energy EFT of QCD can be uniquely
formulated in terms of these degrees of freedom.
One often identifies $\Lambda^2_\chi=(4\pi f^0_\pi)^2\approx 1\GeV^2$, where
$f^0_\pi\simeq0.09\GeV$ is the pion decay constant in the chiral limit.
An alternative is to set the chiral symmetry breaking scale equal to the mass of the
lowest lying non-Goldstone meson, i.e. the $\rho$-meson.
The low energy EFT of QCD is, strictly speaking, non-renormalizable, and an increasing number
of couplings or low energy constants (LECs), in addition to $f^0_\pi$, are required at higher orders to
permit a systematic order-by-order renormalization. The a priori unknown LECs thereby parametrize 
part of the physics of the underlying full theory that is not directly accessible within the low energy EFT.
The predictive power of the EFT-approach is due the fact that the LECs 
are universal and process independent. 

\begin{figure}[t]
  \begin{center}
    \includegraphics[width=0.8\textwidth]{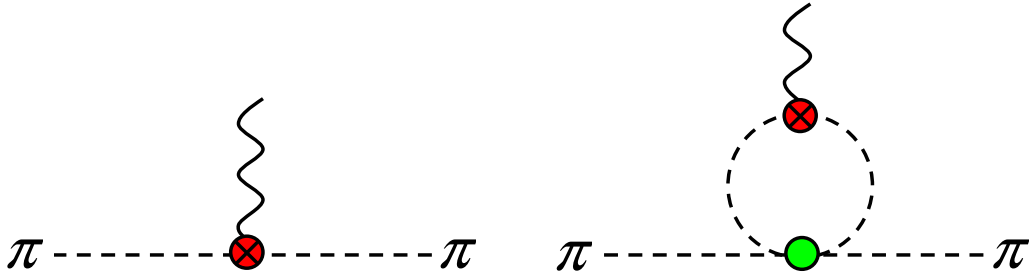}
  \end{center}
  \caption{Illustration of contributions to the pion form factor in ChPT. The coupling to the external vector 
  current is denoted by the crossed circle.}
  \label{ill_ChPT_v1}
 \end{figure}

Terms providing an explicit chiral symmetry breaking 
at non-zero quark masses can be included in the EFT in a perturbative way, giving rise to 
non-vanishing pseudoscalar meson masses. The meson masses are 
related to the quark masses in a non-trivial manner, described by the
Gell-Mann-Oakes-Renner (GMOR) relation at leading order,
\bea
m_\pi^2=-\frac{1}{(f^0_\pi)^2}\langle \overline\psi \psi\rangle \overline m + \mathcal{O}(p/\Lambda_\chi)\, 
  \label{GMOR}
\eea
where $\langle\overline\psi \psi\rangle=
\langle(\overline u u + \overline d d)\rangle$ is the quark condensate, and 
$\overline m=(m_u+m_d)/2$.  
Similar relations hold for the kaons and the $\eta$-meson. 
A power counting scheme has been developed that allows for a systematic expansion in $p/\Lambda_\chi$ to all orders, 
where pion loop diagrams are suppressed by $(p/\Lambda_\chi)^2$ relative to the tree level contributions
\cite{Gasser:1983yg}. 
Such a perturbative expansion is called chiral perturbation theory (ChPT) and predicts in particular  
the form of the quark mass dependence (or pion mass dependence using Eq.(\ref{GMOR}))
of, e.g., pseudoscalar meson masses, decay constants, form factors, etc.
Apart from the analytic quark mass dependence in terms of powers of $m_\pi^2$, 
pion loop diagrams in ChPT lead to the famous (``non-analytic") chiral logarithms, $m_\pi^2/(4\pi f^0_\pi)^2 \ln(m_\pi^2/\lambda^2)$.
Here, we have included a factor $(4\pi)^2$ that is characteristic for the loop integration, and $\lambda$ is the renormalization scale.
An illustration of contributions to the pion form factor 
at leading order ($\mathcal{O}(p^2)$) and 1-loop ($\mathcal{O}(p^4)$) level 
is given in Fig.~\ref{ill_ChPT_v1}.
References to meson structure observables that have been studied in ChPT will be given at the
beginnings of sections \ref{sec:FFs} and \ref{sec:PDFs}.

The inclusion of baryons DOFs into the low energy EFT is formally straightforward, 
but leads to conflicts within the simple power counting scheme used in the pure mesonic sector, 
since baryon momenta in the full relativistic theory 
are generically large, $p_B\simeq m_B\simeq \Lambda_\chi$. 
The power counting scheme can be saved by treating the baryons 
as heavy particles and employing a non-relativistic expansion in $p/m_B$, 
where $p=p_B,p_\pi,m_\pi,\ldots$ \cite{Jenkins:1990jv}.
The result of such a procedure, which can formally be carried out to any order, is called heavy baryon ChPT (HBChPT).
For an early review we refer to \cite{Bernard:1995dp}.
In HBChPT, higher resonances are usually integrated out and consistently included in 
an implicit way in the form of counter-term contributions. 
It has however been noted early on that the $\Delta(1232)$ resonance is special
in at least two ways. First, the nucleon-$\Delta$ mass splitting is less than two times the pion mass
and therefore does not qualify easily as a large scale, and at the same time
the pion-nucleon-$\Delta$ coupling constant is phenomenologically large.
Based on these observations, a formalism called small scale expansion (SSE) 
was developed to consistently include explicit $\Delta(1232)$ degrees of freedom into the HBChPT formalism
\cite{Hemmert:1996xg}.
Within the SSE, the mass splitting $\myDelta m_{N\Delta}\approx 0.27\text{ GeV}$ is treated on the same level as the
pion mass, momentum, etc. as small parameter $\eps=\myDelta m_{N\Delta},p_\pi,m_\pi,\ldots$
We note that a number of nucleon structure calculations have been performed in
a somewhat modified version of the HBChPT-SSE developed in \cite{Hemmert:2002uh}
and further pursued in, e.g., studies of the nucleon mass and of $g_A$
in \cite{Procura:2006bj} and \cite{Procura:2006gq}, respectively.
Another method including explicit $\Delta$-DOFs, called $\delta$-expansion scheme,
where the small expansion parameter is
$\delta=\Delta / \Lambda_{\chi} \sim m_\pi / \Delta$, implying that $\delta^2\sim m_\pi$,
has been developed in \cite{Pascalutsa:2002pi} and, e.g., employed in a relativistic (see below) ChPT-study of the
magnetic moment of the $\Delta$-baryon \cite{Pascalutsa:2004je}.

For many years, due to the lack of a consistent relativistic formulation, hadron structure observables have been 
mostly investigated in the HBChPT (and the SSE, $\delta$-expansion) formalism. 
This is slowly changing since Becher and Leutwyler 
\cite{Becher:1999he} developed a consistent power counting scheme including baryons based 
on the so-called infrared regularization (IR) of loop integrals. 
This has been employed for example already in \cite{Kubis:2000zd} in a slightly modified form for the analysis
of the nucleon electromagnetic form factors.
Since then, further covariant renormalization schemes like the extended on-mass shell (EOMS) \cite{Schindler:2003xv} and
the modified IR ($\overline{\text{IR}}$) \cite{Dorati:2007bk} schemes have been developed, 
so that today solid alternatives to the HBChPT approach are available. 
An approach alternative to the usual dimensional regularization of loop integrals is based on a finite-range regularization 
(FRR) \cite{Donoghue:1998bs,Young:2002ib,Leinweber:2003dg}.

To this day, a large number of baryon structure observables have been studied in covariant and heavy BChPT, 
for example
nucleon form factors, including charge radii, magnetic moments and the axial vector coupling constant, 
moments of polarized and unpolarized nucleon PDFs and GPDs, 
including in particular the form factors of the QCD energy momentum tensor 
nucleon polarizabilities,  
etc.
For a recent review of the investigation of baryons in ChPT we refer to \cite{Bernard:2007zu}.

Depending on the order of the chiral expansion, the pion mass dependence of a given hadronic matrix element
depends in general on several a priori unknown LECs. One can envisage at least two different scenarios: \\
\begin{itemize}
\item In the rare case that all relevant LECs can be determined from experiment and phenomenology,
ChPT gives a parameter-free prediction of the pion mass dependence of the observable under consideration
to a given order, which can be directly compared to results from lattice QCD.
\item In the general case, the numerical values of (some of) the LECs are unknown. The unknown LECs 
may be then be obtained from a fit of the full ChPT result to lattice simulation data at different
$m_\pi$ and results from experiment at the physical pion mass (if available). Alternatively, one may want to
fit to lattice data points alone in order to be able to `predict' (a priori or a posteriori)
the value at the physical point.
\end{itemize}

In either case, it is of crucial importance to study the convergence of the ChPT series,
and to avoid stretching the chiral extrapolation beyond its region of applicability.
These are difficult tasks in praxis since most of the lattice hadron structure
results are available only for pion masses larger than $300$ MeV. The region of applicability 
of ChPT is being studied intensively and will certainly depend to some extent on
the calculational scheme, e.g. HBChPT compared to covariant BChPT, as well as the observables under consideration.
On the lattice side, one has to keep in mind that in most cases finite volume and discretization
effects are not perfectly under control, so that an extrapolation in the pion mass alone based on 
infinite volume, continuum ChPT is subject to systematic uncertainties of largely unknown size.
At this point, it should be stressed that ChPT not only predicts the form of the pion mass dependence of
hadronic observables, but can also be extended to include finite volume effects 
as well as corrections at non-zero lattice spacing (discretization effects).
Chiral perturbation theory calculations in a finite volume have already been carried out 
for different hadron structure observables.
An interesting example is the axial-vector coupling constant $g_A$ that was studied
in HBChPT in finite volume in \cite{Beane:2004rf,Khan:2006de,Procura:2006gq}.
Discretization effects for different (mixed) lattice actions 
have been investigated in particular for hadron masses and decay constants
(see, e.g., \cite{Bar:2003mh,Bar:2005tu}) 
and also for nucleon magnetic moments \cite{Beane:2003xv}.

Chiral perturbation theory calculations at higher orders, including finite volume and
discretization effects, are certainly desirable. However, one has to keep in mind
that the number of unknown LECs increases with the order of the ChPT calculation.

References to results from ChPT for the structure of mesons and baryons in terms
of form factors and moments of PDFs and GPDs will be given 
at the beginnings of the corresponding sections below.

%% file: FFs.tex
\section{Lattice results on form factors}
\label{sec:FFs}
\subsection{Overview of lattice results}
Here we briefly summarize the current standing of each of the topics that will be
discussed in more detail later in this section.

First lattice calculations in the quenched approximation of the 
pion electromagnetic form factor, $F_\pi(Q^2)$, have been published
already in the late 1980's \cite{Martinelli:1987bh,Draper:1988bp}. 
Since then, further quenched and unquenched lattice results for $F_\pi(Q^2)$ have been been presented
\cite{vanderHeide:2003kh,Bonnet:2004fr,Hashimoto:2005am},
and an extensive study in unquenched lattice
QCD has more recently been published in \cite{Brommel:2006ww}. 
Up-to-date calculations in unquenched lattice QCD
employing for example partially twisted boundary 
conditions (pTBCs) to access $F_\pi(Q^2)$
at very low non-zero $Q^2$ and 
all-to-all propagators (see end of previous section \ref{sec:methods})
can be found in \cite{Simula:2007fa,Boyle:2007wg,Boyle:2008yd,JLQCD:2008kx,Frezzotti:2008dr}.
First results for the tensor form factor of the pion, $B^\pi_{T10}$, have been published in 
\cite{Brommel:2007xd}.

A first lattice calculation of the electric form factor of the proton in quenched QCD 
was published in 1989 \cite{Martinelli:1988rr}. 
Since then, several quenched lattice
QCD studies of the nucleon electromagnetic form factors,
including charge radii and magnetic moments, appeared 
\cite{Draper:1989pi,Leinweber:1990dv,Wilcox:1991cq,Gockeler:2003ay,Boinepalli:2006xd}. 
Interestingly, to this date, there is only a small number of peer-reviewed publications 
on the electromagnetic form factors of the nucleon in unquenched lattice QCD available 
\cite{Alexandrou:2006ru,Lin:2008uz}. 
A larger number of unquenched results 
can be found in proceedings 
\cite{Negele:2004iu,Renner:2004ck,Edwards:2005kw,Edwards:2006qx,Gockeler:2006uu,Gockeler:2007hj,Gockeler:2007ir,Ohta:2008kd,Bratt:2008uf}. 
For an overview and a discussion of nucleon electromagnetic form factors in lattice QCD 
we also refer to \cite{Arrington:2006zm}.

Early studies of the axial vector form factors, $G_A(Q^2)$ and $G_P(Q^2)$, 
of the nucleon in quenched QCD have been presented in 
\cite{Liu:1992ab,Liu:1994dr}, and more recent results in unquenched lattice QCD can be found in \cite{Alexandrou:2007zz,Lin:2008uz}.
The nucleon isovector axial vector coupling constant $g_A=G_A(Q^2=0)$ was studied quite intensively in recent years.
A detailed quenched study based on chiral (domain wall) fermions
can be found in \cite{Sasaki:2003jh}, and dynamical lattice results have been published 
in \cite{Khan:2006de,Edwards:2005ym,Yamazaki:2008py}.
Unquenched calculations of the isosinglet axial vector coupling constant 
have been presented in \cite{Fukugita:1994fh,Dong:1995rx,Gusken:1999as}, 
including estimates of contributions from disconnected diagrams. 

The nucleon tensor charge, $g_T$, has been investigated in quenched
QCD for the first time in \cite{Aoki:1996pi} and later using domain wall fermions
in \cite{Orginos:2005uy}. 
More recent results for $g_T$ in full QCD can be found in 
\cite{Dolgov:2002zm,Gockeler:2005cj,Edwards:2006qx,Lin:2008uz}.
First unquenched lattice QCD calculations of the nucleon tensor form factors $g_T(Q^2=-t)=A_{T10}(t)$, 
$B_{T10}(t)$ and $\tilde A_{T10}(t)$ have been presented in \cite{Gockeler:2005cj,Gockeler:2006zu}.

A first direct lattice calculation of $\rho$-meson vector form factors, 
$G_C(Q^2)$, $G_M(Q^2)$ and $G_Q(Q^2)$, in quenched QCD 
was published recently \cite{Hedditch:2007ex} (restricted to $Q^2=0$ and a single non-zero 
momentum transfer, $Q^2\simeq0.1$ GeV$^2$). 
A short while ago, preliminary results for $G_C(Q^2)$, $G_M(Q^2)$ and $G_Q(Q^2)$
for a range of values of $Q^2$ were obtained in full lattice QCD \cite{Gurtler2008}. 
Related results based on density-density correlators in quenched and unquenched lattice QCD
can be found in \cite{Alexandrou:2003qt,Alexandrou:2007pn,Alexandrou:2008ru}.

The electromagnetic form factors of the decuplet baryons, including their quadrupole and octupole
moments, have been studied already some time ago in quenched lattice QCD 
based on results for $Q^2=0$ and a single non-zero momentum transfer $Q^2\sim0.1\GeV^2$ 
\cite{Leinweber:1992hy} and more recently for $Q^2\sim0.23\GeV^2$ in \cite{Boinepalli:2009sq}.
A more detailed analysis of the $Q^2$-dependence of the electromagnetic form factors of the $\Delta$-baryon
in quenched and unquenched lattice QCD has been performed recently \cite{Alexandrou:2008bn}.

Selected results from mostly from unquenched (including hybrid) lattice QCD simulations will 
be discussed in some detail below.

\label{sec:FFsOverview}
\subsection{Results from chiral perturbation theory}
\label{sec:FFsChPT}
\subsubsection*{Pion}
The $m_\pi$- and $Q^2$-dependence of the 
pion vector form factor in infinite volume has already been calculated 25 
years ago at 1-loop level
in ChPT in the seminal paper by Gasser and Leutwyler \cite{Gasser:1983yg}
(see also \cite{Gasser:1984ux}).
A finite volume study of $F_\pi(Q^2)$ to $\mathcal{O}(p^4)$ 
using lattice regularized ChPT was carried out in 
\cite{Borasoy:2004zf}.
Results for the pion form factor in the framework of partially-quenched
ChPT (PQChPT) at one-loop order in a finite volume, including the lattice spacing dependence
for a mixed-action theory of Ginsparg-Wilson valence quarks and staggered sea quarks
were obtained in \cite{Bunton:2006va}.
By now, many observables in the mesonic sector have also been studied at 
NNLO ($\mathcal{O}(p^6)$), including 2-loop diagrams.
An overview can be found in \cite{Bijnens:2006zp}.
Particularly relevant for the structure of the pion is the pion mass dependence 
of $F_\pi(Q^2)$ at 2-loops in infinite volume ChPT \cite{Gasser:1990bv,Bijnens:1998fm}.
First results for the pion mass and $Q^2$-dependence of the 
tensor (chiral-odd ) form factor of the pion, $B^\pi_{T10}(Q^2)$,
were obtained in \cite{Diehl:2006js} at leading 1-loop level in ChPT.
Results for the pion vector and tensor form factor in a finite volume in 
the framework of PQChPT became available soon thereafter \cite{Chen:2006gg}.
Furthermore, due to the reduced space-time symmetries on the lattice, the off-forward pion matrix 
elements of the vector current and the tensor operator receive finite volume corrections 
in form of additional kinematical structures and calculable ``form factors'',
which have been worked out in \cite{Manashov:2007qr}.
Finite volume effects introduced by partially twisted boundary conditions  
are in general exponentially suppressed \cite{Bedaque:2004ax,Sachrajda:2004mi}.
Although probably small in current dynamical calculations of the pion form factor, 
they may be non-negligible in future lattice calculations at 
pion masses $m_\pi<300\MeV$ and for not too large volumes, 
as has been worked out in PQChPT to NLO in \cite{Jiang:2006gna}
for rest frame kinematics, where both isospin and hypercubic symmetry breaking effects
are present. More recently, these calculations
have been revisited in the Breit frame, where isospin and a discrete rotational 
symmetry are preserved \cite{Jiang:2008te}.

\subsubsection*{Nucleon}
As explained in section \ref{sec:ChEFT}, a direct implementation of baryons, in particular the nucleon,
into the relativistic EFT framework 
leads to a conflict with the standard
ChPT-expansion due to lack of a consistent power counting scheme. This has been overcome
in the heavy baryon ChPT (HBChPT) formalism, which has been used in practice from the early 1990's on.
The isovector Dirac and Pauli form factors of the nucleon were studied 
at 1-loop level ($\mathcal{O}(p^4)$) in SU(2) HBChPT in 
\cite{Bernard:1992qa},
and further activities in this framework during the early years have been reviewed in 
\cite{Bernard:1995dp},
including results for the nucleon axial-vector form factor $G_A(Q^2)$ and the 
pseudo-scalar form factor $G_P(Q^2)$.
Systematically taking into account explicit $\Delta$-baryon degrees-of-freedom in the chiral expansion,
the isovector nucleon form factors $G_{E,M}(Q^2)$ and $F_{1,2}(Q^2)$ were revisited 
in the framework of the small scale expansion (SSE) (see section \ref{sec:ChEFT}) in
\cite{Bernard:1998gv}. 
Based on a modified chiral power counting of the isovector 
$\Delta$-nucleon transition coupling constant $c_V$,
a calculation of the nucleon isovector anomalous
magnetic moment $\kappa$ to NLO ($\mathcal{O}(\epsilon^3)$) 
in the framework of a revised version of the SSE HBChPT formalism was presented in \cite{Hemmert:2002uh}. 
These calculations were extended later \cite{Gockeler:2003ay} to include the full $Q^2$-dependence
of the isovector nucleon form factors, $F_{1,2}(Q^2)$, at NLO in SSE HBChPT.
Employing a consistent power counting scheme based on infrared-regularization ($IR$) 
\cite{Becher:1999he}, the nucleon vector form
factors have been investigated in relativistic baryon ChPT (BChPT) in \cite{Kubis:2000zd}, 
where explicit results for the pion mass dependence of the isovector and isoscalar anomalous magnetic moments
and the charge radii of $G_{E,M}(Q^2)$ can be found.
Nucleon electromagnetic form factors were also studied in relativistic BChPT in \cite{Fuchs:2003ir} employing 
the extended on-mass-shell renormalization scheme.
In \cite{Wang:2007iw} and \cite{Wang:2008vb}, chiral extrapolations of the nucleon magnetic form factors 
and octet-baryon charge radii, respectively, have been studied using finite-range regularization.

A detailed study of the nucleon axial vector coupling constant $g_A$ in the SSE approach
to $\mathcal{O}(\epsilon^3)$ has been presented in \cite{Hemmert:2003cb}.
The lattice spacing dependence at $\mathcal{O}(a)$ of $g_A$,
for Wilson fermions and a mixed formulation of
Wilson sea and Ginsparg-Wilson valence fermions, has been studied in the framework 
of PQChPT including explicit $\Delta$-DOFs in \cite{Beane:2003xv}.
In a work by the same authors \cite{Beane:2004rf}, 
the leading finite volume corrections to $g_A$ have been obtained
in HBChPT including explicit $\Delta$-DOFs, in a framework that
differs somewhat from the SSE formalism mentioned above. 
Subsequently, the SSE calculation of \cite{Hemmert:2003cb} was extended to include 
finite volume corrections and applied to unquenched 
lattice QCD results in \cite{Khan:2006de}. 
In addition, a matching and comparison of the employed
SSE-LECs to the LECs used in \cite{Beane:2004rf} was attempted.
Results for the axial vector and pseudoscalar form factors, including the
axial vector coupling constant, have also been obtained 
in relativistic BChPT to $\mathcal{O}(p^4)$ \cite{Schindler:2006it}. Due to
the reformulated IR-scheme employed in this work, it was also possible to
include the axial vector meson $a_1$ explicitly as an additional DOF
in the chiral expansion.
The specific form of $g_A$ at 2-loop level in HBChPT has been worked out in \cite{Bernard:2006te},
where also the coefficient of the double logarithmic term, as well as 
several other numerically important contributions, were obtained.
More recently, lattice spacing artifacts for the nucleon matrix element
of the axial vector current have been investigated to $\mathcal{O}(a^2)$, for 
a mixed action approach of Ginsparg-Wilson and staggered sea quarks, in 
PQChPT with explicit $\Delta$ DOFs \cite{Jiang:2007sn}.

The leading chiral logarithmic contributions in HBChPT to 
the nucleon tensor charge can be be found in \cite{Arndt:2001ye,Chen:2001eg},
and in \cite{Diehl:2006js}, the full set of nucleon tensor form factors have been studied 
in HBChPT at leading 1-loop level. 
A broad study of baryon matrix elements of twist-2 operators 
including vector, axial vector and tensor currents
in the framework of (partially quenched) HBChPT
at finite and infinite volume has been presented in \cite{Detmold:2005pt}.
\subsubsection*{Delta}
The leading non-analytic contributions in HBChPT to the 
magnetic and quadrupole moments of the decuplet baryons, including the $\Delta$ resonance, 
were studied for the first time in \cite{Butler:1993ej}, however no explicit expressions
for the pion mass dependences of $\mu_\Delta$ and $Q_\Delta$ are given in this work.
More explicit results are given in \cite{Banerjee:1995wz}, where 
the HBChPT calculation of the decuplet magnetic moments has
been revisited, differing from \cite{Butler:1993ej} by the inclusion of the Roper
resonance as intermediate state, and in the treatment of $SU(3)$ flavor symmetry.
On the basis of the $\delta$-expansion scheme detailed in \cite{Pascalutsa:2002pi}, 
the $m_\pi$-dependence of the magnetic moment of the $\Delta$-baryon
has been calculated in a covariant way to NLO \cite{Pascalutsa:2004je}.
A study of the magnetic moment of the $\Delta$-baryon using the extended 
on-mass-shell renormalization scheme in relativistic BChPT has been presented
in \cite{Hacker:2006gu}.

Recently, the pion mass dependence of the 
axial vector coupling constant of the $\Delta$-baryon has 
been studied in full and partially quenched HBChPT to NLO \cite{Jiang:2008we}, where also
isospin relations for the underlying $\Delta$-matrix elements of the axial vector current
were derived, which could serve as a useful guide for future lattice calculations of this observable.

Charge radii, magnetic and quadrupole moments of decuplet baryons, with particular emphasis
on a separation of connected and disconnected contributions, were investigated recently in \cite{Tiburzi:2009yd} 
using partially quenched ChPT.

\subsubsection*{Polarizabilities}
The Compton scattering amplitude for the nucleon, parametrized by the nucleon 
electric, magnetic and spin polarizabilities,
has been studied at length over the years in ChPT. 
For an overview, we refer to the review \cite{Bernard:2007zu} 
and the extensive studies in \cite{Hildebrandt:2003fm,Beane:2004ra,Detmold:2006vu}.
The spin-independent nucleon polarizabilities, cf. section \ref{sec:polarizabilities}, Eq.~(\ref{Compton1}),
were studied first in \cite{Bernard:1991rq,Bernard:1991ru} in HBChPT at leading 1-loop level (to $\mathcal{O}(q^3)$).
In \cite{Bernard:1993bg}, this calculation was extended to $\mathcal{O}(q^4)$.
Contributions from kaons and baryon resonances as intermediate 
states were studied in \cite{Butler:1992ci} at leading order in the heavy baryon limit.
A calculation of the polarizabilities in the SSE scheme to $\mathcal{O}(\eps^3)$ including explicit 
$\Delta$-baryon intermediate states  was presented in \cite{Hemmert:1996rw}, and a
similar study in the $\delta$-expansion scheme, also including explicit $\Delta$-DOFs, 
can be found in \cite{Pascalutsa:2002pi}.
More recently, the volume dependence of the polarizabilities has been studied in 
the framework of HBChPT \cite{Detmold:2006vu}, where also results for (partially) 
quenched QCD are provided.

\subsection{Pion form factors}
\label{pionFFs}
\begin{figure}[t]
   \begin{minipage}{0.48\textwidth}
      \centering
          \includegraphics[angle=-90,width=0.98\textwidth,clip=true]{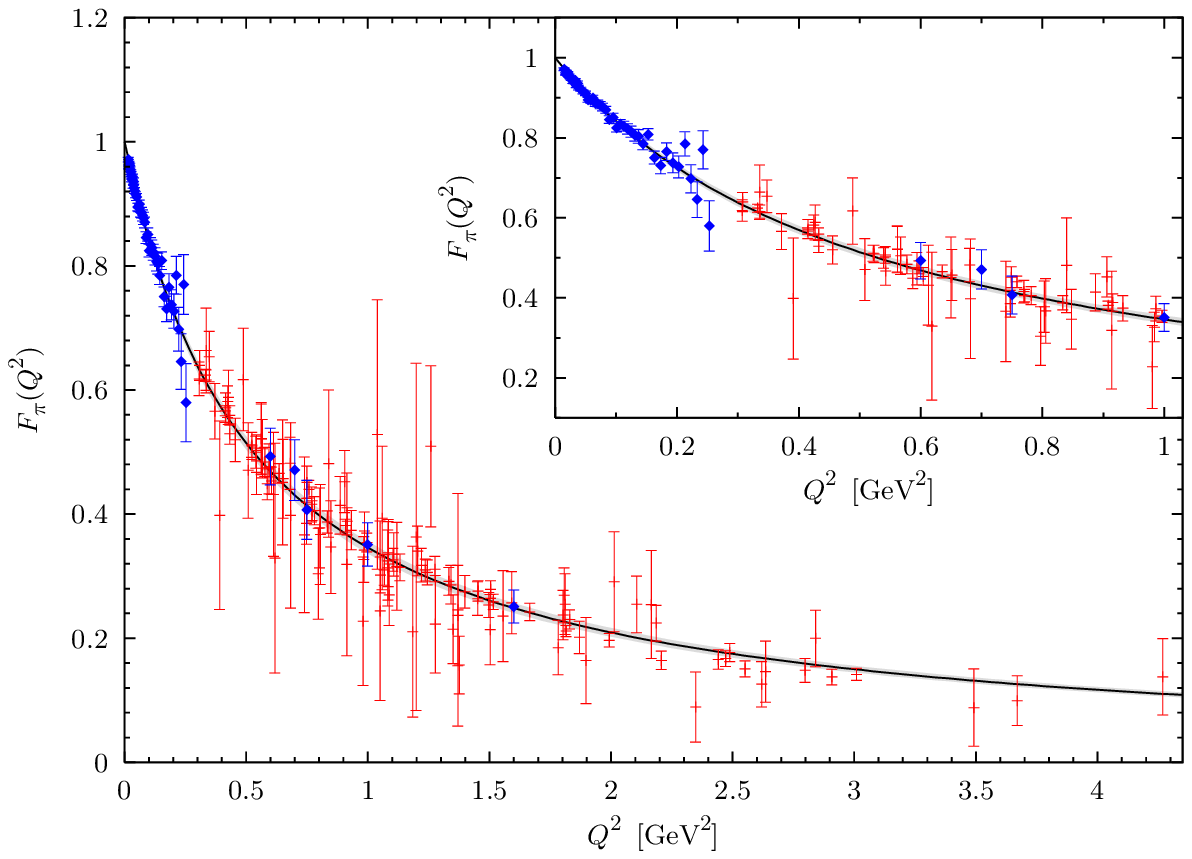}
  \caption{Pion electromagnetic form factor $F_\pi$ in lattice QCD (horizontal dashes)
   and experiment (diamonds) (from \cite{Brommel:2006ww}).}
  \label{FFpion_global_QCDSF}
     \end{minipage}
     \hspace{0.5cm}
    \begin{minipage}{0.48\textwidth}
      \centering
          \includegraphics[angle=-90,width=0.98\textwidth,clip=true]{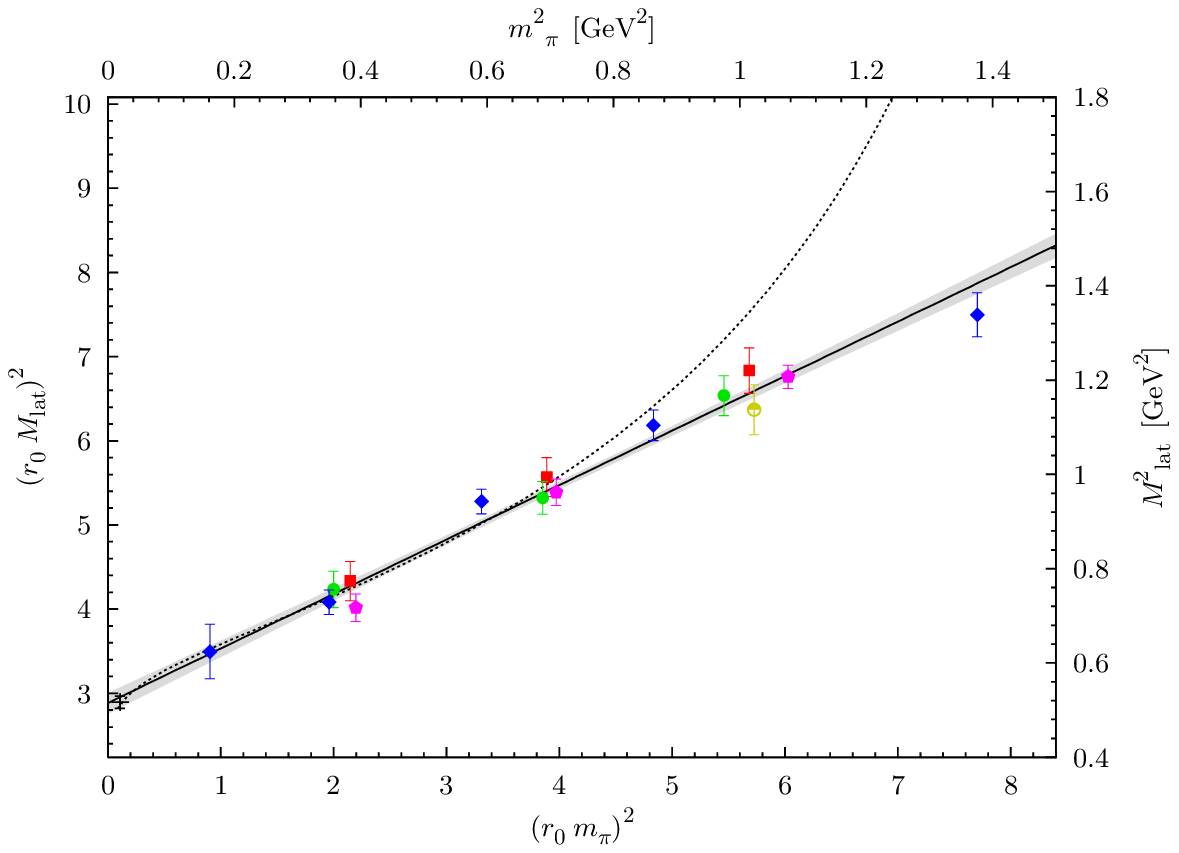}
  \caption{Chiral extrapolation of the monopole mass (from \cite{Brommel:2006ww}).}
  \label{FFpion_extrapol_QCDSF}
     \end{minipage}
 \end{figure}

Instead of going in detail through the history of lattice QCD calculations
of the pion electromagnetic form factor, we will concentrate here on recent 
results for $F_\pi$ obtained in unquenched lattice simulations. 
To begin with, we show in Fig.~\ref{FFpion_global_QCDSF} an overview plot of $F_\pi(Q^2)$ as
a function of $Q^2$, including experimental data points and (chirally extrapolated, see below) lattice
QCD results from QCDSF-UKQCD for $n_f=2$ flavors of
improved Wilson fermions and Wilson glue \cite{Brommel:2006ww}.
The large number of lattice data points have been obtained on the basis
of 15 ensembles with pion masses ranging from $400$ to $1200\text{ MeV}$
and for lattice spacings in the range $a\sim0.07,\ldots,0.12\fm$.
Since the local (non-conserved) vector current has been used, the bare lattice
data has been renormalized to ensure charge conservation, i.e. such that $F_\pi(Q^2=0)=1$. 
A monopole ansatz
\bea
\label{globalmonopole}
F_\pi(Q^2,m_\pi^2)=\frac{F_\pi(0)}{1+\frac{Q^2}{m_{\text{mono}}^2(m_\pi^2)}}\,,
\eea
with $F_\pi(0)=1$ and $m_{\text{mono}}^2(m_\pi^2)=c_0+c_1 m_\pi^2$ 
was fitted to the lattice data to simultaneously parametrize the $Q^2$- and $m_\pi^2$-dependence. 
On the basis of this fit, the lattice results have been shifted
to the physical point by subtracting the difference 
$F_\pi(Q^2,(m^{\lat}_\pi)^2)-F_\pi(Q^2,(m^{\phys}_\pi)^2)$ from the individual data points.
The shifted data points are shown in Fig.~\ref{FFpion_global_QCDSF}, and turn out to be
in very good agreement with the experimental results over the full range
of $Q^2=0,\ldots,4.4\GeV^2$, and in particular at small $Q^2$ as displayed in the inset
in the upper right corner of Fig.~\ref{FFpion_global_QCDSF}.
Notwithstanding the overall successful description of the lattice results based on 
Eq.~\ref{globalmonopole}, it is important to explicitly study the pion mass dependence 
of the monopole masses, as obtained from separate monopole fits at fixed $m_\pi$.
This can be done on the basis of Fig.~\ref{FFpion_extrapol_QCDSF}, displaying
$m_{\text{mono}}^2$ as a function of $m_\pi^2$. Comparing different 
ans\"atze for the pion mass dependence of $m_{\text{mono}}$, it was shown
that the form $m_{\text{mono}}^2(m_\pi^2)=c_0+c_1 m_\pi^2$
that has been used in the global fit above provides the best description 
of the lattice data in terms of $\chi^2/\text{DOF}$.
The resulting chiral extrapolation, represented by the shaded band in 
Fig~\ref{FFpion_extrapol_QCDSF}, gives a value of 
$m_{\text{mono}}=0.727\pm0.016_\text{stat}+0.024_\text{vol}\pm0.046_\text{sys}\GeV$
at the physical pion mass, where also estimates of uncertainties due
to finite volume effects (vol) and the fit- and chiral-extrapolation
ansatz (sys) have been included. This corresponds to a mean square charge radius of 
$\langle r_\pi^2\rangle=6/m_{\text{mono}}^2=0.441\pm0.019_\text{stat}-0.029_\text{vol}\pm0.056_\text{sys}\fm^2$, 
which is in good agreement with 
the experimental value of $\langle r_\pi^2\rangle=0.451(11)\fm^2$ from the PDG \cite{PDG2008}.
Since most of the lattice data points in Fig.~\ref{FFpion_extrapol_QCDSF} 
correspond to rather large pion masses of $m_\pi>590\MeV$, 
and since the only data point with $m_\pi=400\MeV$
has little statistical weight, a theory-based chiral extrapolation
employing results from ChPT discussed in the previous section has not been attempted in this case.
%

\begin{figure}[t]
    \begin{minipage}{0.48\textwidth}
      \centering
\psfrag{xlabel}[c][bc][1][0]{\Huge $Q^2[\GeV^2]$}
\psfrag{ylabel}[c][t][1][0]{\Huge $F_{\pi}(Q^2)$}
\psfrag{exp}[l][lc][1][0]{\LARGE experimental data NA7}
\psfrag{330MeV}[l][lc][1][0]{\LARGE lattice data for $m_\pi=330\MeV$}
\psfrag{NLO330MeV}[l][lc][1][0]{\LARGE $\text{SU}(2)$ NLO lattice-fit; $m_\pi=330\MeV$}
\psfrag{NLO139.57MeV}[l][lc][1][0]{\LARGE $\text{SU}$ NLO lattice-fit;
		$m_\pi=139.57\MeV$}
\psfrag{444OOOOOOOOOOOOOOOOOOOO}[l][lc][1][0]{\LARGE $1+\frac 16\langle r^2_\pi\rangle^{\rm PDG}Q^2$}
 \epsfig{angle=-90,width=0.95\textwidth,file=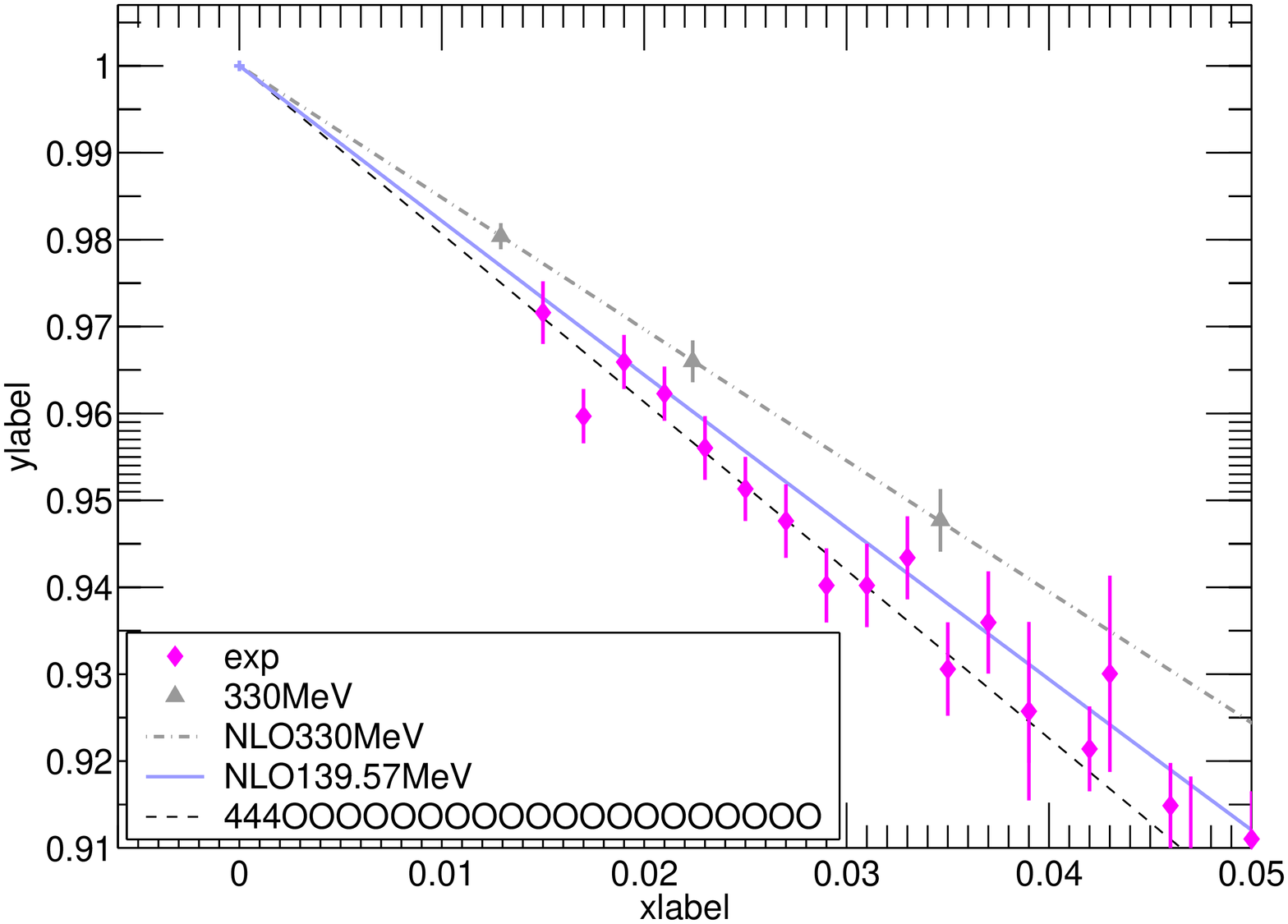}
          %
  \caption{Pion electromagnetic form factor at low $Q^2$ (from \cite{Boyle:2008yd}).}
  \label{FFpion_final_RBCUKQCD}
     \end{minipage}
          \hspace{0.5cm}
        \begin{minipage}{0.48\textwidth}
      \centering
          \includegraphics[angle=0,width=0.92\textwidth,clip=true]{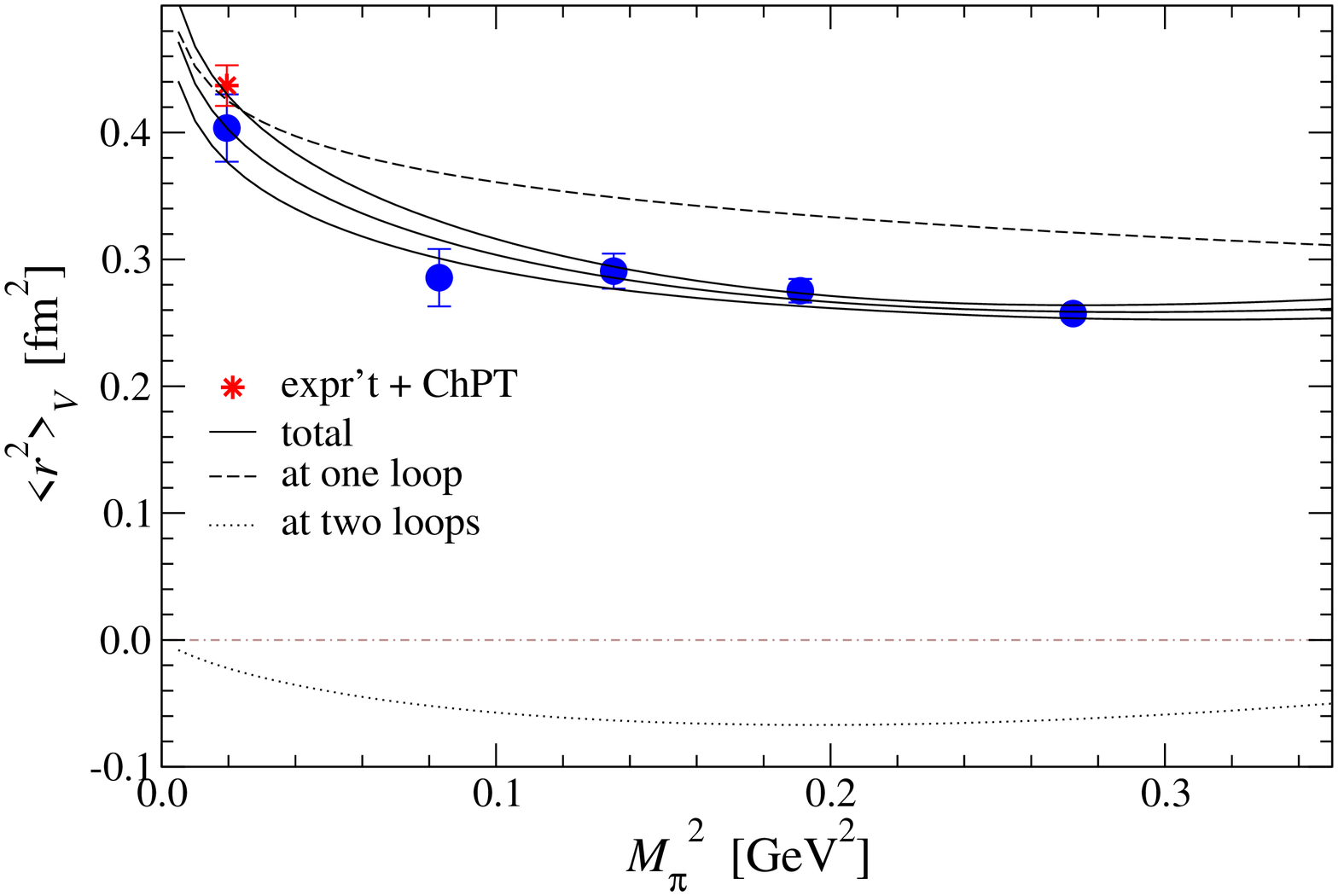}
  \caption{Pion charge radius in experiment and lattice QCD (from \cite{JLQCD:2008kx}).}
  \label{r2_JLQCD08_nnlo}
     \end{minipage}
 \end{figure}

Recently, RBC-UKQCD calculated the pion form factor for a pion mass
of $m_\pi=330\MeV$, employing partially
twisted boundary conditions, as explained at the end of section \ref{sec:methods} above,
to reach very low non-zero values of $Q^2$ \cite{Boyle:2008yd}. The calculation
is based on a simulation with $n_f=2+1$ flavors of domain wall fermions and
the Iwasaki gauge action, for a single lattice spacing of $a\approx0.114\fm$
in a volume of $V\approx(2.74\fm)^3$. 
In contrast to traditional point sources, 
random wall (or noise-) sources with $Z(2)$ noise on a single color- and spin-index combination
have been used for the inversion of the Dirac matrix.
For this particular calculation, it was shown that 
a reduction of the computational cost by a factor of $\approx12$ 
could be achieved compared to a conventional point source computation 
with the similar statistical precision.
In combination with twisted boundary conditions for the valence quarks (pTBCs),
high precision results were obtained for the pion form factor at very small,
non-zero values of the momentum transfer squared, $Q^2=0.013,0.022$ and $0.035\GeV^2$,
as displayed in Fig.~\ref{FFpion_final_RBCUKQCD}.
Quite remarkably, the lowest $Q^2$ accessed in the lattice calculation is 
even below the lowest $Q^2$ reached by experiment,
and the statistical errors of the two leftmost lattice data points in Fig.~\ref{FFpion_final_RBCUKQCD} are 
smaller than for the experimental data points.  
Owing to the very low $Q^2$ that could be accessed, and the reasonably low pion mass of $330\MeV$,
a fit to results from $SU(2)$ ChPT at NLO for the simultaneous $Q^2$- and $m_\pi^2$-
dependence of the pion form factor has been attempted, using the original result by  
Gasser and Leutwyler \cite{Gasser:1983yg}
\bea
\label{NLOChPT_pionFF}
F_\pi(t=-Q^2,m_\pi^2)=1 + \frac{1}{6f_\pi^2} \left(t-4m_\pi^2\right)\overline J(t,m_\pi^2)
+ \frac{t}{6(4\pi f_\pi)^2}\left( \overline l_6^r -\frac{1}{3}\right) \,,
\eea
where $f_\pi=f_\pi^0\approx0.087\GeV$ denotes the pion decay constant in the chiral limit,
$\overline J(t,m_\pi^2)$ is a non-analytic function of $t/m_\pi^2$, and
the (scale independent) LEC $\overline l_6^r$ is related to the scale dependent
LEC (counter-term) $l_6^r(\lambda)$ by
\bea
\label{l6r}
l_6^r(\lambda)=-\frac{1}{96\pi^2} \left( \overline l_6^r + \log\frac{m_\pi^2}{\lambda^2} \right)\,.
\eea
A fit to the lattice data points in Fig.~\ref{FFpion_final_RBCUKQCD}, represented by the 
dashed-dotted line, gives $\langle r_\pi^2\rangle=0.418(31)\fm^2$, and 
$l_6^r(\lambda)=-9.3(1.0)\,10^{-3}$ for $\lambda=m_\rho$,
where the error includes the statistical error
as well as uncertainties in the determination of the lattice spacing, $f_\pi^0$, and
from the continuum extrapolation, all added in quadrature.
Very good agreement of the chirally extrapolated form factor at the physical point,
given by the continuous line, and the experimental data points is found for the lowest 
values of $Q^2$ in Fig.~\ref{FFpion_final_RBCUKQCD}.

%
%

%
An interesting study of the pion form factor for $n_f=2$ overlap fermions with Wilson kernel and 
the Iwasaki gauge action in a fixed topological sector has been presented recently
by the JLQCD collaboration \cite{JLQCD:2008kx}.
Results have been obtained for a lattice spacing of $a=0.118\fm$
in a volume of $V\approx(1.9\fm)^3$, for four different 
pion masses ranging from very low $288$ to $522\MeV$.
Using all-to-all propagators, a remarkable statistical precision was achieved 
with a comparatively small number of only $\mathcal{O}(100)$ 
configurations per ensemble. Contributions to the all-to-all
propagators from the 100 lowest-lying modes of the overlap Dirac operator have been 
treated exactly, while the higher modes were stochastically estimated using 
$Z_2$ noise and the dilution method proposed in \cite{Foley:2005ac}.
The $Q^2$-dependence of the pion vector form factor has been 
fitted with an ansatz based on the $\rho$-meson pole (monopole term)
plus polynomial corrections, and the mean square pion charge radius
and curvature term $c_V$ were extracted from the usual expansion
$F_\pi(Q^2)=1-6^{-1}\langle r_\pi^2\rangle Q^2+c_V(Q^2)^2+\mathcal{O}(Q^6)$.
Figure \ref{r2_JLQCD08_nnlo} displays the resulting values for $\langle r_\pi^2\rangle$ 
as a function of $m_\pi^2$,
which are somewhat low compared to the results discussed above by QCDSF/UKQCD and 
RBC-UKQCD in the overlapping region of $m_\pi\sim300-500\MeV$.  
The result of a simultaneous chiral fit to $\langle r_\pi^2\rangle$
and $c_V$ based on $2$-loop (NNLO) chiral perturbation theory 
\cite{Gasser:1990bv,Bijnens:1998fm}
with four free parameters (LECs) is represented by the solid lines in Fig.~\ref{r2_JLQCD08_nnlo}.
Interestingly, the 2-loop contribution, indicated by the dotted line, is found to be substantial,
of $\mathcal{O}(20\%)$, in the region where lattice data points are available.
Leading finite volume corrections have been included in the analysis of
the pion vector form factor using 
1-loop lattice regularized ChPT \cite{Borasoy:2004zf}, 
from which it is expected that the pion charge radius \emph{decreases}
for $L\rightarrow\infty$.
Finally, from a simultaneous 2-loop ChPT fit additionally including lattice
results for the scalar charge radius, $\langle r_\pi^2\rangle_S$ 
(extracted from the scalar form factor using the ansatz 
$F^S_\pi(Q^2)=F^S_\pi(0)(1-6^{-1}\langle r_\pi^2\rangle_S Q^2+c_S(Q^2)^2)$),
a value of $\langle r_\pi^2\rangle=0.404\pm(22)_{\text{stat}}\pm(22)_{\text{sys}}\fm^2$ 
was found at the physical pion mass, where the systematic error is due to variations
in some LECs and the fitting ranges. 

To further substantiate this interesting analysis, it
would be important to study possible effects from working 
in a fixed (trivial) topological sector in more detail.

%
\begin{figure}[t]
   \begin{minipage}{0.48\textwidth}
      \centering
          \includegraphics[angle=0,width=0.98\textwidth,clip=true]{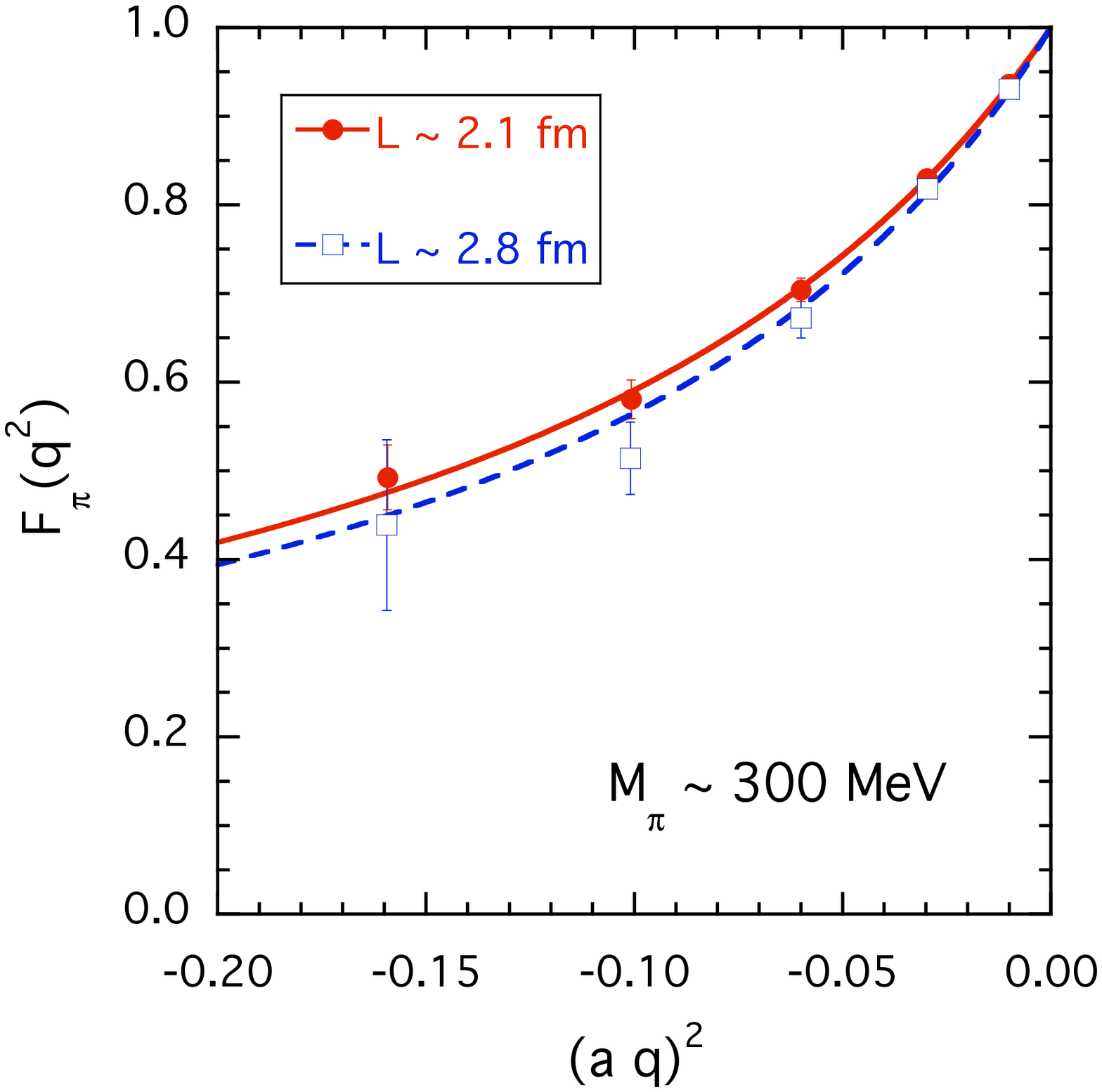}
  \caption{Pion form factor (from \cite{Simula:2007fa}).}
  \label{FFpion_q2_ETMC}
     \end{minipage}
     \hspace{0.5cm}
    \begin{minipage}{0.48\textwidth}
      \centering
          \includegraphics[angle=0,width=0.95\textwidth,clip=true]{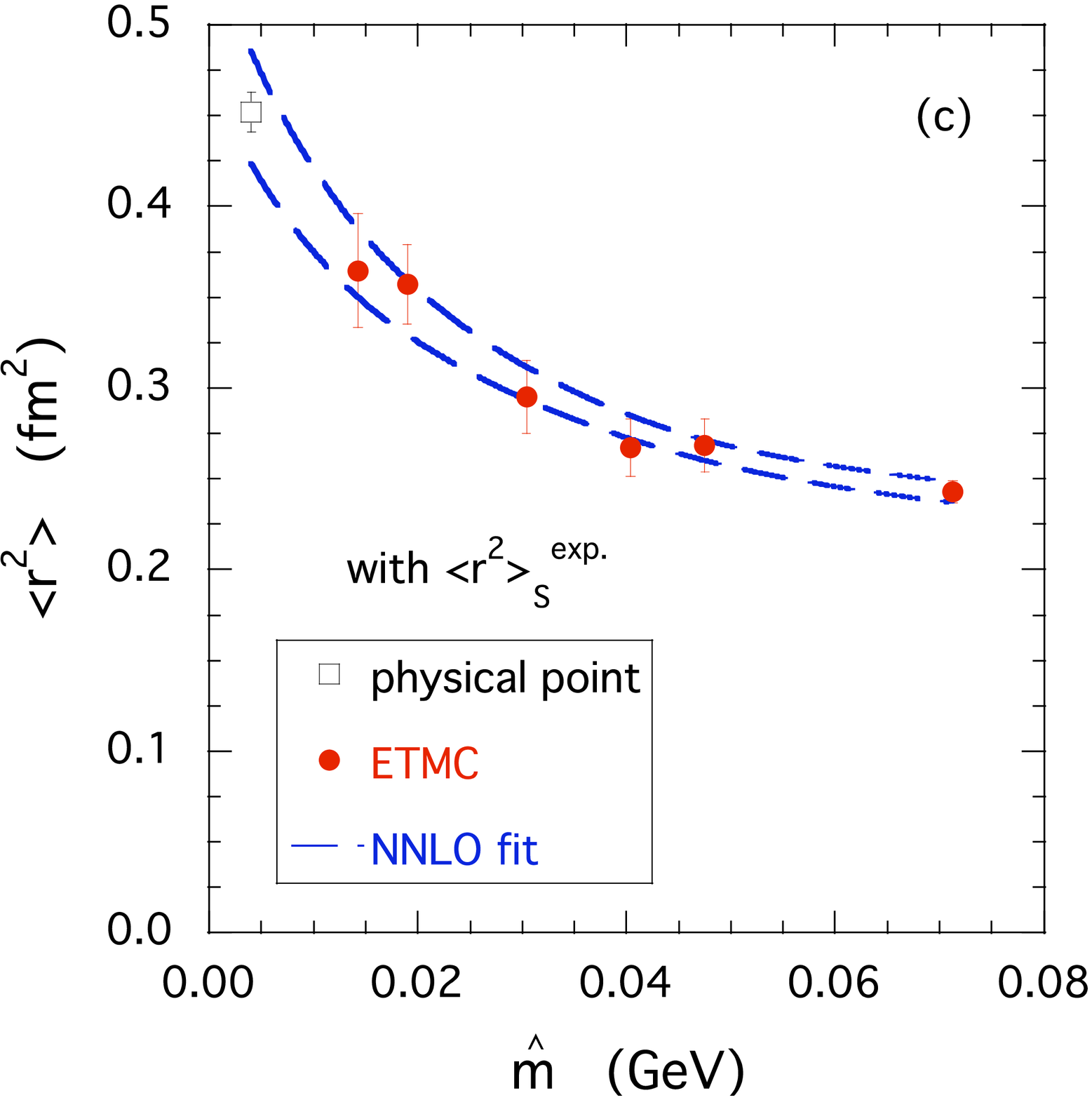}
  \caption{Pion mean square radius as function of the renormalized quark mass (from \cite{Frezzotti:2008dr}).}
  \label{FFpion_r2_ETMC}
     \end{minipage}
 \end{figure}

An extensive and detailed study of the pion form factor was presented recently by ETMC
based on simulations with $n_f=2$ twisted mass fermions (tuned to maximal twist) 
and the tree-level improved Symanzik gauge action \cite{Frezzotti:2008dr}.
Calculations were performed for six pion masses from $m_\pi\simeq260\MeV$ to $m_\pi\simeq580\MeV$,
lattice spacings of $a\simeq0.07\fm$ $a\simeq0.09\fm$ and volumes of $V\approx(2.1\fm)^3$, $V\approx(2.2\fm)^3$ and $V\approx(2.8\fm)^3$.
All-to-all propagators based on the stochastic-source method and the ``one-end-trick'' described above and
in sections \ref{sec:stochastic}, \ref{sec:oneendtrick} were used 
together with partially twisted boundary conditions (pTBCs) (see section \ref{sec:ptbcs})
to compute pion three-point functions in the Breit frame for (non-zero) $Q^2$ as low as $(Q^2)^{\not=0}_{\min}\approx0.05\GeV^2$.
Results for the pion form factor as a function of the squared momentum transfer in lattice units
are displayed in Fig.~\ref{FFpion_q2_ETMC} (note that $a^{-2}\approx4.8\GeV^2$).
A comparison of the lattice results for the two different volumes and with predictions from ChPT 
including pTBCs in a finite volume \cite{Jiang:2008te} indicate that finite size effects are on the few-percent-level,
of about the same size as the statistical errors. 
To minimize systematic uncertainties from FSE, the analysis was restricted to ensembles with $m_\pi L\gtrapprox4$. 
Discretization effects were roughly estimated by comparing the main simulations results for $a\simeq0.09\fm$ 
with a smaller number of results at $a\simeq0.07\fm$ and found to be of the same order as the statistical uncertainties.

The $Q^2$-dependence of the lattice form factor data was fitted with a monopole ansatz, Eq.~(\ref{globalmonopole})
with $m_{\text{mono}}(m_\pi^2)=\text{const}$ as free parameter, as indicated in Fig.~\ref{FFpion_q2_ETMC} by the solid line.
From the fit, the mean square radius and curvature where obtained from $\langle r_\pi^2\rangle=6/m_{\text{mono}}^2$ 
and $c_V=1/m_{\text{mono}}^4$, respectively.
Corresponding results for $\langle r_\pi^2\rangle$ are displayed in Fig.~\ref{FFpion_r2_ETMC}
as a function of the renormalized quark mass $\hat m\propto m_\pi^2 + h.o.$ (see GMOR relation in Eq.~(\ref{GMOR})).
The dashed lines represent the $1\sigma$ error band of a simultaneous chiral fit at NNLO to the lattice results
for $\langle r_\pi^2\rangle$, $m_\pi^2(\hat m)$, $f_\pi$ and $c_V$. 
In order to reduce the substantial
uncertainties in the relevant low energy constants (fit parameters), in particular $\bar{l}_{i=1,\ldots,4}$, in the chiral fit,
the experimental result for the scalar mean square radius of $\langle r_\pi^2\rangle^{\exp}_S=0.61\pm0.04\fm^2$
was included and simultaneously fitted at NNLO together with the lattice results for the other observables mentioned before. 
The fit represented by the error band in Fig.~\ref{FFpion_r2_ETMC} provides a very good description of the lattice data, 
and due to the singnificant upwards bending at low quark masses, the extrapolated result
of $\langle r_\pi^2\rangle^\phys=0.456\pm0.030_{\text{stat}}\pm0.024_{\text{sys}}$ fully overlaps with
the experimental value at the physical point (indicated by the open square). 
Estimates of uncertainties from finite size and discretization effects have been added in quadrature to 
provide the systematic error.
For the details of this extensive analysis, including the values of the numerous fitted low energy constants,
we refer to the original publication \cite{Frezzotti:2008dr}.

An overview and a direct comparison of the recent results for the pion charge radius discussed in this section
will be given below in section \ref{sec:FFsSummary}, cf. Figs.~\ref{overview_r2Pi_v1} and \ref{overview_r2Pi_mPiPhys_v1}.

%
%


First lattice results for the pion tensor form factor $B^\pi_{T10}(Q^2)$
as defined in Eq.~\ref{PionTensorFF} will be discussed in section \ref{PionTensor}
in the context of moments of pion generalized parton distributions.

\subsection{Nucleon form factors}
\subsubsection{Dirac and Pauli form factors}
\begin{figure}[t]
   \begin{minipage}{0.48\textwidth}
      \centering
          \includegraphics[width=0.9\textwidth,clip=true,angle=0]{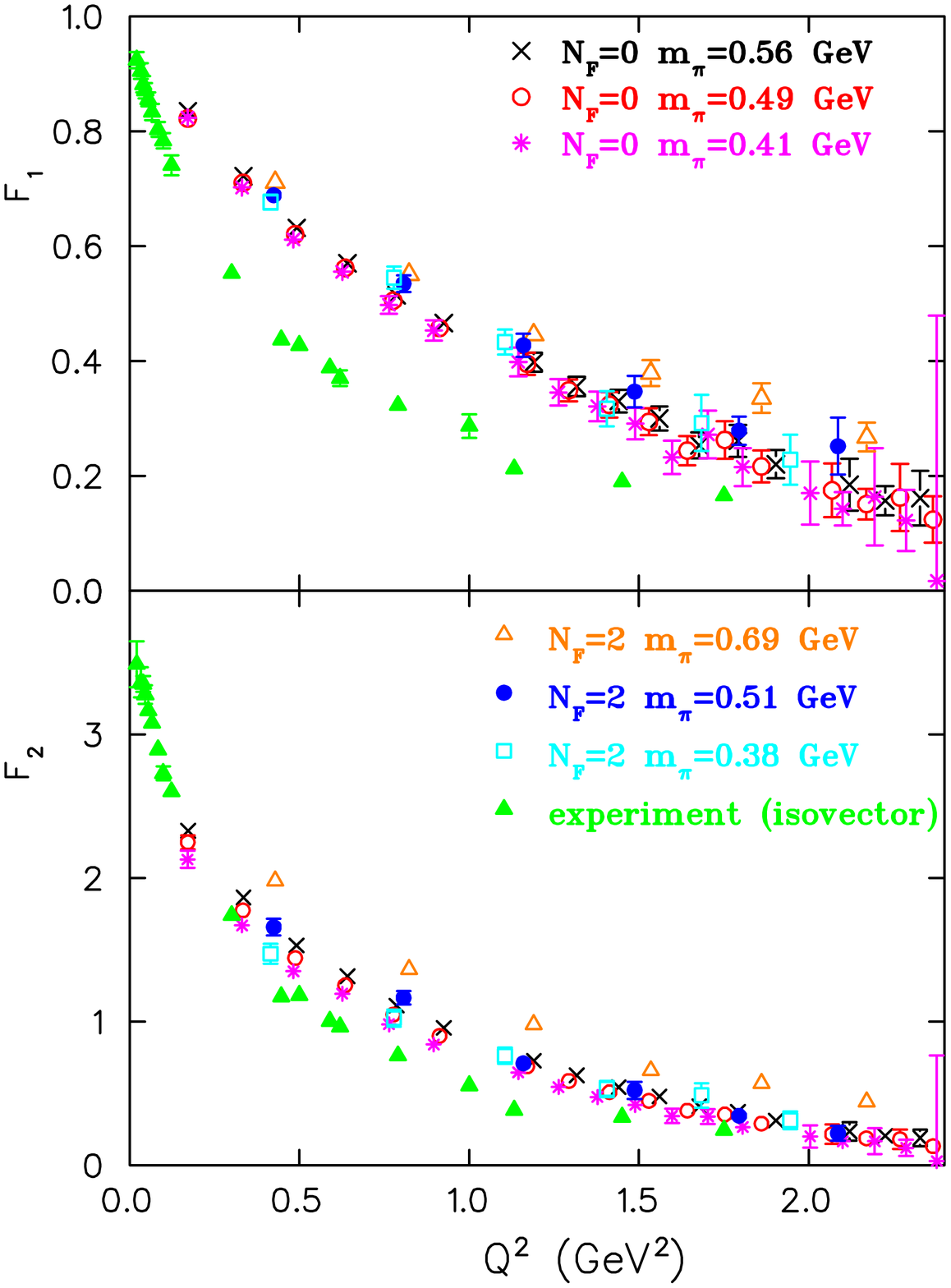}
  \caption{Nucleon form factors $F_1$ (top) and $F_2$ (bottom) in the isovector channel
  compared to experimental data (from \cite{Alexandrou:2006ru}).}
  \label{CyprusMIT_F1F2}
     \end{minipage}
     \hspace{0.5cm}
    \begin{minipage}{0.48\textwidth}
      \centering
           \includegraphics[width=0.9\textwidth,clip=true,angle=0]{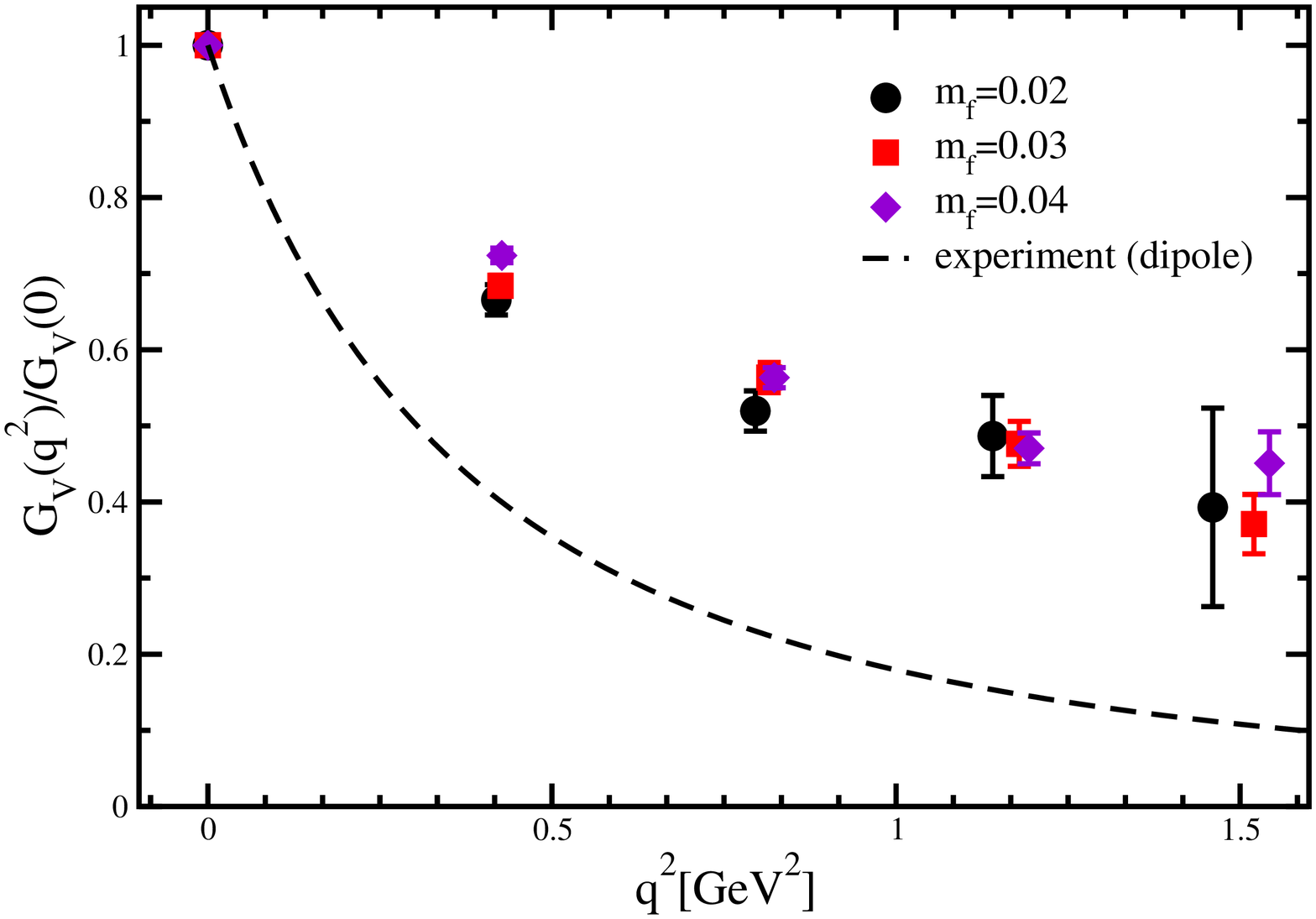}
           \includegraphics[width=0.9\textwidth,clip=true,angle=0]{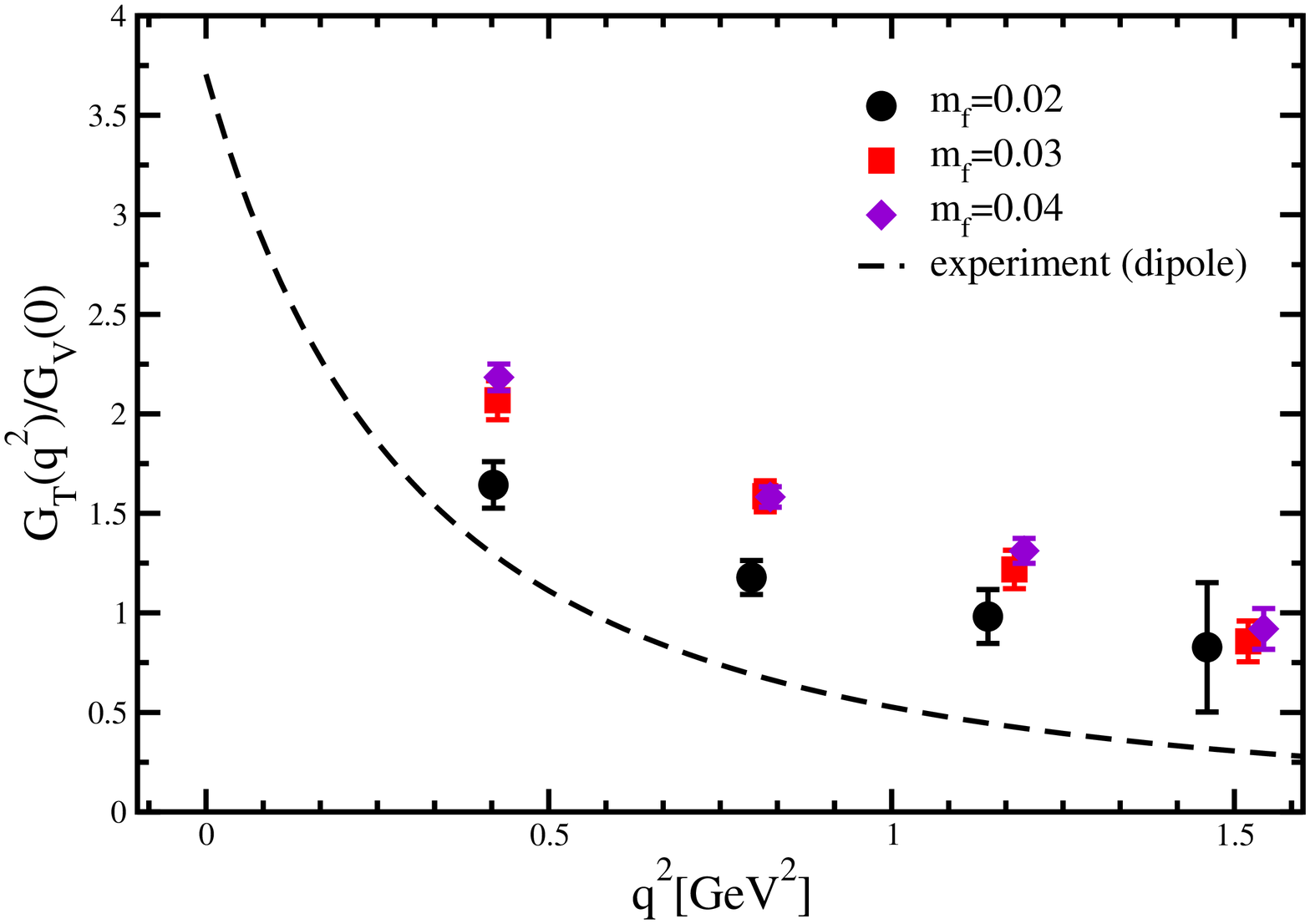}
  \caption{Nucleon form factors $F_1$ (top) and $F_2$ (bottom) in the isovector channel
  compared to dipole parametrizations of experimental data (from \cite{Lin:2008uz}).}
  \label{F1F2_RBCUKQCD}
     \end{minipage}
 \end{figure}

%
Figure \ref{CyprusMIT_F1F2} presents a typical comparison between results from a lattice QCD calculation \cite{Alexandrou:2006ru}
and experiment for the isovector Dirac, $F^{u-d}_1(Q^2)$, and Pauli, $F^{u-d}_2(Q^2)$, form factors.
The lattice results are based on simulations with $n_f=2$ flavors of Wilson fermions at a lattice
spacing of $a\approx0.08$ fm (obtained from an extrapolation of the nucleon mass to the chiral limit) 
in volumes of $V\approx(1.9\text{ fm})^3$.
The local vector current has been renormalized by demanding that $F^{u-d}_1(0)=1$.
Overall, the lattice data shows small statistical errors and a smooth $Q^2$-dependence.
The direct comparison in Fig.~\ref{CyprusMIT_F1F2} exemplifies the substantial 
difference of the slope in $Q^2$ of lattice and experimental results.
Concerning normalization and pion mass dependence 
of $F^{u-d}_2(Q^2)$, we note that the lattice values were obtained using the continuum
parametrization in Eq.~\ref{NuclVec3} with $m_N=m_N^{\text{lat}}$,
i.e. the pion mass, volume, and lattice spacing dependent nucleon mass for the corresponding ensemble.
Empirically, it is known that the lattice nucleon mass drops approximately linearly in $m_\pi$ towards
the physical point (see, e.g., \cite{WalkerLoud:2008bp}), $m^{\text{lat}}_N\sim\text{const}+\alpha m_\pi$ with 
$\alpha\sim1$, so that the choices $m_N=m_N^{\text{lat}}$ or $m_N=m_N^{\text{phys}}$ 
on the right-hand-side in Eq.~\ref{NuclVec3} lead to noticeably different results for $F_2(Q^2)$.
This issue has been discussed in some detail in Ref.~\cite{Gockeler:2003ay} in connection
with the chiral extrapolation of quenched lattice results from QCDSF/UKQCD, and should always
be kept in mind when studying the pion mass dependence of the form factor $F^{u-d}_2(Q^2)$ (for example
in Fig.~\ref{CyprusMIT_F1F2}) and the anomalous magnetic moment $\kappa=F_2(0)$.
Concerning $F^{u-d}_1(Q^2)$, one finds that 
the overall $Q^2$-slope over the full range of $Q^2=0,\cdots,2.5\text{ GeV}^2$
changes only very little going from $m_\pi=0.69\text{ GeV}$ down to $m_\pi=0.38\text{ GeV}$.

%
\begin{figure}[t]
   \begin{minipage}{0.48\textwidth}
      \centering
          \includegraphics[angle=0,width=0.9\textwidth,clip=true]{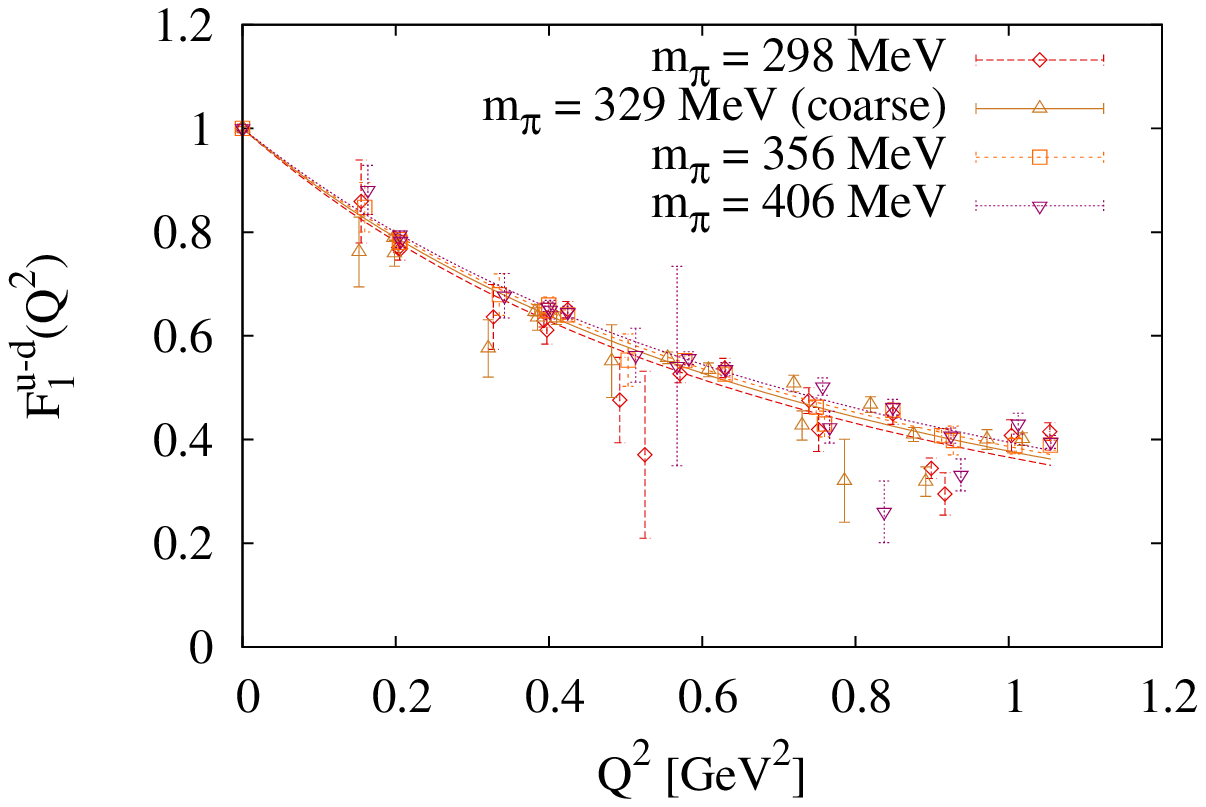}
  \caption{Isovector form factor $F_1$ together with dipole parametrization from 
  (from proceedings \cite{Syritsyn2008}).}
  \label{F1_syritsyn}
     \end{minipage}
     \hspace{0.5cm}
    \begin{minipage}{0.48\textwidth}
      \centering
          \includegraphics[angle=0,width=0.9\textwidth,clip=true]{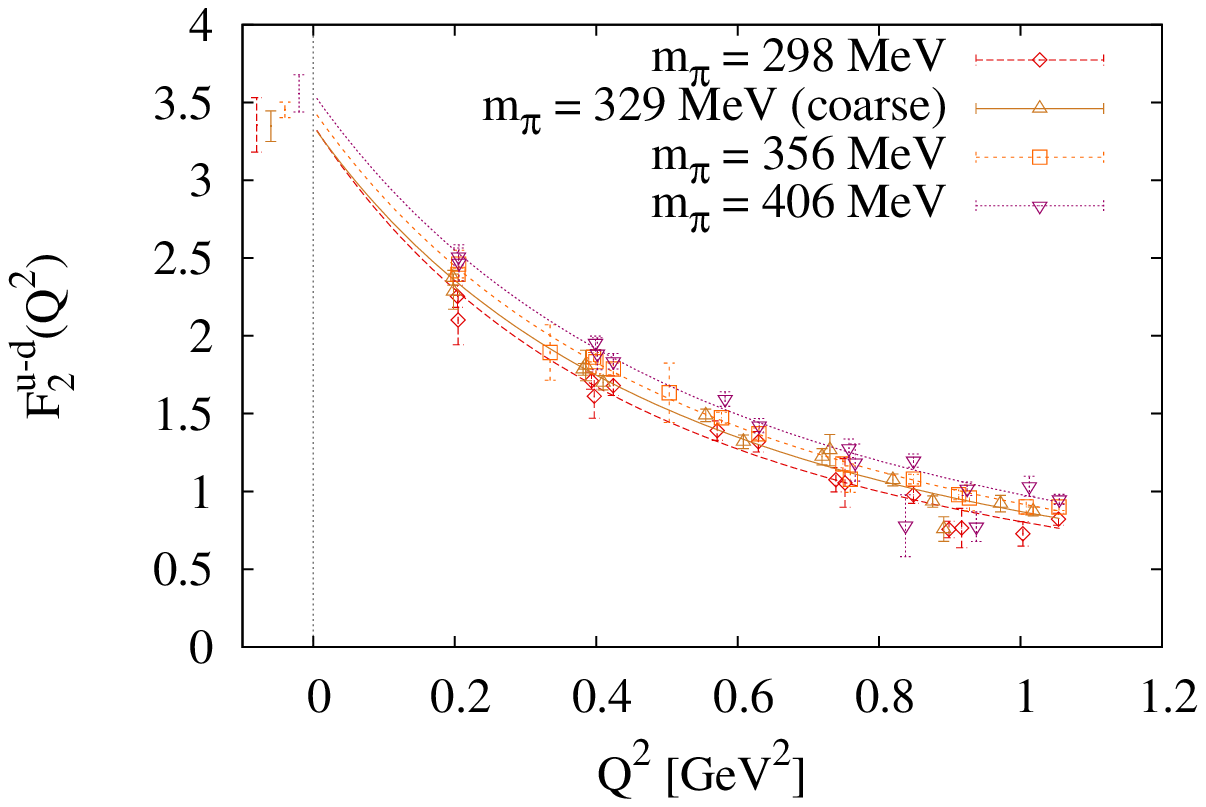}
  \caption{Isovector form factor $F_2$ together with dipole parametrization from 
  (from proceedings \cite{Syritsyn2008}).}
  \label{F2_syritsyn}
     \end{minipage}
 \end{figure}

%
Very similar observations can be made on the basis of recent results from the RBC collaboration
for $n_f=2$ flavors of domain wall fermions and a doubly-blocked Wilson (DBW2) gauge action
shown in Fig.~\ref{F1F2_RBCUKQCD} \cite{Lin:2008uz}.
Results have been obtained for three different pion masses, $m_\pi=0.695,0.607,0.493\text{ GeV}$, 
at a lattice spacing of $a\approx0.11$ fm, determined from a calculation and extrapolation of the
mass of the $\rho$-meson, and for a volume of $V\approx(1.9\text{ fm})^3$.
As above, the lattice nucleon mass has been used for the extraction of $F^{u-d}_2(Q^2)$.
The $Q^2$-slope of $F^{u-d}_1(Q^2)$ from the lattice is again substantially
flatter than for the phenomenological dipole fit describing the experimental data.
At the same time, within statistical errors, there is no systematic pion mass dependence 
visible of the lattice data points for $F^{u-d}_1(Q^2)$.
Corresponding results for the Dirac mean square radius will be
discussed in the following section, cf. Fig.~\ref{msr_v_chi_RBCUKQCD}.
%
Preliminary results for the Dirac and Pauli form factors, 
based on configurations generated by RBC-UKQCD with $n_f=2+1$ flavors of domain wall fermions
and the Iwasaki gauge action, have been presented recently by LHPC \cite{Syritsyn2008}.
Gaussian-smeared sources with APE-smeared links and carefully tuned parameters 
were employed to obtain an optimal overlap with the nucleon ground state, 
supporting the use of a source-sink separation of $\approx1.0\fm$ for the three-point functions.
Coherent sequential propagators based on   
4 nucleon and 4 anti-nucleon sources were employed to increase the statistics. 
Figures \ref{F1_syritsyn} and \ref{F2_syritsyn} show the $Q^2$-dependence of 
$F_1$ and $F_2$ in the isovector channel for pion masses from $\approx298$ to $\approx406\MeV$.
Consistent results were obtained for the ``fine'' lattices, with an estimated lattice spacing
of $\approx0.084\fm$, and the ``coarse'' lattice with $a\approx0.114\fm$. The 
spatial volumes are $V\approx(2.7\fm)^3$ in both cases.
As indicated by the dipole fits, the slope in $Q^2$ of both form factors
slightly increases with decreasing pion masses, but even at the lowest pion mass
the lattice data at non-zero $Q^2$ is still far above the experimental results.
%
%
%
%
%
%
%

A preliminary comparison of a full $n_f=2+1$ DW calculation with a hybrid calculation 
for $n_f=2+1$ flavors of domain wall fermions and staggered Asqtad sea quarks,
based on gauge configurations provided by the MILC collaboration, has been performed recently in
\cite{Bratt:2008uf}.
This careful studied revealed good consistency for $F_1^{u-d}$ of the full DW and the hybrid results
with lattice spacings of $a\approx0.084\fm$ and $a\approx0.124\fm$, respectively,
indicating that discretization effects are small.
Volume effects in the case of the hybrid approach were also analyzed and found to be negligible
within statistical errors.

%
%
\subsubsection{Charge radii and anomalous magnetic moment}
\label{FFradii}
\begin{figure}[t]
    \begin{minipage}{0.48\textwidth}
        \centering
          \includegraphics[angle=-90,width=0.9\textwidth,clip=true,angle=0]{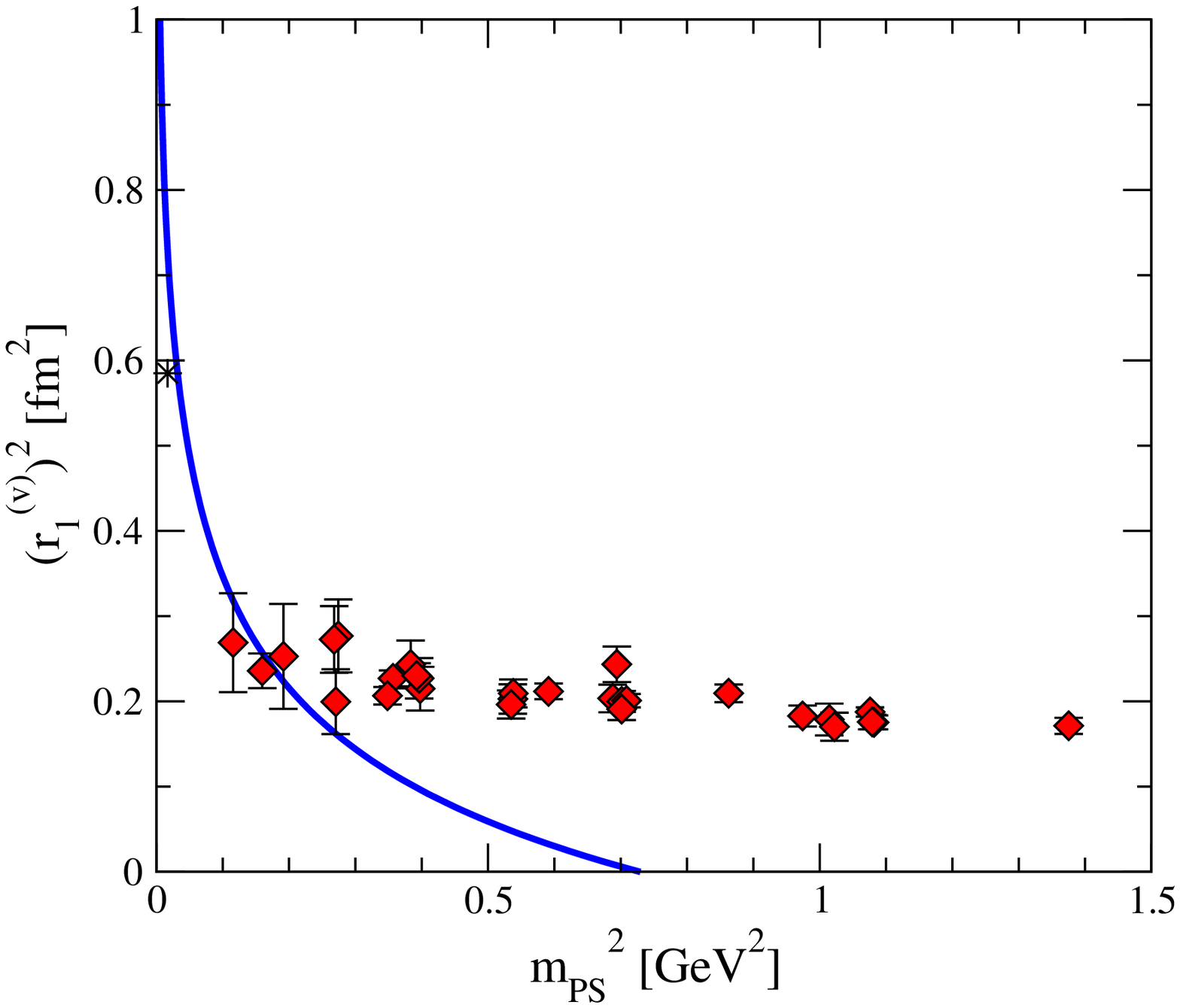}
  \caption{Isovector mean square charge radius for $F^{u-d}_1$ (from proceedings \cite{Gockeler:2006uu}).
  Phenomenological value from \cite{Belushkin:2006qa}.}
  \label{rv1Q_QCDSF}
     \end{minipage} 
         \hspace{0.2cm}
    \begin{minipage}{0.48\textwidth}
      \centering
          \includegraphics[angle=0,width=0.9\textwidth,clip=true,angle=0]{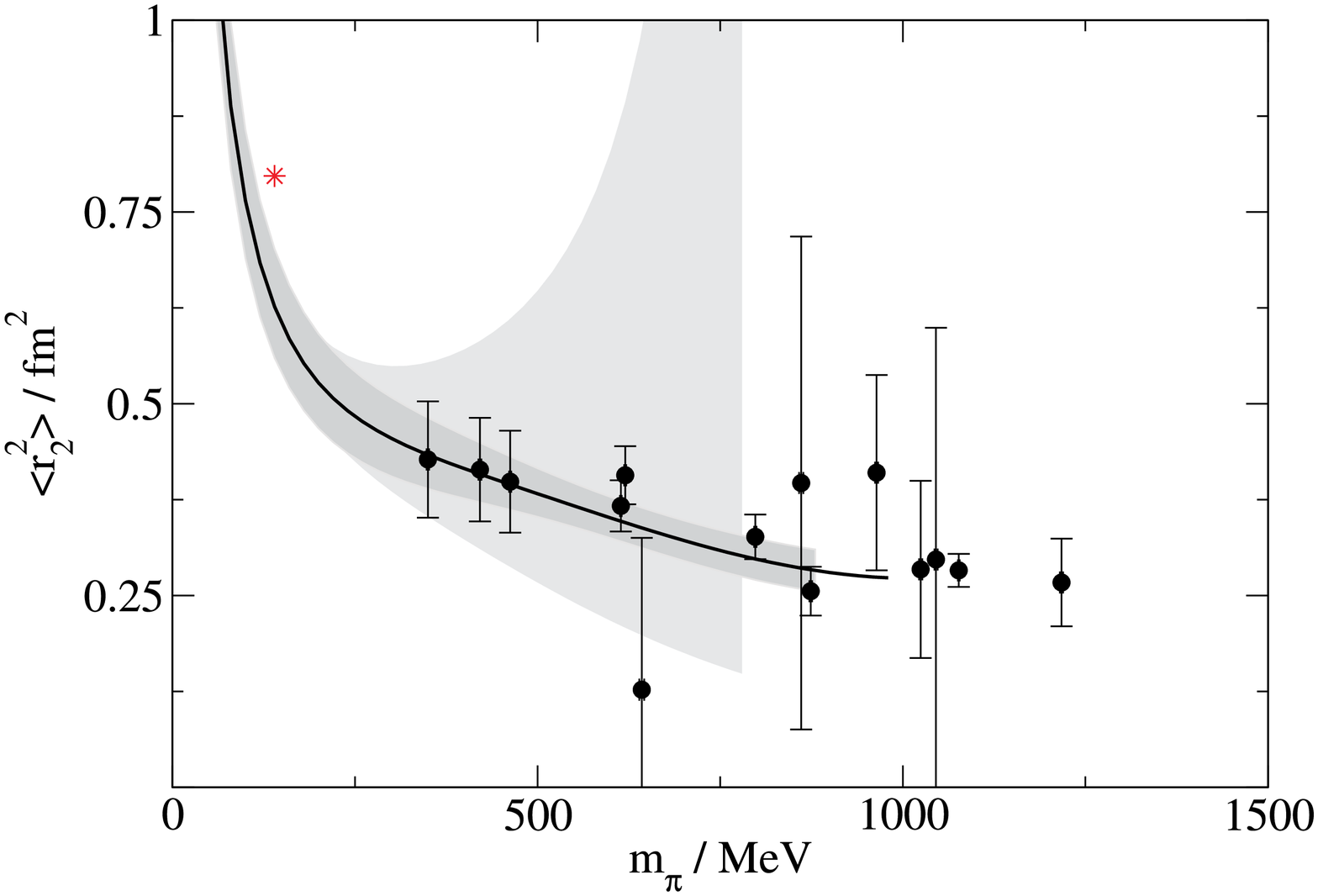}
  \caption{Isovector mean square radius for $F^{u-d}_2$ (from proceedings \cite{Gockeler:2007hj}).}
  \label{rv2Q_QCDSF_2}
     \end{minipage}
 \end{figure}

%
%
Over the last couple of years, the QCDSF-UKCQD collaboration has performed extensive 
hadron structure calculations using $n_f=2$ flavors of clover-improved Wilson fermions and Wilson gluons
for a large number of pion masses, several lattice spacings and volumes.
In \cite{Gockeler:2006uu}, the lattice data for the nucleon Dirac and Pauli 
form factors was fitted with a standard $p$-pole ansatz
\bea
\label{ppole2}
F_i(Q^2)=\frac{F_i(0)}{\left(1+\frac{Q^2}{p\,m_{i}^2}\right)^p}\,,
\eea
for $i=1,2$, choosing $p=2$ for $F_1$ and $p=3$ for $F_2$, and where $F_2(0)$ and $m_{i}$
were treated as free fitting parameters.
The local vector current has been renormalized by demanding that $F^{u-d}_1(0)=1$.
Results for the isovector mean square radius for $F_1$, Eq.~(\ref{radii}), obtained
from $\langle r_1^2\rangle=6/m_1^2$, 
are shown in Fig.~\ref{rv1Q_QCDSF} as a function of $m_\pi^2$.
In \cite{Gockeler:2007hj}, a slightly more general ansatz has been used to 
describe the $Q^2$-dependence of $F_1$ and $F_2$, 
\bea
\label{F1F2para}
F_i(Q^2)=\frac{F_i(0)}{1+c_{i,2}Q^2+c_{i,2i+2}Q^{2i+2}}\,,
\eea
where $c_{i2}$, $c_{i4}$ and $F_2(0)$ are fitting parameters. 
Within the available statistics, both ans\"atze in Eq.~(\ref{ppole2}) and Eq.~(\ref{F1F2para})
provided similarly good descriptions of the lattice data.
Results for the Pauli form factor radius based on the parametrization in Eq.~(\ref{F1F2para}), 
with $\langle r_2^2\rangle=6c_{22}$, are displayed in Fig.~\ref{rv2Q_QCDSF_2} as a function of the pion mass.
We note that the lattice data points in Fig.~\ref{rv1Q_QCDSF} and 
\ref{rv2Q_QCDSF_2} correspond to a range of lattice spacings, 
$a\approx0.67,\ldots,0.85\text{ fm}$ (chirally extrapolated) that have been fixed using the nucleon mass, 
and to different volumes from $V\approx(1.2\fm)^3$ to $V\approx(1.9\fm)^3$, 
with $m_\pi L>3.5$.
%
%
Within statistics, remarkably consistent results were obtained 
for $\langle r_1^2\rangle_{u-d}$ and $\langle r_2^2\rangle_{u-d}$
for the different ensembles (lattice spacings and volumes).
Concerning the pion mass dependence, one finds that 
$\langle r_1^2\rangle_{u-d}$ is very slowly and approximately linearly rising 
over the full range from $m_\pi\sim1.2\GeV$ down to $m_\pi\sim0.330\GeV$,
and that even at the lowest available pion mass the lattice result are roughly
a factor of two below the experimental number 
$\langle r_1^2\rangle_{u-d}=0.635\pm0.007\text{ fm}^2$ (PDG \cite{PDG2008}),
in accordance with the discussion
of Figs.~\ref{CyprusMIT_F1F2} and \ref{F1F2_RBCUKQCD} above.
We note that a recent dispersion relation analysis of experimental data gives 
a smaller value $\langle r_1^2\rangle_{u-d}=0.585\pm0.017$ \cite{Belushkin:2006qa}.

\begin{figure}[t]
      \centering
          \includegraphics[angle=0,width=0.5\textwidth,clip=true,angle=0]{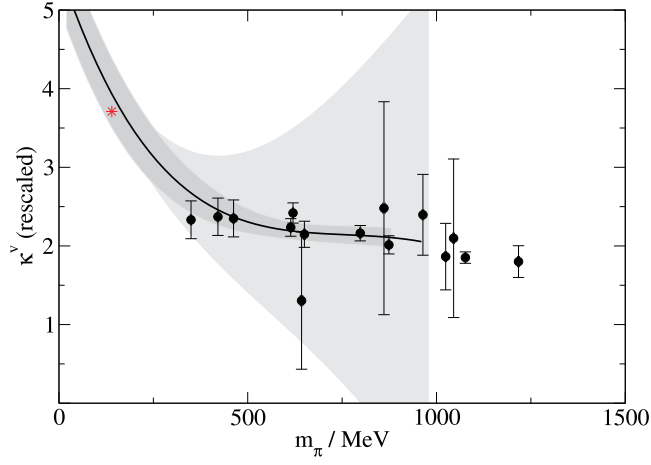}
  \caption{Isovector anomalous magnetic moment $\kappa^{u-d}$ (from proceedings \cite{Gockeler:2007hj}).}
  \label{kappav_QCDSF_2007_2}
 \end{figure}

An explanation for the discrepancy between the lattice results and the experimental
data may be given on the basis of chiral
perturbation theory, which generically predicts that the isovector charge radius diverges 
logarithmically in the chiral limit. 
The chiral extrapolations indicated by the lines and bands in Figs.~\ref{rv1Q_QCDSF}, 
\ref{rv2Q_QCDSF_2} and \ref{kappav_QCDSF_2007_2} were based on the 
results developed in \cite{Hemmert:2002uh,Gockeler:2003ay} 
to $\mathcal{O}(\eps^3)$ in the SSE of HBChPT.
To this order, the pion mass dependence of $\langle r_1^2\rangle_{u-d}$ is given by \cite{Gockeler:2003ay}
\bea
 \langle r_1^2\rangle_{u-d}^{\text{SSE}} &=&
    -  \frac{1}{(4\pi f_\pi)^2}\left\{1+7 g_{A}^2 +
  \left(10 g_{A}^2 +2\right) \ln\left(\frac{m_\pi}{\lambda}\right)\right\} 
     +  \frac{c_A^2}{54\pi^2 f_{\pi}^2}\Bigg\{
        26+30\ln\left(\frac{m_\pi}{\lambda}\right)  \nonumber\\
     &+& 30\frac{\Delta m_{\Delta N}}{\sqrt{\Delta m_{\Delta N}^2-m_{\pi}^2}}
       \ln\left(\frac{\Delta m_{\Delta N}}{m_\pi}
      + \sqrt{\frac{\Delta m_{\Delta N}^2}{m_{\pi}^2}-1}\right) \Bigg\}
      - \frac{12 B_{10}^{(r)}(\lambda)}{(4\pi f_\pi)^2} \,,
\label{F1radius}
\eea
which, in addition to the counter term $B_{10}^{(r)}(\lambda)$ that removes the 
regularization scale dependence, is
governed by four LECs: $f_\pi$, $g_A$, the axial transition pion-nucleon-$\Delta$
coupling constant $c_A=g_{\pi N\Delta}$, and $\Delta m_{\Delta N}$ (all in the chiral limit).
Using their phenomenological values at the physical point as input,
and setting the counter term equal to zero at a scale of $\lambda=600\text{ MeV}$ \cite{Gockeler:2003ay}, 
the SSE result provides a prediction for the $m_\pi$-dependence as indicated by the solid
curve in Fig.~\ref{rv1Q_QCDSF}. 
Clearly, at the given order, the SSE HBChPT formula is maximally applicable up to the lowest
accessible lattice pion masses of $\approx500\MeV$. 
At larger pion masses, the SSE prediction does not flatten off fast enough and therefore 
undershoots the lattice data points. However, it does provide a link between 
the lattice data at the lowest pion masses and the chiral limit.
This indicates that future lattice results for $\langle r_1^2\rangle_{u-d}$
not much below $m_\pi\approx300\text{ MeV}$ may show a strong upwards bending and therefore 
a clear deviation from the linear $m_\pi^2$ dependence. 
However, it should be noted that the SSE prediction also misses the
experimental number by $\approx20\%$, indicating that higher order
corrections may be important already at the physical pion mass. 
A quantitative comparison with lattice results
around or above $m_\pi^{\text{phys}}$ should therefore be considered with some caution.
%
%
%
We now come back to the results from QCDSF-UKQCD for the isovector mean square radius of $F_2$ 
shown in Fig.~\ref{rv2Q_QCDSF_2}, obtained from the parametrization in Eq.~(\ref{F1F2para}).
As for $\langle r_1^2\rangle_{u-d}$, the lattice results at the lowest available
pion masses of $\approx350$ to $\approx400\text{ MeV}$ are $\approx50\%$ below the experimental result,
$\langle r_2^2\rangle_{u-d}=0.776\pm0.011\text{ fm}^2$.
In contrast to the isovector Dirac form factor, the Pauli form factor is
predicted to diverge linearly as $1/m_\pi$ in the chiral limit.
To $\mathcal{O}(\eps^3)$ in SSE HBChPT, 
the pion mass dependence of $\langle r_2^2\rangle_{u-d}$ is given by \cite{Gockeler:2003ay}
\begin{equation} 
\langle r_2^2\rangle_{u-d}^{\text{SSE}} = 
   \frac{m_N}{\kappa_{u-d}(m_\pi)} \Bigg\{
   \frac{g_{A}^2}{8 \pi f_{\pi}^2  m_\pi}
  +\frac{c_A^2}{9 \pi^2 f_{\pi}^2 \sqrt{\Delta m_{\Delta N}^2-m_{\pi}^2}} 
    \ln\left(\frac{\Delta m_{\Delta N}}{m_\pi}+\sqrt{\frac{\Delta m_{\Delta N}^2}{m_{\pi}^2}-1}\right)
  + 24B_{c2} \Bigg\}\,   
\label{F2radius}
\end{equation}
and involves, in addition to the LECs discussed above, also the (pion mass dependent)
isovector anomalous magnetic moment $\kappa_{u-d}(m_\pi)$ and the constant $B_{c2}$. 
The SSE expression for $\kappa_{u-d}$ reads \cite{Gockeler:2003ay}
\bea
 \kappa_{u-d}^{\text{SSE}}&=&
          \kappa_{u-d}^0
          - \frac{g_A^2\,m_\pi m_N}{4\pi F_\pi^2}
          +   \frac{2 c_A^2 \Delta m_{\Delta N} m_N}{9\pi^2 F_\pi^2} 
          \Bigg\{\sqrt{1-\frac{m_\pi^2}{\Delta m_{\Delta N}^2}}
          \ln \left(\frac{\Delta m_{\Delta N}}{m_\pi}+\sqrt{\frac{\Delta m_{\Delta N}^2}{m_{\pi}^2}-1}\right)
          \nonumber\\
          &+& \ln\left(\frac{m_\pi}{2\Delta m_{\Delta N}}\right) \Bigg\}
        +  \frac{4c_A c_V g_A m_N m_\pi^2}{9\pi^2 F_\pi^2} \ln\left(\frac{2\Delta m_{\Delta N}}{\lambda} \right)
        +  \frac{4c_A c_V g_A m_N m_\pi^3}{27\pi F_\pi^2\Delta m_{\Delta N}}
         \nonumber\\
         &-&   \frac{8 c_A c_V g_A \Delta m_{\Delta N}^2 m_N}{27\pi^2 F_\pi^2}
               \Bigg\{\left(1-\frac{m_\pi^2}{\Delta m_{\Delta N}^2}\right)^{3/2} 
               \ln \left(\frac{\Delta m_{\Delta N}}{m_\pi}
               + \sqrt{\frac{\Delta m_{\Delta N}^2}{m_{\pi}^2}-1}\right)
         \nonumber\\
         &+& \left(1-\frac{3m_\pi^2}{2\Delta m_{\Delta N}^2}\right) 
          \ln\left(\frac{m_\pi}{2\Delta m_{\Delta N}}\right) \Bigg\}
             -   8 E_1^{(r)} (\lambda) m_N m_\pi^2\,,
\label{kappa}
\eea
and depends on two more LECs, the isovector anomalous magnetic moment $\kappa^{0}_{u-d}$ and 
the isovector nucleon-$\Delta$ transition coupling constant $c_V=c^0_V$ in the chiral limit, 
as well as an additional scale dependent counter term parameter $E_1^{(r)}(\lambda)$.
They were treated as free parameters in a simultaneous fit
of the SSE HBChPT formulas for the mean square radius $\langle r_2^2\rangle_{u-d}$ (Eq.~\ref{F2radius}) 
and $\kappa_{u-d}$ to the lattice data points 
in Figs.~\ref{rv2Q_QCDSF_2} and \ref{kappav_QCDSF_2007_2},
obtained from the more general ansatz Eq.~(\ref{F1F2para}).
The fit result for $\langle r_2^2\rangle_{u-d}$ is shown in Fig.~\ref{rv2Q_QCDSF_2} 
as smaller error band, where the width of the band represents the statistical errors only.
Although the SSE chiral extrapolation provides some curvature towards the physical point,
it still underestimates the experimental result by $\approx25\%$.
We note that results for $\langle r_2^2\rangle_{u-d}$
and $\kappa^{u-d}$ based on the $p$-pole ansatz in Eq.~(\ref{ppole2})
(presented in \cite{Gockeler:2006uu} but not shown here) lead to visibly different 
chiral fits, with a central extrapolation curve
for $\langle r_2^2\rangle_{u-d}$ that is up to $25\%$ below the 
one in Fig.~\ref{rv2Q_QCDSF_2}.
This reveals that the ansatz used to parametrize the 
form factors can have a significant impact on the resulting chiral
extrapolation. 
Similar results also hold for the anomalous magnetic moment,
as we will discuss in the following.
%
%

\begin{figure}[t]
   \begin{minipage}{0.5\textwidth}
      \centering
          \includegraphics[angle=0,width=0.9\textwidth,clip=true,angle=0]{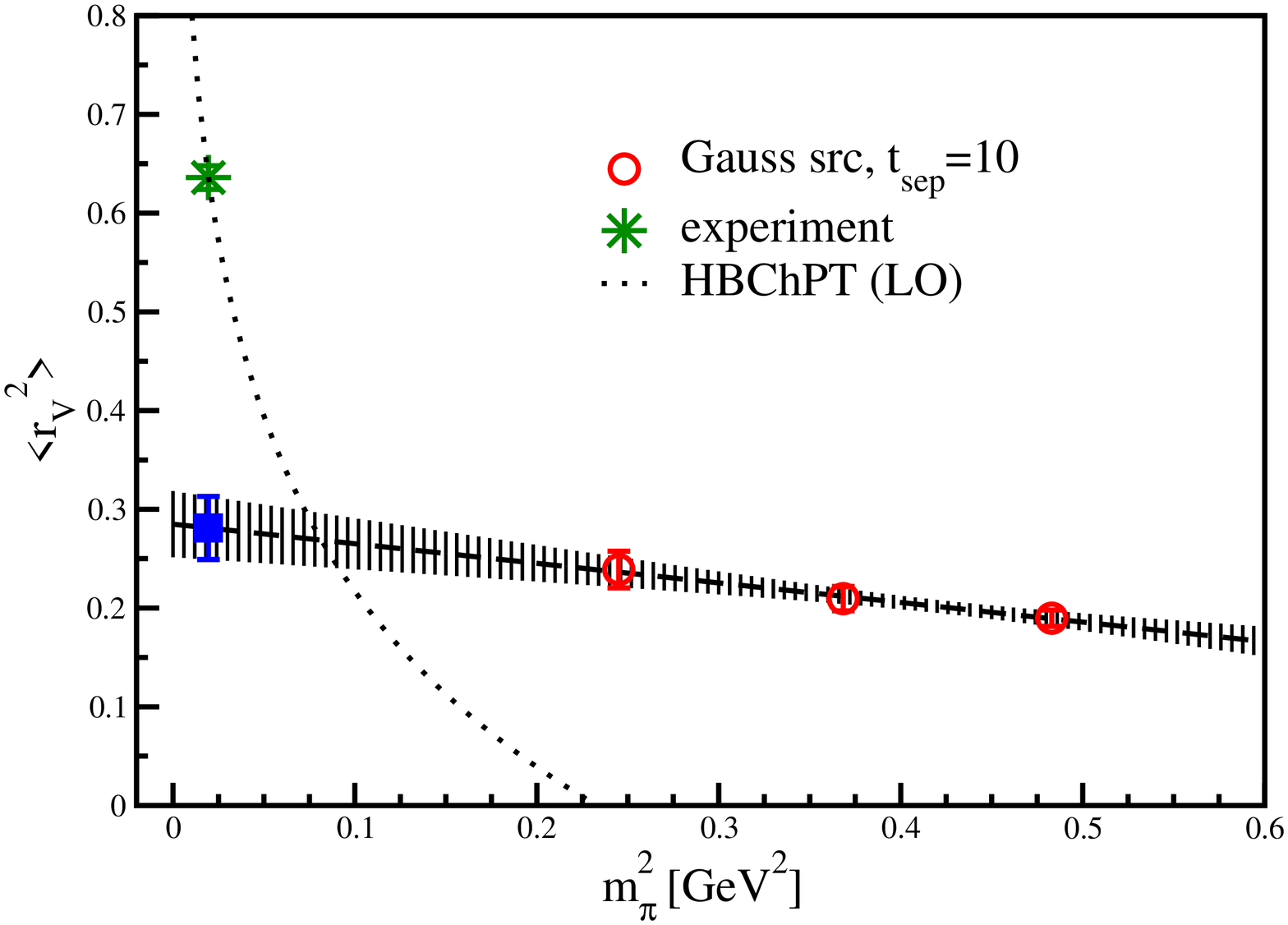}
  \caption{Isovector Dirac rms radius from RBC for $n_f=2$ flavors of DW fermions (from \cite{Lin:2008uz}).
   Experimental value from the PDG \cite{PDG2008}.\newline}
  \label{msr_v_chi_RBCUKQCD}
     \end{minipage}
     \hspace{0.5cm}
    \begin{minipage}{0.46\textwidth}
      \centering
          \includegraphics[angle=0,width=0.9\textwidth,clip=true,angle=0]{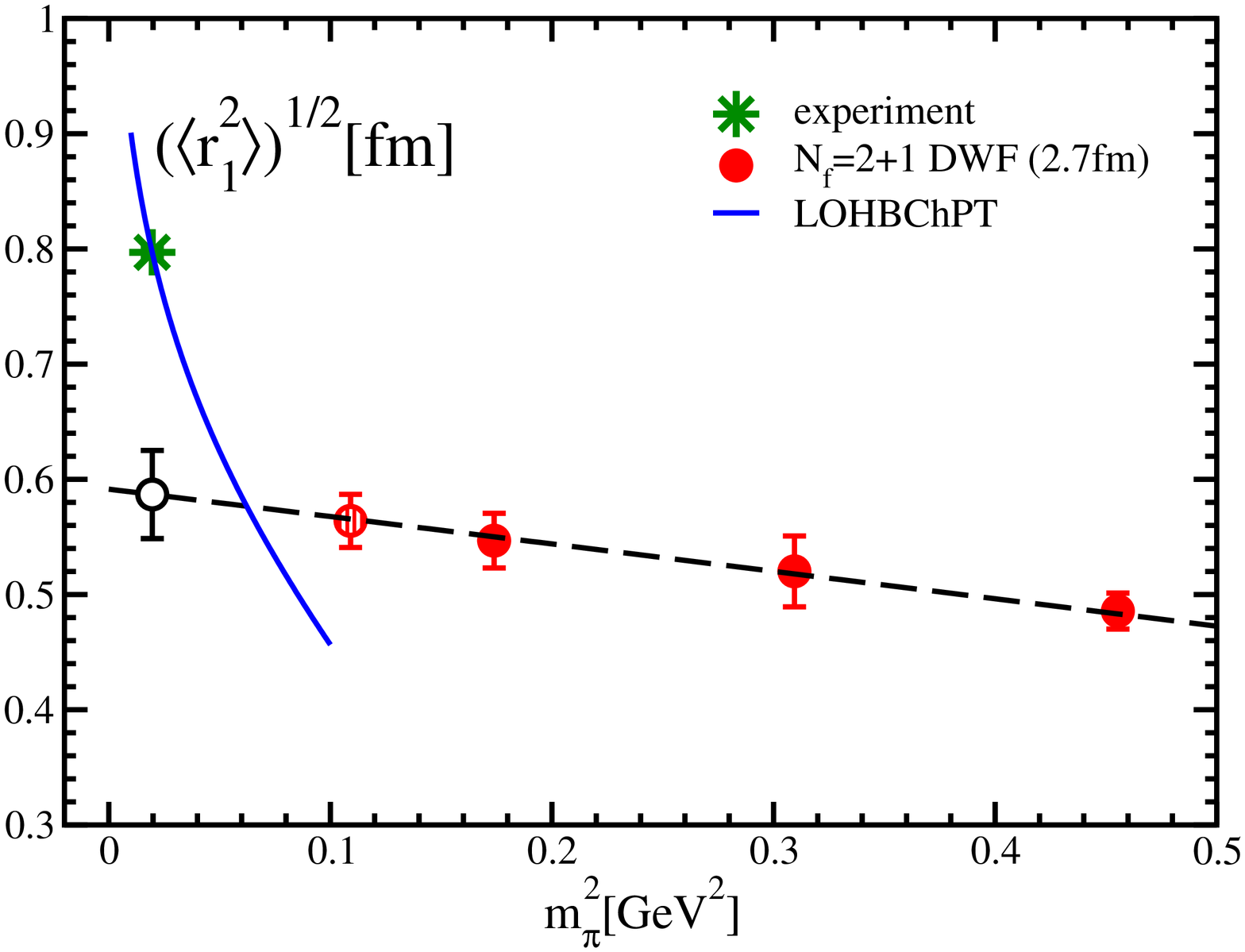}
  \caption{Isovector Dirac rms radius from RBC-UKQCD $n_f=2+1$ flavors of DW fermions 
  with $t_{\text{sep}}=12$ (from lattice proceedings \cite{Ohta:2008kd}). 
  Experimental value from the PDG \cite{PDG2008}.}
  \label{mfr_1_RBCUKQCD_Lat2008}
     \end{minipage}
 \end{figure}
%
%
The lattice data points in Fig.~\ref{kappav_QCDSF_2007_2} were obtained from
the parametrization in Eq.~(\ref{F1F2para}), providing the required 
extrapolations of the Pauli form factor lattice data at non-zero momentum transfer  to $Q^2=0$.
The results for $\kappa_{u-d}$ were scaled with a factor of $m_N^{phys}/m_N^{lat}(m_\pi)$
to remove the additional pion mass dependence that comes in through the use of 
the pion mass dependent nucleon mass in the extraction of $F_2$, as explained above.
Although slowly rising towards lower pion masses, the lattice results still 
lie $\approx 25\%$ below the experimental value at the lowest pion masses of $\approx350$ to $\approx400\text{ MeV}$. 
The chiral extrapolation of the lattice results based 
on a simultaneous fit to $\langle r_2^2\rangle_{u-d}$ and $\kappa_{u-d}$ 
(based on Eq.~(\ref{F2radius}) and Eq.~(\ref{kappa}))
as discussed above is presented by the error band in Fig.~\ref{kappav_QCDSF_2007_2}.
It is promising to see that the chiral fit overlaps within errors with the value from experiment.
Such an agreement within errors between extrapolated lattice results and experiment
could not be observed on the basis of the $p$-pole ansatz in Eq.~\ref{ppole2}
used in \cite{Gockeler:2006uu}. 
We note, however, that the lattice values for $\langle r_2^2\rangle_{u-d}$ and $\kappa_{u-d}$
obtained from the more general ansatz in Eq.~(\ref{F1F2para}) have
larger errors than the results from the $p$-pole parametrization, 
and a comparison of the two approaches is at this point clearly limited by statistics.
Still, we conclude that different methods to parametrize and extrapolate
nucleon form factors to $Q^2=0$ may lead to significantly different final results, in particular 
for $F_2(Q^2)$ and in combination with a chiral extrapolation to the physical point. This is a serious
source of systematic uncertainty, and below in sections 
\ref{sec:DeltaFFs}, \ref{sec:BGfield}, \ref{sec:NuclPTBCs},
we will discuss some alternative and also new approaches, for example
based on background field methods and pTBCs to
access small non-zero $Q^2$, which may help to improve the situation in the near future. 
For a detailed discussion of chiral fits based on Eqs.~(\ref{F1radius}), (\ref{F2radius}) and (\ref{kappa}) 
to quenched lattice data and of the corresponding LECs, we refer to \cite{Gockeler:2003ay}.

Results for the isovector Dirac rms radius from the RBC collaboration, 
based on $n_f=2$ flavors of domain wall fermions
for a lattice spacing of $a\approx0.11$ fm and a volume of $V\approx(1.9\text{ fm})^3$,
are shown in Fig.~\ref{msr_v_chi_RBCUKQCD} \cite{Lin:2008uz}.
The rms radius has been obtained from dipole fits to the form factor data displayed in Fig.~\ref{F1F2_RBCUKQCD}.
The results, which are in overall agreement with the lattice data from QCDSF-UKQCD presented in Fig.~\ref{rv1Q_QCDSF},
are more than a factor of two below the experimental value and only
show a hint of a slope in $m_\pi^2$, so that a linear extrapolation in $m_\pi^2$
to the physical pion mass fails dramatically. As mentioned already above,
LO HBChPT predicts a strong non-linear pion mass dependence at lower pion masses
(represented by the dashed line), clearly indicating that an extrapolation 
linear in $m_\pi^2$ is most probably misleading in this case.

More recently, RBC-UKQCD presented results for
the Dirac and Pauli form factor based on simulations with $n_f=2+1$ flavors of
domain wall fermions and the Iwasaki gauge action \cite{Ohta:2008kd}
(for simulation details, see \cite{Allton:2008pn}).
Calculations were performed for pion masses of $\simeq331$, $\simeq419$, $\simeq557$ and $\simeq672\MeV$,
a lattice spacing of $a\simeq0.114\fm$ and volumes of $\simeq(1.82\fm)^3$ and $\simeq(2.74\fm)^3$.
Bare quark masses and the lattice scale were fixed using $SU(2)$ ChPT fits and
experimental values for the $\pi$, $K$ and $\Omega$ meson masses.
The dynamical lattice strange quark mass turned out to be $\approx12-15\%$ larger than the physical
strange quark mass.
A comparatively large sink-source time separation of $t_{\text{sep}}=12$ units corresponding to $\approx1.37\fm$
was chosen for the nucleon three-point functions in order minimize contaminations from
excited states, which in principle could be very helpful in reducing this sort of systematic error. 
In practice, however, while statistically clean plateaus for the ratio of three- to two-point functions
could be observed for $t_{\text{sep}}=10$ ,
the statistics is notably worse for $t_{\text{sep}}=12$, to an extent that it may even be 
difficult to identify a proper plateau region in the first place, see for example 
Fig.~\ref{Plat10_RBCUKQCD} by RBC related to moments of PDFs.
Furthermore, we note that the nucleon two point functions that enter
the ratio of three- to two-point functions in Eq.~(\ref{ratio1}) in particular at the sink position $t_\snk$, 
may show significant fluctuations at large times. 
Therefore, while excited states may contaminate the plateau for too small $t_\snk$, 
choosing a specific large source-sink separation may introduce an additional systematic uncertainty 
through $C_{\text{2pt}}(t_\snk)$. 
%
%
\begin{figure}[t]
   \begin{minipage}{0.48\textwidth}
      \centering
          \includegraphics[angle=0,width=0.9\textwidth,clip=true,angle=0]{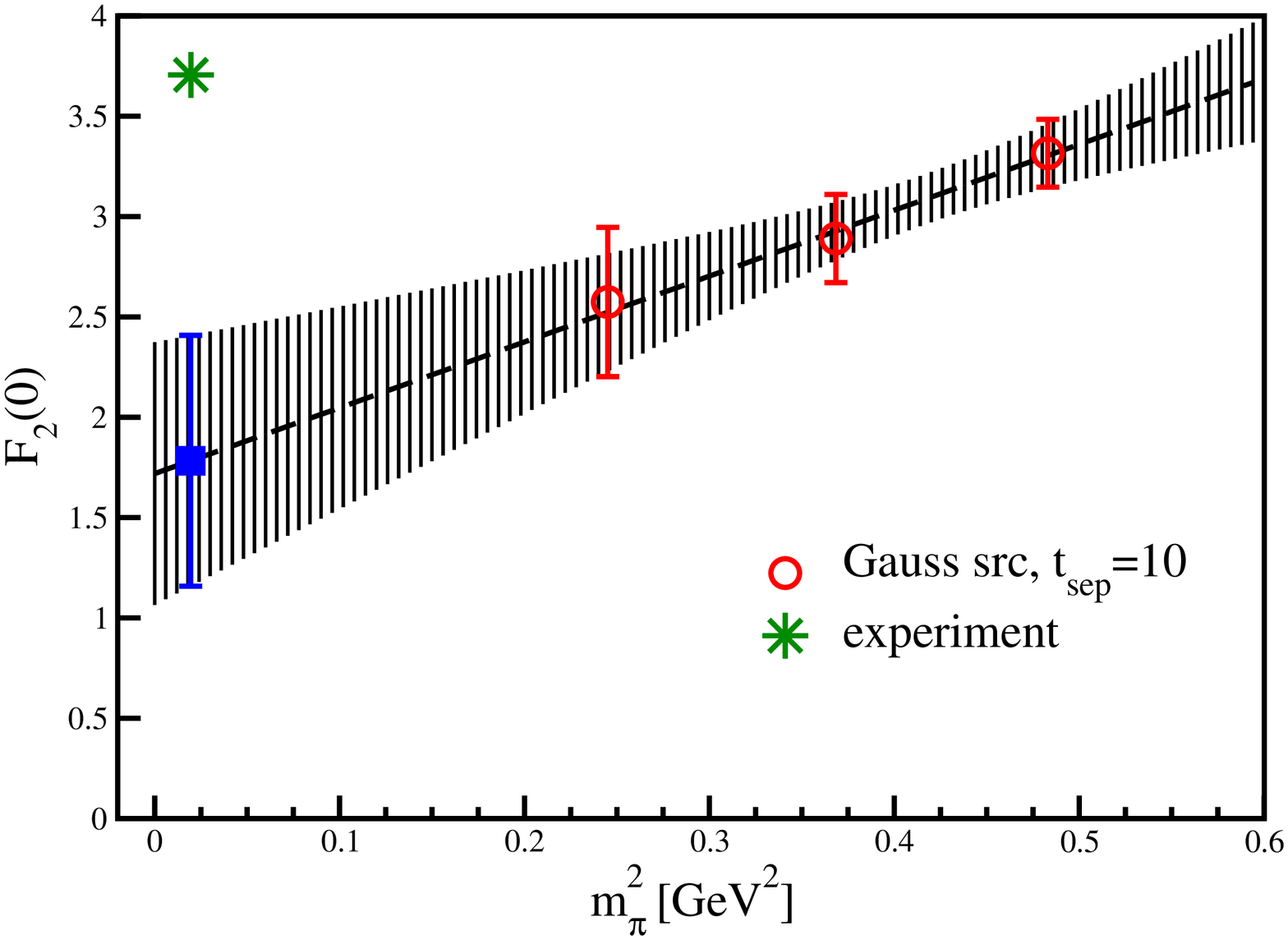}
  \caption{Isovector anomalous magnetic moment for $n_f=2$ flavors of DW fermions 
  (from \cite{Lin:2008uz}).}
  \label{F2_chi_RBCUKQCD}
     \end{minipage}
     \hspace{0.5cm}
    \begin{minipage}{0.48\textwidth}
      \centering
          \includegraphics[angle=0,width=0.9\textwidth,clip=true,angle=0]{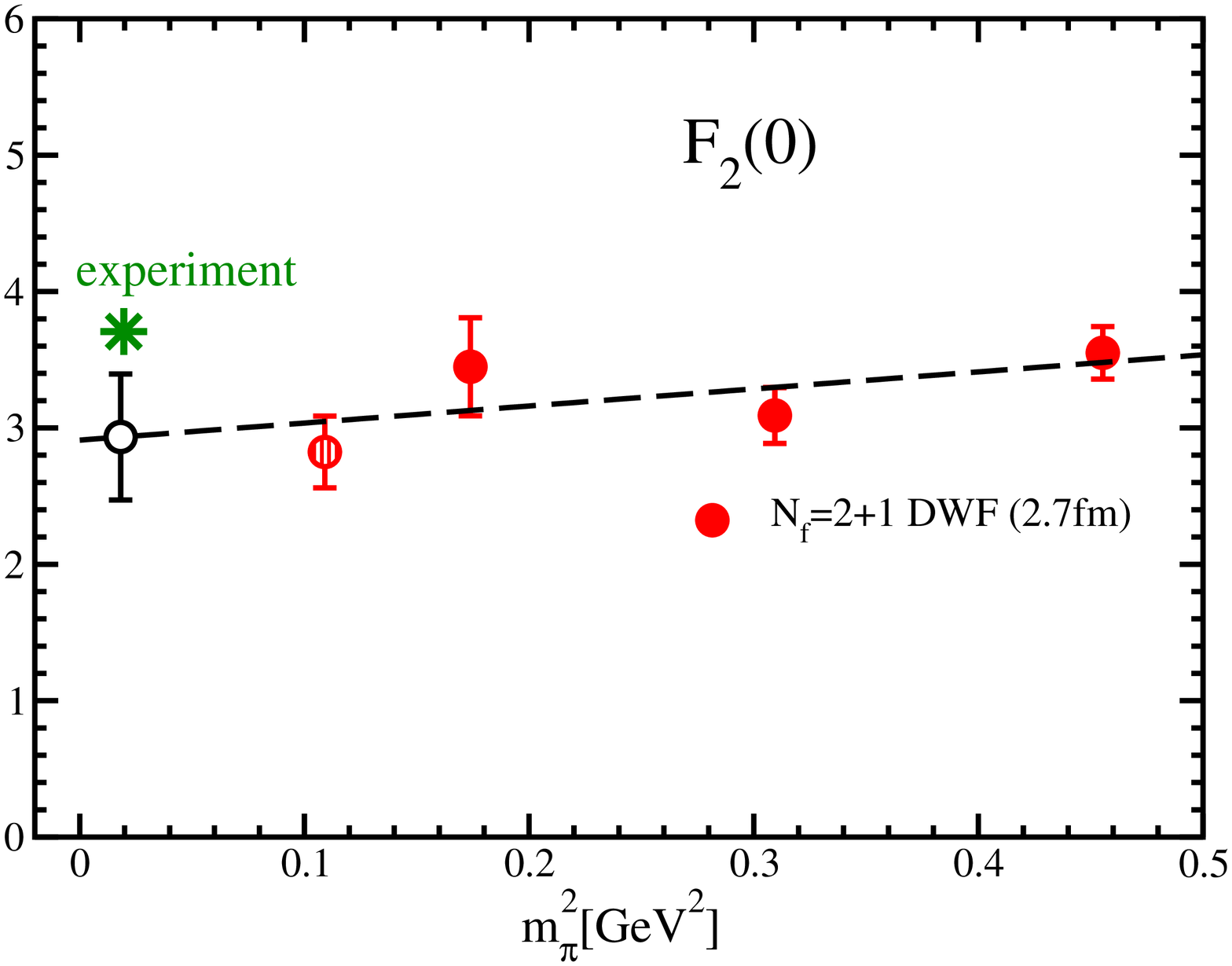}
  \caption{Isovector anomalous magnetic moment for $n_f=2+1$ flavors of 
  DW fermions with $t_{\text{sep}}=12$ (from proceedings \cite{Ohta:2008kd}).}
  \label{mfF_t_RBCUKQCD_Lat2008}
     \end{minipage}
 \end{figure}
%

Figure \ref{mfr_1_RBCUKQCD_Lat2008} 
shows the pion mass dependence of the Dirac rms radius from these $n_f=2+1$ simulations, as obtained from dipole fits to $F_1(Q^2)$.
Even at the lowest accessible pion mass of $\simeq331\MeV$, no sign of an upwards bending
is visible. Consequently, the notorious linear extrapolation to the physical pion mass,
represented by the dashed line, fails.
A comparison of Fig.~\ref{msr_v_chi_RBCUKQCD} and Fig.~\ref{mfr_1_RBCUKQCD_Lat2008}, 
shows that the $n_f=2+1$ results lie systematically above than the values
obtained for $n_f=2$ flavors of DW fermions. Though the effect is not dramatic
within statistical errors, it would be interesting to find out if this
is related to a systematic uncertainty, e.g. the setting of the scale or 
the different source-sink separations, or a genuine effect due to the dynamical strange quark.

Results for the isovector anomalous magnetic moment, $\kappa_{u-d}$, from RBC and RBC-UKQCD
for $n_f=2$ and $n_f=2+1$ flavors of DW fermions are displayed
in Fig.~\ref{F2_chi_RBCUKQCD} and Fig.~\ref{mfF_t_RBCUKQCD_Lat2008}, respectively.
Note that in both cases the pion mass dependent lattice nucleon mass has been used to
extract $F^{u-d}_2(Q^2)$ (cf. Eq.\ref{NuclVec3}), which was then fitted with a dipole ansatz
and extrapolated to $Q^2=0$ to obtain $\kappa_{u-d}=F^{u-d}_2(0)$.
While the linear chiral extrapolation of the $n_f=2+1$ lattice data
to the physical pion mass gives a value that is almost
compatible within errors with the number from experiment,
it clearly fails in the case of the $n_f=2$ results.
The latter shows that a naive linear extrapolation of a small number of
data points lacking sufficient statistical precision may be very misleading.

%

\begin{figure}[t]
   \begin{minipage}{0.48\textwidth}
      \centering
          \includegraphics[angle=0,width=0.9\textwidth,clip=true,angle=0]{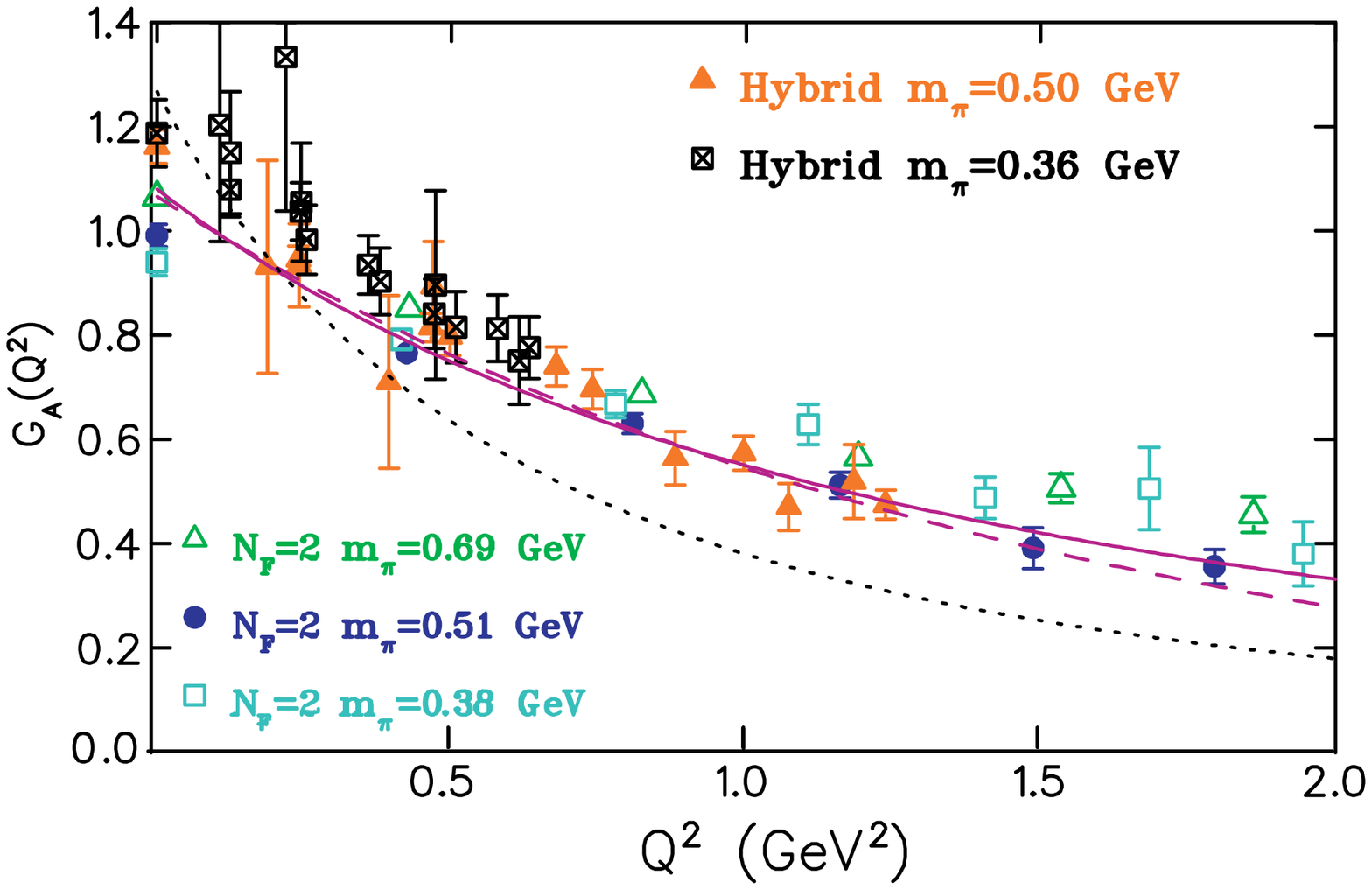}
  \caption{Isovector axial vector form factor $G^{u-u}_A(Q^2)$ (from \cite{Alexandrou:2007zz};
  the figure has been edited for this review).}
  \label{GAHA_LHPC_CyprusMIT}
     \end{minipage}
     \hspace{0.5cm}
    \begin{minipage}{0.48\textwidth}
      \centering
          \includegraphics[angle=0,width=0.9\textwidth,clip=true,angle=0]{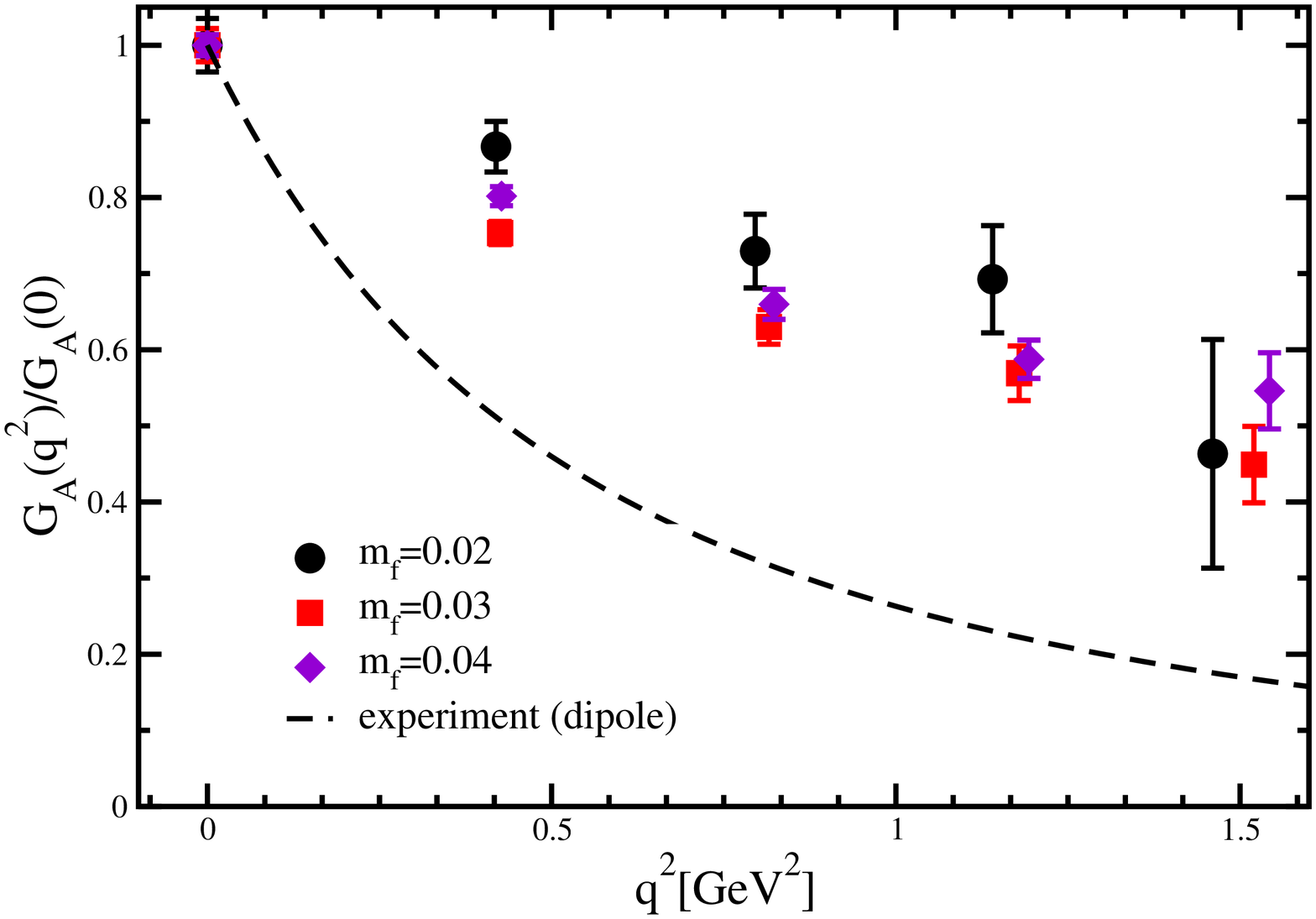}
  \caption{Normalized isovector axial vector form factor $G^{u-d}_A(Q^2)/G^{u-d}_A(0)$ 
  for $n_f=2$ flavors of DW fermions (from \cite{Lin:2008uz}).}
  \label{Ga_mom_RBCUKQCD}
     \end{minipage}
 \end{figure}
%
%

\subsubsection{Axial vector and pseudoscalar form factor}
All results discussed in this section have been non-perturbatively renormalized. 
For the details, we refer to the original works.

Having discussed the vector form factors,
we now turn our attention towards the axial-vector form factors of the nucleon.
Figure \ref{GAHA_LHPC_CyprusMIT} shows results for the proton isovector axial-vector form factor 
$G^{u-d}_A(Q^2)$ from the Athens-Cyprus-MIT collaboration \cite{Alexandrou:2007zz} 
for $n_f=2$ flavors of Wilson fermions together 
with corresponding results based on the $n_f=2+1$ hybrid calculation from LHPC,
for pion masses as indicated in the legends.
The underlying axial-vector currents have been non-perturbatively renormalized 
based on Refs.~\cite{Becirevic:2005ta,Edwards:2005ym}.
We first note that the normalizations of the lattice data, i.e. the forward limit
values $g_A=G_A(Q^2=0)$, do not agree for the different dataset. 
In particular,
the result from the hybrid calculation does not agree with the Wilson fermion result
at similar pion masses, i.e.
for $m_\pi\sim500,510\text{ MeV}$ and $m_\pi\sim360,380\text{ MeV}$. Possible reasons for
such discrepancies, as well as the pion mass dependence of $g_A$ will be discussed
in a separate paragraph below. Apart from the normalization, and considering the scatter 
at the lower pion masses, there is overall good agreement
of the lattice data points within statistical errors . Specifically, there is no clear systematic pion mass 
dependence visible at larger $Q^2$, and the slope in $Q^2$ seems to be approximately
independent of $m_\pi$. 
This observation is in agreement with the $n_f=2$ domain wall fermion data from 
RBC \cite{Lin:2008uz} for $G^{u-d}_A(Q^2)/G^{u-d}_A(0)$ displayed in Fig.~\ref{Ga_mom_RBCUKQCD},
where for example at $Q^2\sim0.4\text{ GeV}$ the lattice value is first decreasing 
going from $m_\pi=695\text{ MeV}$ to $m_\pi=607\text{ MeV}$, only to increase
to an even larger value for $m_\pi=493\text{ MeV}$. It has been speculated 
that this increase at the lowest pion mass is a finite volume effect \cite{Lin:2008uz},
which would be quite remarkable considering that $m_\pi L\approx 4.75$ for this ensemble.
In any case, the $Q^2$-slope of the lattice results shown in Figs.~\ref{GAHA_LHPC_CyprusMIT} and 
\ref{Ga_mom_RBCUKQCD} is consistently much flatter than the one observed in experiment,
represented by the dashed dipole curves. 
%
%
\begin{figure}[t]
   \begin{minipage}{0.48\textwidth}
      \centering
          \includegraphics[angle=0,width=0.95\textwidth,clip=true,angle=0]{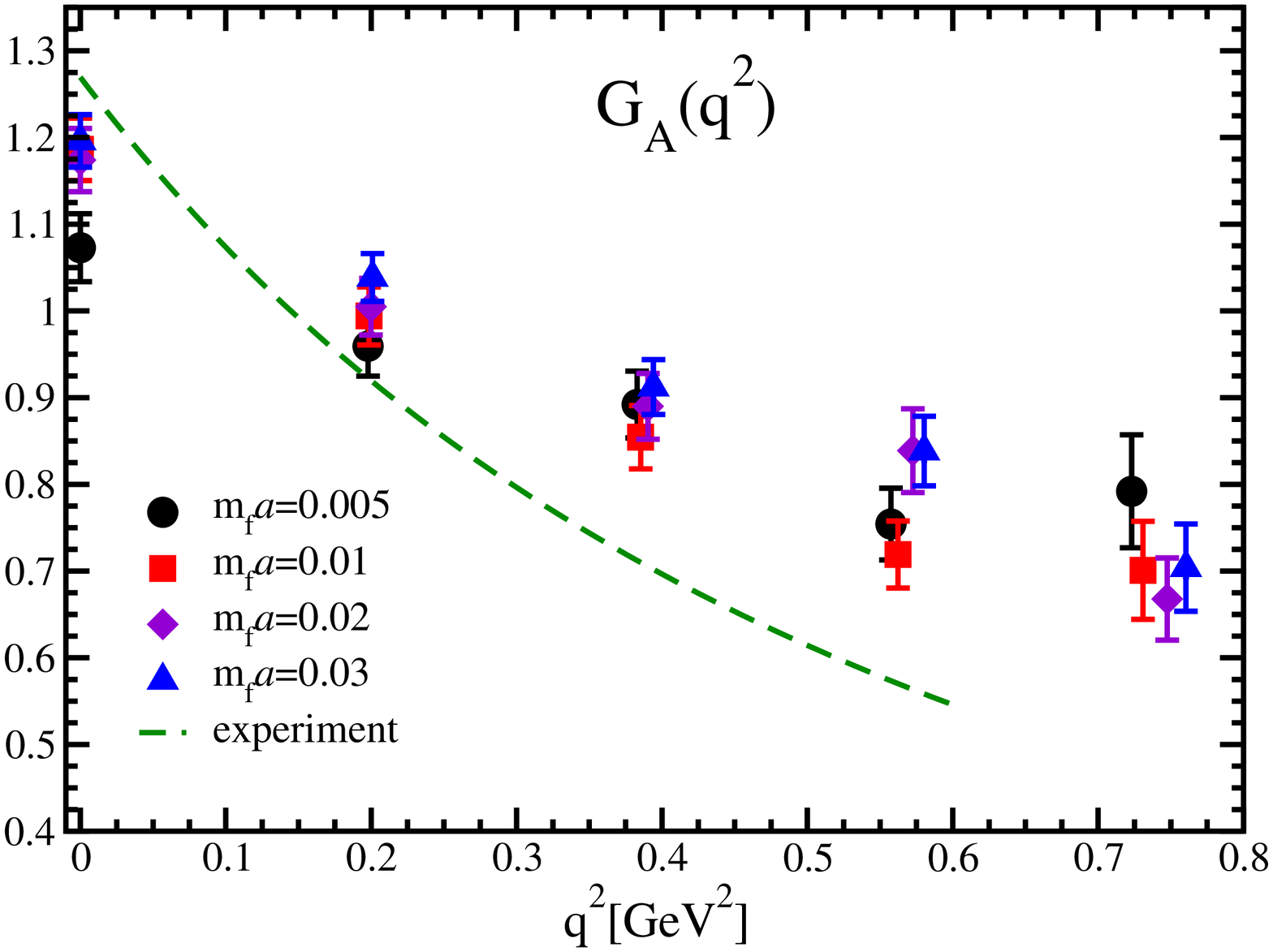}
  \caption{Isovector axial vector form factor $G^{u-u}_A(Q^2)$ for $n_f=2+1$ flavors of 
  DW fermions with (from proceedings \cite{Ohta:2008kd}).}
  \label{F_a_g_v_RBCUKQCD_Lat2008}
     \end{minipage}
     \hspace{0.5cm}
    \begin{minipage}{0.48\textwidth}
      \centering
          \includegraphics[angle=0,width=0.9\textwidth,clip=true,angle=0]{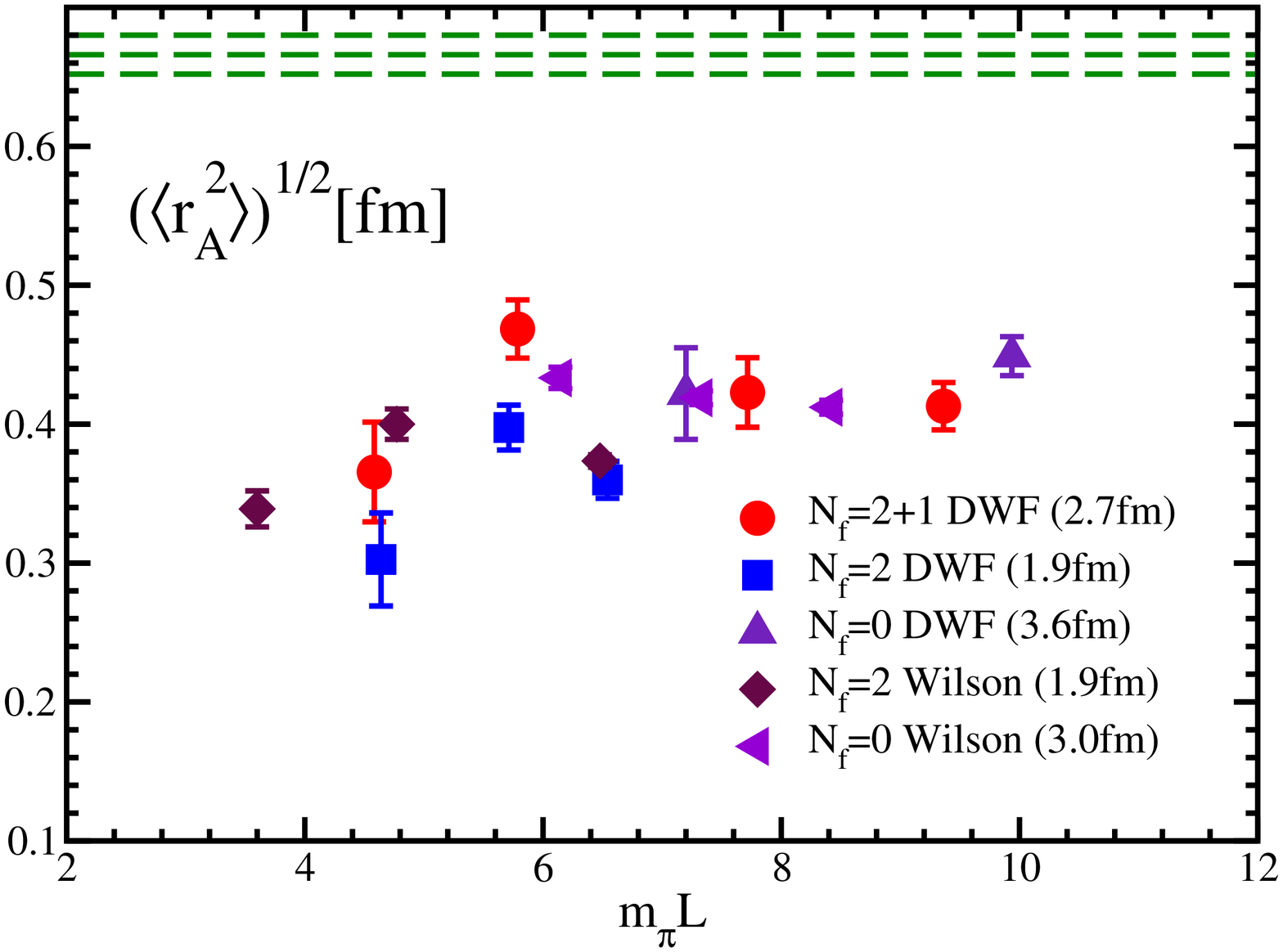}
  \caption{Isovector axial vector rms radius (from proceedings \cite{Ohta:2008kd}).}
  \label{mLr_a_c_RBCUKQCD_Lat2008}
     \end{minipage}
 \end{figure}
%
More recent results from RBC-UKQCD for $G^{u-d}_A(Q^2)$, for $n_f=2+1$ flavors of DW fermions are 
shown in Fig.~\ref{F_a_g_v_RBCUKQCD_Lat2008} \cite{Ohta:2008kd}. 
The different quark masses $m_u=m_d$ in the 
legend correspond to pion masses of $\approx331$, $\approx419$, $\approx557$ and $\approx672\MeV$.
Apart from the results for the lightest quark mass, $ma=0.005$, which may be subject to
finite volume effects, the lattice data points seem to come closer
to the experimental result (indicated by the dashed line) as the quark mass decreases.

An overview of different dynamical and quenched lattice calculations, including the $n_f=2+1$ DW fermion results,
of the axial vector radius $\langle r_A^2\rangle^{1/2}$
obtained from dipole fits to $G^{u-d}_A(Q^2)$, is given 
in Fig.~\ref{mLr_a_c_RBCUKQCD_Lat2008} \cite{Ohta:2008kd}.
The data points lie in a range of $\langle r_A^2\rangle^{1/2}=\approx0.3,\ldots,0.5\fm$
and are 
significantly below experiment indicated by the dashed lines,
where values between $\langle r_A^2\rangle^{1/2}=0.57\fm$ and $0.72\fm$ have been reported
(see, e.g., the discussion in \cite{Schindler:2006it}).
We note that lattice results for $m_\pi L \lessapprox5$ are somewhat lower
than the rest, however contrary to the claim in \cite{Ohta:2008kd},
no clear ``scaling'' in $m_\pi L$ can be observed. In particular the data points
with $m_\pi L \approx6.0,\ldots,6.5$ scatter significantly, indicating that
other systematic effects related to the pion mass dependence, the range 
of $Q^2$ used in the underlying dipole fits, the setting of the scale, etc.,
are relevant in this case.

%
\begin{figure}[t]
    \begin{minipage}{0.48\textwidth}
      \centering
          \includegraphics[angle=0,width=0.95\textwidth,clip=true,angle=0]{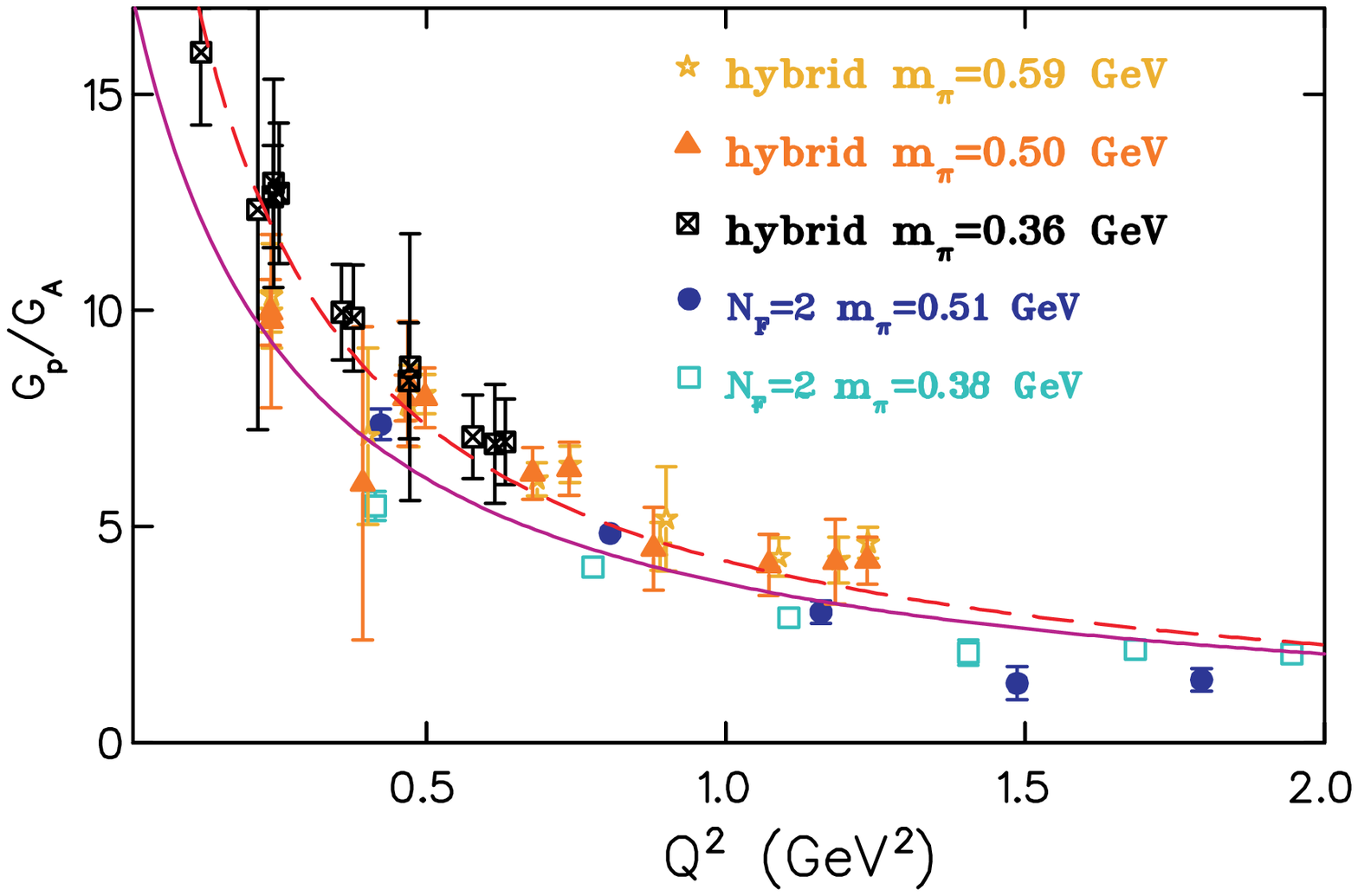}
  \caption{Ratio $\big(G^{u-d}_A/G^{u-d}_P\big)(Q^2)$ (from \cite{Alexandrou:2007zz};
  the figure has been edited for this review).}
  \label{GPoverGA_CyprusMIT}
     \end{minipage}
   \hspace{0.5cm}
   \begin{minipage}{0.48\textwidth}
      \centering
          \includegraphics[angle=0,width=0.9\textwidth,clip=true,angle=0]{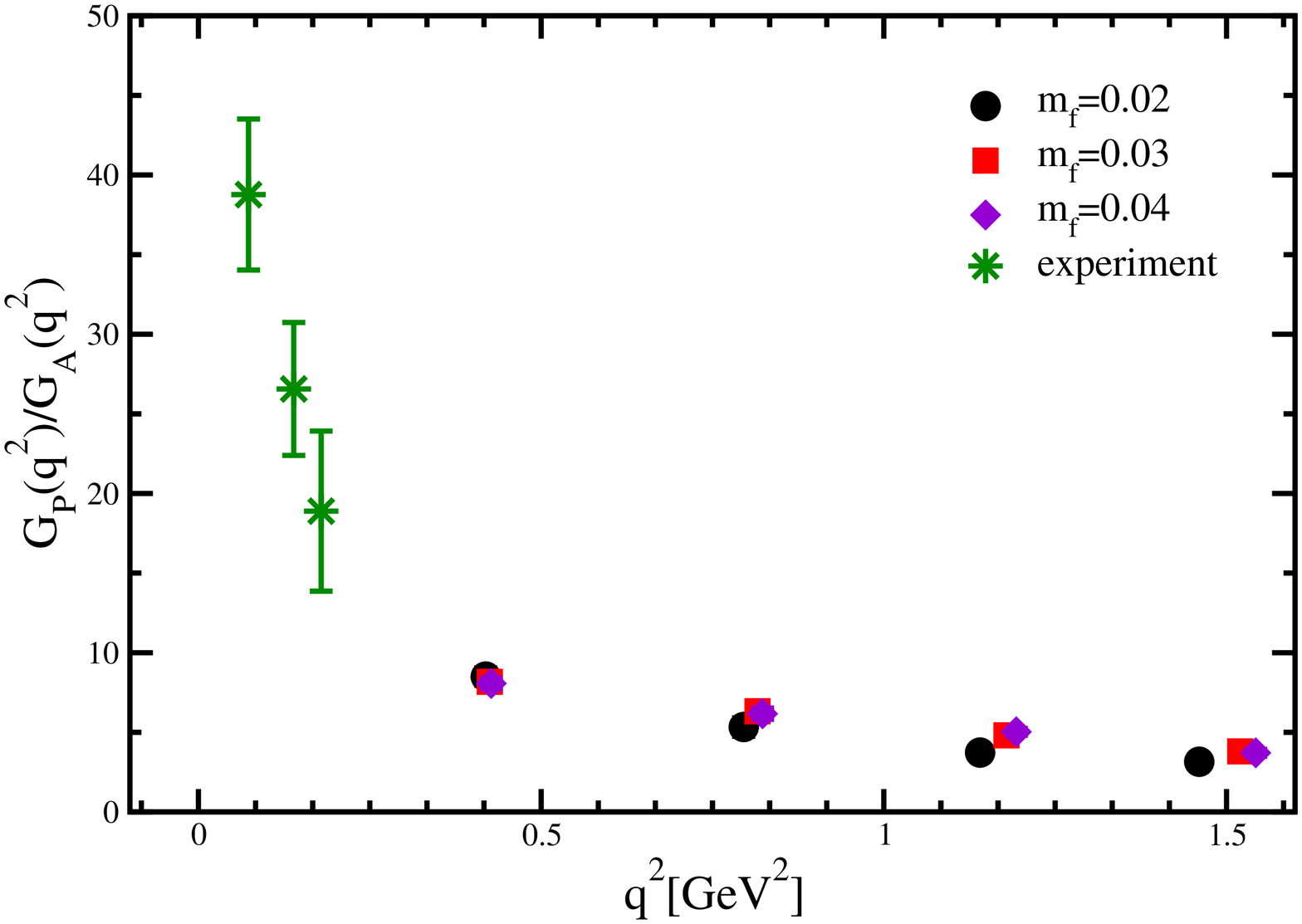}
  \caption{Ratio $\big(G^{u-d}_A/G^{u-d}_P\big)(Q^2)$ (from \cite{Lin:2008uz}).}
  \label{GpGa_mom_RBCUKQCD}
     \end{minipage}
 \end{figure}
%

Results for the induced pseudo-scalar form factor in form
of the ratio $\big(G^{u-d}_P/G^{u-d}_A\big)(Q^2)$ from the Athens-Cyprus-MIT collaboration
are shown in Fig.~\ref{GPoverGA_CyprusMIT}, for the same
ensembles and lattice data as in Fig.~\ref{GAHA_LHPC_CyprusMIT}.
The ratio decreases rapidly for larger $Q^2$ as is expected from
pion pole dominance, which states that 
\bea
\label{pionpole}
\frac{G^{u-d}_P(Q^2)}{G^{u-d}_A(Q^2)}\sim\frac{4m_N^2}{m_\pi^2+Q^2}\,.
\eea
This is plotted for comparison as dashed line in Fig.~\ref{GPoverGA_CyprusMIT} 
for a lattice pion and nucleon mass of $m_\pi=0.411\text{ GeV}$ and $m_N=1.109\text{ GeV}$ 
respectively, and describes the overall trend of the lattice data well.
While the results of the hybrid calculation appear to be overall consistent
within errors, some discrepancies are visible for the $n_f=2$ Wilson fermion results.
Specifically, taking into account that the ratio in Fig.~\ref{GPoverGA_CyprusMIT}
is independent of renormalization issues,
it is quite remarkable that the values obtained from the $n_f=2$ Wilson fermion calculation
are $\approx40\%$ below the result from the $n_f=2+1$ hybrid calculation,
at the lowest pion masses of $m_\pi\approx360,\ldots,370\text{ MeV}$ and
for a momentum transfer squared of $Q^2\approx0.4\text{ GeV}$.
Since for the Wilson fermion data point $m_\pi L\sim 3.5$,
in comparison to a much larger value of $m_\pi L\sim 6.25$ for the hybrid case,
it may be that finite volume effects are at least partially responsible
for the observed discrepancy.

Figure \ref{GpGa_mom_RBCUKQCD} displays results for $\big(G^{u-d}_P/G^{u-d}_A\big)(Q^2)$
from the $n_f=2$ domain wall fermion calculation by RBC \cite{Lin:2008uz}, 
together with experimental data points from \cite{Choi:1993vt}. 
Although the experimental errors 
are rather large, Fig.~\ref{GpGa_mom_RBCUKQCD} still shows exemplary 
the steep rise at lower values of the momentum transfer squared that
may be observed in future lattice studies of $G_P(Q^2)$ at 
lower pion masses and smaller $Q^2$.

\subsubsection{Axial vector coupling constant}
\label{sec:axialvector}
Having discussed the $Q^2$-dependence of the axial-vector and induced pseudo-scalar
form factors, we now turn our attention to the axial-vector coupling, $g_A$,
which we identify with the forward value of the nucleon axial-vector form factor, 
$g_A=G_A(Q^2=0)$, see Eq.~(\ref{BetaDecay1}).
In the following, we will be concentrating on the isovector channel, $g_A=g^{u-d}_A$.
Results for $g^{u+d}_A=\Delta\Sigma^{u+d}=G^{u+d}_A(Q^2=0)$ in the isosinglet channel
will be presented in the framework of the nucleon spin structure in section \ref{sec:SpinStructure} below. 

\begin{figure}[t]
   \begin{minipage}{0.48\textwidth}
      \centering
          \includegraphics[angle=0,width=0.9\textwidth,clip=true,angle=0]{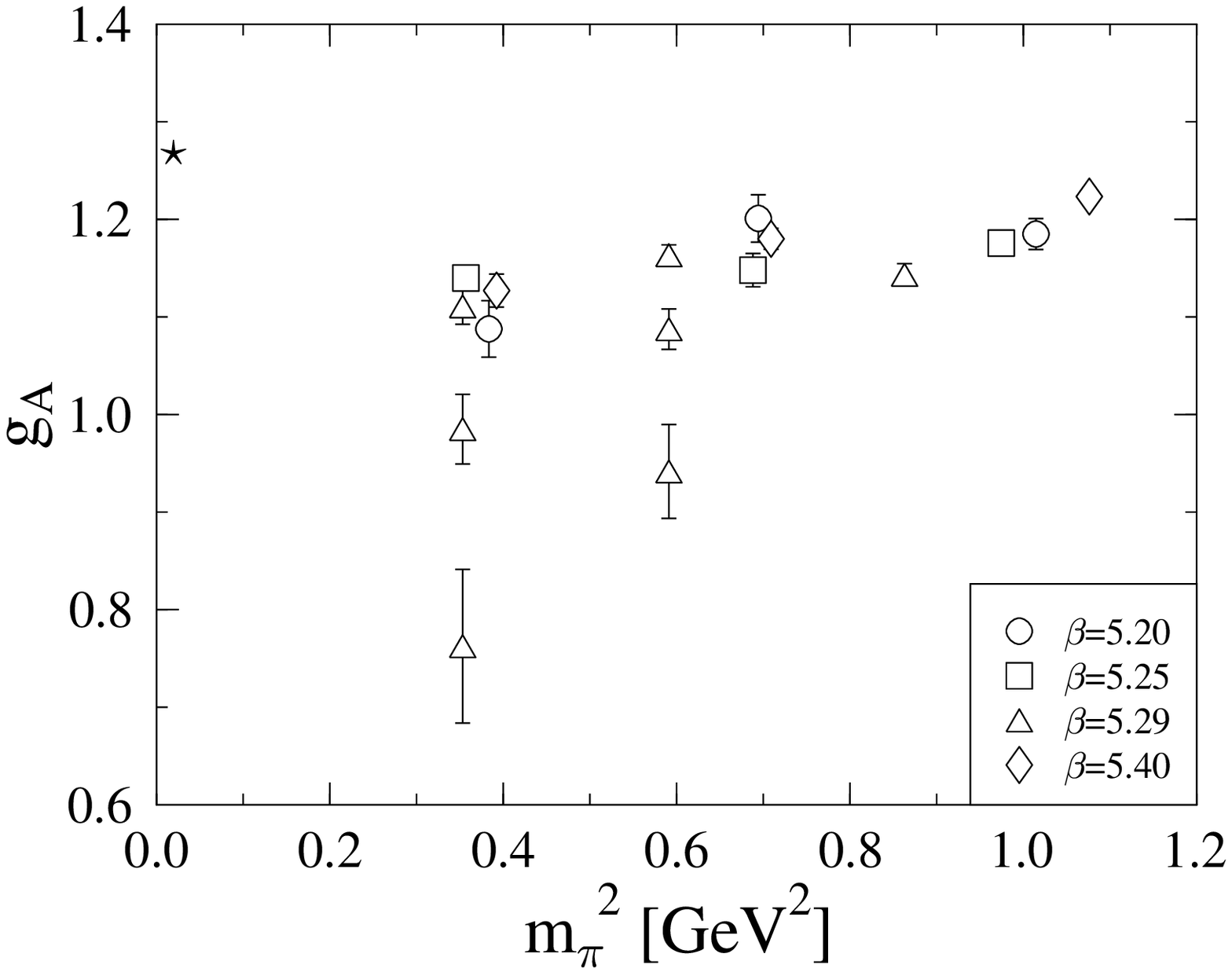}
  \caption{Isovector axial vector coupling constant $g_A$ (from \cite{Khan:2006de}).}
  \label{gadat_QCDSF}
     \end{minipage}
     \hspace{0.5cm}
    \begin{minipage}{0.48\textwidth}
      \centering
          \includegraphics[angle=0,width=0.9\textwidth,clip=true,angle=0]{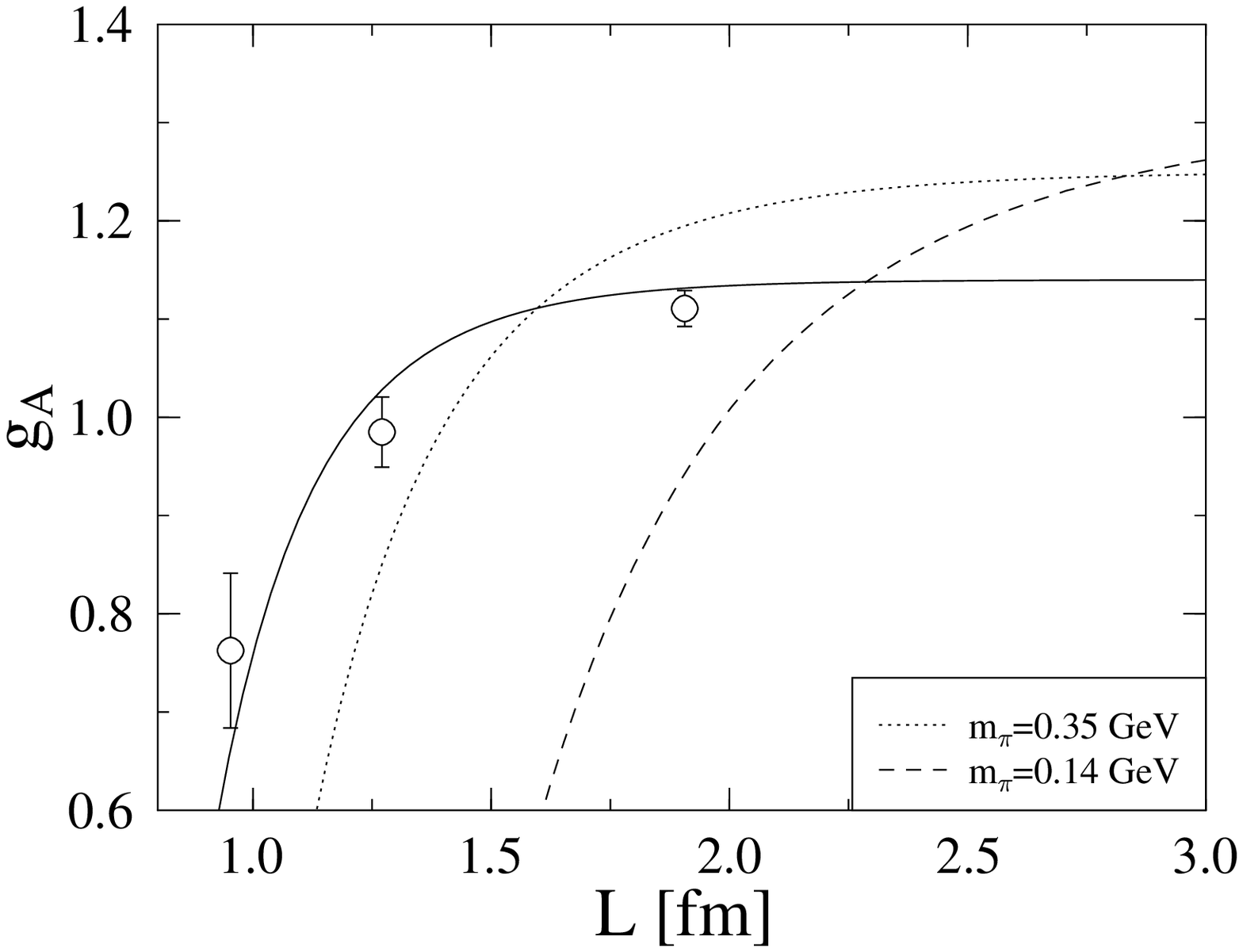}
  \caption{Volume dependence of the isovector axial vector coupling constant $g_A$ (from \cite{Khan:2006de}).}
  \label{ga1L_QCDSF}
     \end{minipage}
 \end{figure}

Noting that the isovector axial vector coupling constant, $g^{u-d}_A$,
plays a central role in QCD low energy dynamics, that it can be 
comparatively easily accessed on the lattice through the local isovector  
axial vector current, and that its experimental value is rather well known, it
may be seen as an important ``benchmark observable'' for lattice QCD calculations \cite{Edwards:2005kw}.
Experimentally, it is known to high statistical precision from neutron beta decay, with 
an average value of $g_A=1.2695(29)$ from the PDG \cite{PDG2008}.
The QCDSF-UKQCD collaboration has studied $g_A$ on the basis of a large number
of ensembles for different pion masses in the range of $m_\pi\approx600\MeV$
to $m_\pi\approx1200\MeV$, lattice spacings from $a\approx0.07\fm$  to $a\approx0.11\fm$\footnote{In this analysis, 
the lattice spacings have not been chirally extrapolated, but taken at the individual quark masses.}
and different volumes from $V\approx(0.95\fm)^3$ to $V\approx(2.0\fm)^3$, 
using $n_f=2$ flavors of improved Wilson fermions \cite{Khan:2006de}. 
An overview of the results
plotted versus $m_\pi^2$ is given in Fig.~\ref{gadat_QCDSF}.
To understand the distribution of the lattice data points in Fig.~\ref{gadat_QCDSF},
it is important to note that the two lowest lying points
correspond to finite volume simulations with smallest lattice volumes of only $V\approx(1.0\text{ fm})^3$.
At least one more low lying point for $\beta=5.29$ can be identified that 
also correspond to a finite volume calculation with $V\approx(1.3\text{ fm})^3$,
while the bulk of the lattice data points in Fig.~\ref{gadat_QCDSF} was obtained for larger volumes.
This strongly indicates that finite volume effects may be significant, i.e. 
larger than $10\%$, for $m_\pi L<4$. 

\begin{figure}[t]
   \begin{minipage}{0.48\textwidth}
      \centering
          \includegraphics[angle=0,width=0.9\textwidth,clip=true,angle=0]{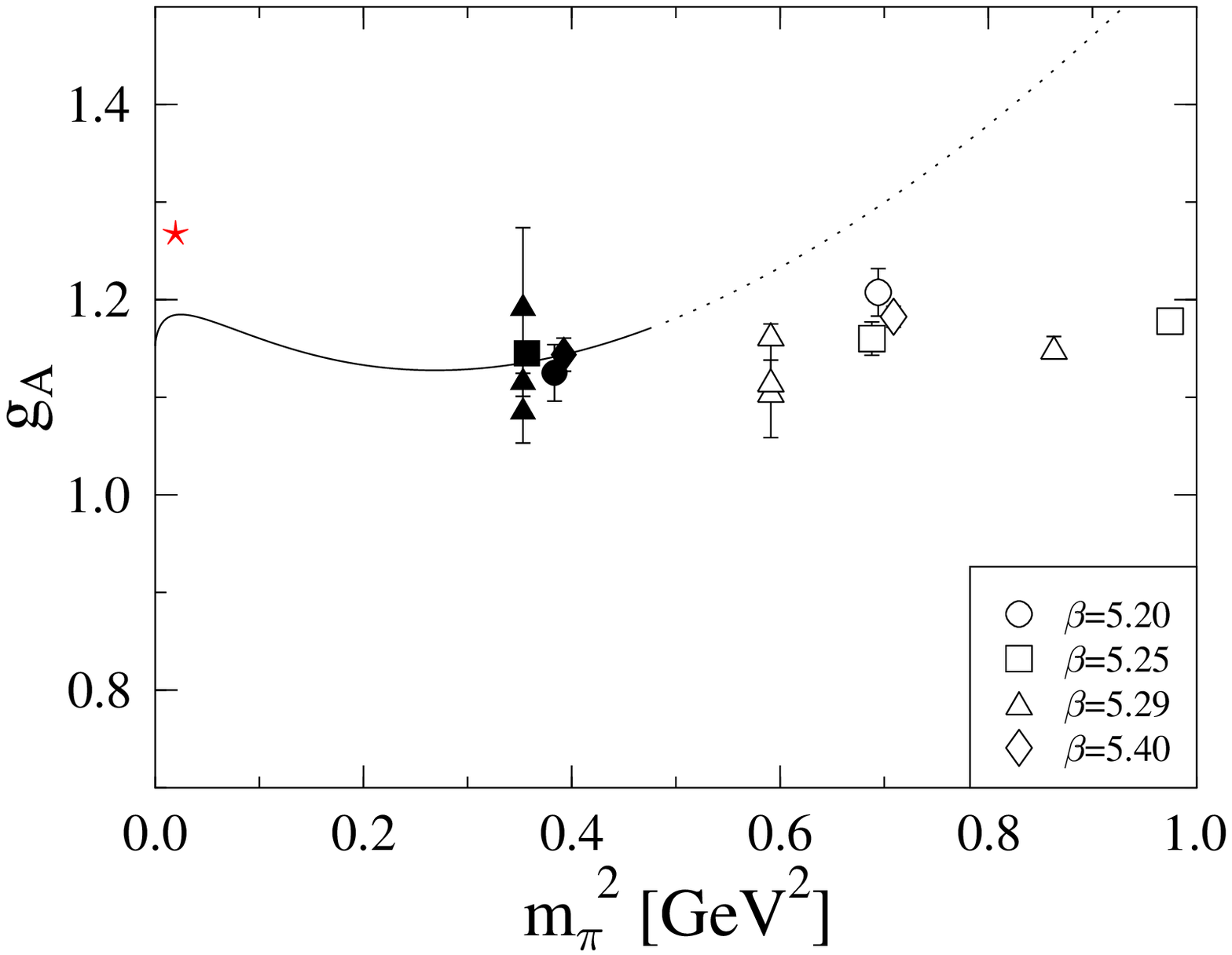}
  \caption{Isovector axial vector coupling constant $g_A$ (from \cite{Khan:2006de}).}
  \label{gafitext1_QCDSF_v2}
     \end{minipage}
     \hspace{0.5cm}
    \begin{minipage}{0.48\textwidth}
      \centering
          \includegraphics[angle=-90,width=0.95\textwidth,clip=true,angle=0]{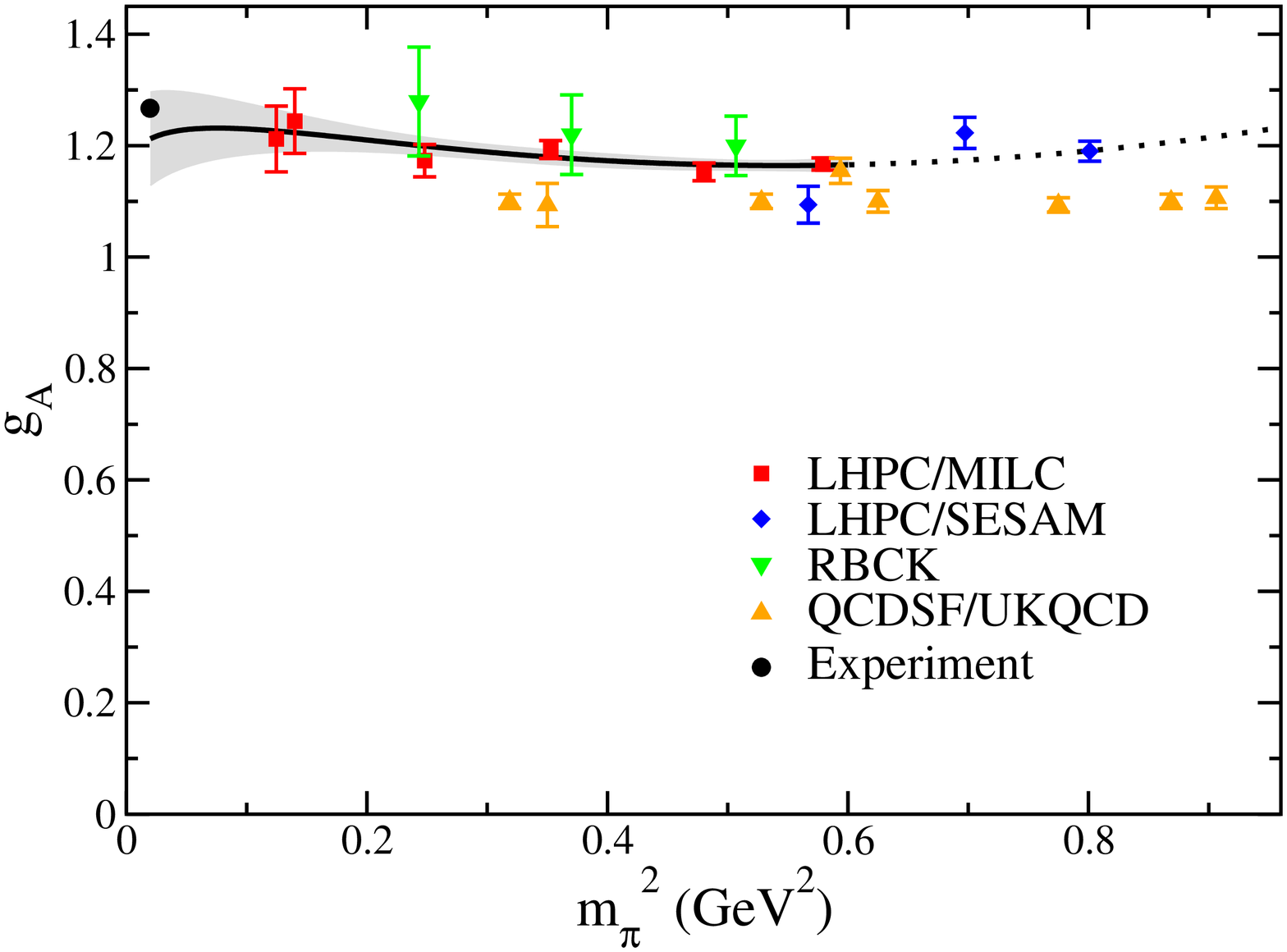}
  \caption{Isovector axial vector coupling constant $g_A$ \cite{Edwards:2005ym} 
  compared to results from QCDSF/UKQCD \cite{Khan:2006de}, LHPC/SESAM \cite{Dolgov:2002zm} 
  and RBC \cite{Ohta:2004mg} (figure from \cite{Edwards:2005ym}).}
  \label{gAfig2_LHPC}
     \end{minipage}
 \end{figure}

Noting that the inclusion of the $\Delta(1232)$-resonance is 
crucial to obtain value of $g_A$ larger than $1$ from the Adler-Weisberger sum rule, 
the pion mass dependence of $g_A$ has been worked out to
$\mathcal{O}(\eps^3)$ SSE-HBChPT (with explicit $\Delta$-intermediate states) in \cite{Hemmert:2003cb}
in infinite volume. This has been extended to finite volume in \cite{Wollenweber:2005,Khan:2006de} 
in the framework of the SSE of HBChPT, and a similar 
finite volume calculation of $g_A$ with explicit $\Delta$-DOFs can also be found in \cite{Beane:2004rf}.
The result can be written as $g_A(L)=g_A(L=\infty)+\Delta g_A(L)$, with 
 \begin{eqnarray}
g^{\text{SSE}}_A(L=\infty)
   &=& g_A^0 + \left\{4\,B_9^r(\lambda) - 8\,g_A^0B_{20}^r(\lambda) - \frac{(g_A^0)^3}{16\pi^2f_\pi^2}
    -\frac{25c_A^2g_1}{324\pi^2f_\pi^2} + \frac{19c_A^2g_A^0}{108\pi^2f_\pi^2}\right\}m_\pi^2 \nonumber \\
    &-& \left((g^0_A)^3 + \frac{1}{2}\,g^0_A\right)
         \frac{m_\pi^2}{4\pi^2 f_\pi^2}\ln{\left(\frac{m_\pi}{\lambda }\right)} 
   +\frac{4c_A^2g_A^0}{27\pi\Delta f_\pi^2}\,m_\pi^3 \nonumber \\
   &+& \left(25c_A^2g_1\Delta^2-57c_A^2g_A^0\Delta^2 - 24c_A^2g_A^0m_\pi^2\right)
      \frac{\sqrt{m_\pi^2-\Delta^2}}{81\pi^2f_\pi^2\Delta}
                  \arccos\left(\frac{\Delta}{m_\pi}\right)\nonumber \\
    &+& \frac{25c_A^2g_1\left(2\Delta^2-m_\pi^2\right)}{162\pi^2f_\pi^2}
                                              \ln\left(\frac{2\Delta}{m_\pi}\right)
    +\frac{c_A^2g_A^0\left(3m_\pi^2-38\Delta^2\right)}{54\pi^2f_\pi^2}
                                             \ln\left(\frac{2\Delta}{m_\pi}\right)
\label{gASSE}
\end{eqnarray}
to $\mathcal{O}(\eps^3)$ in the SSE \cite{Hemmert:2003cb,Khan:2006de}.
The expansion involves the LECs $f_\pi$, $g^0_A$,
$c_A=g_{\pi N\Delta}$, $\Delta m_{N\Delta}$, $g_1$ (all in the chiral limit) and the 
scale dependent counter-terms $B^r_{9,20}(\lambda)$.
For the complete (somewhat lengthy) expression for $\Delta g_A(L)$ at $\mathcal{O}(\eps^3)$,
we refer to \cite{Wollenweber:2005,Khan:2006de}. Since the finite volume formulas are essentially obtained 
from infinite volume expressions by replacing momentum integrals by sums over discrete momentum modes,
$\mbf{p}=2\pi/L \mbf{n}$ with integer $\mbf{n}_i$, $\Delta g_A(L)$ does not depend on any additional LECs.
On the technical side, we note that the counter terms in \cite{Khan:2006de} have been redefined 
compared to \cite{Hemmert:2003cb} in order to
match the definitions in HBChPT (without explicit $\Delta$-DOFs), which allows
for a direct comparison of the LECs in HBChPT and SSE HBChPT to the given order.
Treating the counter term and the couplings $g_A$ and $g_1$ as free parameters,
while fixing the other LECs to their physical or chiral limit values known from the literature,
the finite volume SSE HBChPT result has been fitted to all lattice data points with 
$m_\pi<770\text{ MeV}$ (including all volumes) in Fig.~\ref{gadat_QCDSF}. The resulting
prediction of the volume dependence of $g_A$ is displayed as solid line
and compared to lattice data points in Fig.~\ref{ga1L_QCDSF} for a pion mass of $\approx770\text{ MeV}$.
Keeping in mind that simulation results for rather large values of $m_\pi$
have been taken into account in the fit, the SSE HBChPT result describes
the lattice data surprisingly well within the statistical errors.
In a second step, the fit results were used to shift the lattice data points to
the infinite volume limit by subtracting the finite volume correction, 
$\Delta g_A(L)=g_A(L)-g_A(L=\infty)$, from the 
lattice data points, $g_A^{\text{lat}}(L=\infty)=g_A^{\text{lat}}(L)-\Delta g_A(L)$,
at the given pion masses. The resulting data points together with
the infinite volume chiral extrapolation are shown Fig.~\ref{gafitext1_QCDSF_v2},
where in contrast to Fig.~\ref{ga1L_QCDSF} only lattice data for 
$m_\pi<630\text{ MeV}$, represented by the filled symbols, have been taken into account
in the chiral fit. Compared to Fig.~\ref{gadat_QCDSF} one finds that the 
different finite volume lattice data points for $g_A$ have been shifted to larger values
in the infinite volume limit, where they agree with each other within statistical errors. 
The central value of the infinite volume chiral extrapolation underestimates the experimental
result by $\approx10\%$ at the physical point, but there is a small overlap 
when the extrapolation error of $\approx10\%$ is taken into account.
Despite the overall success, the application of the SSE HBChPT results 
to lattice data with $m_\pi\gtrapprox600\MeV$ should be regarded with caution,
and it would be important to repeat 
this study including lattice results for pion masses of $m_\pi\approx300\MeV$ and below.

%
%

\begin{figure}[t]
      \centering
          \includegraphics[angle=0,width=0.55\textwidth,clip=true,angle=0]{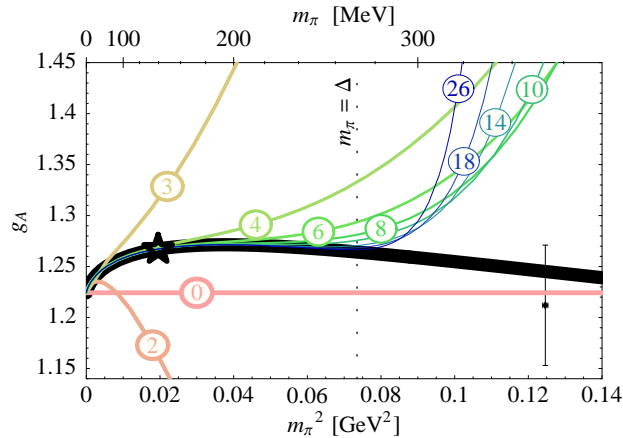}
  \caption{Study of the convergence pattern of the SSE HBChPT extrapolation of $g_A$ to $\mathcal{O}(\eps^3)$ 
  in infinite volume (from \cite{Procura:2006gq}).
  The numbers $n=0,2,3,4,\ldots$ represent the orders at which the expansion in $m_\pi$ was truncated.}
  \label{convergplot_paper}
\end{figure}

A detailed study of the infinite volume SSE HBChPT extrapolation of $g_A$ 
has been presented in \cite{Procura:2006gq}. From a fit to lattice data points
and the experimental value, the convergence pattern of the chiral extrapolation
was numerically investigated by a subsequent series expansion in powers of $m_\pi$.
As displayed in Fig.~\ref{convergplot_paper}, the chiral expansion in $m_\pi$ 
slowly converges only for pion masses below $\approx\myDelta m_{\Delta N}\approx300\MeV$.
This indicates that systematic uncertainties in a conventional HBChPT extrapolation of $g_A$,
where explicit $\Delta$-DOFs have been integrated out and are only included
in the form of LECs (couplings) at higher orders in the chiral Lagrangian, 
may be substantial above $m_\pi\approx300\MeV$. 
An explicit, ``non-perturbative'' inclusion of the $\Delta$-resonance appears to be crucial for $g_A$.

%
%

The results from QCDSF/UKQCD are in overall agreement with calculations performed by LHPC based on the
hybrid approach of domain wall valence and Asqtad sea quarks \cite{Edwards:2005ym}, as can be seen
from Fig.~\ref{gAfig2_LHPC}, where the uncorrected, ``large'' volume results
from Fig.~\ref{gadat_QCDSF} are presented by the upwards pointing triangles. 
In particular for $m_\pi \approx600\text{ MeV}$, 
the QCDSF/UKQCD data points, corresponding to volumes of $V\approx(1.5,\ldots,2\text{ fm})^3$,
would be slightly shifted upwards by finite volume corrections, and therefore agree 
within statistics with the LHPC results, which have been obtained for $V\approx(2.5\text{ fm})^3$
($V\approx(3.5\text{ fm})^3$ for the leftmost data point in Fig.~\ref{gAfig2_LHPC}).
A chiral fit based on results from HBChPT including explicit $\Delta$-DOFs in a finite volume 
 \cite{Beane:2004rf} (very similar to the SSE HBChPT calculation discussed above)
to the LHPC lattice data given by the filled squares is shown in Fig.~\ref{gAfig2_LHPC} as shaded band. 
Finite volume corrections were found to be small and negligible within statistical errors,
as can also be inferred directly from the two leftmost data points in Fig.~\ref{gAfig2_LHPC} 
that were obtained for $V\approx(2.5\text{ fm})^3$ and $V\approx(3.5\text{ fm})^3$, respectively. 
This is in agreement with the results
presented in Fig.~\ref{ga1L_QCDSF}, which point towards small finite
volume corrections for $L\ge2.5\text{ fm}$
even for the lowest accessible pion masses of $m_\pi\approx350\text{ MeV}$, 
i.e. for $m_\pi L\gtrapprox4.5$.
The infinite volume chiral extrapolation based on the LHPC results in
Fig.~\ref{gAfig2_LHPC} is found to be in good agreement with the experimental value
within errors.

\begin{figure}[t]
   \begin{minipage}{0.48\textwidth}
      \centering
          \includegraphics[angle=0,width=0.9\textwidth,clip=true,angle=0]{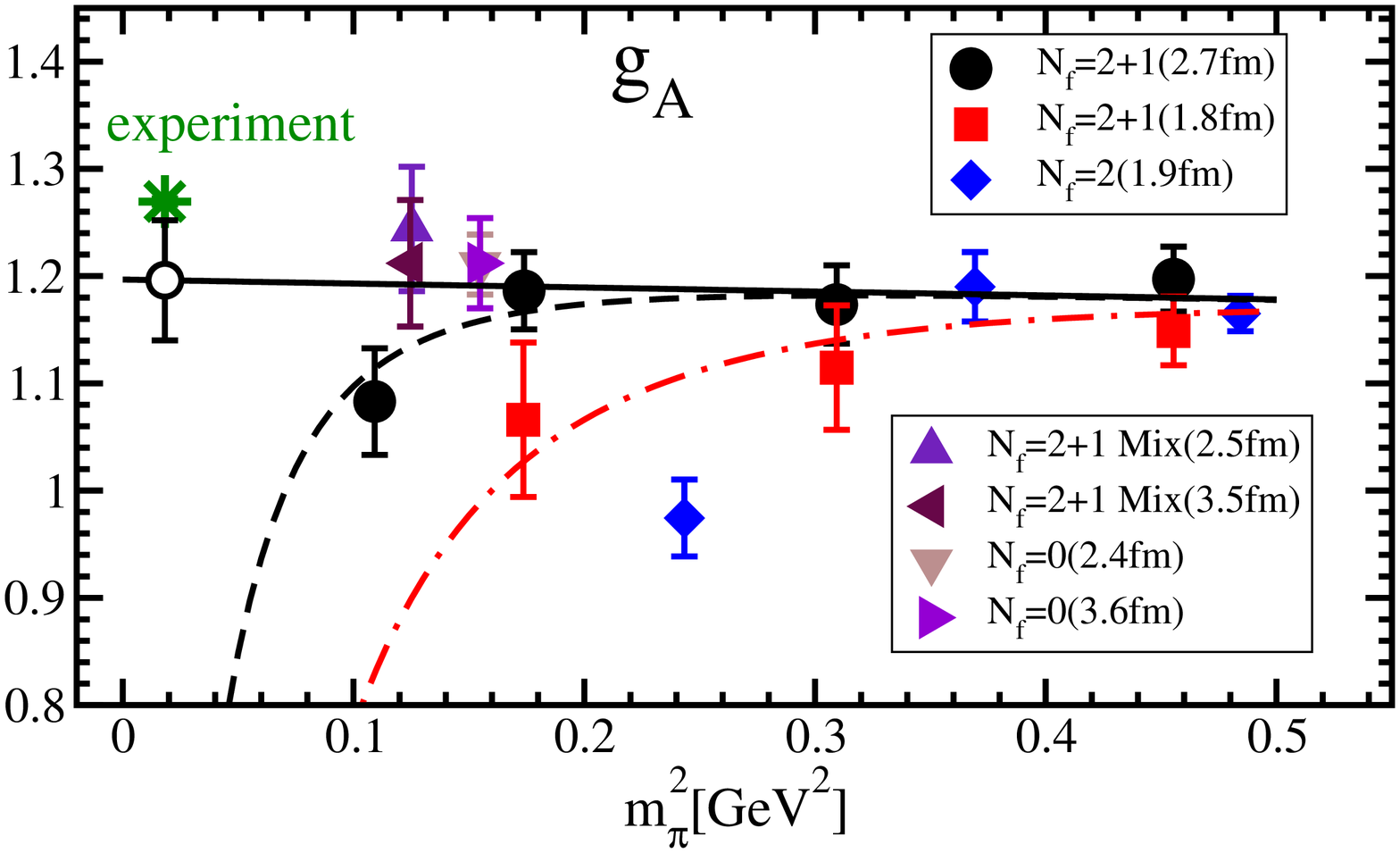}
  \caption{Isovector axial vector coupling constant $g_A$ (from \cite{Yamazaki:2008py}).}
  \label{gA_RBCUKQCD}
     \end{minipage}
     \hspace{0.5cm}
    \begin{minipage}{0.48\textwidth}
      \centering
          \includegraphics[angle=0,width=0.75\textwidth,clip=true,angle=0]{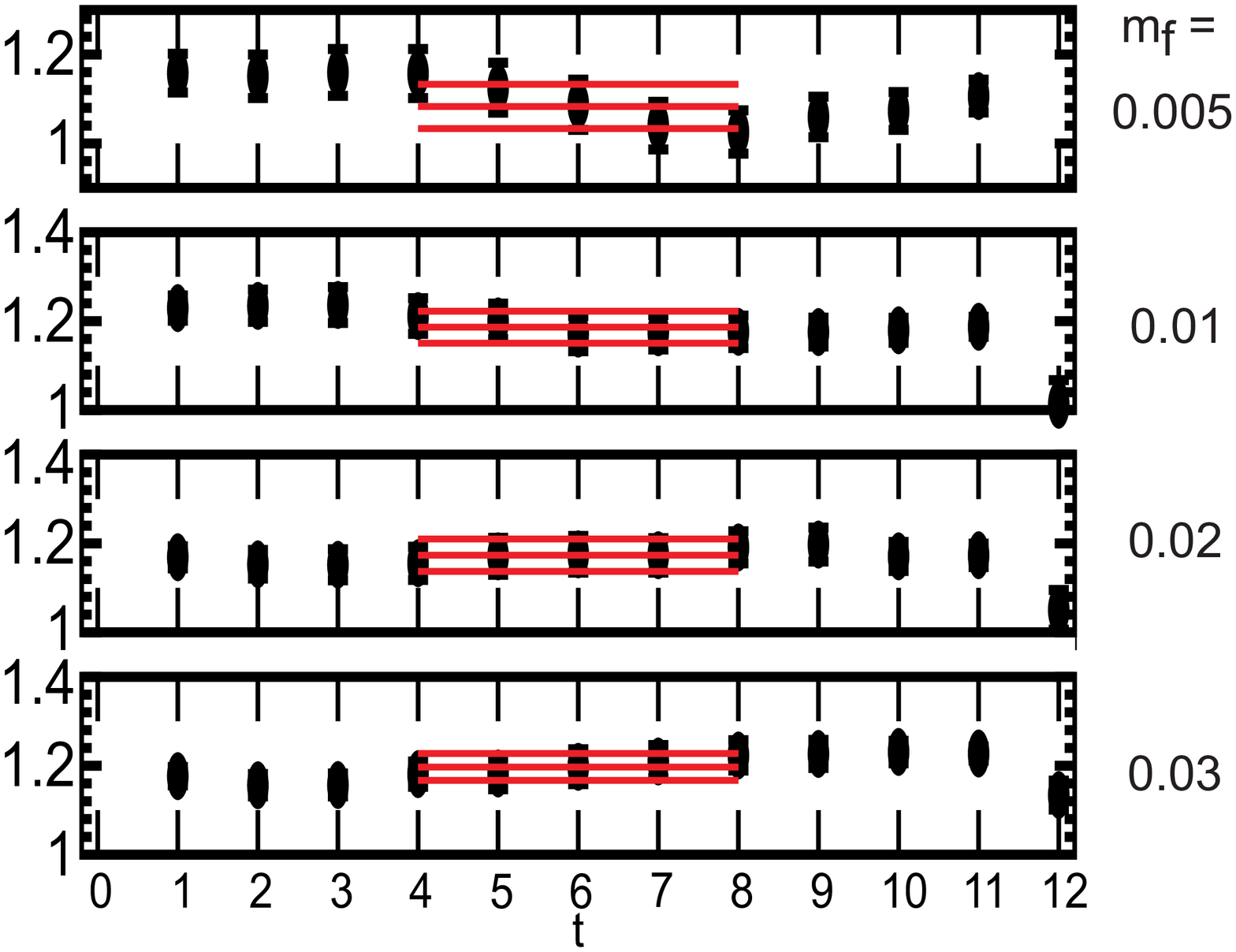}
  \caption{Plateaus for $g_A$ (from \cite{Yamazaki:2008py}).}
  \label{plat_RBCUKQCD}
     \end{minipage}
 \end{figure}

A somewhat different point of view concerning finite volume 
corrections to $g_A$ is expressed in \cite{Yamazaki:2008py}
on the basis of recent results obtained for $n_f=2+1$ flavors
of domain wall fermions displayed in Fig.~\ref{gA_RBCUKQCD}.
It has been noted in particular that the lattice data point for the lowest pion mass
of $m_\pi\approx330\text{ MeV}$ for $V\approx(2.7\text{ fm})^3$ 
in Fig.~\ref{gA_RBCUKQCD} lies $\approx 10\%$ lower than 
the results at $m_\pi\approx420\text{ MeV}$ as well as
the results from LHPC (denoted by ``$N_f=2+1$ Mix'') at $m_\pi\approx350\text{ MeV}$.
Motivated by the empirical observation that
the volume dependence of a number of results for 
$g_A$ from different groups can be approximately described in terms
of the single variable $m_\pi L$, an ansatz of the form
$A+Bm_\pi^2+C\exp(-m_\pi L)$ has been used to simultaneously fit the
volume and pion mass dependence of the lattice data. Results
of such fits are shown by the dashed lines in Fig.~\ref{gA_RBCUKQCD}
for the volumes $V\approx(2.7\text{ fm})^3$ and $V\approx(1.8\text{ fm})^3$.
These phenomenological fits indeed suggest the presence of significant
finite volume corrections in the relevant region of 
$m_\pi\approx330,\ldots,360\text{ MeV}$ and for volumes of $V<(3\text{ fm})^3$,
in contradiction to the SSE HBChPT results discussed above,
as shown in particular in Fig.~\ref{ga1L_QCDSF}.
It may, however, be noted noted that the low-lying, leftmost
lattice data point in Fig.~\ref{gA_RBCUKQCD} has been obtained
from a central average over the somewhat asymmetric plateau 
in the ratio of three- to two-point functions displayed
on the top in Fig.~\ref{plat_RBCUKQCD}. Due to the asymmetrically
distributed data points around $t=6$, the choice of the plateau region
is a potential source of an additional systematic uncertainty. If included,
it may reduce the statistical significance of the observed difference to the other lattice 
data points at $m_\pi\approx350\text{ MeV}$ and $m_\pi\approx420\text{ MeV}$ in Fig.~\ref{gA_RBCUKQCD}.
%
%
%
%
%
\subsubsection{Tensor form factors and tensor charge}
\label{sec:tensorFFs}
As before, all results discussed in this section have been obtained for non-perturbatively renormalized
operators and are given in the $\MSbar$ scheme at a renormalization scale of
$\mu=2\text{ GeV}$, unless stated otherwise. For details, we refer to the original works.

\begin{figure}[t]
      \centering
          \includegraphics[angle=-90,width=0.9\textwidth,clip=true]{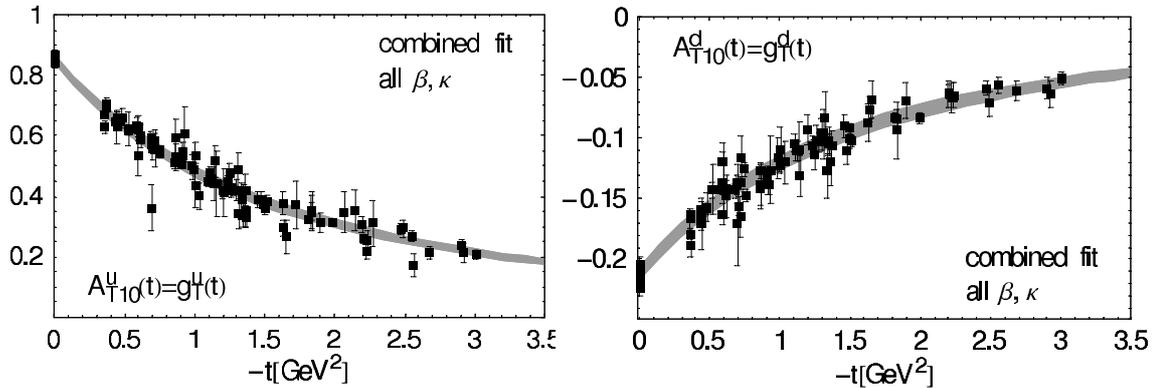}
  \caption{Tensor form factor $G_T(t=Q^2)=A_{T10}(t=Q^2)$ (from \cite{Gockeler:2005cj}).}
  \label{AT10_QCDSF}
 \end{figure}

\begin{figure}[t]
      \centering
         \includegraphics[angle=0,width=0.9\textwidth,clip=true]{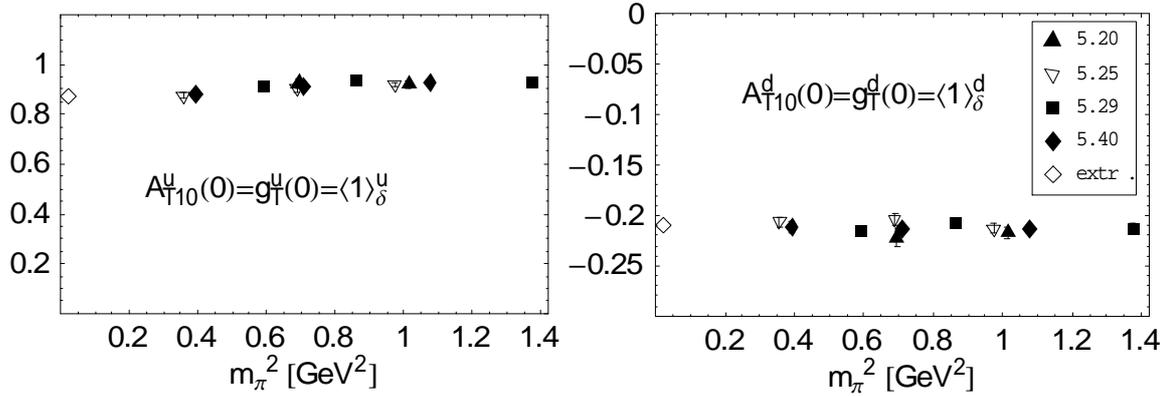}
  \caption{Tensor charge $g_T$ versus $m_\pi^2$ (from \cite{Gockeler:2005cj}).}
  \label{ATn0_m_pi_QCDSF}
 \end{figure}

An overview of lattice results from QCDSF/UKQCD for the tensor 
form factor $G_T(t=-Q^2)=A_{T10}(t=-Q^2)$ for up- and down-quarks 
is given Fig.~\ref{AT10_QCDSF}, obtained for pion masses of
$m_\pi\approx600,\ldots,1200\text{ MeV}$, 
lattice spacings of $a\approx0.07,\ldots,0.11\text{ fm}$ and volumes
of $V\approx(1.4,\ldots,2.0\text{ fm})^3$ \cite{Gockeler:2005cj}.
Contributions from disconnected diagrams, which would only drop out
in the isovector channel, were not included. However, since the
underlying tensor operator $\overline qi\sigma^{\mu\nu}q$ flips the helicity of the quarks,
the disconnected contributions will not contribute in the combined continuum 
and chiral limit, $(a,m_q)\rightarrow0$, and are expected to be suppressed 
even at finite quark masses.
The lattice data points in Fig.~\ref{AT10_QCDSF} have already been extrapolated to the 
continuum limit and the physical pion mass by means of a simultaneous global fit
to the full $Q^2$-, $m_\pi$- and $a$-dependence, including all available ensembles.
A dipole ansatz of the form
\begin{equation}
A_{T10}^{\text{dipole},m_\pi,a} (t) = 
 \frac{A_{T10}^{0}(0)+\alpha_1 m_\pi^2+\alpha_2 a^2}
  {\left( 1 - \frac{t}{(m_D^0+\alpha_3 m_\pi^2)^2}
   \right)^2} \ ,
\label{chiralcontdipole}
\end{equation}
with five fit parameters $A_{T10}^{0}(0)$, $m_D^0$ and $\alpha_1,
\ldots, \alpha_3$, where the $\alpha_i$ describe the
lattice spacing and $m_\pi^2$-dependence of the forward value and
the dipole mass, has been fitted to the lattice data points. 
In a second step, the fit result was used to shift the original 
data points to the continuum limit and the 
physical pion mass by subtracting the difference 
$A_{T10}^{\text{dipole},m_{\pi}^\text{latt},a}(t)-
A_{T10}^{\text{dipole},m_{\pi}^\text{phys},a=0}(t)$.
%
%
The overall agreement of the shifted data points in Fig.~\ref{AT10_QCDSF}
shows that this simplistic treatment of the $Q^2$-, $m_\pi$- and $a$-dependence 
works surprisingly well, but clearly a more sophisticated 
approach will be necessary in the case that lattice data at lower pion masses is included.
From the global simultaneous fit, a tensor charge of $g^u_T=A^u_{T10}(Q^2=0)=0.857(13)$
for up-, and $g^d_T=A^d_{T10}(Q^2=0)=-0.212(05)$ for down-quarks was found.
This agrees within statistical errors with the results from a separate linear chiral extrapolation
of $g^u_T$ and $g^d_T$, as plotted in Fig.~\ref{ATn0_m_pi_QCDSF}. 
The $m_\pi^2$-dependence of the tensor charge is indeed to a very good approximation linear 
and almost flat over the whole range of accessible pion masses.
We also quote the values for the dipole masses obtained from the fit using Eq.~\ref{chiralcontdipole},
which are $m^u_D=1.732(36)\text{ GeV}$ $m^d_D=1.741(56)\text{ GeV}$ at the physical pion mass.
This translates into mean square tensor charge radii of 
$\langle r_T^2\rangle_{u}=0.156(7)\text{ fm}^2$ and $\langle r_T^2\rangle_{d}=0.154(10)\text{ fm}^2$,
which are in the same ballpark as the (isovector) axial-vector radius 
$\langle r_A^2\rangle_{u-d}$ in Fig.~\ref{mLr_a_c_RBCUKQCD_Lat2008}, but clearly smaller than the 
isovector Dirac radius $\langle r_1^2\rangle_{u-d}$ in Fig.~\ref{rv1Q_QCDSF}.

\begin{figure}[t]
   \begin{minipage}{0.48\textwidth}
      \centering
          \includegraphics[angle=0,width=0.9\textwidth,clip=true,angle=0]{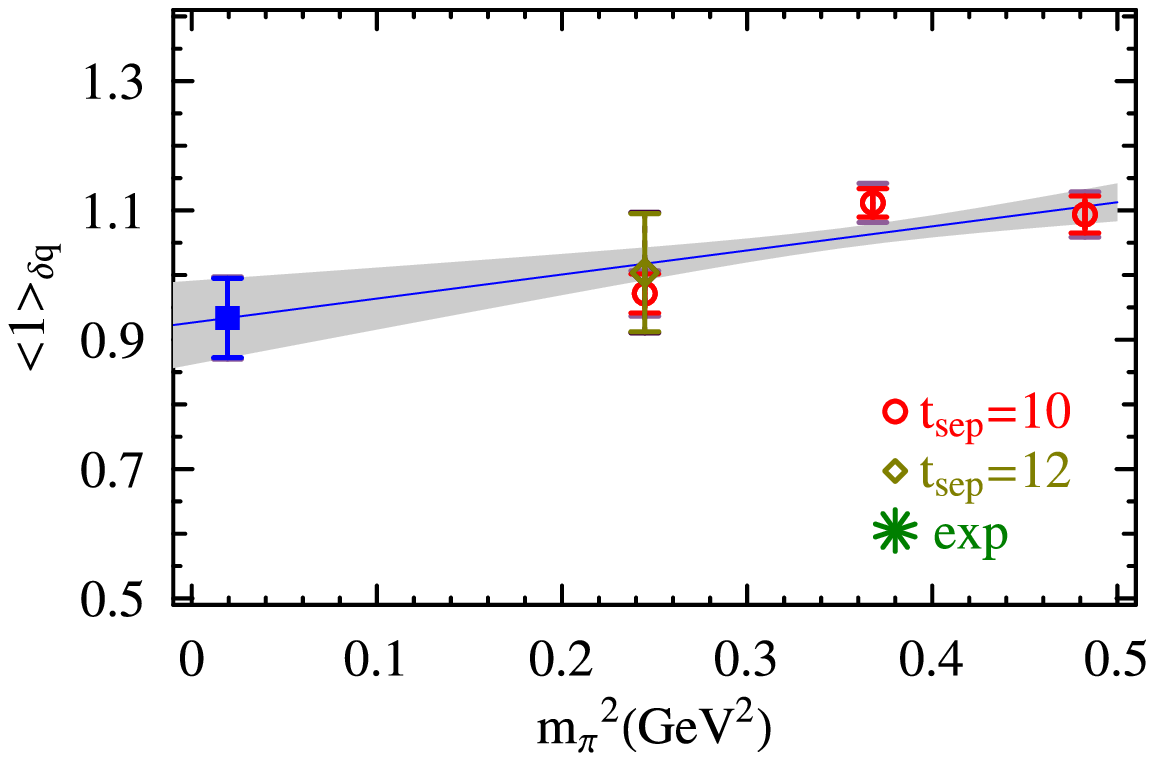}
  \caption{Pion mass dependence of the isovector tensor charge $g_T$ (from \cite{Lin:2008uz}).}
  \label{gt_RBCUKQCD}
     \end{minipage}
     \hspace{0.5cm}
    \begin{minipage}{0.48\textwidth}
      \centering
          \includegraphics[angle=0,width=0.9\textwidth,clip=true]{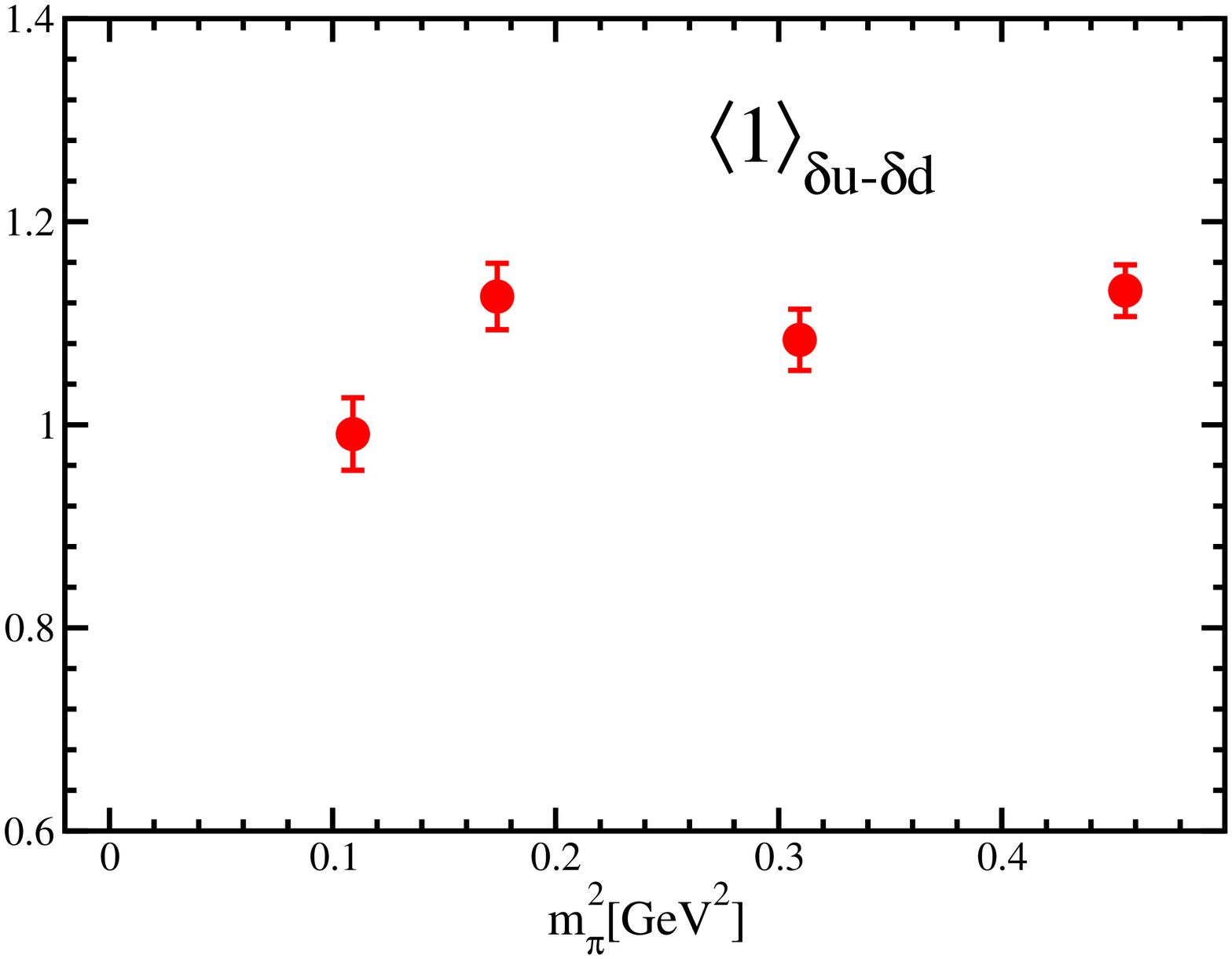}
  \caption{Pion mass dependence of the isovector tensor charge $g_T$ (from proceedings \cite{Ohta:2008kd}).}
  \label{mf1_q_RBCUKQCD_Lat2008}
     \end{minipage}
 \end{figure}

Results for the isovector tensor charge from RBC for $n_f=2$ flavors of domain wall fermions 
are displayed in Fig.~\ref{gt_RBCUKQCD} \cite{Lin:2008uz}. Clearly, the results for the three different
pion masses with $m_\pi>490\text{ MeV}$ do not allow for a systematic extrapolation to the physical point,
so that the value of $g^{u-d}_T=0.93(6)$ obtained from a linear extrapolation in $m_\pi^2$
should, in the best case, be seen as indicative. 
Figure \ref{mf1_q_RBCUKQCD_Lat2008} shows $g^{u-d}_T$ as obtained more recently by RBC-UKQCD
in the framework of simulations with $n_f=2+1$ flavors of domain wall fermions \cite{Ohta:2008kd}
for pion masses between $\approx331$ and $\approx672\MeV$.
The low value of the tensor charge at the smallest pion mass may be an indication
for finite volume effects. 
Overall agreement is observed for the $n_f=2$ and $n_f=2+1$ results 
in Fig.~\ref{gt_RBCUKQCD} and Fig.~\ref{mf1_q_RBCUKQCD_Lat2008}, respectively,
which in both cases were non-perturbatively renormalized and are given in the $\MSbar$ 
scheme at a scale of $\mu=2\text{ GeV}$.
 
\begin{figure}[t]
     \centering
          \includegraphics[angle=-90,width=0.45\textwidth,clip=true]{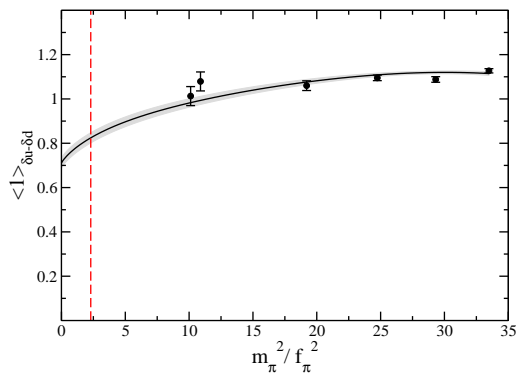}
  \caption{Pion mass dependence of the isovector tensor charge $g_T$ (from proceedings \cite{Edwards:2006qx}).}
  \label{gt_umd_LHPC}
 \end{figure}

Based on the hybrid approach of $n_f=2+1$ domain wall valence Asqtad staggered 
sea quarks, LPHC has computed the isovector tensor charge 
for pion masses in the range of $\approx350\MeV$ to $\approx760\text{ MeV}$, 
as displayed in Fig.~\ref{gt_umd_LHPC} \cite{Edwards:2006qx}. 
In contrast to the corresponding study of the axial-vector
operator discussed in section \ref{sec:axialvector}, 
the tensor operator, $\overline q i\sigma^{\mu\nu}q$, 
has in this case been renormalized and transformed to the $\MSbar$ scheme 
at a scale of $\mu=2\text{ GeV}$ using a non-perturbatively (NP-) improved 
\emph{perturbative} renormalization constant,
\bea
\label{ZLHPC}
Z_\mathcal{O}=Z_\mathcal{O}^{\text{PT}}\frac{Z_A^{\text{NPT}}}{Z_A^{\text{PT}}}\,,
\eea
with $\mathcal{O}=\mathcal{O}_T$, employing the non-perturbative axial-vector
renormalization constant $Z_A^{\text{NPT}}$ in combination with 
$Z_\mathcal{O}^{\text{PT}}$ and $Z_A^{\text{PT}}$ obtained from 1-loop lattice perturbation theory 
\cite{Bistrovic:2005aa}. 
We note that although the perturbative renormalization constants
are already close to unity due to HYP-smearing of the gauge fields, 
it is not guaranteed that the NP-improved renormalization factors in Eq.~\ref{ZLHPC}
agree within statistical errors with the (at this point unknown) 
fully non-perturbative renormalization factors.
This potential systematic uncertainty should be kept in mind when comparing the lattice
data from the mixed-action approach by LHPC, e.g. in Fig.~\ref{gt_umd_LHPC}, 
to other simulation results.
%
%
The shaded band in Fig.~\ref{gt_umd_LHPC} represents a chiral extrapolation
of the lattice data points based on a self-consistent rearrangement
of the leading 1-loop HBChPT formula 
\cite{Arndt:2001ye,Chen:2001eg}, given by \cite{Edwards:2006qx}
\begin{eqnarray}
\label{gtChPT}
g_T=\langle 1 \rangle_{\delta u-\delta d} & = &
\delta g_T^0 \left( 1 - \frac{(4 g_{A,\lat}^2 + 1)}{2(4\pi)^2} \frac{m_{\pi,\lat}^2}{f_{\pi,\lat}^2} \ln \left( \frac{m_{\pi,\lat}^2}{f_{\pi,\lat}^2} \right) \right) + \delta c_0 \frac{m_{\pi,\lat}^2}{f_{\pi,\lat}^2}\,,
\end{eqnarray}
where the LECs in the chiral limit have been replaced by the
respective pion mass dependent values $g_{A,\lat}$ and $f_{\pi,\lat}$ 
obtained in the lattice calculation.
Although chiral extrapolations based on self-consistently improved ChPT 
have shown some success in the extrapolation to the physical point and comparison
with experimental results \cite{Edwards:2006qx}, it is difficult to judge if 
the extrapolation in Fig.~\ref{gt_umd_LHPC} describes the underlying physics correctly.
After all, such a rearrangement of the chiral series cannot possibly 
account for all analytic and non-analytic structures in $m_\pi$ that
are in principle relevant at the currently accessible pion masses.
A value of $g_T=\langle 1 \rangle_{\delta u-\delta d}\approx0.82(2)$ is found at the physical ratio $m_{\pi,\text{phys}}^2/f_{\pi,\text{phys}}^2\approx2.4$. 

%
%
%
%
%

Lattice results for the nucleon tensor form factor $\overline B_{T10}(Q^2)$
are particularly relevant for an understanding of the transverse nucleon spin structure, 
as explained in section \ref{sec:Interpretation}, and will be discussed in relation with transverse 
spin densities of quarks in the nucleon in section \ref{sec:TransverseSpin} below.

\subsection{Electromagnetic form factors of the $\rho$-meson}
\label{sec:rhoFFs}
\begin{figure}[t]
   \begin{minipage}{0.48\textwidth}
      \centering
       \hspace{0pt}\scalebox{0.95}{\input{\myfiguresFFs/FFRhoG1_Cyprus}}
\vspace{-0pt}\captionof{figure}{$G_1$ of the $\rho$-meson as a function of 
 the momentum transfer $Q^2$ for three $\kappa$ values
(from proceedings \cite{Alexandrou:2007pn}).}
  \label{FFRhoG1_Cyprus}
     \end{minipage}
          \hspace{0.5cm}
   \begin{minipage}{0.48\textwidth}
      \centering
          \includegraphics[angle=90,width=0.9\textwidth,clip=true,angle=0]{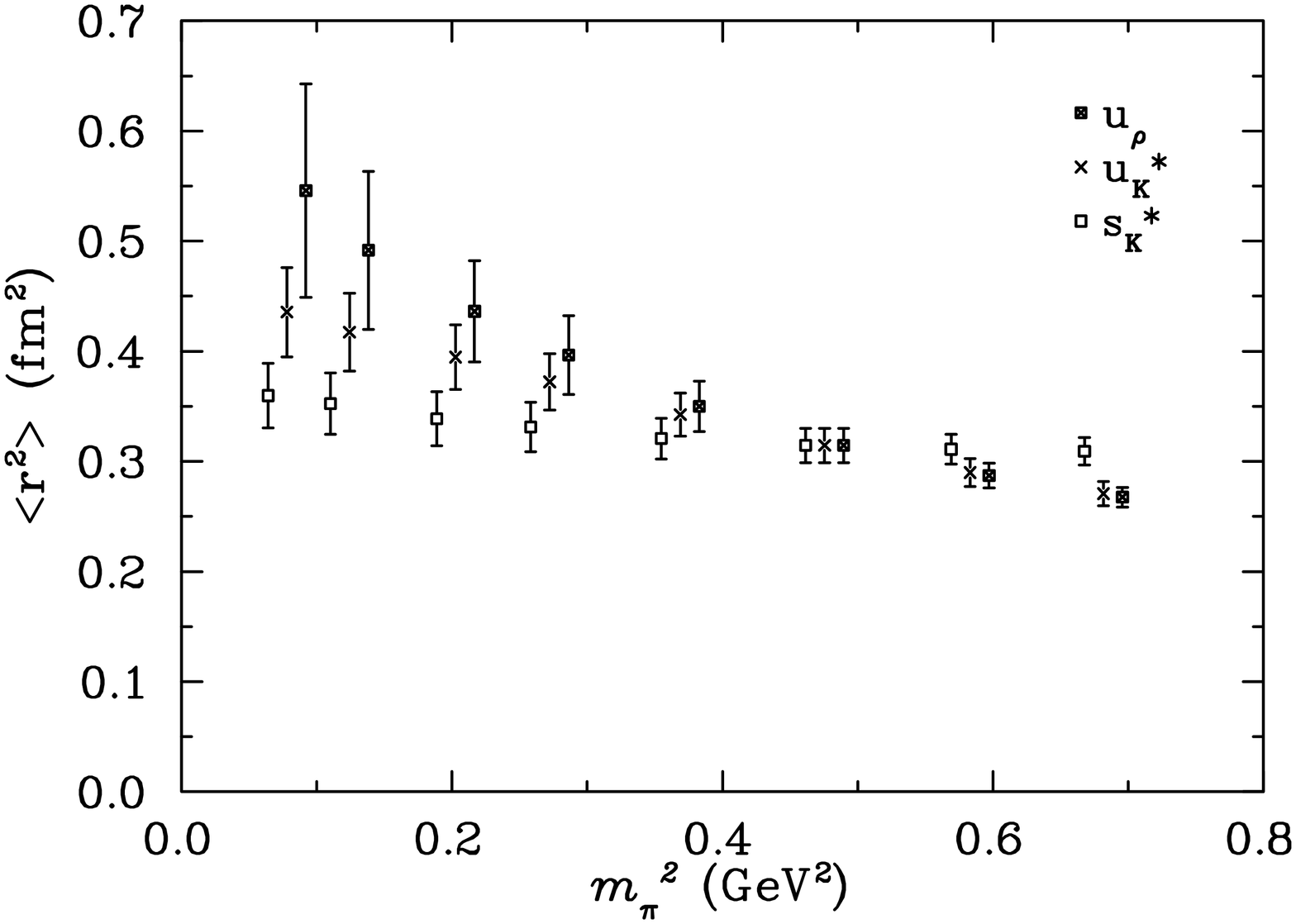}
  \caption{Pion mass dependence of the $\rho$-meson charge radius in the quenched approximation, 
  compared to corresponding up- and strange-quark contributions to the $K^*$ charge radius (from \cite{Hedditch:2007ex}).}
  \label{FFrho_r2_Adelaide}
     \end{minipage}
 \end{figure}

A first preliminary computation of the form factor $G_1(Q^2)$ of the $\rho$-meson in unquenched lattice QCD
has been performed using so-called density-density correlators \cite{Alexandrou:2007pn}
(for a very recent study by the same authors of hadron deformations 
using density-density correlators, we refer to \cite{Alexandrou:2008ru} and section \ref{sec:deformations} below).
For the case of the $\rho$-meson one considers four-point functions of the form
\bea
\label{Rho4pt}
C^{\mu\nu}_{\text{4pt}}(\mbf{x},t_1,t_2)&=&
\int d^3x_1d^3x_2
\langle \rho(\mbf{x}_2,t)|
J^\mu_{u}(\mbf{x}+\mbf{x}_1,t_2) 
J^\nu_{d}(\mbf{x}_1,t_1)
|\rho(\mbf{x}_0,t_0)\rangle\,.
\eea
Fourier-transforming this with respect to $\mbf{x}$, inserting of a complete
set of states, and using isospin symmetry, one finds for e.g. $\mu=\nu=4$
at large time-distances $t_2-t_1,t-t_0,t-t_2\gg0$ that
$C^{44}_{\text{4pt}}(\mbf{P},t_1,t_2)\propto e^{(-E_\rho(\mbf{P})(t_2-t_1))}
e^{(-E_\rho(\mbf{P}')(t-(t_2-t_1)-t_0))}G_1(Q^2)^2$, where $\mbf{P}'=0$, $Q^2=-(P-P')^2$,
from which the (squared) form factor $G_1(Q^2)$ can be extracted by taking appropriate
ratios with $\rho$-meson two-point-functions. Because of the integrals over $x_1$ and $x_2$
in Eq.~(\ref{Rho4pt}), the evaluation of the 4-point correlator requires in general 
all-to-all propagators. It has been shown in \cite{Alexandrou:2007pn}
how the one-end trick that has already been successfully used for the calculation
of pion correlators (see section \ref{sec:methods} above) can also be used for
the evaluation of the four-point functions of the $\rho$-meson, leading to a substantial
reduction of the stochastic noise compared to the standard method using all-to-all
propagators. Gaussian smearing, combined with HYP-smearing, of the interpolating
source and sink fields was shown to be crucial to suppress contaminations from excited states.
Using these methods, calculations have been performed for $n_f=2$ flavors
of Wilson fermions and the Wilson gauge action, for pion masses
of $\approx384$, $\approx510$ and $\approx690\MeV$, a lattice spacing of $a\approx0.08\fm$
and a volume of $V\approx(1.9\fm)^3$.
The results for the $\rho$-meson form factor $G_1(Q^2)$ are displayed in 
Fig.~\ref{FFRhoG1_Cyprus}, showing a very good precision over a wide range of $Q^2$.
Keeping in mind the exploratory character of this calculation, the results in 
Fig.~\ref{FFRhoG1_Cyprus} seem already to indicate that the slope in $Q^2$,
and therefore the charge radius, is increasing for decreasing pion mass.
%
%

\begin{figure}[t]
    \begin{minipage}{0.48\textwidth}
      \centering
          \includegraphics[angle=90,width=0.9\textwidth,clip=true,angle=0]{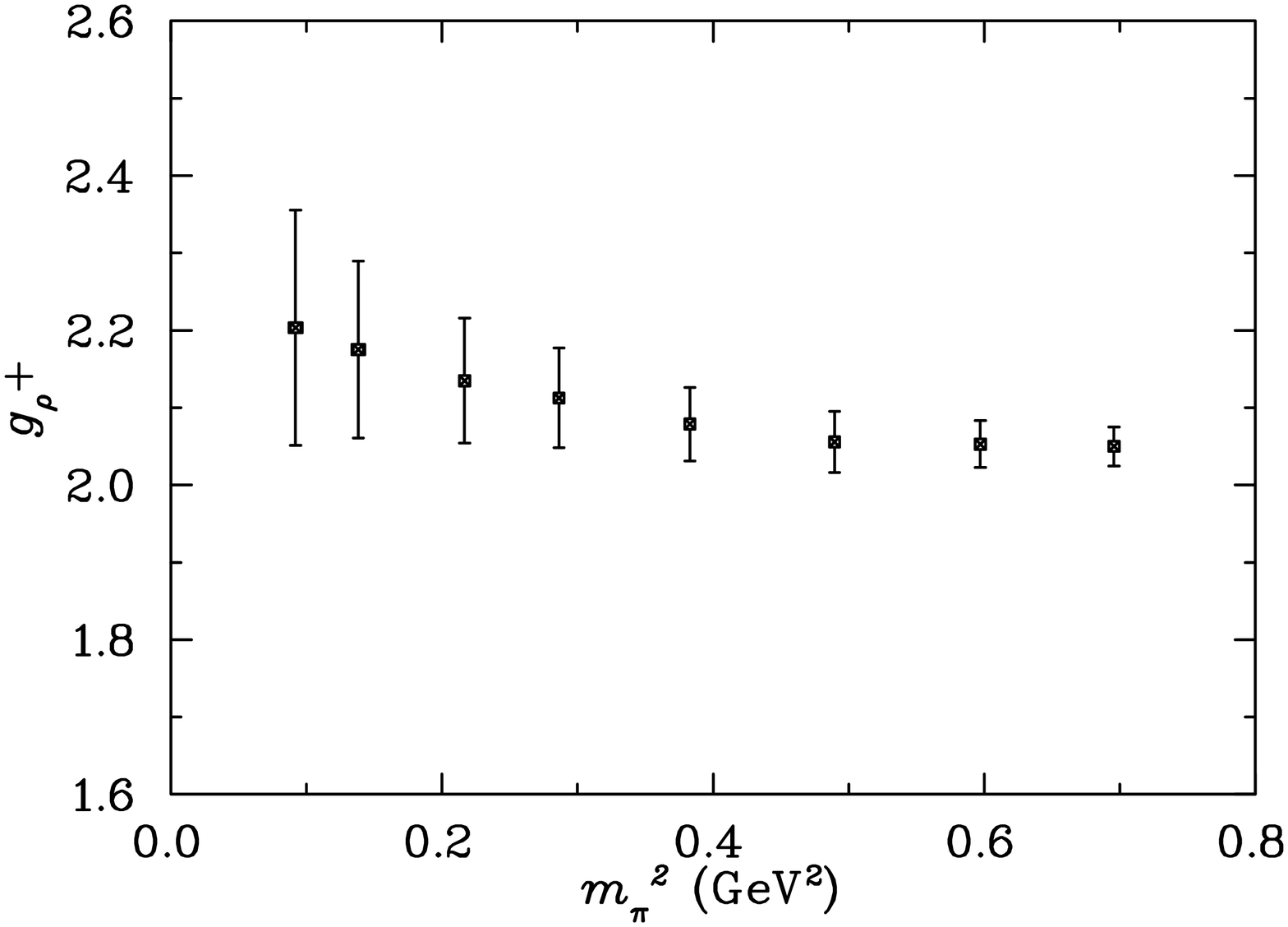}
  \caption{Pion mass dependence of the $\rho$-meson magnetic moment in natural units (g-factor) obtained from
  $\mu_\rho=G_M(0)\widetilde=G_M(0.22\GeV^2)/G_C(0.22\GeV^2)$ in the quenched approximation 
  (from \cite{Hedditch:2007ex}).}
  \label{FFrho_gfactor_Adelaide}
     \end{minipage}
          \hspace{0.5cm}
   \begin{minipage}{0.48\textwidth}
      \centering
          \includegraphics[angle=90,width=0.9\textwidth,clip=true,angle=0]{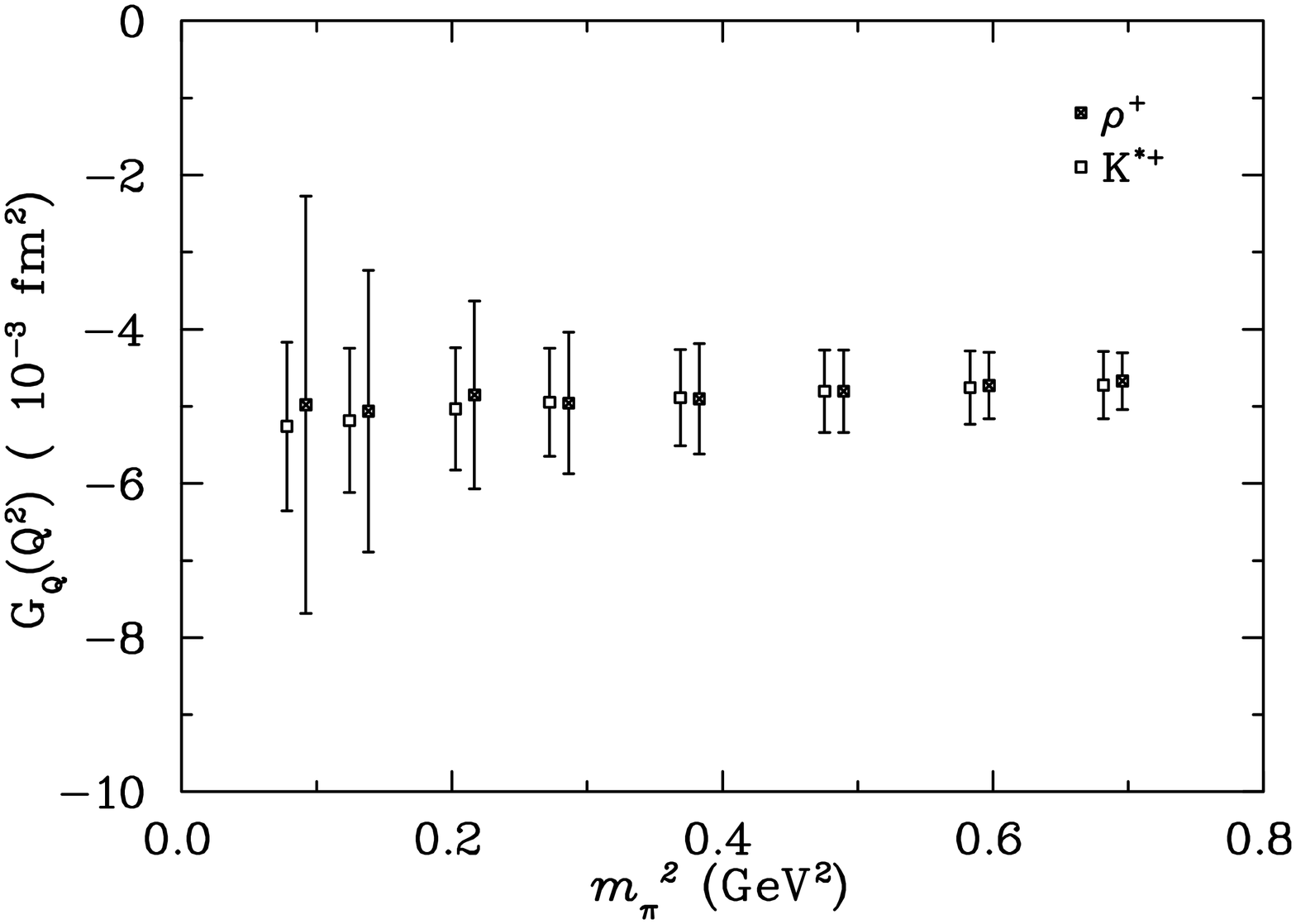}
  \caption{Pion mass dependence of the $\rho$-meson quadrupole moment in the quenched approximation
  (from \cite{Hedditch:2007ex}).}
  \label{FFrho_quad_Adelaide}
     \end{minipage}
 \end{figure}

An extensive study of the charge radii, magnetic and quadrupole moments of
the $\rho$ and the $K^*$ mesons in quenched lattice QCD has been presented recently,
based on fat-link irrelevant clover (FLIC) improved Wilson fermions and 
an $\mathcal{O}(a^2)$ improved gauge action \cite{Hedditch:2007ex}.
Results have been obtained for a lattice spacing of $a\approx0.128\fm$
in a volume of $V\approx(2.6\fm)^3$, with pion masses as low as $\approx300\MeV$.
The local vector current has been renormalized by demanding charge conservation, i.e. $G_C(Q^2\eql0)=1$. 
Since the analysis has been restricted to a single non-zero $Q^2\approx0.22\GeV^2$,
it was assumed in the extraction of the magnetic moment $\mu=G_M(0)$
that the magnetic and charge form factors scale identically at small $Q^2$,
such that $\mu_\rho=G_M(0)\simeq G_M(0.22\GeV^2)/G_C(0.22\GeV^2)$. 
For the same reason, the mean square charge radius was calculated
from $\langle r^2\rangle=6(G_C(Q^2)^{-1}-1)/Q^2$ for $Q^2\approx0.22\GeV^2$,
corresponding to a monopole ansatz for the $Q^2$-dependence of $G_C(Q^2)$.
Results for the charge radius of the $\rho$-meson as a function of the 
pion mass are shown in Fig.~\ref{FFrho_r2_Adelaide}.
As before, the charge radius is increasing at lower pion masses,
with a value of $\langle r_\rho^2\rangle=0.546(97)\fm^2$ at $m_\pi\approx300\MeV$.
Figure \ref{FFrho_gfactor_Adelaide} displays the magnetic moment in natural magnetons 
(also called the $g$-factor), which is approximately constant as a function of $m_\pi^2$.
A value of $g_\rho=\mu_\rho=2.21(15)$ was obtained at the lowest accessible pion mass of $\approx300\MeV$.
The pion mass dependence of the $\rho$-meson quadrupole form factor $G_Q(Q^2)$ at $Q^2\approx0.22\GeV^2$
is displayed in Fig.~\ref{FFrho_quad_Adelaide}.
A small negative value of $G_Q(Q^2\approx0.22\GeV^2)=-0.0050(27)\fm^2$
was found at $m_\pi\approx300\MeV$. Although  
the statistical errors are rather large, these results give a first indication 
that the charge distribution of the $\rho$-meson is oblate.

%
%

\begin{figure}[t]
   \begin{minipage}{0.48\textwidth}
      \centering
          \includegraphics[angle=0,width=0.93\textwidth,clip=true,angle=0]{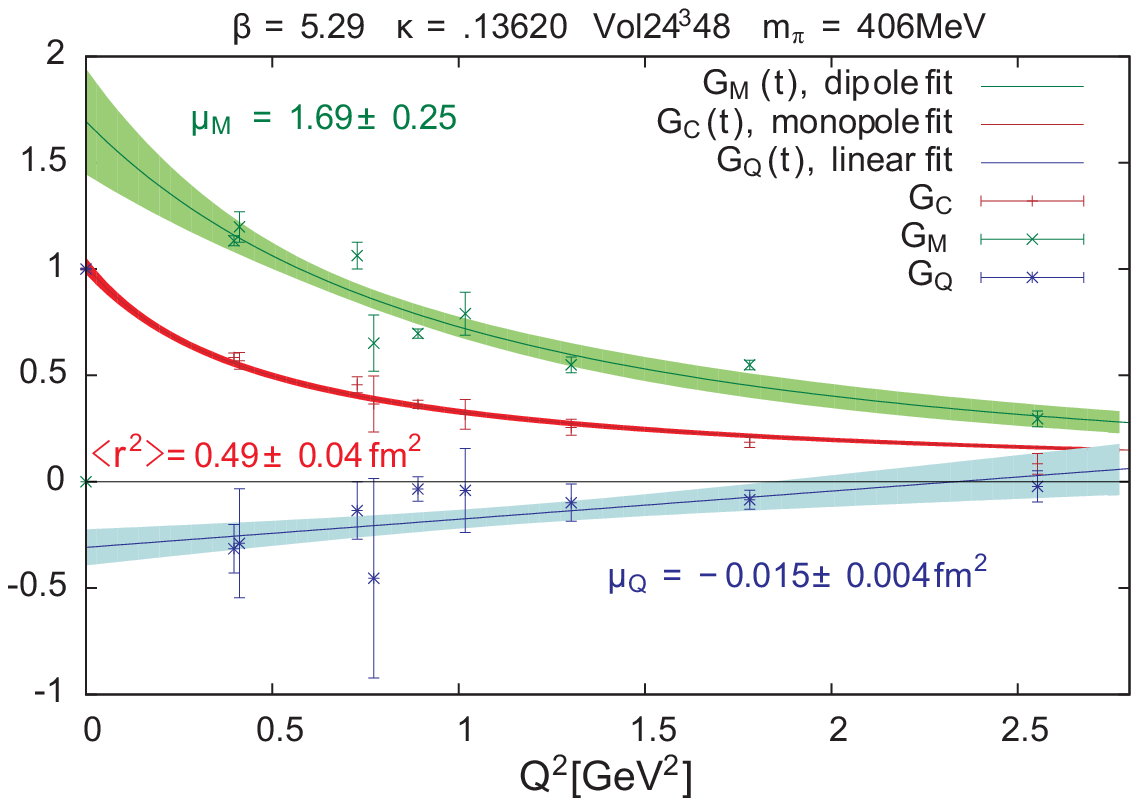}
  \caption{Electromagnetic form factors of the rho meson (from proceedings \cite{Gurtler2008}).}
  \label{FFrho_QCDSF_2}
     \end{minipage}
     \hspace{0.3cm}
    \begin{minipage}{0.48\textwidth}
      \centering
          \includegraphics[angle=0,width=0.93\textwidth,clip=true,angle=0]{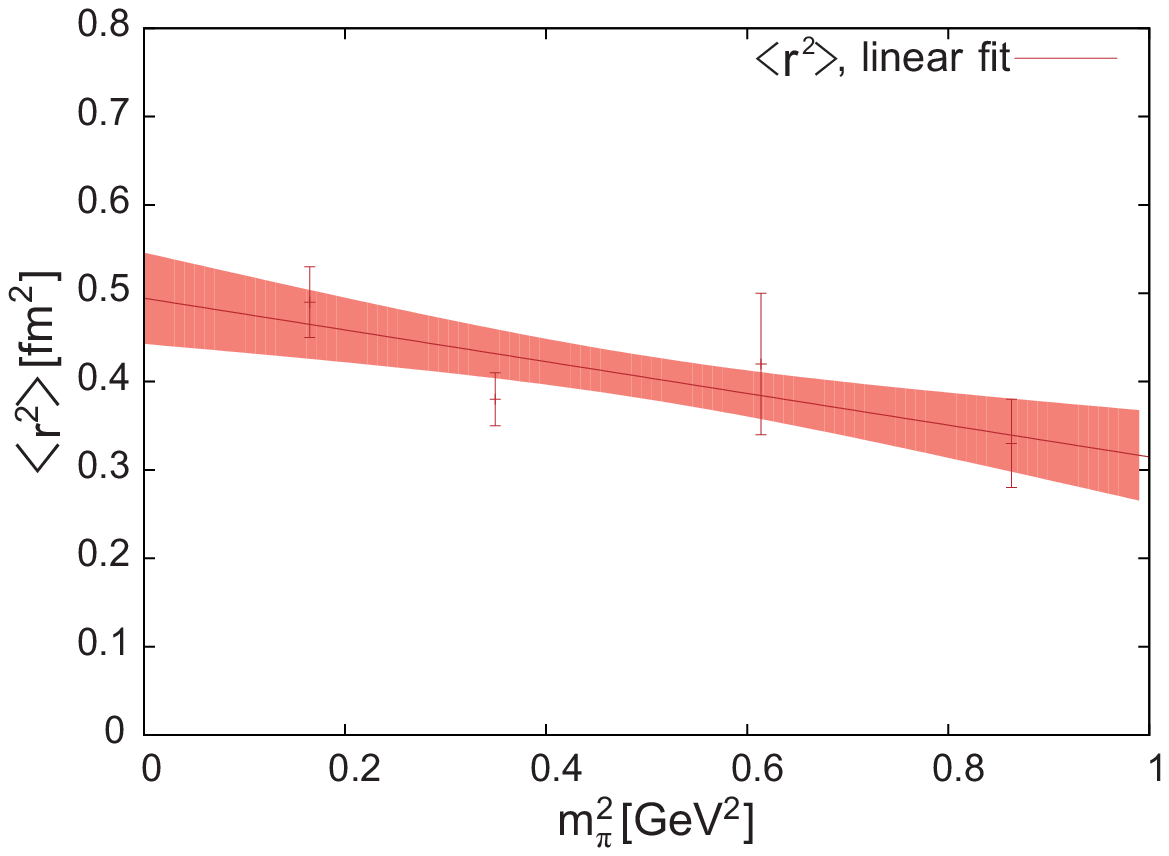}
  \caption{Pion mass dependence of the rho charge radius (from proceedings \cite{Gurtler2008}).}
  \label{FFrho_r2_QCDSF_2}
     \end{minipage}
 \end{figure}

\begin{figure}[t]
   \begin{minipage}{0.48\textwidth}
      \centering
          \includegraphics[angle=0,width=0.93\textwidth,clip=true,angle=0]{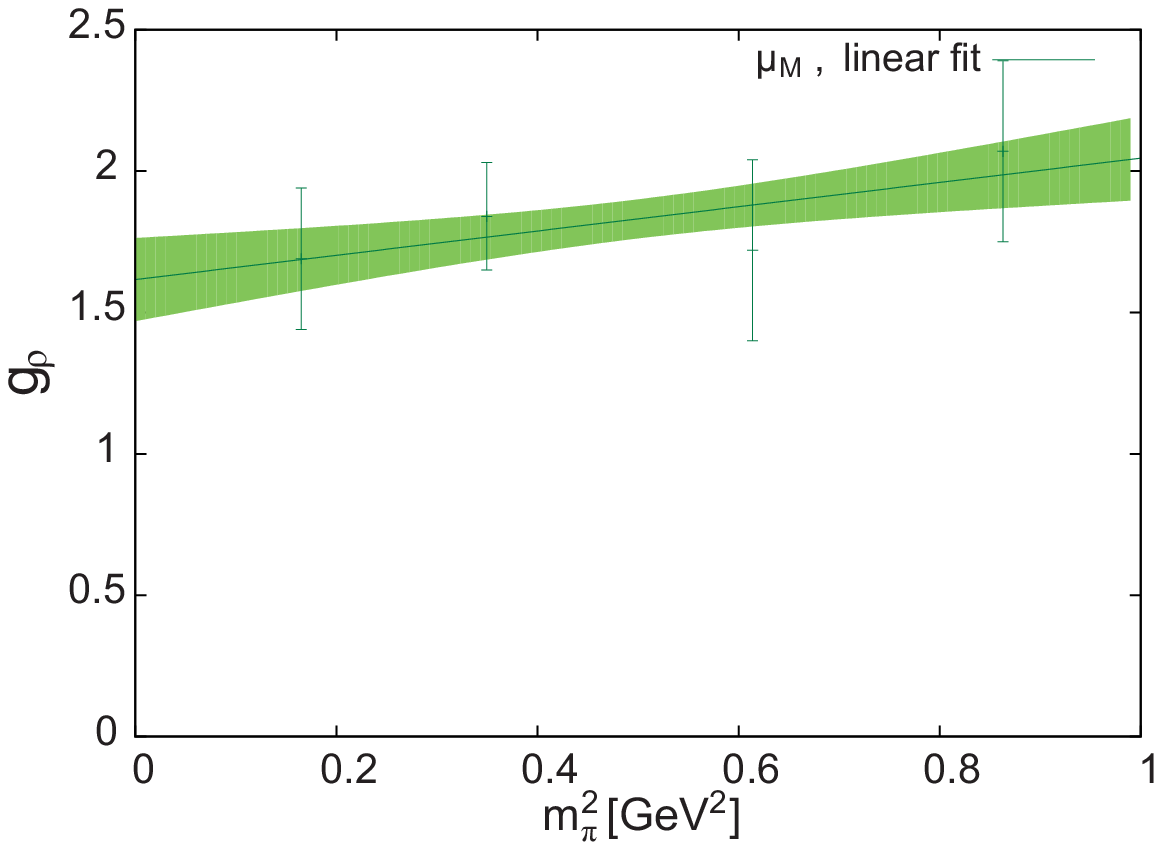}
  \caption{Pion mass dependence of the rho magnetic moment (from proceedings \cite{Gurtler2008}).}
  \label{FFrho_mum_QCDSF_2}
     \end{minipage}
     \hspace{0.3cm}
    \begin{minipage}{0.48\textwidth}
      \centering
          \includegraphics[angle=0,width=0.93\textwidth,clip=true,angle=0]{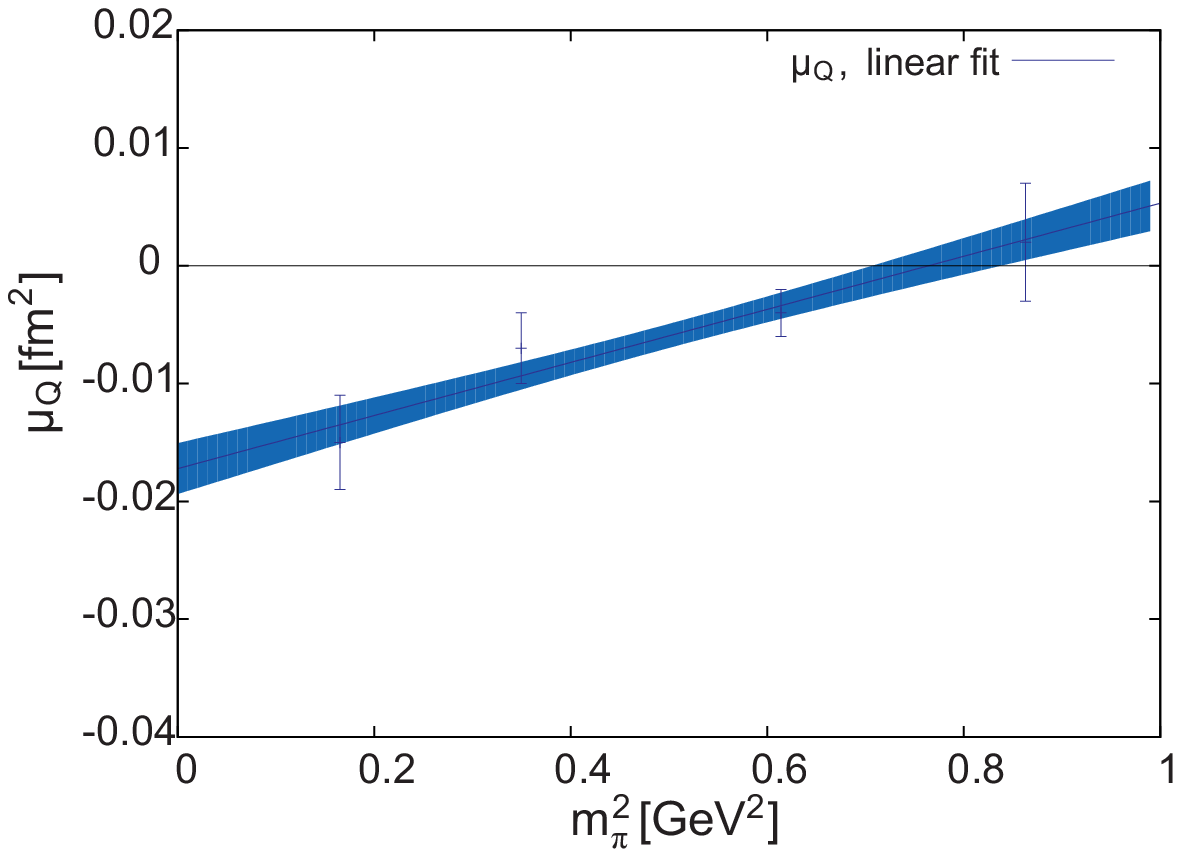}
  \caption{Pion mass dependence of the rho quadrupole moment (from proceedings \cite{Gurtler2008}).}
  \label{FFrho_muq_QCDSF_2}
     \end{minipage}
 \end{figure}

First preliminary results for the electromagnetic
form factors of the rho in unquenched lattice QCD,
based on standard two- and three-point functions,
have recently been obtained by QCDSF/UKQCD \cite{Gurtler2008}.
Calculations have been performed for a
set of ensembles for $n_f=2$ flavors of Wilson fermions,
as described in the previous sections.
Figure \ref{FFrho_QCDSF_2} shows the charge, magnetic and
quadrupole form factors as functions of the momentum transfer squared
for a pion mass of $\approx400\MeV$ in a volume of $V\approx(1.8\fm)^3$.
The local vector current has been renormalized such that $G_C(Q^2=0)=1$.
Monopole, dipole, and linear fits to $G_C(Q^2)$, $G_M(Q^2)$ and $G_Q(Q^2)$,
respectively, are represented by the shaded bands in Fig.~\ref{FFrho_QCDSF_2}.
While the quality of the lattice data for $G_C(Q^2)$ is very good,
the results for $G_M(Q^2)$ in particular show some scatter. 
Many of the lattice data points for the quadrupole form factor
are compatible with zero within errors, but there is a general
trend towards negative values at lower $Q^2$, indicating a non-spherical
charge distribution within the rho.
Since the mass of the $\rho$-meson is in this case close to or even slightly above the 
two-pion threshold, $m_\rho\gtrapprox2m_\pi$, these results have to be considered
with greatest care. For a related preliminary study of the $\rho$-meson mass from a phase-shift analysis
in finite volume, performed in the same simulation framework, we refer to \cite{Gockeler:2008kc}.
The pion mass dependence of the charge radius, which has been obtained from
monopole fits to the lattice data, is displayed in Fig.~\ref{FFrho_r2_QCDSF_2}.
From a linear interpolation in $m_\pi^2$, a value of 
$\langle r_\rho^2\rangle=0.46(3)\fm^2$ was found at $m_\pi\approx450\MeV$.
The magnetic moment has been obtained from an extrapolation of $G_M(Q^2)$ to 
$Q^2=0$ using a dipole parametrization as shown in Fig.~\ref{FFrho_QCDSF_2}, 
and the results for $\mu_\rho$ as a function of $m_\pi^2$ are presented in 
Fig.~\ref{FFrho_mum_QCDSF_2}. Within statistical errors, $\mu_\rho$
is approximately independent of the pion mass, and we note that the central values
are somewhat below the results from \cite{Hedditch:2007ex} displayed in Fig.~\ref{FFrho_gfactor_Adelaide}.
The linear interpolation in $m_\pi^2$ gives $\mu_\rho=1.69(10)$ at $m_\pi\approx450\MeV$,
which is $\approx25\%$ below the corresponding values in Fig.~\ref{FFrho_gfactor_Adelaide}.
Further studies are necessary to see if this is related to the quenched approximation
or the ansatz $\mu_\rho=G_M(0)\simeq G_M(0.22\GeV^2)/G_C(0.22\GeV^2)$ used in \cite{Hedditch:2007ex},
or any other systematic uncertainties affecting the unquenched results in Fig.~\ref{FFrho_QCDSF_2}.
Results for the quadrupole moment, $Q_\rho=G_Q(Q^2=0)/m_\rho^2$, which have been obtained 
from linear extrapolations to $Q^2=0$, are presented in Fig.~\ref{FFrho_muq_QCDSF_2}
versus the pion mass squared.
%
%
%
While the results at larger $m_\pi$ are compatible with zero within errors,
a non-zero, negative value of 
$Q_\rho=-0.0125(15)\fm^2$ at $m_\pi\approx450\MeV$ 
was obtained from a linear interpolation shown by 
the shaded band in Fig.~\ref{FFrho_muq_QCDSF_2}.
This is approximately a factor of two to three larger than the values in
Fig.~\ref{FFrho_quad_Adelaide} above for $G_Q(Q^2\approx0.22\GeV^2)$ obtained
in the quenched approximation.
In summary, to this date all lattice QCD calculations of the quadrupole form factor of
the $\rho$-meson give a non-zero and negative quadrupole moment $Q_\rho$,
indicating that the charge distribution of the rho is oblate.
Further improved studies of the form factors and critical analyses of
the employed assumptions and extrapolations are certainly required 
for a better understanding of the systematic errors and 
in order to confirm these interesting preliminary results.

A study of the magnetic moment of spin-1 mesons 
using the background field method (purely based on hadron two-point functions)
has been presented recently in \cite{Lee:2008qf}.
For a brief introduction to the background field method
and the extraction of, e.g., magnetic moments from 
mass shifts we refer to section \ref{sec:BGfield}.
Calculations were performed in the quenched approximation
using Wilson fermions and the Wilson gauge action for pion masses
ranging from $\approx522\MeV$ to $\approx1015\MeV$, 
a lattice spacing of $a\approx0.1\fm$ and
a volume of $V\approx(2.4\fm)^3$. 
The values for the $g$-factor of the $\rho$-meson
($\mu_\rho$ in natural units) obtained in this study are in remarkably good agreement with the values 
from the conventional approach based on three-point functions display in Fig.~\ref{FFrho_gfactor_Adelaide}
that were also calculated in the quenched approximation.
It is encouraging to see that the two very different methods lead to such consistent results.

%
%

%

\subsection{Form factors of the $\Delta$-baryon}
\label{sec:DeltaFFs}
An extensive first study of the form factors of the decuplet baryons in quenched lattice QCD
was published already in the early 1990's \cite{Leinweber:1992hy}.
The calculations were performed for two values
of the momentum transfer $Q^2=0$ and $Q^2\approx0.16\GeV^2$, for three different pion masses,
on a $24\times12\times12\times24$ lattice, with a lattice spacing of $a\approx0.128\fm$.
Without going into the details of this early study, we mention here only that
results were obtained for the charge radius, magnetic moment, quadrupole and octupole form factor
in particular of the $\Delta$-resonance. From linear chiral extrapolations, 
values for the mean square charge radius of $\langle r_\Delta^2\rangle=0.393(90)\fm^2$,
and for the magnetic form factor of $G^{\Delta^+}_{M1}(0.16\GeV^2)=2.22(44)$ were obtained
for, e.g., the $\Delta^+$.
The central values of the quadrupole, $G^{\Delta^+}_{E2}$, and octupole, $G^{\Delta^+}_{M3}$, 
form factors at $Q^2=0.16\GeV^2$ were found to be negative and positive, respectively, 
but in both cases compatible with zero 
within statistical errors.

\begin{figure}[t]
   \begin{minipage}{0.48\textwidth}
      \centering
          \includegraphics[angle=90,width=0.9\textwidth,clip=true,angle=0]{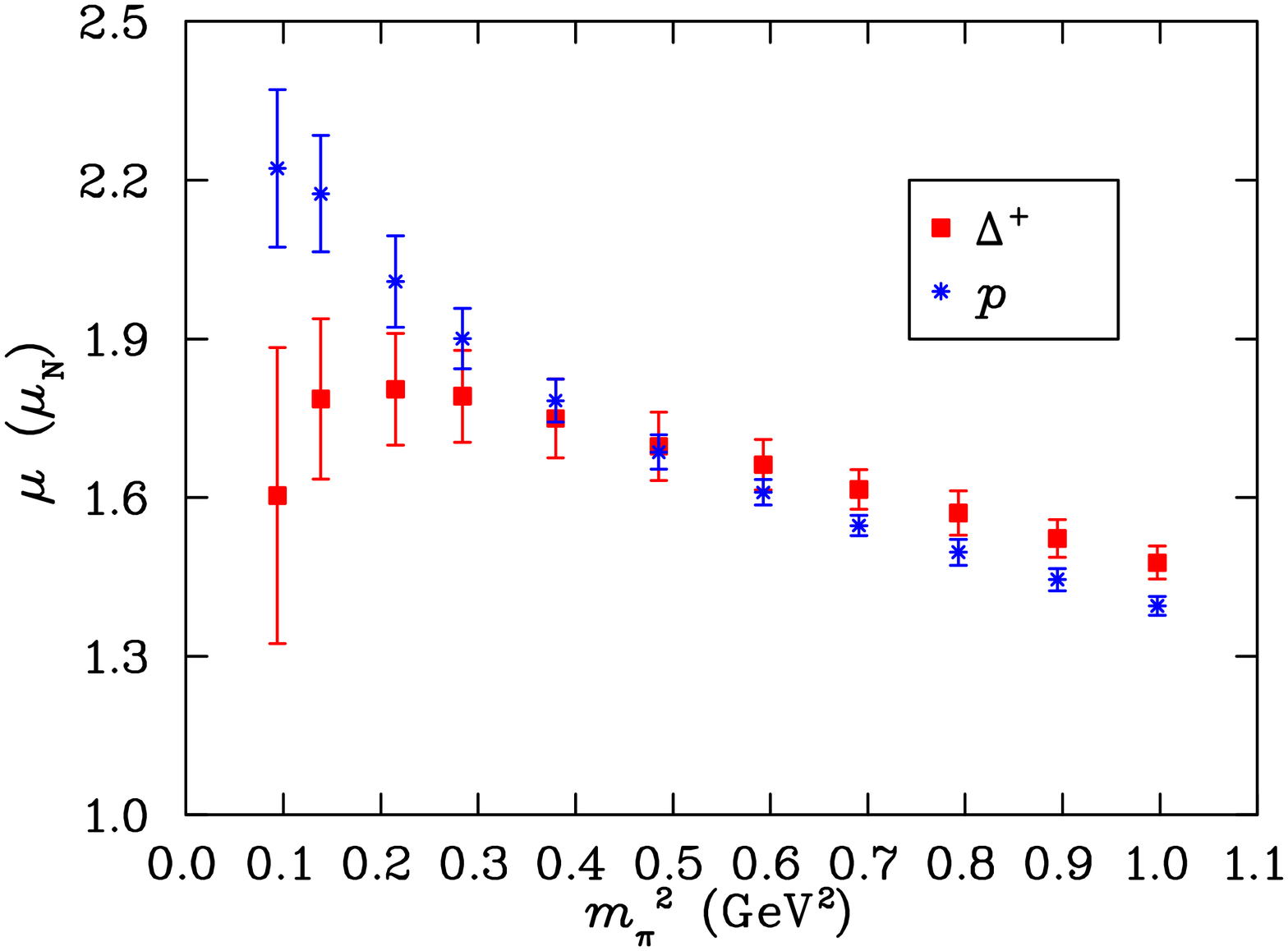}
  \caption{Pion mass dependence of the $\Delta^+$ magnetic moment in the
  quenched approximation (from proceedings \cite{Boinepalli:2006xd}).}
  \label{FFdelta_magmom_PD_Adelaide}
     \end{minipage}
     \hspace{0.5cm}
    \begin{minipage}{0.48\textwidth}
      \centering
          \includegraphics[angle=0,width=0.9\textwidth,clip=true,angle=0]{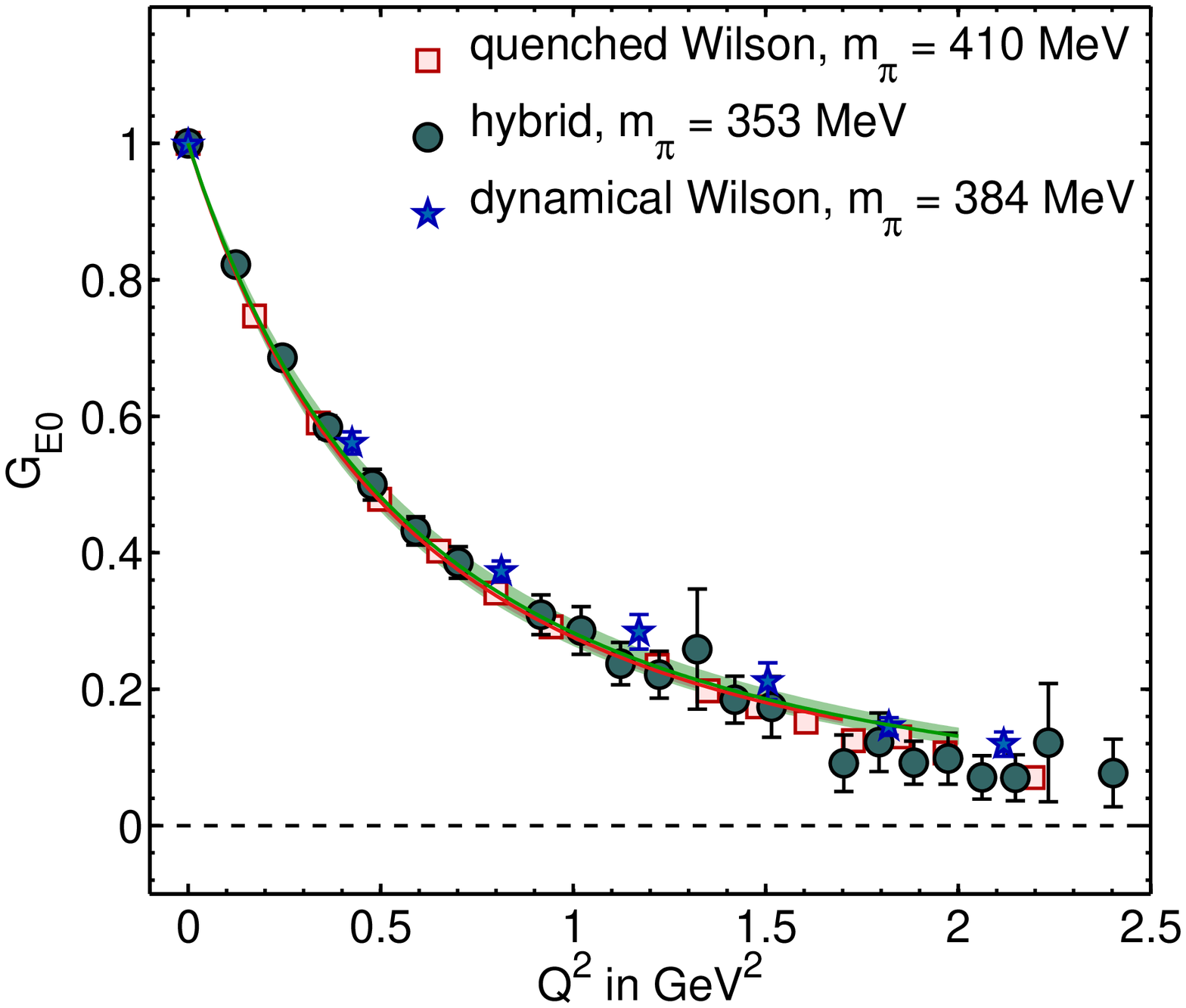}
  \caption{Connected contributions to the electric charge form factor of the $\Delta^+$-baryon
  (from \cite{Alexandrou:2008bn}).}
  \label{FFDelta_ge0_dyn_red}
     \end{minipage}
 \end{figure}

Results for the magnetic moment of the $\Delta^+$-baryon
from a more recent study in quenched lattice QCD using FLIC fermions (with same lattice
parameters as in the study of the $\rho$-meson discussed in the previous section) \cite{Boinepalli:2006ng} 
are displayed in Fig.~\ref{FFdelta_magmom_PD_Adelaide} as a function of $m_\pi^2$, with a 
value of $\mu_{\Delta^+}\approx1.80(15)$ in units of nuclear magnetons
at a pion mass of $\approx372\MeV$.
Notably, the sign of the curvature for the proton and for the $\Delta^+$-state are 
opposite at lower pion masses. This may be understood in the framework
of ChPT, where it has been shown that contributions to $\Delta$-decay that become
increasingly important as one approaches the $\Delta\rightarrow N\pi$ threshold,
have opposite signs in full QCD and QCD in the quenched approximation \cite{Labrenz:1996jy,Leinweber:2003ux}.
The different pion mass dependences of the proton and $\Delta$-baryon magnetic moments
in Fig.~\ref{FFdelta_magmom_PD_Adelaide} have therefore been interpreted as clear signature 
of the quenched approximation \cite{Leinweber:2003ux}. 
This ``quenched oddity'' \cite{Labrenz:1996jy} is interesting by itself but clearly makes a comparison
of lattice results with experimental data in this channel even more delicate
if not impossible.
Magnetic moments of baryons, specifically the $\Delta$ states, were also studied 
in quenched QCD using the background field method (see section \ref{sec:BGfield} below) 
\cite{Lee:2005ds}. The calculations were based on the Wilson gauge and fermion actions,
with a lattice spacing of $a\approx0.1\fm$ for a volume of $\approx(2.4\fm)^3$ and pion masses
in the range of $\approx520\MeV$ to $\approx1010\MeV$. Quantization conditions for the
external magnetic fields were not considered, and it was attempted to keep the
boundary effects small by using Dirichlet boundary conditions,
small values of the external fields, and by placing the sources centrally in the spatial direction.
Four different values of the magnetic field $B$
for the up- and down-quarks were employed in the calculation of the two point functions,
and even-power terms in the expansion Eq.~\ref{massshift} were eliminated by 
also computing the mass shifts $\Delta E$ for the corresponding negative values, $-B$, and then
extracting the magnetic moments from $(\Delta E(B)-\Delta E(-B))/2$.
The results for the magnetic moments in natural units were first converted to nuclear magnetons
with factors of $m_N^{\phys}/m^{\lat}_h$, where $m^{\lat}_h$ is the pion mass dependent hadron
mass measured on the lattice, and then extrapolated to $m^{\phys}_\pi$ using different
ans\"atze for the pion mass dependence. Extrapolations linear in $m_\pi$ gave results
for the baryon octet magnetic moments and the $\Delta^{++}$-baryon
that were $\approx10\%$ to $\approx25\%$ above the experiment 
(in terms of absolute values), with the exception
of the neutron magnetic moment, which came out slightly below the experimental result.
The pion mass dependence of the proton and $\Delta^+$ magnetic moments turns out to be
well compatible with the results in Fig.~\ref{FFdelta_magmom_PD_Adelaide},
and the central values obtained in \cite{Lee:2005ds} using the background field method are
only slightly above the values from the standard form factor calculation of \cite{Boinepalli:2006xd}.
To study possible finite size effects, the background-field calculations were repeated for a 
smaller volume of $V\approx(1.6\fm)^3$. The absolute values for the proton 
and neutron magnetic moment came out significantly lower than for the larger volume,
even for large pion masses.
This may indicate that effects from the boundary, due to abandoning the
quantization condition and using Dirichlet boundary conditions, 
may be non-negligible and affect the results even in the larger volume.

%
%

An extensive study of the form factors of the $\Delta$-baryon for a much larger range of values
of the momentum transfer squared, $Q^2=0,\ldots,2.25\GeV^2$, in quenched and unquenched lattice QCD 
was presented recently \cite{Alexandrou:2008bn}. 
Here we will concentrate on a discussion of the unquenched results. 
Calculations were performed using $n_f=2$ flavors of Wilson
fermions and the Wilson gauge action for a lattice spacing
of $a\approx0.077\fm$ in a volume of $V\approx(1.85\fm)^3$, for three values of 
the pion mass ranging from $\approx384\MeV$ to $\approx691\MeV$. Further results were obtained
in the framework of the mixed action approach with $n_f=2+1$ flavors of DW valence
fermions and Asqtad staggered sea quarks (using MILC configurations), for
a lattice spacing of $a\approx0.124\fm$, a volume of $V\approx(3.5\fm)^3$, and 
a pion mass of $m_\pi\approx353\MeV$.
The analysis was based on 
a standard ratio of $\Delta^+$-three- to two-point functions, where the
three-point functions were calculated using the sequential source technique
with the sink momentum set to zero.
In this study, the (local) charge weighted electromagnetic current, 
$J_{EM}^\mu=\sum_{q=u,d}e_q\overline{q}\gamma^\mu q$, was employed,
and we note that only contributions from quark line connected diagrams were included.
As usual, the current operator was renormalized such that $G^{\Delta^+}_{E0}(0)=1$.
The $Q^2$-dependence of the electric charge form factor $G^{\Delta^+}_{E0}(Q^2)$ is displayed
in Fig.~\ref{FFDelta_ge0_dyn_red} for the lowest accessible pion masses, 
showing a very good precision of the lattice
data points. Although the $n_f=2$ Wilson data points tend to be a bit higher,
overall consistency of the results can be observed within statistical errors. 
A dipole fit to the $n_f=2+1$ mixed action calculation is represented by the lighter
shaded error band, from which a mean square radius of $\langle r_E^2\rangle=0.411(28)\fm^2$
was obtained, compared to $\langle r_E^2\rangle=0.373(21)\fm^2$ for the $n_f=2$ Wilson case,
for pion masses indicated in the figure.

\begin{figure}[t]
   \begin{minipage}{0.48\textwidth}
      \centering
          \includegraphics[angle=0,width=0.9\textwidth,clip=true,angle=0]{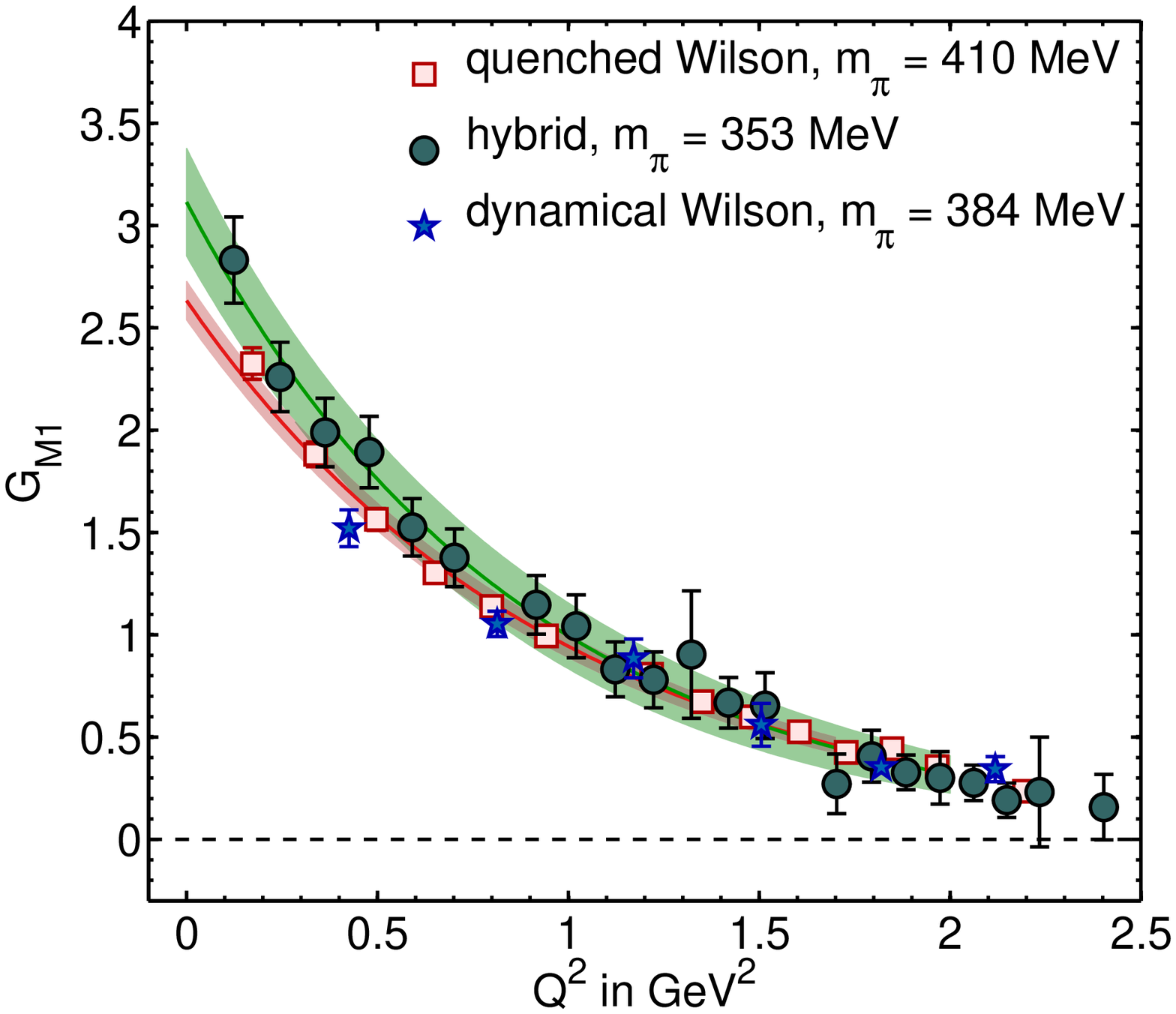}
  \caption{Connected contributions to the magnetic form factor of the $\Delta^+$-baryon
  (from \cite{Alexandrou:2008bn}).}
  \label{FFDelta_gm1_dyn}
     \end{minipage}
     \hspace{0.5cm}
    \begin{minipage}{0.48\textwidth}
      \centering
          \includegraphics[angle=0,width=0.85\textwidth,clip=true,angle=0]{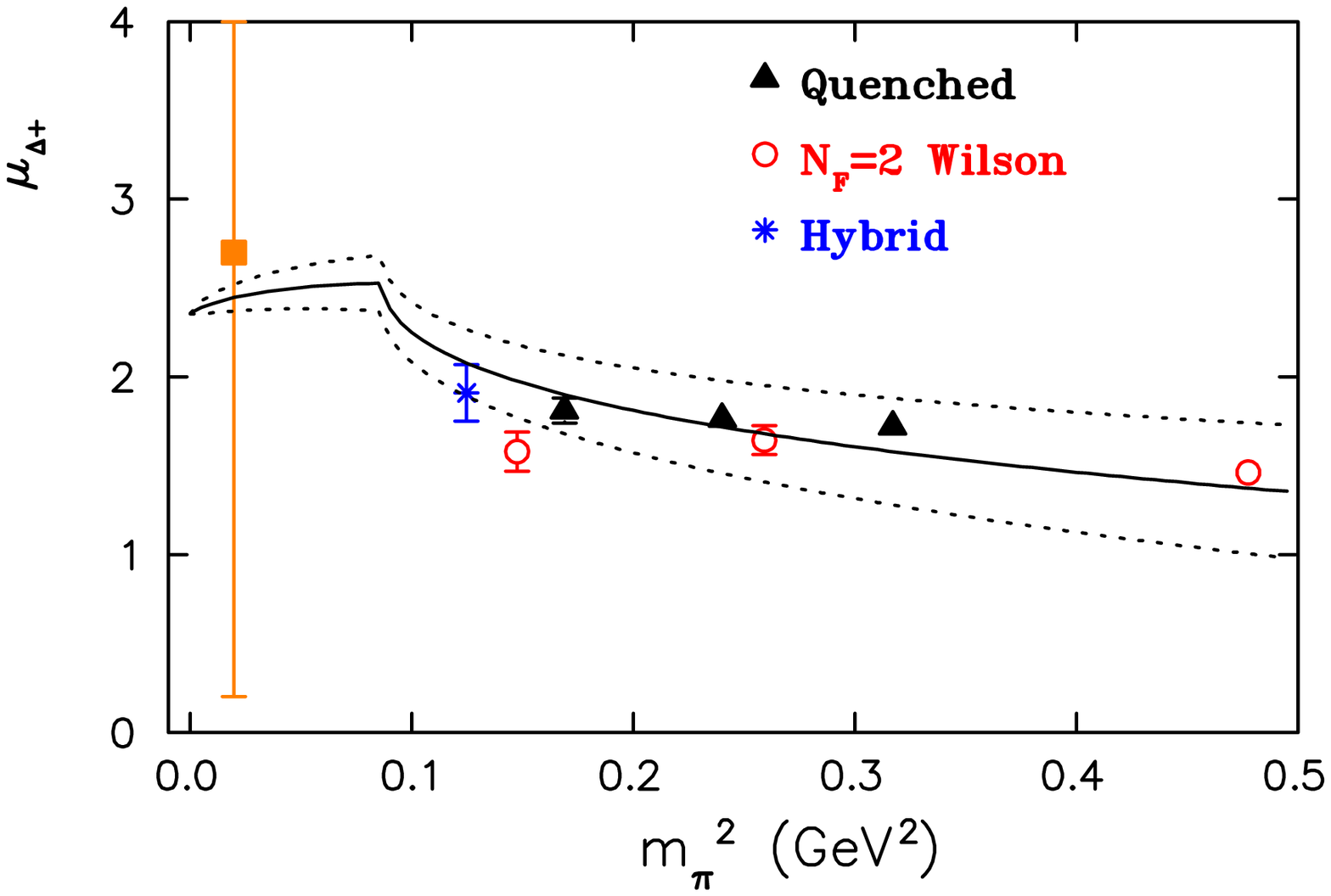}
  \caption{Pion mass dependence of the (real part of the) $\Delta^+$ magnetic form factor in units 
  of nuclear magnetons (from \cite{Alexandrou:2008bn}).}
  \label{FFDelta_mu}
     \end{minipage}
 \end{figure}

Figure \ref{FFDelta_gm1_dyn} shows corresponding results for the
magnetic dipole form factor $G^{\Delta^+}_{M1}(Q^2)$, also with quite
remarkable statistical precision. In this case, an exponential
ansatz, $G^{\Delta^+}_{M1}(Q^2)=G^{\Delta^+}_{M1}(0)\exp(-Q^2/\Lambda^2_{M1})$, 
was used to fit the data and extrapolate to $Q^2=0$. 
The pion mass dependence of the resulting magnetic moment,  $\mu_{\Delta^+}=G^{\Delta^+}_{M1}(0)$
transformed to nuclear magnetons by multiplication with a factor $2m^{\phys}_N/(2m^{\lat}_\Delta)$,
is displayed in Fig.~\ref{FFDelta_mu}.
An extrapolation to the chiral limit based on a relativistic ChPT calculation in the $\delta$-expansion scheme
\cite{Pascalutsa:2004je} is indicated by the curves, clearly showing 
a cusp at the pion production threshold, $m_\delta>m_N+m_\pi$, where  $\mu_{\Delta^+}$
acquires an imaginary part. 
Very good agreement with the central value of the experimental result, represented by the filled square,
is found at the physical pion mass. Clearly, this has to be regarded with great caution
due to the large errors of the experimental data point, as well as potential
systematic uncertainties of the lattice calculation and the chiral extrapolation.
%
%

\begin{figure}[t]
   \begin{minipage}{0.48\textwidth}
      \centering
          \includegraphics[angle=0,width=0.9\textwidth,clip=true,angle=0]{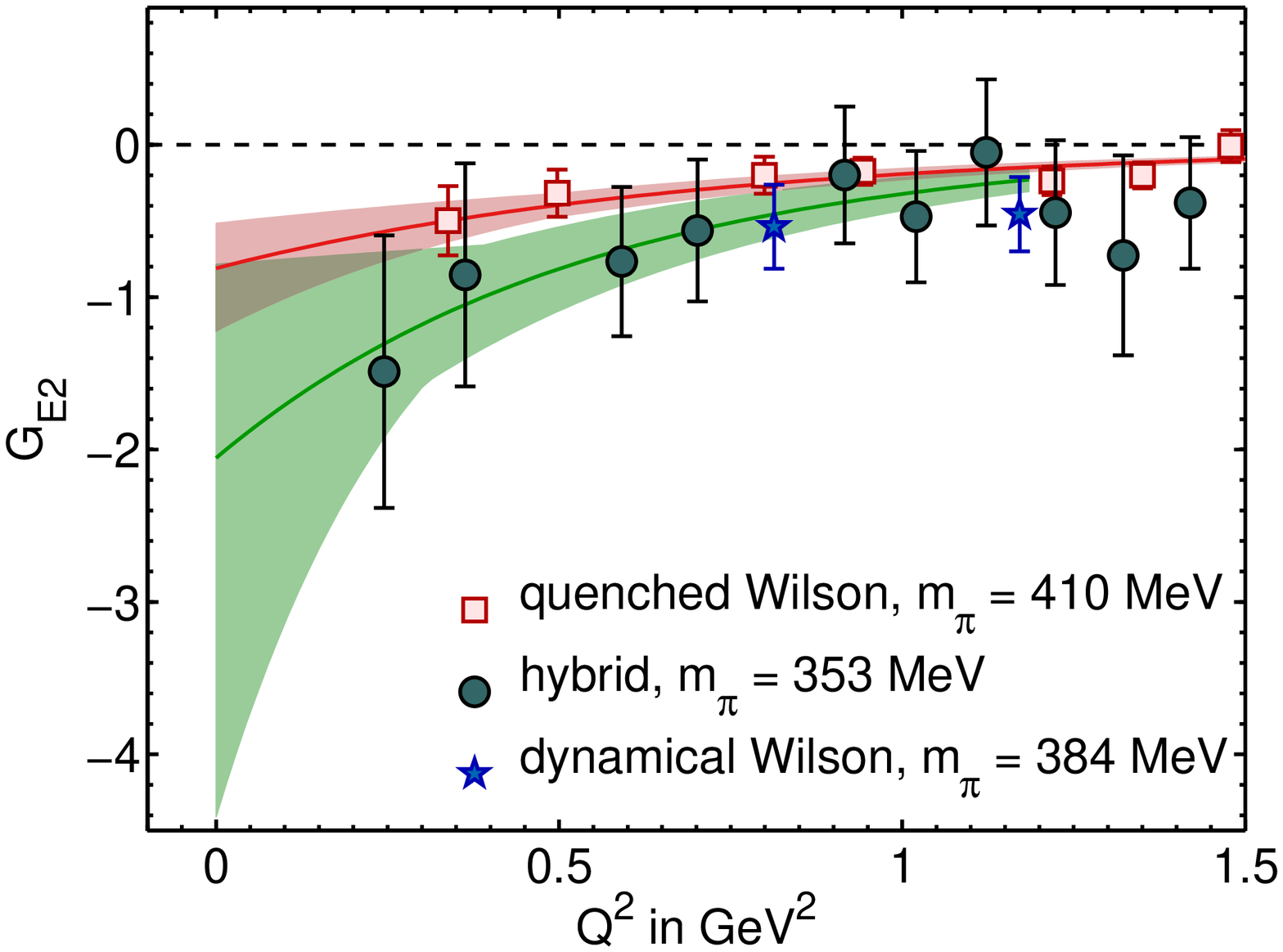}
  \caption{Connected contributions to the quadrupole form factor of the $\Delta^+$-baryon (from \cite{Alexandrou:2008bn}).}
  \label{FFDelta_ge2_dyn}
  \end{minipage}
     \hspace{0.5cm}
    \begin{minipage}{0.48\textwidth}
      \centering
          \includegraphics[angle=0,width=0.9\textwidth,clip=true,angle=0]{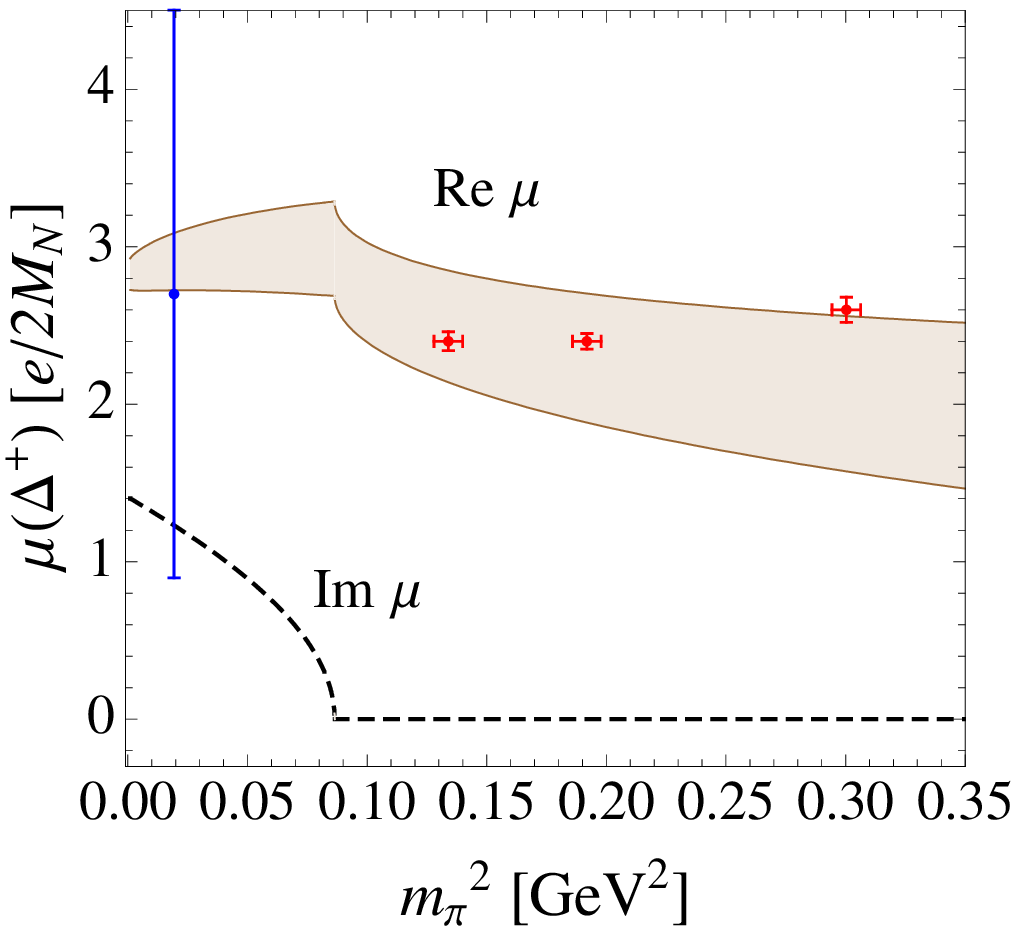}
  \caption{Pion mass dependence of the $\Delta^+$ magnetic moment 
  from a background field calculation (from \cite{Aubin:2008qp}).}
  \label{FFDelta_mu_Aubin}
     \end{minipage}
 \end{figure}

The $Q^2$-dependence of the electric quadrupole form factor, $G^{\Delta^+}_{E2}(Q^2)$, is shown in Fig.~\ref{FFDelta_ge2_dyn},
together with extrapolations based on the exponential ansatz given above.
Although the statistical errors are substantial, the data shows a clear trend towards
sizeable negative values at $Q^2=0$.

A first calculation of the magnetic moments of the $\Delta$-baryons (and the $\Omega^-$)
using background field methods in full QCD has been presented very recently \cite{Aubin:2008qp}.
In this study, the constant external magnetic field, $B$, was implemented using a
``patched'' electromagnetic potential 
(or, in a different terminology, including transverse links \cite{Detmold:2008xk})
as discussed in section \ref{sec:BGfield}, leading to the much less restrictive
quantization condition $qB=2\pi n/L^2$. Calculations were performed for
$n_f=2+1$ flavors of stout-smeared clover-Wilson fermions on anisotropic
lattices with lattice spacings of $a_t\approx0.035\fm$ and $a_s\approx0.122\fm$
in temporal and spatial directions, respectively, spatial volumes
of $V\approx(1.95\fm)^3$ and $V\approx(2.93\fm)^3$, and pion masses
of $\approx438\MeV$ and $\approx548\MeV$ for the smaller, and 
$\approx366\MeV$ for the larger volume.
For the study of the $\Delta$ magnetic moments, three different
$B$-fields, corresponding to $n=\pm1/2,\pm1$ and $\pm2$, were employed.
Boundary effects for the choice $n=\pm1/2$ violating the quantization
condition were expected to be negligible \cite{Aubin:2008qp}.
Mass differences in the presence of the external magnetic fields
were extracted from the $\Delta$- and $\Omega^-$-baryon two-point functions,
from which the corresponding magnetic moments could be obtained
as coefficients of the term linear in $B$, 
$\Delta m_h^{S_z}=m_h^{S_z}(-B)-m_h^{S_z}(B)=2\hat S_z B\mu+\mathcal{O}(B^3)$,
where 
$\hat S_z=S_z/S$ denotes the spin projected in $z$-direction.
Comparing the results for the smaller and larger magnetic fields, 
clear effects from contributions of $\mathcal{O}(B^3)$ 
could be observed in particular for larger values of $B$. 
The data points obtained for the smallest magnetic fields corresponding to $n=\pm1/2$ 
were used as final numbers, and shifts from non-vanishing $\mathcal{O}(B^3)$ terms
were estimated to be at most $5\%$.
The results for magnetic moment of the $\Delta^+$-baryon in units of
nuclear magnetons are displayed in Fig.~\ref{FFDelta_mu_Aubin} 
as a function of $m_\pi^2$, together with the result of
a ChPT calculation in the $\delta$-expansion scheme \cite{Pascalutsa:2004je},
represented by the error band. As already noted above with respect to 
Fig.~\ref{FFDelta_mu}, $\mu_{\Delta^+}$
acquires an imaginary part at the $\Delta \rightarrow \pi N$ threshold, 
i.e. for $m_\delta>m_N+m_\pi$. 
Comparing the results in Fig.~\ref{FFDelta_mu_Aubin} and Fig.~\ref{FFDelta_mu},
we note that the values from the background field calculation are
$\approx25-30\%$ above the data points obtained in the conventional form factor
approach based on three-point functions. It would be important to understand
the origin of this discrepancy and see if it is, e.g., related to the
$Q^2$-extrapolation in the form factor calculation, or other systematic
uncertainties in these studies.
The usual discretization and finite volume effects (not related to the constant magnetic field
and boundary effects) in the background field study 
were estimated to be small, $\approx3\%$ and $\approx2\%$, respectively.
We note that this study is, in the terminology of \cite{Lee:2008qf}, ``$U(1)$-quenched'',
i.e. the constant external magnetic fields were implemented on the level of the 
valence quarks and the quark propagators, but not in the generation of the gauge
configuration, so that the sea quarks effectively carry no electric charges.
Effects from a coupling of the sea quarks to the electromagnetic fields were expected to be at most $1\%$.
Taking all statistical and estimated systematic uncertainties into account, 
a value of $\mu_{\Omega^-}=-1.93(8)_\text{stat}(12)_\text{sys}$ in nuclear magnetons was given
for the magnetic moment of the $\Omega^-$-baryon, obtained directly from a calculation
close to the physical strange quark mass. It is encouraging to see that this is in 
excellent agreement with the average value from the PDG \cite{PDG2008}, $\mu_{\Omega^-}=-2.02(5)_\text{stat+sys}$.

%
%
%
%

%
%

\subsection{Background field methods and static hadron properties}
\label{sec:BGfield}
Static quantities like magnetic moments and polarizabilities can be 
studied using background field methods. Constant electric $\mathbf{E}$ and magnetic $\mathbf{B}$
background fields lead to energy shifts (i.e. shifts in the hadron masses) that can be
expanded for small fields,
\bea
\label{massshift}
\Delta E=-\mathbf{p}\cdot\mathbf{E}-\boldsymbol{\mu}\cdot\mathbf{B} 
-\frac{1}{2}\big(\alpha_E \mathbf{E}^2+\beta_M \mathbf{B}^2\big) + \ldots\,,
\eea
where the electric and magnetic dipole moments are denoted by $\mathbf{p}$ and $\boldsymbol{\mu}$,
respectively, and where $\alpha_E$ is the electric, and $\beta_M$ is the magnetic polarizability.
In a lattice calculation, the mass shifts can be extracted from the exponential decay 
of the corresponding hadron two-point correlators.
Specifically for the magnetic moments $\mu$, background field methods provide an
interesting alternative to the standard approach based on hadron three-point functions
discussed in the previous sections,
where $\mu$ is obtained from an extrapolation of the magnetic form factor to $Q^2=0$, $\mu=G_M(Q^2=0)$. 
A standard way to couple constant external fields, represented by the electromagnetic potential
$A^{EM,\text{ext}}_\mu$, to quark fields in the continuum is to use the minimal coupling prescription, i.e.
a covariant derivative $D_\mu=\partial_\mu-igA_\mu+ie_q A^{EM,\text{ext}}_\mu$, where $e_q$ is the electric
charge of a quark.
In the lattice framework, this corresponds to a replacement of the link variables,
$U_\mu\rightarrow \exp(ie_qA^{EM,\text{ext}}_\mu)U_\mu$, i.e. a multiplication with
a $U(1)$ phase factor. In full QCD, the phase factors have in principle to be
included already in the generation of the gauge configurations, which would, however, be
prohibitively expensive since usually several different values of the external fields need to be 
considered\footnote{Furthermore, due to the different quark charges, 
individual flavors would have to be treated
separately in the Monte Carlo integration.}.
Therefore, to this date, background field calculations have been based on existing configurations
generated for vanishing external fields, and non-zero electric and magnetic fields were
only implemented on the level of the valence quark propagators. This approach has been dubbed
``$U(1)$-quenched'' \cite{Lee:2008qf}, since it corresponds to a situation where 
just the valence quarks carry electric charges and respond to the external fields, while the sea quarks
are treated as uncharged objects.
The effect of an electromagnetic coupling to the sea quarks can, however, be
taken into account perturbatively by expanding the fermion action
in the external fields,.
This leads to insertions of the fields into hadron two-point
correlators and therefore requires the computation of three- and four-point functions.
A recent study and application of this approach to the electric polarizability 
of the neutron by Engelhardt \cite{Engelhardt:2007ub}
will be discussed below in section \ref{sec:Polarizabilities}.

When placing constant external fields on a finite lattice, particular attention has to be paid
to the boundary conditions. Consider for example a constant magnetic field in $z$-direction,
$B=\mbf{B}_z$, obtained from a background potential with non-zero $y$-component,
$A^{\EM,\text{ext}}_{\mu=y}(x,y,z)=Bx$. If we require the corresponding links
$U^{\EM,\text{ext}}_y=\exp(ie_qBx)$ to be continuous across the boundary in the $x$-direction with periodic boundary
conditions, the magnetic field component has to be ``quantized'' in the form $e_qB=2\pi n/L$ with
integer $n$ and where $L$ denotes the spatial lattice extent. Unfortunately for the currently
accessible lattice volumes, even for $n=1$ this corresponds to magnetic fields that are
way too large to justify an expansion as in Eq.~(\ref{massshift}). Clearly, if the quantization
condition is not fulfilled, the (otherwise constant) effective magnetic field 
seen by the charged quarks can strongly fluctuate due to the discontinuity of the link 
(or rather the discontinuity of the plaquette related to the magnetic flux, see below) at the boundary,
which may seriously affect the mass shift.
One approach that has been used in the literature to tame such finite volume effects is to 
employ Dirichlet boundary conditions in (for this example) the $x$-direction in combination with
numerically small magnetic fields independent of any quantization condition (see, e.g., \cite{Lee:2005ds}). 
Boundary effects may then be kept at a minimum by placing the hadron source at maximal distance to the boundary, 
i.e. central in $x$-direction.
A different method that directly relaxes the quantization condition
is based on the realization that it is not the link variable that needs to be continuous
across the boundary, but rather the plaquette 
in the $(x,y)$-plane that corresponds to the magnetic flux \cite{Aubin:2008hz,Detmold:2008xk}.
The continuity of the plaquette can be achieved by setting another component of the 
EM-potential non-zero, this time in $x$-direction,
specifically $A^{\EM,\text{ext}}_{\mu=x}(x,y,z)=-BLy$ at the $x$-boundary, i.e. when $(x,y,z)=(L-1,y,z)$,
but keeping $A^{\EM,\text{ext}}_{\mu=x}(x,y,z)=0$ for $x\not=L-1$.
Then, as long as $y\not=L-1$, the $(x,y)$-plaquette is given by $\exp(-ie_qB)$ on both sides of 
the $x$-boundary and is therefore indeed continuous. However, at the point where the plaquette
crosses the boundaries in $x$- and $y$-direction, i.e. for $(x,y,z)=(L-1,L-1,z)$,
the plaquette changes from $\exp(-ie_qB)$ to $\exp(ie_qB(L^2-1))$. A continuous magnetic
flux then requires a quantized magnetic field given by $e_qB=2\pi n/L^2$. 
Hence, for typical spatial lattice extents of $L\ge16$, the smallest allowed
non-zero values of $B$ are significantly reduced compared to the initial 
constraint ensuring continuous links.
This method of ``patching'' the potential by, e.g., setting $A^{\EM,\text{ext}}_{\mu=x}(x,y,z)=-BLy$ 
at the $x$-boundary has been studied and successfully used in calculations of the magnetic
moment of the $\Delta$-baryon \cite{Aubin:2008hz,Aubin:2008qp} and of the electric
polarizabilities of hadrons \cite{Detmold:2008xk}, where in the latter case
``transverse links'' have been included at the time-boundary to relax the 
quantization condition for the background electric field.
We note that the background field method may also be extended to the calculation
of more general hadronic matrix elements by using exponentiated operators in the action \cite{Detmold:2004kw}.

\subsubsection{Hadron polarizabilities}
\label{sec:Polarizabilities}
%
\begin{figure}[t]
    \begin{minipage}{0.48\textwidth}
      \centering
  \includegraphics[angle=90,width=0.9\textwidth,clip=true,angle=0]{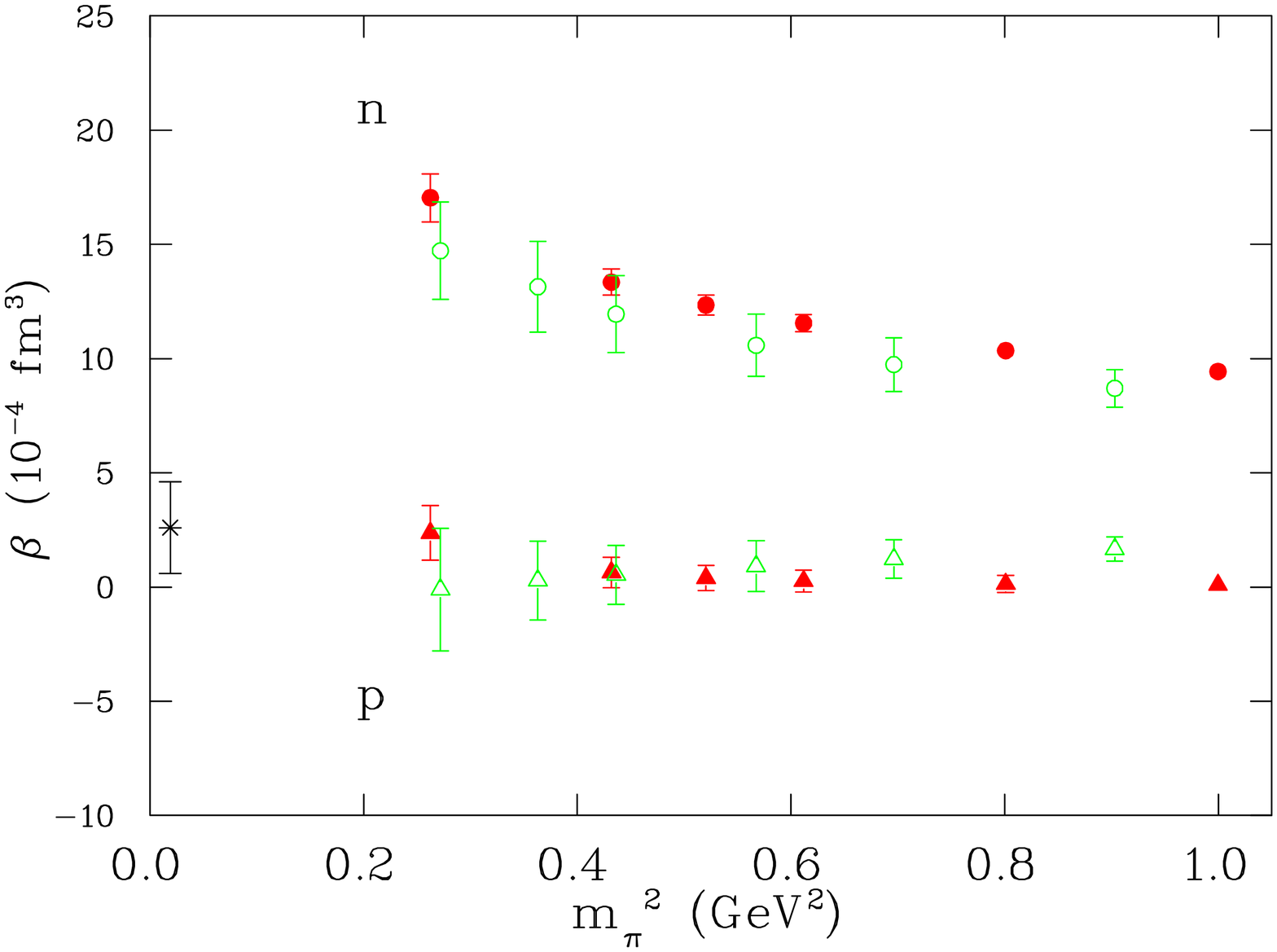}
  \caption{Pion mass dependence of the nucleon magnetic polarizabilities in the quenched approximation (from \cite{Lee:2005ds}).}
  \label{Polariz_nucl_mpol_Wilcox}
     \end{minipage}
            \hspace{0.5cm}
      \begin{minipage}{0.48\textwidth}
      \centering
 \includegraphics[angle=90,width=0.9\textwidth,clip=true,angle=0]{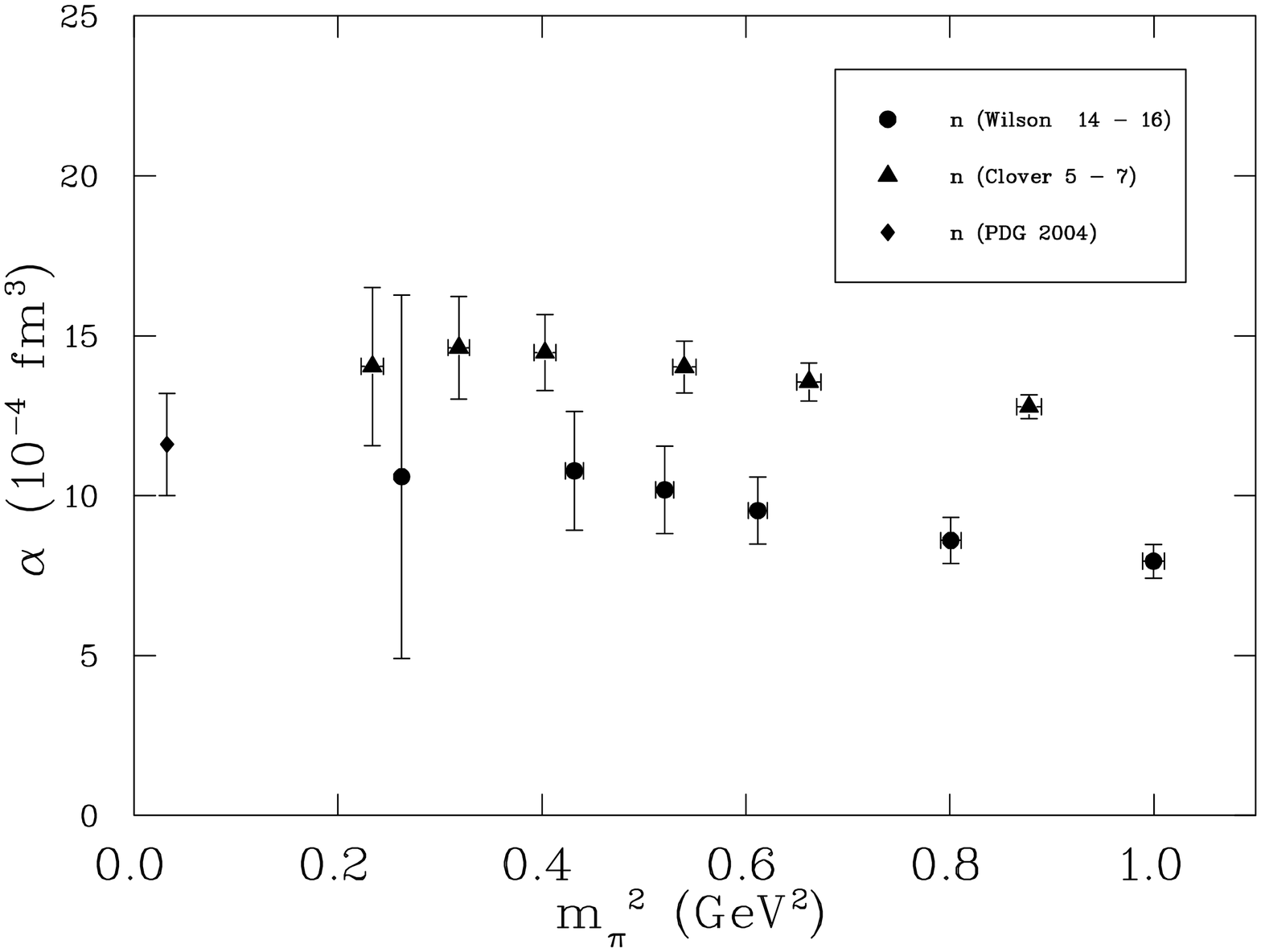}
  \caption{Pion mass dependence of the neutron electric polarizability in the quenched approximation (from \cite{Christensen:2004ca}).
  Note that the overall sign of the lattice data should be inverted for comparison with experiment \cite{Shintani:2006xr}.}
  \label{Polariz_n_epol_wilson_Walter_cl_octnxx}
     \end{minipage}
 \end{figure}
%
Extensive studies of hadron electric and magnetic polarizabilities in quenched lattice QCD have been 
presented in \cite{Christensen:2004ca} and \cite{Lee:2005dq}, respectively. 
As described in sections \ref{sec:BGfield},\ref{sec:polarizabilities}, 
the electric, $\alpha_E$, and magnetic, $\beta_M$, 
polarizabilities lead to hadron mass shifts in the presence of static external
electric, $\mbf{E}$, and magnetic, $\mbf{B}$, fields, i.e. 
$\Delta m(E)= m(E)-m(0)=-\alpha_E\mbf{E}^2/2$ and $\Delta m(B)= m(B)-m(0)=-\beta_M\mbf{B}^2/2$.
Here we have already dropped possible contributions that are odd in the external fields, 
for example a magnetic moment contribution $-\boldsymbol{\mu}\cdot\mbf{B}$,
which are usually removed by considering averages over positive and negative valued fields.
In quenched lattice QCD, the static external fields can be implemented by multiplying
the link variables that enter the calculation of the quark propagators by 
phase factors, for example $U_z\rightarrow \exp(ie_qA_z)U_z$ with a gauge potential $A_z(t)=\mbf{E}_zt$
for a static electric field in $z$-direction, or $U_y\rightarrow \exp(ie_qA_y)U_y$
with a gauge potential $A_y(\mbf{x})=\mbf{B}_z\mbf{x}_x$ for a static magnetic field in $z$-direction.
As discussed above in section \ref{sec:BGfield}, 
such additional $t$- or $\mbf{x}$-dependent phase factors are in general in conflict with 
the (periodic) boundary conditions for the link variables on a finite lattice,
and the electromagnetic fields have in principle to fulfill certain quantization conditions. 

\begin{figure}[t]
      \centering
 \includegraphics[angle=0,width=0.45\textwidth,clip=true,angle=0]{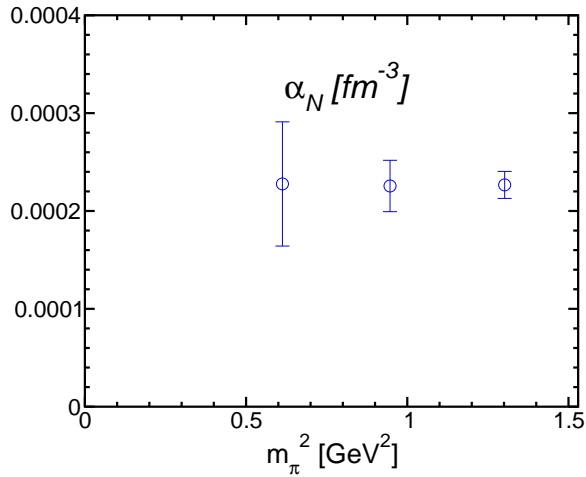}
  \caption{Pion mass dependence of the neutron electric polarizability (from \cite{Shintani:2006xr}).}
  \label{Pol_n_massdep_Aoki}
 \end{figure}

In \cite{Christensen:2004ca} and \cite{Lee:2005dq},
a linearized version of the phase factors, $\exp(iq_fA)U\rightarrow (1+iq_fA)U$, was used
together with Dirichlet boundary conditions in the respective directions. 
Four different values for the electric and for the 
magnetic fields were used, that allowed the extraction of the polarizabilities 
as coefficients of the $E^2$- and $B^2$-dependences from parabolic fits to the effective
mass shifts. The effective mass shifts were obtained in a standard way
from the exponential decay of hadron two-point functions in the presence/absence of the
external fields. Most of the calculations were based on Wilson fermions 
and the Wilson gauge action, with pion masses in the range
$m_\pi\approx512,\ldots,1000\text{ MeV}$, for a lattice spacing of $a\approx0.1\fm$ and a 
volume of $V\approx(2.4\fm)^3$. 
Results for the for the proton and neutron magnetic polarizabilities
as functions of the pion mass are shown in Figure \ref{Polariz_nucl_mpol_Wilcox},
which may be compared to the average values from the PDG of $\beta_M^p=1.9(5)\cdot10^{-4}$
for the proton, and of $\beta_M^n=3.7(2.0)\cdot10^{-4}\fm^{3}$ for the neutron \cite{PDG2008}.
While the lattice data points for $\beta_M^p$ are overall consistent with
the value from experiment, the lattice results for the neutron magnetic polarizability
are surprisingly different and show a trend towards large values of $\beta_M^n$ at
smaller pion masses.
Further calculations for a similar range of pion masses were performed based on clover 
improved Wilson fermions and a L\"uscher-Weisz gauge action, for a lattice 
spacing of $a\approx0.17\fm$ and a volume of $V\approx(2\fm)^3$. 
Corresponding lattice results for the neutron electric polarizability are displayed 
in Fig.~\ref{Polariz_n_epol_wilson_Walter_cl_octnxx} \cite{Christensen:2004ca}, which may be compared to the
PDG average of $\alpha_E^n=11.6(1.5)\cdot10^{-4}\fm^{3}$ \cite{PDG2008}. 
Importantly, it has been noted in \cite{Shintani:2006xr} that 
the real-valued electric field used in \cite{Christensen:2004ca} in the 
(linearized) lattice phase factor, $U_z\rightarrow (1+ie_q\mbf{E}_zt)U_z$, 
corresponds to an imaginary electric field in Minkowski space, $\mbf{E}_{\text{Mink}}=i\mbf{E}$, so that
the sign of the mass shift has to be flipped, 
$\Delta m(E)= m(E)-m(0)=-\alpha_E\mbf{E}_{\text{Mink}}^2/2=\alpha_E\mbf{E}^2/2$\footnote{
Alternatively, an imaginary potential could be used in the first place 
for the Euclidean phase factor, e.g. $U_z\rightarrow \exp(-e_q\mbf{E}_zt)U_z$
with real-valued $\mbf{E}=\mbf{E}_{\text{Mink}}$ \cite{Shintani:2006xr,Alexandru:2008sj}.}.
Consequently, the signs of the lattice data points in Fig.~\ref{Polariz_n_epol_wilson_Walter_cl_octnxx}
should be flipped for comparison with the experimental value, turning
the consistency of the lattice results with experiment that was observed in \cite{Christensen:2004ca}
into a strong disagreement, $\alpha^{n,\exp}_E=11.6(1.5)\cdot10^{-4}\fm^{3}$ compared to
$\alpha^{n,\lat}_E\approx-12(4)\cdot10^{-4}\fm^{3}$ at the lowest accessible pion mass. 
These results have been challenged in a number of recent lattice calculations, which may also
provide possible explanations for the observed striking discrepancy. 
Using a full, non-linearized, (real-valued) phase factor, $U_z\rightarrow \exp(e_q\mbf{E}_zt)U_z$,
together with periodic boundary conditions, calculations in quenched 
lattice QCD of the neutron electric polarizability were performed for 
domain wall as well as clover improved fermions and a renormalization improved gauge action in \cite{Shintani:2006xr}.
For the domain wall calculation with a lattice spacing of $a\approx0.1\fm$ 
in a volume of $V\approx(1.6\fm)^3$, a positive value of $\alpha_E^n=1.32(2)\cdot10^{-4}\fm^{3}$ was 
found at a large pion mass ($m_\pi/m_\rho\approx0.88$).
The clover improved Wilson fermion results for the same lattice spacing, but
a larger volume of $V\approx(2.4\fm)^3$, are shown in 
Fig.~\ref{Pol_n_massdep_Aoki} versus the squared pion mass, also 
providing positive values of the order of $\alpha_E^n\approx2.0\cdot10^{-4}\fm^{3}$.

%
%
%

\begin{figure}[t]
    \begin{minipage}{0.48\textwidth}
      \centering
  \includegraphics[angle=0,width=0.45\textwidth,clip=true,angle=0]{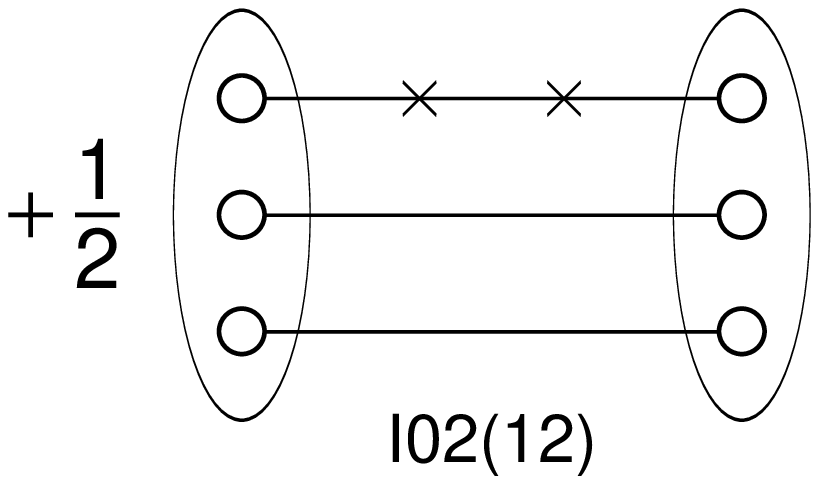}
  \includegraphics[angle=0,width=0.45\textwidth,clip=true,angle=0]{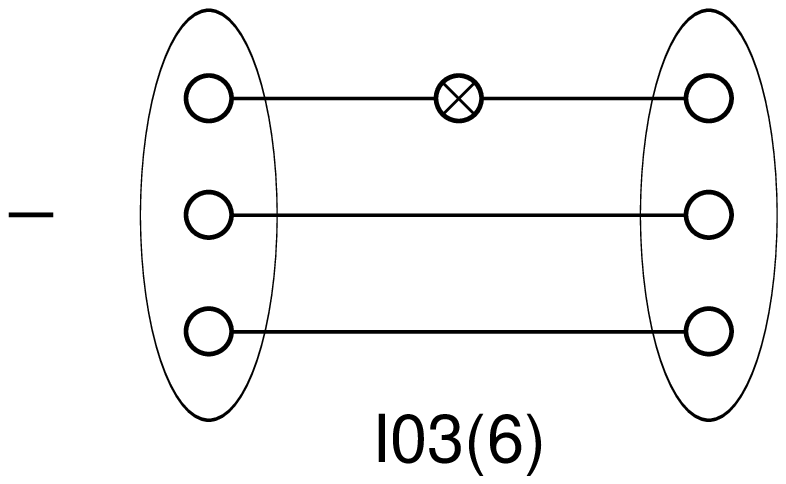}
  \caption{Contributions to the neutron two-point function expanded in the external gauge potential
   to $\mathcal{O}(A_\mu^2)$ (from \cite{Engelhardt:2007ub}).}
  \label{Polariz_n_elec_diagrams_Engelh}
     \end{minipage}
     \hspace{0.5cm}
  \begin{minipage}{0.48\textwidth}
         \centering
         \includegraphics[angle=-90,width=0.9\textwidth,clip=true]{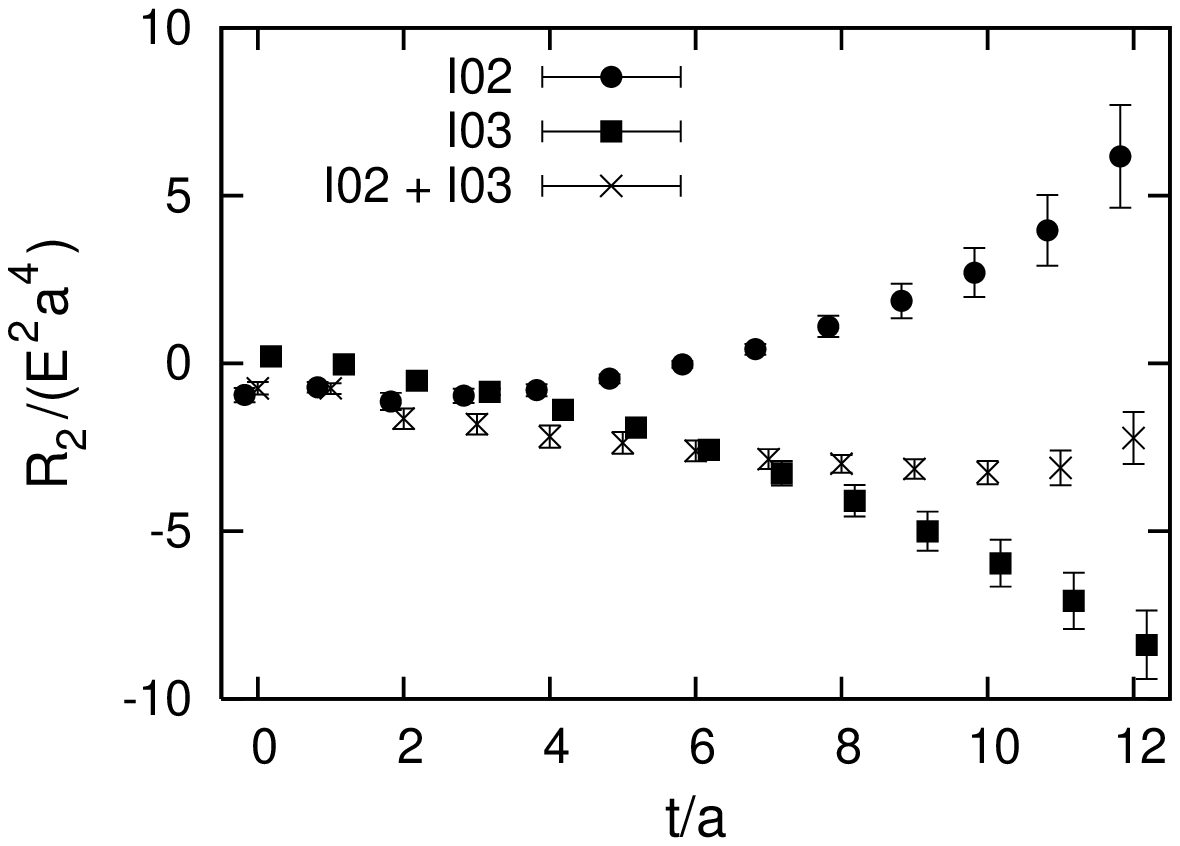}
  \caption{Contributions of diagrams as in Fig.~\ref{Polariz_n_elec_diagrams_Engelh}
  to the ratio $R_2(t)$ (from \cite{Engelhardt:2007ub}).}
  \label{Polariz_n_elec_cancel_Engelh}  
     \end{minipage}
 \end{figure}

More evidence for a small positive value of $\alpha_E^n$ at similar pion masses 
has been found in a first study of the neutron polarizability in 
unquenched lattice QCD \cite{Engelhardt:2007ub}. 
At first sight, such a calculation appears prohibitively expensive, since
the static background fields would have to be included in the generation of the gauge configurations.
The alternative method explored in \cite{Engelhardt:2007ub} is to perturbatively expand 
the neutron two-point function in the electric background field $E$
to the desired order, $\mathcal{O}(E^2)$, resulting in a
set of space-time integrals of correlators, i.e. four-point functions, 
which only depend on the original gauge links but include explicit insertions of the external fields.
Examples for corresponding connected diagram contributions are shown in Fig.~\ref{Polariz_n_elec_diagrams_Engelh}, 
where a cross represents an insertion linear in the background gauge potential $A^{\text{ext}}_\mu$, and the
encircled cross represents a contact term $\propto A^{\text{ext}}_\mu{}^2$.
The relevant four point functions were computed using a standard sequential method,
and contributions from disconnected diagrams were 
evaluated using stochastic sources with complex $Z(2)$ noise, see section \ref{sec:methods}.
The neutron mass shift can then be extracted from the 
$t$-slope of the ratio of the expanded two-point function to the two-point function for vanishing
background fields, $R_2(t)=G_{(E^2)}(t)/G_{(E=0)}(t)$, i.e. $\Delta m=-R'_2(t)$, 
where $t$ is the source-sink separation in the Euclidean time direction. 
In extracting $\Delta m$ from $R_2(t)$, special care has been taken 
with respect to additional contributions to the mass shift
due to a constant background gauge potential $A^{\text{const}}_z=-E_zt_0$ on
a finite lattice, where $t_0$ is the time for which the background gauge 
potential vanishes, $A_z=\mbf{E}_z(t-t_0)$. For the details, we refer to
the corresponding discussions in \cite{Engelhardt:2007ub,Engelhardt:2007sm},
and just note that the final results in \cite{Engelhardt:2007ub}
were obtained by demanding that the mass shift 
is stationary in $A^{\text{const}}_z$ (or $t_0$).
The numerical computations were based on a
hybrid approach of domain wall valence fermions and staggered Asqtad sea
quarks using gauge configurations provided by the MILC collaboration, as described
in \cite{Edwards:2005ym,Hagler:2007xi}, for a lattice spacing of $a\approx0.124\fm$ and
a volume of $V\approx(2.5\fm)^3$. 
A value of $\alpha^n_E=2.0(9)\cdot10^{-4}$ fm$^3$ was finally obtained 
for a pion mass of $\approx760\MeV$\footnote{Here, we have already reversed
the sign of $\alpha^n_E$ as discussed above and also explained at the end of Ref.~\cite{Engelhardt:2007ub}.}.
Interestingly, this number is well compatible with the results of the quenched 
calculation displayed in Fig.~\ref{Pol_n_massdep_Aoki}. The question then
arises why the results from \cite{Christensen:2004ca}, shown in
Fig.~\ref{Polariz_n_epol_wilson_Walter_cl_octnxx}, are so different
in magnitude and sign. Possible explanations are the use of Dirichlet boundary
conditions in \cite{Christensen:2004ca}, the issue of a constant background gauge potential
on a finite lattice
\cite{Engelhardt:2007ub,Engelhardt:2007sm}, and the linearization of
the phase factor employed in \cite{Christensen:2004ca}. 
The use of Dirichlet boundary conditions has been studied in \cite{Shintani:2006xr},
and the last two points have been investigated in \cite{Engelhardt:2007ub}. 
Concerning the latter, it has been observed in particular that a linearization of the phase factor
excludes contributions with insertions $\propto (A^{\text{ext}}_\mu{})^2$ 
as represented on the right hand side in Fig.~\ref{Polariz_n_elec_diagrams_Engelh}.
These are indeed irrelevant in the naive continuum limit, but in general are required to
renormalize propagators with two insertions linear in the background gauge field
as shown on the left hand side in Fig.~\ref{Polariz_n_elec_diagrams_Engelh}.
Figure \ref{Polariz_n_elec_cancel_Engelh} explicitly shows that both
types of diagrams give substantial contributions to $R_2$, which however 
cancel to a large extent in the sum. 
In an attempt to understand the results of \cite{Christensen:2004ca},
$\alpha^n_E$ was calculated again for values of 
$t_0$ (related to the constant background potential $A^{\text{const}}_z$) 
resembling the choices in \cite{Christensen:2004ca}, and at the same time dropping the contributions
from quadratic insertions $\propto A^{\text{ext}}_\mu{}^2$ to reproduce
the situation of a linearized phase factor. 
Remarkably, these results turned out to be compatible with the findings in \cite{Christensen:2004ca},
i.e. $\alpha^n_E$ was found to be significantly larger and 
of opposite (negative) sign, albeit with larger statistical errors \cite{Engelhardt:2007ub}.

The issues of real versus imaginary potentials in the phase factors and 
their linearization have also been studied recently in detail by Alexandru and Lee \cite{Alexandru:2008sj}.
In the framework of a quenched study, they confirmed what has been
discussed above, in particular that the use of a proper Euclidean phase factor
gives small positive values for $\alpha^n_E$, in excellent quantitative agreement 
with the result from \cite{Engelhardt:2007ub}, and only slightly below the values in Fig.~\ref{Pol_n_massdep_Aoki} from 
\cite{Shintani:2006xr}.
Furthermore, it was shown that a linearization of the phase factor leads to large negative values
for the polarizability, fully consistent with the earlier results from \cite{Christensen:2004ca}
displayed in Fig.~\ref{Polariz_n_epol_wilson_Walter_cl_octnxx}
(when using the appropriate sign).

In any case, the results for the neutron electric polarizability 
from quenched lattice QCD in Fig.~\ref{Pol_n_massdep_Aoki}
and from the unquenched calculation discussed above \cite{Engelhardt:2007ub} at pion masses
around $600\MeV$ are approximately a factor of $5$ below the experimental average.
This may be expected from ChPT calculations, which predict that $\alpha^n_E$ 
diverges as $1/m_\pi$ to $+\infty$ in the chiral limit \cite{Bernard:1992qa}. 
Calculations at lower pion masses with better 
statistical precision have to be performed before a fully quantitative 
comparison of lattice results with experiment may be attempted using ChPT.
%

%
%

%

The electric polarizabilities of neutral and \emph{charged} hadrons have very recently
also been studied in unquenched lattice QCD in a mixed action approach based
on the $n_f=2+1$ domain wall configuration from the RBC-UKQCD collaboration and
tadpole-improved Clover-Wilson fermions for the valence quarks \cite{Detmold:2008xk}.
In contrast to the unquenched study in \cite{Engelhardt:2007ub} discussed above, the calculation
was mainly based on the energy shifts extracted directly from the corresponding hadron two-point
functions. Effects from a coupling of the sea quarks to the electric background field
were not included (``$U(1)$-quenched'' \cite{Lee:2008qf}).
As usual, the electric field was implemented by multiplying the 
gauge links by $U(1)$ links of the form $U^{\EM,\text{ext}}_{\mu=z}=\exp(-ie_qE_\text{\EM}t)$ 
with real valued $E_\text{EM}$
(requiring an analytic continuation to Minkowski space).
Particular attention has been paid to the quantization conditions on a periodic lattice (see 
our discussion in section \ref{sec:BGfield}).
It was argued that by including additional transverse links at the time boundary, the
electromagnetic flux represented by the plaquette can be rendered continuous across
the boundary if the electric field obeys the relaxed quantization condition
$qE=2\pi n/(L_tL_s)$ with integer $n$ and temporal and spatial lattice extents 
$L_t$ and $L_s$, respectively. Comparing effective masses in the presence of 
constant external electric fields, calculated with and without
the transverse link, and with integer and non-integer $n$, the effectiveness of the
method was also impressively verified on the numerical side.
Values for the electric polarizabilities of different charge-neutral hadrons were obtained
from fits to the particle energies as functions of the external field,
corresponding to $n=1,-2,3,4,5,-6,7$.
A preliminary value of $\alpha^n_E=3.6(1.3)\cdot10^{-4}\fm^3$ was obtained for, e.g.,
the neutron electric polarizability at a pion mass of $m_\pi\approx420\MeV$, which
is in good agreement with the results from \cite{Shintani:2006xr,Engelhardt:2007ub} discussed above.
Concerning charged particles, it is clear that they will be accelerated in the presence of
external electric fields, and at first sight it is not obvious how the two-point functions
precisely depend on the Euclidean time, the particle momentum and
the constant background field in this case. 
This has been studied in the non-relativistic limit  in an 
effective field theory framework for a spin-$1/2$ particle coupled to a classical gauge field
\cite{Detmold:2006vu}, where it was found that the exponential fall-off
gets modified for charged particles:
$C^h_\text{2pt}(t)\stackrel{NR}{\rightarrow}\exp(-E_ht-q^2E^2_\text{EM}t^3/(6m_h))$,
where $E_h$ is the hadron energy depending on the background field $E_\text{EM}$, and where
the hadron mass is given by $m_h=E_h(E_\text{EM}=0)$.
The energies $E_h$ of the charged hadrons in the presence
of external fields were extracted from fits to 
the respective two-point functions using the full functional form of the 
$t$-dependence of $C^h_\text{2pt}(t)$. Finally, the 
electric polarizabilities $\alpha^h_E$ were obtained from fits to
the $E_\text{EM}$-dependence of $E_h(E_\text{EM})=\alpha^h_E E^2_\text{EM}/2+\mathcal{O}(E^4_\text{EM})$. 
Examples for some of the preliminary results given in \cite{Detmold:2008xk}
are $\alpha^{\pi^+}_E=3.4(0.4)\cdot10^{-4}\fm^3$ the pion,
and $\alpha^{p}_E=8.8(5.9)\cdot10^{-4}\fm^3$ for the proton, for $m_\pi\approx420\MeV$.
It would be highly interesting to extend and improve this promising study
with respect to the statistics and systematics, for example by
investigating potential finite volume and discretization effects, 
and in particular the pion mass dependence of the results in greater detail.

\subsection{Discussion and summary}
\label{sec:FFsSummary}

\begin{figure}[t]
      \centering
 \includegraphics[angle=0,width=0.6\textwidth,clip=true,angle=0]{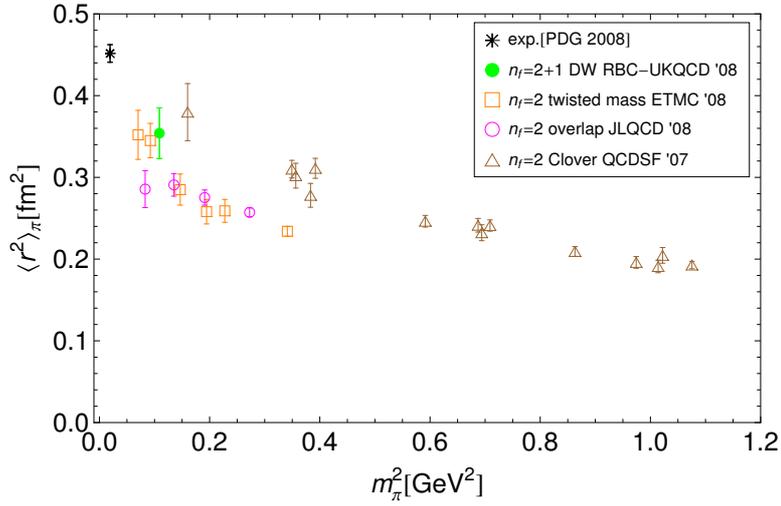}
  \caption{Overview of dynamical lattice QCD results for the pion mean square radius as a function
  of $m_\pi^2$.}
  \label{overview_r2Pi_v1}
 \end{figure}

\begin{figure}[t]
      \centering
 \includegraphics[angle=0,width=0.6\textwidth,clip=true,angle=0]{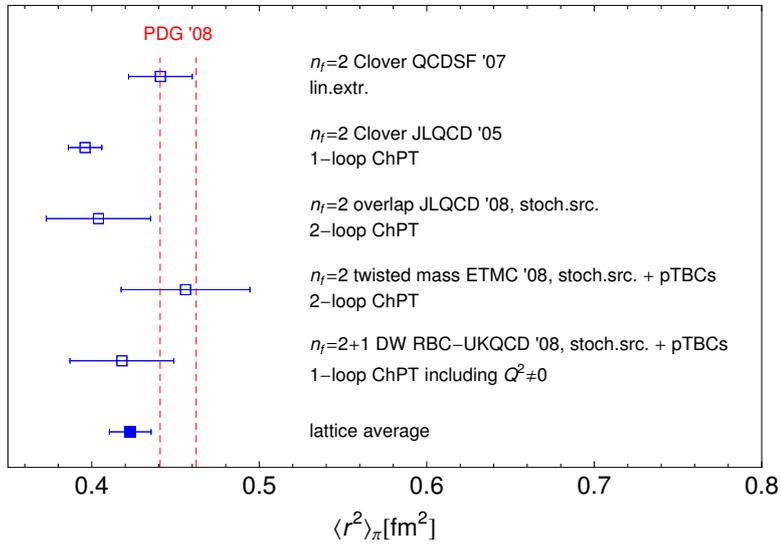}
  \caption{Overview of dynamical lattice QCD results for the chirally extrapolated pion mean square radius
  at the physical pion mass.}
  \label{overview_r2Pi_mPiPhys_v1}
 \end{figure}

\begin{figure}[t]
      \centering
 \includegraphics[angle=0,width=0.6\textwidth,clip=true,angle=0]{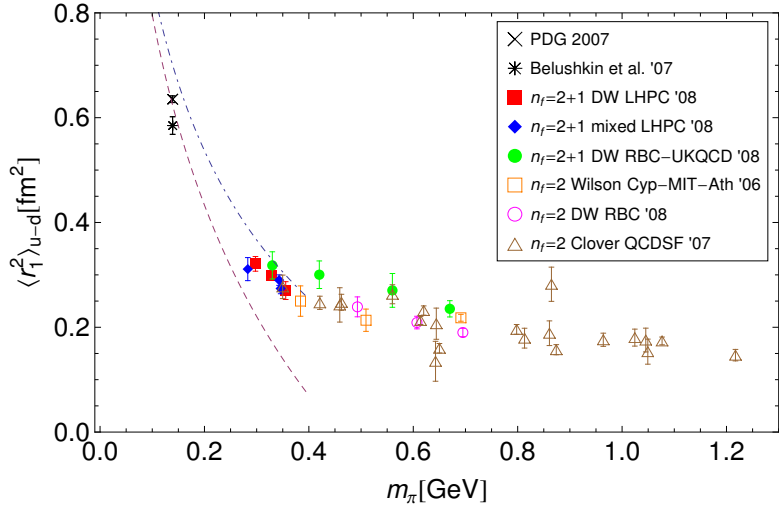}
  \caption{Overview of dynamical lattice QCD results for the isovector Dirac mean square radius. 
  The dashed curve represents the leading 1-loop HBChPT prediction \cite{Bernard:1992qa}, 
  and the dotted-dashed curve the result obtained in the SSE \cite{Gockeler:2003ay}, Eq.~(\ref{F1radius}).}
  \label{overview_r1_v2}
  \end{figure}

\begin{figure}[t]
      \centering
 \includegraphics[angle=0,width=0.6\textwidth,clip=true,angle=0]{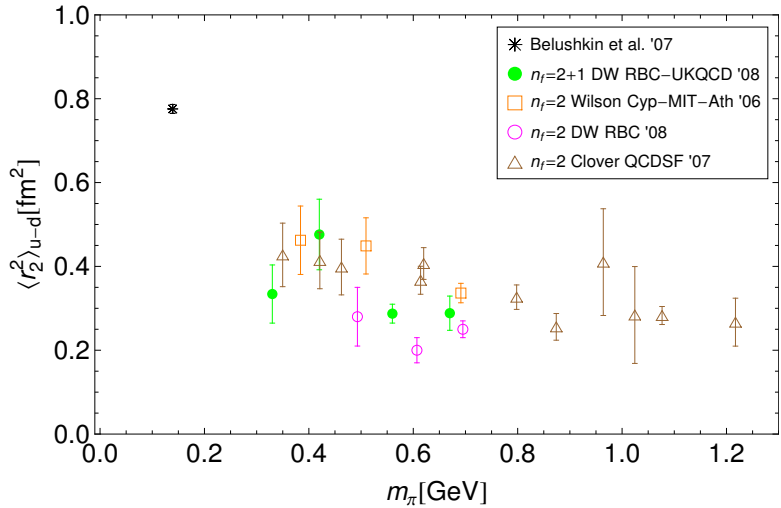}
  \caption{Overview of dynamical lattice QCD results for the isovector Pauli mean square radius.}
  \label{overview_r2_v2}
 \end{figure}

\begin{figure}[t]
      \centering
  \includegraphics[angle=0,width=0.6\textwidth,clip=true,angle=0]{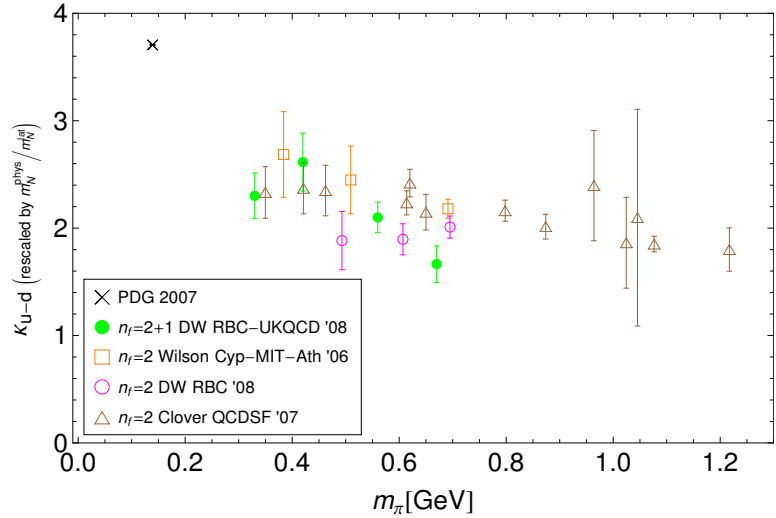}
  \caption{Overview of dynamical lattice QCD results for the isovector anomalous magnetic moment.}
  \label{overview_kappa_v2} 
\end{figure}

We have seen substantial progress in hadron form factor calculations
in recent years, in particular related to dynamical $n_f=2$ and $n_f=2+1$ lattice QCD simulations.
An overview of results from different collaborations for the pion and nucleon form factors 
is given in Fig.~\ref{overview_r2Pi_v1} and \ref{overview_r2Pi_mPiPhys_v1} for pion charge radius, in 
Figs.~\ref{overview_r1_v2} and \ref{overview_r2_v2} for the nucleon isovector Dirac and Pauli mean square radii,
and in Fig.~\ref{overview_kappa_v2} for the isovector anomalous magnetic moment.
It is remarkable that calculations are carried out routinely with pion masses as low as $300\MeV$ 
with excellent statistical precision for the pion form factor and in some cases very good precision
for the nucleon form factors. The statistics are in general lower for quantities like 
magnetic moments that have to be extracted from extrapolations of form factors to $Q^2=0$. 
%
%
%
%
%
%

Thanks to the use of advanced methods involving all-to-all propagators, the one-end-trick and 
partially twisted boundary conditions,
great progress has been reported by several collaborations with respect to
the pion form factor, where high precision results have been obtained in unquenched lattice QCD for
pion masses below $300\MeV$ and very small, non-zero values of the momentum transfer
down to $Q^2\approx0.01\GeV^2$. 
An overview of corresponding results for 
the mean square pion charge radius as a function of $m_\pi^2$ is displayed in Fig.~\ref{overview_r2Pi_v1}.
While an overall good consistency can be observed for most of the data points
in the range of $m_\pi^2\approx0.1,\ldots,0.4\GeV^2$, we note, however, 
that the data points from the $n_f=2$ clover improved Wilson fermion study by
QCDSF/UKQCD lie significantly above the rest. This discrepancy may be at least partially due to the
different ways $\langle r_\pi^2\rangle$ has been extracted from the $Q^2$-dependence
of the form factor data:
QCDSF/UKQCD fitted a monopole ansatz with the monopole mass as free parameter to lattice form factor data
in a wide range
of $Q^2=0,\ldots,\approx4\GeV^2$.
The RBC-UKQCD data point was obtained using 1-loop
ChPT to describe simultaneously the $m_\pi$- and $Q^2$-dependence of $F_\pi(Q^2)$ (calculated with pTBCs).
JLQCD employed an ansatz based on the $\rho$-meson pole with polynomial corrections, while ETMC 
also used a monopole ansatz to fit the form factor data in a smaller range
of $Q^2=0,\ldots,\approx1.25\GeV^2$ (employing pTBCs).
In addition, finite volume effects (although they have been included
in several cases), discretization effects, and the setting of the scale
may help to explain the observed differences.

The lattice results for the pion charge radius at the physical pion mass presented
in Fig.~\ref{overview_r2Pi_mPiPhys_v1} mostly agree within errors, 
which is remarkable by itself since the chiral extrapolations differ
significantly in the details. Most of the central values, as well as the lattice average,
are $\approx10\%$ below the experimental result.
It will be an interesting task to work out the precise origin of this discrepancy 
in future studies of the pion form factor.
%

%
%

%
%
Moving on to the nucleon form factors, we note that although there is
some scatter visible in the case of the Dirac mean square radius in Fig.~\ref{overview_r1_v2},
a general trend towards the experimental value at small pion masses can still be observed. 
Chiral perturbation theory indeed predicts a steep rise of $\langle r_1^2\rangle$ towards
the chiral limit as indicated by the dashed and dotted-dashed curves, but it seems that for a
fully quantitative extrapolation to the physical pion mass,
more precise lattice data is needed at pion masses below $300\MeV$.
The statistical uncertainty of the lattice results for $\langle r_2^2\rangle$ and
the anomalous magnetic moment in Figs.~\ref{overview_r2_v2} and \ref{overview_kappa_v2}, respectively, 
is $\mathcal{O}(10\%)$ at the lowest pion masses,
which must be significantly reduced before a quantitatively meaningful chiral extrapolation
and comparison with the very precise experimental values can be performed.
Partially twisted boundary conditions could help in these cases as they promise access to very small, non-zero
values of the momentum transfer, and thereby significantly reduce the errors
induced by the extrapolations based on, e.g., dipole fits to $Q^2=0$.
For the presently accessible pion masses and statistics, a clear, systematic difference
between the $n_f=2$ and $n_f=2+1$ cannot be observed for any of these observables.

Discrepancies between lattice data points that are observed even when statistical
uncertainties are taken into account indicate the presence of systematic effects. 
Interestingly, in several cases finite volume and also discretization effects 
have been checked at least semi-quantitatively and were found to be smaller than the statistical errors.
It may, however, be deceptive to assume that systematic errors are completely negligible in such cases.
It would be very helpful if at least rough quantitative estimates of systematic uncertainties,
including errors due to the setting of the lattice scale in physical units, would be routinely provided.
From the overview plots in
Figs.~\ref{overview_r1_v2}, \ref{overview_r2_v2} and \ref{overview_kappa_v2},
and the comparisons with the experimental data points it is in any case
clear that a strong pion mass dependence is expected in the region below $\approx300\MeV$.
Obviously, while there is at least qualitative agreement with results from ChPT,
naive linear extrapolations in $m^2_\pi$ from the currently accessible pion masses to the 
physical pion mass fail in general and should be avoided.

To this date, most lattice results for nucleon form factors are given in the isovector channel, where
contributions from quark line disconnected diagrams drop out because of isospin symmetry.
Estimates of disconnected contributions, even if only in the form of upper limits,
would be very useful and allow to provide first quantitative results
for the equally important isosinglet sector and the contributions from individual flavors.

\begin{figure}[t]
      \centering
  \includegraphics[angle=0,width=0.8\textwidth,clip=true,angle=0]{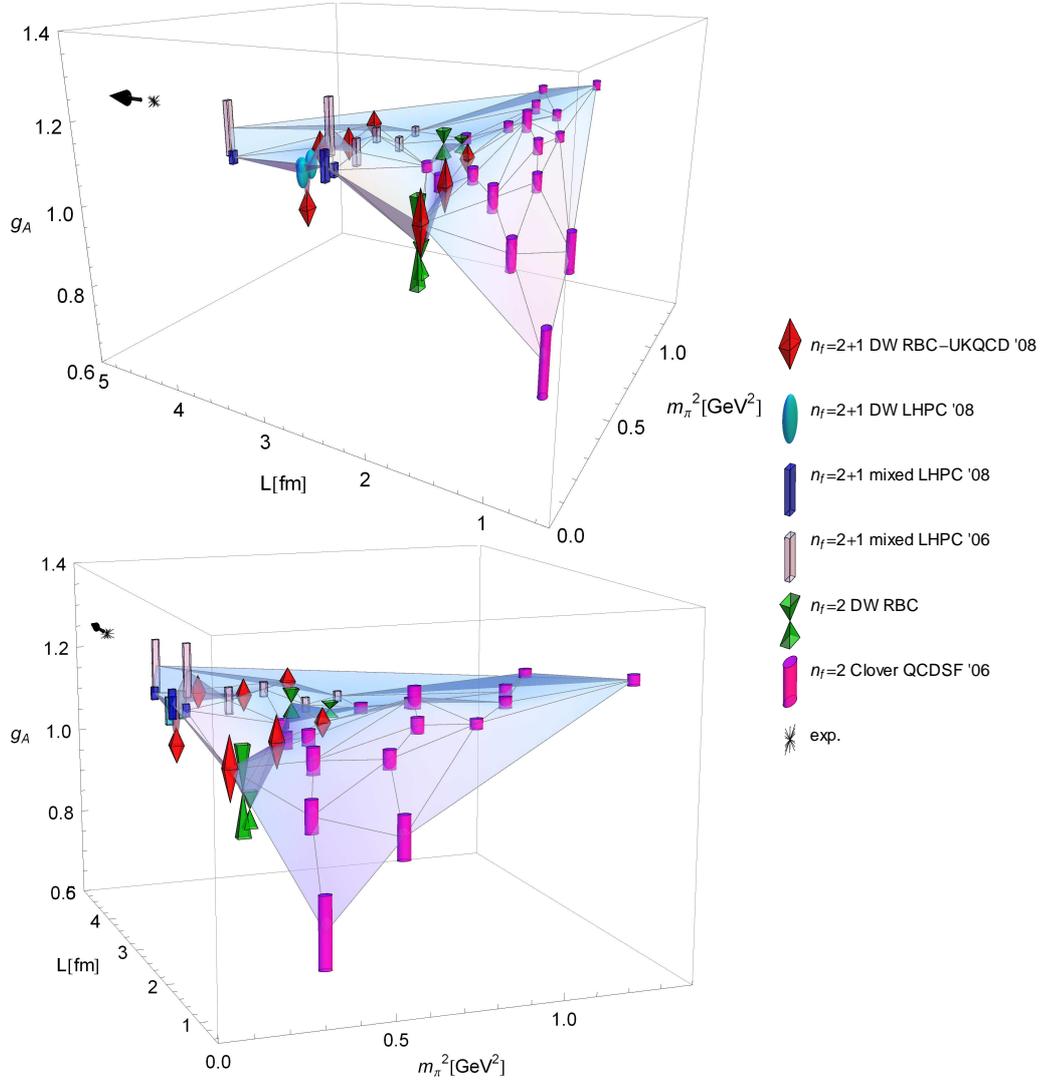}
  \caption{The ``$g_A$-landscape'', giving an overview of dynamical 
  lattice QCD results for the nucleon isovector axial vector coupling constant.}
  \label{overview_gA}
 \end{figure}

An overview of lattice results for the nucleon axial vector coupling constant
as a function of $m_\pi$ and the spatial box size $L$
is given in Fig.~\ref{overview_gA}. The required renormalization of the axial vector 
current has been carried out in all cases non-perturbatively, where simulations with domain wall
fermions have the clear advantage that $g_A/g_V$ is automatically renormalized
due to the lattice chiral symmetry.
Although the $g_A$-landscape in $m_\pi^2$ and box length $L$ is not completely
smooth, in particular due to the low lying data points from RBC-UKQCD at
low pion masses, it clearly shows that $g_A$ tends towards lower values
as the box length decreases.
In this case, finite volume effects are no longer necessarily seen as 
systematic uncertainty: The volume dependence has been
included and studied in the framework of ChPT \cite{Beane:2004rf,Khan:2006de}, and
as no new low energy constants have to be introduced at finite volume,
the $L$-dependence of $g_A$ provides an additional handle
in the matching of ChPT predictions and lattice results. 
A fit to the simultaneous $m_\pi$- and $L$-dependence of the
lattice data for $g_A$ based on ChPT may therefore 
help to better constrain the universal low energy constants and
provide a more accurate and precise chiral and infinite volume extrapolation.
This has already been attempted quite successfully in \cite{Khan:2006de}, however
the application of results from ChPT to the rather large pion masses of $m_\pi\gtrapprox600\MeV$
accessible in this work should be regarded with some caution.

%
%
%
In addition to the Dirac and Pauli form factors and the axial vector coupling of the nucleon,
a broad range of interesting observables like tensor form factors of the pion and the nucleon, and
electromagnetic form factors of the $\rho$-meson and the $\Delta$-baryon have by now been studied
in dynamical lattice QCD.
Many of these quantities are very difficult to access in experiment, and
lattice calculations therefore often provide at least qualitatively important insights into the charge distributions,
magnetization densities, possible deformations and the spin structure of hadrons.

\begin{figure}[t]
      \centering
  \includegraphics[angle=0,width=0.6\textwidth,clip=true,angle=0]{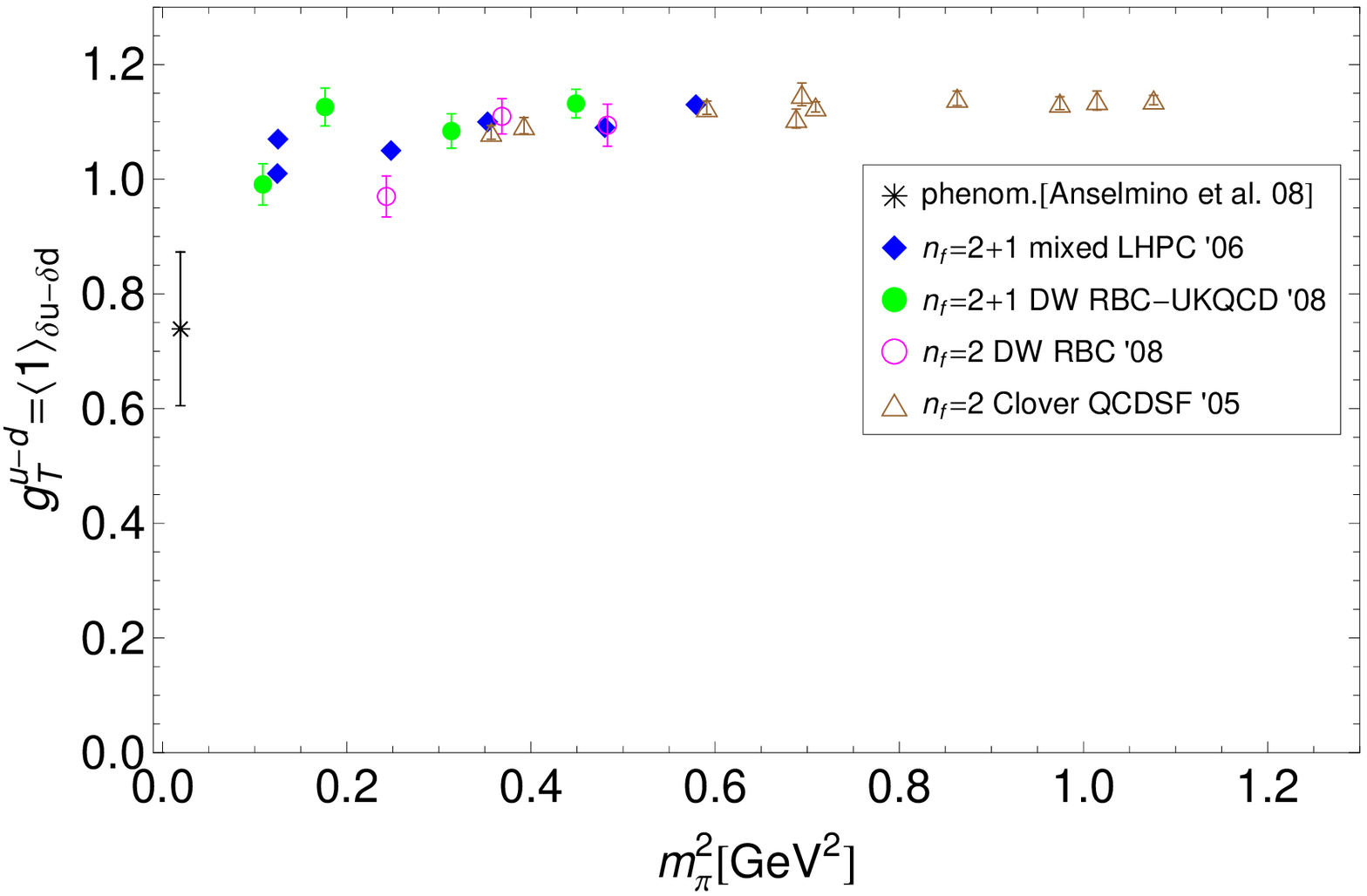}
  \caption{Overview of dynamical lattice QCD results for the nucleon isovector tensor charge $g^{u-d}_T=A^{u-d}_{T10}(\t0)$, 
  see Eq.~(\ref{NuclTensor3}).}
  \label{overview_gT}
 \end{figure}

An example of an observable that can be accessed in lattice QCD 
with a precision similar to $g_A$, but is difficult to determine directly in experiment,
is the tensor charge $g_T$ of the nucleon. An overview of non-perturbatively renormalized
lattice results for $g^{u-d}_T$ in the $\MSbar$ scheme at a scale of 
$4\GeV^2$ is given in Fig.~\ref{overview_gT}. 
The lattice data points, which show a remarkable statistical
precision, may be compared to results for the valence quark contribution to $g^{u-d}_T$
obtained from a phenomenological description of SIDIS data \cite{Anselmino:2008sj} as
indicated by the star.

%

%
\begin{figure}[t]
      \centering
  \includegraphics[angle=0,width=0.6\textwidth,clip=true,angle=0]{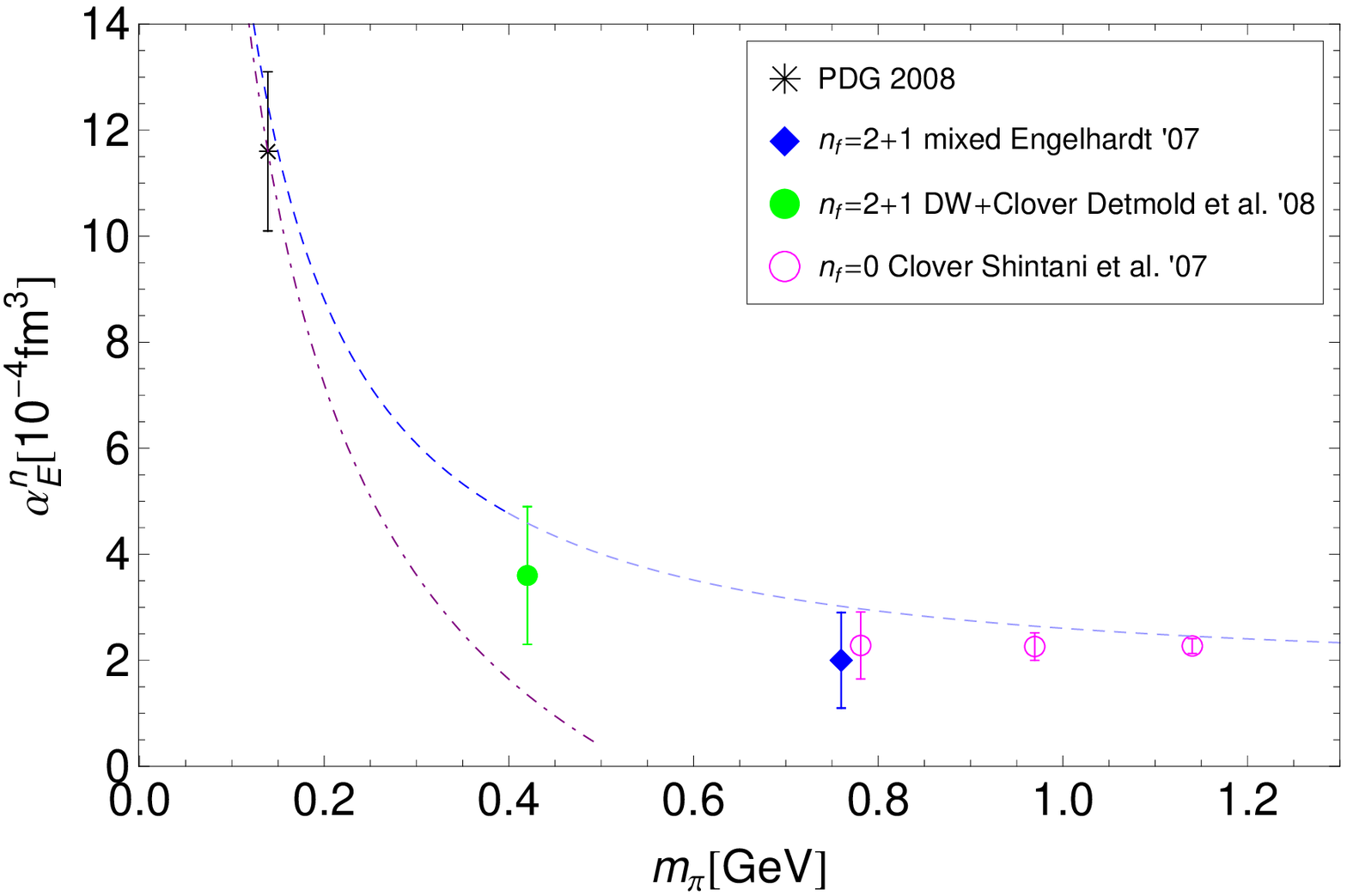}
  \caption{Overview of lattice QCD results for the electric polarizability of the neutron $\alpha^n_E$.
 Predictions from 1-loop HBChPT \cite{Bernard:1995dp} and SSE \cite{Hildebrandt:2003fm}
  are represented by the dashed and dot-dashed curves, respectively, which have been continued to larger pion masses
  only to guide the eye.}
  \label{overview_alphaN}
 \end{figure}

%

In particular with the help of improved background field methods, very interesting studies
have been performed of the magnetic moment of the $\Delta$-baryon and nucleon polarizabilities. 

Concerning the electric polarizability of the neutron, there is mounting evidence from different 
lattice studies that $\alpha^n_E\approx2\ldots3\cdot10^{-4}\fm^3$ for pion masses $m_\pi\gtrapprox400\MeV$,
as shown in the overview plot Fig.~\ref{overview_alphaN}.
In view of the different techniques that were used for the extraction of the polarizabilities 
(though all were based on the background field method),
it is gratifying to see that the results not only agree in sign but also in magnitude within errors.
As we have noted before, comparatively low values at unphysically large pion masses were to
be expected from ChPT, predicting that $\alpha^n_E$ diverges as $1/m_\pi\rightarrow+\infty$ in the chiral limit.
This is adequately illustrated by the dashed curve in Fig.~\ref{overview_alphaN}, representing a 1-loop HBChPT calculation \cite{Bernard:1995dp}. 
It would be very interesting to improve the statistics of the lattice calculations, extend them to lower pion masses 
and investigate if a quantitatively reliable chiral extrapolation to the physical pion can be performed.

%% file: figures/FFs/FFRhoG1_Cyprus.tex
\begingroup%
  \makeatletter%
  \newcommand{\GNUPLOTspecial}{%
    \@sanitize\catcode`\%=14\relax\special}%
  \setlength{\unitlength}{0.1bp}%
\begin{picture}(2339,1836)(0,0)%
{\GNUPLOTspecial{"
/gnudict 256 dict def
gnudict begin
/Color true def
/Solid true def
/gnulinewidth 5.000 def
/userlinewidth gnulinewidth def
/vshift -33 def
/dl {10.0 mul} def
/hpt_ 31.5 def
/vpt_ 31.5 def
/hpt hpt_ def
/vpt vpt_ def
/Rounded false def
/M {moveto} bind def
/L {lineto} bind def
/R {rmoveto} bind def
/V {rlineto} bind def
/N {newpath moveto} bind def
/C {setrgbcolor} bind def
/f {rlineto fill} bind def
/vpt2 vpt 2 mul def
/hpt2 hpt 2 mul def
/Lshow { currentpoint stroke M
  0 vshift R show } def
/Rshow { currentpoint stroke M
  dup stringwidth pop neg vshift R show } def
/Cshow { currentpoint stroke M
  dup stringwidth pop -2 div vshift R show } def
/UP { dup vpt_ mul /vpt exch def hpt_ mul /hpt exch def
  /hpt2 hpt 2 mul def /vpt2 vpt 2 mul def } def
/DL { Color {setrgbcolor Solid {pop []} if 0 setdash }
 {pop pop pop 0 setgray Solid {pop []} if 0 setdash} ifelse } def
/BL { stroke userlinewidth 2 mul setlinewidth
      Rounded { 1 setlinejoin 1 setlinecap } if } def
/AL { stroke userlinewidth 2 div setlinewidth
      Rounded { 1 setlinejoin 1 setlinecap } if } def
/UL { dup gnulinewidth mul /userlinewidth exch def
      dup 1 lt {pop 1} if 10 mul /udl exch def } def
/PL { stroke userlinewidth setlinewidth
      Rounded { 1 setlinejoin 1 setlinecap } if } def
/LTw { PL [] 1 setgray } def
/LTb { BL [] 0 0 0 DL } def
/LTa { AL [1 udl mul 2 udl mul] 0 setdash 0 0 0 setrgbcolor } def
/LT0 { PL [] 1 0 0 DL } def
/LT1 { PL [4 dl 2 dl] 0 1 0 DL } def
/LT2 { PL [2 dl 3 dl] 0 0 1 DL } def
/LT3 { PL [1 dl 1.5 dl] 1 0 1 DL } def
/LT4 { PL [5 dl 2 dl 1 dl 2 dl] 0 1 1 DL } def
/LT5 { PL [4 dl 3 dl 1 dl 3 dl] 1 1 0 DL } def
/LT6 { PL [2 dl 2 dl 2 dl 4 dl] 0 0 0 DL } def
/LT7 { PL [2 dl 2 dl 2 dl 2 dl 2 dl 4 dl] 1 0.3 0 DL } def
/LT8 { PL [2 dl 2 dl 2 dl 2 dl 2 dl 2 dl 2 dl 4 dl] 0.5 0.5 0.5 DL } def
/Pnt { stroke [] 0 setdash
   gsave 1 setlinecap M 0 0 V stroke grestore } def
/Dia { stroke [] 0 setdash 2 copy vpt add M
  hpt neg vpt neg V hpt vpt neg V
  hpt vpt V hpt neg vpt V closepath stroke
  Pnt } def
/Pls { stroke [] 0 setdash vpt sub M 0 vpt2 V
  currentpoint stroke M
  hpt neg vpt neg R hpt2 0 V stroke
  } def
/Box { stroke [] 0 setdash 2 copy exch hpt sub exch vpt add M
  0 vpt2 neg V hpt2 0 V 0 vpt2 V
  hpt2 neg 0 V closepath stroke
  Pnt } def
/Crs { stroke [] 0 setdash exch hpt sub exch vpt add M
  hpt2 vpt2 neg V currentpoint stroke M
  hpt2 neg 0 R hpt2 vpt2 V stroke } def
/TriU { stroke [] 0 setdash 2 copy vpt 1.12 mul add M
  hpt neg vpt -1.62 mul V
  hpt 2 mul 0 V
  hpt neg vpt 1.62 mul V closepath stroke
  Pnt  } def
/Star { 2 copy Pls Crs } def
/BoxF { stroke [] 0 setdash exch hpt sub exch vpt add M
  0 vpt2 neg V  hpt2 0 V  0 vpt2 V
  hpt2 neg 0 V  closepath fill } def
/TriUF { stroke [] 0 setdash vpt 1.12 mul add M
  hpt neg vpt -1.62 mul V
  hpt 2 mul 0 V
  hpt neg vpt 1.62 mul V closepath fill } def
/TriD { stroke [] 0 setdash 2 copy vpt 1.12 mul sub M
  hpt neg vpt 1.62 mul V
  hpt 2 mul 0 V
  hpt neg vpt -1.62 mul V closepath stroke
  Pnt  } def
/TriDF { stroke [] 0 setdash vpt 1.12 mul sub M
  hpt neg vpt 1.62 mul V
  hpt 2 mul 0 V
  hpt neg vpt -1.62 mul V closepath fill} def
/DiaF { stroke [] 0 setdash vpt add M
  hpt neg vpt neg V hpt vpt neg V
  hpt vpt V hpt neg vpt V closepath fill } def
/Pent { stroke [] 0 setdash 2 copy gsave
  translate 0 hpt M 4 {72 rotate 0 hpt L} repeat
  closepath stroke grestore Pnt } def
/PentF { stroke [] 0 setdash gsave
  translate 0 hpt M 4 {72 rotate 0 hpt L} repeat
  closepath fill grestore } def
/Circle { stroke [] 0 setdash 2 copy
  hpt 0 360 arc stroke Pnt } def
/CircleF { stroke [] 0 setdash hpt 0 360 arc fill } def
/C0 { BL [] 0 setdash 2 copy moveto vpt 90 450  arc } bind def
/C1 { BL [] 0 setdash 2 copy        moveto
       2 copy  vpt 0 90 arc closepath fill
               vpt 0 360 arc closepath } bind def
/C2 { BL [] 0 setdash 2 copy moveto
       2 copy  vpt 90 180 arc closepath fill
               vpt 0 360 arc closepath } bind def
/C3 { BL [] 0 setdash 2 copy moveto
       2 copy  vpt 0 180 arc closepath fill
               vpt 0 360 arc closepath } bind def
/C4 { BL [] 0 setdash 2 copy moveto
       2 copy  vpt 180 270 arc closepath fill
               vpt 0 360 arc closepath } bind def
/C5 { BL [] 0 setdash 2 copy moveto
       2 copy  vpt 0 90 arc
       2 copy moveto
       2 copy  vpt 180 270 arc closepath fill
               vpt 0 360 arc } bind def
/C6 { BL [] 0 setdash 2 copy moveto
      2 copy  vpt 90 270 arc closepath fill
              vpt 0 360 arc closepath } bind def
/C7 { BL [] 0 setdash 2 copy moveto
      2 copy  vpt 0 270 arc closepath fill
              vpt 0 360 arc closepath } bind def
/C8 { BL [] 0 setdash 2 copy moveto
      2 copy vpt 270 360 arc closepath fill
              vpt 0 360 arc closepath } bind def
/C9 { BL [] 0 setdash 2 copy moveto
      2 copy  vpt 270 450 arc closepath fill
              vpt 0 360 arc closepath } bind def
/C10 { BL [] 0 setdash 2 copy 2 copy moveto vpt 270 360 arc closepath fill
       2 copy moveto
       2 copy vpt 90 180 arc closepath fill
               vpt 0 360 arc closepath } bind def
/C11 { BL [] 0 setdash 2 copy moveto
       2 copy  vpt 0 180 arc closepath fill
       2 copy moveto
       2 copy  vpt 270 360 arc closepath fill
               vpt 0 360 arc closepath } bind def
/C12 { BL [] 0 setdash 2 copy moveto
       2 copy  vpt 180 360 arc closepath fill
               vpt 0 360 arc closepath } bind def
/C13 { BL [] 0 setdash  2 copy moveto
       2 copy  vpt 0 90 arc closepath fill
       2 copy moveto
       2 copy  vpt 180 360 arc closepath fill
               vpt 0 360 arc closepath } bind def
/C14 { BL [] 0 setdash 2 copy moveto
       2 copy  vpt 90 360 arc closepath fill
               vpt 0 360 arc } bind def
/C15 { BL [] 0 setdash 2 copy vpt 0 360 arc closepath fill
               vpt 0 360 arc closepath } bind def
/Rec   { newpath 4 2 roll moveto 1 index 0 rlineto 0 exch rlineto
       neg 0 rlineto closepath } bind def
/Square { dup Rec } bind def
/Bsquare { vpt sub exch vpt sub exch vpt2 Square } bind def
/S0 { BL [] 0 setdash 2 copy moveto 0 vpt rlineto BL Bsquare } bind def
/S1 { BL [] 0 setdash 2 copy vpt Square fill Bsquare } bind def
/S2 { BL [] 0 setdash 2 copy exch vpt sub exch vpt Square fill Bsquare } bind def
/S3 { BL [] 0 setdash 2 copy exch vpt sub exch vpt2 vpt Rec fill Bsquare } bind def
/S4 { BL [] 0 setdash 2 copy exch vpt sub exch vpt sub vpt Square fill Bsquare } bind def
/S5 { BL [] 0 setdash 2 copy 2 copy vpt Square fill
       exch vpt sub exch vpt sub vpt Square fill Bsquare } bind def
/S6 { BL [] 0 setdash 2 copy exch vpt sub exch vpt sub vpt vpt2 Rec fill Bsquare } bind def
/S7 { BL [] 0 setdash 2 copy exch vpt sub exch vpt sub vpt vpt2 Rec fill
       2 copy vpt Square fill
       Bsquare } bind def
/S8 { BL [] 0 setdash 2 copy vpt sub vpt Square fill Bsquare } bind def
/S9 { BL [] 0 setdash 2 copy vpt sub vpt vpt2 Rec fill Bsquare } bind def
/S10 { BL [] 0 setdash 2 copy vpt sub vpt Square fill 2 copy exch vpt sub exch vpt Square fill
       Bsquare } bind def
/S11 { BL [] 0 setdash 2 copy vpt sub vpt Square fill 2 copy exch vpt sub exch vpt2 vpt Rec fill
       Bsquare } bind def
/S12 { BL [] 0 setdash 2 copy exch vpt sub exch vpt sub vpt2 vpt Rec fill Bsquare } bind def
/S13 { BL [] 0 setdash 2 copy exch vpt sub exch vpt sub vpt2 vpt Rec fill
       2 copy vpt Square fill Bsquare } bind def
/S14 { BL [] 0 setdash 2 copy exch vpt sub exch vpt sub vpt2 vpt Rec fill
       2 copy exch vpt sub exch vpt Square fill Bsquare } bind def
/S15 { BL [] 0 setdash 2 copy Bsquare fill Bsquare } bind def
/D0 { gsave translate 45 rotate 0 0 S0 stroke grestore } bind def
/D1 { gsave translate 45 rotate 0 0 S1 stroke grestore } bind def
/D2 { gsave translate 45 rotate 0 0 S2 stroke grestore } bind def
/D3 { gsave translate 45 rotate 0 0 S3 stroke grestore } bind def
/D4 { gsave translate 45 rotate 0 0 S4 stroke grestore } bind def
/D5 { gsave translate 45 rotate 0 0 S5 stroke grestore } bind def
/D6 { gsave translate 45 rotate 0 0 S6 stroke grestore } bind def
/D7 { gsave translate 45 rotate 0 0 S7 stroke grestore } bind def
/D8 { gsave translate 45 rotate 0 0 S8 stroke grestore } bind def
/D9 { gsave translate 45 rotate 0 0 S9 stroke grestore } bind def
/D10 { gsave translate 45 rotate 0 0 S10 stroke grestore } bind def
/D11 { gsave translate 45 rotate 0 0 S11 stroke grestore } bind def
/D12 { gsave translate 45 rotate 0 0 S12 stroke grestore } bind def
/D13 { gsave translate 45 rotate 0 0 S13 stroke grestore } bind def
/D14 { gsave translate 45 rotate 0 0 S14 stroke grestore } bind def
/D15 { gsave translate 45 rotate 0 0 S15 stroke grestore } bind def
/DiaE { stroke [] 0 setdash vpt add M
  hpt neg vpt neg V hpt vpt neg V
  hpt vpt V hpt neg vpt V closepath stroke } def
/BoxE { stroke [] 0 setdash exch hpt sub exch vpt add M
  0 vpt2 neg V hpt2 0 V 0 vpt2 V
  hpt2 neg 0 V closepath stroke } def
/TriUE { stroke [] 0 setdash vpt 1.12 mul add M
  hpt neg vpt -1.62 mul V
  hpt 2 mul 0 V
  hpt neg vpt 1.62 mul V closepath stroke } def
/TriDE { stroke [] 0 setdash vpt 1.12 mul sub M
  hpt neg vpt 1.62 mul V
  hpt 2 mul 0 V
  hpt neg vpt -1.62 mul V closepath stroke } def
/PentE { stroke [] 0 setdash gsave
  translate 0 hpt M 4 {72 rotate 0 hpt L} repeat
  closepath stroke grestore } def
/CircE { stroke [] 0 setdash 
  hpt 0 360 arc stroke } def
/Opaque { gsave closepath 1 setgray fill grestore 0 setgray closepath } def
/DiaW { stroke [] 0 setdash vpt add M
  hpt neg vpt neg V hpt vpt neg V
  hpt vpt V hpt neg vpt V Opaque stroke } def
/BoxW { stroke [] 0 setdash exch hpt sub exch vpt add M
  0 vpt2 neg V hpt2 0 V 0 vpt2 V
  hpt2 neg 0 V Opaque stroke } def
/TriUW { stroke [] 0 setdash vpt 1.12 mul add M
  hpt neg vpt -1.62 mul V
  hpt 2 mul 0 V
  hpt neg vpt 1.62 mul V Opaque stroke } def
/TriDW { stroke [] 0 setdash vpt 1.12 mul sub M
  hpt neg vpt 1.62 mul V
  hpt 2 mul 0 V
  hpt neg vpt -1.62 mul V Opaque stroke } def
/PentW { stroke [] 0 setdash gsave
  translate 0 hpt M 4 {72 rotate 0 hpt L} repeat
  Opaque stroke grestore } def
/CircW { stroke [] 0 setdash 
  hpt 0 360 arc Opaque stroke } def
/BoxFill { gsave Rec 1 setgray fill grestore } def
/BoxColFill {
  gsave Rec
  /Fillden exch def
  currentrgbcolor
  /ColB exch def /ColG exch def /ColR exch def
  /ColR ColR Fillden mul Fillden sub 1 add def
  /ColG ColG Fillden mul Fillden sub 1 add def
  /ColB ColB Fillden mul Fillden sub 1 add def
  ColR ColG ColB setrgbcolor
  fill grestore } def
%
%
/PatternFill { gsave /PFa [ 9 2 roll ] def
    PFa 0 get PFa 2 get 2 div add PFa 1 get PFa 3 get 2 div add translate
    PFa 2 get -2 div PFa 3 get -2 div PFa 2 get PFa 3 get Rec
    gsave 1 setgray fill grestore clip
    currentlinewidth 0.5 mul setlinewidth
    /PFs PFa 2 get dup mul PFa 3 get dup mul add sqrt def
    0 0 M PFa 5 get rotate PFs -2 div dup translate
	0 1 PFs PFa 4 get div 1 add floor cvi
	{ PFa 4 get mul 0 M 0 PFs V } for
    0 PFa 6 get ne {
	0 1 PFs PFa 4 get div 1 add floor cvi
	{ PFa 4 get mul 0 2 1 roll M PFs 0 V } for
    } if
    stroke grestore } def
/Symbol-Oblique /Symbol findfont [1 0 .167 1 0 0] makefont
dup length dict begin {1 index /FID eq {pop pop} {def} ifelse} forall
currentdict end definefont pop
end
gnudict begin
gsave
0 0 translate
0.100 0.100 scale
0 setgray
newpath
1.300 UL
LTb
300 415 M
28 0 V
1862 0 R
-28 0 V
1.300 UL
LTb
300 635 M
28 0 V
1862 0 R
-28 0 V
1.300 UL
LTb
300 855 M
28 0 V
1862 0 R
-28 0 V
1.300 UL
LTb
300 1075 M
28 0 V
1862 0 R
-28 0 V
1.300 UL
LTb
300 1296 M
28 0 V
1862 0 R
-28 0 V
1.300 UL
LTb
300 1516 M
28 0 V
1862 0 R
-28 0 V
1.300 UL
LTb
300 1736 M
28 0 V
1862 0 R
-28 0 V
1.300 UL
LTb
300 200 M
0 28 V
0 1508 R
0 -28 V
1.300 UL
LTb
458 200 M
0 14 V
0 1522 R
0 -14 V
615 200 M
0 28 V
0 1508 R
0 -28 V
1.300 UL
LTb
773 200 M
0 14 V
0 1522 R
0 -14 V
930 200 M
0 28 V
0 1508 R
0 -28 V
1.300 UL
LTb
1088 200 M
0 14 V
0 1522 R
0 -14 V
1245 200 M
0 28 V
0 1508 R
0 -28 V
1.300 UL
LTb
1403 200 M
0 14 V
0 1522 R
0 -14 V
1560 200 M
0 28 V
0 1508 R
0 -28 V
1.300 UL
LTb
1718 200 M
0 14 V
0 1522 R
0 -14 V
1875 200 M
0 28 V
0 1508 R
0 -28 V
1.300 UL
LTb
2033 200 M
0 14 V
0 1522 R
0 -14 V
2190 200 M
0 28 V
0 1508 R
0 -28 V
1.300 UL
LTb
1.300 UL
LTb
300 200 M
1890 0 V
0 1536 V
-1890 0 V
300 200 L
0.900 UP
LTb
LTb
0.900 UP
1.300 UL
LT0
LTb
LT0
1834 1623 M
256 0 V
-256 31 R
0 -62 V
256 62 R
0 -62 V
300 1494 M
0 43 V
-31 -43 R
62 0 V
-62 43 R
62 0 V
557 1095 M
0 28 V
-31 -28 R
62 0 V
-62 28 R
62 0 V
776 884 M
0 31 V
745 884 M
62 0 V
-62 31 R
62 0 V
1147 676 M
0 32 V
-31 -32 R
62 0 V
-62 32 R
62 0 V
1310 606 M
0 39 V
-31 -39 R
62 0 V
-62 39 R
62 0 V
1741 499 M
0 48 V
-31 -48 R
62 0 V
-62 48 R
62 0 V
98 -58 R
0 40 V
-31 -40 R
62 0 V
-62 40 R
62 0 V
92 -87 R
0 85 V
-31 -85 R
62 0 V
-62 85 R
62 0 V
300 1516 BoxF
557 1109 BoxF
776 899 BoxF
1147 692 BoxF
1310 626 BoxF
1741 523 BoxF
1870 509 BoxF
1993 485 BoxF
1962 1623 BoxF
0.900 UP
1.300 UL
LT1
LTb
LT1
1834 1523 M
256 0 V
-256 31 R
0 -62 V
256 62 R
0 -62 V
300 1485 M
0 61 V
-31 -61 R
62 0 V
-62 61 R
62 0 V
550 1035 M
0 41 V
-31 -41 R
62 0 V
-62 41 R
62 0 V
757 800 M
0 46 V
726 800 M
62 0 V
-62 46 R
62 0 V
1098 608 M
0 36 V
-31 -36 R
62 0 V
-62 36 R
62 0 V
116 -82 R
0 41 V
-31 -41 R
62 0 V
-62 41 R
62 0 V
1632 380 M
0 186 V
1601 380 M
62 0 V
-62 186 R
62 0 V
300 1516 Crs
550 1055 Crs
757 823 Crs
1098 626 Crs
1245 582 Crs
1632 473 Crs
1962 1523 Crs
0.900 UP
1.300 UL
LT2
LTb
LT2
1834 1423 M
256 0 V
-256 31 R
0 -62 V
256 62 R
0 -62 V
300 1474 M
0 84 V
-31 -84 R
62 0 V
-62 84 R
62 0 V
546 992 M
0 43 V
515 992 M
62 0 V
-62 43 R
62 0 V
744 730 M
0 56 V
713 730 M
62 0 V
-62 56 R
62 0 V
1067 599 M
0 62 V
-31 -62 R
62 0 V
-62 62 R
62 0 V
1206 523 M
0 137 V
1175 523 M
62 0 V
-62 137 R
62 0 V
1567 405 M
0 160 V
1536 405 M
62 0 V
-62 160 R
62 0 V
76 -130 R
0 76 V
-31 -76 R
62 0 V
-62 76 R
62 0 V
300 1516 TriUF
546 1014 TriUF
744 758 TriUF
1067 630 TriUF
1206 592 TriUF
1567 485 TriUF
1674 473 TriUF
1962 1423 TriUF
1.300 UL
LTb
300 200 M
1890 0 V
0 1536 V
-1890 0 V
300 200 L
0.900 UP
stroke
grestore
end
showpage
}}%
\put(1784,1423){\makebox(0,0)[r]{$m_\rho$ = 0.853(37)~GeV}}%
\put(1784,1523){\makebox(0,0)[r]{$m_\rho$ = 0.910(38)~GeV}}%
\put(1784,1623){\makebox(0,0)[r]{$m_\rho$ = 1.002(41)~GeV}}%
\put(1275,-10){\makebox(0,0){Q$^2$ (GeV$^2$)}}%
\put(500,1626){\makebox(0,0){G$_1($Q$^2)$}}%
\put(2190,100){\makebox(0,0){3.0}}%
\put(1875,100){\makebox(0,0){2.5}}%
\put(1560,100){\makebox(0,0){2.0}}%
\put(1245,100){\makebox(0,0){1.5}}%
\put(930,100){\makebox(0,0){1.0}}%
\put(615,100){\makebox(0,0){0.5}}%
\put(300,100){\makebox(0,0){0.0}}%
\put(250,1736){\makebox(0,0)[r]{1.2}}%
\put(250,1516){\makebox(0,0)[r]{1.0}}%
\put(250,1296){\makebox(0,0)[r]{0.8}}%
\put(250,1075){\makebox(0,0)[r]{0.6}}%
\put(250,855){\makebox(0,0)[r]{0.4}}%
\put(250,635){\makebox(0,0)[r]{0.2}}%
\put(250,415){\makebox(0,0)[r]{0.0}}%
\end{picture}%
\endgroup
 

%% file: PDFs.tex
\section[Lattice results on PDFs and GPDs]{Lattice results on parton distribution functions and generalized parton distributions}
\label{sec:PDFs}
\subsection{Overview of lattice results}
\label{sec:PDFsOverview}
\subsubsection{Pion}
First calculations of the lowest two moments of the pion quark distribution function, 
$\langle x\rangle^\pi$ and $\langle x^2\rangle^\pi$,
in quenched lattice QCD have been presented in the pioneering works by Martinelli and Sachrajda
in the late 1980's \cite{Martinelli:1987zd,Martinelli:1987bh}.
An extensive study in the quenched approximation, including the next higher moment $\langle x^3\rangle^\pi$,
was presented many years later in \cite{Best:1997qp}.
The ZeRo 
collaboration has studied the momentum fraction of 
quarks in the pion in some detail in quenched QCD, employing non-perturbative renormalization of the corresponding operators and 
including an investigation of finite size effects \cite{Guagnelli:2003hw,Guagnelli:2004ww,Guagnelli:2004ga}.
A detailed study of $\langle x\rangle^\pi$ using twisted mass fermions in the 
quenched approximation has been presented by the $\chi$LF collaboration 
\cite{Capitani:2005jp}.
Preliminary (unrenormalized) results for the momentum fraction of quarks in the pion
in unquenched lattice QCD, based on twisted mass fermions, were obtained by the
European Twisted Mass
collaboration \cite{Baron:2007ti}.
A detailed study of the lowest moments, $\langle x^{n-1}\rangle^\pi$ with $n=2,\ldots,4$, of the pion PDF in unquenched lattice
QCD has been presented by QCDSF 
\cite{Brommel:2006zz}.
First results for moments of the pion vector GPD, in particular for the generalized form 
factor (GFF) $A^\pi_{20}(t=q^2)$ in Eq.~(\ref{PionVectorGFFn2}), in unquenched lattice QCD 
were presented earlier by QCDSF in \cite{Brommel:2005ee}, and a
calculation of the lowest moments moments of the pion tensor GPD in the same simulation framework
was published recently \cite{Brommel:2007xd}.
The momentum fraction carried by gluons in the pion has been investigated for
the first time in \cite{Meyer:2007tm} in the quenched approximation.

\subsubsection{Nucleon}
The lowest two moments of the unpolarized nucleon PDF
were calculated already twenty years ago in the quenched approximation \cite{Martinelli:1988rr}.
Thereafter, moments of unpolarized and polarized nucleon PDFs have been investigated
in quenched lattice QCD in a number of works by G\"ockeler \emph{et al.}
\cite{Gockeler:1995wg,Gockeler:1997zr,Gockeler:2000ja}.
A first detailed study of moments of unpolarized, polarized and tensor (transversity) PDFs
in unquenched and quenched lattice QCD by LHPC/SESAM can be found in \cite{Dolgov:2002zm}.
Preliminary results for the unpolarized quark momentum fraction 
in quenched lattice QCD using overlap fermions were presented in \cite{Gurtler:2004ac},
and a quenched calculation of moments of unpolarized and polarized 
PDFs employing domain wall fermions is described in \cite{Orginos:2005uy}.
A careful investigation of the unpolarized momentum fraction, $\langle x\rangle$, of quarks in the nucleon
in the quenched approximation,
using an improved action and employing non-perturbative renormalization,
was presented by QCDSF \cite{Gockeler:2004wp}.
The quark momentum fraction has also been studied by LPHC in the framework of a lattice investigation of GPDs
with dynamical quarks in \cite{Hagler:2007xi}, where also numerical results for the 
higher moment, $\langle x^2\rangle$, and the moments of the polarized PDFs can be found.
Recently, the isovector polarized and unpolarized quark momentum fractions 
have been computed by RBC and UKQCD in unquenched lattice QCD using domain wall fermions 
\cite{Lin:2008uz,Ohta:2008kd}.
A number of more recent results on moments of quark PDFs obtained in dynamical lattice QCD appeared in proceedings, see, e.g.,
\cite{Edwards:2006qx,Renner:2007pb,Gockeler:2007rx,Brommel:2007sb}.
An attempt to extract the unpolarized PDF as a function of the momentum fraction $x$
from lattice results for the lowest $x$-moments has been presented in \cite{Detmold:2001dv}.

Angular momentum contributions to the nucleon spin have been first investigated
in quenched lattice QCD in \cite{Mathur:1999uf} and, based on a different method 
avoiding extrapolations in the momentum transfer, in \cite{Gadiyak:2001fe}, 
where in both cases also contributions from disconnected diagrams have been evaluated.
First studies of moments of unpolarized nucleon generalized parton distributions, i.e.
the generalized form factors $A_{20}(t=q^2)$, $B_{20}(t)$
and $C_{20}(t)$ in Eq.~(\ref{NuclVec5}), appeared in 2003 in the quenched approximation by QCDSF \cite{Gockeler:2003jf}
and in unquenched lattice QCD by LHPC/SESAM\cite{Hagler:2003jd}.
Higher moments of GPDs were investigated in unquenched lattice QCD in \cite{LHPC:2003is}.
More recently, an extensive study of the polarized and unpolarized moments of GPDs with dynamical quarks 
in a mixed action approach has been presented by LHPC \cite{Hagler:2007xi}.
Nucleon tensor (generalized) form factors $g_T(t)=A_{T10}(t)$ and $A_{T20}(t)$ in Eqs.~(\ref{NuclTensor3}) and (\ref{NuclTensor5}), 
including the $x$-moment of the transversity distribution $\langle x\rangle_\delta=A_{T20}(\t0)$, 
have been calculated in unquenched lattice QCD by QCDSF 
\cite{Gockeler:2005cj}.
The transverse spin structure of the nucleon, including in particular the tensor 
(generalized) form factors $\overline B_{T10}=B_{T10}+2\widetilde A_{T10}$ in Eq.~(\ref{NuclTensor3}), was studied in the 
same simulation framework by QCDSF in \cite{Gockeler:2006zu}.
As before, a number of results on moments of GPDs in unquenched lattice QCD can be found in proceedings, see, e.g.,
\cite{Gockeler:2004mn,Gockeler:2005cd,Gockeler:2005aw,Gockeler:2006ui,Brommel:2007sb,Schroers:2003mf,Negele:2004iu,Edwards:2006qx,Bratt:2008uf}.

Turning to the gluon structure of the nucleon, we note that
the first and only attempt to extract the gluon contribution to the nucleon
spin, $\Delta g$, in quenched lattice QCD has been made almost 20 years
ago in \cite{Mandula:1990ce} (see also the pertinent comments in \cite{Efremov:1990ga}
and a reply in \cite{Mandula:1991ex}).
The unpolarized gluon momentum fraction in the nucleon has been 
investigated so far only in \cite{Gockeler:1996zg} more than 10 years ago in 
the quenched approximation.
\subsubsection{$\rho$-meson}
The only lattice QCD calculation of moments of unpolarized and 
polarized PDFs of the $\rho$-meson in the quenched approximation 
was published more than ten years ago in \cite{Best:1997qp}.
So far, there are no lattice results available for moments of GPDs of the $\rho$-meson,
apart from the lowest $n=1$-moments corresponding to the $\rho$-meson electromagnetic FFs
discussed above in section \ref{sec:rhoFFs}.
\subsection{Results from chiral perturbation theory}
\label{sec:PDFsChPT}
Moments of unpolarized PDFs have been studied at leading 1-loop level in ChPT
for the pion and the nucleon (in HBChPT) by Arndt and Savage in \cite{Arndt:2001ye},
where also contributions from explicit $\Delta$-DOFs were included.
Leading non-analytic corrections for the complete set of twist-2 nucleon matrix elements, 
i.e. the moments of the unpolarized, polarized and transversity PDFs,
were independently obtained in HBChPT by Chen and Ji \cite{Chen:2001eg}.
The inclusion of $\Delta(1232)$-degrees of freedom ($\Delta$-DOFs) was also studied in \cite{Detmold:2002nf}
for moments of the isovector twist-2 PDFs of the nucleon.
A partially-quenched ChPT study of isovector twist-2 matrix elements of baryons
including $\Delta$-DOFs can be found in \cite{Chen:2001yi,Beane:2002vq}.
More recently, finite volume corrections to twist-2 nucleon matrix elements were investigated in some detail
in \cite{Detmold:2005pt} to leading 1-loop order in partially-quenched HBChPT.
Leading chiral corrections to the angular momentum contributions of quarks and gluons to the 
proton spin, $J_{q,g}$ have been calculated in \cite{Chen:2001pv}, in the framework of
HBChPT including $\Delta$ intermediate states.
Subsequently, the pion mass and momentum dependence of form factors of the energy momentum tensor, i.e.
the gravitational form factors, for the nucleon in the isovector and isosinglet channel has been worked out
to leading 1-loop order in HBChPT \cite{Belitsky:2002jp}.
The complete set of moments of the nucleon vector, axial-vector and tensor GPDs has been investigated
in \cite{Ando:2006sk} in HBChPT to leading 1-loop order.
Moments of the unpolarized and polarized (vector and axial-vector) nucleon GPDs in the 
isosinglet channel were also calculated in \cite{Diehl:2006ya} to leading 1-loop order in HBChPT.
Corresponding results in ChPT including the isovector case and moments of tensor GPDs 
of the pion and the nucleon were presented by the same authors in \cite{Diehl:2006js}.
A calculation of the isovector and isosinglet gravitational form factors of the nucleon,
including a study of the quark angular momentum contribution to the nucleon spin, 
was also recently performed in covariant BChPT to $\mathcal{O}(p^2)$ 
based on a modified infrared regularization scheme \cite{Dorati:2007bk}.
Leading non-analytic chiral corrections to the moments of the pion vector GPDs were already studied 
in \cite{Kivel:2002ia} and later revisited in \cite{Diehl:2005rn}.
More recently, finite volume corrections to pion matrix elements of vector and tensor operators
were calculated in \cite{Manashov:2007qr}.
\subsection{Moments of PDFs}
\label{sec:MomentumFractions}
In this section, we are mainly concerned with the lowest $x^{n-1}$-moments of the quark PDFs of the pion, nucleon and $\rho$-meson,
see Eqs.~(\ref{VecMoments1},\ref{AxVecMoments1},\ref{TensorMoments1}) and adjacent discussion, 
which can be directly obtained from corresponding forward hadron matrix elements of the local operators in Eq.~(\ref{localOps}).
\subsubsection{Pion}
\label{sec:pionPDFs}
%
\begin{figure}[t]
   \begin{minipage}{0.48\textwidth}
      \centering
          \includegraphics[angle=-90,width=0.99\textwidth,clip=true]{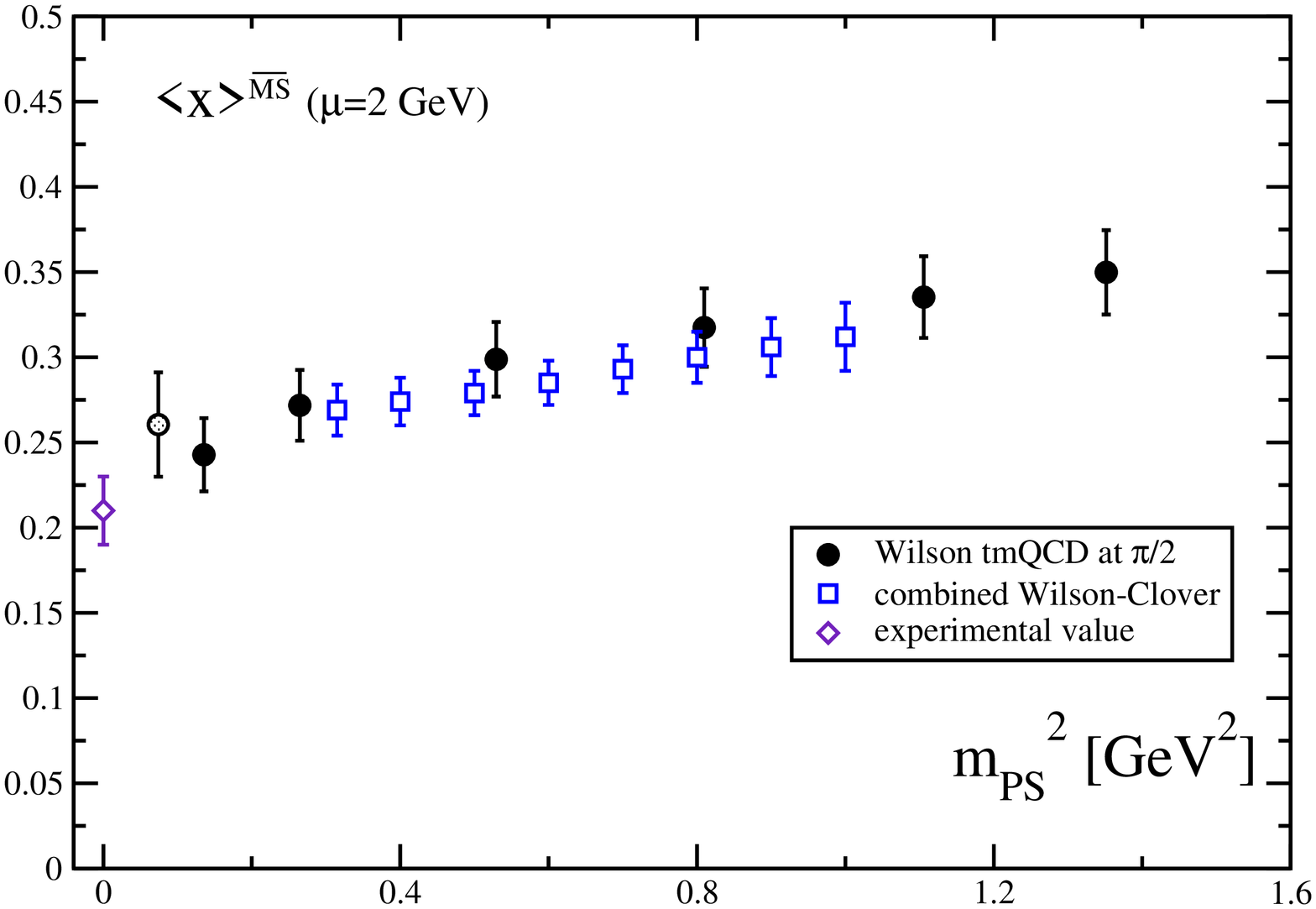}
  \caption{The momentum fraction of up-quarks in a $\pi^+$ (from \cite{Capitani:2005jp}).}
  \label{Pion_x_ChiLF}
     \end{minipage}
     \hspace{0.3cm}
    \begin{minipage}{0.48\textwidth}
      \centering
          \includegraphics[angle=0,width=0.9\textwidth,clip=true]{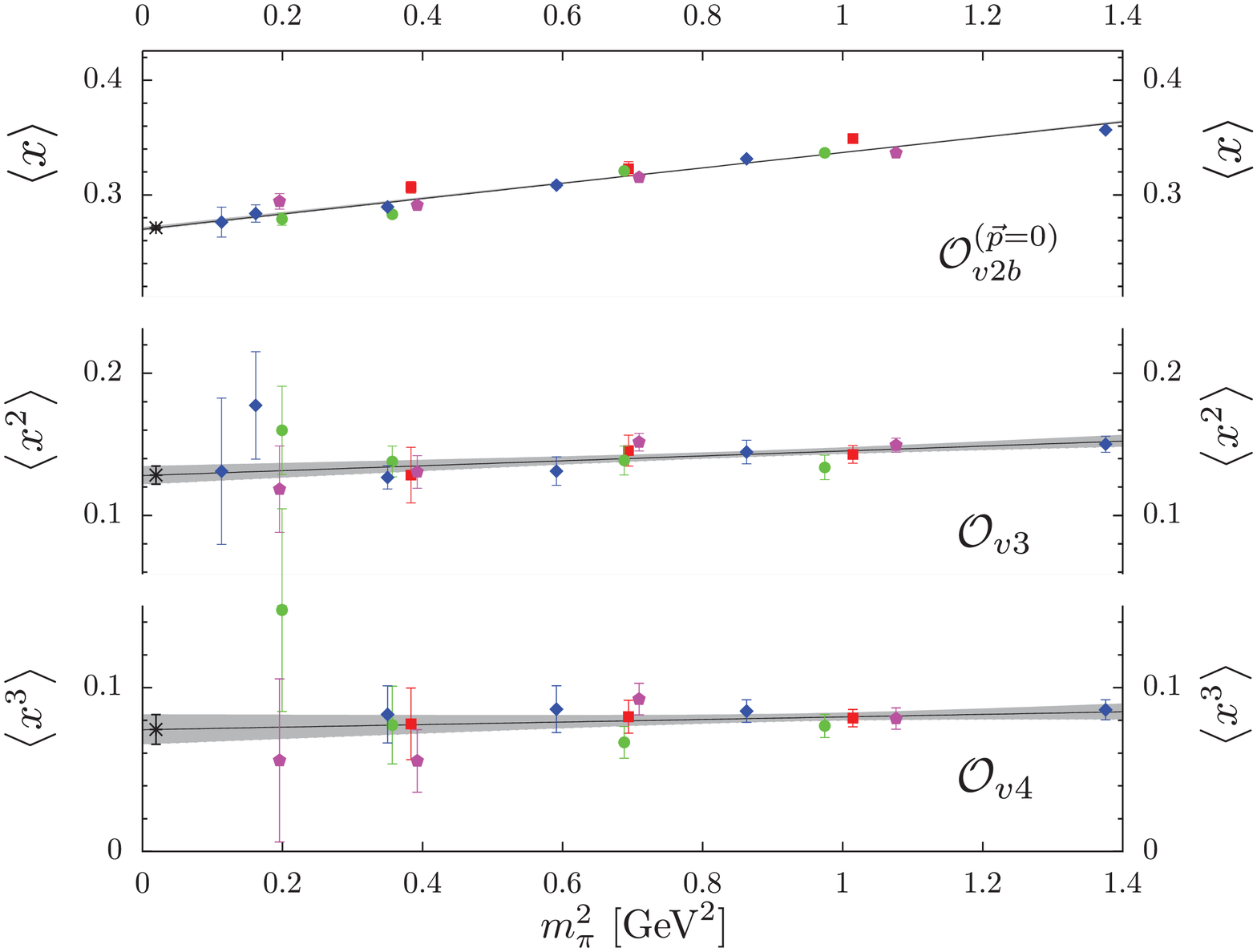}
  \caption{The momentum fraction of up-quarks in a $\pi^+$ (from proceedings \cite{Brommel:2006zz}).}
  \label{Pion_x-moms_QCDSF07}
     \end{minipage}
 \end{figure}
%
The $\chi_\text{L}\text{F}$ collaboration has calculated the momentum fraction of quarks
in the pion, $\langle x\rangle^\pi$, in the quenched approximation using
Wilson twisted mass fermions and the standard Wilson gauge action for a wide range
of pion masses from $\approx270\MeV$ to $\approx1160\MeV$, for up to five different lattice spacings
from $\approx0.123\fm$ down to $\approx0.048\fm$ and volumes in the range of $\approx(1.5,\ldots,2.0\fm)^3$ \cite{Capitani:2005jp}.
Employing the Schr\"odinger functional renormalization scheme, the underlying 
twist-2 operator
\bea
\label{op22}
\mathcal{O}^{n=2}_{v}=(\mathcal{O}_{44}-1/3(\mathcal{O}_{11}+\mathcal{O}_{22}+\mathcal{O}_{33}))\,,
\eea
a linear combination of the operators listed in Eq.~(\ref{op2}),
has been non-perturbatively renormalized, and the results have been transformed to the $\MSbar$-scheme
at a scale of $\mu=2\GeV$. Finite volume effects, which are non-negligible for some of the
accessed pion masses and lattice volumes, have been taken into account following the 
study in \cite{Guagnelli:2004ww}, where good agreement between a power-law, $\propto1/L^{p}$,
and an exponential, $\propto \exp(c L)$, ansatz 
for the volume dependence (with constants $p,c$) was found.
Results for different lattice spacings were utilized in a careful study
of the continuum limit, exhibiting sizeable $\mathcal{O}(a^2)$ discretization effects
at low pion masses in the case where the vanishing of the pion mass was used for the definition
of the critical quark mass $m_c$ (critical hopping parameter $\kappa_c$), and much smaller scaling violations 
for the PCAC definition of $m_c$. A linear extrapolation in $a^2$ to the continuum limit
was performed, and the corresponding results for $\langle x\rangle^u_\pi$ as a function of 
$m_\pi^2$, including finite volume corrections, 
are shown in Fig.~{\ref{Pion_x_ChiLF}} as filled circles. The lattice data point
at the lowest pion mass of $\approx270\MeV$, denoted by the
shaded circle, was not extrapolated to the infinite volume limit.
Apart from the lattice data point at $m_\pi\approx370\MeV$, the results 
appear, to a good approximation, to be linear in $m_\pi^2$.

Although a direct quantitative comparison with results from experiment and phenomenology
is not possible due to the nature of the quenched approximation,
we note that the central values of the lattice data points at the lowest 
accessible pion masses lie approximately 15\% above the value of
$\langle x\rangle^{u,v}_\pi=0.217(11)$ obtained for the valence quark sector 
(not including possible contributions from anti-quarks)
in a recent phenomenological analysis of Drell-Yan scattering data \cite{Wijesooriya:2005ir}.
It is reassuring that the lattice data points of the
twisted mass approach agree well within errors with the results of
a comparable lattice study by the ZeRo collaboration based on 
clover-improved Wilson fermions \cite{Guagnelli:2003hw,Guagnelli:2004ww,Guagnelli:2004ga}, 
denoted by the open squares in Fig.~{\ref{Pion_x_ChiLF}}.

We now turn to a discussion of results from QCDSF obtained in the framework
of lattice simulations with $n_f=2$ flavors of clover-improved Wilson quarks and the Wilson gauge action \cite{Brommel:2006zz}.
Calculations have been performed for pion masses ranging from 
$\approx350\MeV$ to $\approx1200\text{ MeV}$, lattice spacings of $a\approx0.07,\ldots,0.12\fm$, and
volumes of $\approx(1.3,\ldots,2.5\fm)^3$, including dedicated finite volume runs.
Local lattice operators with one, two and three covariant derivatives,
needed for the computation of the $x^{n-1}$-moments $\langle x\rangle_\pi$,
$\langle x^2\rangle_\pi$ and $\langle x^3\rangle_\pi$, 
have been non-perturbatively renormalized using the Rome-Southampton 
method, and the results were given in the $\MSbar$ scheme at a scale of $\mu=2\GeV$.
While the extraction of $\langle x\rangle^\pi$ based on the operator in Eq.~\ref{op22}
can be done for zero pion momentum, $\mbf{P}=0$, 
a non-zero $\mbf{P}$ is require to obtain non-vanishing
pion matrix matrix elements for the higher moments based on the operators 
\bea
\label{op4}
\mathcal{O}^{n=3}_{v}&=&\mathcal{O}_{144}-1/2(\mathcal{O}_{224}+\mathcal{O}_{334})\,,\\
\label{op5}
\mathcal{O}^{n=4}_{v}&=&\mathcal{O}_{1122}+\mathcal{O}_{1133}+\mathcal{O}_{2244}+\mathcal{O}_{3344}
 - 2\left(\mathcal{O}_{1144}+\mathcal{O}_{2233}\right)\,.
\eea
Consequently, the higher moments are subject to larger statistical noise.
Figure \ref{Pion_x-moms_QCDSF07} displays the corresponding  lattice
results for $\langle x^{n-1}\rangle^u_{\pi^+}$ with $n=2,\ldots,4$ versus $m_\pi^2$, 
together with linear chiral extrapolations represented by the shaded bands.
A study of the volume dependence showed that finite size 
effects may be of the order of a few percent already for
$m_\pi L\approx 5$ and up to 10\% for the lowest pion masses
included in Fig.~\ref{Pion_x-moms_QCDSF07}. For comparison, see also the results in \cite{Guagnelli:2004ww}. 
Finite volume effects (FVEs) are therefore significant with respect to the statistical errors for $\langle x\rangle_\pi$. 
In an attempt to correct for FVEs, an ansatz
\bea
\label{Pion_x_fit}
\langle x^{n-1}\rangle^u_{\pi^+}=c_{n,0}+c_{n,1} m_\pi^2+c_{n,2} m_\pi^2 e^{-m_\pi L}\,,
\eea
including a term with an exponential dependence on the lattice extent $L$, was
used to fit the lattice data points. A value of 
$\langle x^{}\rangle^u_{\pi^+}=0.271\pm0.002_{(\text{stat})}\pm0.010_{(\text{ren})}$ was obtained 
for the momentum fraction of up-quarks in the $\pi^+$ at the physical point and in the infinite
volume, where the second error represents uncertainties from the non-perturbative renormalization.
For a comparison of the $(n=2,\ldots,4)$-moments in Fig.~\ref{Pion_x-moms_QCDSF07}
with quenched lattice calculations and results 
from phenomenological studies of the pion PDFs
in \cite{Sutton:1991ay,Gluck:1999xe}, we refer to \cite{Brommel:2006zz}.
We note, however, that such a direct comparison 
has to be considered with caution, since a number of different 
approximations and assumptions have been employed in either of these studies.
In particular, contributions from disconnected diagrams, which are present for even $n$, 
have not been included in the lattice calculations.

Recently, Negele and Meyer performed a first calculation of the momentum fraction
carried by gluons in the pion, $\langle x\rangle^\pi_g$, using the Wilson action 
in the quenched approximation \cite{Meyer:2007tm}.
Noting that $\langle \pi|\bar T^g_{00}|\pi\rangle=N\langle x\rangle^\pi_g$,
where $N$ is a kinematic factor and $\bar T^g_{\mu\nu}$ is the
traceless part of the gluon energy momentum tensor, the numerical 
studies were based on plaquette and clover discretizations
of the operator $\bar T^g_{00}=(-\mbf{E}^2+\mbf{B}^2)/2$, where, e.g., $\mbf{E}^2=\sum_a\mbf{E}^a\cdot\mbf{E}^a$. 
HYP-smearing of the gauge fields was employed to
reduce the statistical noise of the correlators due to 
short-range fluctuations of the gauge fields, which represents a major challenge 
in calculations of gluonic contributions to hadron structure observables.
To preserve locality as much as possible, the plaquette operator 
was used together with HYP-smearing for the final calculations, 
providing a reduction of the variance by a factor of $\approx40$ 
compared to the unsmeared case.
The gluon operator, which mixes with the quark singlet operator $\bar T^q_{00}$
under renormalization, was (partially) non-perturbatively renormalized,
employing results from \cite{Guagnelli:2003hw,Guagnelli:2004ga} for 
the quark non-singlet renormalization constant and the bare quark momentum fraction.
Computations were performed for three different pion masses
from  $\approx620\MeV$ to $\approx1060\MeV$, giving a value of
$\langle x\rangle^\pi_g=0.37\pm0.08_{(\text{stat})}\pm0.12_{(\text{ren})}$
in the $\MSbar$ scheme at a scale of $\mu=2\GeV$, for a 
pion mass of $m_\pi\approx890\MeV$.
Since the value of one of the renormalization constants was not known beyond
its trivial tree-level value, a systematic error of $\pm0.12$ was added
to account for higher order contributions.
The result indicates that even for such large pion masses, the gluons
carry a substantial fraction of the total momentum of the parent hadron.
It would be highly interesting to repeat this study in full QCD
and to reduce in particular the systematic uncertainties due to the 
operator renormalization.
\subsubsection{Nucleon}
\label{sec:nuclPDFs}
In the following, we will be concentrating on the isovector case, 
for which only contributions from connected diagrams are relevant. 
The computationally much more demanding disconnected contributions
are, to this date, not routinely included in lattice hadron structure 
calculations. Still, a large number of interesting results have been
obtained over the years for the \emph{connected parts} of
isosinglet quark contributions to observables. 
Some of those will be discussed in particular
in relation with the spin structure of the nucleon below in section \ref{sec:SpinStructure}.

\subsubsection*{Unpolarized quark momentum fraction}

Already the pioneering work by Martinelli and Sachrajda in the late 1980's
indicated that the isovector momentum fraction 
of quarks in the nucleon, as obtained from lattice QCD simulations 
in the quenched approximation, is approximately $\approx50\%$ larger than the 
result from experiment and phenomenology, i.e. $\langle x\rangle_{u-d}^\text{lat}\approx0.25$ 
compared to $\langle x\rangle_{u-d}^\pheno\approx0.16$ (in the $\MSbar$ scheme at a scale of $\mu=2\GeV$). 
Recent global PDF-analyses of DIS, Drell-Yan, and 
other scattering data give, e.g., $\langle x\rangle_{u-d}^\text{CTEQ6.6}\simeq0.154$ \cite{Nadolsky:2008zw}
and $\langle x\rangle_{u-d}^\text{MRST06}\simeq0.158$ \cite{Martin:2007bv}, 
see also \cite{Martin:2002aw,Martin:2003sk,Alekhin:2006zm}.

Years later, the initial quenched lattice results were confirmed with higher precision 
and for pion masses down to $\mathcal{O}(600\MeV)$ in \cite{Gockeler:1995wg}.
Since then, a number of possible explanations and solutions for the
discrepancy have been suggested, as summarized by the following keywords:
\begin{itemize}
\item quenched approximation
\item renormalization of the underlying lattice operators
\item discretization effects
\item finite volume effects
\item phenomenological analysis of experimental data giving $\langle x\rangle^{u-d}_\text{exp}$
\item chiral extrapolation
\end{itemize}
A first extensive study based on simulations with $n_f=2$ dynamical Wilson fermions
by the LHPC-SESAM collaboration \cite{Dolgov:2002zm} confirmed again the earlier
findings in the quenched approximation at similar pion masses.
Several years later, many of the above items were looked at in great detail 
in the framework of a quenched lattice QCD study using clover-improved Wilson fermions \cite{Gockeler:2004wp}.
First, a reanalysis of moments of structure functions, which in contrast
to the PDFs are directly accessible in experiment, revealed 
good agreement with the results obtained from global PDF-analyses.
On the side of the lattice calculation,
operator improvement terms were studied and found to be numerically small, and
different perturbative and non-perturbative renormalization procedures
were investigated and compared, including the possibility of operator mixing.
Non-perturbatively renormalized operators were then used to obtain the final results   
for the momentum fraction, which were found to be practically independent of 
the pion mass in the accessible range of $m_\pi\approx550$ to $m_\pi\approx1250\MeV$. 
Separate linear extrapolations in $m^2_\pi$ to the chiral limit were performed
for three different values of the lattice spacing,
and these values were finally extrapolated 
linearly in $a^2$ to the continuum limit. All these efforts 
culminated in a value of $\langle x\rangle^{u-d}_\text{lat}=0.245(9)$
in the $\MSbar$ scheme at a scale of $2\GeV$ \cite{Gockeler:2004wp}, in perfect agreement
with the results obtained in the years before in quenched and unquenched lattice QCD, where
only perturbatively renormalized operators have been employed.
Although new discrepancies within different lattice approaches emerged
more recently, as will be discussed below, it is likely that a proper, non-linear,
chiral extrapolation of $\langle x\rangle^{u-d}$ to the physical
pion mass is the clue for bridging at least part of the
large gap of $\approx50\%$ between the lattice results and the experimentally observed value.
Some of the attempts to reach agreement at the physical point 
based on various forms of chiral extrapolations will be discussed in the following.
Results for the quark momentum fractions obtained in the framework of 
lattice studies of moments of GPDs will be discussed in some detail in 
section \ref{sec:EMT} further below.

%
\begin{figure}[t]
    \begin{minipage}{0.5\textwidth}
      \centering
          \includegraphics[angle=0,width=1.\textwidth,clip=true]{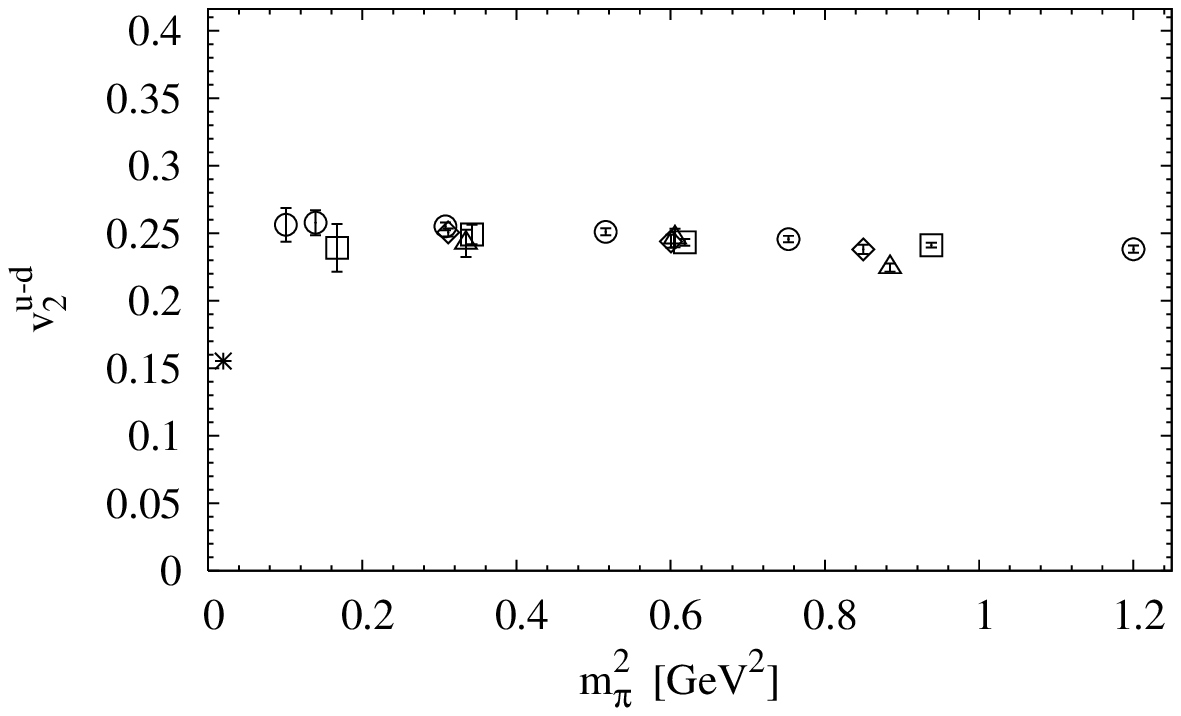}
      \vspace*{-0.cm}
  \caption{The momentum fraction of quarks in the isovector channel, $\text{v}_2^{u-d}=\langle x\rangle^{u-d}$ 
  (from proceedings \cite{Gockeler:2006ui}).}
  \label{x_v2b_QCDSF_2006}
     \end{minipage}
     \hspace{0.5cm}
   \begin{minipage}{0.48\textwidth}
      \centering
          \includegraphics[angle=0,width=0.9\textwidth,clip=true]{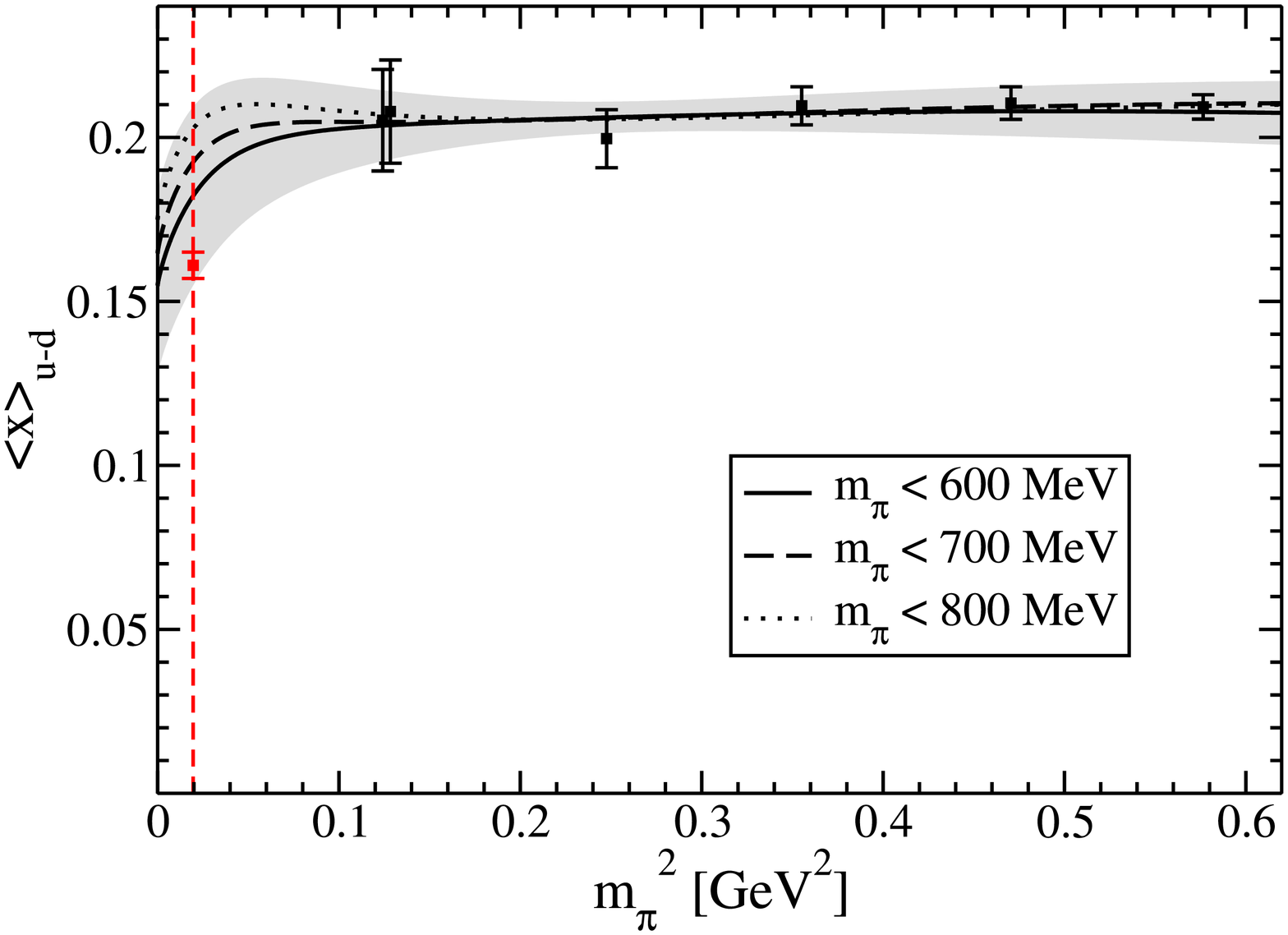}
  \caption{The isovector momentum fraction of quarks (from proceedings \cite{Renner:2007pb}).}
  \label{x_LHPC}
     \end{minipage}
 \end{figure}
%

The situation just described is reflected in the more recent results for $\langle x\rangle^{u-d}$
from QCDSF displayed in Fig.~\ref{x_v2b_QCDSF_2006} as a function of the pion mass \cite{Gockeler:2006ui}, 
for $n_f=2$ flavors of clover-improved Wilson quarks and the Wilson gauge action.
The lattice data points, which have been obtained
using the non-perturbatively renormalized lattice operator $\mathcal{O}_{v2b}=\mathcal{O}^{n=2}_{v}$
in Eq.~\ref{op22},
are not only practically flat in $m_\pi^2$ over the full range of 
$m_\pi\approx350,\ldots,1400\text{ MeV}$,
they also agree within statistics for all the different ensembles with
lattice spacings of $a\approx0.07,\ldots,0.12\fm$ and 
volumes of $\approx(1.3,\ldots,2.5\fm)^3$, with $m_\pi L\gtrsim3.5$.
In particular, there are no significant discretization effects visible, 
which was already observed for a similar set of ensembles in \cite{Gockeler:2004vx}
and more recently confirmed in \cite{Brommel:2008tc}.
Finite size effects have been analyzed in a similar lattice framework for a pion mass of $\approx770\MeV$ 
and were found to be small compared to the statistical errors \cite{Gockeler:2004vx},
although they have been predicted to become more severe at light quark masses \cite{Detmold:2005pt}.

Results for the isovector momentum fraction of quarks in the proton by LHPC, 
based on a hybrid (mixed action) approach of $n_f=2+1$ flavors of domain wall valence and 
staggered Asqtad sea quarks, with gauge configurations provided by the 
MILC collaboration \cite{Bernard:2001av,Aubin:2004wf}, 
are presented in Fig.~\ref{x_LHPC} \cite{Renner:2007pb}.
Calculations were performed for pion masses in the range of 
$m_\pi\approx350$ to $m_\pi\approx760\text{ MeV}$, a lattice spacing of $a\approx0.12\text{ fm}$ 
and a volume of $V\approx(2.5\text{ fm})^3$.
The analysis was repeated for the lowest pion mass in a larger volume of 
$V\approx(3.5\text{ fm})^3$ to check for possible finite size effects.
A perturbatively calculated renormalization constant, including a non-
perturbative correction factor as given in Eq.~\ref{ZLHPC}, was employed 
for the renormalization of the operator in Eq.~\ref{op22} \cite{Bistrovic:2005aa,Hagler:2007xi}.
HYP-smearing of the gauge fields in the valence quark action
proved to be beneficial by suppressing large contributions from loop integrals
in the perturbative renormalization, leading to perturbative renormalization
factors that differ from unity by only a few percent.
As can be seen from Fig.~\ref{x_LHPC}, the two lattice data points
at the lowest pion mass, obtained for the two different volumes
of  $\approx(2.5\text{ fm})^3$ and  $\approx(3.5\text{ fm})^3$, agree well, indicating
that potential finite volumes effects are smaller than the statistical uncertainties.
Similar to the results from QCDSF discussed above, the lattice data points
are remarkably constant in $m^2_\pi$ over the full range of accessible pion masses.
Most notable, however, is that the overall normalization of the lattice results in 
Fig.~\ref{x_LHPC} turned out to be approximately 20\% lower than
for the corresponding results in Fig.~\ref{x_v2b_QCDSF_2006}.
Unfortunately, the origin of this discrepancy could not be tracked down to this date.
Apart from the usual systematic uncertainties due to finite volume and discretization effects that
have already been addressed to some extent, but still may have been underestimated, 
further possible causes include the renormalization procedure (NP-improved
perturbatively calculated renormalization constants) and potential
discretization errors in the framework of the mixed action approach by LHPC,
the missing dynamical strange quark in the case of QCDSF ($n_f=2$ versus $n_f=2+1)$,
as well as potential contaminations from excited states of the plateaus 
in the ratio of three- to two-point functions on either side.
The latter was, however, studied already in some detail in the case of the
hybrid calculation by LHPC in \cite{Renner:2007pb}, where no indications
for uncontrolled contributions from excited states were observed.

In the same spirit as in the case of the tensor charge shown
in Fig.~\ref{gt_umd_LHPC} and discussed in the context of
Eq.~\ref{gtChPT}, the isovector momentum fraction
in Fig.~\ref{x_LHPC} has been chirally extrapolated employing
a self-consistent rearrangement of the leading 1-loop HBChPT formula 
\cite{Arndt:2001ye,Chen:2001eg}, given by \cite{Edwards:2006qx,Renner:2007pb}
\begin{eqnarray}
\label{xLHPCChPT}
\langle x\rangle^{u-d} & = &
\langle x\rangle^{u-d,0} \left( 1 - \frac{(3 g_{A,\lat}^2 + 1)}{(4\pi)^2} \frac{m_{\pi,\lat}^2}{f_{\pi,\lat}^2} \ln \left( \frac{m_{\pi,\lat}^2}{f_{\pi,\lat}^2} \right) \right) + c_0 \frac{m_{\pi,\lat}^2}{f_{\pi,\lat}^2}\,.
\end{eqnarray}
As in Eq.~\ref{gtChPT}, the LECs in the chiral limit, 
$g_A^0$ and $f_\pi^0$, have been replaced by the respective 
pion mass dependent values $g_{A,\lat}$ and $f_{\pi,\lat}$ 
obtained in the lattice calculation. The result of a fit
with free parameters $\langle x\rangle^{u-d,0}$ and $c_0$ to the lattice data points
is represented by the shaded band in Fig.~\ref{x_LHPC}. Owing to the
overall low normalization of the LHPC lattice data,
the chiral extrapolation band even marginally overlaps with the phenomenological value.
However, we note again that although chiral extrapolations based on 
self-consistently improved ChPT have shown some phenomenological success in the past,
it is difficult to judge if the underlying chiral dynamics
is properly included in this framework over such a wide range of pion masses.

%
\begin{figure}[t]
   \begin{minipage}{0.48\textwidth}
      \centering
          \includegraphics[angle=0,width=0.99\textwidth,clip=true]{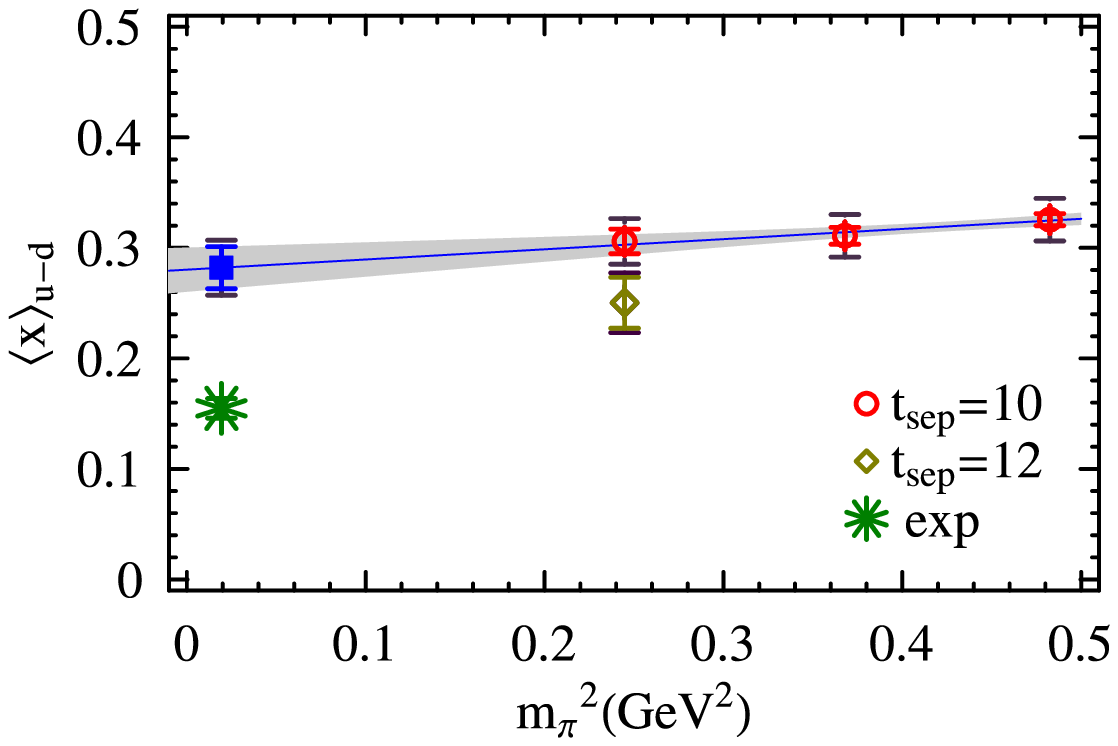}
          \includegraphics[angle=0,width=0.85\textwidth,clip=true]{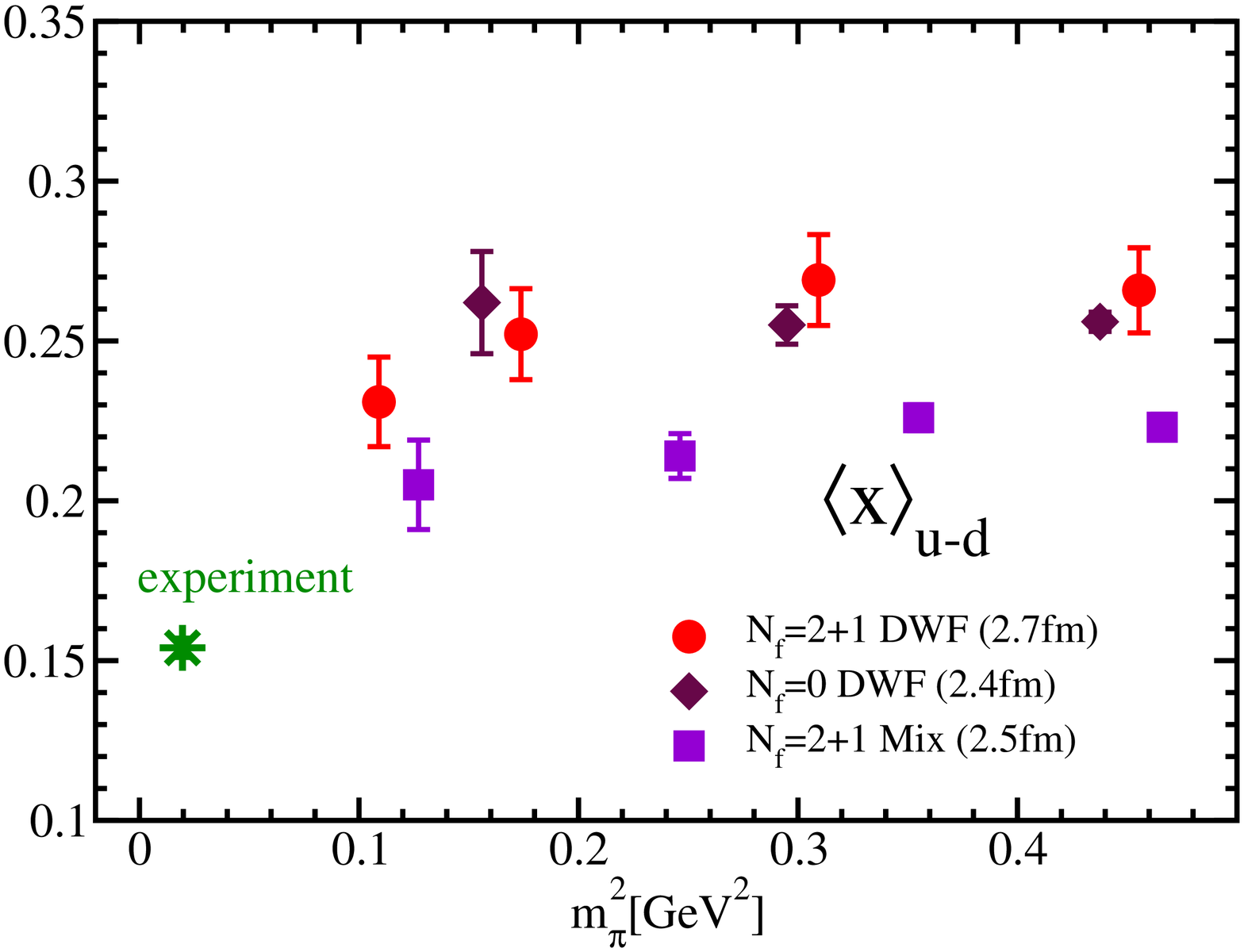}
  \caption{The momentum fraction of quarks in the isovector channel, from \cite{Lin:2008uz} (upper figure)
  and recent proceedings \cite{Ohta:2008kd} (lower figure).}
  \label{x_RBCUKQCD}
     \end{minipage}
     \hspace{0.5cm}
   \begin{minipage}{0.48\textwidth}
      \centering
          \includegraphics[angle=0,width=0.93\textwidth,clip=true]{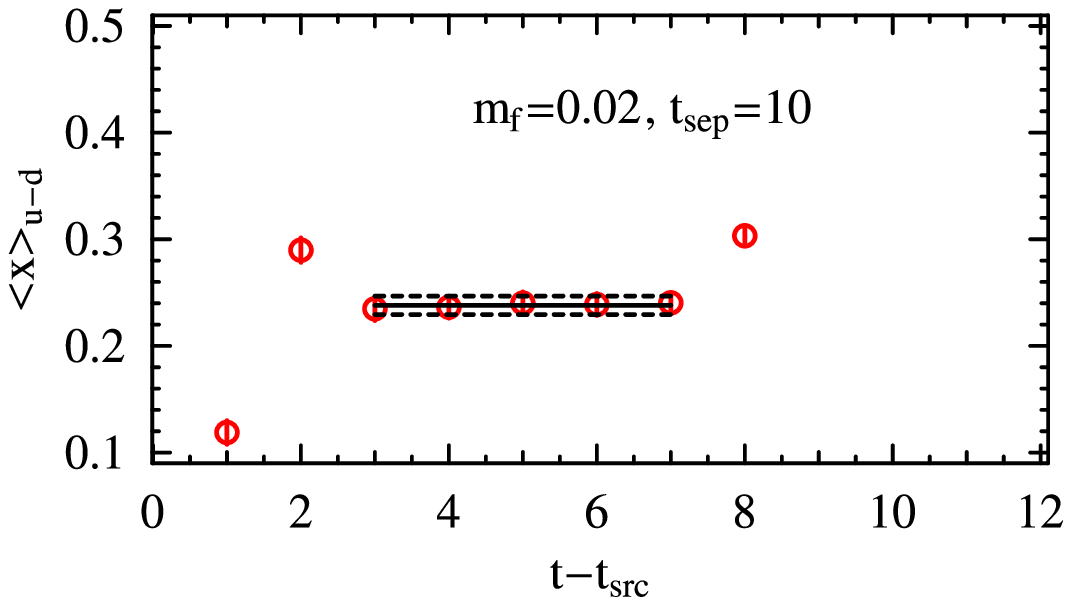}
          \includegraphics[angle=0,width=0.93\textwidth,clip=true]{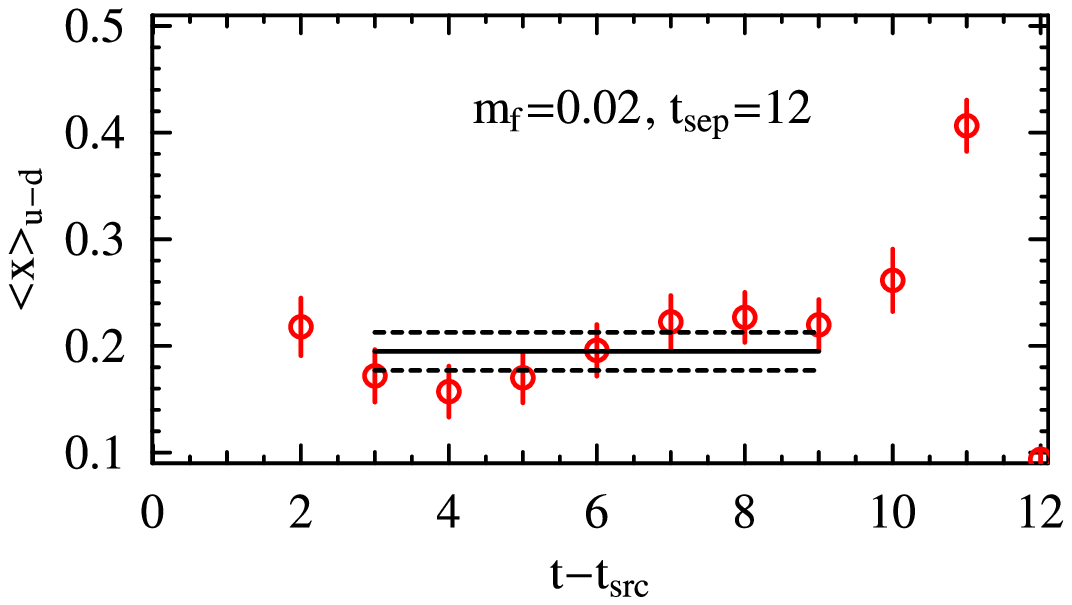}
  \caption{Plateau plots for the bare (unrenormalized) 
  momentum fraction of quarks in the isovector channel (from \cite{Lin:2008uz}).}
  \label{Plat10_RBCUKQCD}
     \end{minipage}
\end{figure}
%

The situation is becoming even more challenging as soon as recent results from
RBC, based on simulations with $n_f=2$ flavors of domain wall fermions and the
DBW2 gauge action, are taken into consideration \cite{Lin:2008uz}.
Results for the momentum fraction in the isovector channel that were obtained using a non-perturbatively
renormalized operator, for three different pion masses of $m_\pi\approx 493\MeV, 607\MeV$ and $\approx695\text{ MeV}$, 
a lattice spacing of $a\approx0.11$ fm and a volume of $V\approx(1.9\text{ fm})^3$, 
are displayed in the upper part of Fig.~\ref{x_RBCUKQCD}.
The lattice data points for a sink-source time separation of $(at)_\text{sep}=10$
(corresponding to a physical distance of $t_\text{sep}\sim1.1\fm$)
at $m_\pi\approx493,607\MeV$ correspond to values of $\langle x\rangle^{u-d}\approx0.31(3)_\text{stat+ren}$,
and are therefore $\approx50\%$ higher than the $n_f=2+1$ mixed action results
from LHPC (Fig.~\ref{x_LHPC}), and $\approx25\%$ higher than the $n_f=2$ clover-improved Wilson Fermion
results by QCDSF (Fig.~\ref{x_v2b_QCDSF_2006}) at similar pion masses. Contaminations from excited states
have been offered as possible explanation for this dramatic discrepancy \cite{Lin:2008uz}. 
At first sight, the bare momentum fraction 
in the upper part of Fig.~\ref{Plat10_RBCUKQCD}
exhibits a perfectly flat plateau between $(at)=3$ and $(at)=7$
for the sink-source separation time of $at_\text{sep}=10$, thereby giving 
no direct indication for contributions from excited states. 
Nevertheless, the calculation has been repeated for a larger $at_\text{sep}=12$,
shown in the lower part of Fig.~\ref{Plat10_RBCUKQCD}.
In this case, there is no clear plateau visible due to larger
statistical fluctuations. However, an average over the
range from $at=3$ to $at=9$ does indeed lead to a lower value for the bare
momentum fraction. Consequently, a substantially lower value of
$\langle x\rangle^{u-d}\approx0.25(3)_\text{stat+ren}$,
this time in good agreement with the results from QCDSF in Fig.~\ref{x_v2b_QCDSF_2006}, 
is obtained for $(at_\text{sep})=12$, as represented by the open diamond in 
the upper part of Fig.~\ref{x_RBCUKQCD} at a pion mass of $m_\pi\approx493\MeV$.
It should be noted, however, that taking a simple average in 
the absence of a clear plateau as in, e.g., the lower part of Fig.~\ref{Plat10_RBCUKQCD},
may introduce an additional systematic uncertainty.
For this reason, no strong conclusions can be drawn from the above analysis at this point.

More recent results by RBC-UKQCD based on $n_f=2+1$ flavors of domain wall fermions and
the Iwasaki gauge action, for a lattice spacing of $a\approx0.114\fm$
in a volume of $V\approx(2.74\fm)^3$ and pion masses ranging from 
$m_\pi\approx331\MeV$ to $\approx672\text{ MeV}$ are shown in 
the lower part of Fig.~\ref{x_RBCUKQCD} \cite{Ohta:2008kd}.
In the range of $m_\pi^2\approx0.2,\ldots,0.5\GeV^2$, 
these results are, in turn, $\approx17\%$ below the $n_f=2$ DW results in the upper part of Fig.~\ref{x_RBCUKQCD}, 
and therefore on the same level as the $n_f=2$ clover-improved Wilson Fermion results by QCDSF/UKQCD in Fig.~\ref{x_v2b_QCDSF_2006}. 
A possible explanation is the larger source-sink separation
of $t_{\text{sep}}=1.4\fm$ used in the case of the 
$n_f=2+1$ results in the lower part of Fig.~\ref{x_RBCUKQCD},
leading to a stronger suppression of contributions from excited states,
which may have affected the $n_f=2$ DW calculation with $at_{\text{sep}}=1.1\fm$ as discussed above.
One should keep in mind, however, that the quality of the signal of the two-point function 
at the sink, $C_{2pt}(t_\snk)$, which is used in the calculation of $\langle x\rangle$,
deteriorates for large $t_\snk$ in the case of light quark masses. 
The choice of a particular large source-sink separation, $t_\snk$,
may therefore be a source of an additional systematic uncertainty 
(see also the corresponding discussion in section \ref{FFradii}).
Notably, the data point at the lowest pion mass in Fig.~\ref{x_RBCUKQCD}
exhibits a somewhat lower central value. Once the statistics has been improved,
and possible systematic uncertainties like finite volume effects and fluctuations 
of the two-point function at large sink times are under control, this may be 
regarded as an encouraging sign of a bending over towards the experimental value.
The $n_f=2+1$ mixed action results by LHPC that are displayed in Fig.~\ref{x_RBCUKQCD}
will be discussed in detail in section \ref{sec:EMT} below, cf. Fig.~\ref{A20umd_LHPC}.

For results on individual up- and down-quark momentum fractions
in quenched QCD we refer to, e.g., 
\cite{Gockeler:1995wg,Dolgov:2002zm,Gockeler:2004wp}
and references therein.
Connected diagram contributions to individual up- and down-quark
momentum fractions from simulations with dynamical quarks
were presented in, e.g., \cite{Dolgov:2002zm}
More recent results from LHPC \cite{Hagler:2007xi}
on connected contributions to $\langle x\rangle^{u+d}$ 
will be discussed in some more detail below in section \ref{sec:SpinStructure}.
Interesting attempts to finally get quark line disconnected contributions to
hadron structure observables under control will be presented separately in section \ref{sec:disconnected}.

\subsubsection*{Unpolarized gluon momentum fraction}

A first preliminary lattice study of the longitudinal momentum carried by the gluons in the
nucleon was already presented more that 10 years ago in the quenched approximation 
using the Wilson action \cite{Gockeler:1996zg}.
The gluon momentum fraction in the nucleon is given by
$\langle P|\mathcal{O}_g^{\mu\nu}|P\rangle=2P^\mu P^\nu\langle x\rangle^g$ with
$\mathcal{O}_g^{\mu\nu}=\mytr F^{\mu\alpha}(0)F^{\alpha\nu}(0)$,
which corresponds directly to the pure gauge part of the QCD energy momentum tensor, $T_g^{\mu\nu}$.
Results were presented for a perturbatively renormalized, 
traceless symmetric operator similar to Eq.~(\ref{op22}),
given by $\mathcal{O}_b=\mytr\{\mbf{E}^2-\mbf{B}^2\}$.
In order to overcome the expected noise from short-range fluctuations,
a large number of up to $3500$ quenched Wilson gauge configurations were employed.
Using a standard ratio of nucleon three- to two-point functions,
averaged over the operator insertion time $\tau$ 
(keeping $\tau$ at a distance of $\delta t=4$ to source and sink),
clearly non-vanishing contributions were observed for the individual chromo-electric and
magnetic contributions, which however canceled out to a large extent in $\mathcal{O}_b$.
An average value of $\langle x\rangle^g\approx0.5$ with an error of $\approx 50\%$
was obtained for the three different hopping parameters $\kappa=0.1550,0.1530,0.1515$ at 
a lattice scale of $a^{-1}\approx2\GeV$, corresponding to pion masses of $m_\pi\approx600,850,1000\MeV$.

More recently, the $\chi$QCD collaboration extended their studies of
disconnected contributions in the quark sector to 
include the gluon momentum fraction in the nucleon \cite{Doi:2008hp}.
Some details of their approach to so-called disconnected insertions (DIs)
based on all-to-all quark propagators obtained from stochastic sources
will be presented below in section \ref{sec:disconnected} 
(for a brief introduction to stochastic methods, see section \ref{sec:methods}).
In order to reduce the short-range fluctuations of the gluon operator,
the overlap Dirac operator, $D_{\text{ov}}$, which is non-ultralocal and therefore may 
help to efficiently filter the ultraviolet modes, was used in the construction of
a discretized field strength tensor, $F^{\text{ov}}_{\mu\nu}\propto \mytr[\sigma_{\mu\nu}D_{\text{ov}}]$.
That this is indeed a valid identification has been shown explicitly before 
in \cite{Liu:2007hq}.
The overlap Dirac operator was estimated numerically using space-time diluted $Z(4)$ noise vectors.
Based on only $500$ quenched Wilson configurations, but including a large number of nucleon sources, 
and employing (unrenormalized) non-diagonal operators of the type $\mathcal{O}_g^{4i}$, 
a non-zero signal was observed for $\langle x\rangle^g$ for a pion mass of $\approx650\MeV$.
Once the gluon operators have been (non-perturbatively) renormalized, and
the signals improved with the help of noise reduction techniques, these 
calculations may provide very interesting insights into the nucleon momentum sum rule Eq.~\ref{MomSumrule1}.
\subsubsection*{Higher moments of the unpolarized quark PDF}
Results for higher moments of the unpolarized quark distribution, 
$\langle x^{n-1}\rangle_{q}$, with $n=3,4$, obtained in the quenched approximation,
can be found in \cite{Gockeler:1995wg,Dolgov:2002zm} based on perturbatively, 
and in \cite{Gockeler:2004wp} based on non-perturbatively renormalized operators.
The LHPC-SESAM collaboration has calculated higher moments based on
$n_f=2$ flavors of Wilson fermions, for rather large pion masses in the range of
$m_\pi=\approx700\MeV$ to $\approx900\text{ MeV}$, a lattice spacing of $\approx0.1\fm$
and a volume of $V\approx(1.6\fm)^3$ \cite{Dolgov:2002zm}.
Results from unquenched simulations at lower pion masses were to this date only 
presented in \cite{Gockeler:2004vx} for $\langle x^{2}\rangle_{u-d}$
and $\langle x^{3}\rangle_{u-d}$, and data for $\langle x^{2}\rangle_{u-d,u+d}$
(only including contributions from connected diagrams) can be found in \cite{Hagler:2007xi}.
The former were obtained for $n_f=2$ flavors of clover-improved Wilson fermions
and pion masses down to $\approx560\text{ MeV}$, while the latter were calculated
in the framework of a hybrid calculation of $n_f=2+1$ flavors of domain wall valence quarks
on the Asqtad MILC ensembles for $m_\pi\sim350,\ldots,760\MeV$. 
We note that Ref.~\cite{Gockeler:2004vx} 
gives $\langle x^{2}\rangle_{u-d}$ and $\langle x^{3}\rangle_{u-d}$
only in the renormalization group invariant (RGI) scheme, and
that the results in \cite{Hagler:2007xi} were obtained using (non-perturbatively
improved) perturbatively renormalized operators as described in the previous section, see Eq.~\ref{ZLHPC}
and adjacent discussion.
Since the evaluation of the nucleon three-point functions for $n=3,4$
requires non-zero hadron momenta, the signals are in general much noisier than
for $n=2$ (corresponding to the momentum fractions). 
With this in mind, we just note that the values for $\langle x^{2}\rangle_{u-d}$
reported in quenched \cite{Gockeler:2004wp} and unquenched \cite{Gockeler:2004vx} lattice QCD 
for $m_\pi\ge550\MeV$ are $\approx30-40\%$ larger than results from 
global PDF analyses, $\langle x^{2}\rangle^{\text{CTEQ6.6}}_{u-d}\simeq0.054$ \cite{Nadolsky:2008zw}
and $\langle x^{2}\rangle^{\text{MRST06}}_{u-d}\simeq0.057$ \cite{Martin:2007bv}.
From the same references, the lattice results for the next higher 
moment, $\langle x^{3}\rangle_{u-d}$, are approximately a factor of two larger in  
the quenched, and $\approx30\%$ larger in the unquenched 
case compared to the value of $\langle x^{3}\rangle^{\pheno}_{u-d}\simeq0.023$
from the global PDF-analyses.

\subsubsection*{Moments of longitudinally polarized PDFs}
In this section we review lattice calculations of the lowest moments of polarized PDFs
of quarks in the nucleon, denoted by $\langle x^{n-1}\rangle_{\Delta q}$ with $n=1,2,\ldots$.
The lowest moment, $n=1$, in the isovector channel is identified
with the axial-vector coupling constant, $\langle x^{n-1}\rangle_{\Delta u-\Delta d}=g_A$,
and was discussed above in section \ref{sec:axialvector} in the framework of nucleon form factors.

Early studies of the polarized momentum fraction in quenched lattice QCD gave 
a value of $\langle x\rangle_{\Delta u-\Delta d}\approx0.246(9)$
\cite{Gockeler:1997zr}, which is, similar to the situation for 
the unpolarized case discussed above, $\approx 25\%$ larger than
results from phenomenology, $\langle x\rangle^{\pheno}_{\Delta u-\Delta d}\simeq0.20$ 
in the $\MSbar$-scheme at a scale of $\mu=2\GeV$ \cite{deFlorian:2008mr}.
This has been confirmed in unquenched Wilson action simulations \cite{Dolgov:2002zm},
where a value of $\langle x\rangle_{\Delta u-\Delta d}\approx0.271(25)$
was found. Both calculations were based on perturbatively 
renormalized operators of the form $\overline q\gamma_5\gamma_{\{\mu}\Dlr_{4\}}$, 
with $\mu=2$ or $\mu=3$, and the cited values 
were obtained from a linear chiral extrapolation in $m_\pi^2$ to the physical pion mass
in the $\MSbar$-scheme at a scale of $\mu=2\GeV$.

\begin{figure}[t]
    \begin{minipage}{0.48\textwidth}
      \centering
          \includegraphics[angle=0,width=0.99\textwidth,clip=true]{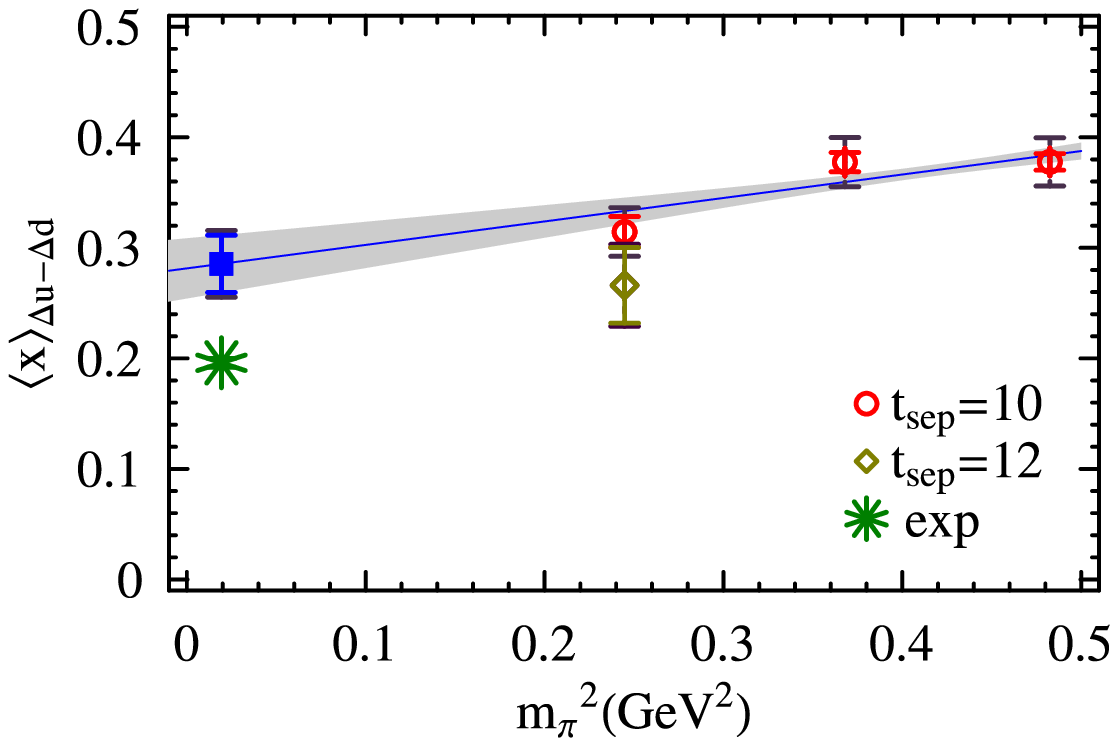}
          \includegraphics[angle=0,width=0.85\textwidth,clip=true]{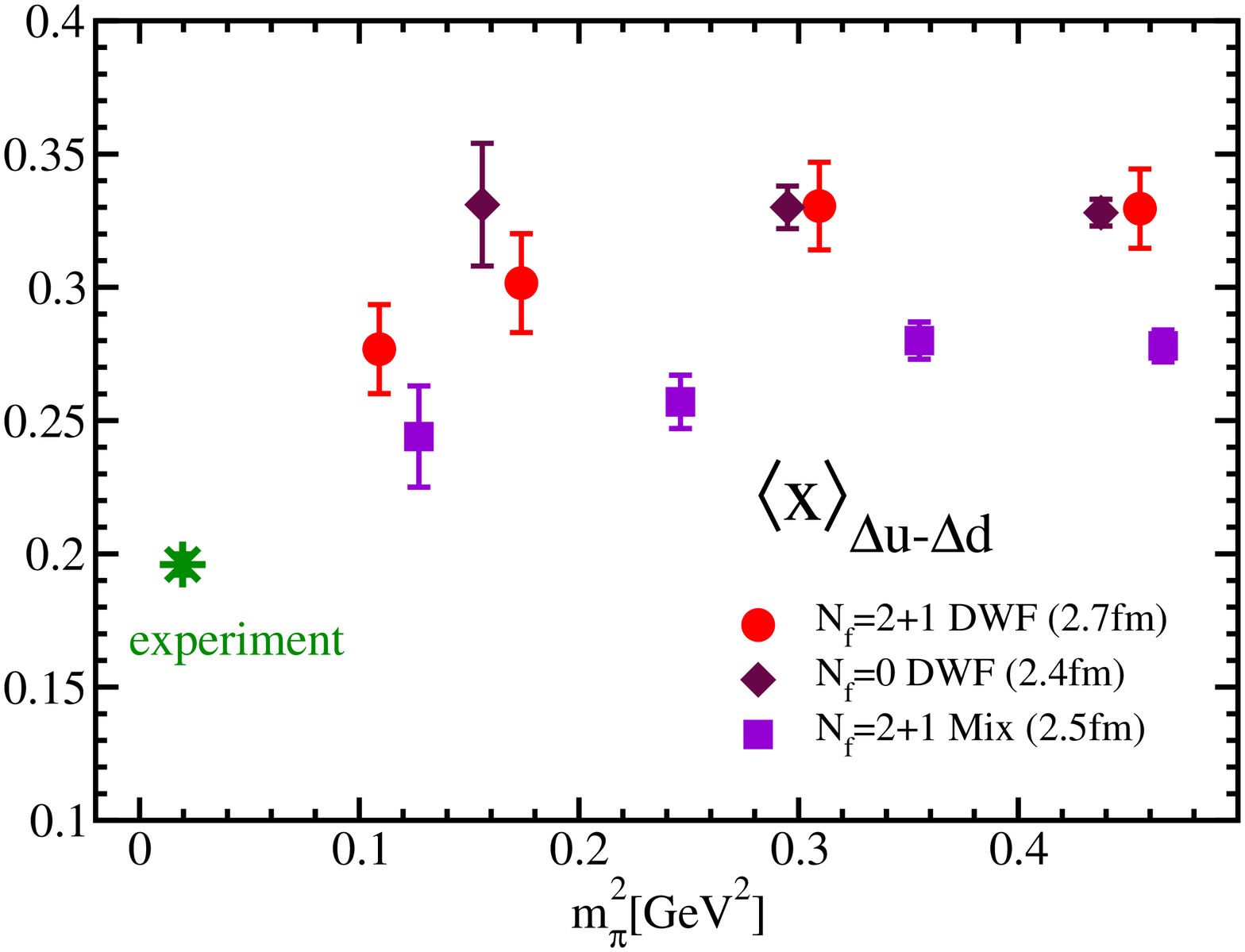}
  \caption{The polarized momentum fraction of quarks in the isovector channel from \cite{Lin:2008uz}
  (upper part) and the proceedings \cite{Ohta:2008kd} (lower part).}
  \label{Deltax_RBCUKQCD}
     \end{minipage}
          \hspace{0.3cm}
    \begin{minipage}{0.48\textwidth}
      \centering
          \includegraphics[angle=0,width=0.9\textwidth,clip=true]{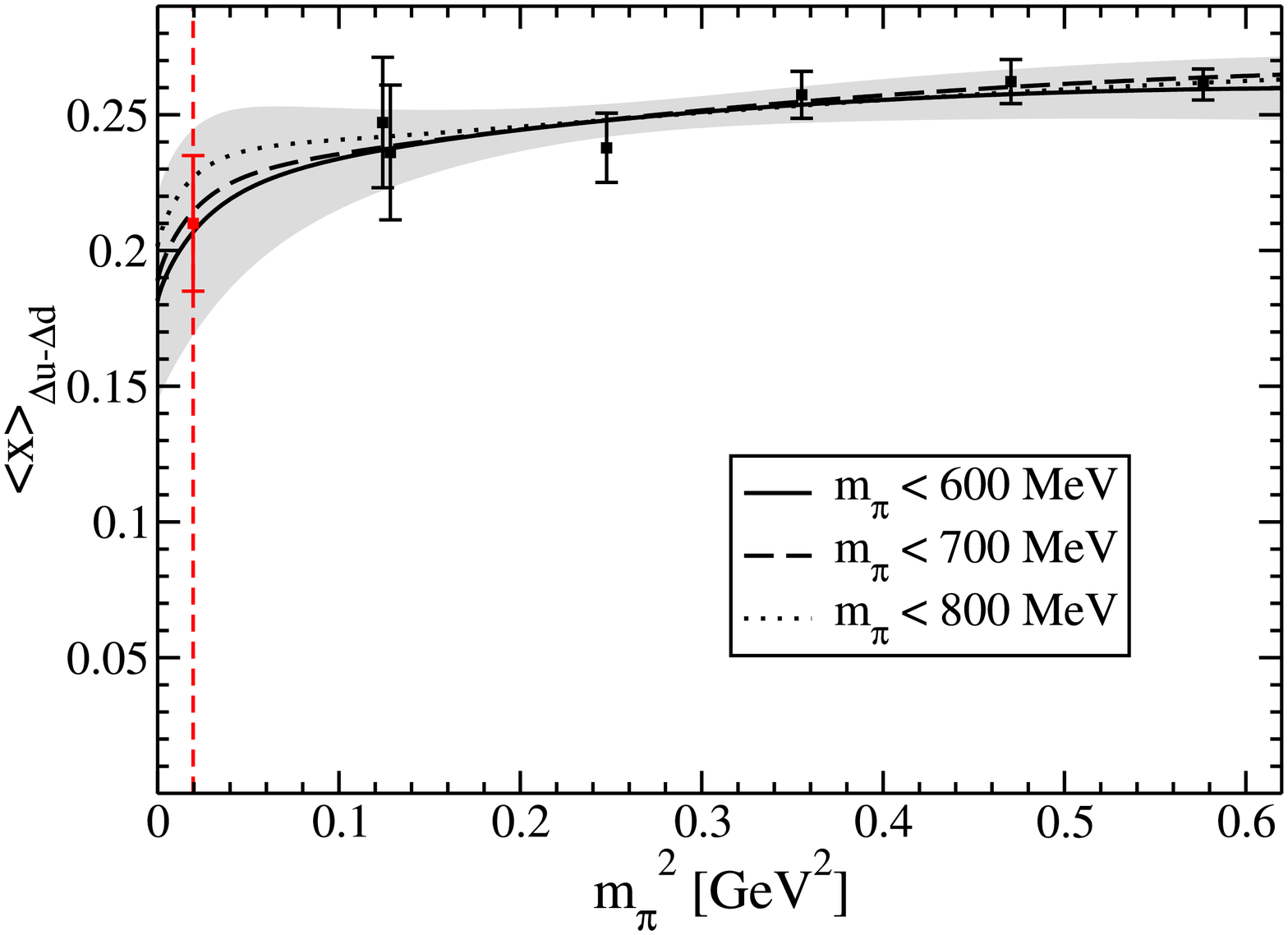}
  \caption{The polarized momentum fraction of quarks in the isovector channel (from proceedings \cite{Renner:2007pb}).}
  \label{Deltax_LHPC}
     \end{minipage}
\end{figure}

Recently, RBC has published results for the polarized momentum fraction
based on $n_f=2$ flavors of domain wall fermions \cite{Lin:2008uz}, obtained in the same framework as
for the unpolarized momentum fraction shown in Fig.~\ref{x_RBCUKQCD}.
The non-perturbatively renormalized operator $\overline q\gamma_5\gamma_{\{3}\Dlr_{4\}}$
was employed for the analysis of  $\langle x\rangle_{\Delta u-\Delta d}$, and the 
results, in the $\MSbar$-scheme at a scale of $\mu=2\GeV$, 
are displayed in the upper part of Fig.~\ref{Deltax_RBCUKQCD} 
as a function of $m_\pi^2$. 
As has already been observed in the previous section for
$\langle x\rangle_{u-d}$, the lattice data points 
for a source-sink separation of $t_\text{sep}=10$ 
are found to be substantially larger than the values that were obtained in 
previous unquenched calculations, e.g. in \cite{Dolgov:2002zm}. 
Contributions from excited states
have been cited as a possible cause, and the calculation has been repeated
for a larger source-sink separation of $t_\text{sep}=12$ at the 
lowest pion mass of $\approx493\MeV$. The result is represented by the
open diamond in the upper part of Fig.~\ref{Deltax_RBCUKQCD}, with a central value
that is indeed substantially lower than for $t_\text{sep}=10$. 
However, the underlying plateau is, similar to what has been discussed
in relation with $\langle x\rangle_{u-d}$ in the upper part 
of Fig.~\ref{x_RBCUKQCD} and with Fig.~\ref{Plat10_RBCUKQCD}, 
subject to much larger statistical noise. 
Furthermore, the two-point function at the sink in the ratio of three- to two-point functions
may show fluctuations for large sink times. It would be important to study these 
issues in some more detail before more definite conclusions can be drawn from these results.

The lower part of Fig.~\ref{Deltax_RBCUKQCD} shows very recent results from RBC-UKQCD
for $n_f=2+1$ flavors of domain wall fermions, obtained using a large
source-sink separation of $at_\text{sep}=1.4\fm$ \cite{Ohta:2008kd}. At larger pion masses, these
results are $\approx13\%$ below the $n_f=2$ results discussed above.
We note that both the $n_f=2$ and $n_f=2+1$ DW results for $\langle x\rangle_{\Delta u-\Delta d}$
from RBC and RBC-UKQCD are substantially above the early unquenched results from \cite{Dolgov:2002zm},
which may however be in parts related to the perturbative renormalization
employed in the latter case.
Further investigations are necessary to see if the apparent bending towards the experimental
value at lower pion masses in the lower part of Fig.~\ref{Deltax_RBCUKQCD} is a 
physical effect due to the chiral dynamics, or rather related to the finite volume 
or other systematic effects.

%
%

Figure \ref{Deltax_LHPC} displays results from LHPC calculated in the framework of
a hybrid approach with $n_f=2+1$ flavors of domain wall quarks and Asqtad staggered
sea quarks \cite{Renner:2007pb}, for a (non-perturbatively improved)
perturbatively renormalized operator (see Eq.~(\ref{ZLHPC}) and corresponding discussion), 
transformed to the $\MSbar$-scheme at a scale of $\mu=2\GeV$. 
Notably, the overall normalization,
$\langle x\rangle_{\Delta u-\Delta d}\approx0.24,\ldots,0.26$,
is on the same level as the first unquenched calculations, 
and therefore closer to the phenomenological value
than the result from RBC-UKQCD discussed above. 
We note, however, that the lower overall normalization may be related to the 
specific (partially perturbative) operator renormalization employed in this calculation.
An extrapolation to the physical point, using self-consistently improved 1-loop HBChPT
based on an equation similar to Eq.~\ref{xLHPCChPT}, with a different
coefficient in front of the chiral logarithm, 
$(3 g_{A,\lat}^2 + 1)\rightarrow (2 g_{A,\lat}^2 + 1)$, was attempted. The 
chiral fit to the lowest four lattice data points with $m_\pi<600\MeV$ is 
represented by the shaded band in Fig.~\ref{Deltax_LHPC} and
broadly overlaps within errors with the phenomenology result.

Apparently, the normalization 
issues in the unpolarized and polarized momentum
fractions are very similar. It has been noted 
in this respect that a number of normalization 
uncertainties may be attenuated in the ratio
$R=\langle x\rangle_{u-d}/\langle x\rangle_{\Delta u-\Delta d}$
\cite{Orginos:2005uy}.
This ratio is not only naturally renormalized 
for chiral, e.g. domain wall, fermions
since the renormalization constants
for vector and axial-vector operators are identical
in this case due to (lattice) chiral symmetry, but also
the leading chiral logarithms cancel out to some extent.
The latter may be, however, only relevant at very small pion masses.
Interestingly, quenched and unquenched lattice results
for $R$ show in general 
good agreement over a wide range of pion masses
with the corresponding ratio from experiment and phenomenology
of $R^{\exp}\approx0.8$ at the physical point, see, e.g.,
\cite{Orginos:2005uy,Lin:2008uz,Edwards:2005kw}.

\begin{figure}[t]
    \begin{minipage}{0.48\textwidth}
      \centering
          \includegraphics[angle=0,width=0.95\textwidth,clip=true]{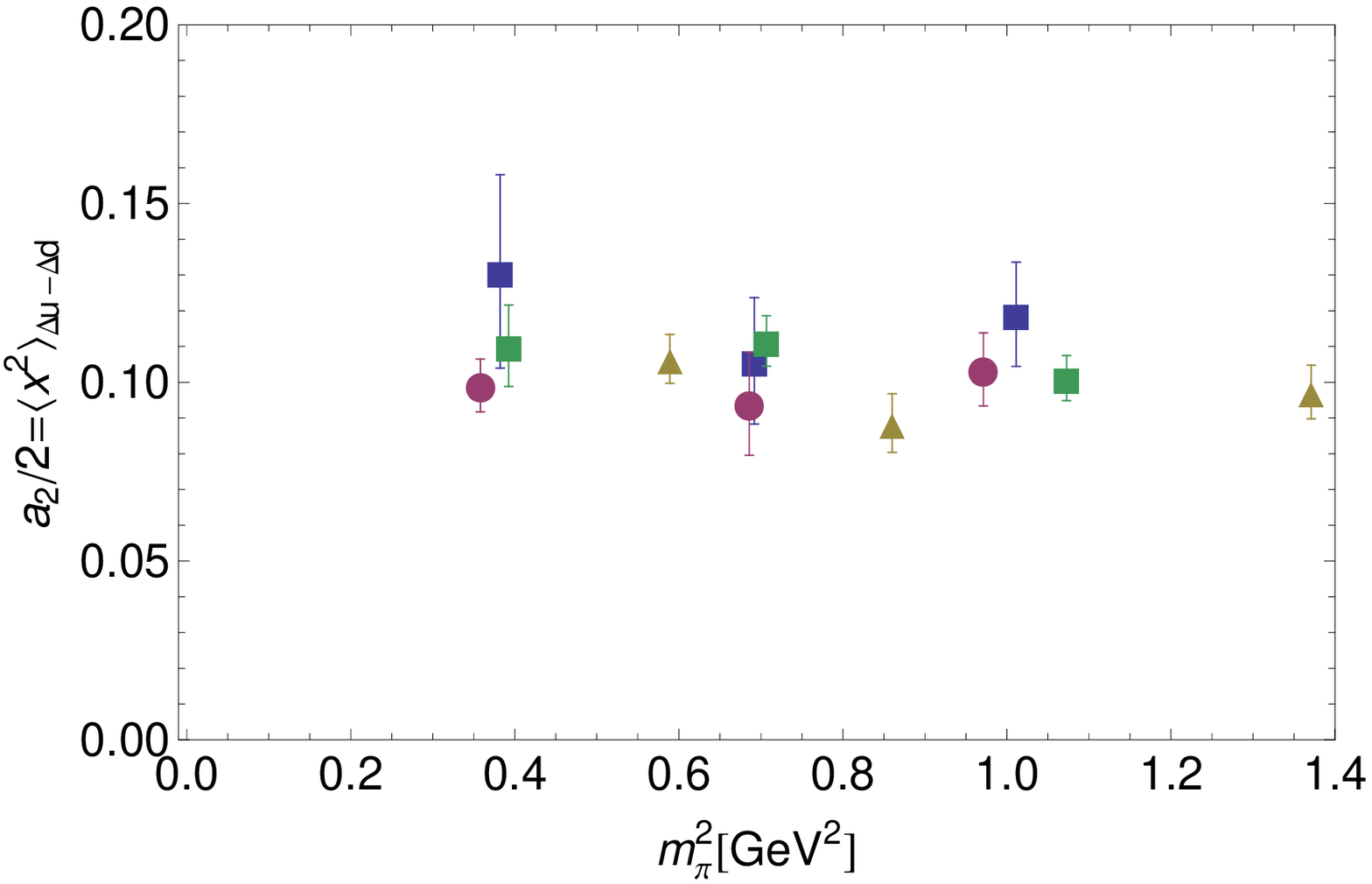}
  \caption{Higher moment of the isovector polarized quark distribution (from \cite{Gockeler:2005vw}).}
  \label{Deltax2_QCDSF}
     \end{minipage}
          \hspace{0.3cm}
    \begin{minipage}{0.48\textwidth}
      \centering
          \includegraphics[angle=-90,width=0.95\textwidth,clip=true]{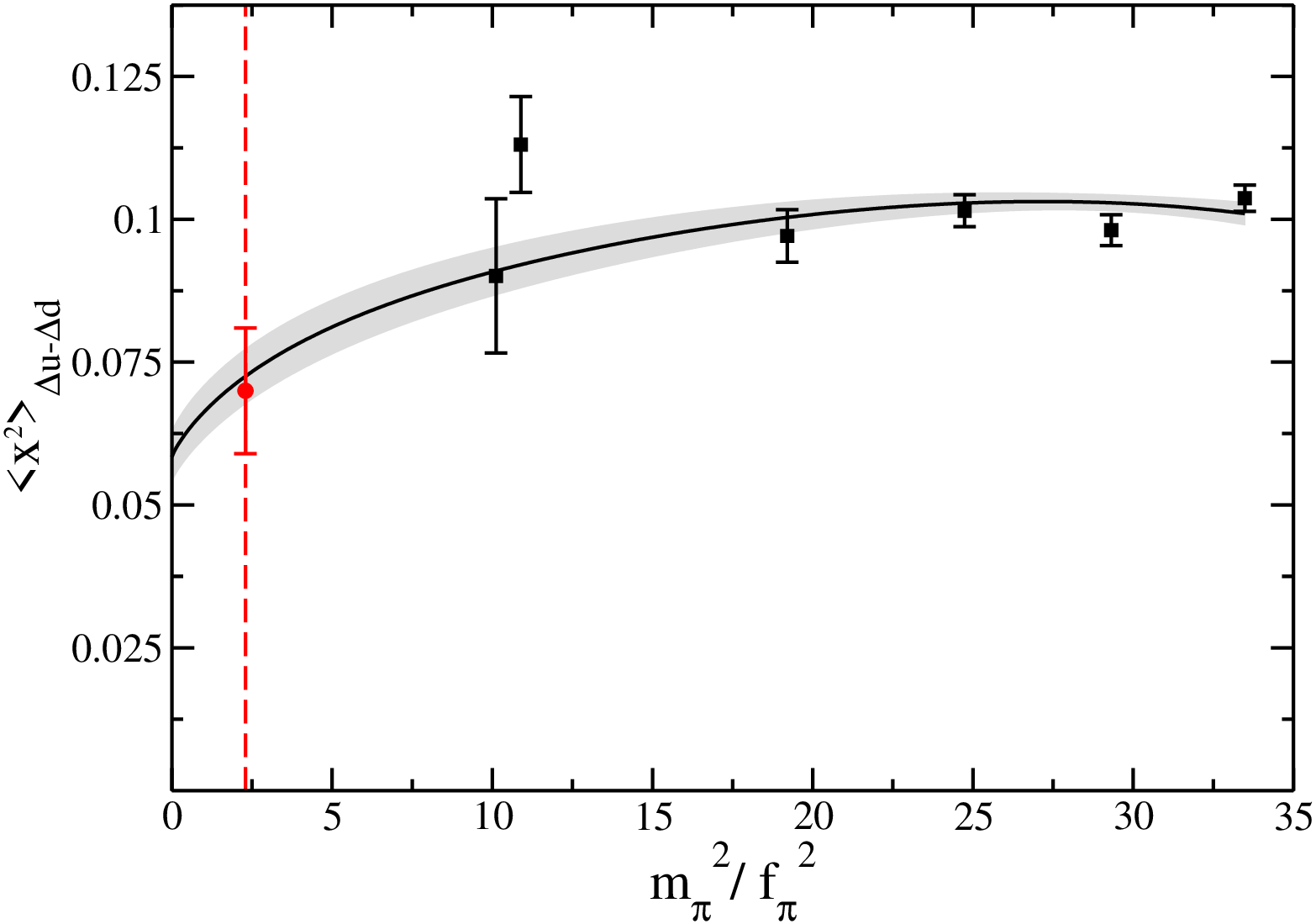}
  \caption{Higher moment of the isovector polarized quark distribution (from proceedings \cite{Edwards:2006qx}).}
  \label{Deltax2_LHPC}
     \end{minipage}
\end{figure}

We now turn to a brief discussion of the next higher
moment of the polarized quark distribution,
$\langle x^2\rangle_{\Delta q}$.
QCDSF has published a dedicated study of moments of the
spin-dependent nucleon structure functions 
in the framework of simulations with $n_f=2$ flavors of
clover-improved Wilson fermions \cite{Gockeler:2005vw}.
A non-perturbatively renormalized operator 
\bea
 \mathcal{O}^5_{214}=\overline q\gamma_5\gamma_{\{2}\Dlr_1\Dlr_{2\}}q \,, 
 \label{oppol1}
\eea
where the curly brackets denote symmetrization over the respective indices,
was used for the extraction of $\langle x^2\rangle_{\Delta q}$\footnote{Denoted 
by $a_2\equiv2\langle x^2\rangle$ in Ref.~\cite{Gockeler:2005vw}},
which is given in the $\MSbar$ scheme at a scale of $5\GeV^2$.
Results for the isovector channel are presented in Fig.~\ref{Deltax2_QCDSF}
for pion masses in the range of $\approx600\MeV$ to $\approx1200\text{ MeV}$. 
Although there are some fluctuations visible, the lattice data points agree for all different
ensembles and are practically flat in $m_\pi^2$ within statistical errors.
They are on average approximately $35\%$ above the phenomenological value 
of $\langle x^2\rangle^{\pheno}_{\Delta u - \Delta d}\simeq0.065$ \cite{deFlorian:2008mr}. 

These results are well compatible with the numbers obtained
by LHPC based on the hybrid approach of 
$n_f=2+1$ flavors of domain wall quarks and Asqtad staggered sea quarks,
displayed in Fig.~\ref{Deltax2_LHPC} as a function of the ratio $(m_\pi/f_\pi)^2$.
As explained already several times before, the lattice operators have been
renormalized using a non-perturbatively improved perturbative renormalization factor, cf. Eq.~(\ref{ZLHPC}), 
and transformed to the $\MSbar$ scheme at a scale of $2\GeV$ \cite{Edwards:2006qx}.
A chiral extrapolation based on a self-consistently improved
1-loop HBChPT formula, as discussed above in relation with $\langle x\rangle_{\Delta u-\Delta d}$,
is represented by the shaded band, and agreement within errors is observed with the phenomenological value 
of $\langle x^2\rangle^{\pheno}_{\Delta u - \Delta d}\simeq0.068$  \cite{deFlorian:2008mr}
at the physical ratio $(m^\phys_\pi/f^\phys_\pi)^2\approx2.4$.

%

%
%

\subsubsection*{Moments of the transversity distribution}
Lattice QCD results for the lowest moment of the transversity
distribution, the tensor charge $\langle 1\rangle_{\delta q}=\delta q$,
were discussed above in section \ref{sec:tensorFFs}.
First lattice calculations of the $n=2$-moment, $\langle x^{}\rangle_{\delta q}$,
in quenched QCD and for $n_f=2$ Wilson fermions, based on a perturbatively renormalized 
operator of the form $\overline q\gamma_5\sigma_{3\{4}\Dlr_{1\}}q$,
were presented in \cite{Dolgov:2002zm}.
Values of $\langle x^{}\rangle_{\delta u-\delta d}\approx0.5$ with errors of $10\%-20\%$ were reported,
in the $\MSbar$-scheme at a scale of $\mu=2\GeV$
for pion masses of $m_\pi\approx700$ and higher. 
We note that this is about a factor of two larger
than the respective moments of the unpolarized and polarized PDFs, which
are $\langle x^{}\rangle_{u-d}\approx\langle x^{}\rangle_{\Delta u-\Delta d}\approx0.25$
as obtained in the same lattice study. 

\begin{figure}[t]
      \centering
         \includegraphics[angle=0,width=0.9\textwidth,clip=true]{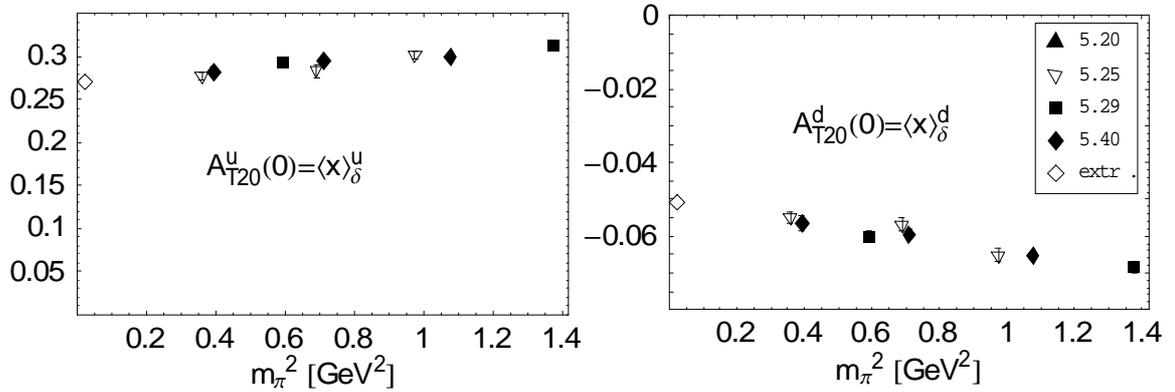}
  \caption{Up- and down-quark (connected) contributions 
  to the $x$-moment of the transversity distribution (from \cite{Gockeler:2005cj}).}
  \label{xdelta_AT20_m_pi_v2_QCDSF}
 \end{figure}

Substantially lower values for $\langle x^{}\rangle_{\delta u-\delta d}$ 
have been reported by QCDSF in an unquenched study with
$n_f=2$ flavors of Clover-improved Wilson quarks for pion masses
in the range of $\approx600$ to $\approx1200\text{ MeV}$ \cite{Gockeler:2005cj}, where special care was
taken that the normalization and parametrization of the corresponding
nucleon matrix elements is fully consistent with the
definition of moments of the transversity distribution $\delta q(x)$ \cite{Jaffe:1991kp}. 
These calculations have been performed in the framework of 
an analysis of tensor GPDs and were based on the full set of non-perturbatively
renormalized operators from the two 8-dimensional multiplets 
transforming according to the $H(4)$ representations $\tau_1^{(8)}$ and $\tau_2^{(8)}$, with
typical members $\overline q(\sigma_{12}\Dlr_{2}-\sigma_{13}\Dlr_{3})q$ 
and $\overline q\sigma_{1\{2}\Dlr_{4\}}q$.
Results for individual up- and down-quark contributions, where only 
contributions from connected diagrams were taken into account,
are displayed in Fig.~\ref{xdelta_AT20_m_pi_v2_QCDSF} for the $\MSbar$-scheme at a scale of $\mu=2\GeV$.
The lattice data points show a remarkable statistical precision over the full range of accessible pion masses,
and are overall consistent for the different underlying ensembles, corresponding to lattice spacings 
in the range of $a\approx0.07$ to $\approx0.11\fm$ and volumes of $\approx(1.4-2.0 \fm)^3$. 
In the isovector channel, a value of $\langle x^{}\rangle_{\delta u-\delta d}=0.322(6)$ was obtained
from a linear chiral extrapolation in $m_\pi^2$ to the physical point \cite{Gockeler:2005cj}.
It has been speculated in \cite{Gockeler:2005cj} that the noticeable difference 
between this value and the results of \cite{Dolgov:2002zm}
may be related to a different normalization of the corresponding
nucleon matrix elements employed in the latter work.
%
%
\subsubsection*{Moments of PDFs of the $\rho$-meson}
The first and only lattice QCD study of the moments of unpolarized and polarized
PDFs of the $\rho$-meson was presented more than ten years ago in \cite{Best:1997qp}. 
Calculations were performed in the quenched approximation with Wilson fermions, 
for three pion masses of $\approx0.71$, $1.0$ and $1.2\GeV$,
with a lattice spacing of $a\approx0.082\fm$, calculated using the $\rho$-mass and 
extrapolated to the chiral limit. 
The study includes the moments $\langle x^{n-1}\rangle^\rho_{q}$ 
and $H_5^{n}$ of the unpolarized distributions $H_1(x)$ 
in Eq.~\ref{RhoVec2} and $H_5(x)$ in Eq.~\ref{RhoVec2b}, respectively,
for $n=2,3,4$, as well as the lowest three moments 
$\langle x^{n-1}\rangle^\rho_{\Delta q}$, with $n=1,2,3$,
of the polarized distribution $\widetilde H_1(x)$ in Eq.~\ref{RhoAxial2}\footnote{We note that the moments of the
unpolarized and polarized PDFs of the $\rho$-meson in \cite{Best:1997qp} are denoted by $a_n$, $d_n$ and $r_n$.}.
Several perturbatively renormalized operators were used for the extraction of 
these lowest moments. Possible effects from operator mixing
for $n=3,4$, corresponding to operators with two and three
covariant derivatives, were not included in the analysis.
The total momentum fraction carried by 
quarks in the rho was found to be $2\langle x^{}\rangle^{\rho^+}_{u}=0.706(19)$ for $m_\pi\approx1\GeV$
in the $\MSbar$ scheme at a scale of $\mu\approx2.4\GeV$, 
approximately of the same size as for the nucleon \cite{Dolgov:2002zm}
and somewhat larger than for the pion \cite{Best:1997qp} in quenched QCD 
at similar pion masses. 
At the same large pion mass, a value of $\frac{1}{2}\;2\;\langle 1\rangle^{\rho^+}_{\Delta u}=0.702(20)$
was obtained for the total contribution of the quark spin to the spin of the rho, $S=1$.
This is close to results for the nucleon in the
quenched approximation \cite{Gockeler:1995wg,Dolgov:2002zm},
where quarks carry approximately 65\% to 70\% percent of the total
spin of the nucleon $S=1/2$ (see also discussion below in section \ref{sec:SpinStructure}).
\subsection{Moments of GPDs}
\subsubsection{Moments of unpolarized GPDs of the pion}
\label{sec:PionGPDs}
Calculations of moments of unpolarized (vector) and tensor GPDs of the pion
have been performed over the last couple of years by QCDSF in unquenched lattice 
QCD \cite{Brommel:2005ee,Brommel:2006zz,Brommel:2007xd}, however very little has so far been published 
in particular for the unpolarized GFFs $A^\pi_{ni}(t)$ for $n\ge2$, $i=0,2,\ldots\le n$, see, e.g., Eq.~\ref{PionGPDsn2}.
We note again that the unpolarized GFF for $n=1$ is equal to the pion form factor, $A^{\pi^{+}}_{u,10}(t)=F_\pi(t=-Q^2)$, 
for which a number of lattice results are available, see section \ref{pionFFs}.
For a preliminary analysis of the GFFs $A^\pi_{20}(t)$ and $A^\pi_{22}(t)$ 
in unquenched lattice QCD, we refer to the PhD thesis by Br\"ommel \cite{Brommel:2007zz}.
Lattice QCD results for the tensor GPD of the pion, $E^\pi_T(x,\xi,t)$, Eq.~(\ref{PionTensor1}), 
in the form of the lowest two moments $B_{T10}^{\pi}(t)$ and $B_{T20}^{\pi}(t)$,
will be presented in the framework of a discussion of the 
transverse pion spin structure below in section \ref{PionTensor}.
\subsubsection{Form Factors of the Nucleon Energy Momentum Tensor}
\label{sec:EMT}
The form factors of the energy momentum tensor for quarks in the nucleon,
denoted by $A^q_{20}(t)$, $B^q_{20}(t)$ and $C^q_{20}(t)$ (see Eqs.~(\ref{EMT1}) and (\ref{EMTFFsGFFs1}))
have been studied for the first time in the quenched approximation 
with Wilson fermions by QCDSF \cite{Gockeler:2003jf} for pion masses
in the range of $\approx600$ to $\approx1000\MeV$, 
and at about the same time by LHPC-SESAM \cite{Hagler:2003jd} using $n_f=2$ flavors
of Wilson fermions and the Wilson gauge action 
for a pion mass of $\approx900\MeV$.

\begin{figure}[t]
    \begin{minipage}{0.48\textwidth}
      \centering
          \includegraphics[angle=0,width=0.99\textwidth,clip=true]{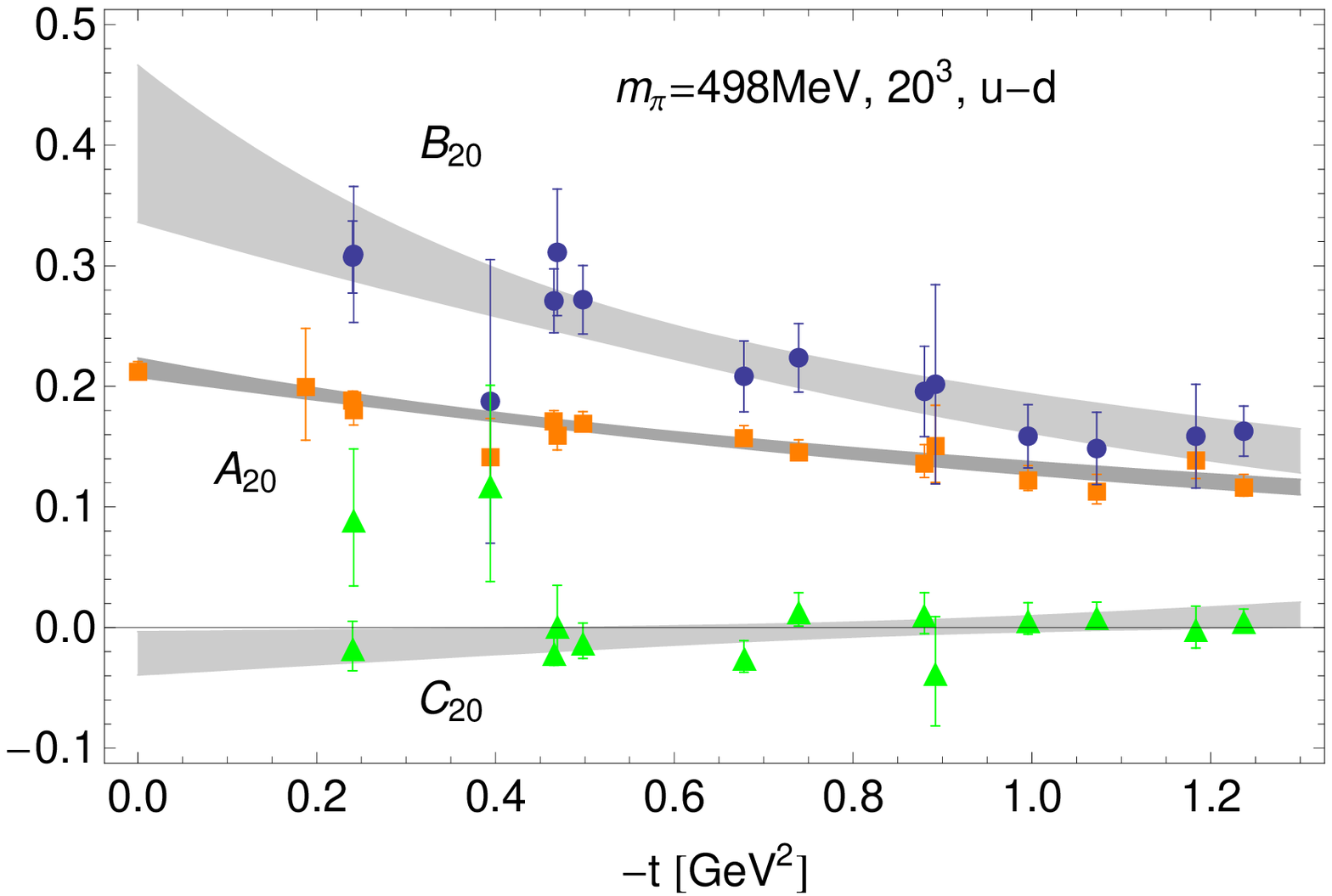}
  \caption{Form factors of the energy momentum tensor in the isovector channel (from \cite{Hagler:2007xi}).}
  \label{A20umd_LHPC}
     \end{minipage}
          \hspace{0.3cm}
    \begin{minipage}{0.48\textwidth}
      \centering
          \includegraphics[angle=-90,width=0.85\textwidth,clip=true]{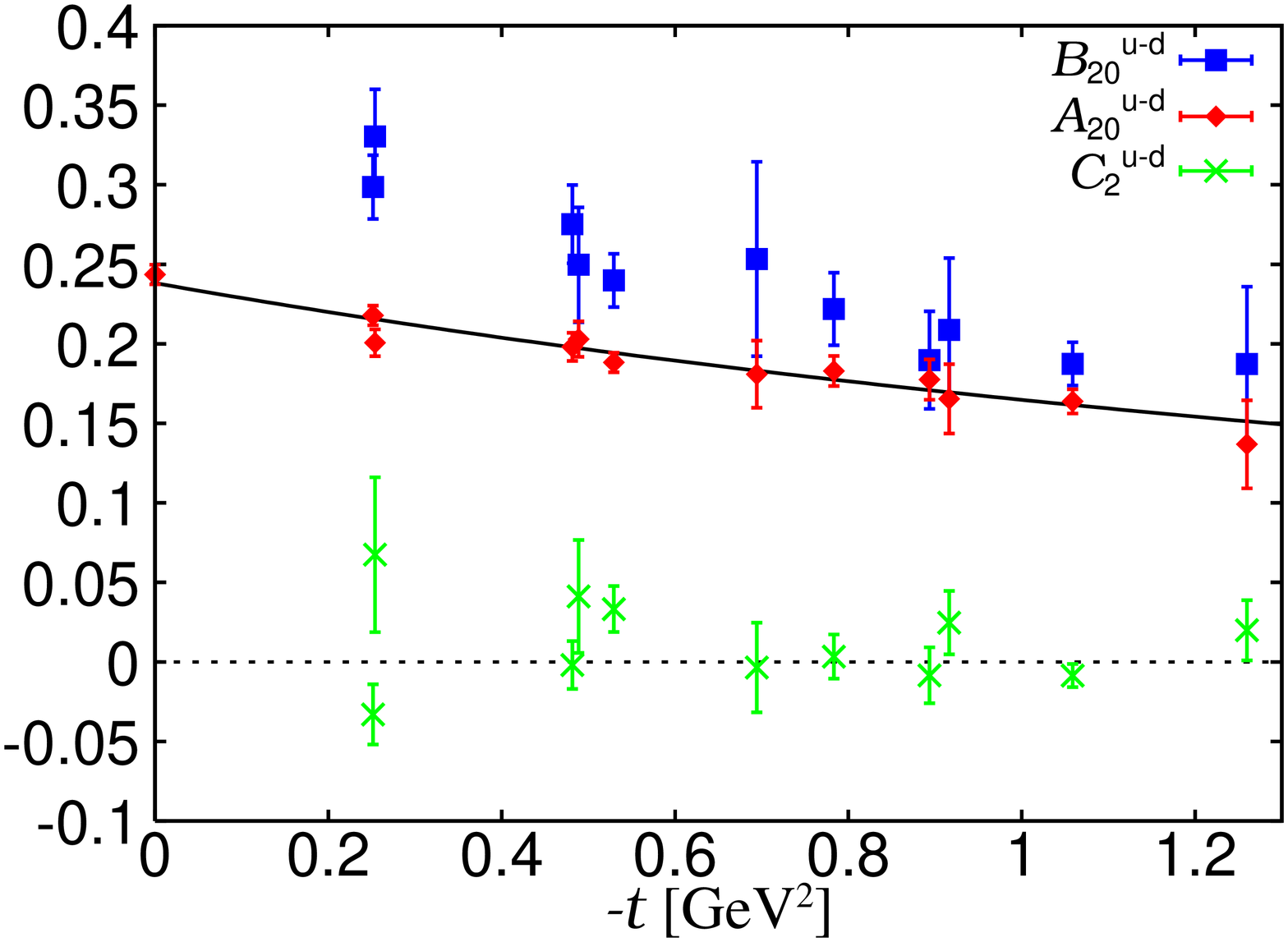}
  \caption{Form factors of the energy momentum tensor in the isovector channel 
  (from proceedings \cite{Brommel:2007sb}).}
  \label{GFF20v_QCDSF07}
     \end{minipage}
\end{figure}

Figure \ref{A20umd_LHPC} displays as an example results 
for the generalized form factors (GFFs) $A^{u-d}_{20}(t)$, $B^{u-d}_{20}(t)$ and $C^{u-d}_{20}(t)$ 
as functions of the momentum transfer squared, $t$,
for $m_\pi\approx498\MeV$ obtained recently by LHPC in a hybrid approach based on $n_f=2+1$
flavors of domain wall valence and Asqtad staggered sea quarks \cite{Hagler:2007xi}.
As has been discussed several times before, the underlying lattice
operators in this analysis were renormalized perturbatively,
supplemented by a non-perturbative improvement factor as given
in Eq.~\ref{ZLHPC}, and all results were transformed to the $\MSbar$ scheme
at a scale of $\mu=2\GeV$. The shaded bands in Fig.~\ref{A20umd_LHPC}
represent dipole fits to the lattice data points for 
$A^{u-d}_{20}(t)$ and $B^{u-d}_{20}(t)$, and a linear fit in $t$ to $C^{u-d}_{20}(t)$.

Similar results obtained by QCDSF for $n_f=2$ flavors of
clover-improved Wilson fermions are shown for comparison in Fig.~\ref{GFF20v_QCDSF07}
for a pion mass of $m_\pi\approx350\MeV$ \cite{Brommel:2007sb}.
Non-perturbatively renormalized lattice operators were employed for the extraction
of the GFFs, which are also given in the $\MSbar$ scheme at a scale of $\mu=2\GeV$.
Although the corresponding pion masses are different,
the results in Fig.~\ref{A20umd_LHPC} and Fig.~\ref{GFF20v_QCDSF07}
are in general compatible, apart from a difference in the overall
normalization of the lattice data points for $A^{u-d}_{20}(t)$ and $B^{u-d}_{20}(t)$,
where the results from QCDSF are $\approx10$ to $25\%$ above the
values from LHPC. This discrepancy has already been discussed
in connection with the forward value 
$A^{u-d}_{20}(t=0)=\langle x\rangle^{u-d}$ in Figs.~\ref{x_v2b_QCDSF_2006} and 
\ref{x_LHPC} in section \ref{sec:nuclPDFs} above.
We note in particular that the GFF $B^{u-d}_{20}(t)$ is sizeable and
larger than $A^{u-d}_{20}(t)$ over the full range  $-t\le1.3\GeV^2$.
At the same time, one finds that the GFF $C^{u-d}_{20}(t)$ is 
compatible with zero within statistical errors
for all accessible values of $-t\approx0.25,\ldots,1.3\GeV^2$.
%
%
%

\begin{figure}[t]
    \begin{minipage}{0.48\textwidth}
      \centering
          \includegraphics[angle=0,width=0.99\textwidth,clip=true]{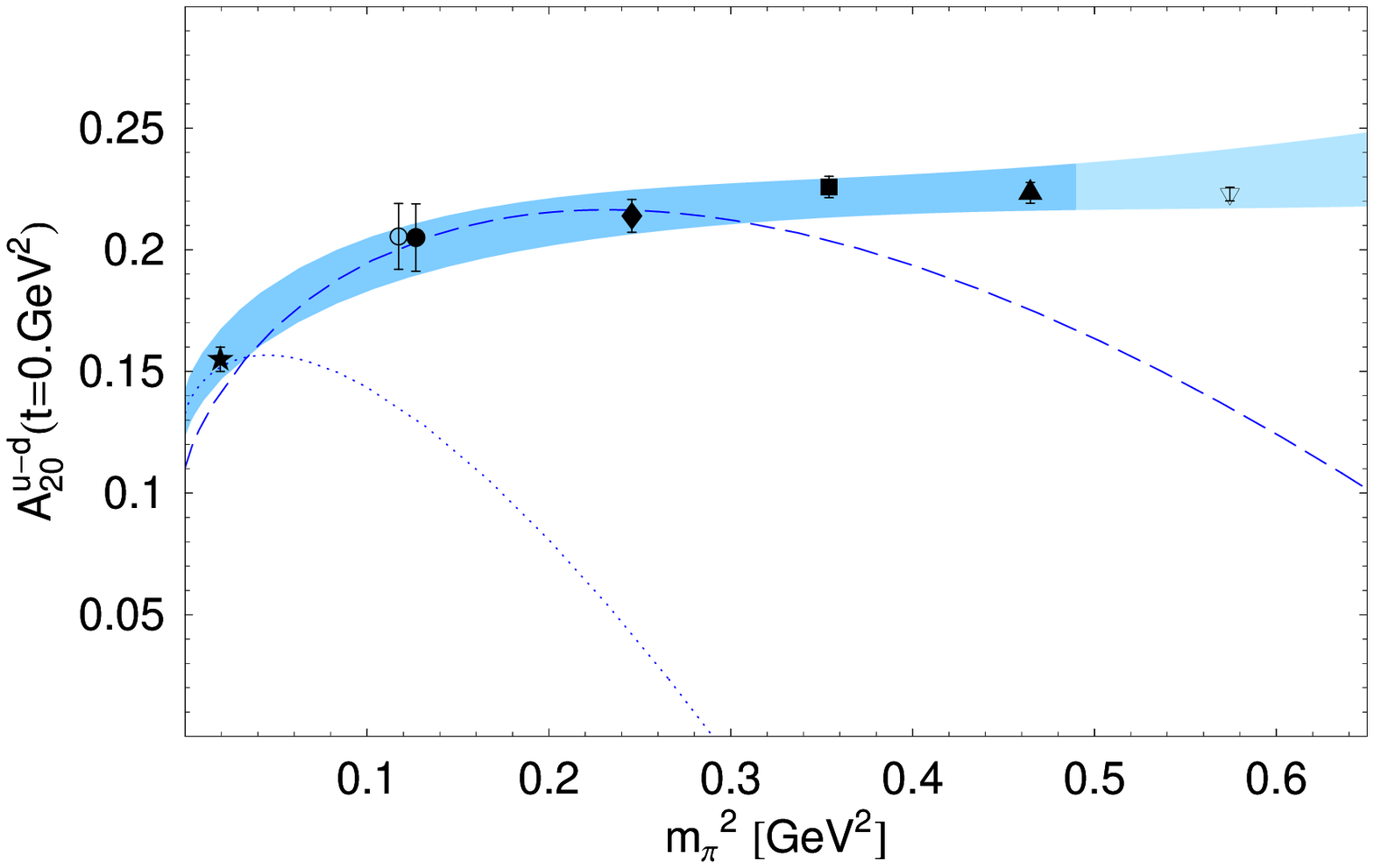}
  \caption{Pion mass dependence and chiral extrapolation of 
  $\langle x\rangle_{u-d}=A_{20}^{u-d}(t=0)$ (from \cite{Hagler:2007xi}).}
  \label{A20_umd_SimulFit}
     \end{minipage}
          \hspace{0.3cm}
    \begin{minipage}{0.48\textwidth}
      \centering
          \includegraphics[angle=0,width=0.99\textwidth,clip=true]{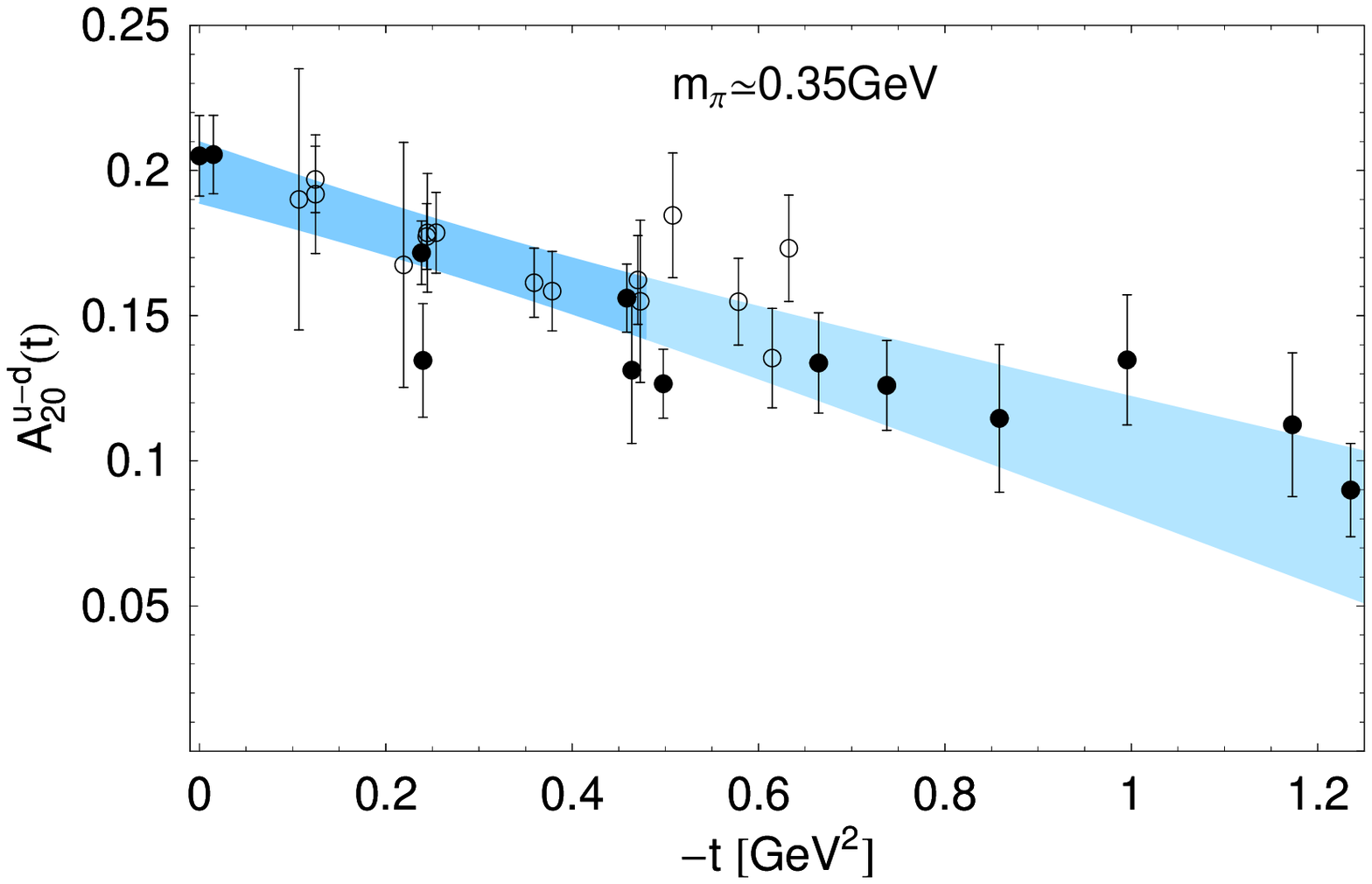}
  \caption{Dependence of $A_{20}^{u-d}(t)$ on the squared momentum transfer $t$ for a pion mass of
  $\approx350\MeV$ (from \cite{Hagler:2007xi}).}
  \label{A20_umd_mPip35_SimulFit}
     \end{minipage}
\end{figure}

In the following, we discuss the results by LHPC and QCDSF for the form
factors of the energy momentum tensor in some more detail.
LHPC has calculated $A^{q}_{20}(t)$, $B^{q}_{20}(t)$ and $C^{q}_{20}(t)$ 
for six different pion masses from $\approx355\MeV$ to $\approx760\MeV$, with a lattice
spacing of $a\approx0.124\fm$, for a volume of $\approx(2.5\fm)^3$ \cite{Hagler:2007xi}. The calculations
at the lowest pion mass were repeated in a larger volume of $\approx(3.5\fm)^3$,
allowing to check for possible finite volume effects.
The forward value $A^{u-d}_{20}(t=0)=\langle x\rangle^{u-d}$
is shown in Fig.~\ref{A20_umd_SimulFit} as a function of $m_\pi^2$.
These results differ from the results for the momentum fraction of $u-d$ quarks
presented in Fig.~\ref{x_LHPC} only in the number of lattice operators 
that were included in the respective analyses. While the lattice data points
in Fig.~\ref{x_LHPC} were obtained using the operator in Eq.~\ref{op22},
the results in Fig.~\ref{A20_umd_SimulFit} were based on the complete set
of symmetric and traceless lattice operators transforming according
to the irreducible representations $\tau_1^{(3)}$ and $\tau_1^{(6)}$ 
of dimensions three and six, respectively \cite{Gockeler:1996mu}.  
The good agreement of the lattice data points at $m_\pi\approx355\MeV$
for the two different volumes, represented by the open and filled circles in Fig.~\ref{A20_umd_SimulFit},
indicates that 
finite volume effects are
small compared to the statistical errors, at least at this particular pion mass.

A chiral extrapolation of the GFFs  $A^{u-d}_{20}(t)$, $B^{u-d}_{20}(t)$ and $C^{u-d}_{20}(t)$
was attempted using recent results obtained in the framework
of a covariant baryon ChPT (CBChPT) calculation to $\mathcal{O}(p^2)$ \cite{Dorati:2007bk}.
The $m_\pi$- and $t$-dependences of the GFFs are given by
\bea
A_{20}^{u-d}(t,m_\pi)&=&A_{20}^{0,u-d}\bigg(f_A^{u-d}(m_\pi) 
+ \frac{(g^0_A)^2}{192 \pi^2(f^0_\pi)^2}h_A(t,m_\pi)\bigg) 
 + \langle x\rangle^{0}_{\Delta u-\Delta d} j_A^{u-d}(m_\pi)\nonumber\\
  &+& \delta_{A}^{t,u-d}\,t 
 + \delta_{A}^{m_\pi,u-d}\,m_\pi^2\,,
\label{ChPTA20umdp4}
\eea
\begin{equation}
B_{20}^{u-d}(t,m_\pi)=\frac{m_N(m_\pi)}{m^0_N} B_{20}^{0,u-d} + A_{20}^{0,u-d} h_B^{u-d}(t,m_\pi) 
+ \frac{m_N(m_\pi)}{m^0_N}\bigg(\delta_{B}^{t,u-d}\,t + \delta_{B}^{m_\pi,u-d}\,m_\pi^2  \bigg)\,,
\label{ChPTB20umdp4}
\end{equation}
\begin{equation}
C_{20}^{u-d}(t,m_\pi)=\frac{m_N(m_\pi)}{m^0_N} C_{20}^{0,u-d} + A_{20}^{0,u-d} h_C^{u-d}(t,m_\pi) 
+ \frac{m_N(m_\pi)}{m^0_N}\bigg(\delta_{C}^{t,u-d}\,t + \delta_{C}^{m_\pi,u-d}\,m_\pi^2 \bigg)\,.
\label{ChPTC20umdp4}
\end{equation}
Apart from the axial-vector coupling and the pion decay constant in the chiral limit,
which were fixed to $g_A^0=1.2$ and $f_\pi^0=0.092\GeV$,
Eqs.~(\ref{ChPTA20umdp4},\ref{ChPTB20umdp4},\ref{ChPTC20umdp4}) depend on 10 additional LECs: The chiral limit values 
of the GFFs at $t=0$, $F_{20}^{0,u-d}=F_{20}^{u-d}(t\eql0,m_\pi\eql0)$, the counter-terms 
$\delta_{F}^{t,u-d}$ and $\delta_{F}^{m_\pi,u-d}$, all with $F=A,B,C$, and the polarized momentum fraction
in the chiral limit, $\langle x\rangle^{0}_{\Delta u-\Delta d}$. 
The latter was fixed to $\langle x\rangle^{0}_{\Delta u-\Delta d}=0.17$ as obtained from the
self-consistently improved chiral extrapolation in Fig.~\ref{Deltax_LHPC} \cite{Edwards:2006qx}.
Explicit expressions for the functions $f_A^{u-d}(m_\pi)$, $h_F(t,m_\pi)$ and $j_A^{u-d}(m_\pi)$, 
which contain the non-analytic dependences on the momentum transfer squared and the pion mass, can
be found in \cite{Dorati:2007bk}. 
It turns out that the functions
$h_F(t,m_\pi)$ are practically independent of $t$ in the accessible ranges of $t$ and $m_\pi$.
The counter terms $\delta_{F}^{t,u-d}$ and $\delta_{F}^{m_\pi,u-d}$ for $F=B,C$
included in the fit are formally of $\mathcal{O}(p^3)$ and therefore represent parts
of higher order corrections. Since the extraction of the GFFs was based on 
the parametrization in Eq.~\ref{EMT1} using the lattice values for the nucleon mass $m_N$,
the ratio $m_N(m_\pi)/m^0_N$ of the pion mass dependent nucleon mass to
the nucleon mass in the chiral limit, which has been fixed to
$m^0_N=0.89\GeV$, appears in Eqs.~(\ref{ChPTB20umdp4}) and (\ref{ChPTC20umdp4}).

We note in particular that Eqs.~(\ref{ChPTA20umdp4},\ref{ChPTB20umdp4}) and (\ref{ChPTC20umdp4})
depend simultaneously on the momentum fraction in the isovector channel in the chiral limit,
$A_{20}^{0,u-d}=\langle x\rangle^{0}_{u-d}$. For this reason, a global simultaneous chiral
fit to the $t$- and $m_\pi$-dependence of the GFFs $A^{u-d}_{20}(t)$, $B^{u-d}_{20}(t)$ and $C^{u-d}_{20}(t)$
was performed including nine free parameters: The LECs $F_{20}^{0,u-d}$, and the counter-terms 
$\delta_{F}^{t,u-d}$ and $\delta_{F}^{m_\pi,u-d}$, for $F=A,B,C$.
To improve the statistics, all lattice results for $-t<0.48\GeV^2$ and $m^2_\pi\le500^2\MeV^2$
were included in the fit, amounting to more than 120 lattice data points. 
The maximal values for the pion mass and the momentum transfer that were taken into
account in the chiral fit appear to be rather large and are probably beyond 
the commonly anticipated region of applicability of a $\mathcal{O}(p^2)$-CBChPT calculation.
The stability of the fit has already been checked in \cite{Hagler:2007xi} to some extent 
by lowering the maximal values of included pion masses down to $500\MeV$, and a consistent description
of $A^{u-d}_{20}(t=0)$ within errors was obtained in all studied cases. Still, it would
be important to investigate the issue of convergence of the chiral expansions
of all three GFFs in greater detail in future studies.

\begin{figure}[t]
    \begin{minipage}{0.48\textwidth}
      \centering
          \includegraphics[angle=0,width=0.99\textwidth,clip=true]{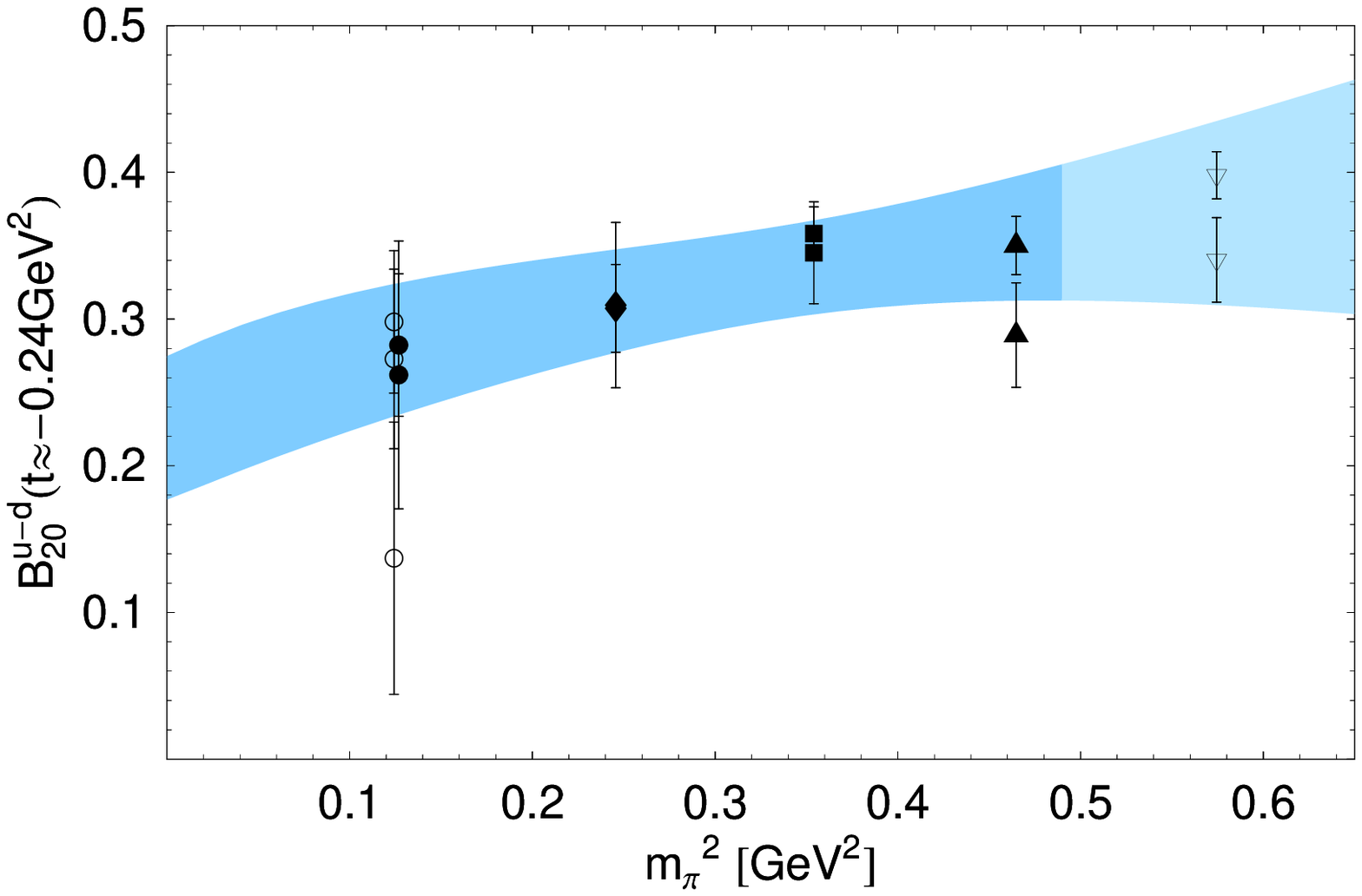}
  \caption{Pion mass dependence and chiral extrapolation of $B_{20}^{u-d}(t\simeq0.24\GeV^2)$ (from \cite{Hagler:2007xi}).}
  \label{B20_umd_tp24_SimulFit}
     \end{minipage}
          \hspace{0.3cm}
    \begin{minipage}{0.48\textwidth}
      \centering
          \includegraphics[angle=0,width=0.99\textwidth,clip=true]{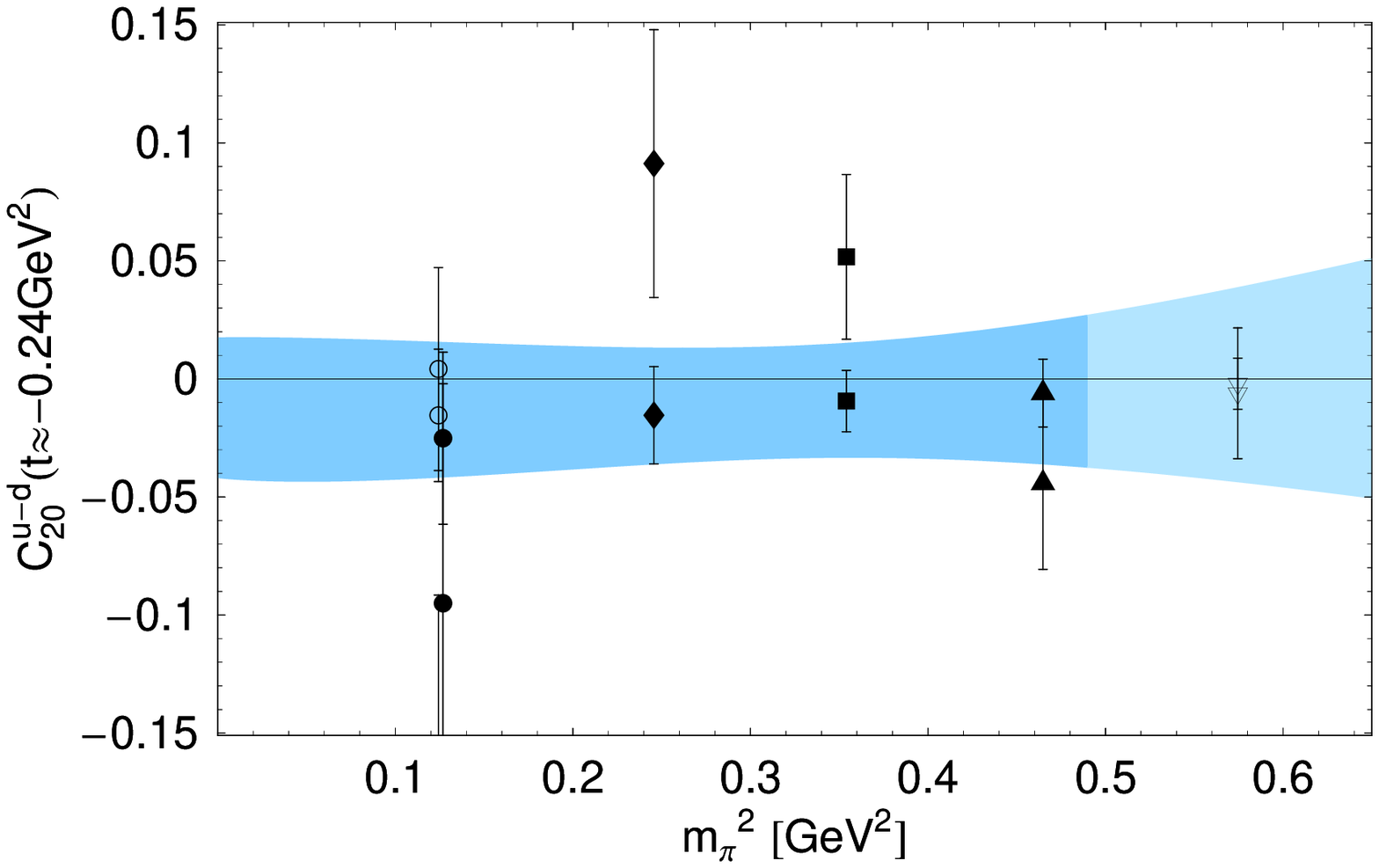}
  \caption{Pion mass dependence and chiral extrapolation of $C_{20}^{u-d}(t\simeq0.24\GeV^2)$ (from \cite{Hagler:2007xi}).}
  \label{C20_umd_tp24_SimulFit}
     \end{minipage}
\end{figure}

The result of the global simultaneous chiral fit just described 
is represented by the shaded bands in
Figs.~\ref{A20_umd_SimulFit}, \ref{A20_umd_mPip35_SimulFit}, 
\ref{B20_umd_tp24_SimulFit} and \ref{C20_umd_tp24_SimulFit}.
A value of $A_{20}^{u-d}(\t0)=\langle x\rangle_{ u- d}=0.157(10)$
was obtained at the physical point, in very good agreement
with results from global PDF-analyses, e.g. the MRST and CTEQ
PDF parametrizations, $\langle x\rangle^{\text{CTEQ6.6}}_{u-d}=0.154$ 
and $\langle x\rangle^{\text{MRST06}}_{u-d}=0.158$ 
\cite{Martin:2007bv,Nadolsky:2008zw}, 
represented by the star in Fig.~\ref{A20_umd_SimulFit}.

The dotted line in Fig.~\ref{A20_umd_SimulFit} represents the heavy baryon limit,
$m_N\rightarrow\infty$, of the CBChPT fit. If the CBChPT fit describes the
physics correctly, then the early breakdown of the heavy-baryon limit curve
strongly suggest that the tower of terms $\propto(m_\pi/m_N)^n$
included in the covariant approach is essential for a 
successful description of the lattice data beyond the physical point.
However, it is interesting to note that a direct fit of the 
corresponding HBChPT formula for $\langle x\rangle_{ u-d}$ \cite{Arndt:2001ye,Chen:2001eg}
to the lattice data for $m_\pi<500\MeV$, as represented by the dashed line,
misses the value from phenomenology\&experiment only by about $10\%$.

%
%

Figure \ref{A20_umd_mPip35_SimulFit} shows the lattice results for $A_{20}^{u-d}(t)$
at a pion mass of $\approx355\MeV$, together with the chiral extrapolation,
versus the momentum transfer squared. Apparently, the almost linear
dependence on $t$ obtained from the chiral fit is in good agreement
with the lattice results for $-t<1\GeV^2$ within statistical errors.

The pion mass dependences of the GFFs $B_{20}^{u-d}(t)$ and $C_{20}^{u-d}(t)$ 
for fixed $t\simeq-0.24\GeV^2$ are displayed in Figs.~\ref{B20_umd_tp24_SimulFit}
and \ref{C20_umd_tp24_SimulFit}, respectively. 
In both cases, the lattice data points are described reasonably well by the CBChPT fit, 
albeit the rather large statistical errors preclude any strong conclusions.
A clearly non-zero value is obtained for $B_{20}^{u-d}$ in the
forward limit at the physical pion mass, $B_{20}^{u-d}(\t0)=0.273(63)$,
while the chiral fit gives a value for $C_{20}^{u-d}$ that is compatible 
with zero within errors over the full range of accessible pion masses
and values of the momentum transfer, 
with $C_{20}^{u-d}(\t0)=-0.017(41)$ at $m_\pi=0.140\MeV$.
%
%

\begin{figure}[t]
    \begin{minipage}{0.48\textwidth}
      \centering
          \includegraphics[angle=-90,width=0.9\textwidth,clip=true]{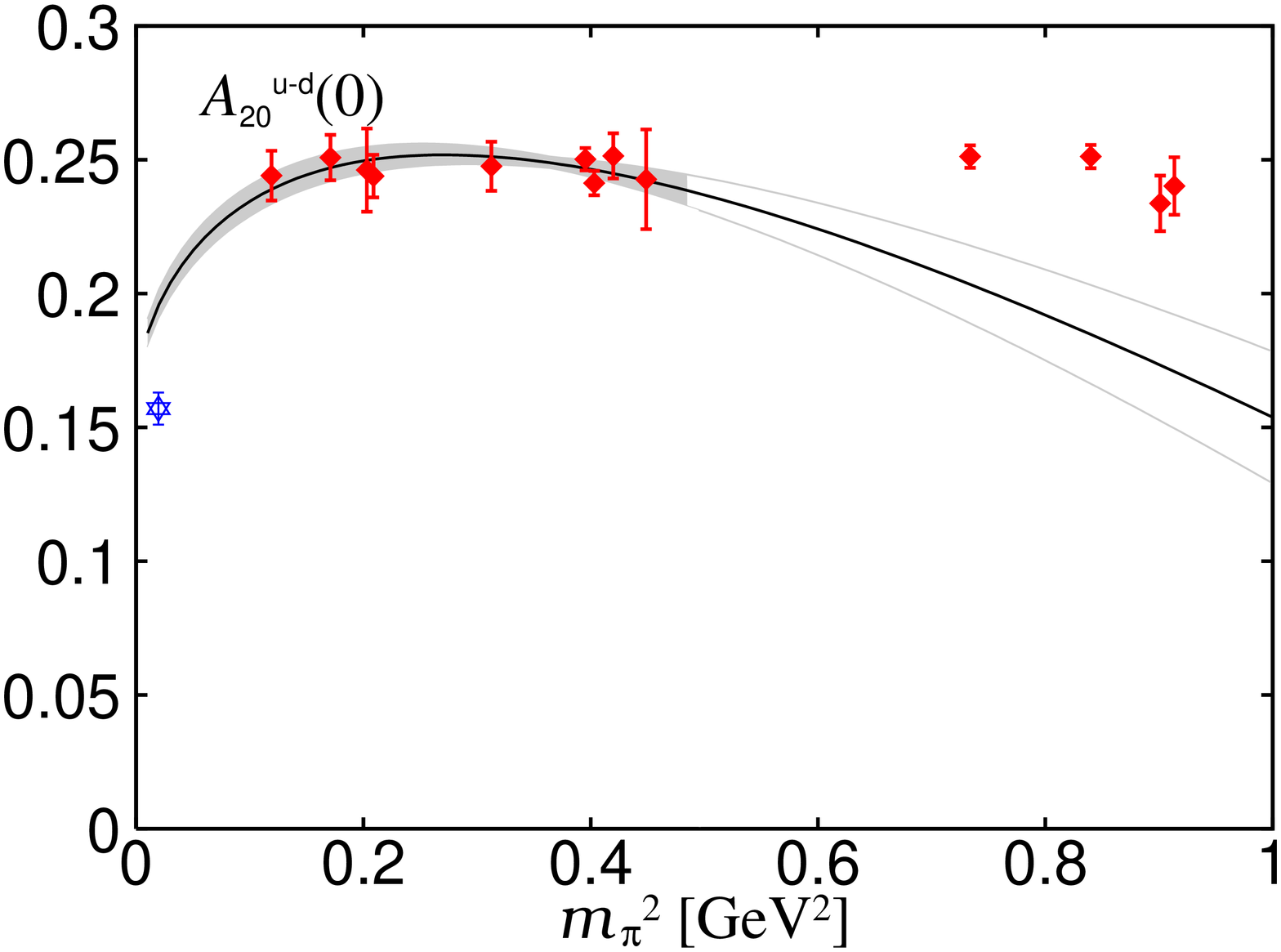}
  \caption{Pion mass dependence and chiral extrapolation of 
  $\langle x\rangle_{u-d}=A_{20}^{u-d}(\t0)$ (from proceedings \cite{Brommel:2007sb}).}
  \label{xv_QCDSF07}
     \end{minipage}
          \hspace{0.5cm}
    \begin{minipage}{0.48\textwidth}
      \centering
          \includegraphics[angle=-90,width=0.9\textwidth,clip=true]{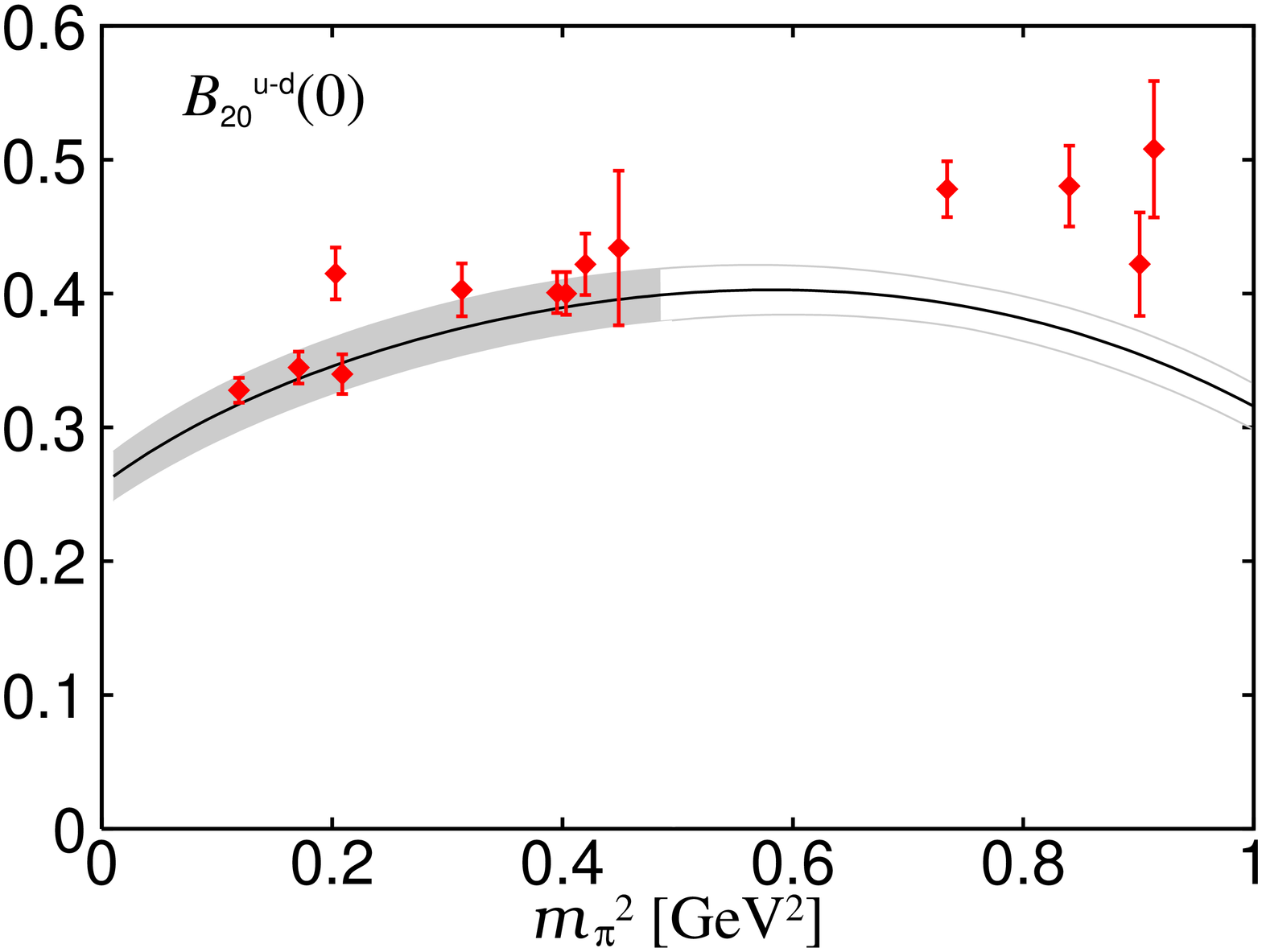}
  \caption{Pion mass dependence and chiral extrapolation of  $B_{20}^{u-d}(\t0)$ (from proceedings \cite{Brommel:2007sb}).}
  \label{b20v_QCDSF07}
     \end{minipage}
\end{figure}

Corresponding results for $A^{u-d}_{20}(\t0)$ and $B^{u-d}_{20}(\t0)$ 
from the QCDSF collaboration for $n_f=2$ flavors of
clover-improved Wilson fermions are presented in Figs.~\ref{xv_QCDSF07}
and \ref{b20v_QCDSF07} as functions of the squared pion mass \cite{Brommel:2007sb}.
The shaded band in Fig.~\ref{xv_QCDSF07} 
represents a chiral fit based on Eq.~(\ref{ChPTA20umdp4}) at fixed $t=0$
with two parameters $A_{20}^{0,u-d}=\langle x\rangle^{0}_{ u- d}$ and $\delta_{A}^{m_\pi,u-d}$
to the lattice data points for $m_\pi\le700\MeV$.
Due to the higher normalization of the data points compared to 
Fig.~\ref{A20_umd_SimulFit}, a value of 
$A_{20}^{u-d}(\t0)=\langle x\rangle_{u-d}=0.198(8)$ was found at the 
physical pion mass, which is $\approx25\%$ larger than the 
phenomenological PDF-parametrizations would suggest.

Employing Eq.~(\ref{ChPTB20umdp4}), a chiral fit
with three parameters $B_{20}^{0,u-d}$, $\delta_{B}^{t,u-d}$ 
and $\delta_{B}^{m_\pi,u-d}$ was performed to the lattice data for $B^{u-d}_{20}(t)$ 
for $m_\pi\le700\MeV$.
Since the GFFs $A^{u-d}_{20}$ and $B^{u-d}_{20}$  were not fitted simultaneously,
the LEC $A_{20}^{0,u-d}(\t0)$ in Eq.~(\ref{ChPTB20umdp4}) was fixed to the
value obtain from the fit in Fig.~\ref{xv_QCDSF07}.
Figure \ref{b20v_QCDSF07} shows the results of the fit for 
vanishing momentum transfer as a function of $m_\pi^2$ as shaded band, 
and the displayed data points are the result of an extrapolation of
the lattice data to $t=0$ using the fitting formula.
A value of $B^{u-d}_{20}(\t0)=0.269(20)$ was found at the physical point, 
which is in good agreement with the results from LHPC discussed above.

While the description of the lattice data points for 
$A^{u-d}_{20}(\t0)$ and $B^{u-d}_{20}(\t0)$ by the CBChPT extrapolations
in Figs.~\ref{xv_QCDSF07} and \ref{b20v_QCDSF07} is quite satisfactory
at lower pion masses, the fit bands start to bend down and thus do not
follow the trend of the lattice results for $m_\pi>700\MeV$. 
It would be interesting to investigate the stability of these 
chiral extrapolations by restricting the ranges of $m_\pi$ and $t$ included in the fit,
and by considering simultaneous global fits to all three
GFFs $A^{q}_{20}(t)$, $B^{q}_{20}(t)$ and $C^{q}_{20}(t)$ as discussed above.
%
%

\begin{figure}[t]
    \begin{minipage}{0.48\textwidth}
      \centering
          \includegraphics[angle=0,width=0.99\textwidth,clip=true]{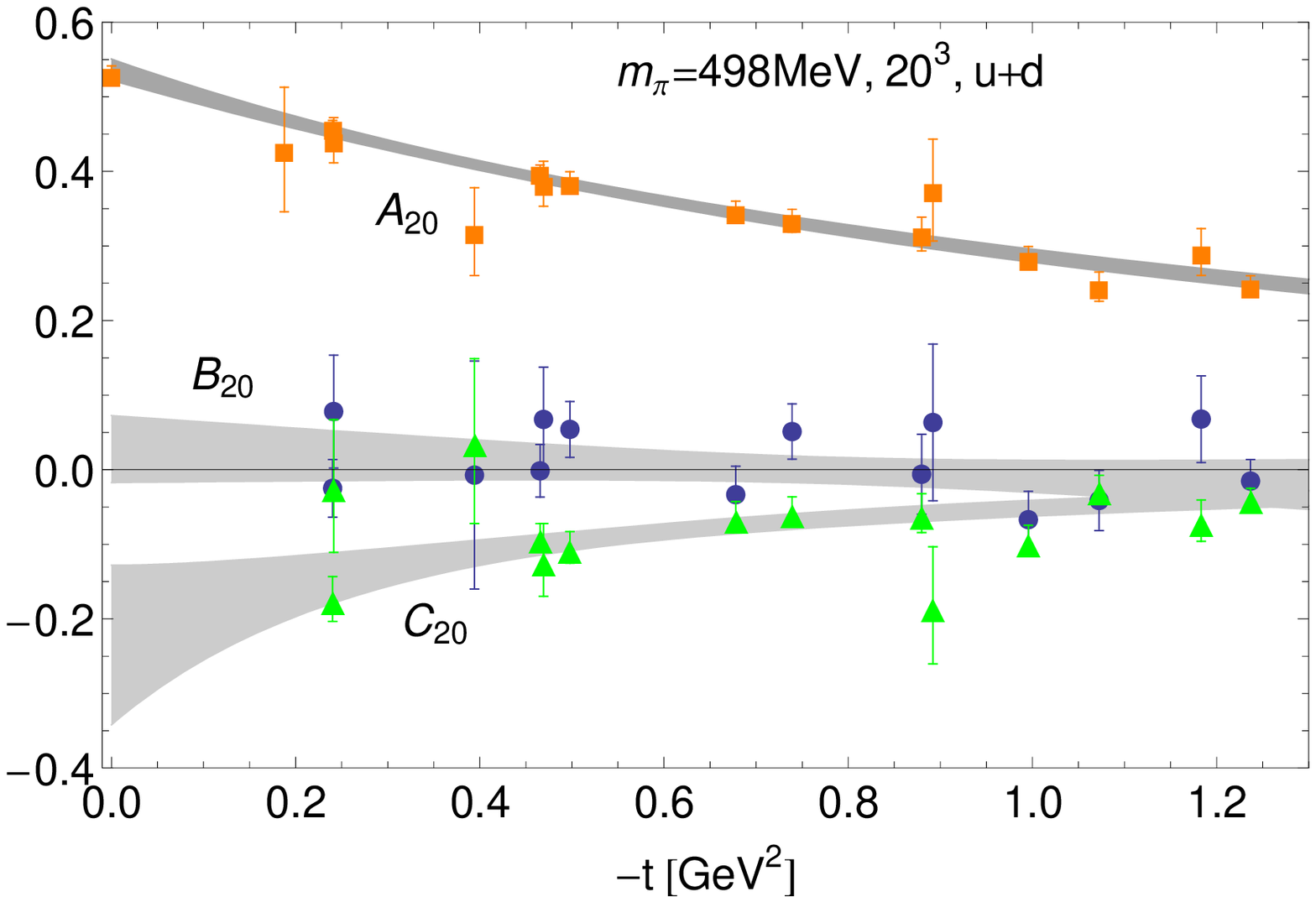}
  \caption{Connected contributions to the form factors of the energy momentum tensor in the isosinglet channel
   (from \cite{Hagler:2007xi}).}
  \label{ABCupd_LHPC}
     \end{minipage}
          \hspace{0.3cm}
    \begin{minipage}{0.48\textwidth}
      \centering
          \includegraphics[angle=0,width=0.99\textwidth,clip=true]{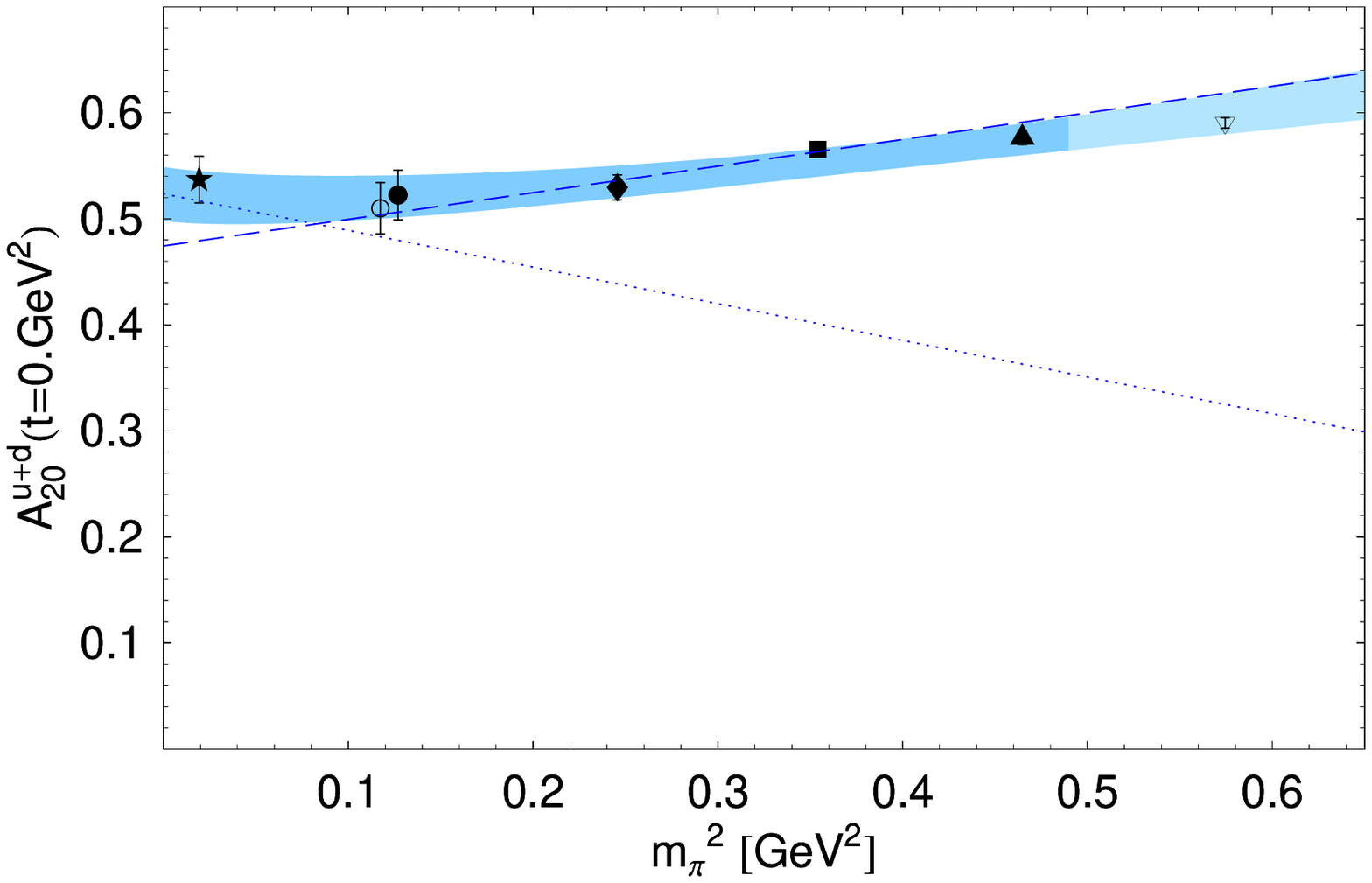}
  \caption{Pion mass dependence and chiral extrapolation of connected contributions to
  $\langle x\rangle_{u+d}=A_{20}^{u+d}(\t0)$ (from \cite{Hagler:2007xi}).}
  \label{A20_upd_SimulFit}
     \end{minipage}
\end{figure}

We now turn to a discussion of the energy momentum tensor form factors in the isosinglet channel.
All results presented in the remainder of this section correspond only to quark
line connected diagrams. Disconnected contributions,
which are required to obtain the full isosinglet results, were not included so far.
Exemplary results for $A^{u+d}_{20}(t)$, $B^{u+d}_{20}(t)$ and $C^{u+d}_{20}(t)$
from LHPC for a pion mass of $\approx498\MeV$, based on the same lattice approach and ensembles as 
discussed at the beginning of this section, is presented in Fig.~\ref{ABCupd_LHPC} \cite{Hagler:2007xi}. 
While the GFF $A^{u+d}_{20}(t)$ is rather large and positive over the full range of $t$,
the values for $C^{u+d}_{20}(t)$ are clearly negative (albeit with larger statistical fluctuations),
and one finds in particular that the GFF $B^{u+d}_{20}(t)$ is compatible with zero within statistical
errors for all accessible values of $t$.
In comparison with the isovector channel displayed in Fig.~\ref{ABCupd_LHPC},
we note 
that $A^{u+d}_{20}>A^{u-d}_{20}$, $B^{u-d}_{20}\gg B^{u+d}_{20}$,
and $|C^{u+d}_{20}|\gtrapprox |C^{u-d}_{20}|$. These results are in overall agreement
with counting rules for GPDs obtained in the framework of an expansion in $1/N_c$ \cite{Goeke:2001tz},
which predict that 
$(A^{u-d}_{20}/A^{u+d}_{20},B^{u+d}_{20}/B^{u-d}_{20},C^{u-d}_{20}/C^{u+d}_{20})\sim 1/N_c$.

The dependence on the pion mass and the momentum transfer of the
GFFs in the isosinglet channel in CBChPT \cite{Dorati:2007bk} is given by
\begin{equation}
A_{20}^{u+d}(t,m_\pi)=A_{20}^{0,u+d}\bigg(f^{u+d}_A(m_\pi) - \frac{g_A^2}{64 \pi^2f_\pi^2}h_A(t,m_\pi)\bigg)  + 
 \delta_{A}^{t,u+d}\,t + \delta_{A}^{m_\pi,u+d}\,m_\pi^2 + \Delta A_{20}^{u+d}(t,m_\pi)\,,
\label{ChPTA20umdp4new}
\end{equation}
\bea
B_{20}^{u+d}(t,m_\pi)&=&\frac{m_N(m_\pi)}{m_N} B_{20}^{0,u+d} + A_{20}^{0,u+d} h_B^{u+d}(t,m_\pi) 
+ \Delta B_{20}^{u+d}(t,m_\pi)\nonumber\\
&+& \frac{m_N(m_\pi)}{m_N}\bigg\{\delta_{B}^{t,u+d}\,t 
+ \delta_{B}^{m_\pi,u+d}\,m_\pi^2\bigg\} \,,
\label{ChPTB20updp4}
\eea
\bea
C_{20}^{u+d}(t,m_\pi)&=&\frac{m_N(m_\pi)}{m_N} C_{20}^{0,u+d} + A_{20}^{0,u+d} h_C^{u+d}(t,m_\pi)  + 
\Delta C_{20}^{u+d}(t,m_\pi) 
\nonumber\\
&+& \frac{m_N(m_\pi)}{m_N}\bigg\{\delta_{C}^{t,u+d}\,t 
+ \delta_{C}^{m_\pi,u+d}\,m_\pi^2\bigg\} \,,
\label{ChPTC20updp4}
\eea
where the functions $f_A^{u+d}(m_\pi)$, $h_F(t,m_\pi)$ and $\Delta F_{20}^{u+d}(t,m_\pi)$, with $F=A,B,C$,
contain the non-analytic dependences on the momentum transfer squared and the pion mass.
We note that the terms $\Delta F_{20}^{u+d}(t,m_\pi)$, $\delta_{B}^{t,u+d}\,t$, $\delta_{B}^{m_\pi,u+d}\,m_\pi^2$
represent only specific parts of the full $\mathcal{O}(p^3)$ corrections \cite{Dorati:2007bk}, and that the counter-terms
$\delta_{C}^{t,u+d}\,t$ and $\delta_{C}^{m_\pi,u+d}\,m_\pi^2$ formally 
appear only at $\mathcal{O}(p^4)$ in the chiral expansion and were added by hand in Eq.~(\ref{ChPTC20updp4}).
The chiral extrapolation formulas depend on the 
LECs $F_{20}^{0,u+d}=F_{20}^{u+d}(\t0,m_\pi\eql0)$ and the counter-terms 
$\delta_{F}^{t,u+d}$ and $\delta_{F}^{m_\pi,u+d}$. Furthermore, the $\mathcal{O}(p^3)$-correction 
terms $\Delta F_{20}^{u+d}(t,m_\pi)$ are all proportional to another LEC, 
the total momentum fraction of quarks in the pion in the chiral limit, $\langle x\rangle^{\pi,0}_{u+d}$,
which was fixed to $0.5$, as suggested by phenomenology and the lattice studies
mentioned in section \ref{sec:pionPDFs}.

Preliminary fits to the lattice data for $B_{20}^{u+d}$ turned out to be unstable
due to unreasonably large contributions from the term $\Delta B_{20}^{u+d}(t,m_\pi)$ in
Eq.~\ref{ChPTB20updp4}, which subsequently was excluded from the full chiral analysis \cite{Hagler:2007xi}.
Noting that $A_{20}^{0,u+d}=\langle x\rangle^{0}_{u+d}$ is a common LEC 
in the chiral expansions in Eqs.~(\ref{ChPTA20umdp4new},\ref{ChPTB20updp4},\ref{ChPTC20updp4}),
LHPC has performed a simultaneous global chiral fit to more than 120 lattice data points for
$A^{u+d}_{20}(t)$, $B^{u+d}_{20}(t)$ and $C^{u+d}_{20}(t)$, with $m^2_\pi<500\MeV^2$ and
$|t|<0.48\GeV^2$, including the 9 free parameters $\Delta F_{20}^{u+d}(t,m_\pi)$, 
$\delta_{F}^{t,u+d}\,t$ and $\delta_{F}^{m_\pi,u+d}\,m_\pi^2$.

Figure \ref{A20_upd_SimulFit} displays the lattice data from LHPC for $A^{u+d}_{20}(\t0)$
versus $m_\pi^2$ together with the result of the chiral fit represented by the shaded band. 
An upwards bending of the chiral extrapolation band is visible at lower pion masses,
which leads to a value of $A_{20}^{u+d}(\t0)=\langle x\rangle_{u+d}=0.520(24)$
at the physical pion mass that is in a good agreement with 
global PDF-parametrizations, $\langle x\rangle^{\text{MRST06}}_{u+d}=0.545$ 
and $\langle x\rangle^{\text{CTEQ6.6}}_{u+d}=0.535$
\cite{Martin:2007bv,Nadolsky:2008zw}.
It has been noted, however, that the term 
$\Delta A_{20}^{u+d}(t,m_\pi)$ in Eq.~\ref{ChPTA20umdp4new},
which represent only part of the full $\mathcal{O}(p^3)$ correction,
is mainly responsible for the observed upward bending. Furthermore,
the inclusion of disconnected contributions could not only change the 
normalization but also the pion mass and $t$-dependence of the lattice data, 
and therefore could lead to a different chiral fit. 
Hence the extrapolation in Fig.~\ref{A20_upd_SimulFit} and in particular
the good agreement with the PDF parametrizations should be considered 
with great caution, at least until the complete $\mathcal{O}(p^3)$ ChPT results 
become available and can be used in fits to lattice data, including 
contributions from disconnected diagrams.

\begin{figure}[t]
    \begin{minipage}{0.48\textwidth}
      \centering
          \includegraphics[angle=0,width=0.99\textwidth,clip=true]{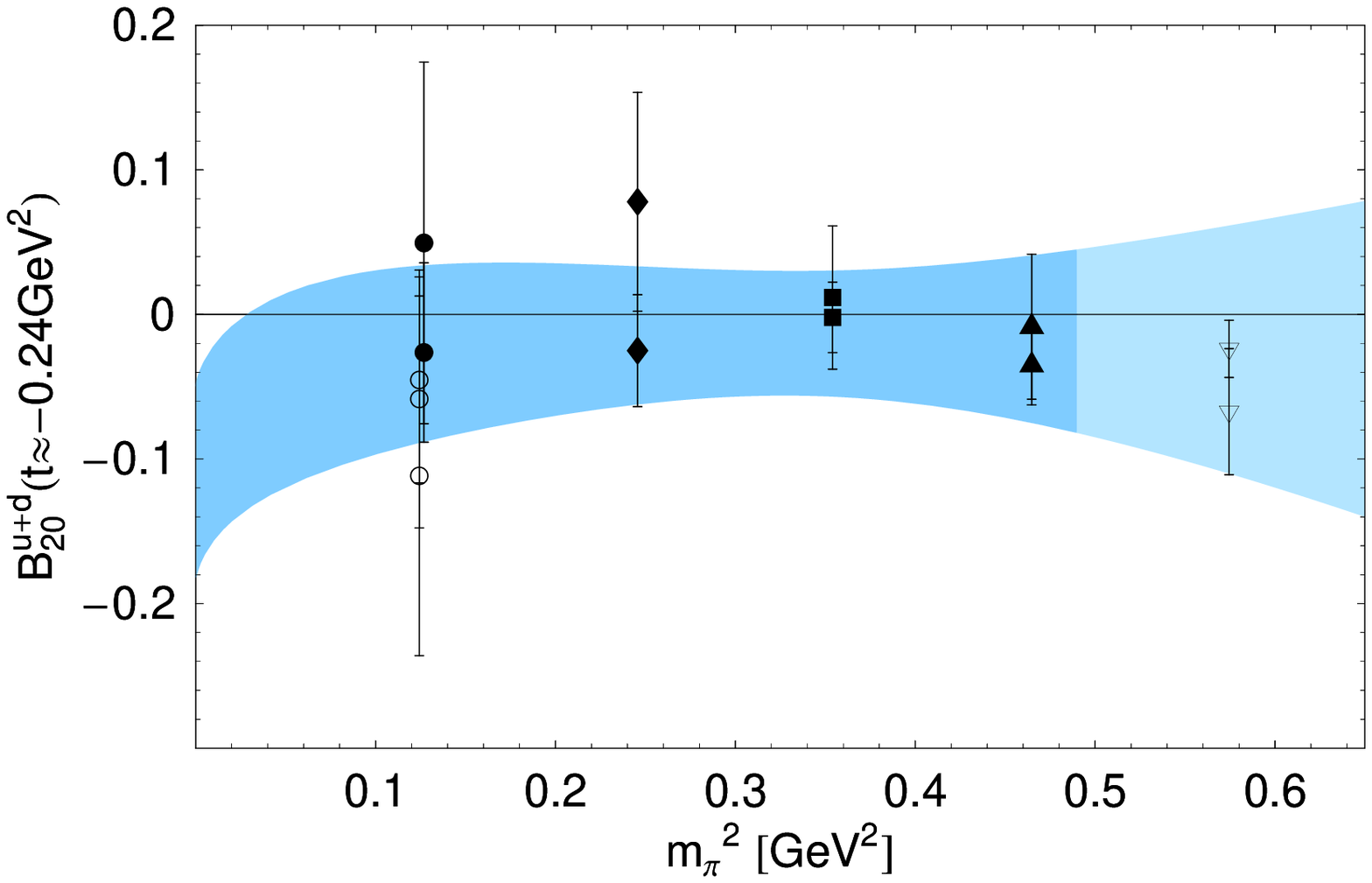}
  \caption{Pion mass dependence and chiral extrapolation of connected contributions to $B_{20}^{u+d}(t\simeq-0.24\GeV^2)$  (from \cite{Hagler:2007xi}).}
  \label{B20_upd_tp24_SimulFit}
     \end{minipage}
          \hspace{0.3cm}
    \begin{minipage}{0.48\textwidth}
      \centering
          \includegraphics[angle=0,width=0.99\textwidth,clip=true]{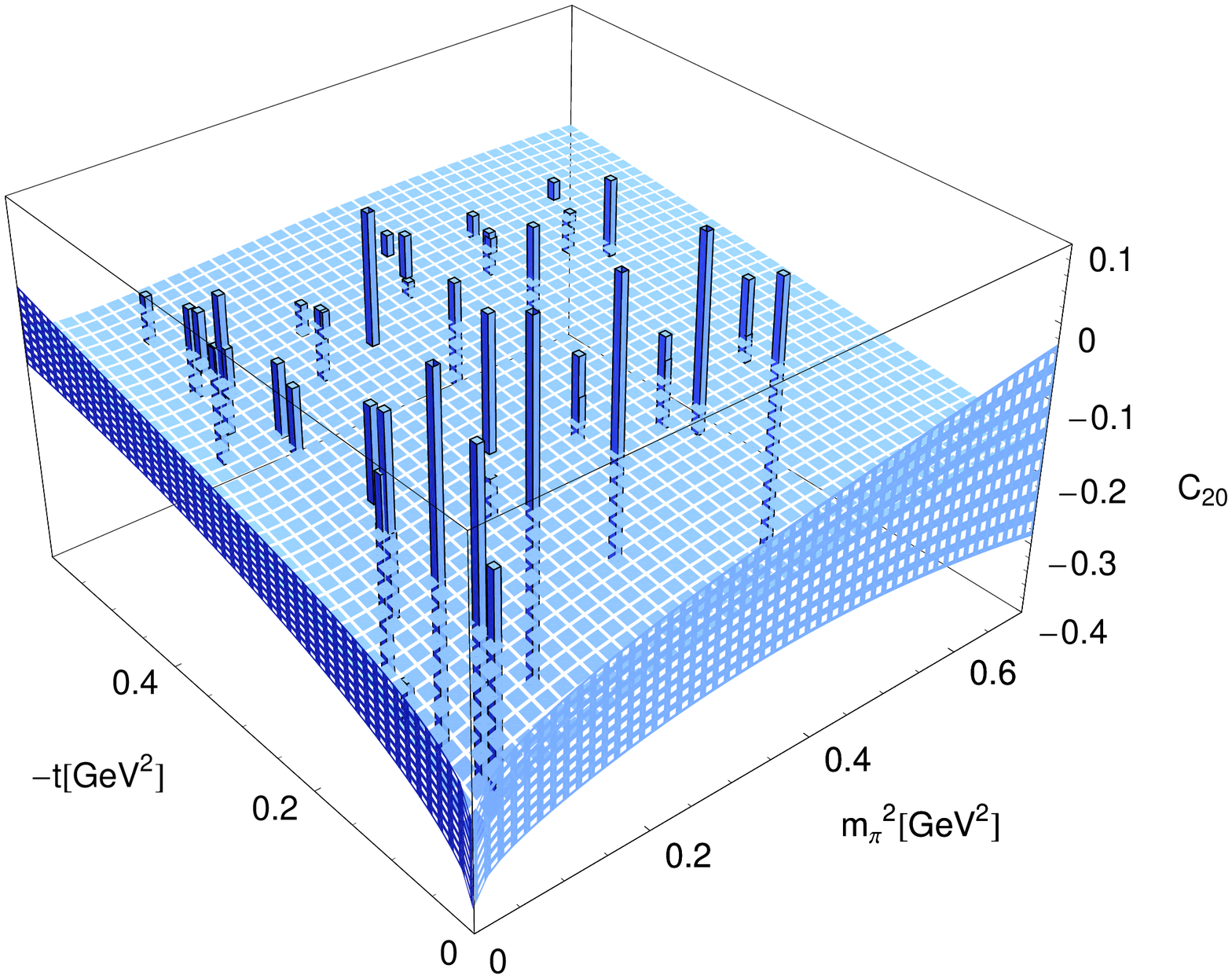}
  \caption{Pion mass and $t$-dependence together with a chiral extrapolation of connected contributions 
  to $C_{20}^{u+d}(t)$ (from \cite{Hagler:2007xi}).}
  \label{C20_upd_3d_SimulFit}
     \end{minipage}
\end{figure}

Similar concerns apply to the results for the GFF $B_{20}^{u+d}$, which
are shown in Fig.~\ref{B20_upd_tp24_SimulFit} for fixed $t\simeq-0.24\GeV^2$
versus $m_\pi^2$. The lattice data points are well described 
by the chiral extrapolation band, which bends down towards
negative values at small pion masses. In the forward limit, the fit gives 
$B^{u+d}_{20}(\t0)=-0.140(84)$ at the physical pion mass. 
As before, the full $\mathcal{O}(p^3)$ corrections on the side
of ChPT and disconnected contributions on the side of the lattice calculation
must be included before this can be regarded as a solid result.
%
%
%

Figure \ref{C20_upd_3d_SimulFit} displays the simultaneous pion mass 
and $t$-dependence of the lattice results for the GFF $C_{20}^{u+d}$, together with 
the simultaneous global chiral fit. The lattice data points
and their statistical errors are represented by the elongated cuboids,
the central result from the chiral extrapolation is given by the surface, and the
statistical errors from the fit are illustrated by the error bands
at $t=0$ and $m_\pi=0$.
From the chiral fit, a sizeable negative value of $C^{u-d}_{20}(\t0)=-0.267(62)$
was found at the physical pion mass. Similar comments as made above for 
$A^{u+d}_{20}$ and $B^{u+d}_{20}$ regarding the
reliability of the chiral extrapolation apply also in this case.

\begin{figure}[t]
    \begin{minipage}{0.48\textwidth}
      \centering
          \includegraphics[angle=-90,width=0.9\textwidth,clip=true]{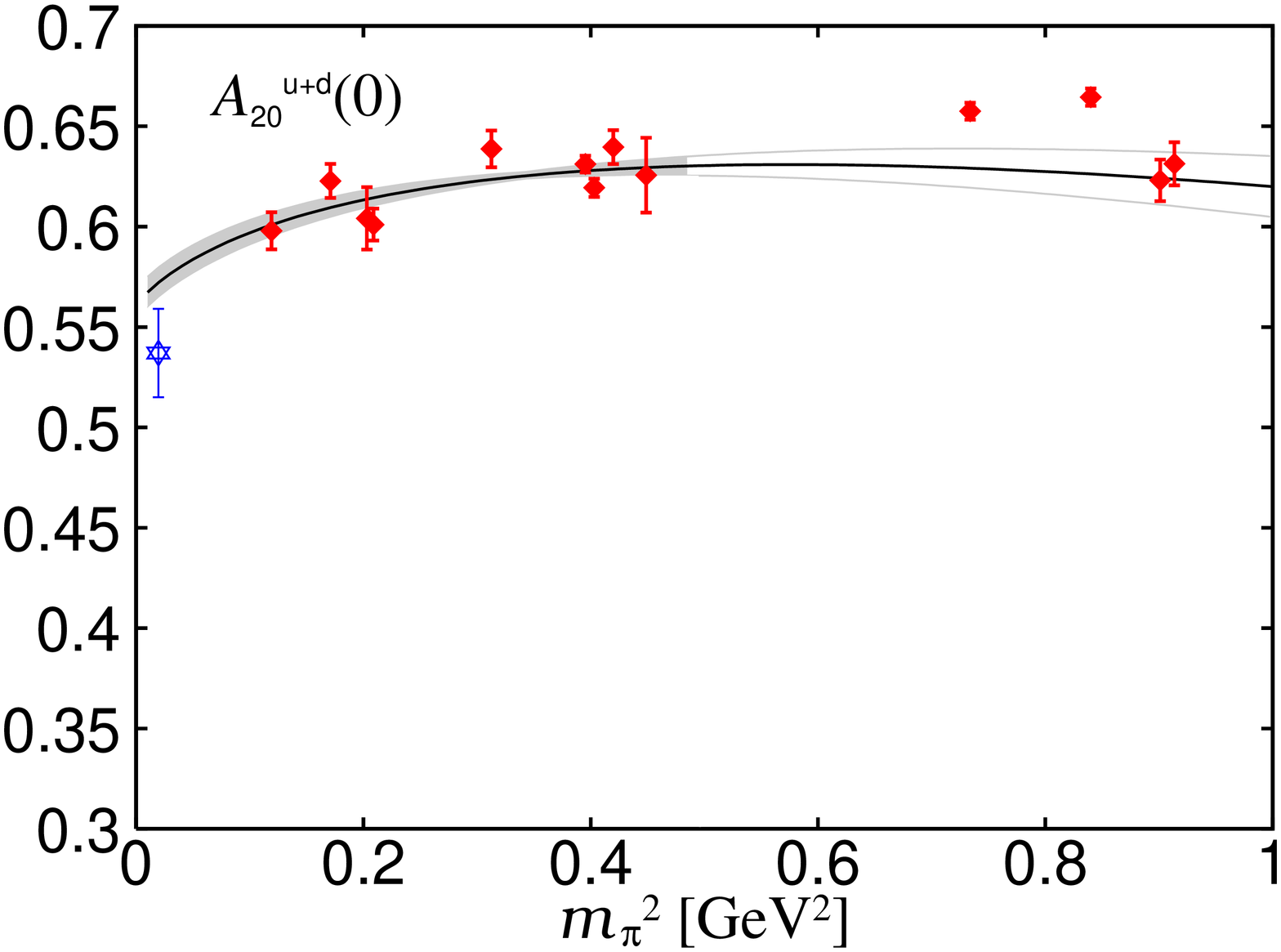}
  \caption{Pion mass dependence and chiral extrapolation of connected contributions to
  $\langle x\rangle_{u+d}=A_{20}^{u+d}(\t0)$  (from proceedings \cite{Brommel:2007sb}).}
  \label{xs_QCDSF07}
     \end{minipage}
          \hspace{0.5cm}
    \begin{minipage}{0.48\textwidth}
      \centering
          \includegraphics[angle=-90,width=0.9\textwidth,clip=true]{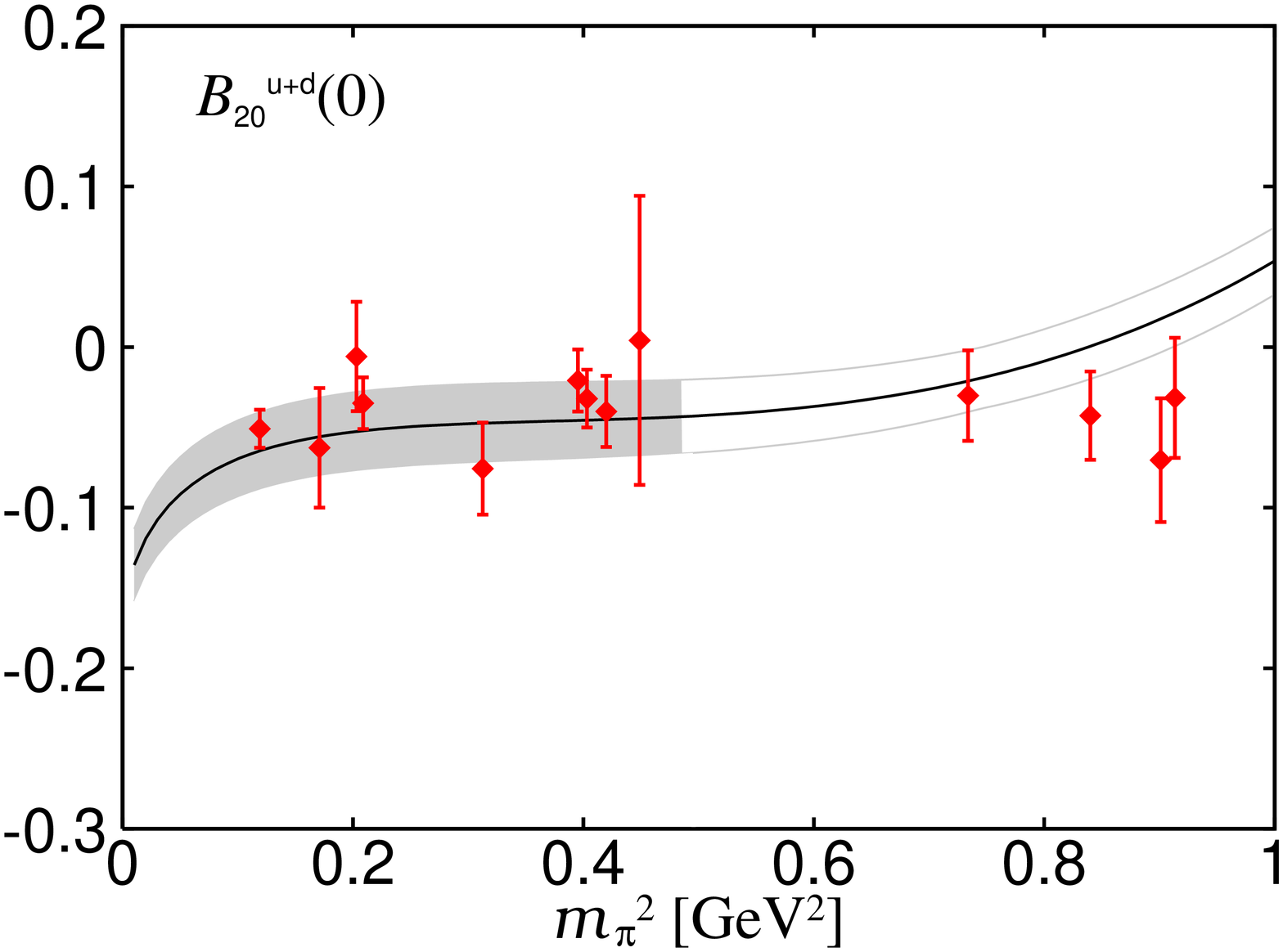}
  \caption{Pion mass dependence and chiral extrapolation of connected contributions to $B_{20}^{u+d}(\t0)$ 
  (from proceedings \cite{Brommel:2007sb}).}
  \label{b20s_QCDSF07}
     \end{minipage}
\end{figure}

Results from QCDSF for the isosinglet GFFs $A^{u+d}_{20}$ and $B^{u+d}_{20}$
are presented in Figs.~\ref{xs_QCDSF07} and \ref{b20s_QCDSF07}, based on the 
same simulations and ensembles as discussed above in relation with
Figs.~\ref{xv_QCDSF07} and \ref{b20v_QCDSF07} \cite{Brommel:2007sb}. 
As in the case of the LHPC simulations,
contributions from disconnected diagrams were not yet included.
Notably, the normalization of the lattice data points 
for $A^{u+d}_{20}$ in Fig.~\ref{xs_QCDSF07} is $\approx10\%$ higher
than of the corresponding results from LHPC in Fig.~\ref{A20_upd_SimulFit}.
A chiral fit based on Eq.~\ref{ChPTA20umdp4new} at fixed $t=0$
and \emph{excluding} the $\mathcal{O}(p^3)$ correction term $\Delta A_{20}^{u+d}(t,m_\pi)$,
with two free parameters $A_{20}^{0,u+d}=\langle x\rangle^{0}_{u+d}$ and 
$\delta_{B}^{m_\pi,u+d}$, is represented by the solid lines and the shaded band.
The chiral fit smoothly bends down as $m_\pi\rightarrow 0$,
giving $A_{20}^{u+d}=\langle x\rangle_{u+d}=0.572(12)$
at the physical point, which is approximately $6\%$ above the 
values from phenomenological PDF parametrizations.
Since the term $\Delta A_{20}^{u+d}(t,m_\pi)$
was not included in the fit, there is no upwards bending visible
at low pion masses as seen in Fig.~\ref{A20_upd_SimulFit}. This confirms
that the full $\mathcal{O}(p^3)$ corrections have to be worked out
in CBChPT and applied to the lattice results before any definite
conclusions can be drawn. 
%
%
%

Equation (\ref{ChPTB20updp4}), with $\Delta B_{20}^{u+d}(t,m_\pi)$ set to zero, 
was used to fit the lattice data for $B^{u+d}_{20}(t)$, 
treating $B^{0,u+d}_{20}$, $\delta_{B}^{t,u+d}$ and $\delta_{B}^{m_\pi,u+d}$ as free parameters,
and setting the LEC $A_{20}^{0,u+d}=\langle x\rangle^{0}_{u+d}$ to the value obtained from the fit 
in Fig.~\ref{xs_QCDSF07}. 
The result is shown as shaded band in Fig.~\ref{b20s_QCDSF07}, 
where the lattice data points have already been extrapolated to $t=0$ at fixed $m_\pi$ based on the chiral fit. 
From the chiral extrapolation, a value of $B^{u+d}_{20}(\t0)=-0.120(23)$ was obtained at the physical
pion mass \cite{Brommel:2007sb}, 
in good agreement with the result from LHPC based on the simultaneous global fit discussed above.
Since contributions from disconnected diagrams and the full $\mathcal{O}(p^3)$ CBChPT corrections
were not included, these results should be considered with appropriate caution. 
%
%

\subsubsection{Nucleon spin structure}
\label{sec:SpinStructure}
A central role in our understanding of the spin structure of the nucleon
is played by the total quark spin contribution $\Delta\Sigma=\sum_{q=u,d,s}\Delta q$, cf. Eq.~(\ref{SpinSumrule2}).
Early experimental results by EMC suggested a very small value (albeit with rather large errors) of 
$\Delta\Sigma\simeq0.098\pm0.136$ \cite{Ashman:1989ig} at a scale of $\approx10\GeV^2$, 
which triggered the famous ``proton spin crisis''.
From a recent precise measurement of the spin structure functions at HERMES \cite{Airapetian:2007aa},
a value of $\Delta\Sigma=\sum_{q=u,d,s}\Delta q=0.330(11)_{\text{th}}(25)_{\text{exp}}(28)_{\text{evo}}$
has been extracted for the $\MSbar$ scheme at a scale of $5\GeV^2$.
These results strongly indicate that the major fraction of the total nucleon spin is indeed carried
by the gluon spin and the orbital angular momentum contributions of quarks and gluons. 

\begin{figure}[t]
    \begin{minipage}{0.48\textwidth}
      \centering
          \includegraphics[angle=-90,width=0.9\textwidth,clip=true]{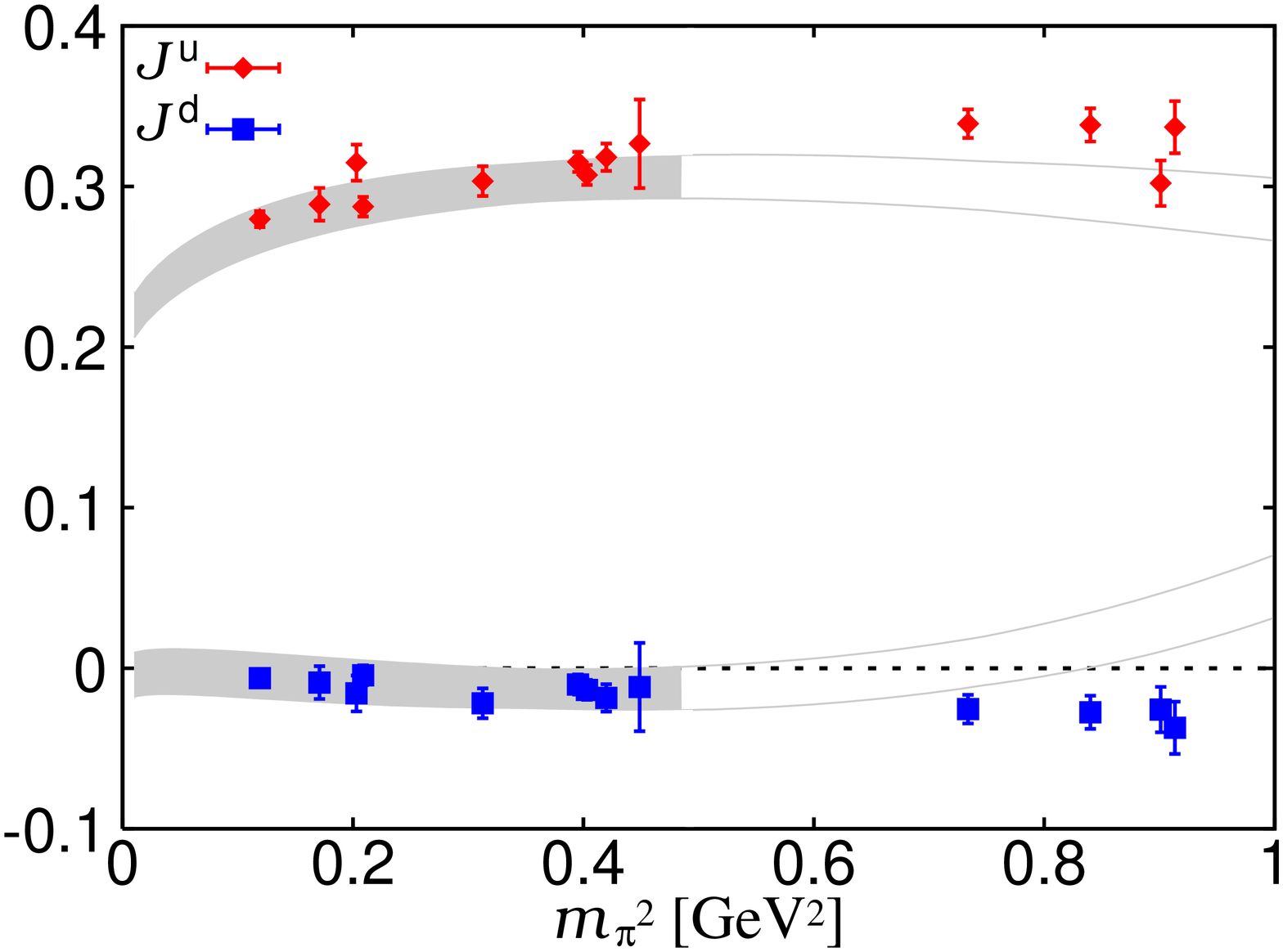}
  \caption{Angular momentum contributions to the nucleon spin (from proceedings \cite{Brommel:2007sb}).}
  \label{Jq_QCDSF07_Ohtani}
     \end{minipage}
          \hspace{0.5cm}
   \begin{minipage}{0.48\textwidth}
      \centering
          \includegraphics[angle=-90,width=0.9\textwidth,clip=true,angle=0]{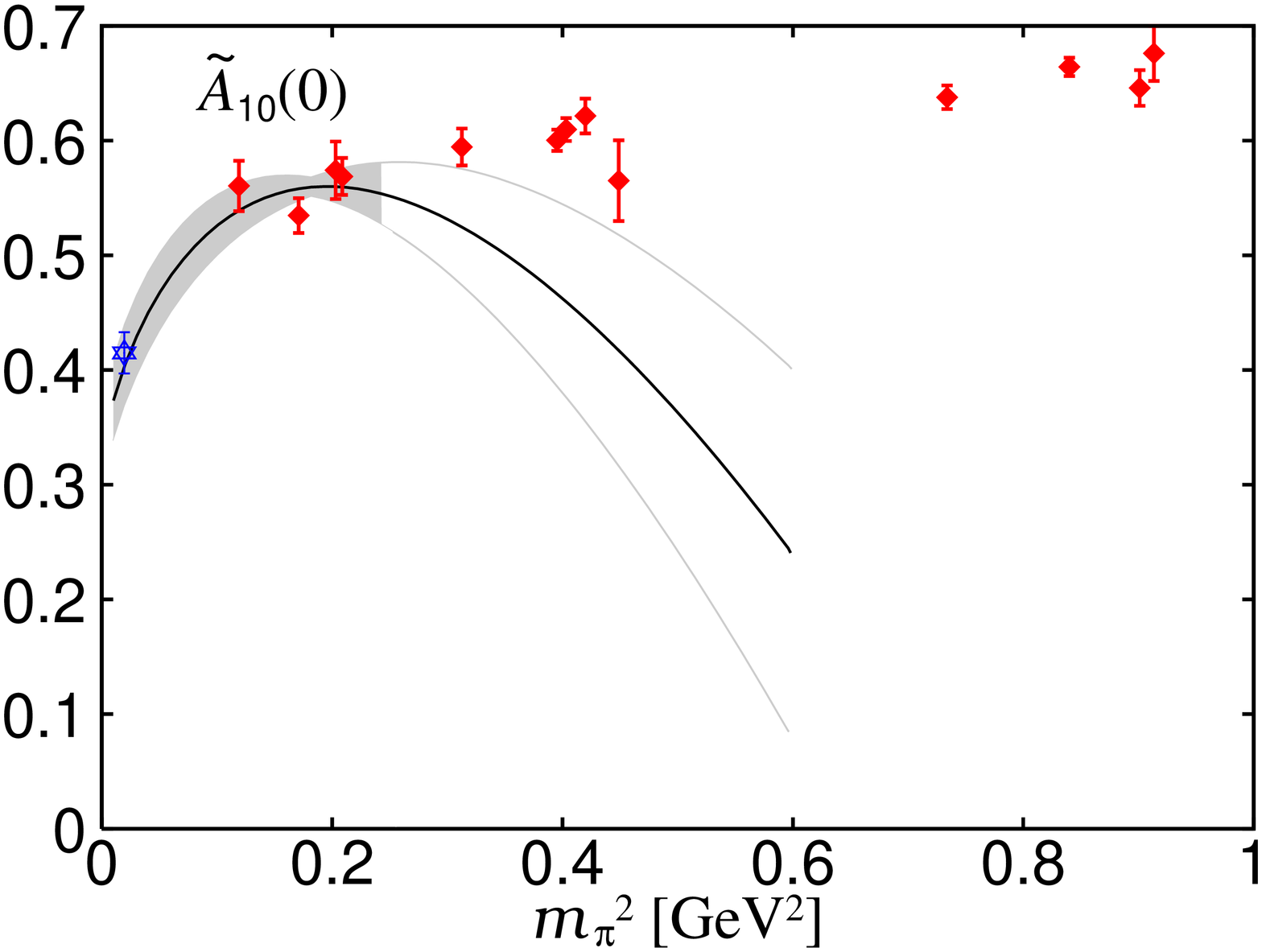}
  \caption{Isosinglet quark spin fraction $\Delta\Sigma^{u+d}$ (from proceedings \cite{Brommel:2007sb}).}
  \label{DxS_QCDSF07_Ohtani}     
\end{minipage}
\end{figure}

\begin{figure}[t]
    \begin{minipage}{0.48\textwidth}
      \centering
          \includegraphics[angle=-90,width=0.9\textwidth,clip=true]{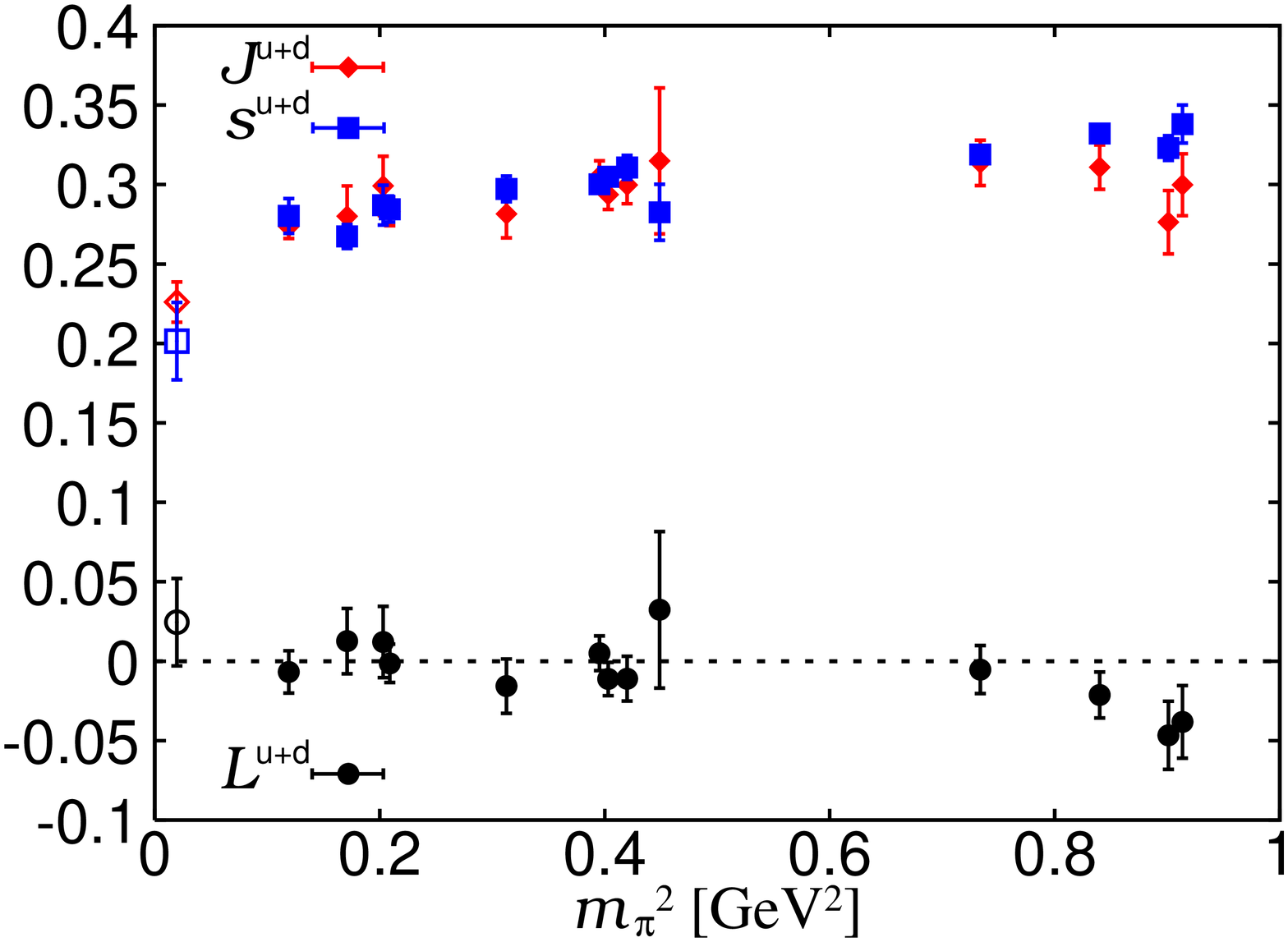}
  \caption{Spin and orbital momentum contributions to the nucleon spin (from proceedings \cite{Brommel:2007sb}).}
  \label{JsL_QCDSF07_Ohtani}
     \end{minipage}
          \hspace{0.1cm}
    \begin{minipage}{0.48\textwidth}
      \centering
          \includegraphics[angle=0,width=0.99\textwidth,clip=true]{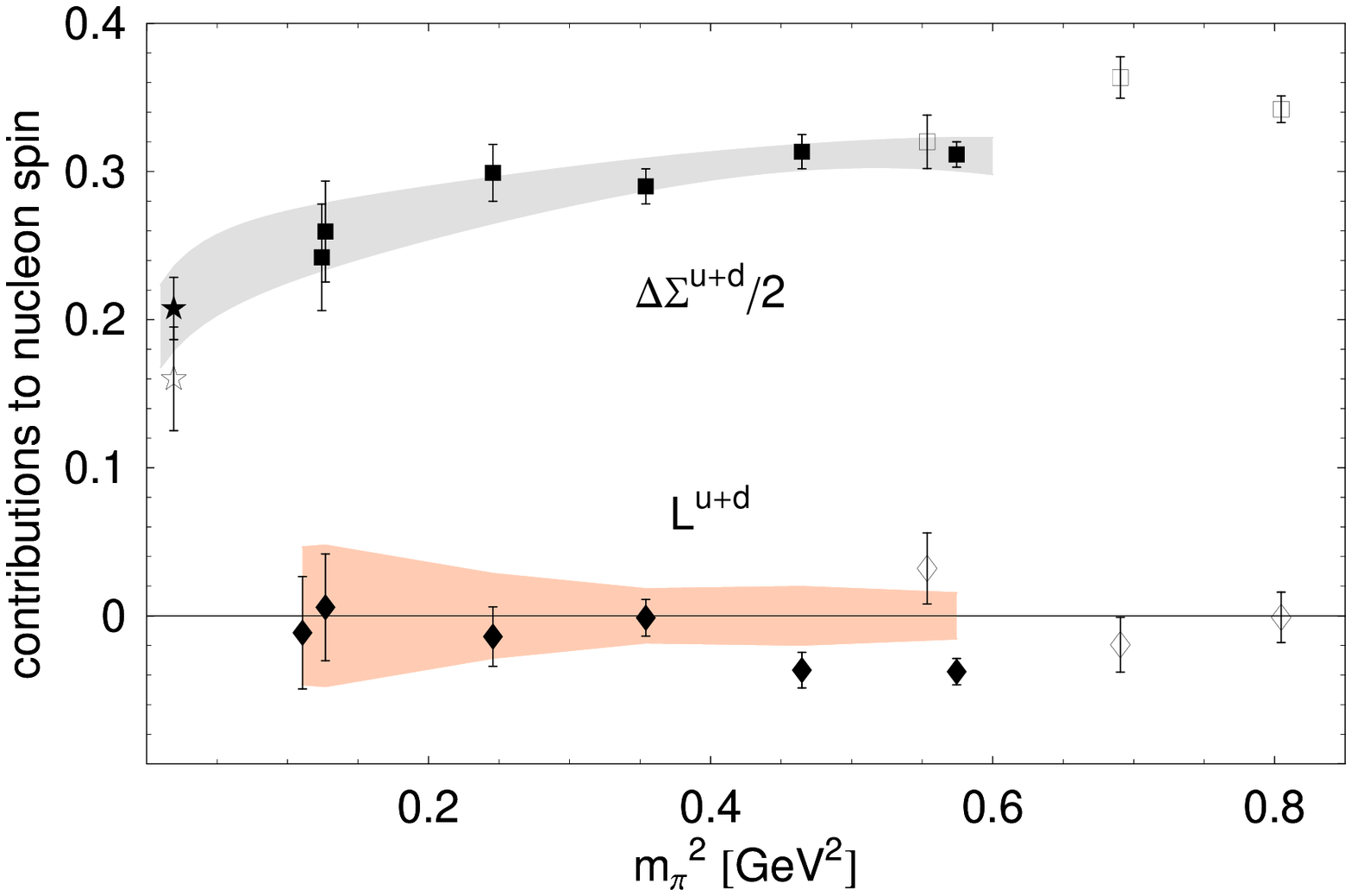}
  \caption{Spin and orbital angular momentum contributions to the nucleon spin (from \cite{Hagler:2007xi}).}
  \label{Lq_Sq_u_plus_d_vs_mPi2_v2_LHPC}
     \end{minipage}
\end{figure}

\begin{figure}[t]
      \centering
          \includegraphics[angle=0,width=0.55\textwidth,clip=true]{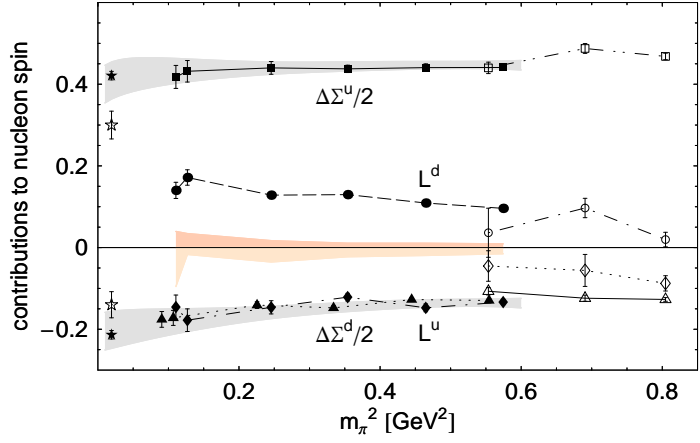}
  \caption{Contributions to the nucleon spin (from \cite{Hagler:2007xi}).}
  \label{Lq_Sq_u_d_vs_mPi2_v2_LHPC}
\end{figure}

Pioneering studies of $\Delta\Sigma$ in quenched lattice QCD were presented in \cite{Fukugita:1994fh,Dong:1995rx}.
In both cases, perturbatively renormalized operators were employed, and
values of $\Delta\Sigma=0.18(10)$ \cite{Fukugita:1994fh} and $\Delta\Sigma=0.25(12)$ \cite{Dong:1995rx} (see also Table \ref{tabMathur} below)
were cited at lattice scales of $\mu=1/a\approx1.4\GeV$, \emph{including} estimates of the disconnected contributions for $q=u,d,s$.
A similar value of $\Delta\Sigma=0.20(12)$ was presented in \cite{Gusken:1999as},
for $n_f=2$ flavors of Wilson fermions and the Wilson gauge action, and based on a perturbatively
renormalized operator. As in the earlier quenched studies, estimates of disconnect contributions 
using stochastic noise methods were also included in this case.
More recent results for the connected contributions to $\Delta\Sigma^{u+d}$ from unquenched lattice QCD will
be presented and discussed further below in this section, cf. Figs.~\ref{DxS_QCDSF07_Ohtani} to \ref{Lq_Sq_u_d_vs_mPi2_v2_LHPC}.

Together with the relations discussed in section \ref{sec:Interpretation}, in particular
Eq.~(\ref{J1}), the results for the GFFs $A^q_{20}$ and $B^q_{20}$ in the isovector and isosinglet channel
from the previous section can now be used to study the quark angular momentum, $J^q$, contributions to the 
nucleon spin $1/2$, Eqs.~(\ref{SpinSumrule1},\ref{SpinSumrule2}). 
Furthermore, taking into account the results for the quark spin contributions, $\Delta \Sigma^q$,
which was for the isovector case (identifying $\Delta \Sigma^{u-d}=g^{(3)}_A$) 
already discussed in section \ref{sec:axialvector},
we also gain access to the orbital angular momentum, $L^q$, carried by the quarks in the nucleon,
as defined in Eq.~(\ref{Jq}).

Before going into the details of the recent results obtained by LHPC and QCDSF in unquenched lattice QCD, 
we note that a first study of the angular and orbital angular momentum contributions to
the spin of the nucleon was performed in the quenched approximation in \cite{Mathur:1999uf}.
In this work, the sum $T^q(t)\equiv (A^q_{20}(t)+B^q_{20}(t))/2$, for three different
non-zero values of $t$, was calculated for up to four different pion masses 
from $\approx535\MeV$ to $\approx1200\MeV$ in a volume of $V\approx(1.7\fm)^3$. 
The corresponding lattice operators were perturbatively
renormalized and transformed to the $\MSbar$ scheme.
Results for the connected contributions, i.e. \emph{connected insertions} (CI),  
were extrapolated to $t=0$ based on a dipole ansatz
to obtain the quark angular momentum, $J^q_{CI}=T_{CI}^q(t=0)$.
Contributions from disconnected diagrams, i.e. \emph{disconnected insertions} (DI),
were evaluated using stochastic sources with complex $Z(2)$ noise, in combination with 
an unbiased variational subtraction scheme as suggested in \cite{Thron:1997iy}
to reduce the error. 
A monopole ansatz
was employed to obtain the DI-contributions to the angular momentum, $J^q_{DI}=T_{DI}^q(t=0)$,
from an extrapolation in $t$. The final results, which have been linearly
extrapolated in the quark mass to the chiral limit (critical quark mass),
are given in Table \ref{tabMathur}, together with
corresponding values for the quark spin fractions, $\Delta\Sigma^q$ \cite{Dong:1995rx,Gusken:1999as,Fukugita:1994fh},
at the lattice scale of $\mu=a^{-1}=1.76\GeV$.

\begin{table}
\begin{center}
\begin{tabular}{|l|l|l|l|l|l|}
\hline
&$J^{q}$&$\frac{1}{2}\Sigma^q$&$L^{q}$\\
\hline
$(u + d)_{CI}$&$0.44(7)$&$0.31(4)$& $0.13(7)$\\
$(u,d)_{DI} $&$-0.047(12)$&$-0.062(6)$& $0.015(12)$\\
\hline
$(u+d)_{CI+DI}$ & $0.35(7)$ & $0.19(4)$ & $0.16(7)$ \\
\hline
$s$&$-0.047(12)$&$-0.058(6)$& $0.011(12)$\\
\hline
$(u+d)_{CI+DI}+s$ & $0.30(7)$&$0.13(6)$&$0.17(6)$\\
\hline
\end{tabular}
\caption{\label{tabMathur}Quark contributions to the nucleon spin in the quenched approximation (from \cite{Mathur:1999uf}).
  CI and DI refer to (quark line) connected and disconnected insertions, respectively, and $s$ denotes strange quark contributions
  obtained from linear extrapolations in the light (valence) quark masses to the chiral limit at fixed 
  $\kappa_s=0.154$ (corresponding to $m_s\approx124\MeV$).}
\end{center}
\end{table}

The total angular momentum of quarks is found to be $\approx60\%$ of $1/2$, while
the quark orbital angular momentum contributions amount to $\approx34\%$.
We note that the DI-contributions to $\Sigma^q$ in particular are surprisingly large,
i.e. of the order of $50\%$ when summed over up-, down- and strange quarks.
It would be important to perform a similar calculation in unquenched lattice
QCD at lower pion masses, and to study the chiral extrapolations more carefully using ChPT.

Another study of quark angular momentum in quenched lattice QCD was
performed in \cite{Gadiyak:2001fe}, with special emphasis on avoiding extrapolations of
$A^q_{20}(t)+B^q_{20}(t)$ in the momentum transfers to $t=q^2=0$. These extrapolations, which 
not only depend on the chosen ansatz, e.g. monopole or dipole, are even more problematic
in small volumes since the lowest non-zero values of $|t|$ are quite large, e.g. in this case
$|t|^{\not=0}_{\text{min}}\approx0.5\GeV^2$, so that a large gap to $t=0$ would have to be bridged,
introducing additional systematic uncertainties. 
We remark that very small $|t|^{\not=0}_{\text{min}}$ may be accessed using (partially) twisted boundary
conditions as discussed at the end of section \ref{sec:methods} and in sections \ref{pionFFs}, \ref{sec:NuclPTBCs}.
In \cite{Gadiyak:2001fe}, an extrapolation in $t$ was avoided by 
replacing the standard lattice operator in Eq.~\ref{op1},
$\mathcal{O}_{\mu\nu}(x)\rightarrow x_\tau\mathcal{O}_{\mu\nu}(x)$,
including an explicit factor of the position vector $x_\tau$.
Inserted into a nucleon matrix element and integrated over $x$, 
this factor can be translated into a derivative with respect to the 
momentum transfer $q=\Delta$ at $q=0$, acting on the matrix element 
of the Fourier-transformed original operator $\mathcal{O}_{\mu\nu}$.
Choosing appropriate components $\mu,\nu,\tau$, the derivative of this
matrix element is then found to be proportional to $A^q_{20}(0)+B^q_{20}(0)$
and thereby gives direct access to the quark angular momentum.
The lattice calculations based on this direct method gave a value of 
$J_{CI}^{u+d}=0.47(7)$ for the connected, and $J_{DI}^{u+d}=-0.12(6)$
for the disconnected insertions, which were computed using random
sources with $Z(2)$ noise and the unbiased subtraction method mentioned above.
These results, obtained for $m_\pi\approx850\MeV$, are in very good
agreement with the chirally extrapolated values in Table \ref{tabMathur}.

It has, however, been noted already in the framework of a lattice study 
of the nucleon form factors \cite{Wilcox:1991cq} that the direct method to 
compute the nucleon magnetic moment $\mu=F_1(0)+F_2(0)$ based on a current operator $J_\mu(x)$
multiplied by a factor of $x$ (very similar to what has been used to access $J^q$)
may be problematic. In simple terms, the corresponding discrete lattice derivative
with respect to the momentum transfer $q$ is in general a bad approximation 
to the continuum derivative in small volumes due to the size of the lowest non-zero
components $\vec q^{\;\;\not=0}_{i,\text{min}}\approx0.7\GeV$. It is therefore
possible that the direct method to calculate $\mu$ and $J^q$ is 
much more susceptible to finite volume effects.
Numerical indications for such large finite volume effects may have been observed already in 
\cite{Gadiyak:2001fe} for the case of the nucleon magnetic moment.

%
%
%

We now come back to the unquenched lattice QCD studies of LHPC and QCDSF discussed in the previous section.
Figure \ref{Jq_QCDSF07_Ohtani} shows the angular momentum of up- and down quarks, $J^{u,d}=(A^{u,d}_{20}(0)+B^{u,d}_{20}(0))/2$, as a function of $m_\pi^2$, computed on the basis of
the results from QCDSF  \cite{Brommel:2007sb} 
in Figs.~\ref{xv_QCDSF07}, \ref{b20v_QCDSF07}, \ref{xs_QCDSF07} and 
\ref{b20s_QCDSF07}, together with the corresponding chiral extrapolations represented
by the shaded bands. It is quite remarkable that the total angular momentum
(from connected insertions) is mainly carried by the up quarks, while the
down quark contribution is very small and tends to zero towards the chiral limit.
From the CBChPT extrapolations, the values $J^{u}=0.230(8)$ and $J^{d}=-0.004(8)$
were obtained at the physical pion mass \cite{Brommel:2007sb}.

This is in good agreement within errors with the results from LHPC \cite{Hagler:2007xi} based on the 
simultaneous global chiral fits in Figs.~\ref{A20_umd_SimulFit} to ~\ref{C20_umd_tp24_SimulFit}
for the isovector, and Figs.~\ref{A20_upd_SimulFit} to ~\ref{C20_upd_3d_SimulFit}
for the isosinglet channel, giving $J^{u}=0.214(27)$ and $J^{d}=-0.001(27)$.

The total angular momentum contribution from $u+d$ quarks to the spin of the nucleon (obtained
from connected diagrams) is therefore $J^{u+d}\approx40-50\%$ of $1/2$,
which is significantly below the values given in Table \ref{tabMathur} and obtained in \cite{Gadiyak:2001fe}
for the connected contributions, which are of the order $J^{u+d}\approx80-90\%$ of $1/2$. This 
discrepancy is certainly to a large extent the result of the different chiral
extrapolations, i.e. the linear extrapolation in $m_\pi^2$ employed in \cite{Mathur:1999uf}
compared to the CBChPT extrapolations used in \cite{Brommel:2007sb} and \cite{Hagler:2007xi}.
The band in Fig.~\ref{Jq_QCDSF07_Ohtani} shows that a substantial chiral
curvature may be expected at smaller pion masses, which cannot be provided
by a linear fit to results at larger pion masses.

%
%

An analysis of the quark orbital angular momentum contributions, $L^q$,
requires, in addition to the total angular momenta $J^q$, also the quark spin fractions $\Delta\Sigma^{q}$. 
For the isovector channel, they were discussed in some detail in section \ref{sec:axialvector}
in terms of the axial vector coupling constant, which we identify with the 
quark spin fraction, $\Delta\Sigma^{u-d}=\langle 1\rangle_{\Delta u-\Delta d}=g^{u-d}_A$. 
We refer in particular to the results from QCDSF in Fig.\ref{gafitext1_QCDSF_v2} \cite{Khan:2006de} and 
from LPHC in Fig.\ref{gAfig2_LHPC} \cite{Edwards:2005ym}.
Corresponding results for the spin fraction of quarks in the isosinglet channel, obtained
by QCDSF on the basis of the same lattice approach and similar ensembles as discussed before, are
displayed in Fig.\ref{DxS_QCDSF07_Ohtani} versus $m_\pi^2$.
The shaded band represents a chiral extrapolation based on a
leading 1-loop HBChPT calculation \cite{Diehl:2006ya}, which gives
\bea
\label{DeltaSigmaHBChPT}
\Delta\Sigma^{u+d} =
\Delta\Sigma^{0,u+d}  \left( 1 - \frac{3 g_{A}^2 m_{\pi}^2}{(4\pi)^2f_{\pi}^2}
 \left\{\ln \frac{m_{\pi}^2}{\lambda^2}  + 1 \right\} \right)
  + c_0 m_{\pi}^2\,.
\eea
A value of $\tilde A^{u+d}_{10}(0)=\Delta\Sigma^{u+d}=0.402(24)$ was obtained at the physical pion mass
from a fit with two free parameters, $\Delta\Sigma^{0,u+d}$ and $c_0$, to the 
lattice data points with $m_\pi<500\MeV$. The strong downwards bending
observed below $m_\pi\sim350\MeV$ leads to a very good agreement with the 
recent result from HERMES \cite{Airapetian:2007aa}, $\Delta\Sigma^{u+d,\exp}=0.415(20)$, represented by the open star. 
We note, however, that the extrapolation not only breaks down quickly 
for pion masses above $500\MeV$, but that leading 1-loop HBChPT is in general not
expected to be applicable for pion masses of $300\MeV$ and larger. 
The good agreement of the chiral extrapolation with the result from experiment in 
Fig.\ref{DxS_QCDSF07_Ohtani} should therefore be considered with caution and may well be accidental.

Together with the results for $J^{u+d}$ discussed above, the total quark orbital angular momentum
contribution was found to be small and compatible with zero within errors,
$L^{u+d}=J^{u+d}-\Delta\Sigma^{u+d}/2=0.025(27)$, at the physical pion mass \cite{Brommel:2007sb}. 
Figure \ref{JsL_QCDSF07_Ohtani} shows
that this remarkable cancellation of $J^{u+d}$ and $\Delta\Sigma^{u+d}/2$
also holds over the full range of accessible pion masses up to $m_\pi\approx900\MeV$.
%
%
%

These findings are in good agreement with the results from LHPC \cite{Hagler:2007xi} 
in Fig.~\ref{Lq_Sq_u_plus_d_vs_mPi2_v2_LHPC},
showing the spin and OAM contributions in the isosinglet channel as functions of the squared pion mass. The 
OAM for $u+d$-quarks, which was in this case obtained from dipole extrapolations of the
GFF $B^{u+d}_{20}(t)$ to $t=0$ at fixed pion masses, 
is also found to be compatible with zero within errors for $m\lesssim600\MeV$.
The lattice data points for the total quark momentum fraction were extrapolated to the
chiral limit based on a self-consistently modified result from leading 1-loop HBChPT,
\begin{eqnarray}
\label{DeltaSigmaHBChPT2}
\Delta\Sigma^{u+d} =
\Delta\Sigma^{0,u+d} \left( 1 - \frac{3 g_{A,\lat}^2}{(4\pi)^2} 
\frac{m_{\pi,\lat}^2}{f_{\pi,\lat}^2} \left(\ln  \frac{m_{\pi,\lat}^2}{f_{\pi,\lat}^2} + 1\right) \right) + c_0 \frac{m_{\pi,\lat}^2}{f_{\pi,\lat}^2}\,.
\end{eqnarray}
with free parameters $\Delta\Sigma^{0,u+d}$ and $c_0$, see also Eq.~\ref{xLHPCChPT} and corresponding discussion.
The chiral fit, which is represented by the 
error band in Fig.~\ref{Lq_Sq_u_plus_d_vs_mPi2_v2_LHPC}, gives  
$\Delta\Sigma^{u+d}=0.415(56)$, also in very good agreement with recent results by HERMES \cite{Airapetian:2007aa}.

Together with the results from LHPC for $J^{u,d}$ discussed above,
one obtains for the total $u+d$ quark OAM contribution to the nucleon spin a value
of $L^{u+d}=J^{u+d}-\Delta\Sigma^{u+d}/2=0.005(52)$ at the physical pion mass \cite{Hagler:2007xi}, 
perfectly consistent with zero and compatible with the results from QCDSF within errors.

Further insight into the nucleon spin structure may be obtained
by decomposing the total quark angular momentum into the spin and orbital angular momentum 
contributions of up- and down-quarks separately. Such a decomposition is displayed
in Fig.~\ref{Lq_Sq_u_d_vs_mPi2_v2_LHPC}, showing lattice results from LHPC for 
$\Delta\Sigma^{u,d}$ and $L^{u,d}$ versus $m_\pi^2$ \cite{Hagler:2007xi}. 
They were
obtained by combining the results for the GFFs $A^{u,d}_{20}(t)$ and $B^{u,d}_{20}(t)$,
where the $B^{u,d}_{20}(t)$ were extrapolated to $t=0$ using dipole fits, 
with the isovector axial vector coupling discussed in section \ref{sec:axialvector}, cf. Fig.~\ref{gAfig2_LHPC}, 
providing $\Delta\Sigma^{u-d}$, and for $\Delta\Sigma^{u+d}$ 
shown in Fig.~\ref{Lq_Sq_u_plus_d_vs_mPi2_v2_LHPC}.

The most remarkable feature of this analysis is that the \emph{individual} up- and 
down-quark OAM contributions are quite substantial, with absolute values $|L^u|\sim|L^d|$
of the order of $20-30\%$ of $1/2$, but opposite in sign, so that they 
cancel almost exactly in the sum, $L^{u+d}\sim0$. 
The values at the physical pion mass,
obtained from the self-consistently modified HBChPT extrapolation of
$\Delta\Sigma^{u+d}$, the SSE extrapolation of $\Delta\Sigma^{u-d}=g_A^{u-d}$ \cite{Edwards:2005ym},
and the CBChPT extrapolated GFFs $A^{u\pm d}_{20}(t)$ and $B^{u\pm d}_{20}(t)$
are $L^{u}=-0.195(44)$ and $L^{d}=0.200(44)$ \cite{Hagler:2007xi}.
Similarly, the individual spin- and OAM contributions of the down-quarks are
of the same size, $|L^d|\sim|\Delta\Sigma^d/2|\sim0.1,\ldots0.15$, but also
opposite in sign, so that the down-quark angular momentum is zero within errors
due to a cancellation of spin and OAM, $J^d=L^d+\Delta\Sigma^d/2\sim0$.

The observation that the total light quark angular momentum
is compatible with zero in lattice QCD, $L^{u+d}\sim0$, seems at first sight to 
be at odds with expectations from relativistic quark models, where the quark OAM contribution is in general substantial, 
$L_{\text{rel}}^{u+d}\sim30,\ldots,40\%$ of $1/2$.
Such a strong discrepancy could in principle point towards significant deficiencies of the
model, or substantial systematic effects in the lattice calculation.
At the same time, the quark spin contribution $\Delta\Sigma_{\text{rel}}^{u+d}/2$ is 
approximately $60,\ldots,70\%$ of $1/2$, in overall agreement with 
lattice results.
A likely explanation that can at least partially account for these observations has been given in 
\cite{Wakamatsu:2007ar,Thomas:2008ga,Bratt:2008uf} :
While the lattice results are given in the $\MSbar$ scheme at a scale of $4\GeV^2$,
the model calculations generically correspond to a much lower ``hadronic'' scale $\mu_{\text{had}}\ll 1\GeV$,
so that the respective numbers may not be compared directly in the first place.
At this point it is interesting to note that the evolution equations
for spin and OAM in the leading-logarithmic approximation \cite{Ji:1995cu}, 
although strictly speaking not applicable at very low scales, predict a strong change 
in magnitude and even the sign of the quark OAM contributions at scales $\mu^2\lessapprox 0.4\GeV^2$.
Depending on the initial conditions and the hadronic starting scale, relativistic quark model results 
were even found to be in rough quantitative agreement with the lattice results when evolved up to the common
scale of $4\GeV^2$ \cite{Thomas:2008ga}.
Similarly, the overall consistency between the model and the lattice results for
$\Delta\Sigma_{\text{rel}}^{u+d}$
may be understood by noting that the quark spin is conserved at leading-log order.

We note again that many of the lattice results presented in this section correspond to the quark line
connected contributions, and that a complete calculation including 
disconnected diagrams could in principle lead to a significantly different decomposition of the nucleon spin.

A summary of the more recent lattice QCD results for quark spin-, OAM- and 
total angular momentum contributions
to the nucleon spin will be given below in Table \ref{slj}.

%
\subsubsection{Higher moments of unpolarized and polarized GPDs}
\label{sec:GPDsMoments}
\subsubsection{Transverse nucleon structure}
\label{sec:TransverseStructure}
Higher moments of the unpolarized nucleon GPDs $H$ and $E$ are
given by linear combinations of the the GFFs $A_{ni}(t)$, $B_{ni}(t)$
with $n=1,2,3,\ldots$ and $i=0,2,4,\ldots\le (n-1)$, and $C_{n0}$ with $n=2,4,6,\ldots$, cf. Eqs.~(\ref{Hn},\ref{En}).
They coincide with the form factors of the energy momentum tensor for $n=2$ that
we discussed in detail in the previous sections. 
Moments of the polarized
GPDs $\widetilde H$ and $\widetilde E$ are parametrized by the GFFs 
$\widetilde A_{ni}(t)$, $\widetilde B_{ni}(t)$ with
$n=1,2,3,\ldots$ and $i=0,2,4,\ldots\le n$ \cite{Hagler:2004yt}.

First results for the $n=2$ moments of the polarized GPDs were
presented in \cite{Schroers:2003mf} in unquenched lattice QCD.
The unpolarized and polarized moments with $n=1,2,3$ and $i=0$ 
have been investigated for the first time by LHPC/SESAM 
in \cite{LHPC:2003is} for $n_f=2$ flavors of Wilson fermions in a
``heavy pion world'' with pion masses of $\approx744\MeV$ and $\approx897\MeV$. In this study,
the focus has been on the change of the $t$-dependence (the slope) of the GFFs $A^q_{n0}(t)$ 
and $\widetilde A^q_{n0}(t)$ when going from the lowest moment $n=1$ (the form factors) 
to the highest accessible moment, $n=3$. The physics behind these results
will be discussed below.
Since then, a number of results, mostly for $A^q_{n0}(t)$, $B^q_{n0}(t)$,
$\widetilde A^q_{n0}(t)$ and $\widetilde B^q_{n0}(t)$ for the lowest $n$,
obtained in unquenched QCD, were presented in proceedings, see, e.g., 
\cite{Gockeler:2004mn,Gockeler:2005cd,Gockeler:2005aw,Gockeler:2006ui,Brommel:2007sb,Schroers:2003mf,Negele:2004iu,Edwards:2006qx,Bratt:2008uf}.
The most comprehensive lattice study to this date of moments of unpolarized and 
polarized GPDs for $n=1,2,3$ and $i=0$ was published more recently by LHPC, in the framework of the hybrid approach
with $n_f=2+1$ flavors of domain wall valence and Asqtad staggered sea quarks \cite{Hagler:2007xi}.
For a discussion of the corresponding results for the unpolarized $n=2$-moments,
we refer to section \ref{sec:EMT}, cf. Figs.~\ref{A20umd_LHPC}, \ref{A20_umd_SimulFit} to \ref{C20_umd_tp24_SimulFit}, 
and \ref{ABCupd_LHPC} to \ref{C20_upd_3d_SimulFit}.
%
\begin{figure}[t]
    \begin{minipage}{0.48\textwidth}
      \centering
\includegraphics[angle=0,width=0.99\textwidth,clip=true]{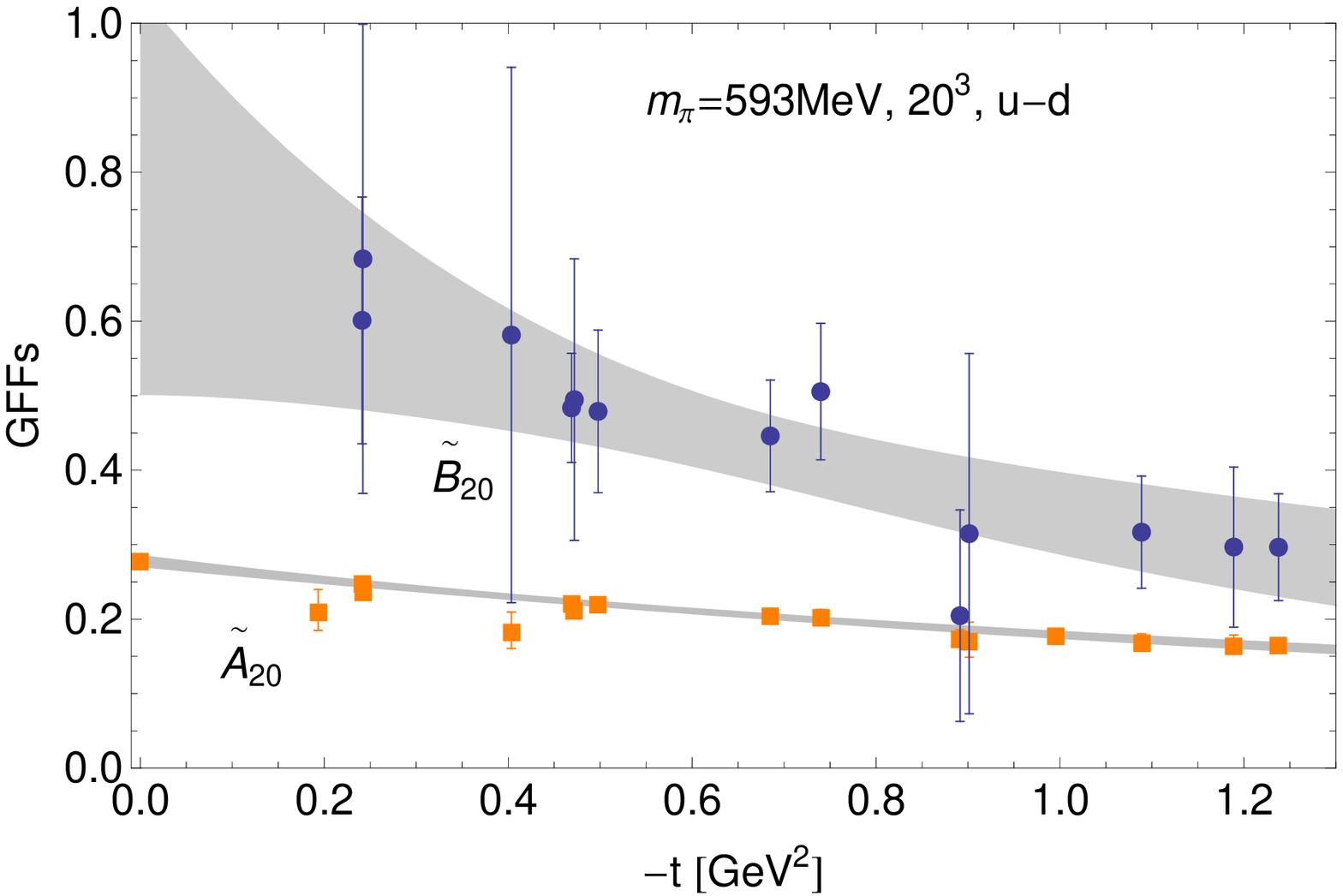}
  \caption{The GFFs $\widetilde A^{u-d}_{20}(t)$ and $\widetilde B^{u-d}_{20}(t)$ at a 
  pion mass of $595\MeV$ (from \cite{Hagler:2007xi}).}
  \label{ABpol_umd_LHPC}
     \end{minipage}
          \hspace{0.3cm}
    \begin{minipage}{0.48\textwidth}
      \centering
          \includegraphics[angle=0,width=0.99\textwidth,clip=true]{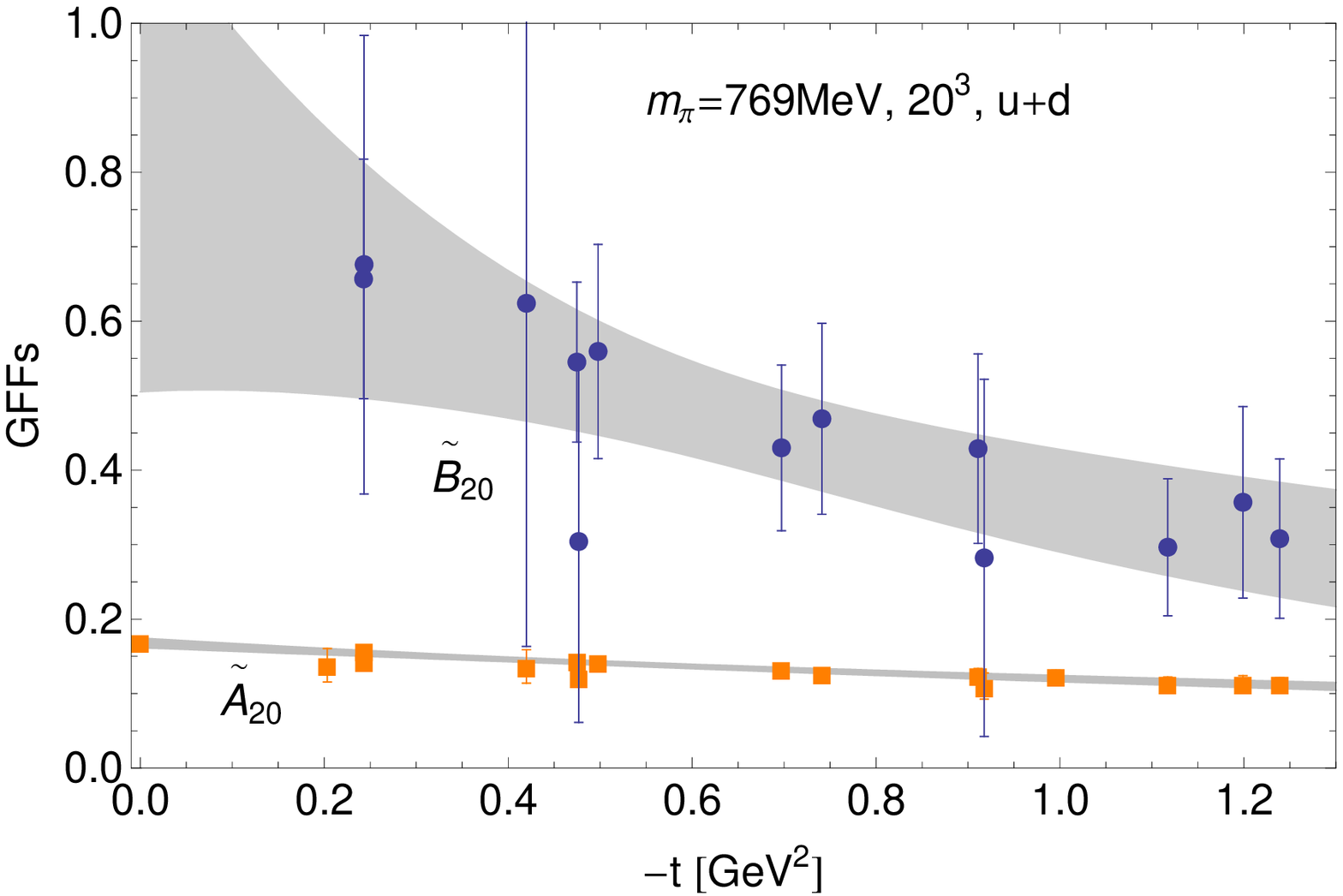}
  \caption{Connected contributions to the GFFs $\widetilde A^{u+d}_{20}(t)$ and $\widetilde B^{u+d}_{20}(t)$ at a 
  pion mass of $758\MeV$ (from \cite{Hagler:2007xi}).}
  \label{ABpol_upd_LHPC}
     \end{minipage}
\end{figure}
%
%

An example for the polarized GFFs $\widetilde A_{20}(t)$ and $\widetilde B_{20}(t)$
is given in Figs.~\ref{ABpol_umd_LHPC} and \ref{ABpol_upd_LHPC} 
for the isovector channel at a pion mass of $\approx595\MeV$,
and for the isosinglet channel at a pion mass of $\approx758\MeV$, respectively. 
As for the
unpolarized case, the underlying axial vector lattice operators were
perturbatively renormalized including a non-perturbative improvement
factor as given in Eq.~\ref{ZLHPC}, and all results are given in the
$\MSbar$ scheme at a scale of $\mu=2\GeV$. 
While the statistical precision for the GFF $\widetilde A_{20}(t)$ is very good, 
the signal for $\widetilde B^{u+d}_{20}(t)$ in particular
is rather noisy with errors of $\mathcal{O}(50\%)$
and larger at pion masses of $\approx595\MeV$ and below. 
Still, within errors the GFF $\widetilde B_{20}(t)$ 
turns out to be large compared to $\widetilde A_{20}(t)$,
with central values of 
$\widetilde B^{u-d}_{20}(t)\sim\widetilde B^{u+d}_{20}(t)\sim0.6$
at the lowest accessible values of $-t\simeq0.25\GeV^2$. 
We note that in particular the pion mass dependence of the forward values
$\langle x\rangle_{\Delta u-\Delta d}=\widetilde A^{u-d}_{20}(\t0)$
(obtained in this lattice simulation framework) was presented and discussed in section \ref{sec:nuclPDFs}, 
Fig~\ref{Deltax_LHPC}. 
Further results for $\widetilde A_{20}(t)$ and $\widetilde B_{20}(t)$
for pion masses down to $\approx350\MeV$ can be found in \cite{Hagler:2007xi}.

\begin{figure}[t]
    \begin{minipage}{0.48\textwidth}
      \centering
\includegraphics[angle=0,width=0.99\textwidth,clip=true]{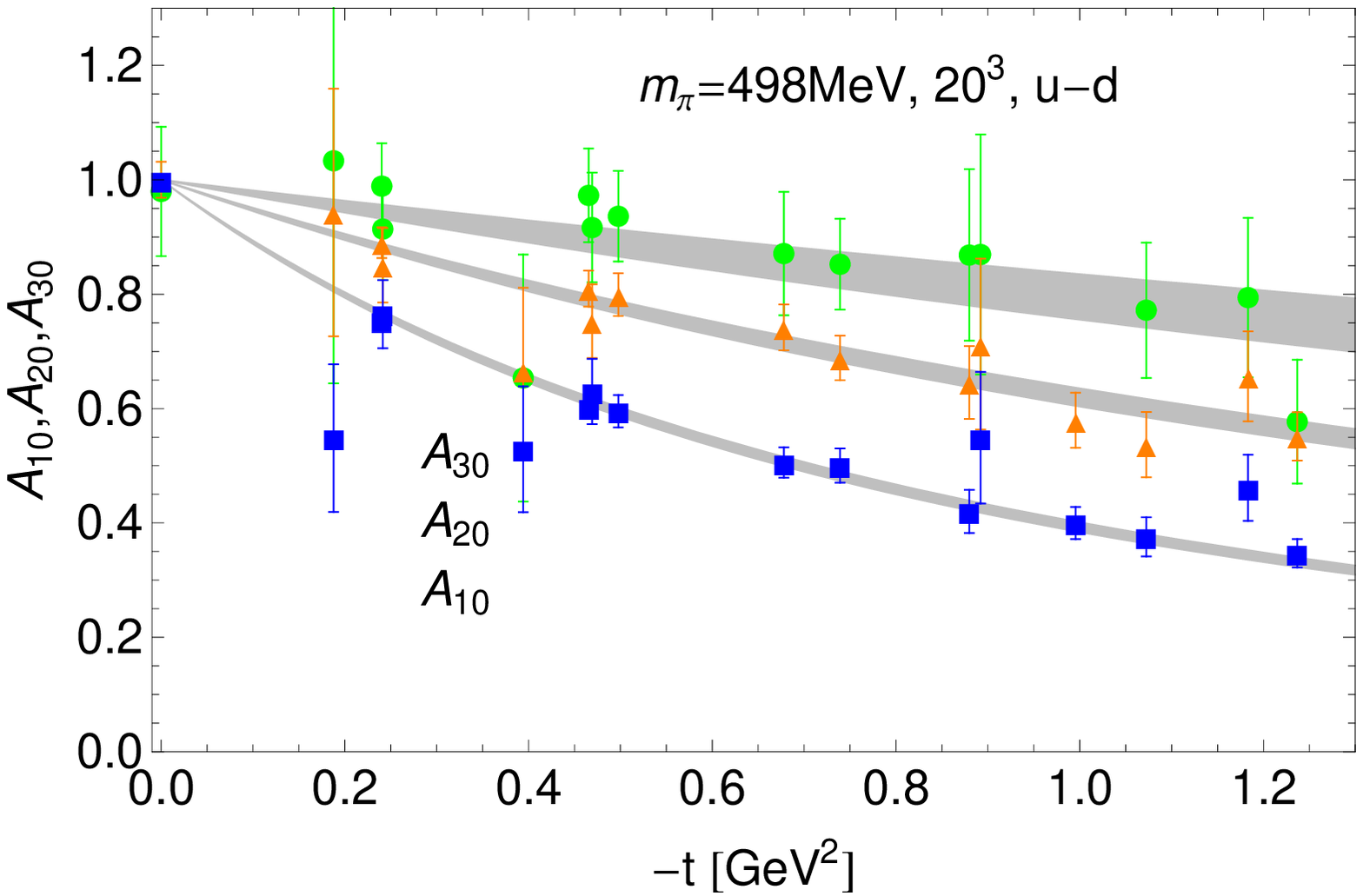}
  \caption{Lowest three moments of the unpolarized GPD $H(x,\xi=0,t)$ for $u-d$ (from \cite{Hagler:2007xi}).}
  \label{A123_umd_LHPC}
     \end{minipage}
          \hspace{0.3cm}
    \begin{minipage}{0.48\textwidth}
      \centering
          \includegraphics[angle=0,width=0.99\textwidth,clip=true]{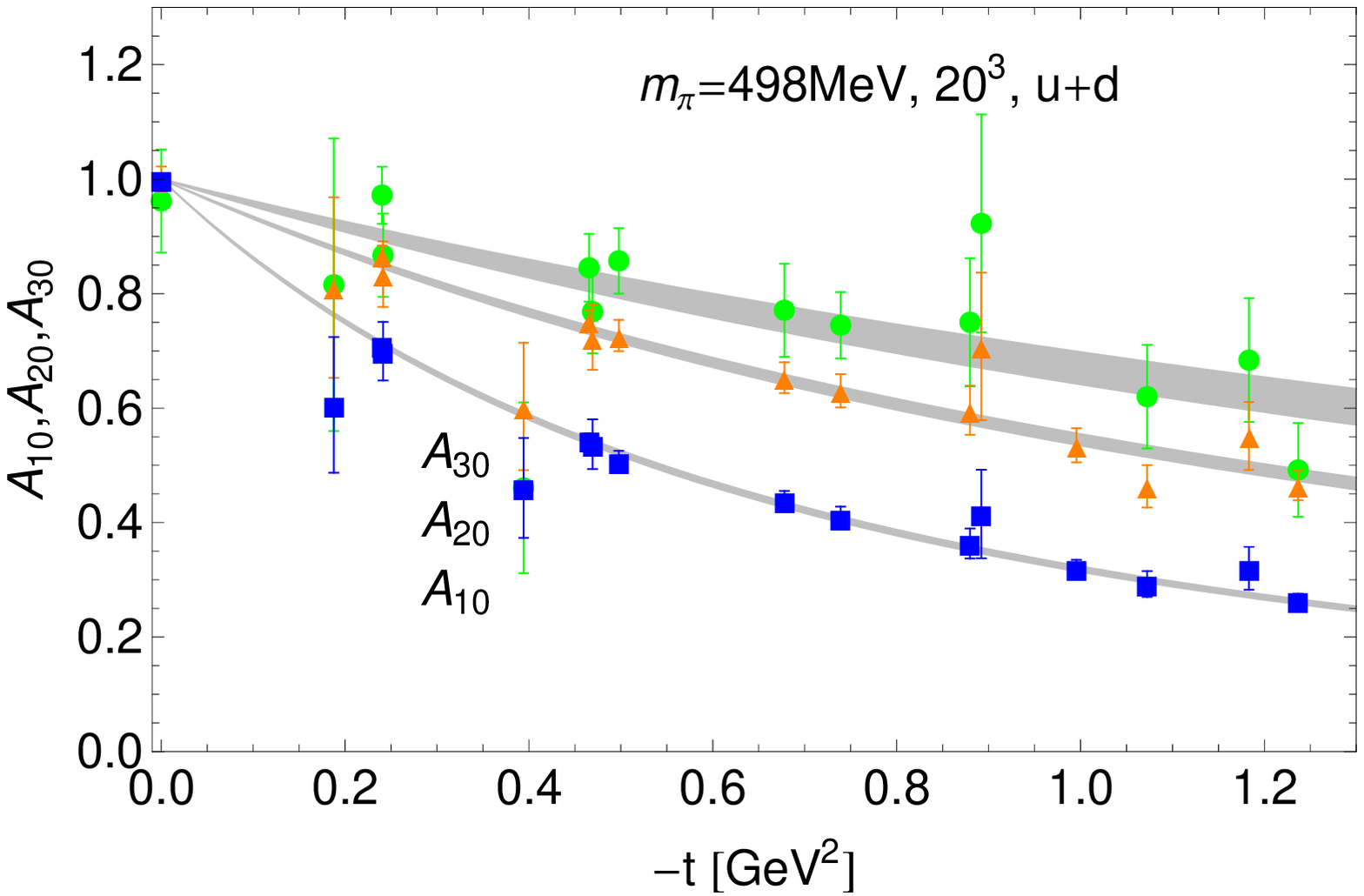}
  \caption{Lowest three moments of the unpolarized GPD $H(x,\xi=0,t)$ for $u+d$ (from \cite{Hagler:2007xi}). 
  Contributions from disconnected diagrams are not included.}
  \label{A123_upd_LHPC}
     \end{minipage}
\end{figure}

A direct comparison of the $t$-dependences of the lowest three moments 
of the unpolarized GPD $H$, $A_{(n=1,2,3)0}(t)$, in the isovector and isosinglet channel
is provided in Figs.~\ref{A123_umd_LHPC} and \ref{A123_upd_LHPC} for $m_\pi\approx496\MeV$,
where the GFFs have been normalized to unity at $t=0$. Dipole fits with
fixed forward values, $A_{(n=1,2,3)0}(\t0)=1$,
to the lattice data points are represented by the shaded bands.
In both the isovector and isosinglet case, the slope in $t$ is found to flatten significantly 
for increasing $n$, confirming the results of the first unquenched lattice QCD
study by LHPC/SESAM \cite{LHPC:2003is} of $A^{u\pm d}_{n0}(t)$ 
and $\widetilde A^{u-d}_{n0}(t)$ at larger pion masses.
%
%
%

An simple interpretation of this interesting observation can be given 
in the framework of the light-cone wave function (LCWF) or overlap representations of the GPDs
in \cite{Brodsky:2000xy,Diehl:2000xz}).
There, one finds that the dependence of the LCWF of the outgoing hadron on the quark transverse momentum, 
upon interaction with an external current, 
is given by $k^{\text{LCWF}}_{f,\perp}=k^{\text{LCWF}}_{i,\perp} - (1-x)\Delta_\perp$,
where $x$ is the longitudinal momentum fraction carried by the active quark,
$k^{\text{LCWF}}_{i,\perp}$ is the initial transverse momentum in the LCWF, and $\Delta_\perp\sim |t|^{1/2}$ 
is the transferred transverse momentum (in the frame specified in \cite{Brodsky:2000xy}).
Hence at small $x$, the intrinsic transverse momentum dependence of the LCWF
changes strongly for larger values of the momentum transfer $\Delta_\perp$, leading to a small wavefunction overlap.
Physically, the large transferred momentum would have to be redistributed among all constituents
to prevent a break-up of the bound state. 
As a result, the coupling to the external current, parametrized by the GPDs, is suppressed. 
For $x\rightarrow1$, $k^{\text{LCWF}}_{f,\perp}\sim k^{\text{LCWF}}_{i,\perp}$, the wavefunction overlap is large,
corresponding to a higher probability that the hadron stays intact, even for larger $\Delta_\perp^2= -t^2$.
In this case, a stronger coupling to the hadron will be observed.
Concerning moments of PDFs and GPDs, we note that the limit $n\rightarrow\infty$
corresponds to $x\rightarrow1$ due to the weighting with high powers $x^{n-1}$, Eq.~(\ref{moments1}).
From the above discussion, we therefore expect that the (relative) coupling to the
hadron at larger $t$, as parametrized by the generalized form factors, 
\emph{increases} for \emph{increasing} $n$. 
More precisely, it is predicted that the GFFs
approach a constant in $t$ for infinitely high moments, 
\be
\label{largen}
\frac{A_{n0}(t)}{A_{n0}(\t0)}\xrightarrow{n\rightarrow\infty}1\;.
\ee
%
\begin{figure}[t]
    \begin{minipage}{0.48\textwidth}
      \centering
\includegraphics[angle=0,width=0.99\textwidth,clip=true]{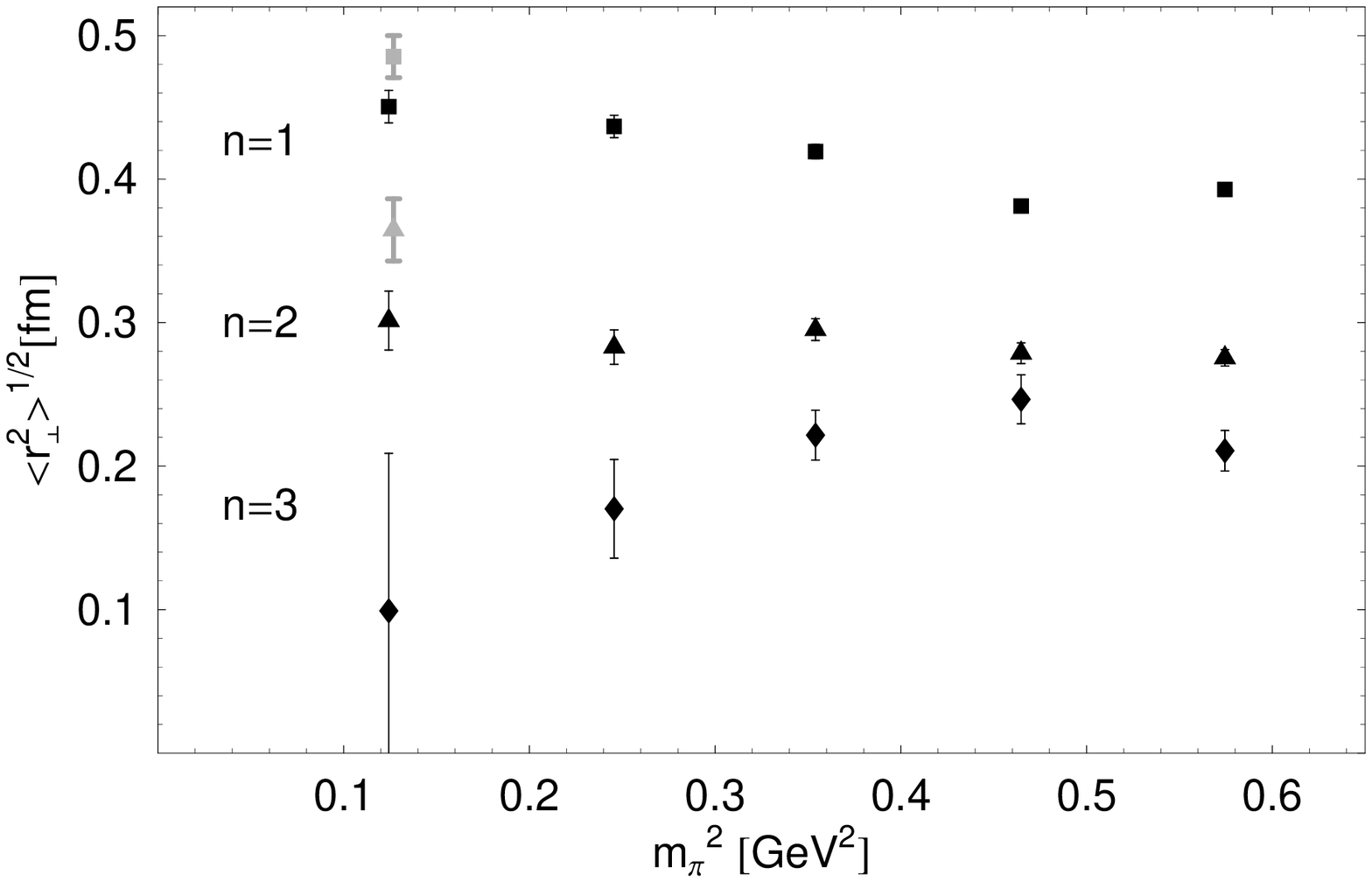}
  \caption{Isovector (generalized) transverse charge radii for $n=1,2,3$ (from \cite{Hagler:2007xi}).}
  \label{Charge_radii_vector_v2_LHPC}
     \end{minipage}
          \hspace{0.3cm}
    \begin{minipage}{0.48\textwidth}
      \centering
          \includegraphics[angle=0,width=0.9\textwidth,clip=true]{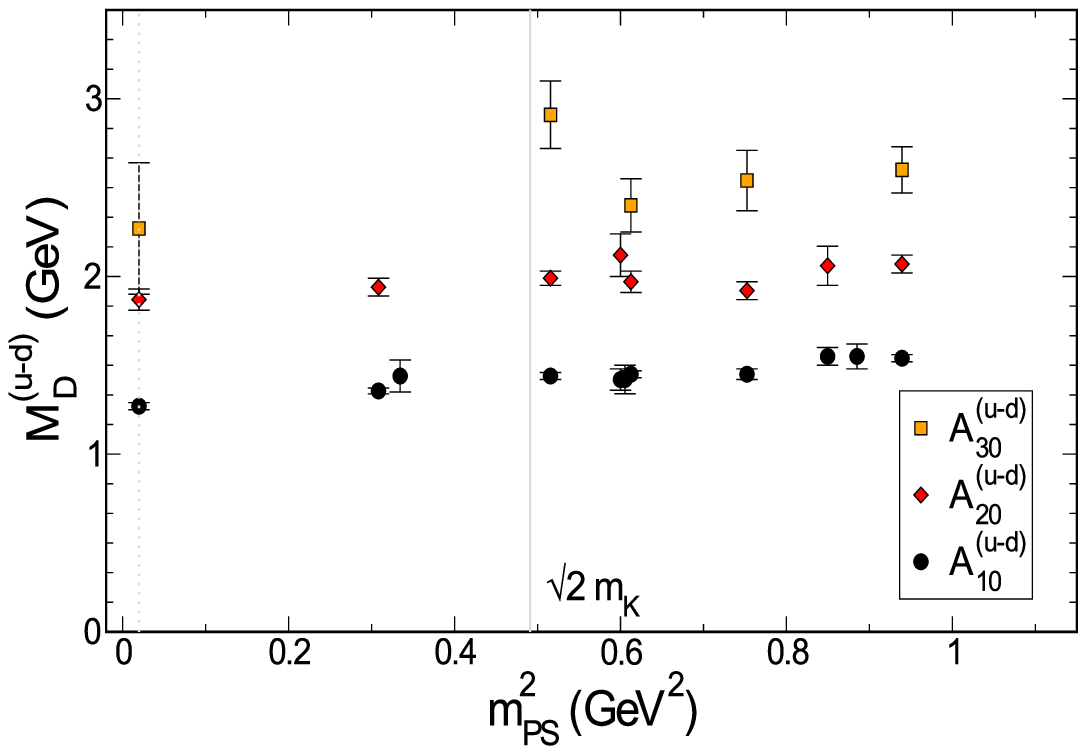}
  \caption{Dipole masses of the GFFs $A^{u-d}_{n0}$ for $n=1,2,3$ (from proceedings \cite{Gockeler:2005aw}).}
  \label{mD_Ai0_umd_QCDSF05}
     \end{minipage}
\end{figure}
%

This is qualitatively confirmed in Fig.~\ref{A123_umd_LHPC},
which shows that the relative coupling 
for $n=3$ is already a remarkable factor of 2 larger than the coupling for $n=1$,
i.e. $A^{u-d}_{30}(t)/A^{u-d}_{30}(0)\sim2A^{u-d}_{10}(t)/A^{u-d}_{10}(0)$ 
for, e.g., a squared momentum transfer of $-t\approx1\GeV^2$,
at the given pion mass. The effect is also clearly visible
but less pronounced for the polarized case \cite{LHPC:2003is,Hagler:2007xi}.

An analogous and possibly even more intuitive interpretation of these
observations may be given in impact parameter (transverse coordinate) space \cite{Burkardt:2000za}.
The impact parameter dependent GPDs (and GFFs) are obtained
by Fourier-transforming the momentum space GPDs (GFFs) 
with respect to the transverse momentum transfer, $t=-\Delta_\perp^2$,
as given in Eq.~\ref{Impact1}, for $\xi=0$. 
The impact parameter, $b_\perp$,
specifies the distance of the active quark to the center of
momentum (COM) of the parent hadron, 
$R_\perp=\sum_ix_ir_{i,\perp}/\sum_ix_i=\sum_ix_ir_{i,\perp}$
where the sums run over all partons. In the 
limit $x\rightarrow1$, the active parton carries
all the momentum and therefore represents the COM.
Hence it is expected that the impact parameter distribution
of partons peaks around $b_\perp=0$ as $x$ approaches unity.
Formally, using, e.g., light cone hadron wave functions,
it can be shown that the GPD $H(x,\xi=0,t)$, transformed to impact parameter
space, is proportional to a delta function in the limit $x\rightarrow1$,
$q(x,b_\perp)=H(x,b_\perp)\overset{x\rightarrow1}{\propto}\delta^2(b_\perp)$ \cite{Diehl:2002he}.
This is of course in one-to-one correspondence with the prediction of
the $t$-\emph{independence} of the GFFs in momentum space as $n\rightarrow\infty$, cf. Eq.~(\ref{largen}).

A measure of the transverse size of a hadron is given by the
(generalized) transverse mean squared radius \cite{LHPC:2003is}, 
\begin{equation}
\label{genRad}
\langle r_\perp^2 \rangle_{n} 
= \frac{\int d^2b_\perp b^2_\perp \int dx x^{n-1} q(x,\vec b_\perp)}
{\int d^2b_\perp  \int dx x^{n-1} q(x,\vec b_\perp)}
=\frac{\int d^2b_\perp b^2_\perp A_{n0}(b_\perp)}
{\int d^2b_\perp A_{n0}(b_\perp)}\, ,
\end{equation}
which coincides with the standard (transverse) mean square charge radius for $n=1$, 
$\langle r_\perp^2 \rangle_{n=1}=\langle r_\perp^2 \rangle=\frac{2}{3}\langle r^2 \rangle$,
defined through the Dirac form factor $F_1(Q^2)=A_{10}(t)$ as in Eq.~(\ref{radii}).
From what has been said above, the transverse size
of the hadron as measured by $\langle r_\perp^2 \rangle_{n}$ is expected
to vanish in the limit $n\rightarrow\infty$.
We also note that the generalized mean square charge radii are directly related
to dipole masses, $\langle r_\perp^2 \rangle_{n}=8/m^2_{D,n}$, 
obtained from, e.g., dipole fits to the lattice data for the GFFs $A_{(n=1,2,3)0}(t)$, 
as represented by the bands in Figs.~\ref{A123_umd_LHPC} and \ref{A123_upd_LHPC}. 
Figure \ref{Charge_radii_vector_v2_LHPC} shows the pion mass dependence of 
isovector transverse RMS radii $\langle r_\perp^2 \rangle^{1/2}_{n}$
for $n=1,2,3$, obtained by LHPC from dipole fits to the lattice data
for $|t|\lessapprox1\GeV^2$ \cite{Hagler:2007xi}.
As the pion mass decreases, the significant spread between $\langle r_\perp^2 \rangle^{1/2}_{n=1}$
and $\langle r_\perp^2 \rangle^{1/2}_{n=3}$  seems to increase. 
However, the statistical errors are quite large at the lowest pion mass,
and the data points in Fig.~\ref{Charge_radii_vector_v2_LHPC} are
in addition subject to systematic uncertainties related to the 
dipole ansatz used to parametrize the GFFs, as well as the region
of $t$ included in the dipole fits. More precise lattice results are needed
before fully quantitative predictions about the decrease of the transverse size
of the nucleon for increasing $x$ at the physical pion mass can be made.

Corresponding results for the generalized axial-vector charge radii
also show a decrease from $n=1$ to $n=3$ \cite{Hagler:2007xi}, which is however not
as pronounced as for the vector case in Fig.~\ref{Charge_radii_vector_v2_LHPC}.

Similar conclusions about the decrease of the nucleon transverse size 
were reached independently by QCDSF based on simulations with
$n_f=2$ flavors of clover-improved Wilson fermions, see, e.g., \cite{Gockeler:2005aw}. 
A summary of the results is displayed in Fig.~\ref{mD_Ai0_umd_QCDSF05},
showing the dipole masses $m_D^{u-d}$, obtained from fits to the lattice
data for the GFFs $A^{u-d}_{(n=1,2,3)0}(t)$, as functions of $m_\pi^2$.
As before, a significant increase of the dipole masses is observed, going from $n=1$ to $n=3$. 
We note that the results in Fig.~\ref{mD_Ai0_umd_QCDSF05}
for $n=1$ correspond to the results for the Dirac radii shown in Fig.~\ref{rv1Q_QCDSF}. 

As we have discussed in section \ref{FFradii}, the lattice QCD results for the Dirac radius
lie approximately a factor of two below the value from experiment at the
lowest accessible pion masses. We therefore would like to stress that
a proper chiral extrapolation of the lattice data points in Fig.~\ref{mD_Ai0_umd_QCDSF05}
and also in Fig.~\ref{Charge_radii_vector_v2_LHPC} will be, at least for $n=1$, highly non-linear
and may lead to significantly different conclusions about the
transverse nucleon structure at the physical pion mass.

\subsubsection{Tensor GPDs and the transverse spin structure of hadrons}
\label{sec:TransverseSpin}
Having discussed lattice QCD results for moments of unpolarized and polarized (vector and axial-vector)
GPDs, we now turn our attention to the tensor (also quark helicity flip or transversity) GPDs of the pion,
$H^\pi_T(x,\xi,t)$ (Eq.~\ref{PionTensorFF}), and the nucleon, $H_T(x,\xi,t)$, $E_T(x,\xi,t)$, $\widetilde H_T(x,\xi,t)$ 
and $\widetilde E_T(x,\xi,t)$ (Eq.~\ref{NuclTensor1}). 
The corresponding $x^{n-1}$-moments are parametrized by the the GFF $B^\pi_{Tni}(t)$ for the pion, 
see Eqs.~(\ref{PionTensorFF},\ref{PionTensorGFFn2}),
and the tensor GFFs $A_{Tni}(t)$, $\overline B_{Tni}(t)$, $\widetilde A_{Tni}(t)$ and $\widetilde B_{Tn(i+1)}(t)$ for the nucleon, 
with $n=1,2,3,\ldots$ and $i=0,2,4,\ldots\le (n-1)$, and where $\widetilde B_{T11}(t)=0$ due to time
reversal symmetry constraints, see Eqs.~(\ref{NuclTensor3},\ref{NuclTensor5}) \cite{Hagler:2004yt}.

\subsubsection{Pion}
\label{PionTensor}
\begin{figure}[t]
   \begin{minipage}{0.48\textwidth}
      \centering
          \includegraphics[width=0.9\textwidth,clip=true]{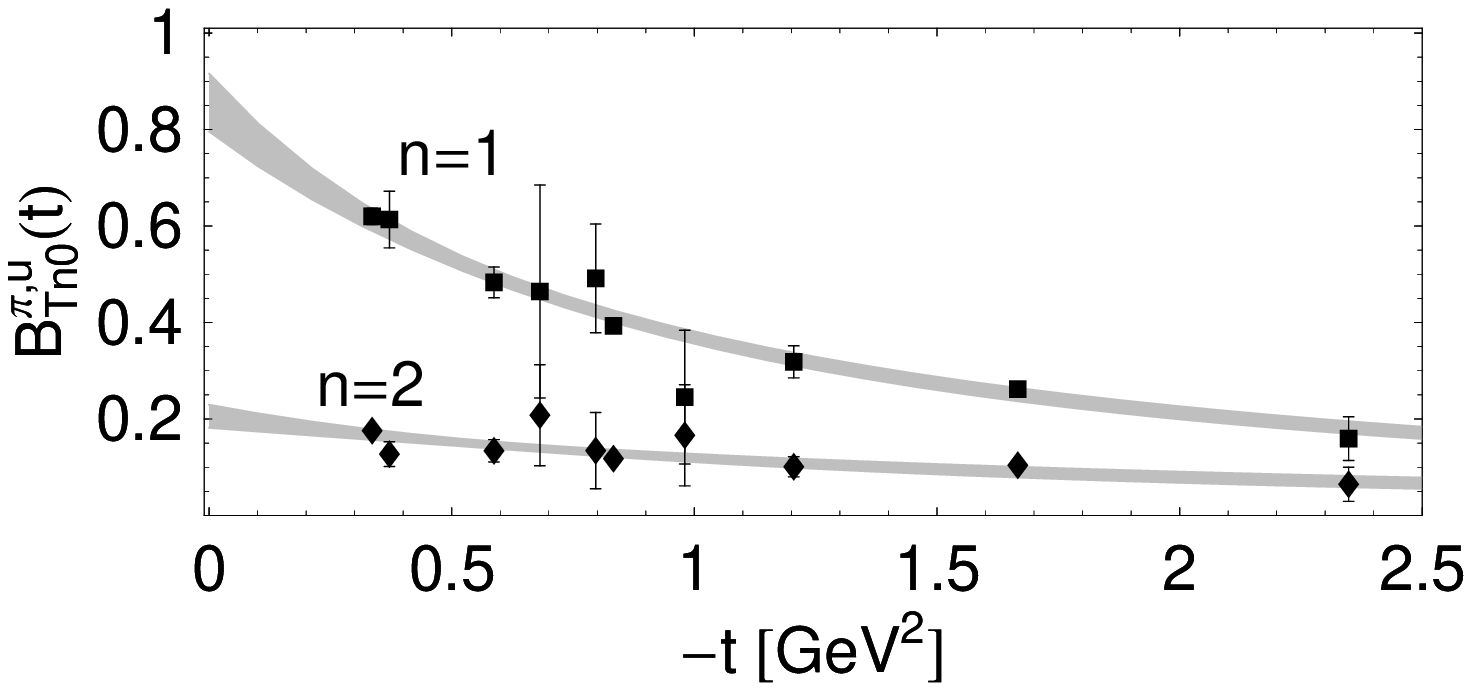}
  \caption{Tensor (generalized) form factors $B^\pi_{Tn0}(t)$ for up-quarks in a $\pi^+$ 
  with $m_\pi\sim600\MeV$ (from \cite{Brommel:2007xd}).}
  \label{BTn0_pion_b5p29k13590_v2_QCDSF}
     \end{minipage}
     \hspace{0.5cm}
    \begin{minipage}{0.48\textwidth}
      \centering
          \includegraphics[width=0.9\textwidth,clip=true]{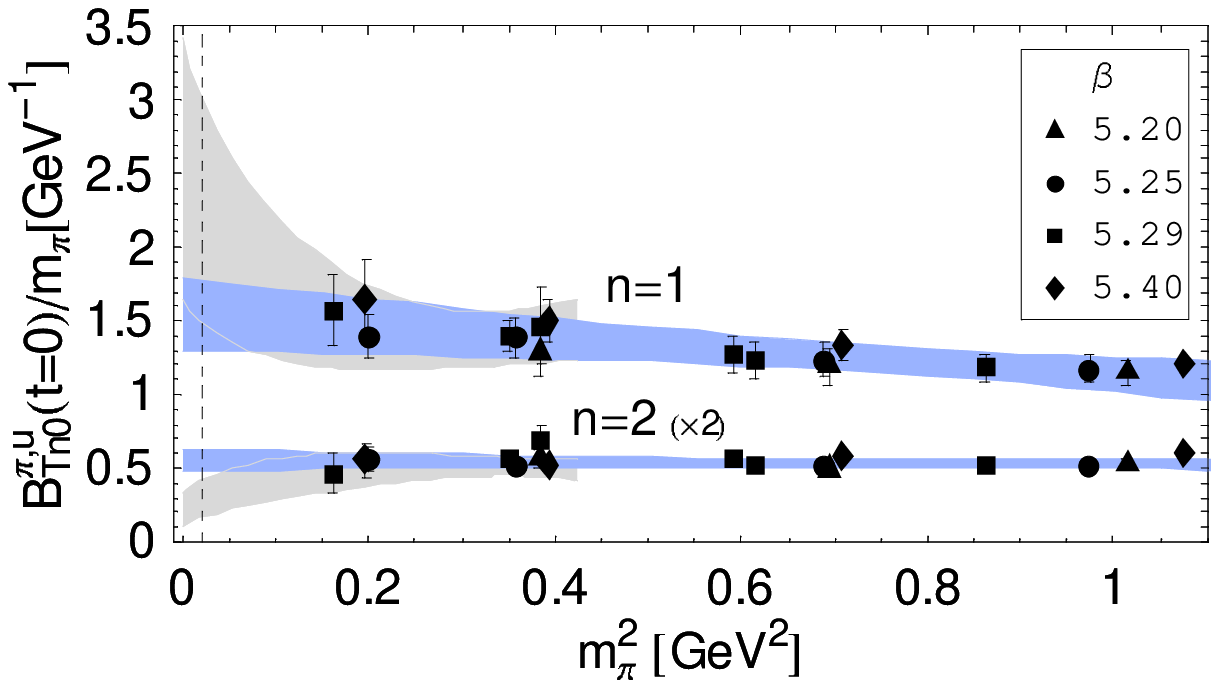}
  \caption{Forward values of tensor (generalized) form factors for up-quarks in a $\pi^+$ as a 
  function of the pion mass squared (from \cite{Brommel:2007xd}).}
  \label{BTn0_pion_over_mPi_t0_linear_log_v5_QCDSF}
     \end{minipage}
 \end{figure}

In this section, we present first lattice QCD results for the pion tensor form factor $B^\pi_{T10}(t=-Q^2)$
as defined in Eq.~\ref{PionTensorFF} and the corresponding next highest moment, $B^\pi_{T20}(t=-Q^2)$, Eq.~(\ref{PionTensorGFFn2}).
Based on the same set of $n_f=2$ improved Wilson fermion and Wilson gauge action 
ensembles that were used for the study of the pion form factor
discussed in section \ref{pionFFs}, Figs.~\ref{FFpion_global_QCDSF} and \ref{FFpion_extrapol_QCDSF}, 
QCDSF/UKQCD recently computed $B^\pi_{Tn0}(t)$ 
for $n=1,2$ in a range of $-t\approx 0.3,\ldots,3\GeV^2$ \cite{Brommel:2007xd}. 
The underlying local tensor operators
have been non-perturbatively renormalized, and the results were transformed to the
$\MSbar$ scheme at a scale of $4\GeV^2$. A typical result for the $t$-dependence
of the pion tensor form factor $B^\pi_{T10}(t)$ 
and the GFF $B^\pi_{T20}(t)$ is shown in Fig.~\ref{BTn0_pion_b5p29k13590_v2_QCDSF}
for a pion mass of $\approx 600\MeV$. Using a $p$-pole ansatz of the form
\bea
\label{ppole}
F(t)=\frac{F(0)}{\left(1-\frac{t}{p\,m_p^2}\right)^p}\,,
\eea
with free parameters $F(0)$ and $m_p^2$, the results for $B^\pi_{Tn0}(t)$ have been fitted and extrapolated
to $t=0$. Since the pion tensor GFFs $B^\pi_{Tn0}(t)$ are defined (see, e.g., Eq.~\ref{PionTensorFF}) including a
normalization factor of $1/m_\pi$, they have to be proportional to $m_\pi$, and therefore vanish in the 
chiral limit,
as predicted by ChPT \cite{Diehl:2006js}. 
It is therefore sensible to study the chiral
extrapolations of $B^\pi_{Tn0}(\t0)/m_\pi$, which are presented in 
Fig.~\ref{BTn0_pion_over_mPi_t0_linear_log_v5_QCDSF} as functions of $m_\pi^2$, obtained for $p=1.6$.
The lattice results are, for the accessible pion masses from $\approx400$ to $\approx1000\MeV$,
within errors well compatible with a linear dependence on $m_\pi^2$. 
In order to account for
uncertainties due to possible finite volume effects, an ansatz of the form 
$B^\pi_{Tn0}(\t0)/m_\pi=c_0+c_1m_\pi^2+c_2m_\pi^2\exp(-m_\pi L)$, where $L$
is the spatial lattice extent, with free parameters $c_i$, has been used to fit the lattice data points 
separately for $n=1,2$ and extrapolate to the physical point. This fit is represented by the dark shaded band
in Fig.~\ref{BTn0_pion_over_mPi_t0_linear_log_v5_QCDSF}. 

Although predictions from 1-loop ChPT \cite{Diehl:2006js} are strictly speaking not applicable at 
the accessible pion masses and volumes,
they still might give a qualitative idea about the uncertainties related to the chiral extrapolation.
To this extent, additional fits based on 1-loop ChPT,
\begin{eqnarray}
\label{BTn0ChPT}
\hat B^\pi_{Tn0}(\t0,m_\pi)& = &
\hat B^{\pi,0}_{Tn0} 
\left( 1 + c_1^{(n)}\frac{m_{\pi}^2}{(4\pi)^2f_{\pi}^2}
 \left\{\ln \frac{m_{\pi}^2}{\lambda^2}  + 1 \right\} \right)
  + c_2^{(n)} m_{\pi}^2 + c_3^{(n)} m_\pi^2 e^{-m_\pi L}\,,
\end{eqnarray}
with chiral coefficients $c_1^{(1)}=1/2$, $c_1^{(2)}=-3/2$, and
where $\hat B^\pi_{Tn0}=B^\pi_{Tn0}/m_\pi$ and $\hat B^{\pi,0}_{Tn0}=\hat B^\pi_{Tn0}(m_\pi\eql0)$, 
including finite volume correction terms $c_3^{(n)}m_\pi^2\exp(-m_\pi L)$ that
were added by hand, have been performed to the lattice data points for $m_\pi<650\MeV$.
%
%
The results are represented by the lighter shaded band in Fig.~\ref{BTn0_pion_over_mPi_t0_linear_log_v5_QCDSF}.
From the linear chiral extrapolation in Fig.~\ref{BTn0_pion_over_mPi_t0_linear_log_v5_QCDSF},
already a rather large value of $B^\pi_{T10}(\t0)/m_\pi=1.54(24)\GeV^{-1}$ was found at the physical point,
while the extrapolation based on 1-loop ChPT tends to give an even larger central value.
Analogously to the case nucleon tensor GFF $\overline B_{T10}$ introduced in section \ref{sec:FFsPDFsGPDs},
one may define a tensor anomalous magnetic moment of the pion, $\kappa_T^\pi=B^\pi_{T10}(\t0)$, 
for which a value of $\kappa_T^\pi=0.215(33)$ is obtained at the physical pion mass
from the linear chiral extrapolation in Fig.~\ref{BTn0_pion_over_mPi_t0_linear_log_v5_QCDSF}.
Also based on a linear chiral extrapolation in $m_\pi^2$, a value of $m_p = 0.756(95)\GeV$ 
was obtained for the corresponding $p$-pole mass at the physical point, with $p=1.6$.

In the case of $B^\pi_{T20}(\t0)$, the result of the linear extrapolation (represented by the dark shaded band),
is $B^\pi_{T20}(0)/m_\pi=0.277(71)\GeV^{-1}$ at the physical pion mass 
and the infinite volume limit. A value of $m_p = 1.130(265)\GeV$ was found for the corresponding 
$p$-pole mass with $p=1.6$, obtained from a linear chiral extrapolation in $m_\pi^2$ to the physical
pion mass.
The fit based on 1-loop ChPT, Eq.~\ref{BTn0ChPT},
shown by the
light shaded band in Fig.~\ref{BTn0_pion_over_mPi_t0_linear_log_v5_QCDSF},
clearly gives a much smaller value for $B^\pi_{T20}(\t0)$ at the physical point, nearly
compatible with zero within errors. We note again, however, that the results from
a 1-loop ChPT fit at such large pion masses cannot be regarded as reliable, 
and only provide an indication for uncertainties in the chiral extrapolation.
%
%
\begin{figure}[t]
   \begin{minipage}{0.62\textwidth}
      \centering
           \includegraphics[angle=0,width=0.9\textwidth,clip=true]{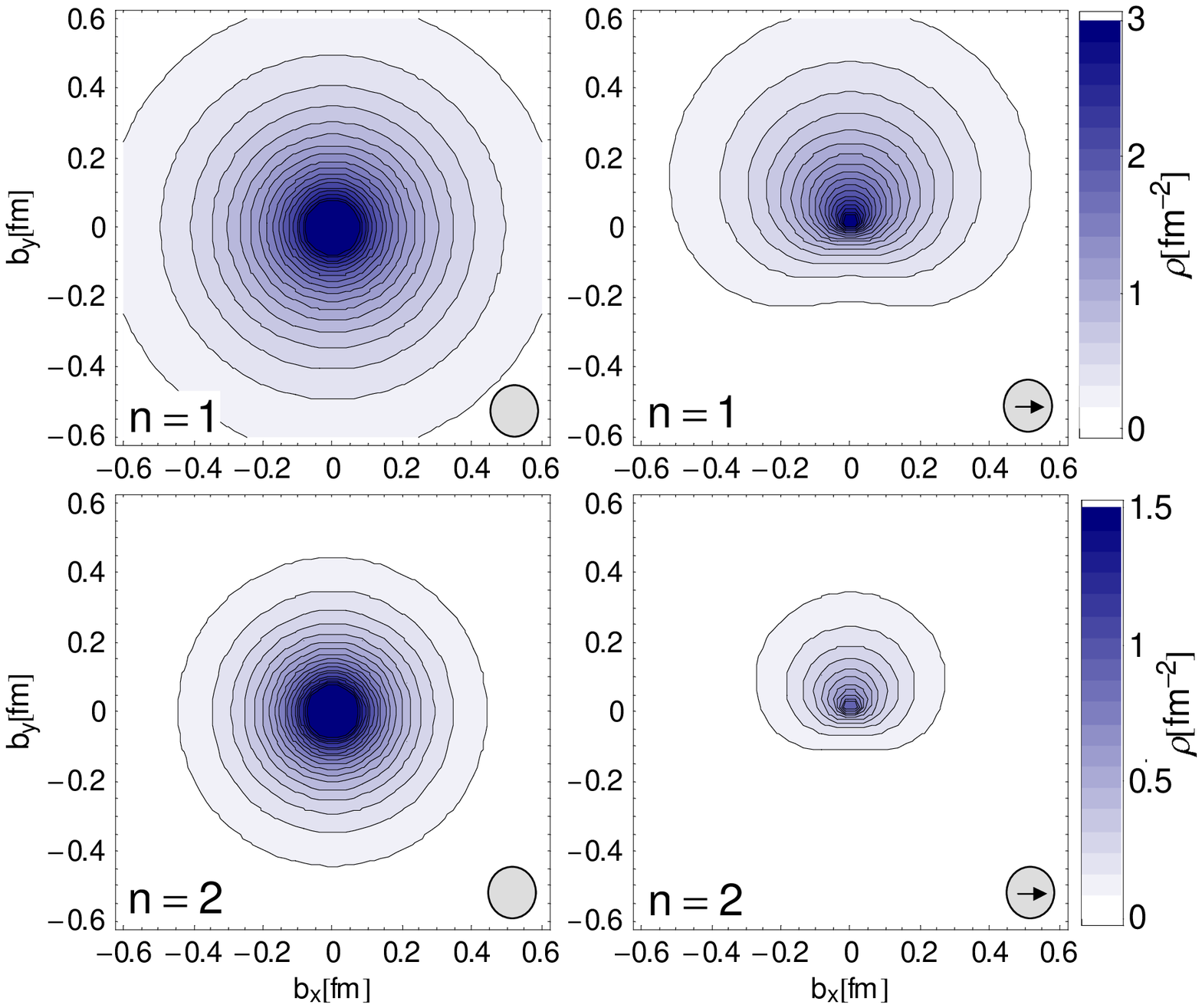}
  \caption{Density of transversely polarized up-quarks in the $\pi^+$ (from \cite{Brommel:2007xd}).}
  \label{Pion_densities_v3}
     \end{minipage}
     \hspace{0.5cm}
    \begin{minipage}{0.34\textwidth}
      \centering
          \includegraphics[width=0.9\textwidth,clip=true]{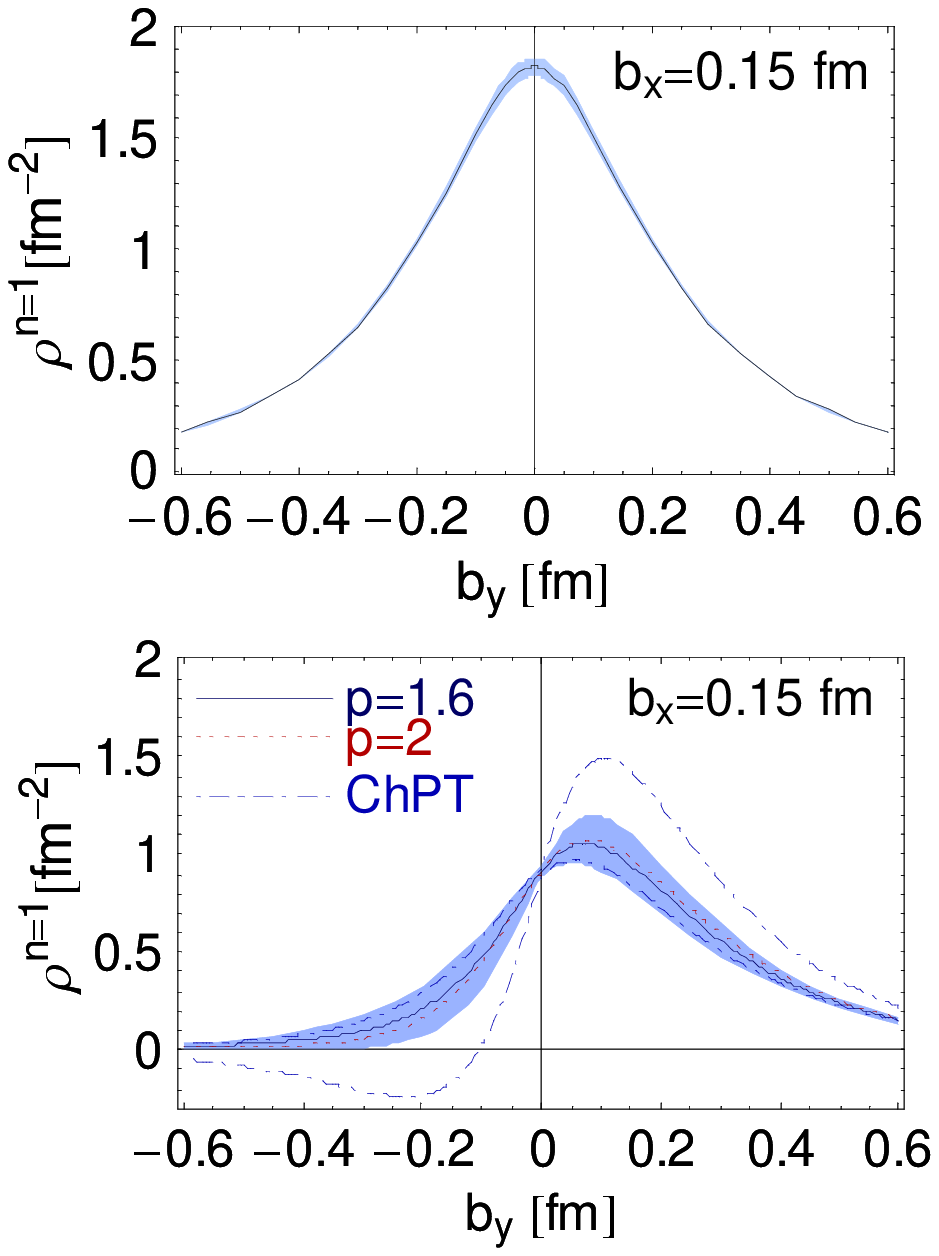}
  \caption{Density profiles for transversely polarized up-quarks in the $\pi^+$ (from \cite{Brommel:2007xd}).}
  \label{Pion_profile_v2}
     \end{minipage}
 \end{figure}
%
%
As we will explain in the following, the lattice results for the moments of the 
pion vector and tensor GPDs may be used for a first study of the spin structure of the pion.

It has been noted in \cite{Brommel:2007xd} that the $x^{n-1}$-moments of the density 
of transversely polarized quark with transverse spin $s_\perp$ in a pion is given by
\bea
\rho^{n}(b_\perp,s_\perp)
 = \int_{-1}^{1} dx\, x^{n-1} \rho(x,b_\perp,s_\perp) 
= \frac{1}{2}\left(\ms A^\pi_{n0}(b_\perp^2) 
  - \frac{s_\perp^i \epsilon^{ij}\ms b_\perp^j}{m_\pi}\,
    \frac{\partial}{\partial b_\perp^2}\, B_{Tn0}^\pi(b_\perp^2)
  \,\right) \,,
\label{PionDensity}
\eea
where the GFFs $A^\pi_{n0}(b_\perp^2)$ and $B_{Tn0}^\pi(b_\perp^2)$ 
in impact parameter space are related to the momentum-space 
GFFs $A^\pi_{n0}(t)$ and $B_{Tn0}^\pi(t)$ by a Fourier-transformation as in Eq.~(\ref{Impact1}).
A numerical evaluation of the density $\rho^{n}(b_\perp,s_\perp)$ using the lattice results
requires representations of the GFFs as functions of $b_\perp$. To this end, 
the $p$-pole parametrization in Eq.~(\ref{ppole}) was Fourier-transformed
to impact parameter space and used together with the numerical lattice
results for the forward values, $F(0)$, and the respective $p$-pole masses, $m_p$,
obtained from $p$-pole fits to the lattice data points and subsequent chiral
extrapolations to the physical pion mass.
For the pion form factor $A^\pi_{10}(t)=F_\pi(t)$, the results from \cite{Brommel:2006ww}, 
discussed in section \ref{pionFFs}, Figs~\ref{FFpion_global_QCDSF} and \ref{FFpion_extrapol_QCDSF},
were used, giving a monopole mass of $m_\text{mono}=0.727\pm0.054_\text{stat+vol+sys}\GeV$,
in addition to the renormalization condition $F_\pi(\t0)=1$ as required by charge conservation.
The lattice results for the GFF $A_{20}^\pi(t)$ were also parametrized using a monopole
($p=1$) ansatz, with a forward value of $A_{20}^\pi(\t0)\approx0.26$
and a monopole mass of $m_\text{mono}\approx1.2\GeV$ \cite{Brommel:2005ee}.
For the GFFs $B_{Tn0}^\pi$, the results of the linear chiral extrapolations 
of $B_{Tn0}^\pi(t=0)/m_\pi$ and the corresponding $p$-pole masses,
presented above, were employed \cite{Brommel:2006zz,Brommel:2007xd}. 

Figure \ref{Pion_densities_v3} shows the final results for the $n=1$ (upper part of figure)
and $n=2$ (lower part of figure) moments of the density of up-quarks in a $\pi^+$.
In contrast to the unpolarized, symmetric densities on the left, the
densities for up-quarks with transverse spin in the $x$-direction, $s_\perp=(1,0)$,
on the right in Fig.~\ref{Pion_densities_v3} are significantly deformed
due to the dipole-terms $\propto s_\perp^i \epsilon^{ij} b_\perp^j {\partial}_{b_\perp^2}B_{Tn0}^\pi(b_\perp^2)$ 
in Eq.~\ref{PionDensity} in combination with the large non-zero values that 
were found in particular for the tensor GFFs $B_{Tn0}^\pi$.

In order to see how the various statistical and systematic uncertainties
of the analysis affect the deformation, Fig.~\ref{Pion_profile_v2}
shows profile plots of the $n=1$-densities for $b_x=0.15\fm$ as
functions of $b_y$. While the shaded bands represent the statistical uncertainties
from the linear chiral extrapolations in $m_\pi^2$,
the dash dotted lines represent the results for the minimal
and maximal values obtained for $B_{T10}^\pi(t=0)$
from the 1-loop ChPT extrapolation based on Eq.~\ref{BTn0ChPT} and displayed 
in Fig.~\ref{BTn0_pion_over_mPi_t0_linear_log_v5_QCDSF}.
Even for the smallest values of $B_{T10}^\pi(t=0)$, there is still
a deformation visible. On the other hand, for the maximal values 
the density becomes negative for $b_y\lessapprox-0.1\fm$, which is unphysical.
This demonstrates that 1-loop ChPT applied to pion 
masses of $m_\ge400\MeV$ is quantitatively unreliable
and in the best case only gives a rough idea about the 
trend of the chiral extrapolation.

In summary, the large values obtained by QCDSF \cite{Brommel:2007xd} in particular for the
tensor anomalous magnetic moment of the pion,
$k^\pi_T=B_{T10}^\pi(\t0)$, as displayed in Fig.~\ref{BTn0_pion_over_mPi_t0_linear_log_v5_QCDSF},
point towards a surprisingly non-trivial transverse spin structure of the pion,
as demonstrated by the deformed densities in Fig.~\ref{Pion_densities_v3}.

Clearly, more precise data from lattice QCD at lower pion masses
and in larger volumes, together with improved results from ChPT, 
are required before quantitative predictions can be made for the densities at the physical point.

For a discussion of 
a possible relevance of these lattice results 
for the transverse momentum dependent Boer-Mulders function 
$h_1^{\perp,\pi}(x,k_\perp)$ for pions and
azi\-muthal asymmetries in semi-inclusive
deep inelastic scattering and Drell-Yan production, we refer to \cite{Brommel:2007xd} and references therein.

\subsubsection{Nucleon}
%
\begin{figure}[t]
      \centering
          \includegraphics[angle=-90,width=0.9\textwidth,clip=true,angle=0]{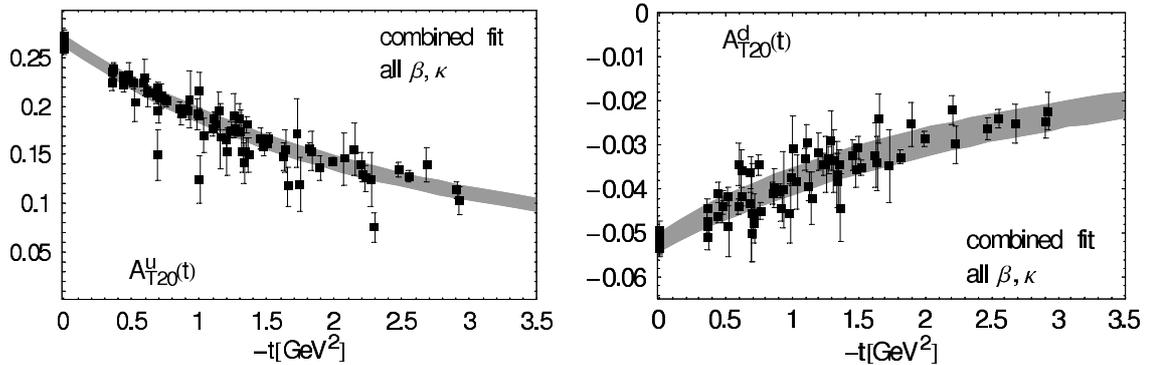}
  \caption{Tensor form factor $A_{T20}(t)$ (from \cite{Gockeler:2005cj}).}
  \label{AT20_QCDSF}
 \end{figure}
In the same framework of simulations with $n_f=2$ flavors of clover-improved Wilson fermions,
QCDSF/UKQCD has also performed calculations of the tensor GFFs of the nucleon \cite{Gockeler:2005cj}.
The relevant lattice tensor operators have been non-perturbatively renormalized,
and all results were transformed to the $\MSbar$ scheme at a scale of $\mu^2=4\GeV^4$.
An overview of the up- and down-quark contributions to the GFF $A_{T20}(t)$
is given in Fig.~\ref{AT20_QCDSF}, based on the same lattice parameters and ensembles
as discussed in relation with the corresponding results for $A_{T10}(t)$ in Fig.~\ref{AT10_QCDSF}.
Using a generalized dipole ansatz of the form Eq.~\ref{chiralcontdipole}, 
a simultaneous global fit to the full $t$-, $m_\pi$- and $a$-dependence of the lattice
results was performed, including all available ensembles.
Based on the fit, the lattice data points were shifted to $a=0$ and $m_\pi=140\MeV$,
and the result is given in Fig.~\ref{AT20_QCDSF}.
We note that the forward values for $A^q_{T20}(\t0)=\langle x\rangle_{\delta q}$ 
obtained in the framework of this analysis have already been presented above at the end of section \ref{sec:nuclPDFs},
Fig.~\ref{xdelta_AT20_m_pi_v2_QCDSF}.
From the simultaneous global fit, represented by the error band in Fig.~\ref{AT20_QCDSF},
the values $A^u_{T20}(t=0)=\langle x\rangle_{\delta u}=0.268(6)$ with $m^u_D=2.31(7)\GeV$
and $A^d_{T20}(t=0)=\langle x\rangle_{\delta d}=-0.052(2)$ with $m^d_D=2.45(17)\GeV$
were found at the physical point.
%

%
\begin{figure}[t]
   \begin{minipage}{0.48\textwidth}
      \centering
          \includegraphics[angle=0,width=0.9\textwidth,clip=true,angle=0]{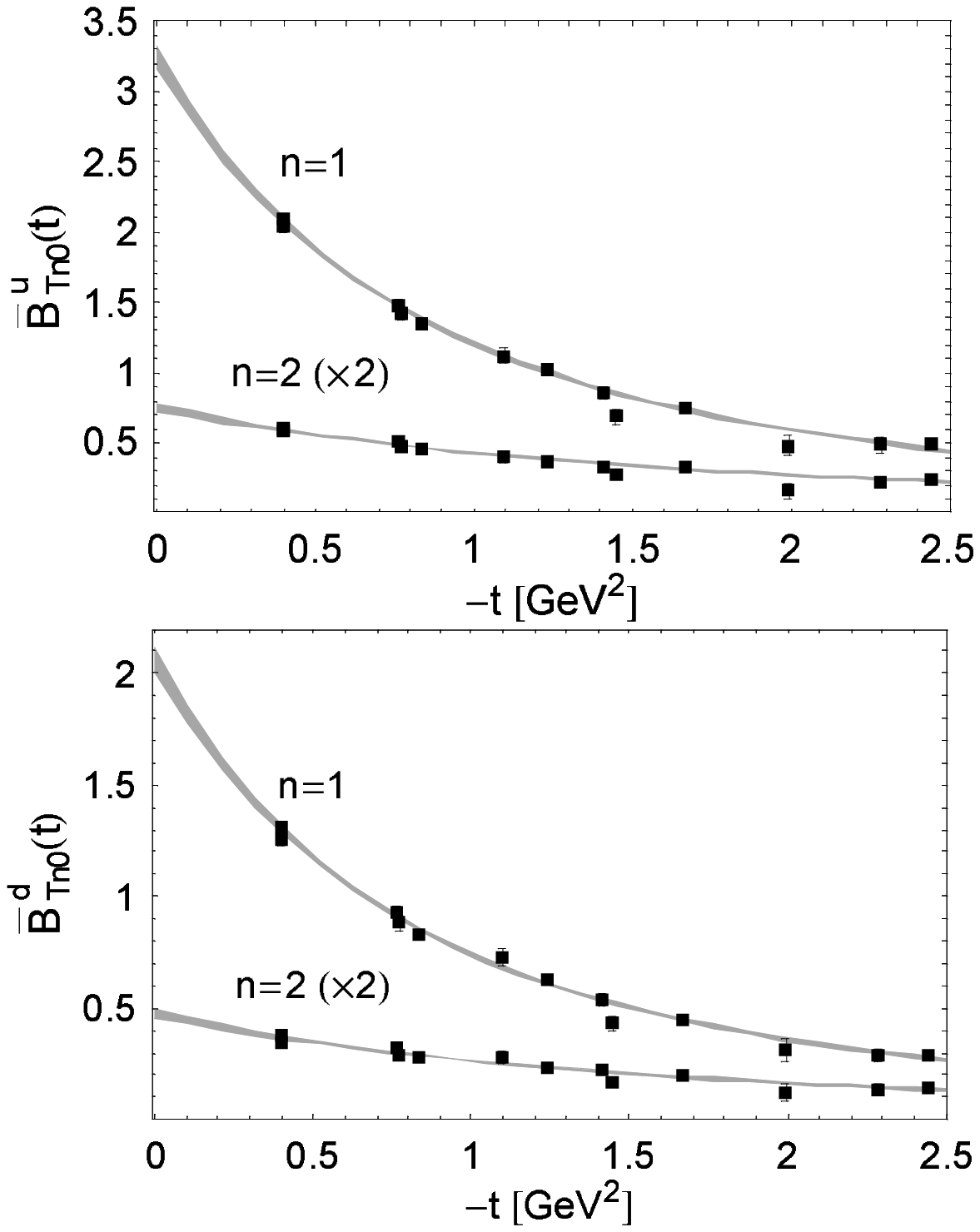}
  \caption{Tensor form factor $\overline B_{T10}(t=Q^2)$ (from \cite{Gockeler:2006zu}).}
  \label{BbarT1020_ud_b5p29k13590_QCDSF}
    \vspace{0.5cm}
     \end{minipage}
     \hspace{0.5cm}
    \begin{minipage}{0.48\textwidth}
      \centering
          \includegraphics[angle=0,width=0.9\textwidth,clip=true,angle=0]{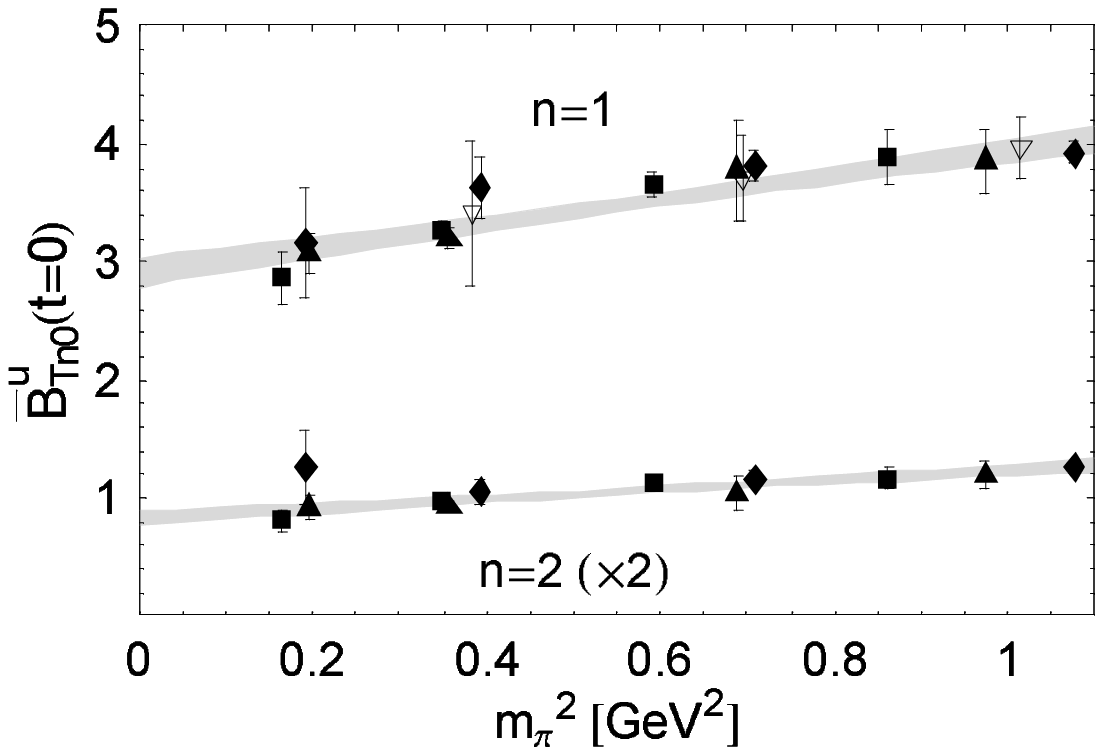}
  \caption{Pion mass dependence of the form factor $\overline B_{T10}(Q^2=0)$ (from \cite{Gockeler:2006zu}).}
  \label{BbarT1020_up_vs_mPi2_QCDSF}
     \end{minipage}
 \end{figure}
%

%

First results for the nucleon tensor form factor $\overline B_{T10}(t=-Q^2)$, Eq.~\ref{NuclTensor3},
and the GFF $\overline B_{T20}(t)$, Eq.~\ref{NuclTensor5},
in the same lattice framework as before, i.e. for simulations with $n_f=2$ flavors of clover-improved Wilson 
fermions, were presented by the QCDSF-UKQCD collaboration in \cite{Gockeler:2006zu}.
Examples for the $t$-dependences of the tensor form factor $\overline B_{T10}(t)$ and the
GFF $\overline B_{T20}(t)$ for up- and down-quarks are presented in
Fig.~\ref{BbarT1020_ud_b5p29k13590_QCDSF} for a pion mass of $\approx600\MeV$
and a coupling of $\beta=5.29$ \cite{Gockeler:2006zu}, 
showing a quite remarkable statistical precision.
The lattice data points for $\overline B_{Tn0}(t)$ were fitted with a $p$-pole ansatz, 
Eq.~\ref{ppole}, with $p=2.5$, as given by the shaded bands, and extrapolated to $t=0$.
Interestingly, the extrapolated forward values $\overline B_{Tn0}(\t0)$ drop substantially going from $n=1$ to $n=2$,
with a ratio of $\overline B_{T20}(\t0)/\overline B_{T10}(\t0)\approx0.12$ for both flavors.

Figure \ref{BbarT1020_up_vs_mPi2_QCDSF} shows the pion mass dependence
of the $\overline B^u_{Tn0}(t=0)$, together with linear extrapolations in $m_\pi^2$ to the chiral limit
represented by the error bands. 
At this point we note that the pion mass dependent nucleon mass $m_N(m_\pi)$ has been used in the
extraction of $\overline B_{Tn0}$ based on the parametrization in Eq.~\ref{NuclTensor3},
which clearly influences the slope in $m_\pi^2$ in Fig.~\ref{BbarT1020_up_vs_mPi2_QCDSF}.
Although results from HBChPT to leading 1-loop order 
for the pion mass and $t$-dependence of the tensor GFFs are available  
\cite{Diehl:2006ya,Diehl:2006js,Ando:2006sk}, they are most likely
not applicable at the accessible pion masses of $\approx400\MeV$ and larger, and
were therefore not utilized for the chiral extrapolation of the lattice data.

Concerning the form factor $\overline B_{T10}$, it has been suggested \cite{Burkardt:2005hp} 
to identify its forward value with a tensor anomalous magnetic moment, $\kappa_T=\overline B_{T10}(\t0)$, 
in analogy to the standard nucleon anomalous magnetic moment $\kappa=F_2(\t0)$.
The linear chiral extrapolation in $m_\pi^2$, represented by the shaded band in 
Fig.~\ref{BbarT1020_up_vs_mPi2_QCDSF}, gives for up-quarks $\kappa^u_T=2.93(13)$ 
and for down-quarks $\kappa^d_T=1.90(9)$. These values, which may be compared to
the up- and down quark contributions to the 
anomalous magnetic moments, $\kappa^u_{\text{exp}}\approx1.67$ and
$\kappa^d_{\text{exp}}\approx-2.03$, are remarkably large and, in contrast
to $\kappa^{u,d}$, also of the same sign.
In the case of the GFF $\overline B_{T20}$, the linear extrapolation of the forward values
to the physical pion mass gives $\overline B^u_{T20}(t=0)=0.420(31)$ and $\overline B^d_{T20}(t=0)=0.260(23)$. 
%
%
%

%
%

%

%

%
\begin{figure}[t]
   \begin{minipage}{0.48\textwidth}
      \centering
          \includegraphics[angle=0,width=0.99\textwidth,clip=true,angle=0]{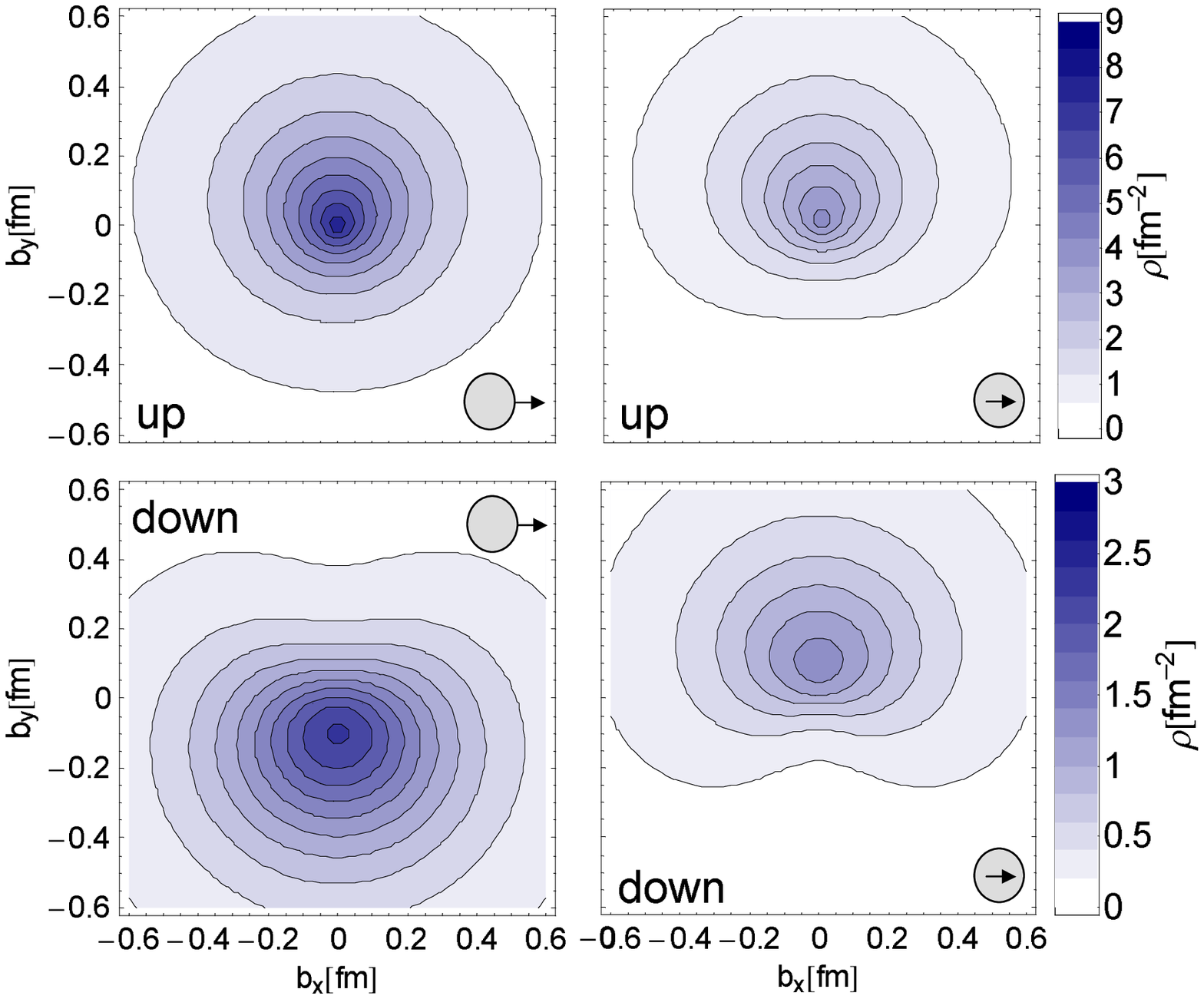}
  \caption{Transverse spin densities of quarks in the nucleon for $n=1$ (from \cite{Gockeler:2006zu}).}
  \label{densities_n1_ud_QCDSF_v2}
     \end{minipage}
     \hspace{0.3cm}
    \begin{minipage}{0.48\textwidth}
      \centering
          \includegraphics[angle=0,width=0.99\textwidth,clip=true,angle=0]{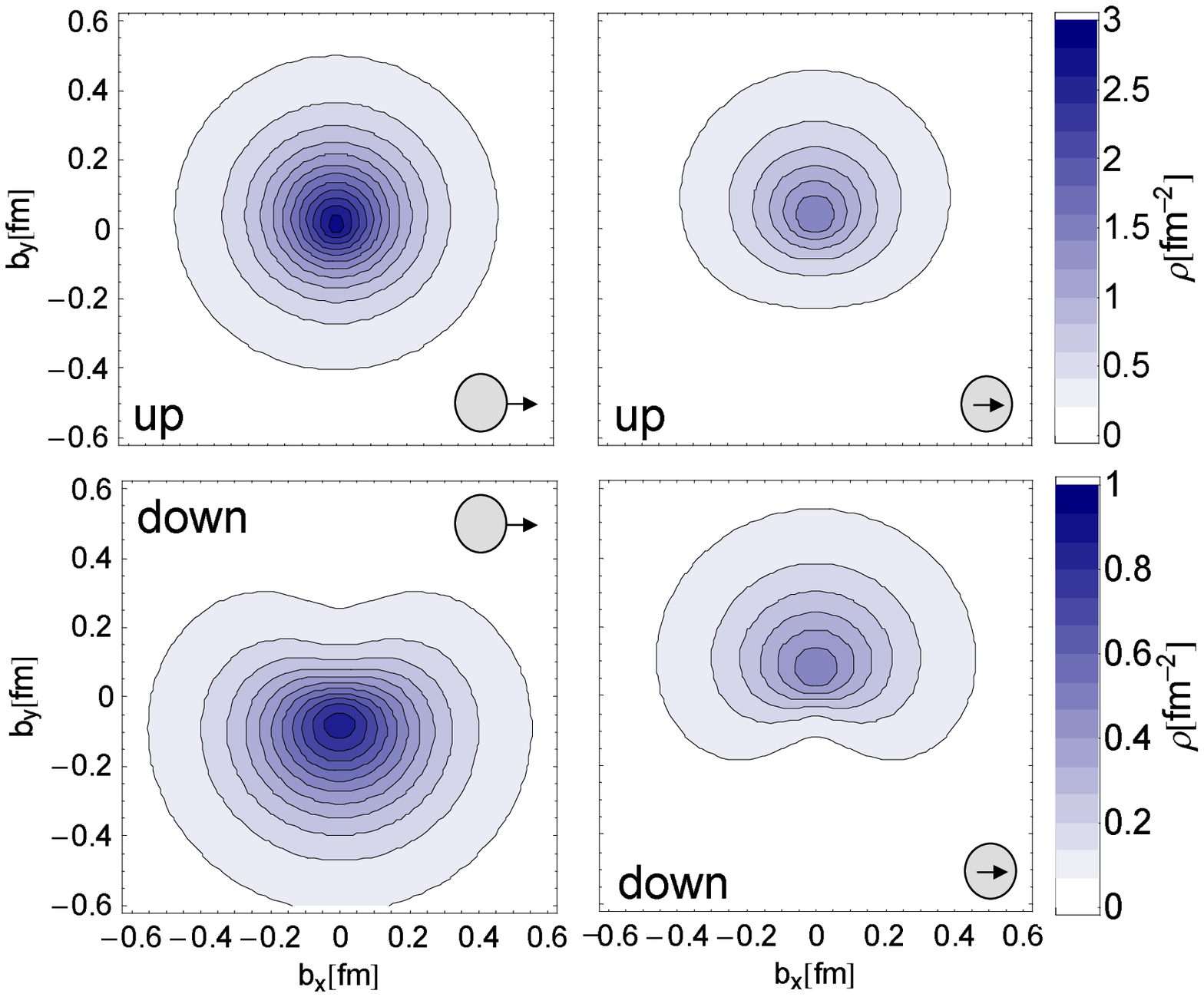}
  \caption{Transverse spin densities of quarks in the nucleon for $n=2$ (from \cite{Gockeler:2006zu}).}
  \label{densities_n2_ud_QCDSF_v2}
     \end{minipage}
 \end{figure}

The tensor generalized form factors in particular provide important information about
the nucleon transverse spin structure. To see this more clearly, we note that the $x^{n-1}$-moments of the spin
density of transversely polarized quarks in the nucleon \cite{Diehl:2005jf}, Eq.~\ref{density1},
are given by
\bea
\rho^{n}(b_\perp,s_\perp,S_\perp)&=&\int_{-1}^{1}dx\,x^{n-1}\rho(x,b_\perp,s_\perp,S_\perp) \nonumber\\
&=& \frac{1}{2}\left\{A_{n0}(b_\perp^2)
+ s_\perp^i S_\perp^i \left( A_{Tn0}(b_\perp^2)
- \frac{1}{4m_N^2} \Delta_{b_\perp} \widetilde{A}_{Tn0}(b_\perp^2) \right) \right.  \nonumber \\
&+& \frac{ b_\perp^j \eps ^{ji}}{m_N} \left( S_\perp^i B_{n0}'(b_\perp^2)
+ s_\perp^i \overline{B}_{Tn0}'(b_\perp^2) \right)
\nonumber \\
&+& \left. s_\perp^i ( 2 b_\perp^i b_\perp^j
- b_\perp^2 \delta^{ij} ) S_\perp^j \frac{1}{m_N^2} \widetilde{A}_{Tn0}''(b_\perp^2)
\right\},\,
\label{density2}
\eea
where $f'(b_\perp^2)=\partial_{b_\perp^2}f(b_\perp^2)$, and where $s_\perp$ and $S_\perp$ denote
the transverse spin vectors of the quark and nucleon, respectively.
For fixed $n$, the moment $\rho^{n}(b_\perp,s_\perp,S_\perp)$ of the density is therefore
fully determined by the five unpolarized (vector) and tensor GFFs
$A_{n0}(b_\perp^2)$, $B_{n0}(b_\perp^2)$, $A_{Tn0}$, $\overline{B}_{Tn0}(b_\perp^2)$ and $\widetilde{A}_{Tn0}(b_\perp^2)$ and their respective derivatives in impact parameter space.

To make use of the lattice results obtained in momentum ($t$-) space, the $p$-pole parametrization
in Eq.~(\ref{ppole}) was Fourier-transformed to impact parameter space and used in 
combination with the numerical lattice results for the forward values of the GFFs, 
$F(\t0)$, and the respective $p$-pole masses, $m_p$, as obtained obtained from the $p$-pole fits.
The lattice results obtained by QCDSF for the relevant (generalized) form factors for $n=1$
and $n=2$, that were (in parts) discussed in 
section \ref{FFradii} for $A_{10}(t=-Q^2)=F_1(Q^2)$, $B_{10}(t=-Q^2)=F_2(Q^2)$,
section \ref{sec:tensorFFs} for $A_{T10}(t)$,
section \ref{sec:EMT} for $A_{20}(t)$, $B_{20}(t)$, and finally
in this section for $A_{T20}(t)$ and $\overline{B}_{Tn0}(t)$,
all linearly extrapolated in $m_\pi^2$ to the physical pion mass,
were employed for a numerical study of the density of 
quarks in the nucleon \cite{Gockeler:2006zu}.

Densities for $n=1$ are displayed in Fig.~\ref{densities_n1_ud_QCDSF_v2} and 
for $n=2$ in Fig.~\ref{densities_n2_ud_QCDSF_v2}\footnote{
Note that $\rho^{n=1}$ corresponds to the difference of quark- and anti-quark-densities,
while $\rho^{n=2}$ is given by a sum and therefore must be strictly positive 
(see Eqs.~(\ref{VecMoments1},\ref{TensorMoments1})).}.
For the case of unpolarized quarks in a transversely polarized nucleon with transverse
spin $S_\perp=(1,0)$ on the left in Fig.~\ref{densities_n1_ud_QCDSF_v2}, the densities
are deformed in positive and negative $b_y$-direction for up- and down quarks, respectively.
This is a consequence of the dipole-term $\propto b_\perp^j \eps ^{ji}S_\perp^i B_{10}'(b_\perp^2)$
in Eq.~\ref{density2} and the large positive and negative values
for the anomalous magnetic moment contributions of up- and down-quarks,
being of the order of 
$\kappa^u=B^u_{10}(t=0)\approx1.3$ and $\kappa^d=B^d_{10}(t=0)\approx-1.5$.
in this lattice study (somewhat below the
values from experiment, $\kappa^u\approx1.67$ and $\kappa^d\approx-2.03$).

This type of dipole distortion of quark densities in a transversely
polarized nucleon has first been noted and numerically studied
in the framework of a GPD model by Burkardt \cite{Burkardt:2002ks,Burkardt:2003uw},
where also a relation to measurable transverse single spin asymmetries in semi-inclusive deep inelastic scattering
was established. 

An even stronger deformation of the density is observed for the case
of transversely polarized quarks in an unpolarized nucleon, as shown
on the right hand side in Fig.~\ref{densities_n1_ud_QCDSF_v2} for $s_\perp=(1,0)$, $S_\perp=(0,0)$.
The deformation, which can be traced back to the dipole-term
$b_\perp^j \eps ^{ji}s_\perp^i \overline{B}_{Tn0}'(b_\perp^2)$ in Eq.~\ref{density2},
is in this case directed in positive $b_y$-direction both for up- and
for down-quarks due to the large and positive values that were obtained
for the tensor form factor $\overline{B}^{u,d}_{T10}$ in the lattice calculation, as shown in 
Figs.~\ref{BbarT1020_ud_b5p29k13590_QCDSF} and \ref{BbarT1020_up_vs_mPi2_QCDSF}.
The effect of the dipole-like distortions for transversely polarized up-quarks in a transversely
polarized nucleon has also been discussed in \cite{Gockeler:2005cd}.

\subsection{Discussion and summary}
\label{sec:PDFsSummary}

Since the advent of lattice QCD studies of hadron structure in the
1980's, a large number of studies of the lowest 
moments of PDFs of the pion, $\rho$-meson and nucleon have been presented.
By now, a large number of quenched, and an increasing number of unquenched, 
lattice results are available for
the unpolarized momentum fraction, $\langle x^{}\rangle_{q}$, 
carried by the quarks in the 
pion, $\rho$-meson and nucleon, the polarized/helicity quark momentum fractions
$\langle x^{}\rangle_{\Delta q}$ of the $\rho$ and nucleon, and the $x$-moment of 
the transversity distribution, $\langle x^{}\rangle_{\delta q}$, 
of the nucleon.

\begin{figure}[t]
      \centering
  \includegraphics[angle=0,width=0.6\textwidth,clip=true,angle=0]{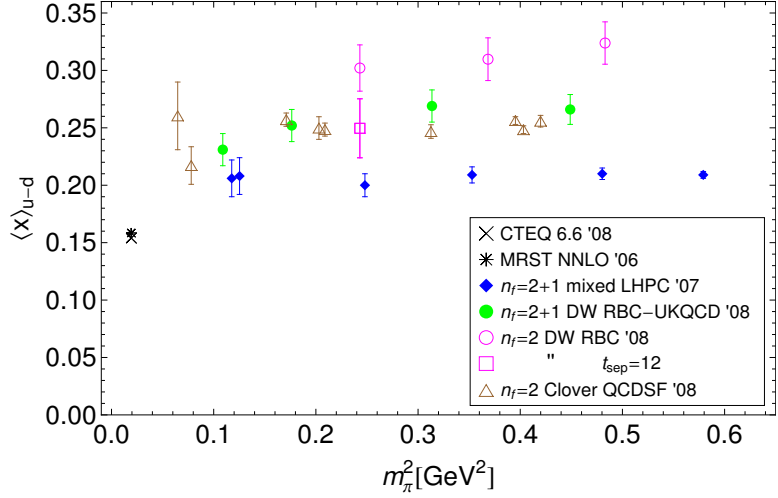}
  \caption{Overview of dynamical lattice QCD results for the isovector quark momentum fraction
  in the $\MSbar$ scheme at a scale of $4\GeV^2$.}
  \label{overview_x} 
\end{figure}

Despite the enormous efforts by many groups during the last 20 years,
one of the biggest puzzles in lattice hadron structure calculations is
still unsolved: In most of the past and current lattice simulations,
the momentum fraction carried by up- minus down-quarks in
the nucleon is approximately $50\%$ above the value from global PDF-analyses,
as shown in the overview plot of dynamical lattice results, Fig.~\ref{overview_x}.

Many possible causes have been separately studied and ruled out or excluded over the years,
including the quenched approximation, perturbative (versus non-perturbative)
operator renormalization, and also in many cases discretization errors and finite volume effects. 
Although pion masses as low as $350\MeV$ have been reached recently, the most likely solution is that 
they are still too large for any chiral dynamics to play a significant role.
It has to be expected that a chiral bending towards the experimental number will only occur
for $m_\pi\lessapprox250,\ldots,300\MeV$.  
In any case, it is remarkable how little the lattice results for $\langle x^{}\rangle_{u-d}$ 
(in a given simulation framework) depend at all on the pion mass in the region above $\approx350\MeV$.
%

Notably, there are substantial discrepancies visible in the overall normalizations of some of the
lattice results in Fig.~\ref{overview_x}. In particular, the $n_f=2$ domain wall results
from RBC are on average above most other data points,
and the results from the $n_f=2+1$ mixed action calculation by LHPC lie exceptionally low
compared to the rest. Possible explanations that were proposed in the literature are
related to the semi-non-perturbative operator
renormalization in the case of the LHPC analysis, and contaminations from excited states.
Although the chiral extrapolation of the LHPC lattice data in Fig.~\ref{A20_umd_SimulFit}, section \ref{sec:EMT},
based on covariant baryon ChPT looks very promising, the results have to be viewed with great caution
as long as the normalization issues have not been resolved.

Furthermore, for a fully quantitative comparison with experimental results,
the simulations have to be pushed to even lower pion masses, $m_\pi\lessapprox300\MeV$, 
where results from chiral perturbation theory can be more safely applied and used for 
extrapolations to the physical point.

The lowest moments of unpolarized, polarized and tensor/transversity generalized parton distributions 
of the nucleon have been studied quite extensively in unquenched lattice QCD in recent years. 
This includes in particular the nucleon form factors of the energy momentum tensor, which are essential for
an understanding of the nucleon spin structure and the decomposition of the nucleon spin 
in form of Ji's sum rule, Eqs.~(\ref{SpinSumrule1},\ref{SpinSumrule2}).

An overview of results for quark spin and (orbital) angular momentum contributions
from $n_f=2$ clover-improved Wilson fermion simulations by QCDSF \cite{Brommel:2007sb}
and $n_f=2+1$ mixed action calculations by LHPC \cite{Hagler:2007xi} is given in Table \ref{slj}, 
in the $\MSbar$-scheme for $\mu^2=4\GeV^2$.
\begin{table}[t]
\begin{minipage}{0.48\textwidth}
{\rule[-3mm]{0mm}{8mm}LHPC \cite{Edwards:2005ym,Hagler:2007xi}} 
\centering
  \begin{tabular}{|c|c|c|c|}
       \hline 
      & $\frac{1}{2}\Delta\Sigma$ & $L$ & $J=\frac{1}{2}\Delta\Sigma+L$        
     \\ \hline
     $u$ &               $0.409(34)$ & $-0.195(44)$ & $0.214(27)$ 
     \\ \hline
     $d$ &               $-0.201(34)$ & $0.200(44)$ & $-0.001(27)$
     \\ \hline
     $u+d$ &              $0.207(28)$ & $0.005(52)$  & $0.213(44)$ 
    \\ \hline
  \end{tabular}
\end{minipage}
     \hspace{0.1cm}
\begin{minipage}{0.48\textwidth}
{\rule[-3mm]{0mm}{8mm}QCDSF-UKQCD \cite{Khan:2006de,Brommel:2007sb}} 
\centering
  \begin{tabular}{|c|c|c|}
       \hline 
             $\frac{1}{2}\Delta\Sigma$ & $L$ & $J=\frac{1}{2}\Delta\Sigma+L$          
     \\ \hline
              $0.428(31)$ & $-0.198(32)$ & $0.230(8)$ 
     \\ \hline
              $-0.227(31)$ & $0.223(32)$ & $-0.004(8)$
     \\ \hline
              $0.201(24)$ & $0.025(27)$  & $0.226(13)$ 
    \\ \hline
  \end{tabular}
\end{minipage}
\caption{Overview of spin and angular momentum contributions of quarks to the 
  spin of the proton. All results have been extrapolated to the physical pion mass using results from ChPT
  and are given in the $\MSbar$-scheme at $\mu=2\GeV$. The numbers have been collected from
  \cite{Edwards:2005ym,Hagler:2007xi} for the $n_f=2+1$ mixed action calculation of LHPC
  (table on the left) and \cite{Khan:2006de,Brommel:2007sb} for the 
  $n_f=2$ clover-improved Wilson fermion simulations by QCDSF/UKQCD (table on the right).}
 \label{slj}
\end{table}

These results certainly represent a first major step towards a quantitative decomposition of 
the nucleon spin in full QCD. 
In reviewing them, we note, however, that several items should be kept in mind:
\begin{itemize}
\item only connected contributions are included, disconnected parts contributing in the singlet sector are missing
\item since $J_q=(\langle x^{}\rangle_{q}+B^q_{20}(0))/2$, the results are in principle affected by the discrepancy 
in the normalization of the quark momentum fraction
of results from QCDSF and LHPC (see Fig.~\ref{overview_x} and discussion above)
\item different types of chiral extrapolations were used:
leading order (self-consistently improved) HBChPT for $\Delta\Sigma^{u+d}$, HBChPT with explicit $\Delta$-DOFs
(SSE) for $g_A=\Delta\Sigma^{u-d}$, and covariant BChPT at leading order $\mathcal{O}(p^2)$ (partially including NLO $\mathcal{O}(p^3)$ corrections)
for the form factors of the energy-momentum-tensor
\item the lowest pion masses are $\gtrapprox350\MeV$, where the applicability of ChPT is not established
\end{itemize}

Nevertheless, the present results show some highly remarkable features.
Most prominently, the total (up- plus down-) quark orbital angular momentum contribution 
to the nucleon spin turns out to be very small and fully compatible with zero within statistical errors. 
Interestingly, the individual OAM contributions from up- and down-quarks are sizeable but of 
opposite sign and only cancel in the sum.
These findings are at first sight completely at odds with expectations from relativistic quark models. 
Noting that the quark models generically live at a low hadronic scales,
the results may however be reconciled if the scale evolution of OAM is taken into account.

Secondly, the down-quark angular momentum contribution is also very small and
zero within errors. Again, this can be seen as the result of a precise cancellation, in this case 
of the individual spin and OAM contributions, which are separately non-zero and of the same
size, but opposite in sign. 
Another interesting outcome is the observed near cancellation of the individually sizeable 
up- and down-quark contributions to the anomalous gravitomagnetic moment $B_{20}(\t0)$.
If confirmed in future studies, this would not only be theoretically very interesting, but
also have far reaching consequences for the gluon contributions to the nucleon spin, since
from Noethers theorem $\sum_qB^q_{20}(\t0)+B^g_{20}(\t0)=0$.

In addition to the form factors of the energy momentum tensor that coincide with the
$x$-moments of the unpolarized GPDs, also the next higher moments of GPDs have been studied. 
From a comparison of the lowest three $x^{n-1}$-moments, i.e. for $n=1,2,3$, it was found that the 
the slopes of the corresponding generalized form factors as functions of the momentum transfer squared $t$
decrease with increasing $n$, see, e.g., Fig.~\ref{A123_umd_LHPC}. 
Transformed to coordinate (impact parameter) space,
this corresponds to a mean square radius of the nucleon that is rapidly decreasing as the
longitudinal quark momentum fraction increases, i.e. as $x\rightarrow1$, see Fig.~\ref{Charge_radii_vector_v2_LHPC}.
Since the absolute values of the generalized form factors do not enter in this case, this result is also 
unaffected by the normalization issue discussed above.


From first lattice studies of moments of tensor generalized parton distributions 
of the pion and the nucleon, qualitatively important insights were gained with respect to the distribution of 
transversely polarized quarks inside hadrons.
The observed large values for some of the tensor GFFs point towards strong correlations 
between transverse spin and coordinate degrees of freedom, which in turn lead to 
characteristically deformed impact parameter densities of quarks with 
transverse polarization in unpolarized and transversely polarized hadrons, 
cf. Figs.~\ref{Pion_densities_v3} and \ref{densities_n1_ud_QCDSF_v2}.
Specifically, it was found that the (spin-$0$) pion has a quite non-trivial transverse spin structure.

Finally, in order to be able to provide quantitatively reliable results and solid predictions for experiment,
contributions from quark line disconnected diagrams will have to be consistently included in unquenched
lattice calculations of flavor singlet quantities.
This is a real challenge, and first promising attempts based on stochastic methods and including
noise reduction techniques will be discussed below in section \ref{sec:disconnected}.

In the course of full calculations in the flavor singlet sector, it will also be important and necessary
because of operator mixing to study the corresponding gluon observables, 
in particular the momentum fraction carried by gluons, 
as well as their spin and total angular momentum contributions to the hadron spin.
First attempts in this direction have been reported already more than 10 years ago 
\cite{Mandula:1990ce,Gockeler:1996zg}, but only recently 
some groups show renewed interest in this subject \cite{Meyer:2007tm,Doi:2008hp}.

%% file: DAs.tex
\section{Lattice results on hadronic distribution amplitudes}
\label{sec:DAs}
\subsection{Moments of meson distribution amplitudes}
\label{sec:DAsMesons}
In this section, we briefly report on lattice QCD results for
the lowest moments of meson and nucleon distribution amplitudes (DAs).

Moments of meson distribution amplitude were first studied 
in the late 1980's in quenched \cite{Gottlieb:1986ie,Martinelli:1987si,DeGrand:1987vy}, and already a few years later in unquenched 
lattice QCD \cite{Daniel:1990ah}.
The early calculations of the $\xi^2$-moment of, e.g., the pion distribution amplitude, 
$\langle \xi^2\rangle_\pi$ (see Eq.~\ref{PionDA3}), 
were mostly based on lattice operators of the form 
\bea
\mathcal{O}^{5DD}_{\{\mu\nu\rho\}}=\overline q\gamma_5\gamma_{\{\mu}\Dlr_{\nu}\Dlr_{\rho\}}q-\text{traces}
\label{O5DD}
\eea
with $\nu=\rho$, $\mu\neq\nu$, transforming according to the 8-dimensional representation $\tau_2^{(8)}$ of $H(4)$.
Operators of the type $\mathcal{O}^{5DD}_{\{\mu\nu\rho\}}$
with $\mu$, $\nu$ and $\rho$ all different, belonging to $\tau_3^{(4)}$ were also investigated
in some cases (concerning the classification of the operators, see \cite{Gockeler:1996mu}).
Results of a calculation of $\langle \xi^2\rangle_\pi$ in the quenched approximation
purely based on operators of type $\mathcal{O}^{5DD}_{\{\mu\nu\rho\}}$ (with all indices different)  were presented in 
\cite{DelDebbio:2002mq}.
The mixing properties of these operators were studied more recently in some detail in 
\cite{Gockeler:2004xb},
where also the mixing coefficients were calculated in lattice perturbation theory at the one-loop level.
It was found in particular that the operator of type $\mathcal{O}^{5DD}_{\{\mu\nu\nu\}}$ mixes with
a lower dimensional operator of type 
\bea
\mathcal{O}^{D}_{\mu\nu\omega}=\overline q\sigma_{\mu\nu}\Dlr_{\omega}q\,,
\eea
giving rise to power divergences of the form $1/a$ in the renormalization procedure, which would
have to be subtracted non-perturbatively.
On the other hand, the operators $\mathcal{O}^{5DD}_{\{\mu\nu\rho\}}$ (with all indices different) 
only mix with operators of the same dimension of type 
\bea
\mathcal{O}^{5\partial\partial}_{\{\mu\nu\rho\}}=\partial_{\{\nu}\partial_\omega(\overline q\gamma_{\mu\}}\gamma_5q)
\eea
\cite{Gockeler:2006nb}.
Although at least two non-zero components of the hadron momentum are required for the evaluation
of matrix elements of these operators, leading in general to more noisy
correlators, they probably still represent the safer alternative for the extraction of, e.g., 
$\langle \xi^2\rangle_\pi$, due to the absence of mixing with lower-dimensional operators.

\subsubsection{Pseudoscalar mesons}
\label{psedoscalarDAs}
An extensive study of the lowest moments of the pion and kaon DAs, Eqs.~(\ref{PionDA1}) to (\ref{PionDA4}), has been performed by QCDSF
for $n_f=2$ flavors of clover-improved Wilson fermions, for pion masses in the
range of $\approx 600\MeV$ to $\approx 1200\MeV$, up to four different values of the coupling
$\beta=5.20,5.25,5.29,5.40$ corresponding to lattice spacings from
$\approx0.07\fm$ to $\approx0.12\fm$, and volumes of $V\approx(1.5,\ldots,2.2\fm)^3$ 
\cite{Braun:2006dg}.
The calculations were based on the axial-vector counterparts of the
lattice operators in Eqs.~(\ref{op2}) and (\ref{op3})
for the  $\xi$-moment of the kaon DA, $\langle \xi\rangle_K$,
and on an operator of the type $\mathcal{O}^{5DD}_{\{\mu\nu\rho\}}$ as discussed above
for the $\xi^2$-moments of the pion and kaon DAs.
All lattice operators were non-perturbatively renormalized, apart from the mixing
with the operator $\mathcal{O}^{5\partial\partial}_{\{\mu\nu\rho\}}$, for which
a perturbatively calculated mixing coefficient was used \cite{Gockeler:2006nb}.
Both types of pion (and kaon) interpolating fields as given in Eq.~(\ref{interpol1}) with
$\gamma_5$ and $\gamma_4\gamma_5$ were employed in the numerical calculations.
%
%
\begin{figure}[t]
   \begin{minipage}{0.48\textwidth}
      \centering
          \includegraphics[angle=0,width=0.99\textwidth,clip=true,angle=0]{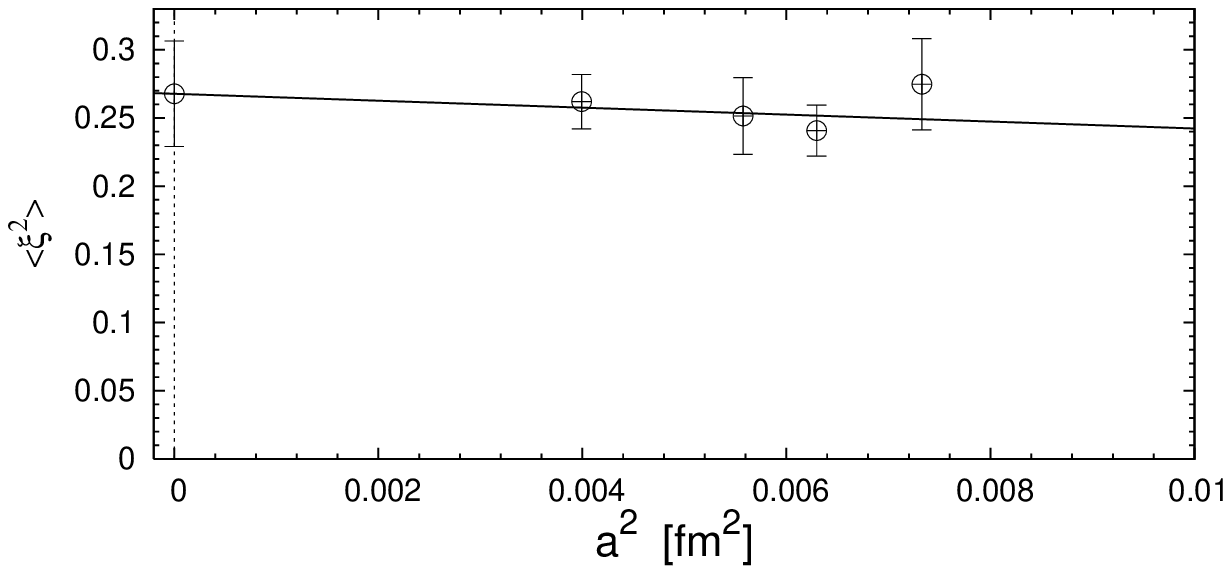}
  \caption{Lattice spacing dependence of the $\xi^2$-moment of the pion distribution amplitude (from \cite{Braun:2006dg}).}
  \label{Pion5_DA2a_chiral_QCDSF06}
     \end{minipage}
     \hspace{0.3cm}
    \begin{minipage}{0.48\textwidth}
      \centering
          \includegraphics[angle=0,width=0.99\textwidth,clip=true,angle=0]{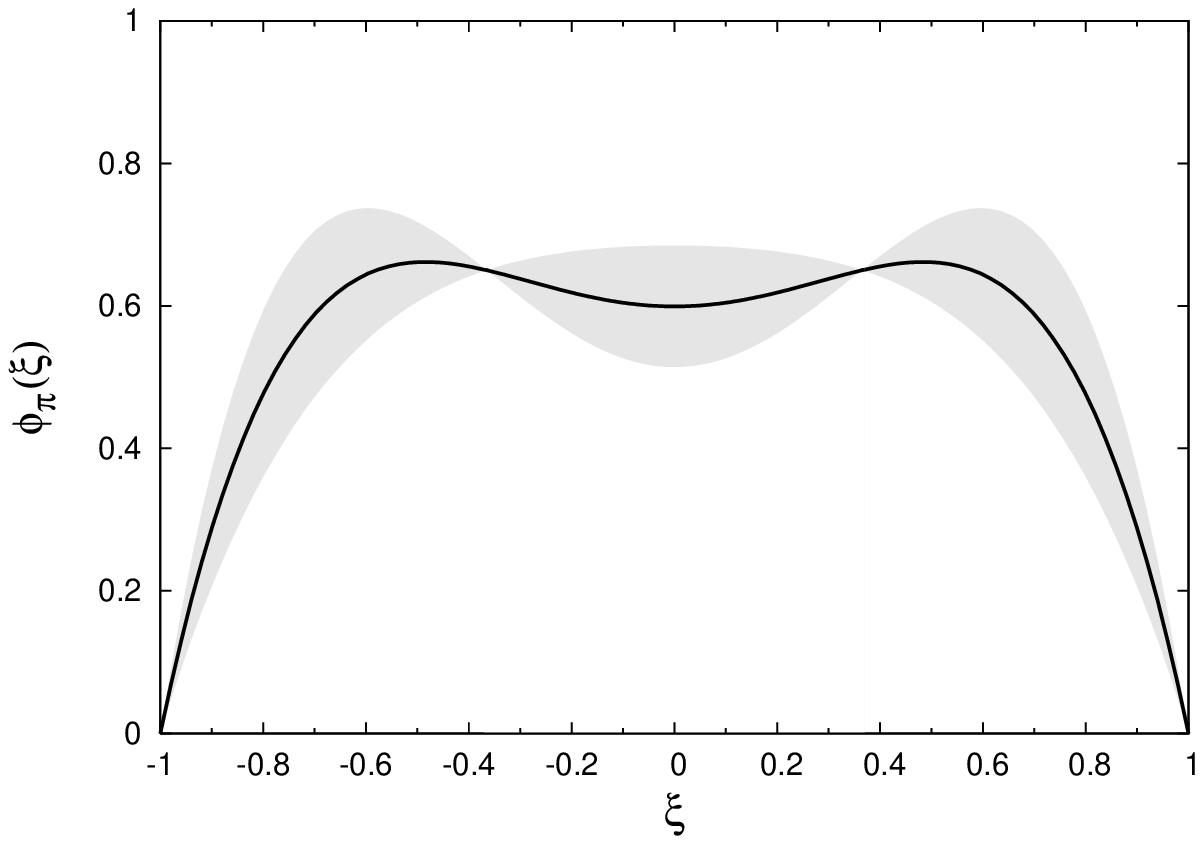}
  \caption{The pion distribution amplitude $\phi_\pi(\xi)$ (from \cite{Braun:2006dg}).}
  \label{phi_pi_a4eq0_QCDSF06}
     \end{minipage}
 \end{figure}

For the case of the pion, the moments $\langle \xi^2\rangle_{\pi}$ were first
linearly extrapolated in $m_\pi^2$ to the physical pion mass for each $\beta$.
The results are plotted in Fig.~\ref{Pion5_DA2a_chiral_QCDSF06}
versus $a^2$, where the lattice spacings of the four data points correspond to chirally
extrapolated values at the physical pion mass.
Within statistical uncertainties, there is no clear lattice spacing dependence visible,
and from a linear extrapolation in $a^2$ to the continuum limit, a value of 
$\langle \xi^2\rangle_{\pi}=0.269(39)$ was obtained in the $\MSbar$ scheme at a scale of $\mu=2\GeV$,
which is somewhat larger than the asymptotic value $\langle \xi^2\rangle^{\mu\rightarrow\infty}=0.2$.
An illustration of the pion distribution amplitude $\phi_\pi(x)$, based on this numerical result and
the Gegenbauer-expansion in Eq.~(\ref{Gegenbauer1}) in combination with the relation 
$a_{(n=2)}^\pi=35\langle \xi^2\rangle_{K}/12-7/12$, is shown in Fig.~\ref{phi_pi_a4eq0_QCDSF06},
where the higher Gegenbauer moments have been set to zero by hand.
The shaded band represents the statistical error for $\langle \xi^2\rangle_{\pi}$,
and it should be noted that contributions from higher moments $a_{(n>2)}^\pi$ may
change the shape of the DA significantly.

\begin{figure}[t]
   \begin{minipage}{0.48\textwidth}
      \centering
     \includegraphics[angle=0,width=0.99\textwidth,clip=true,angle=0]{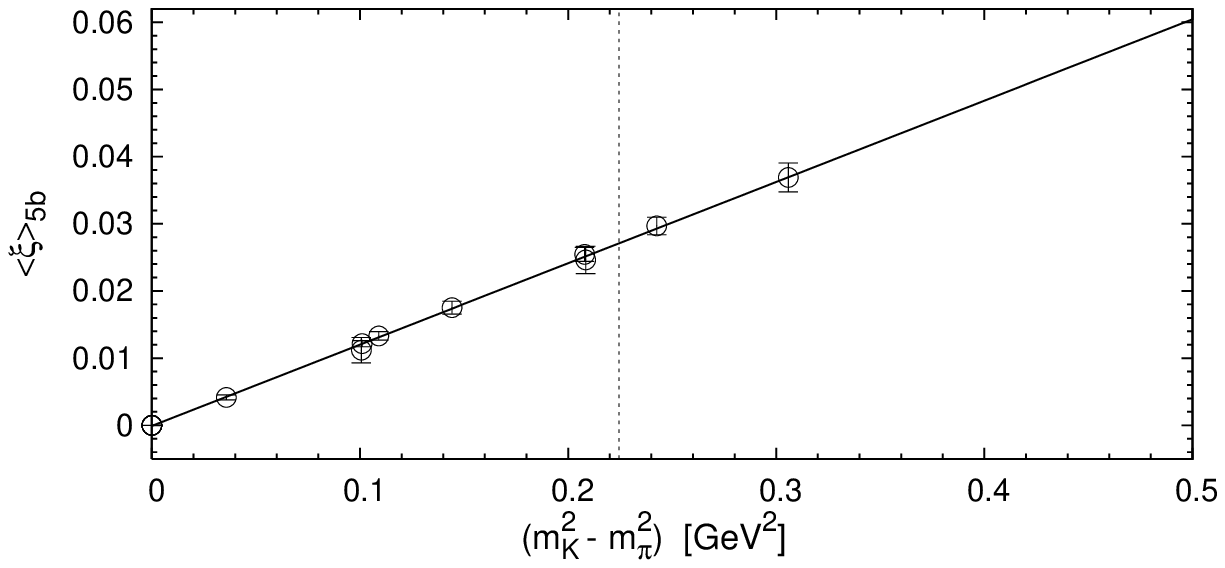}
  \caption{Moment of the kaon distribution amplitude $\langle \xi\rangle_{K}$ at for fixed sea quark mass (from \cite{Braun:2006dg}).}
  \label{Pion5_DA1bPQ_b5p29kp13590_QCDSF06}
     \end{minipage}
     \hspace{0.3cm}
    \begin{minipage}{0.48\textwidth}
      \centering
          \includegraphics[angle=0,width=0.99\textwidth,clip=true,angle=0]{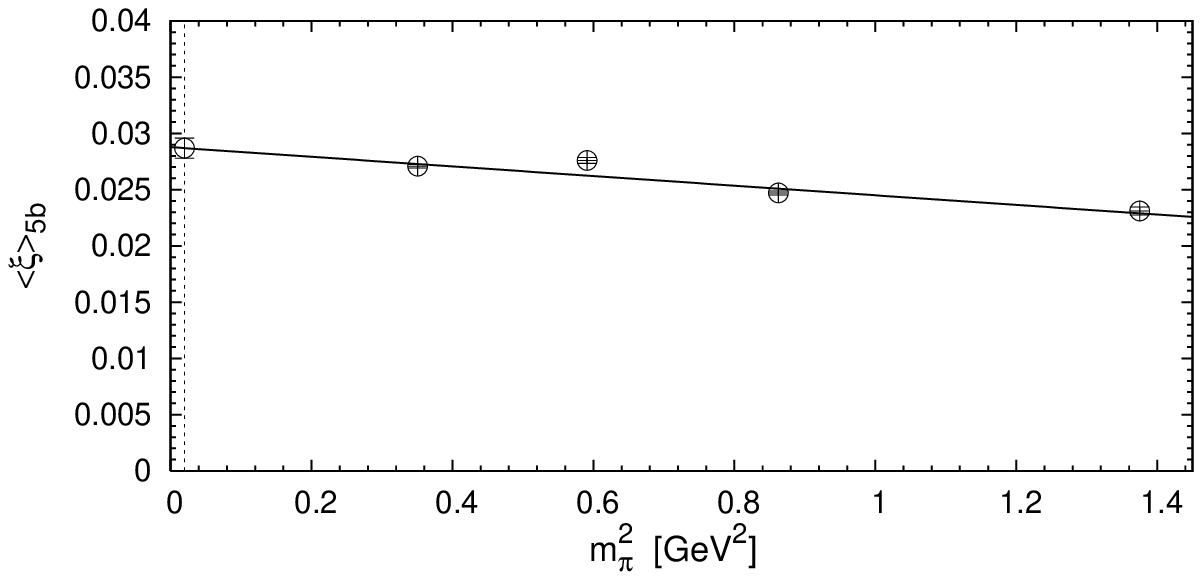}
  \caption{Sea quark mass dependence of $\langle \xi\rangle_{K}$ (from \cite{Braun:2006dg}).}
  \label{Slope5_DA1bPQ_b5p29_QCDSF06}
     \end{minipage}
 \end{figure}

The moments of the kaon distribution amplitude $\langle \xi\rangle_{K}$ and $\langle \xi^2\rangle_{K}$
were also computed in the framework of this analysis for a fixed coupling of $\beta=5.29$.
Results for $\langle \xi\rangle_{K}$ are given in Fig.~\ref{Pion5_DA1bPQ_b5p29kp13590_QCDSF06}
for the axial-vector counterpart of the operator $\mathcal{O}^b_{1234}$ in Eq.~(\ref{op2}), as a function 
of the difference $m_K^2-m_\pi^2$, for a fixed sea quark mass corresponding to a hopping
parameter of $k_\text{sea}=0.13500$. The lattice data points in
Fig.~\ref{Pion5_DA1bPQ_b5p29kp13590_QCDSF06} were obtained for a range of values
of smaller ($\hat= m^\text{val}_{u,d}$) and larger ($\hat= m^\text{val}_{s}$) valence quark masses, 
for which also the corresponding pion masses, $m_\pi^2$, and kaon masses, $m_K^2$, were calculated. 
Chiral perturbation theory predicts
that $\langle \xi\rangle_{K}\propto m_s-m_{u,d}$ \cite{Chen:2003fp}, which is according
to the Gell-Mann-Oakes-Renner relations, Eq.~(\ref{GMOR}), proportional to the difference
$m_K^2-m_\pi^2$. The lattice data points in Fig.~\ref{Pion5_DA1bPQ_b5p29kp13590_QCDSF06}
should therefore lie on a straight line, which is indeed the case to very good approximation.
From a linear interpolation to the physical difference $(m^\phys_K)^2-(m^\phys_\pi)^2\approx0.22\GeV^2$, 
indicated by the vertical line, values for $\langle \xi\rangle_{K}$ were inferred at the given
sea quark mass, i.e. in this case for $k_\text{sea}=0.13500$.
The sea-quark mass dependence in terms of $m_\pi^2$ is displayed in Fig.~\ref{Slope5_DA1bPQ_b5p29_QCDSF06}, 
together with a linear extrapolation to the physical point, as represented by the straight line.
A final value of $\langle \xi\rangle_{K}=0.0272(5)$ in the $\MSbar$ scheme
for $\mu=2\GeV$ was obtained by averaging over the results 
obtained for different lattice operators and kaon interpolating fields.

\begin{figure}[t]
   \begin{minipage}{0.48\textwidth}
      \centering
        \includegraphics[angle=0,width=0.99\textwidth,clip=true,angle=0]{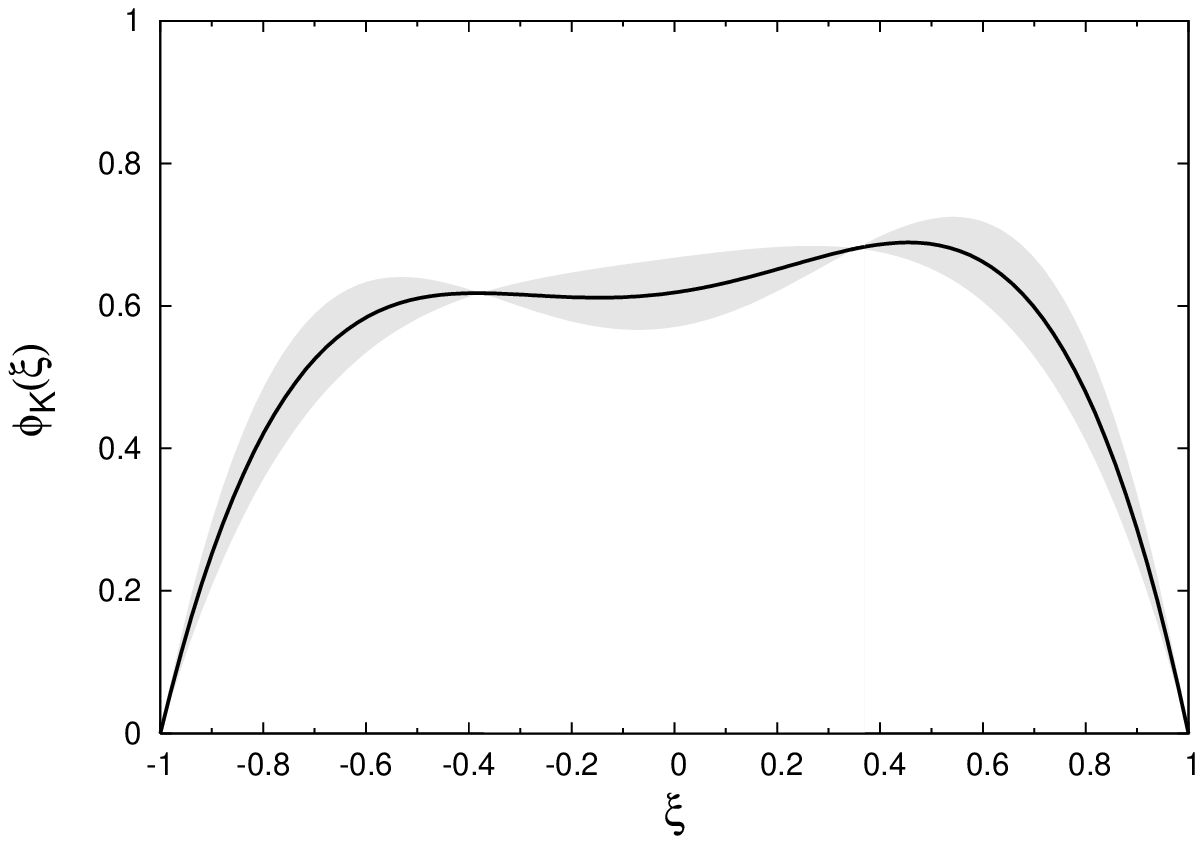}
  \caption{Illustration of the kaon distribution amplitude (from \cite{Braun:2006dg}).}
  \label{phi_K_QCDSF06}
     \end{minipage}
     \hspace{0.3cm}
    \begin{minipage}{0.48\textwidth}
      \centering
  \psfrag{mqplusmres}[t][c][1][0]{\Huge $am_q+am_{\rm res}$}
  \psfrag{fstmom}[c][t][1][0]{\Huge $\langle \xi\rangle_{K}^{\textrm{bare}}$}
  \psfrag{Legendmass1}[c][c][1][0]{\Huge $am_{ud}=0.01$}
  \psfrag{Legendmass2}[c][c][1][0]{\Huge $am_{ud}=0.02$}
  \psfrag{Legendmass3}[c][c][1][0]{\Huge $am_{ud}=0.03$}
  \epsfig{angle=270,width=0.99\textwidth,file=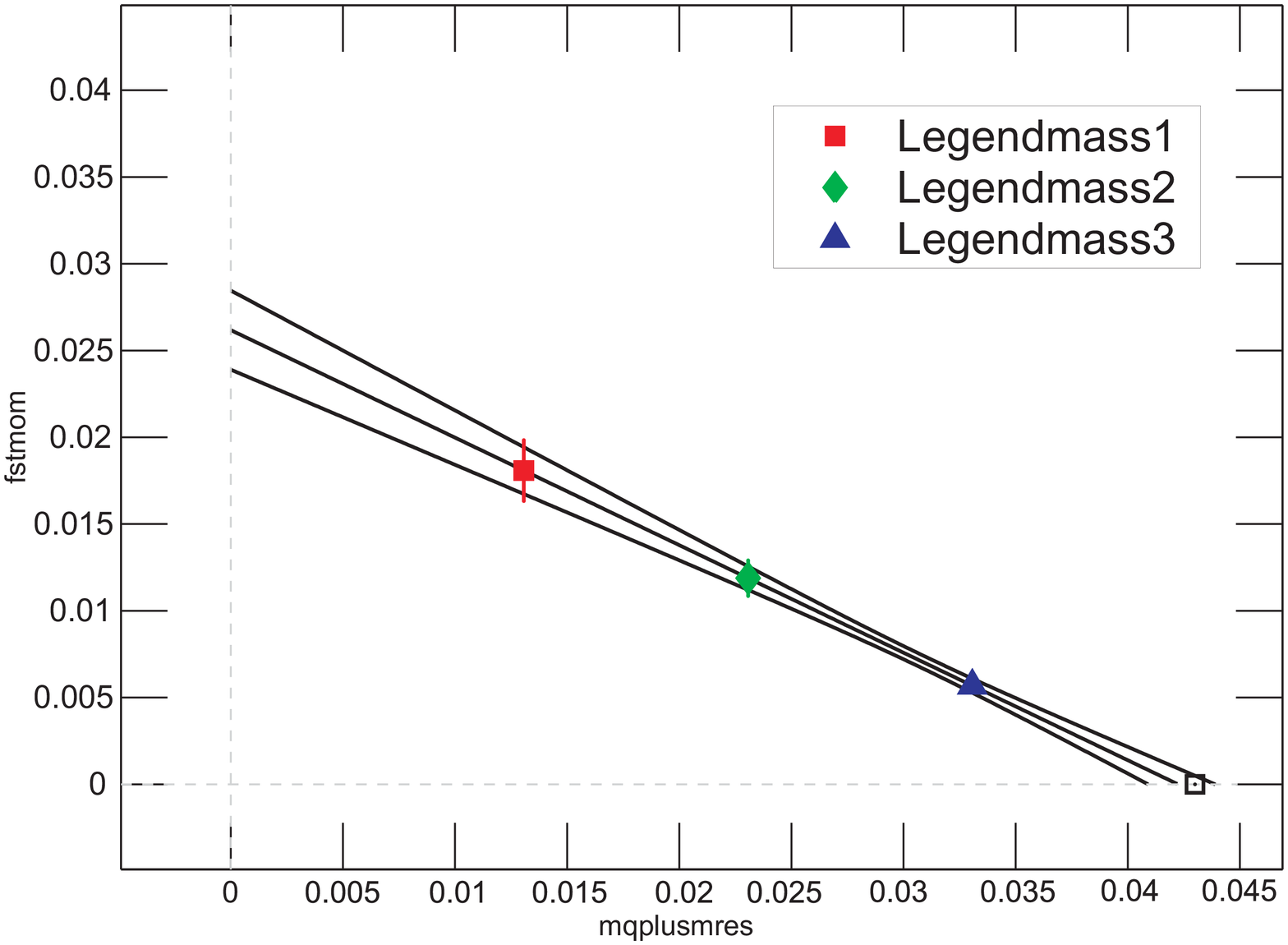}
  %
  \caption{Chiral extrapolation of $\langle \xi\rangle_{K}$ (from \cite{Boyle:2006pw}).}
  \label{kaonDA_first_RBCUKQCD}
     \end{minipage}
 \end{figure}

Similar calculations of the $\xi^2$-moment of the pion DA, based on the operator $\mathcal{O}^{5DD}_{\{\mu\nu\rho\}}$
discussed above, were also performed for $\langle \xi^2\rangle_{K}$. 
The lattice results for the different combinations of valence, $m^\text{val}_{u,d}$ and $m^\text{val}_{s}$, 
and sea-quark masses at fixed $\beta=5.29$ 
were fitted using the global ansatz $\langle \xi^2\rangle_{K}=c_0+c_1m_\pi^2+c_2m_K^2$,
where $m_K$ corresponds to the valence quarks masses, $m_q^\text{val1}\ge m_q^\text{val2}$, 
and $m_\pi$ to the sea-quark mass.
From the global fit, a value of $\langle \xi^2\rangle_{K}=0.260(6)$
was obtained at the physical pion and kaon masses, for a scale of $\mu=2\GeV$ in the $\MSbar$ scheme. 
Figure \ref{phi_K_QCDSF06} shows an illustration of the kaon DA $\phi_K(\xi)$, based on the numerical lattice results
for $\langle \xi\rangle_{K}$,  $\langle \xi^2\rangle_{K}$
and the Gegenbauer-expansion in Eq.~(\ref{Gegenbauer1}) with $a_{(n=1)}^K=5\langle \xi^2\rangle_{K}/3$
and $a_{(n=2)}^K=35\langle \xi^2\rangle_{K}/12-7/12$, and where all higher moments were set to zero by hand.

%

Results for the $\xi$-moment of the kaon DA in the framework of lattice simulations
with $n_f=2+1$ flavors of domain wall fermions and the Iwasaki gauge action
have been published by RBC-UKQCD \cite{Boyle:2006pw}.
The computations were based on non-diagonal operators of the type 
$\mathcal{O}^{5D}_{\{\mu\nu\}}=\overline q\gamma_5\gamma_{\{\mu}\Dlr_{\nu\}}q$
with $\mu\neq\nu$ that belong to the $H(4)$ representation $\tau_4^{(6)}$ \cite{Gockeler:1996mu}
and that do not mix with other operators of the same or lower dimension.
Both types of pion interpolating fields in Eq.~(\ref{interpol1}) were used for the calculation of the correlators.
Numerical results have been obtained for a lattice spacing of $a\approx0.12\fm$,
a volume of $V\approx(2\fm)^3$, a strange quark mass that was tuned 
approximately to the physical value, and
three different light quark masses corresponding to pion masses
of $\approx390\MeV$, $\approx520\MeV$ and $\approx620\MeV$ and kaon masses
of $\approx570\MeV$, $\approx620\MeV$ and $\approx667\MeV$.
Results for the unrenormalized (bare) moment, $\langle \xi\rangle^{\text{bare}}_{K}$,
are shown in Fig.~\ref{kaonDA_first_RBCUKQCD} versus the bare light quark mass plus
the residual mass in lattice units, $a(m_{u,d}+m_\text{res})$, with a residual
mass of $am_\text{res}=0.003$ (the length of the fifth dimension
in the DW simulations was set to $L_s=16$). 
The quark mass dependence of the lattice data points is well described by an ansatz
linear in $m_{u,d}$, in agreement with the predictions from ChPT \cite{Chen:2003fp}
and with the results by QCDSF displayed in Fig.~\ref{Pion5_DA1bPQ_b5p29kp13590_QCDSF06}.
In the $SU(3)$ symmetric case, $\langle \xi\rangle_{K}\rightarrow0$, which is indeed
observed in Fig.~\ref{kaonDA_first_RBCUKQCD} where the fitting curve
intersects the $x$-axis close to the expected mass of the strange quark,
i.e. where $m_{u,d}\sim m_s$.
Perturbatively calculated renormalization constants were employed to
transform the results of the linear fit, represented by the solid lines in 
Fig.~\ref{kaonDA_first_RBCUKQCD}, to the $\MSbar$ scheme at a scale of $\mu=2\GeV$, giving 
$\langle \xi\rangle_{K}=0.032(3)$.
This value is in reasonable agreement with the result of the $n_f=2$ Wilson fermion
study by QCDSF/UKQCD discussed above, taking into account potential discretization 
errors and finite size effects on both sides.

Preliminary results for $\langle \xi\rangle_{K}$ and $\langle \xi^2\rangle_{\pi,K}$,
obtained in the framework of simulations with $n_f=2+1$ flavors of domain wall fermions,
for a lattice spacing of $a=1.729(28)\fm$, a larger volume of $V\approx(2.7\fm)^3$, and including a lower pion mass of
$m_\pi\sim330\MeV$, were presented more recently by RBC/UKQCD in \cite{Boyle:2008nj}
(see also \cite{Donnellan:2007xr}).
All operators were perturbatively renormalized, including the mixing of operators
of type $\mathcal{O}^{5DD}_{\{\mu\nu\nu\}}$ with the operators $\mathcal{O}^{5\partial\partial}$, 
as required for the calculation of the $\xi^2$-moments and discussed at the beginning of this section.
The preliminary results, which were linearly extrapolated to the physical masses, are 
$\langle \xi\rangle_{K}=0.0289(19)$ and $\langle \xi^2\rangle_{K}=0.267(17)$ 
for the moments of the kaon DA, and 
$\langle \xi^2\rangle_{\pi}=0.272(20)$
for the moment of the pion DA in the $\MSbar$ scheme at a scale of $\mu=2\GeV$,
again in good agreement with the $n_f=2$ Wilson fermion results from
QCDSF/UKQCD\cite{Braun:2006dg}.

\subsubsection{Vector mesons}
First results for the $\xi$-moment of the (spin-1) $K^*$-meson DAs,
$\phi_\Vert(\xi)$ and $\phi_\perp(\xi)$, see section \ref{secDAs},
calculated in the framework of simulations with $n_f=2$ flavors of 
clover-improved Wilson fermions, were presented by QCDSF/UKQCD in \cite{Braun:2007zr}.
This preliminary study was restricted to a small subset of all available
ensembles, specifically to a fixed coupling of $\beta=5.29$,
(sea quark) pion masses in the range of $\approx630\MeV$ to $\approx1100\MeV$, and various valence quark masses.
For the study of $\langle \xi\rangle^{\Vert}_{K^*}$, 
operators analogous to (\ref{op2})
and (\ref{op3}) were employed, and the transverse moment $\langle \xi\rangle^{\perp}_{K^*}$
was extracted using tensor operators (corresponding to two different $H(4)$ representations) of the form
$\mathcal{O}^{\perp D}_{ij4}+\mathcal{O}^{\perp D}_{i4j}-\mathcal{O}^{\perp D}_{4ij}-\mathcal{O}^{\perp D}_{4ji}$
with $i\neq j$, and $\mathcal{O}^{\perp D}_{i44}-\sum_{j\not=i}\mathcal{O}^{\perp D}_{ijj}/2$,
where
$\mathcal{O}^{\perp D}_{\mu\nu\tau}=\overline q\gamma_{\mu}\gamma_{\nu}\Dlr_{\tau}u$.
They were non-perturbatively renormalized, and the final results were transformed to 
the $\MSbar$ scheme at a scale of $\mu=2\GeV$.
Analogously to the pseudoscalar kaon results displayed in 
Fig.~\ref{Pion5_DA1bPQ_b5p29kp13590_QCDSF06}, the lattice results for 
$\langle \xi\rangle^{\Vert,\perp}_{K^*}$ were first interpolated 
to the physical value of the difference of the squared \emph{pseudoscalar} kaon and pion masses,
$m_K^2-m_\pi^2$, for a fixed sea quark mass but varying valence quark masses, 
and then linearly extrapolated in $m_\pi^2$ (corresponding to the sea quarks)
to the chiral limit, giving 
$\langle \xi\rangle^{\Vert}_{K^*}\approx0.033\pm0.005_{\text{stat+sys}}$
and 
$\langle \xi\rangle^{\perp}_{K^*}\approx0.030\pm0.008_{\text{stat+sys}}$,
where the errors represent statistical and systematic uncertainties of
the calculation.

Corresponding results for the lowest two moments $\langle \xi^{(n=1,2)}\rangle^{\Vert}$ of $K^*$-, $\rho$- and $\phi$-vector meson DAs
were more recently presented by RBC-UKQCD \cite{Boyle:2008nj}. 
The calculations were performed in the same lattice framework (with $n_f=2+1$ flavors of domain wall fermions)
as already discussed in the context of pseudoscalar mesons in the previous section.
Employing perturbatively renormalized operators, values of 
$\langle \xi\rangle^{\Vert}_{K^*}\approx0.0324\pm0.0016_{\text{stat}}\pm0.0021_{\text{sys}}$
and 
$\langle \xi^2\rangle^{\Vert}_{K^*}\approx0.248\pm0.017_{\text{stat}}\pm0.012_{\text{sys}}$
were obtained from linear extrapolations to the physical masses, in the $\MSbar$ scheme at $\mu=2\GeV$
(for a volume of $V\approx(2.7\fm)^3$).
We note that the result for $\langle \xi\rangle^{\Vert}_{K^*}$ is, within errors, perfectly compatible
with the value from QCDSF/UKQCD \cite{Braun:2007zr} discussed above.

\subsection{Moments of nucleon distribution amplitudes}
\label{sec:DAsNucleon}
Recently, the lowest moments of the leading twist-2 nucleon distribution amplitude $\varphi(x_1,x_2,x_3)$,
see section \ref{secDAs} and Eq.~(\ref{ProtonDA2}),
were studied for the first time in lattice QCD by QCDSF UKQCD 
\cite{Gockeler:2008xv}.\footnote{A more extensive study by the same authors has been presented very recently in \cite{Braun:2008ur}.}
Calculations were performed using $n_f=2$ flavors of clover-improved Wilson fermions, for
two couplings $\beta=5.29$ and $\beta=5.40$, corresponding to chirally extrapolated
lattice spacings of $\approx0.075\fm$ and $\approx0.067\fm$, respectively, pion
masses in the range of $\approx380\MeV$ to $\approx1200\MeV$, and 
volumes $V=L^3$ with $m_\pi L\ge3.7$.
Special care was taken to properly account for possible mixings of the relevant three-quark lattice operators.
These baryonic operators of half-integer spin can be classified according to the so-called spinorial hypercubic group $\overline H(4)$, 
as compared to the hypercubic group $H(4)$ that is relevant for the quark- anti-quark lattice operators of integer spin,
defining the moments of PDFs, GPDs and meson DAs.
As in the case of the $\overline q q$-operators, three-quark lattice operators that transform identically
with respect to $\overline H(4)$ can and do mix in general under renormalization. 
The transformation properties of the three-quark operators 
with up to two derivatives 
were studied in detail in \cite{Kaltenbrunner:2008pb}, where the multiplets of operators
belonging to the irreducible representations $\tau_{1,2}^{\underline 4}$ of dimension 4, $\tau_{}^{\underline 8}$ of dimension 8, and 
$\tau_{1,2}^{\underline 12}$ of dimension 12, have been worked out explicitly.
These results were exploited in the lattice study \cite{Gockeler:2008xv} by choosing sets of lattice operators 
that keep operator-mixing at a minimum and in particular avoid mixing with lattice operators
of lower mass dimension, which would introduce power divergences of the form $1/a$.
The corresponding twist-$2$ operators with maximally two covariant derivatives
were non-perturbatively renormalized, and the results for the moments of the proton DA have been 
transformed to the $\MSbar$ scheme at a scale of $\mu=2\GeV$ \cite{Gockeler:2008we}.

Instead of the moments of the standard nucleon DA $\varphi^{klm}$ in Eq.~(\ref{ProtonDA2}) \cite{Dziembowski:1987es}, 
a different linear combination of moments of the vector, axial-vector and tensor DAs, denoted by 
$\phi^{klm}$, with normalization $\phi^{000}=1$, was used, which is related to the
standard DA by $\varphi^{klm}=2\phi^{klm}-\phi^{mlk}$.
To reduce the substantial statistical noise for the higher moments, it turned out
to be beneficial to construct normalized lattice moments (ratios) of the form
$R^{klm}=\phi^{klm}/\overline\phi_{k+l+m}$ with 
$\overline\phi_1=2\phi^{100}+\phi^{010}+\phi^{001}$ and
$\overline\phi_2=2(\phi^{110}+\phi^{101}+\phi^{011})+(\phi^{200}+\phi^{020}+\phi^{002})$.

\begin{figure}[t]
   \begin{minipage}{0.48\textwidth}
      \centering
        \includegraphics[angle=0,width=0.99\textwidth,clip=true,angle=0]{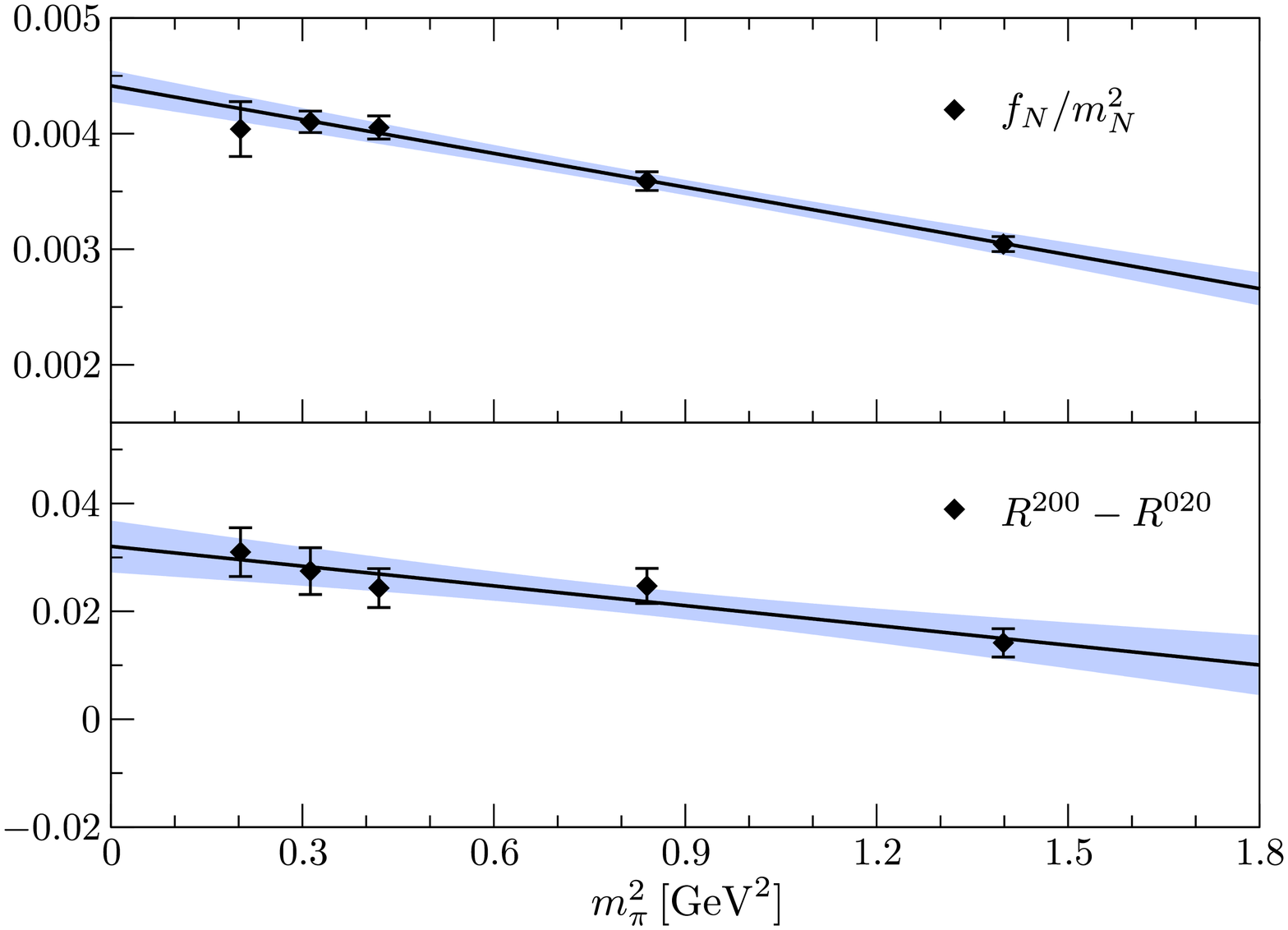}
  \caption{Decay constant and second moment of the nucleon distribution amplitude (from \cite{Gockeler:2008xv}).}
  \label{nDA_QCDSF08}
     \end{minipage}
     \hspace{0.3cm}
    \begin{minipage}{0.48\textwidth}
      \centering
          \includegraphics[angle=0,width=0.99\textwidth,clip=true,angle=0]{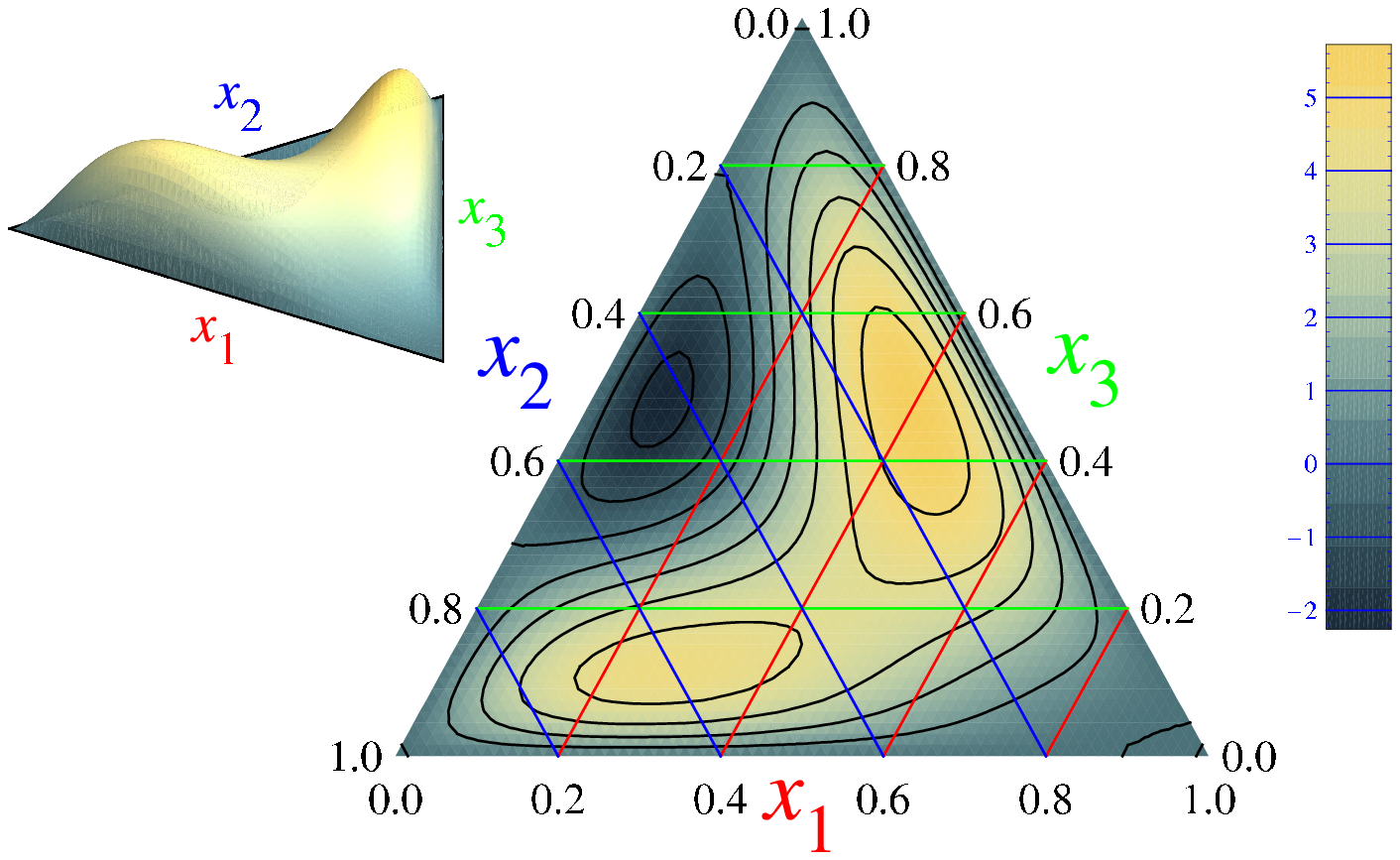}
  \caption{Illustration of $\varphi(x_1,x_2,x_3)$ as obtained from the lattice results for 
  the lowest moments of the nucleon distribution amplitude (from \cite{Gockeler:2008xv}).}
  \label{nDA_contour_QCDSF08}
     \end{minipage}
 \end{figure}

Exemplary pion mass dependences of bare (unrenormalized) lattice results for
the ``decay constant'', $f_N$, and
an ``asymmetry'' constructed from the normalized $(k+l+m=2)$-moments,  
$R^{200}-R^{020}=(\phi^{200}-\phi^{020})/\overline\phi_2$
are displayed in Fig.\ref{nDA_QCDSF08}.
In the limit of infinitely large scale, 
$\phi(x_1,x_2,x_3)\stackrel{\mu\rightarrow\infty}{\longrightarrow}120x_1x_2x_3$,
giving, e.g.,  $\phi^{200}=\phi^{020}=\phi^{002}=1/7$, so that the clearly non-zero asymmetry 
$\propto(\phi^{200}-\phi^{020})$ in Fig.~\ref{nDA_QCDSF08} provides non-trivial information
about deviations from the asymptotic form of the DA.
Within statistics, the lattice data points lie to a good approximation on straight lines.
Fits linear in $m_\pi^2$, as represented by the error
bands in Fig.~\ref{nDA_QCDSF08}, were performed and used in particular to
extrapolate all lattice results for the ratios $R^{100},\ldots,R^{001}$,
$R^{110},R^{101},R^{011}$ and $R^{200},\ldots,R^{002}$ to the physical pion mass.

Finally, the moments $\phi^{klm}$ with $(k+l+m=1,2)$
were extracted from the chirally extrapolated ratios
by demanding momentum conservation for the renormalized,
extrapolated values, $\overline\phi^{\text{ren,extr}}_{1,2}\stackrel{!}{=}1$,
i.e. by setting $\phi^{klm}_{\text{ren,extr}}=R^{klm}_{\text{ren,extr}}$.
Explicit results for the these moments with $(k+l+m=1,2)$ can be found in \cite{Gockeler:2008xv,Braun:2008ur}.
%
%

%
To get a first impression about possible deviations of the nucleon DA $\varphi(x_1,x_2,x_3)$ from its asymptotic shape,
an expansion of the form
\begin{equation}
 \varphi(x_1,x_2,x_3,\mu)=120 x_1 x_2 x_3 \sum_{n=0}^N c_n(\mu_0) P_n(x_1,x_2,x_3)
\left(\frac{\alpha_s(\mu)}{\alpha_s(\mu_0)}\right)^{\frac{3}{11N_c-2n_f}\gamma_n}
\label{NuclDAExp}
\end{equation}
was used, where the anomalous dimensions $\gamma_n$ and the polynomials $P_n(x_1,x_2,x_3)$
are given by the eigenvalues and eigenfunctions, respectively,
of the three-quark-operator evolution matrix \cite{Braun:1999te}.
Setting higher order contributions with $n>2$ to zero, the
non-perturbative coefficients $c_n(\mu_0=2\GeV)$ with $n\le2$ in Eq.~(\ref{NuclDAExp}) were
calculated from the lattice data for the moments of the DA, $\phi^{klm}_{\text{ren,extr}}$
with $(k+l+m=1,2)$, obtained for the ensemble with $\beta=5.40$.
An illustration of the resulting $x_i$-dependencies of $\varphi(x_1,x_2,x_3)$
from the truncated expansion in Eq.~(\ref{NuclDAExp}) is displayed in Fig.~\ref{nDA_contour_QCDSF08}
in form of a barycentric contour plot (note that $\sum x_i=1$).
In contrast to the asymptotic case, where the DA has a single maximum at $x_i=1/3$,
the inclusion of the  $(k+l+m=2)$-moment in particular leads the two offset local maxima
and thereby a non-trivial distribution of the quark momentum. 

%% file: InProgress.tex
\section{Recent developments and work in progress}
\label{sec:InProgress}

\subsubsection{Deformation of hadrons from density-density correlators}
\label{sec:deformations}

\begin{figure}[t]
    \begin{minipage}{0.48\textwidth}
      \centering
          \includegraphics[angle=0,width=0.85\textwidth,clip=true]{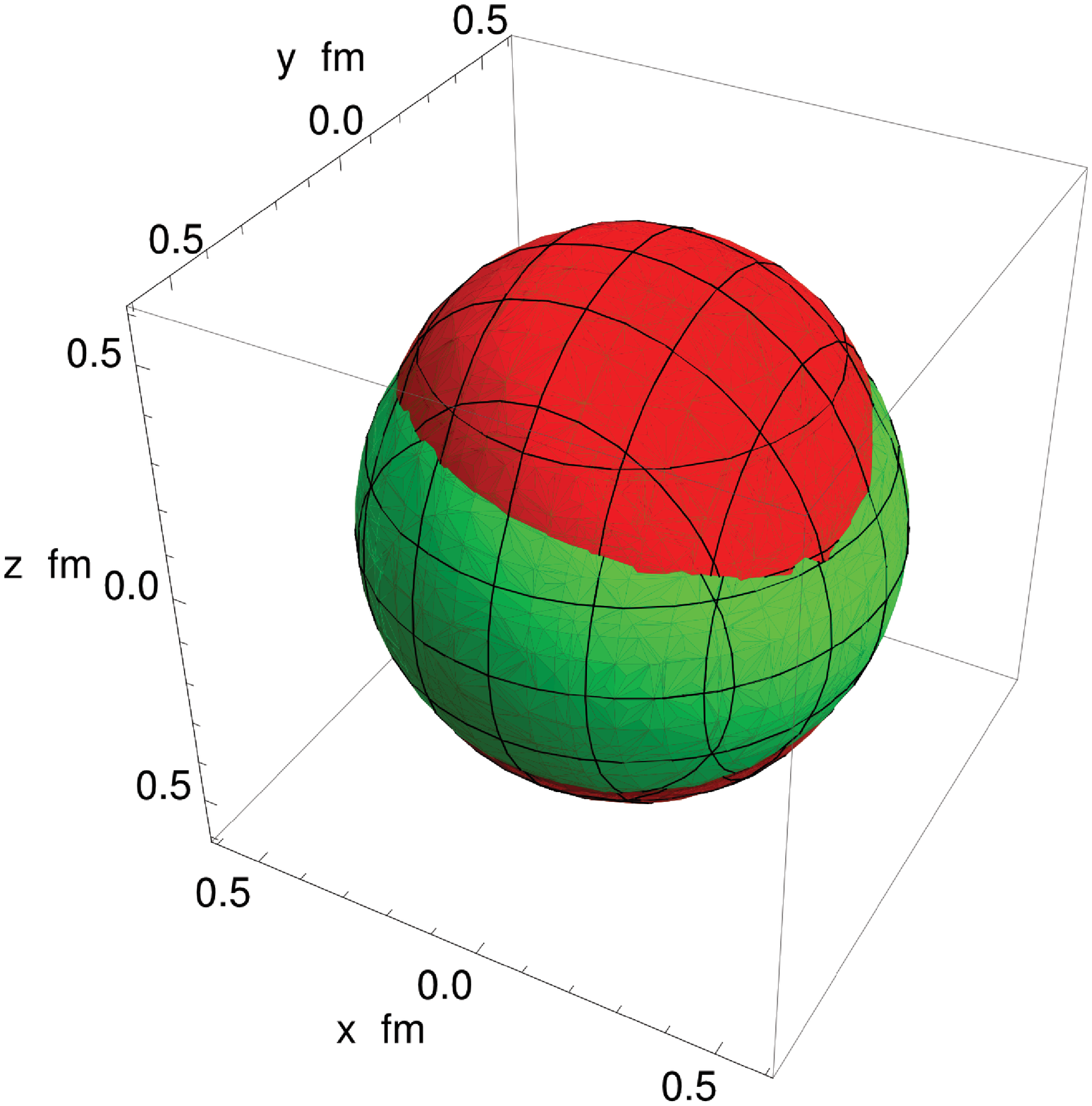}
  \caption{Surface plot of the density-density correlator $C_{\rho}(\mbf{r})$
  for the $\rho$-meson with spin projection $s_z=0$ for $m_\pi=509\MeV$ (from \cite{Alexandrou:2008ru}).
  The darker shaded surface corresponds to all $\mbf{r}$ with $C_{\rho}(\mbf{r})= C_{\rho}(\mbf{r}=0)/2$, intersected
  by the lighter shaded sphere with a radius of $|\mbf{r}|\approx0.5\fm$.}
  \label{cont3d_imgc_rho0_k0p1580}
    \end{minipage}
          \hspace{0.1cm}
    \begin{minipage}{0.48\textwidth}
      \centering
          \includegraphics[angle=0,width=0.85\textwidth,clip=true]{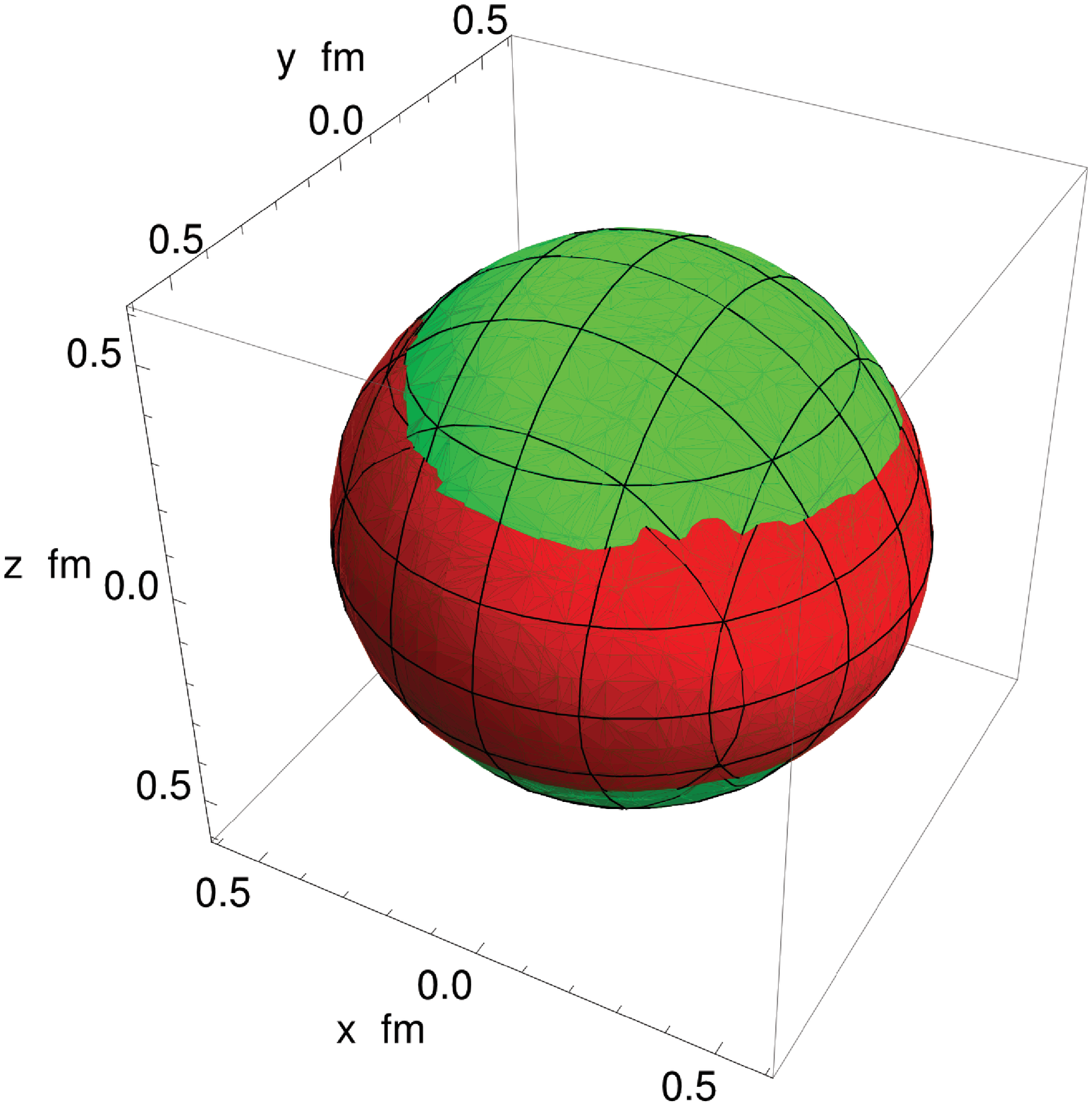}
  \caption{Surface plot of the density-density correlator of the $\rho$-meson with spin projection $s_z\pm1$
  for $m_\pi=509\MeV$ (from \cite{Alexandrou:2008ru}). Surfaces are as described in the caption of Fig.~\ref{cont3d_imgc_rho0_k0p1580}.}
  \label{cont3d_imgc_rho1_k0p1580}
    \end{minipage}
\end{figure}

An investigation of the deformation of hadrons based on density-density correlators
was recently presented by the Cyprus group \cite{Alexandrou:2008ru}. 
The study was based on equal-time four-point correlators as in Eq.~\ref{Rho4pt}, 
$C(\mbf{r}=\mbf{x},t_1)=C^{00}_{\text{4pt}}(\mbf{x},t_1=t_2)$,
for the pion, the $\rho$-meson, the nucleon, and the $\Delta$-baryon.
All-to-all propagators and the ``one-end-trick'' (see section \ref{sec:methods}), which was extended to
four-point-functions, were used for the first time for the calculation of four-point correlators.
In what was called the ``direct'' method, the all-to-all propagators were replaced by their
stochastic estimates, Eq.~\ref{stochastic1}, based on $Z(2)$ noise vectors.
Dilution in color, spin, and space (even-odd dilution) was employed to reduce the stochastic noise.
For the meson correlators, the direct method was compared to calculations based on the ``one-end-trick''.
Calculations were performed for $n_f=2$ flavors of Wilson fermions for pion masses of 
$m_\pi\approx384\MeV, 509\MeV$ and $\approx691\MeV$, lattice spacing of $a\approx0.077\fm$ and a volume of $V\approx(1.85\fm)^3$.
Gaussian smeared sources with HYP smeared gauge links were employed to improve 
the ground state signals and suppress contributions from excited states.
The density-density operator insertion time $t_1$ was chosen to be centered between sink and source times,
$\Delta t=t_1-t_\src=t_\snk-t_1$, allowing to extract a time-independent correlator
at large $\Delta t$, $C(\mbf{r})=\lim\displaylimits_{\Delta t\to\infty}C(\mbf{r},t_1)$.
In a first step, it was shown that the use of the ``one-end-trick'' leads indeed to a significant improvement in the 
statistical precision of $C_{\pi,\rho}(\mbf{r})$ compared to the direct method.
However, it was noted that the density-density correlators could be rather sensitive to finite size effects.
Due to the finite box size and the periodic boundary conditions, the correlator represents a 
sum over all periodic images, $C(\mbf{r})=\sum_{\mbf{n}}C_0(\mbf{r}+L\mbf{n})$. This  
must be taken into account in the interpretation of the results particularly for large distances $|\mbf{r}|$.
In a first attempt to correct for these finite size effects, the nearest neighbor contributions 
$C_0(\mbf{r}+L\mbf{n})$ with $|\mbf{n}|_\max=\sqrt{3}$ were included.
The lattice correlator $C(\mbf{r})$ was fitted using spherically symmetric 
(for the pion and nucleon) and non-symmetric (for the $\rho$ and $\Delta$) ans\"atze 
for the $\mbf{r}$-dependence of $C_0(\mbf{r})$ with up to five free parameters, and
corrected correlators were finally obtained by a subtraction of the nearest neighbor periodic images.
Results for the $\rho$-meson are displayed in Figs.~\ref{cont3d_imgc_rho0_k0p1580} 
and \ref{cont3d_imgc_rho1_k0p1580} for the spin projections 
$s_z=0$ and $s_z=\pm1$, respectively, in form of surface-plots with $\mbf{r}$ fixed by $C_{\rho}(\mbf{r})= C_{\rho}(\mbf{r}=0)/2$.

%
%

Compared to the lighter (green) shaded spheres, the darker shaded (red) surfaces 
are clearly deformed in $z$-direction, pointing towards a prolate shape for $s_z=0$ 
and oblate shape for $s_z=\pm1$.

In the case of the $\Delta$-baryon, where the one-end-trick could not be applied, no clear spatial deformations 
could be observed within the available statistics. 

The results presented in Figs.~\ref{cont3d_imgc_rho0_k0p1580} 
and \ref{cont3d_imgc_rho1_k0p1580} are certainly very promising.
However, the procedure to remove the periodic images introduces a systematic uncertainty 
due to fit-ans\"atze for the $\mbf{r}$-dependence of $C_0(\mbf{r})$.
It would therefore be very interesting to consider larger volumes, and to
investigate the influence of these ans\"atze on the final results in some more detail.

\subsubsection{Disconnected contributions}
\label{sec:disconnected}
Quark line disconnected contributions to three-point functions, as illustrated in Fig.~\ref{ill_3pt_disc_v1},
have been discussed already several times throughout this report in sections \ref{sec:Polarizabilities},
\ref{sec:nuclPDFs} and \ref{sec:SpinStructure}. 
They have been estimated, e.g., in the context of calculations of 
strange quark contributions to nucleon form factors
\cite{Dong:1997xr,Lewis:2002ix}, the quark momentum fraction 
and 
quark contributions to the nucleon spin 
\cite{Fukugita:1994fh,Dong:1995rx,Gusken:1999as,Mathur:1999uf,Gadiyak:2001fe,Deka:2008xr}, and 
polarizabilities \cite{Engelhardt:2007ub}.
Disconnected diagram contributions to quark three-point functions, also referred to as 
disconnected insertions (DI), are of the form 
$\langle G_h\times \mytr\overline{\tilde{\mathcal{O}}}\rangle_U$ and measure gauge field correlations 
between the (vacuum subtracted) trace of an operator,
$\mytr\overline{\tilde{\mathcal{O}}}=\mytr\tilde{\mathcal{O}}-\langle \mytr\tilde{\mathcal{O}}\rangle_U$,
and the hadron propagator $G_h$, see, e.g., the last line of Eq.~(\ref{C3ptProps}). 
These correlations turn out to
be in general small in magnitude compared to contributions from
quark line connected diagrams. 
As has been discussed in section \ref{sec:methods},
the DIs involve all-to-all quark propagators, and stochastic methods are required for their calculation.
Because of the additional stochastic noise and the relative smallness of the signal,
it is a challenge to obtain statistically significant results for the disconnected contributions.
Here, we briefly review some of the more recent efforts to improve the 
computation of disconnected diagram contributions in unquenched lattice QCD.

The Kentucky group recently reported promising results in particular for disconnected
contributions to the quark momentum fraction $\langle x\rangle_q$ for 
up-, down- and strange-quarks in the nucleon \cite{Deka:2008xr}.
Calculations were performed in the quenched approximation 
using 500 Wilson gauge configurations for a coupling of $\beta=6.0$ on a lattice of dimensions $16^3\times24$.
Contributions from quark line disconnected diagrams involving all-to-all propagators,
specifically the traces of the operators 
$\mytr\tilde{\mathcal{O}}(\tau)=\mytr\{ M^{-1}\tilde{K}^{\mathcal{O}}(\tau)\}$ 
(i.e. quark loops with operator insertion at time $\tau$, see Fig.~\ref{ill_3pt_disc_v1}),
were estimated using stochastic noise methods as discussed in section \ref{sec:stochastic}.
Up to 400 complex $Z_2$ noise estimators were used per gauge configuration,
and a noise reduction was performed by emplyoing an
unbiased subtraction based on the hopping parameter expansion \cite{Thron:1997iy}.
A further reduction of the noise was achieved by considering the 
parity, charge and $\gamma_5$-hermiticity transformation properties
of the nucleon propagator $G_N$ and the relevant quark loops
and by subsequently combining only those parts that are expected to give a non-zero
contribution. 
For example, the disconnected contribution to $\langle x\rangle_q$ can be obtained 
from a combination of nucleon three-point functions at zero momentum transfer, 
for a quark operator of non-diagonal type $\mathcal{O}^a_{4i}$, $i=1,\ldots,3$ (see Eq.~(\ref{op3}))
of the form
\bea
&&C^{N,\mathcal{O}^a_{4i}}_{\text{3pt};\Gamma}(\tau,\mbf{P}_i) 
- C^{N,\mathcal{O}^a_{4i}}_{\text{3pt};\Gamma}(\tau,-\mbf{P}_i)=\nonumber\\
&&\quad\quad\quad\quad-\big\langle\myIm\big( \mytr \big\{\Gamma G_N(t_\snk,\mbf{P}_i)\big\}-\mytr \big\{\Gamma G_N(t_\snk,-\mbf{P}_i)\big\}\big)
   \times \myIm\big(\mytr\tilde{\mathcal{O}}(\tau)\big)\big\rangle_U\,,
   \label{DI3pt}
\eea
involving only imaginary parts, while the real parts, which would contribute to the noise,
have been dropped explicitely.

An important observation with respect to disconnected contributions is that 
the sink-time $t_\snk$, e.g. in Eq.~(\ref{DI3pt}), may be easily varied,
in contrast to a conventional sequential-source calculation of the connected contribution,
which requires fixing $t_\snk$, see, e.g., Eq.~(\ref{C3pt2_seq}) and related discussion.
Therefore, in order to increase the statistics,
one may sum the three-point functions in Eq.~(\ref{DI3pt})
over the insertion time $\tau$, while studying the residual dependence on $t_\snk$ 
\cite{Maiani:1987by,Viehoff:1997wi,Gusken:1999as,Deka:2008xr}.
Introducing a fixed source-time $t_\src$ (which may be set to zero), one finds 
\bea
&&\frac{\displaystyle\sum\limits_{\tau=t_\src+1}^{t_\snk-1}C^{N,\mathcal{O}^a_{4i}}_{\text{3pt};\Gamma}(\tau,t_\snk,\mbf{P}_i)}
       {C^N_{\text{2pt};\Gamma}(t_\snk,\mbf{P}_i)} = c(\mbf{P})\, \langle x\rangle_q t_\snk \,+\, \text{const} \,+\, \cdots\,,
   \label{sumDIratio}
\eea
which depends linearly on $t_\snk$, while contributions involving excited (positive and negative parity) states
with energies $E',E'',\ldots$ indicated by the dots are exponentially suppressed by factors 
$e^{-(E'-E) (t_\snk-t_\src)}$. Since $c(\mbf{P})$ is a known kinematical coefficient,
the momentum fraction $\langle x\rangle_q$ may therefore be extracted from the slope in $t_\snk$ of Eq.~(\ref{sumDIratio})
in the limit $t_\snk-t_\src\gg(E'-E)^{-1}$.
This approach is different from the standard method where a plateau in the
ratio of three- to two-point functions is sought at operator insertion times far
away from sink and source, $t_\snk-\tau\gg(E'-E)^{-1}$, $\tau-t_\src\gg(E'-E)^{-1}$,
and where contributions from excited states are suppressed by factors of
$e^{-(E'-E) (t_\snk-\tau)}$ and $e^{-(E'-E) (\tau-t_\src)}$, see, e.g., Eq.~(\ref{ratio1}).
Contributions from excited states close to source and sink are included
in the sum over $\tau$ in Eq.~(\ref{sumDIratio}), and in practice it should be checked
if they are sufficiently suppressed for the accessible source-sink separations $t_\snk-t_\src$.

Calculations were performed for three different quark masses (hopping parameters) with $m_u=m_d$,
corresponding to pion masses of $\approx478\MeV$, $\approx538\MeV$ and $\approx650\MeV$, with
a lattice spacing of $a\approx0.11\fm$ that was set using the nucleon mass.
The lattice operators were perturbatively renormalized and transformed to the $\MSbar$-scheme 
at a scale of $\mu=2\GeV$.

\begin{figure}[t]
    \begin{minipage}{0.48\textwidth}
      \centering
          \includegraphics[angle=-90,width=1.\textwidth,clip=true]{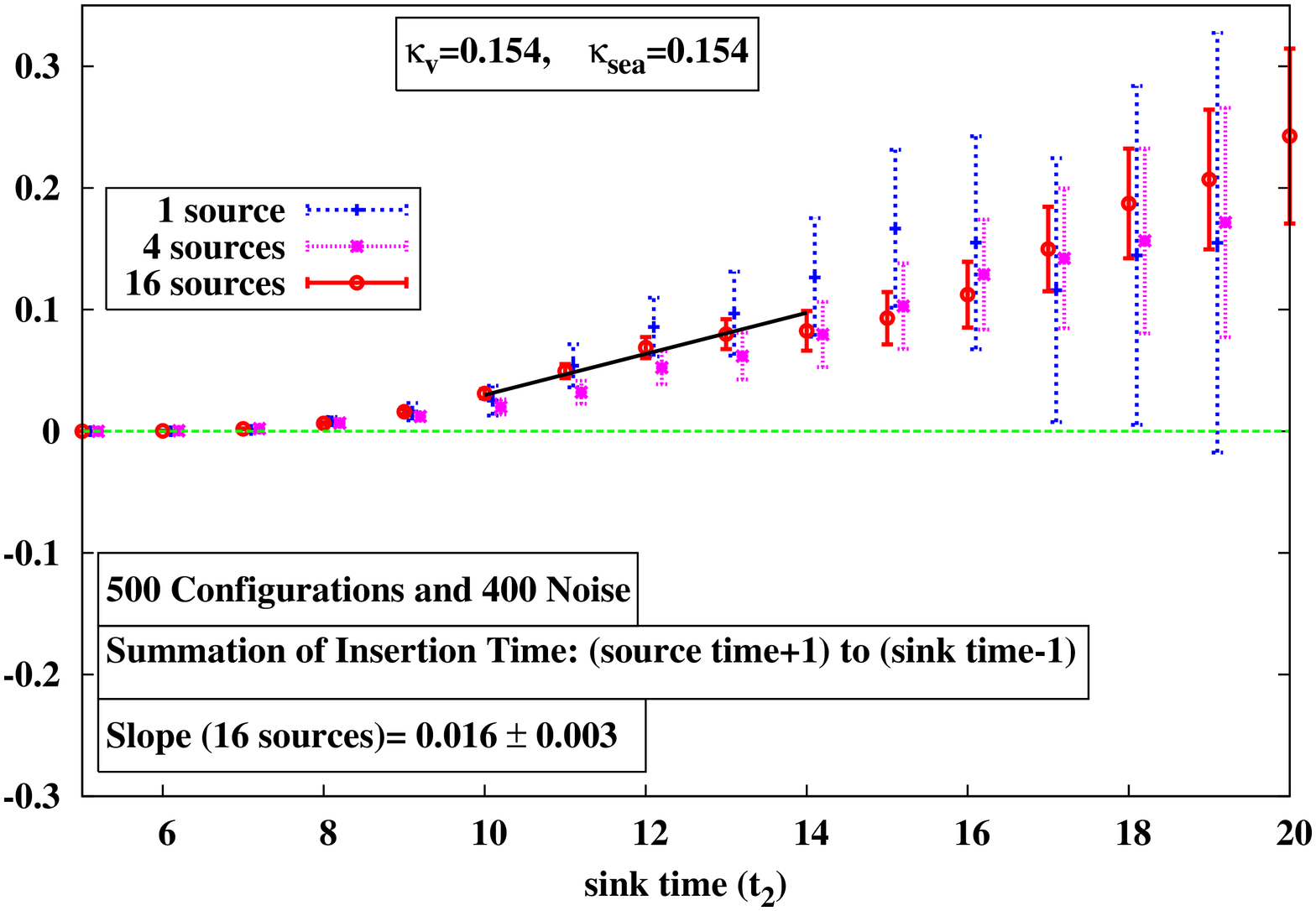}
  \caption{Summed ratio of three- to two-point functions for the disconnected insertion
  as a function of the sink time (from \cite{Deka:2008xr}).}
  \label{ratio_Deka}
  \end{minipage}
          \hspace{0.1cm}
    \begin{minipage}{0.48\textwidth}
        \centering
          \includegraphics[angle=-90,width=1.\textwidth,clip=true]{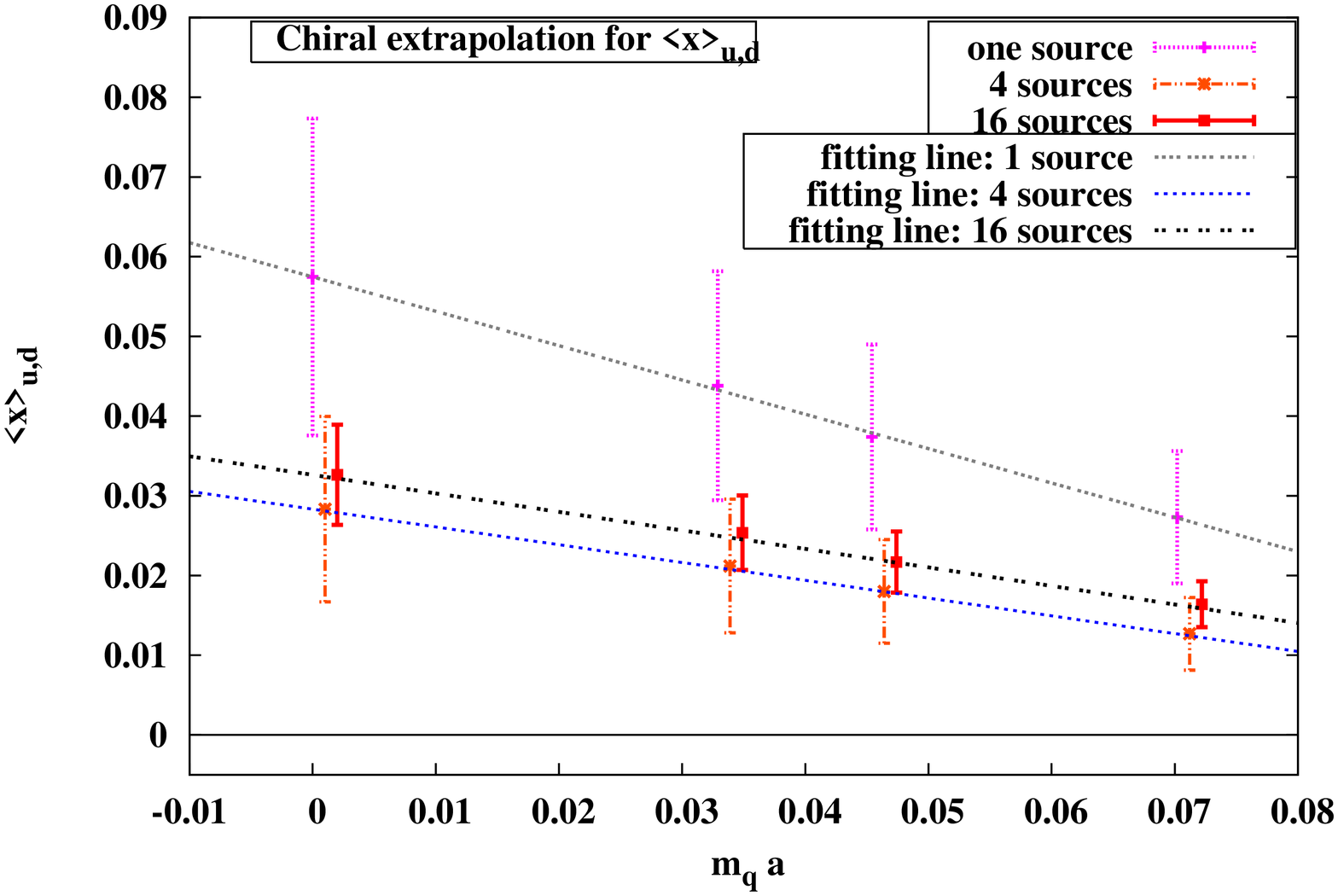}
  \caption{Quark mass dependence of $\langle x\rangle^{\text{disc}}_{u,d}$ (from \cite{Deka:2008xr}).}
  \label{DI_xud_vs_mq_Deka}
  \end{minipage}
\end{figure}

Results for the summed ratio, Eq.~(\ref{sumDIratio}), as a function of $t_\snk$ are
displayed in Fig.~\ref{ratio_Deka} for a pion mass of $\approx650\MeV$.
Unbiased subtraction and the use of up to 16 nucleon sources was essential to obtain
a clean signal and to increase the statistics. 
The slope, i.e. the momentum fraction, was extracted from a correlated fit to the
lattice data points (open circles) in the range of $t_\snk=10,\ldots,14$
as indicated by the solid line in Fig.~\ref{ratio_Deka}, giving a 
clearly non-zero value of $\langle x\rangle^{\text{disc}}_q=0.016\pm0.003$.

Figure ~\ref{DI_xud_vs_mq_Deka} shows the quark mass ($m_\pi^2$-) dependence of $\langle x\rangle^{\text{disc}}_{u,d}$
together with linear extrapolations in $m_q$ to the chiral limit, represented by the dotted lines,
from which a final value of $\langle x\rangle^{\text{disc,extr}}_{u,d}=0.032\pm0.006$ was obtained 
for the case of 16 sources.
By keeping the quark mass in the loop fixed at the largest value (corresponding to $\kappa=0.154$)
and only varying the valence quark masses, the strange quark contribution to the momentum fraction in the nucleon
was extracted in a similar way, giving $\langle x\rangle^{\text{extr}}_{s}=0.027\pm0.006$.
From these results, a ratio of strange to (disconnected) up quark contributions
of $(\langle x\rangle_{s}/\langle x\rangle_{u})^{\text{disc,extr}}=0.88\pm0.07$ was found.

In addition, calculations of the connected contributions to the momentum fractions were performed employing the conventional
sequential source technique. From plateaus in $\tau$ of ratios of (connected) three- to two-point functions
and linear extrapolations to the chiral limit, values of
$\langle x\rangle^{\text{con,extr}}_{u}=0.408\pm0.038$ and $\langle x\rangle^{\text{con,extr}}_{d}=0.148\pm0.019$ were obtained,
in overall agreement with earlier results in the quenched approximation, see, e.g., \cite{Dolgov:2002zm}. 

In summary, this study gives strong indications that the often neglected contributions from disconnected diagrams 
to, e.g., the total light quark momentum fraction in the nucleon, $\langle x\rangle_{u+d}$, may be as large as $\approx10-20\%$. 
If correct, many lattice studies of isosinglet ovservables,
such as quark angular momentum contributions to the nucleon spin in section \ref{sec:SpinStructure}, 
in which only the connected contributions were taken into account,
could be affected by substantial systematic uncertainties and should therefore be re-assessed.

The above results should of course be considered with caution in particular because of
the use of the quenched approximation and the linear extrapolations to the chiral limit.
With respect to the extraction of the disconnected parts from slopes in $t_\snk$, cf. Eq.~(\ref{sumDIratio}) and Fig.~\ref{ratio_Deka},
it would also be important to check for possible contaminations from excited states.

%
%
%

For preliminary results by the Kentucky group on disconnected and connected contributions 
to the quark momentum fraction obtained for $n_f=2+1$ Clover Wilson fermions, we
refer to the proceedings \cite{Doi:2008hp,Mankame:2008gt}.

Progress in the stochastic computation of disconnected contributions 
based on a combination of various noise reduction methods has been reported recently in \cite{Bali:2008sx}.
Partitioning in time has been employed by setting the noise vectors equal to zero except on a single time-slice.
The idea of the hopping parameter expansion (see section \ref{sec:methods}) has been implemented by noting that
the lowest terms in the HPE for the Wilson action do not contribute to the trace, i.e. one can replace
$\mytr(\Gamma M^{-1})=\mytr(\Gamma[1-\kappa\Dslash]^{-1})=\mytr[\Gamma(\kappa\Dslash)^nM^{-1}]$,
with, e.g., $n=8$ for the axial vector current, $\Gamma=\gamma_\mu\gamma_5$.
In some cases, a further reduction of the noise has been achieved using the truncated eigenmode
approach, as briefly discussed in section \ref{sec:methods}.

Finally, a new method called the truncated solver method (TSM) introduced in \cite{Collins:2007mh},
was used to improve the computation of the stochastic estimates. 
In its basic form, it is based on two independent sets (set $1$ and set $2$)
of noise sources with numbers of sources
equal to $N_{\eta,1}$ and $N_{\eta,2}$. The idea is to stop the iterative solver 
used to solve $M\phi^{}=\eta_{1}^{(j)}$ for set $1$ 
\emph{before} convergence is achieved, i.e. after $n_\text{trunc}<<n_\text{conv}$ iterations.  
Subsequently, the difference to the full converged solution can be stochastically estimated
based on set $2$, giving an unbiased estimate of the propagator in the form 
$E[M^{-1}_\text{conv}]=E_1[M^{-1}_\text{trunc}]+E_2[M^{-1}_\text{conv}-M^{-1}_\text{trunc}]$.
If the convergence behavior is such that already for small $n_\text{trunc}$ a result close to the
converged answer is obtained, then the stochastic error may be significantly reduced 
by using a large number of sources in the truncated inversion, and a small
number of sources for the expensive estimate of the correction to the converged results,
i.e.  $N_{\eta,1}>>N_{\eta,2}$. 

The numerical studies in \cite{Collins:2007mh} were based on a mixed action approach using the Wilson action 
for the valence fermions together with gauge configurations for $n_f=2+1$ stout-link improved (rooted) staggered sea quarks 
and a Symanzik improved gauge action, for a lattice spacing of $a\approx0.127\fm$, and a physical volume
of $V\approx(2.0\fm)^3$.
Using a standard ratio of the disconnected parts of the three-point to the two-point function,
the 
disconnected contribution to the nucleon spin, $\Delta q^{\text{disc}}$, was
estimated for three different quark masses in the quark loop and 2 different quark masses 
in the nucleon 2-point correlators, corresponding to pion masses in the range of $\approx300\MeV$ to $\approx600\MeV$. 
All results for $\Delta q^{\text{disc}}$ turned out to be compatible with zero.
Compared to a sole partitioning in time, the use of the additional noise reduction
techniques lead to absolute errors that were smaller by up to a factor of $3$,
while the central values moved even closer to zero.
%
%
\begin{figure}[t]
       \centering
          \includegraphics[angle=0,width=0.45\textwidth,clip=true]{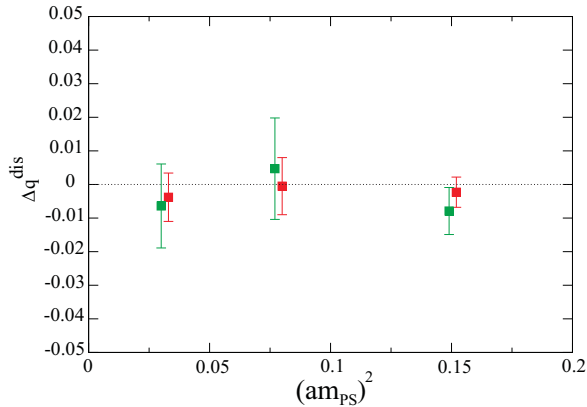}
  \caption{Unrenormalized disconnected quark contribution to the nucleon spin (from proceedings \cite{Bali:2008sx}).}
  \label{DeltaS_extrap_loop}
\end{figure}
%

Figure \ref{DeltaS_extrap_loop} displays $\Delta q^{\text{disc}}$
as a function of the squared pion mass in lattice units (corresponding to the three different
quarks masses in the loop). 
Within errors, no systematic variation of the results with the quark mass could be observed.
An already quite impressive upper limit of $|\Delta s|<0.022$ at $95\%$ confidence level
is found for the unrenormalized strange quark contribution to the nucleon spin at the intermediate nucleon mass.
This may be compared to recent results from 
HERMES of $\Delta s=-0.085(13)_\text{th}(8)_\text{exp}(9)_\text{evo}$
obtained from inclusive DIS and using $SU(3)$ symmetry \cite{Airapetian:2007aa}, and 
$\Delta s^{[x=0.02-0.4]}=0.037(19)_\text{stat}(27)_\text{sys}$ 
from charged kaon production in semi-inclusive DIS \cite{Airapetian:2008qf}, both 
for the $\MSbar$-scheme at a scale of $5\GeV^2$.
Keeping in mind that the experimental results generically suffer from 
uncertainties due to lack of data in the region of very low $x$, further improved lattice
calculations (including a proper operator renormalization) 
might indeed have a significant impact and help to get a better quantitative understanding of
strange, and possibly also the light, quark contributions to the nucleon spin in the near future.

Further preliminary results for strange quark contributions to nucleon form factors
using ``extremely diluted'' stochastic sources have been reported by Babich et al. \cite{Babich:2007jg,Babich:2009rq}.

\subsubsection{Operator product expansion on the lattice}
\label{sec:OPE}
Many hadron structure studies in lattice QCD focus on hadronic matrix elements
of local operators, providing, e.g., moments of PDFs.
They are related to the cross section of deep inelastic scattering (DIS),
i.e. the structure functions, by the operator product expansion (OPE).
The OPE of the hadronic current tensor given by the product of two electromagnetic currents
can be written in momentum space as
\bea
W_{\mu\nu}=\langle h(P)|J_\mu(q)J^\dagger_\nu(q)|h(P)\rangle
=\sum_{i,n} C_{\mu\nu,\{i,n\}}(q) \langle h(P)|\mathcal{O}_{\{i,n\}}|h(P)\rangle \,,
\label{OPE}
\eea
where in DIS $q$ is the momentum transferred by the virtual
photon to the hadron $h$, $C_{\mu\nu,\{i,n\}}(q)$ are the Wilson coefficients, and $\mathcal{O}_{\{i,n\}}$ 
are local quark and gluon operators. Concentrating on the non-singlet sector, 
the sum in Eq.~(\ref{OPE}) runs over an infinite set of local quark operators,
\bea
\mathcal{O}_{\{i,n\}}=\mathcal{O}_{\Gamma_i\mu_1\ldots\mu_n}=\bar q\Gamma_i\Dlr_{\mu_{1}}\cdots\Dlr_{\mu_{n}}q\,,
\label{OPE2}
\eea
where $\Gamma_i$ is an element of the Dirac algebra, $i=1,\ldots,16$,
and $\{i,n\}$ is shorthand for the Lorentz indices and Dirac structure.
The hadronic matrix elements of the local operators are parametrized in terms of
moments of leading and higher twist PDFs.
In the usual phenomenological approach,
the Wilson coefficients $C_{\mu\nu,\{i,n\}}(q)$ are calculated to a given order in QCD perturbation theory, and are
used to extract the PDFs from the measured structure functions using the OPE, Eq.~\ref{OPE}.
The moments of the PDFs obtained in this way can finally be compared to corresponding lattice results.
It has, however, been noted that also the Wilson coefficients $C_{\mu\nu,\{i,n\}}(q)$ 
can be calculated non-perturbatively in lattice QCD 
\cite{Capitani:1998fe,Capitani:1999fm,Detmold:2005gg,Bietenholz:2007wa,Bietenholz:2008fe}. 
This is feasible
because they are independent of the hadron $h$ and its momentum $P$, so that
off-shell quark states $|h(P)\rangle=|q(P)\rangle$ can be used to
calculate the corresponding matrix elements on the left and right hand sides of Eq.~(\ref{OPE}).
To get a non-zero result, this clearly has to be done in a fixed gauge, for example Landau gauge. 
A non-perturbative calculation of Wilson coefficients is not only interesting by itself,
but could be used together with lattice determinations of nucleon
matrix elements of local quark operators, discussed in section \ref{sec:nuclPDFs}, to obtain
fully non-perturbative results for moments of nucleon structure functions.  
If the Wilson coefficients and the nucleon matrix elements are calculated within
the same lattice framework, their scale and scheme dependences will cancel out
in the calculation of the hadronic tensor, Eq.~\ref{OPE}. Bare lattice results
may therefore be used, avoiding the 
costly non-perturbative renormalization of local lattice operators.
Furthermore, higher twist contributions to moments of nucleon structure functions,
implicitly included in the OPE in Eq.~(\ref{OPE})\footnote{Note that the operators in Eq.~\ref{OPE2} are neither traceless nor symmetrized.},
may be directly accessed in the framework of a non-perturbative calculation of the Wilson-coefficients
and the corresponding nucleon matrix elements of local operators \cite{Capitani:1999fm}.

In practice, only a finite number of operators can be considered,
so that the expansion in Eq.~\ref{OPE} has to be truncated. 
A truncation of terms of higher dimension, i.e. at large $n$, can be justified if
the involved scales are well separated, that is if $|P^2|<<|q^2|$. At the same time,
in order to keep discretization effects in a lattice calculation under control, $|q^2|<<(\pi/a)^2$.
So far, operators with up to three covariant derivatives, $n=1,\ldots,n_{\max}=3$, have been included in
lattice QCD calculations \cite{Capitani:1998fe,Capitani:1999fm,Bietenholz:2007wa,Bietenholz:2008fe}. 
The most recent study of Wilson coefficients in lattice QCD was done in the quenched approximation,
using the L\"uscher-Weisz gauge action with a lattice spacing of 
$a\approx0.075\fm$ and a volume of $V\approx(1.8\fm)^3$, and overlap valence quarks 
with a bare quark mass of $m_q\approx73\MeV$ \cite{Bietenholz:2008fe}.
The large number of $1360$ operators for $n_{\max}=3$ could be reduced to only $67$
independent operators by choosing an isotropic momentum transfer, $q\propto(1,1,1,1)$,
and using symmetry arguments \cite{Bietenholz:2007wa}.
To determine the $N=67$ Wilson coefficients, off-shell quark matrix elements 
of the product of two electromagnetic currents, and of the local operators, 
were calculated for up to $M=25$ different lattice momenta $P_{i=1,\ldots,M}$.
For fixed external indices, for example $\mu=\nu=4$, this leads to linear equations of the form
$W_{44}(P_i,q)=\sum_{m=1}^{67}C^m_{44}(q)M_m(P_i)$,
where $M_m(P)=\langle q(P)|\mathcal{O}_m|q(P)\rangle$, the index $m$ specifies
the 67 independent operators, and $i=1,\ldots,M$. Taking also into account
all $4\times4=16$ combinations of the spinor-indices of the initial and final quark states,
this provides a system of $16\times M$ equations. In the case that this system
is overdetermined, i.e. that the number of linearly independent equations is larger than 
the number of independent Wilson coefficients, an approximate solution can be constructed
using standard methods, for example the singular value decomposition.

\begin{figure}[t]
      \centering
          \includegraphics[angle=-90,width=0.5\textwidth,clip=true]{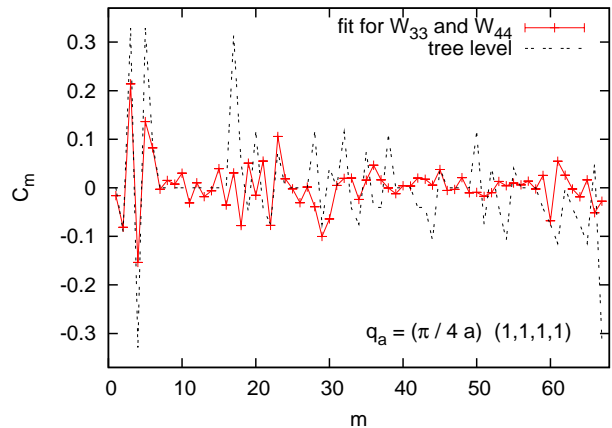}
  \caption{Wilson coefficients for operators with up to three derivatives (from proceedings \cite{Bietenholz:2008fe}).}
  \label{C_W3344q2}
\end{figure}

Numerical results for the $m=1,\ldots,67$ Wilson coefficients obtained using 
$\mu=\nu=3$ and $\mu=\nu=4$, for a momentum transfer squared of $q^2\approx17\GeV^2$, 
are displayed in Fig.~\ref{C_W3344q2}. For orientation, we note that 
$m=1$ correspond to operators with zero, $m=2,\ldots,6$ with one,
$m=7,\ldots,16$ with two, and $m=17,\ldots,67$ with three derivatives.
Typical choices for independent operators are $\mathcal{O}_{m=1}=\overline q\mathds{1}q$ for zero derivatives, and 
$\overline q \sum_{i=1}^{3}\gamma_iD_iq$ for the one-derivative case.
Because of chiral symmetry, the Wilson coefficients for $\mathcal{O}_{m=1}$ and for
any other of the 67 operators with an even number of derivatives must vanish in the combined continuum and chiral limit.
In this study, due to the use of overlap fermions,
which provide an exact lattice chiral symmetry, the corresponding Wilson coefficients were expected to be small.
This was indeed numerically confirmed as shown in Fig.~\ref{C_W3344q2}, where the corresponding coefficients turn out to be suppressed.
%
%
In particular for low $m$, the lattice data points follow a pattern similar to the tree-level results
given by the dashed line. This is somewhat different for larger $m$, where clear
differences between the perturbative and non-perturbative results can be observed.
It would be very interesting to further improve these preliminary studies by
using larger sets of lattice momenta $P_{i=1,\ldots,M}$, additional momentum transfers $q$,
and by including all possible combinations of the indices $\mu,\nu$. Lower $P^2$, and therefore
a better scale separation, will be achieved by using partially twisted boundary conditions 
\cite{Bietenholz:2008fe}.
\subsubsection{Nucleon structure with partially twisted boundary conditions}
\label{sec:NuclPTBCs}
\begin{figure}[t]
    \begin{minipage}{0.48\textwidth}
      \centering
\includegraphics[angle=0,width=0.9\textwidth,clip=true]{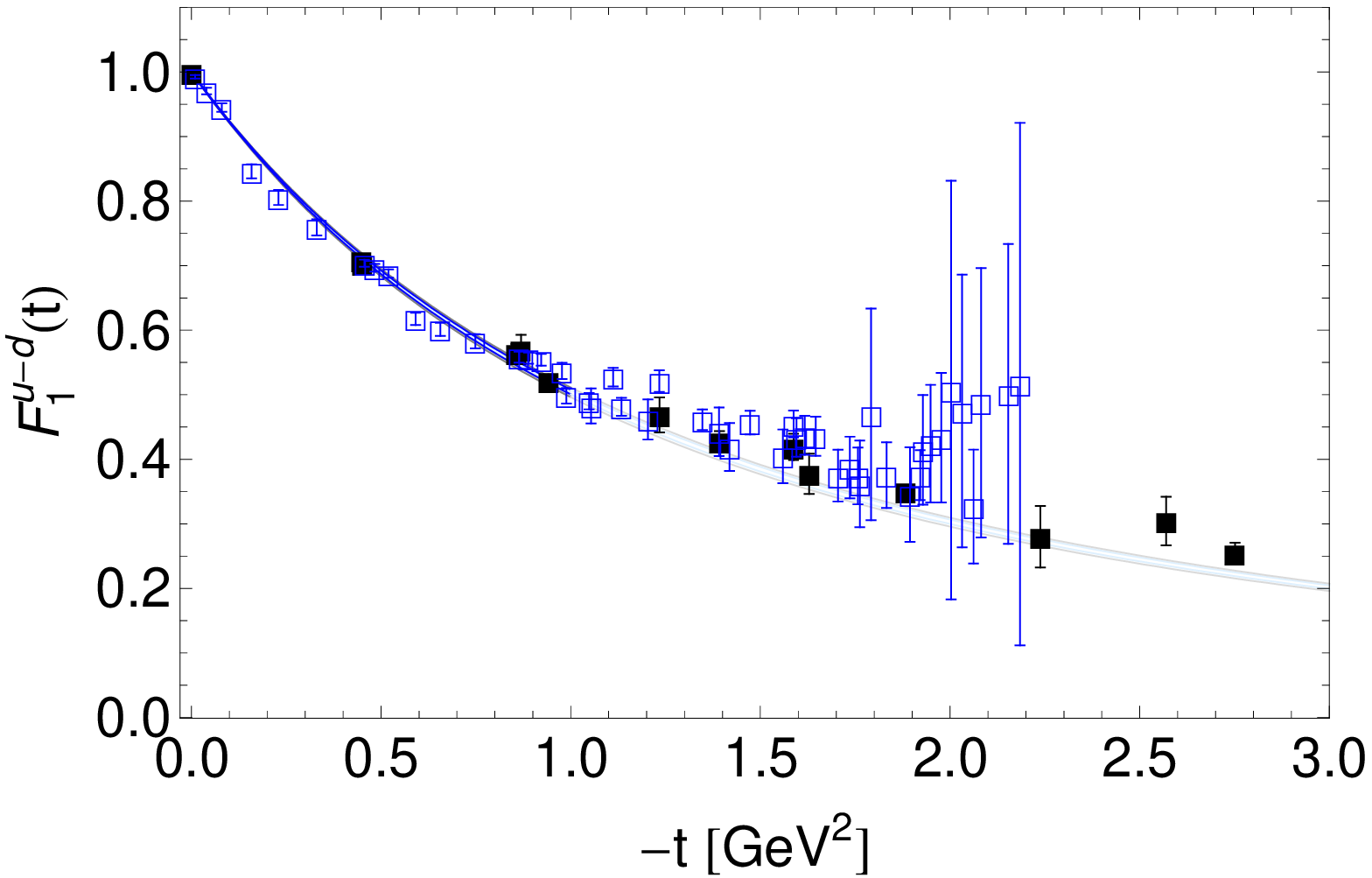}
  \caption{The Dirac form factor in the isovector channel using pTBCs (from proceedings \cite{Hagler2008}).}
  \label{F1_umd_13590}
     \end{minipage}
          \hspace{0.5cm}
    \begin{minipage}{0.48\textwidth}
      \centering
          \includegraphics[angle=0,width=0.9\textwidth,clip=true]{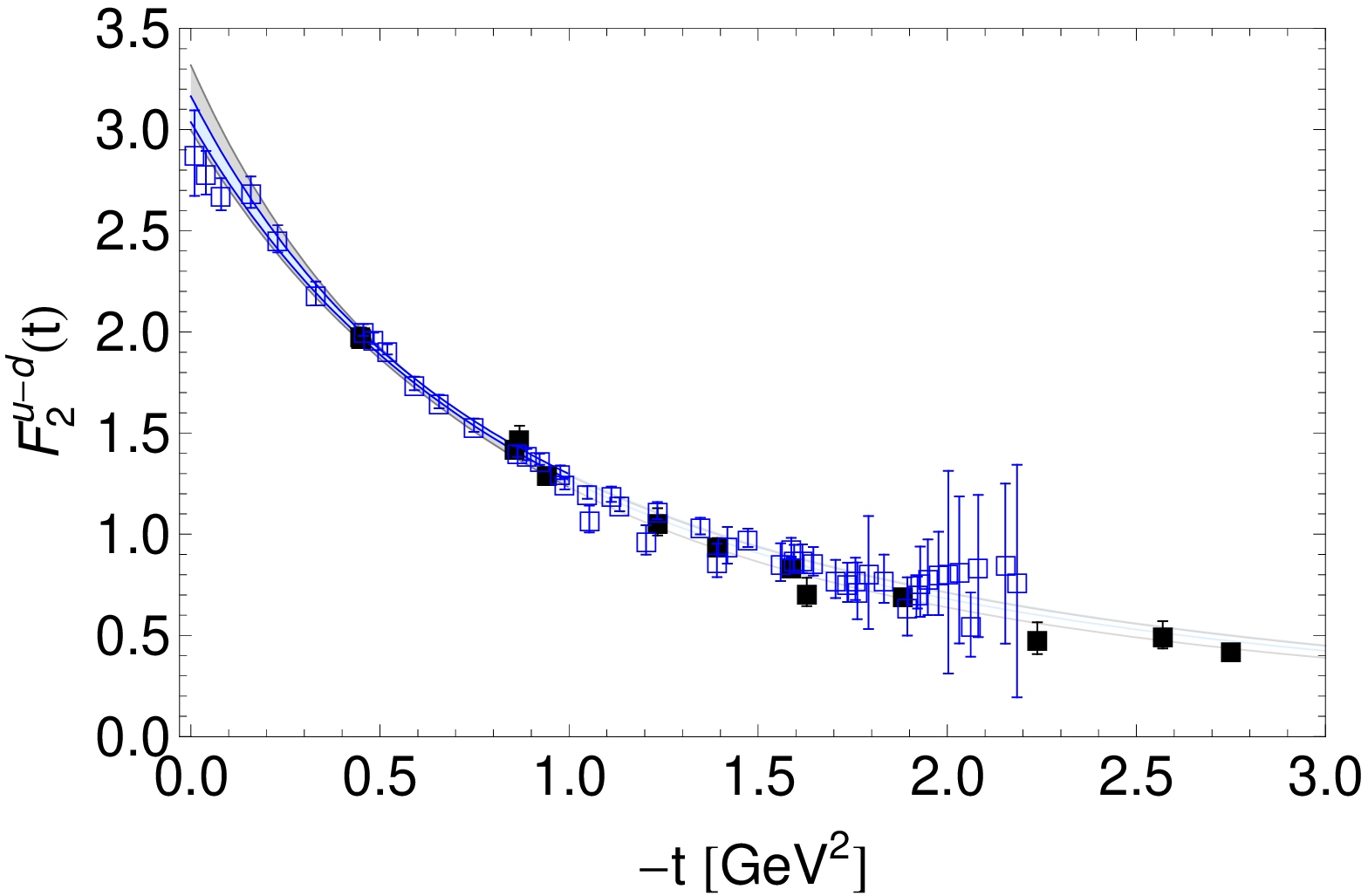}
  \caption{The Pauli form factor in the isovector channel using pTBCs (from proceedings \cite{Hagler2008}).}
  \label{F2_umd_13590}
     \end{minipage}
\end{figure}

Partially twisted boundary conditions (pTBCs), introduced at the end of section \ref{sec:methods}, 
have already been successfully employed in recent years 
for calculations in the meson sector, for example the kaon to pion transition
form factor and the pion form factor \cite{Boyle:2007wg,Boyle:2008yd}, as discussed in
section \ref{pionFFs}.
More recently, pTBCs have been used for the first time in an exploratory 
study of the nucleon structure \cite{Hagler2008}.
Calculations were performed in the framework of simulations with $n_f=2$ flavors
of non-perturbatively clover-improved Wilson fermions and the Wilson gauge action.
Preliminary results were obtained for an ensemble with $\beta=5.29$ and $\kappa=0.13590$, 
corresponding to a lattice spacing of $a\approx0.075\fm$, a pion mass of $m_\pi\approx630\MeV$,
for a spatial volume of $V=L^3\approx(1.8\fm)^3$, with $m_\pi L\approx5.7$.
The twisting angles were chosen such that very small, non-zero values of the
squared momentum transfer could be reached, $|t^{\not=0}_{\min}|\simeq0.01\GeV^2$,
and that the large gaps of $\approx0.5\GeV^2$ between the accessible Fourier momenta, 
$\mbf{p}^F=(2\pi/(aL))\mbf{n}$, could be filled.
Figures \ref{F1_umd_13590} and \ref{F2_umd_13590} show the $t$-dependence
of the isovector Dirac and the Pauli form factors, respectively.
Results for standard periodic boundary conditions (``Fourier-momenta'') 
are given by the filled squares, while the larger number of open
squares represent the lattice data points obtained with pTBCs.
Dipole fits to the filled squares, and to the combined data points 
for $|t|\le1\GeV^2$ are shown by the gray and the blue error bands, respectively.
The statistical precision of the results at low momentum transfers
is quite remarkable. However, the pTBC data points above $t\sim-1\GeV^2$
scatter significantly, and sets of data points below
$t\sim-1\GeV^2$ seem to be systematically lower than the average. 
A likely explanation are currently uncontrolled systematic uncertainties, 
in particular discretization effects and finite volume effects related to the pTBCs. 
As has been observed in \cite{Tiburzi:2006px,Jiang:2008ja} for the isovector magnetic moment
in the framework of partially quenched ChPT, finite volume effects 
from pTBCs may indeed be sizeable in the nucleon sector. 

From the dipole fit to the combined data points for the Dirac form factor in Fig.~\ref{F1_umd_13590},
a mean square radius of $\langle r_1^2\rangle^{u-d}=12/m_D^2=0.191(2)\fm^2$ was obtained,
showing a substantially reduce statistical error compared to
$\langle r_1^2\rangle^{u-d}=0.191(5)\fm^2$
for the Fourier momentum (periodic BCs) based results.
A similar improvement in statistics by using pTBCs was observed for the Pauli form factor 
displayed in Fig.~\ref{F2_umd_13590}.
Dipole fits to the combined lattice data give a mean square radius of 
$\langle r_2^2\rangle^{u-d}=0.259(10)\fm^2$
and an anomalous magnetic moment of $\kappa^{u-d}=3.101(64)$,
compared to $\langle r_2^2\rangle^{u-d}=0.272(26)\fm^2$ and $\kappa^{u-d}=3.158(160)$
from the fit to the filled data points for the periodic boundary conditions.
Even more promising, a value of $\kappa^{u-d}\approx F^{u-d}_2(t\approx-0.01\GeV^2)=2.88(21)$
with a statistical error below $10\%$ could be 
directly obtained very close to the forward limit with pTBCs.
However, as has been noted above, finite volume effects may be significant
in this case, so that the results should be regarded with due caution.
%
%
%
%

\begin{figure}[t]
    \begin{minipage}{0.48\textwidth}
      \centering
\includegraphics[angle=0,width=0.9\textwidth,clip=true]{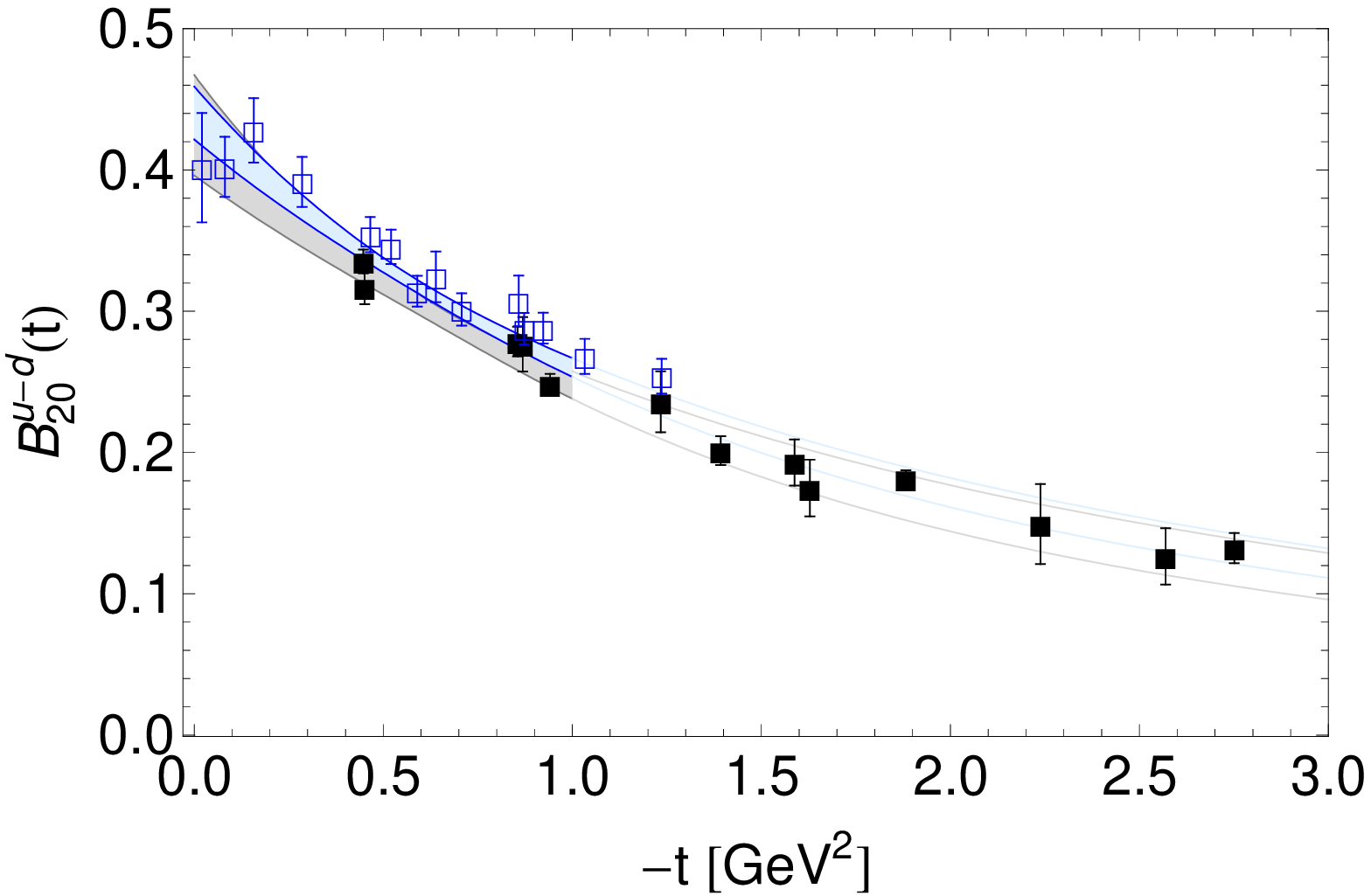}
  \caption{The form factor $B_{20}(t)$ in the isovector channel using pTBCs (from proceedings \cite{Hagler2008}).}
  \label{B20_umd_13590}
     \end{minipage}
          \hspace{0.5cm}
    \begin{minipage}{0.48\textwidth}
      \centering
          \includegraphics[angle=0,width=0.9\textwidth,clip=true]{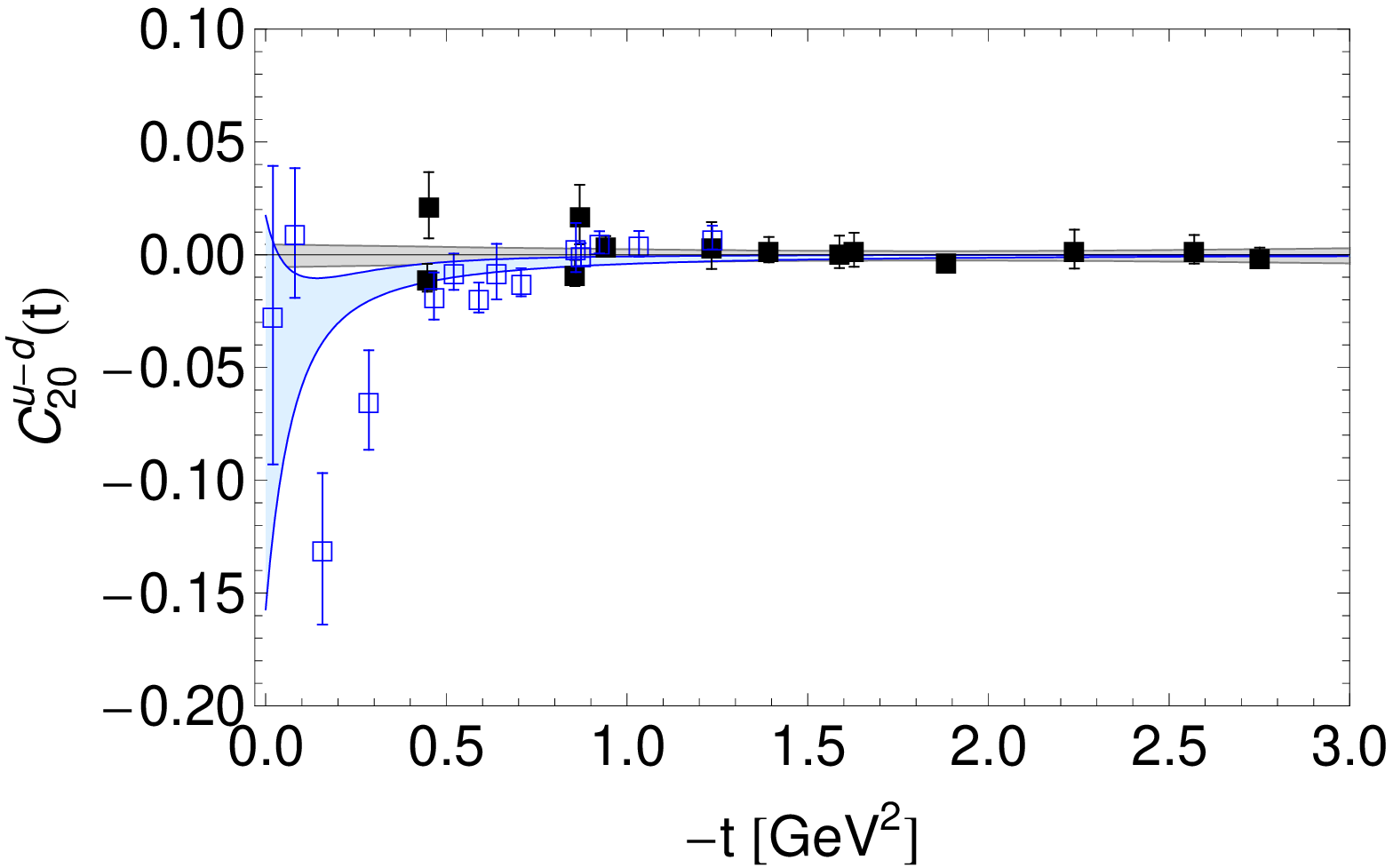}
  \caption{The form factor $C_{20}(t)$ in the isovector channel using pTBCs (from proceedings \cite{Hagler2008}).}
  \label{C20_umd_13590}
     \end{minipage}
\end{figure}

Preliminary results for the nucleon generalized form factors
$B_{20}(t)$, which plays a central role for the nucleon spin structure, Eq.~(\ref{SpinSumrule1}), and $C_{20}(t)$,
obtained using pTBCs are shown in Figs.~\ref{B20_umd_13590} and \ref{C20_umd_13590} in the isovector channel.
A calculation of these GFFs employing pTBCs is particularly
promising, since neither $B_{20}(t)$ nor $C_{20}(t)$ can be directly obtained
at zero momentum transfer, as discussed at the end of section \ref{sec:methods}.

As before, the filled squares represent the results for standard periodic BCs,
and the data points obtained with pTBCs are given by the open squares.
Dipole fits to the periodic BC and to the combined data points are represented
by the error bands.
A direct calculation of $B^{u-d}_{20}(t)$ very close to the 
forward limit gives $B^{u-d}_{20}(-0.02\GeV^2)=0.402(39)$ with 
good statistical precision of $\sim10\%$. 
Within errors, this is compatible with the values from
the dipole fits of $B^{u-d}_{20}(0)=0.440(19)$ for the combined, and 
$B^{u-d}_{20}(0)=0.432(35)$, for the periodic BC
(Fourier momentum) lattice data.
Qualitatively new insights may be obtained for $C^{u-d}_{20}(t)$.
In this case, the lattice data points for the periodic BCs are
compatible with zero over the full range of $|t|\simeq0.5,\ldots,3\GeV^2$.
However, for the low values of $|t|\le0.5\GeV^2$ that could be reached with pTBCs,
a trend towards non-zero, negative values can be observed in Fig.~\ref{C20_umd_13590},
albeit with rather large statistical errors.
Provided that systematic uncertainties such as finite volume effects 
can be brought under control, partially twisted boundary conditions hold great promise
for the understanding of a large number of key nucleon structure observables.

%

%
%

\subsubsection{Nucleon form factors at high momentum transfer on anisotropic lattices}
\label{sec:highQ}
Lattice calculations of nucleon form factors at large values of the momentum transfer
are interesting for a number of reasons.
First, they might provide a better understanding of the large $Q^2$-scaling behavior
in particular in comparison with respective predictions from perturbative QCD.
Moreover, %
they could provide interesting predictions for planned experiments in the high-$Q^2$ region at, e.g., JLab.
Specifically, such calculations would allow to directly investigate a possible zero-crossing of the Sachs 
electric form factor $G_E^p(Q^2)$ somewhere above $Q^2\gtrapprox6\GeV^2$ \cite{Gayou:2001qd}.

Lattice calculations of nucleon form factors at large $Q^2$ 
are challenging since they suffer in general from poor signal-to-noise ratios.
They require high-momentum nucleon states, but the smearing 
of the sources and sinks is usually tuned to provide an 
optimal overlap with the nucleon ground state \emph{at rest} and therefore
potentially provides a rather poor overlap with a fast-moving nucleon.
Furthermore, 
the signal-to-noise ratio of the correlators deteriorates quickly 
for large nucleon momenta and large times.
This is particularly a problem in calculations
based on standard ratios of three- to two-point functions (cf. Eq.~\ref{ratio1})
since this requires accurate knowledge of the
nucleon two-point function for large source and/or sink momentum at large sink-times.
%

\begin{figure}[t]
      \centering
          \includegraphics[angle=0,width=0.5\textwidth,clip=true]{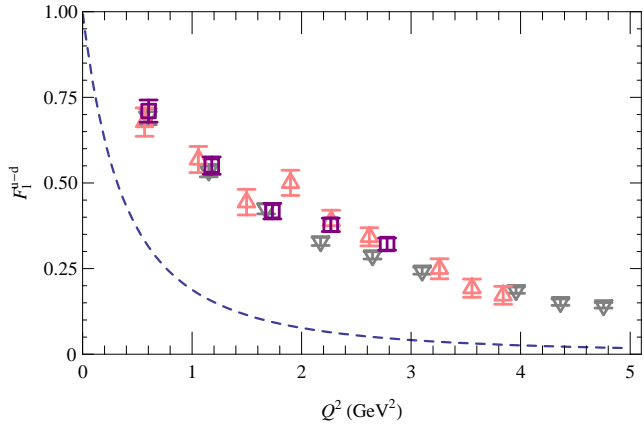}
  \caption{Isovector Dirac form factor for pion masses of $\approx480\MeV$ (triangles)
  $\approx720\MeV$ (squares) and $\approx1100\MeV$ (downward triangles) for a large range of $Q^2$
  (from proceedings \cite{Lin:2008gv}).}
  \label{F1-iso-sim_Lin2008}
\end{figure}

A new strategy to improve the calculation of form factors at large $Q^2$ on the lattice
has been proposed and tested recently \cite{Lin:2008gv}.
It is based on the idea of using a set of differently Gaussian smeared \cite{Alexandrou:1992ti} 
interpolating fields, including those that have a better overlap with high-$P$ and excited nucleon states,
on \emph{anisotropic} lattices with $a_s>a_t$.
This then provides for a better resolution in the time direction 
and hence allows to study the exponential decay of excited and high momentum states more precisely.
In this study, three different Gaussian smearing strengths have been used, and the nucleon 
two-point functions were calculated for all 9 possible source-sink smearing combinations.
From a diagonalization of the correlators, the energies and overlap factors of the ground and first excited state
could be extracted.
These results were then used to extract the relevant matrix elements (and from them finally the form factors) 
from simultaneous fits to the three-point functions, which have also been calculated for all 
9 source-sink smearing combinations.
Preliminary results were obtained in the quenched approximation using the Wilson
gauge action for three pion masses of $\approx480\MeV$, $\approx720\MeV$ and $\approx1100\MeV$,
a spatial lattice spacing of $a_s\approx0.1\fm$, and with $a_s=3a_t$. 
As displayed in Fig.~\ref{F1-iso-sim_Lin2008}, an impressively clean signal was found for the Dirac form factor
$F_1$ in the isovector channel up to rather large momentum transfers $Q^2\approx5\GeV^2$.
Similar results have been obtained for the Pauli form factor, and from the combination
of $F_1$ and $F_2$ it was found that the Sachs proton electric form factor does not
cross zero in the accessible region of $Q^2$.
Further results were obtained for nucleon excited-ground state transition form factors.
It would be very interesting to further improve this promising approach and to extend
the calculations to unquenched QCD and even higher $Q^2$.

%
%

\subsubsection{Transverse momentum dependent parton distribution functions in lattice QCD}
\label{sec:TMDPDFs}
We now turn to a brief discussion of another class of interesting observables,
the so-called transverse momentum dependent parton distribution functions (tmdPDFs).
These provide important information about hadron structure that is to a large extent 
complementary to the physics content of PDFs and GPDs, which have already been discussed
in some detail in this report.
The tmdPDFs depend, in addition to the longitudinal momentum 
fraction $x$, on the intrinsic transverse momentum $k_\perp$ carried by the quarks and gluons
in the hadron, and have in general a probability density interpretation,
similar to the generalized parton distributions (GPDs), 
which provide information about the distribution of quarks and gluons in transverse coordinate 
space (impact parameter-, $b_\perp$-space, see Eq.~(\ref{Impact1}) and illustration Fig.~\ref{ill_impact_1}).
We stress, however, that the transverse momentum, $k_\perp$, and coordinate, $b_\perp$, dependent distribution functions are 
not directly related by a Fourier-transformation and provide in general complementary information 
about the structure of hadrons.

Transverse momentum dependent PDFs play a central role in the description
of semi-inclusive deep inelastic scattering (SIDIS) and the Drell-Yan-process (DY-process).
For a recent review, we refer to \cite{DAlesio:2007jt}.
In such processes, correlations between the intrinsic transverse momenta of the partons, 
the hadron momentum, and their spins can lead to very interesting asymmetries.
Prominent examples are single spin azimuthal asymmetries in SIDIS generated by the
Collins- and Sivers-effect, which have been studied at HERMES, COMPASS and BELLE \cite{Airapetian:2004tw,Alexakhin:2005iw,Abe:2005zx}.
Despite the fact that tmdPDFs are phenomenologically relevant and provide important information about 
structure of hadrons, they received to this date very little attention in the framework of lattice QCD.
One of the reasons for this may be that there is no unique and broadly accepted definition 
of tmdPDFs
in terms of matrix elements of quark and gluon operators available. 
A range of definitions, some of which were proposed in the context
of SIDIS-factorization, can be found in Refs. \cite{Collins:1999dz,Collins:2003fm,Ji:2004wu,Collins:2004nx,Cherednikov:2008ua}.

For the exploratory lattice studies in \cite{Musch:2007ya,Musch:2008jd}, a basic definition
in terms of the correlators (see, e.g., \cite{Mulders:1995dh} and references therein)
\begin{equation}
	\Phi_{\Gamma}(x,k_\perp;P,S) \equiv
	\frac{1}{2} \int d(\bar n\cdot k) \int \frac{d^4l}{(2\pi)^4}
	e^{-ik \cdot l}\ 
	\langle P,S| \bar q(l)\, \Gamma \mathcal{U}_{\mathcal{C}(l,0)}\ q(0)| P,S\rangle
	\label{tmdPDFs1}
\end{equation}
was employed, were $|P,S\rangle$ denotes a nucleon state, $\bar n$ is a lightlike
vector with $\bar n\cdot n=1$, and where the Wilson line $\mathcal{U}_{\mathcal{C}(l,0)}$ 
guarantees the gauge invariance
of the bilocal quark operator (for the notation, see also section \ref{sec:observables}). 
These correlators are, for example, parametrized by the tmdPDF
$f_1(x,k_\perp^2)$ in the unpolarized case, $\Gamma=\;\not \hspace{-0.7ex}n$, and by
$g_{1L}(x,k_\perp^2)$ and $g_{1T}(x,k_\perp^2)$ in the polarized case, $\Gamma=\;\not \hspace{-0.4ex}n \gamma_5$, 
\bea
	\Phi_{\not n}(x, k_\perp;P,S) & =& f_1(x,k_\perp^2) \nonumber\\
	\Phi_{\not n \gamma_5}(x, k_\perp;P,S) &=& \frac{m_N}{n\cdot P} n\cdot S g_{1L}(x,k_\perp^2)\ 
	+\ \frac{k_\perp \cdot S_\perp}{m_N}\ g_{1T}(x,k_\perp^2).
	\label{tmdPDFs2}
\eea
Formally integrating over the transverse momentum, the operator in Eq.~\ref{tmdPDFs1} reduces
to the light-cone operator in Eq.~(\ref{QuarkOp}), and one finds a direct relation to the
standard PDFs, 
\bea
	q(x)&=&f_1(x)= \int d^2k_\perp f_1(x,k_\perp^2)\;,\nonumber\\
	\Delta q(x)&=&g_1(x) = \int d^2k_\perp g_{1L}(x,k_\perp^2)\;.
	\label{tmdPDFsInt}
\eea
Note that the $k_\perp$-integrals in Eqs.~(\ref{tmdPDFsInt}) are logarithmically divergent, and that the
upper integration limit is related to the scale-, $\mu$-, dependence of the PDFs described by the
well-known DGLAP evolution equations.

The gauge link path $\mathcal{C}(l,0)$ in Eq.~\ref{tmdPDFs1} is a crucial ingredient
in the definition of tmdPDFs, and is given by the combination of three straight lines of the form
$\mathcal{C}(l,0)=[l,l\pm\infty n][l\pm\infty n,\pm\infty n][\pm\infty n,0]$, 
where the plus and minus signs correspond to the cases of SIDIS and DY-scattering, respectively.
It turns out, however, that the gauge links on the light cone lead to so-called
light cone singularities in (continuum) perturbation theory\footnote{The light-cone singularities cancel 
out for the $k_\perp$-integrated quantities.}, 
so that in fact a modified definition of the tmdPDFs is needed
\cite{Collins:1999dz,Collins:2003fm,Ji:2004wu,Collins:2004nx,Cherednikov:2008ua}.

To simplify the problem, a straight Wilson line, $\mathcal{U}_{[l,0]}$, directly connecting
the quark field operators, was used for the numerical studies in \cite{Musch:2007ya,Musch:2008jd}. 
It is important to note that the first lattice results are therefore \emph{not} directly related
to the phenomenologically relevant tmdPDFs in SIDIS and DY-processes.
The nucleon matrix elements in Eq.~\ref{tmdPDFs1} can be parametrized by complex valued, invariant 
amplitudes $\tilde A_i(l^2,l\cdot P)$. In the conventions of \cite{Musch:2008jd} one has for example 
\bea
		\langle P,S| \bar q(l)\gamma_\mu\mathcal{U}_{\mathcal{C}(l,0)} q(0)| P,S\rangle
		& =& 4 \tilde{A}_2 P_\mu
		+ 4i m_N^2 \tilde{A}_3 l_\mu \;,\nonumber\\
	\langle P,S| \bar q(l) \gamma_\mu\gamma_5\mathcal{U}_{\mathcal{C}(l,0)} q(0)| P,S\rangle
		& = &
		- 4 m_N \tilde{A}_6 S_\mu
		- 4im_N \tilde{A}_7 P_\mu l \cdot S
		+ 4 m_N^3 \tilde{A}_8 l_\mu l \cdot S\,.
	\label{tmdPDFs3}
	\eea
For the lattice calculation of the nucleon matrix elements, standard
nucleon two-point and three-point functions (obtained using the sequential source technique)
have been employed. The distance between the quark fields in the time direction was set to zero,
$l_0=0$, and in the cases that the spatial separations were not directed along a single axis,
the straight lines in the continuum were approximated by zigzag (step-like) paths on the lattice.
Products of link variables were used to represent 
the Wilson lines $\mathcal{U}_{[l,0]}$ in the discretized non-local operators.

It is well known \cite{Dorn:1986dt,Eichten:1989kb,Maiani:1991az} 
that calculations in a cutoff regularized theory, for example in
lattice QCD with cutoff $\Lambda=a^{-1}$, involving Wilson lines are plagued by potential ultraviolet divergences
in $\Lambda \mathcal{L}$, where $\mathcal{L}$ is the length of the path.
To give a physical meaning to the $l$-dependence of the correlators, the potential
power divergence has to be removed by a renormalization of the lattice operator.
This can be achieved by multiplication with a factor $\exp(-\delta m \mathcal{L})$ \cite{Dorn:1986dt},
where $\delta m\propto a^{-1}$. 
The renormalization constant $a\delta m$ 
was estimated using leading order lattice perturbation theory, and also non-perturbatively
based on a comparison of Wilson lines directed along an axis and along other directions represented
by zigzag paths on the lattice (`taxi driver method') \cite{Musch:2007ya,Musch:2008jd}.
More recently, a different renormalization procedure, employing a specific renormalization condition
that involves the string potential, has been used \cite{Musch:2009xy}.
Since the renormalization is still preliminary and not free of ambiguities, all results below have to be regarded with due caution.

\begin{figure}[t]
    \begin{minipage}{0.48\textwidth}
      \centering
\includegraphics[angle=0,width=0.99\textwidth,clip=true]{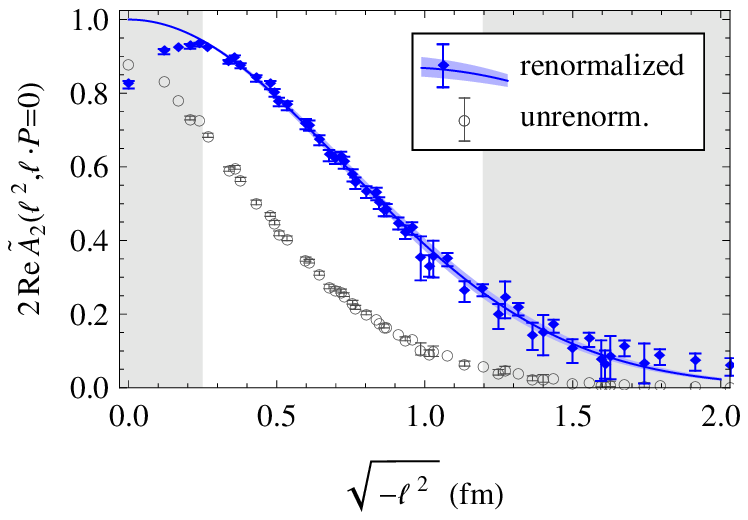}
  \caption{Real part of the amplitude $\tilde A^{u-d}_2$ as a function of the distance $|l|$, obtained for a renormalization condition
  based on the string potential (from \cite{Musch:2009xy}). 
  Similar results based on a different renormalization procedure were presented in the proceedings \cite{Musch:2008jd}.}
  \label{ReA2}
     \end{minipage}
          \hspace{0.3cm}
    \begin{minipage}{0.48\textwidth}
      \centering
          \includegraphics[angle=0,width=0.99\textwidth,clip=true]{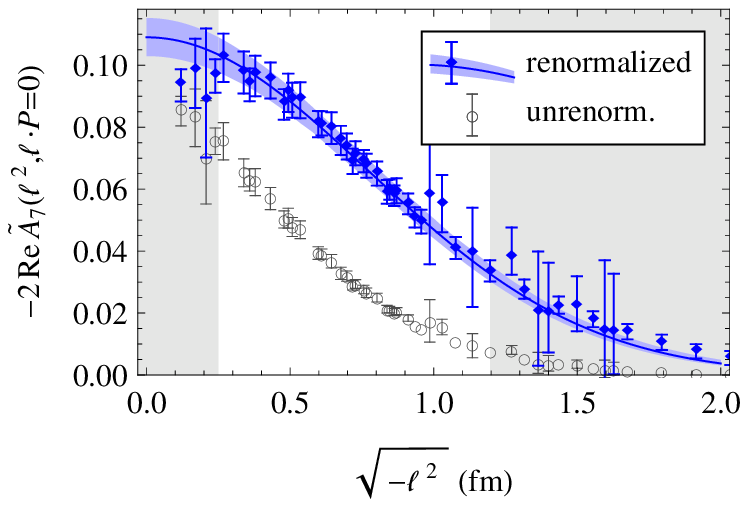}
  \caption{Real part of the amplitude $\tilde A_7^{u-d}$ as a function of the distance $|l|$, obtained for a renormalization condition
  based on the string potential (from \cite{Musch:2009xy}). 
  Similar results based on a different renormalization procedure were presented in the proceedings \cite{Musch:2008jd}.}
  \label{m020_zZ_U_ren_ReA7_vs_l}
     \end{minipage}
\end{figure}

The numerical computations were based on nucleon two- and three-point functions
calculated by LHPC (see, e.g., \cite{Hagler:2007xi}) for $n_f=2+1$ flavors of
domain wall valence fermions and Asqtad staggered sea quarks (MILC configurations), 
for a lattice spacing of $a\approx0.125\fm$, a volume of $V\approx(2.5\fm)^3$,
and three pion masses of $m_\pi\approx496\MeV$, $m_\pi\approx596\MeV$ and $m_\pi\approx759\MeV$.
First lattice results at the lightest pion mass for the real parts of the 
amplitudes $\tilde{A}_2(l^2,l\cdot P\eql0)$ and $\tilde{A}_7(l^2,l\cdot P\eql0)$ for
up-quarks as functions of the distance $|l|$ are displayed in Figs.~\ref{ReA2} 
and \ref{m020_zZ_U_ren_ReA7_vs_l}.
%
%
The statistical errors of the lattice data points for $\tilde{A}_2(l^2,l\cdot P\eql0)$ are
remarkably small. Both amplitudes show a Gaussian like fall-off and can be phenomenologically well described
using a double Gaussian fit ansatz of the form $\tilde A(l^2)=A \exp(-l^2/\sigma_A^2)+B \exp(-l^2/\sigma_B^2)$, as
represented by the solid curves and shaded error bands. 

From Eqs.~(\ref{tmdPDFs1}), (\ref{tmdPDFs2}) and (\ref{tmdPDFs3}), and by noting that $l\cdot P=0$ corresponds
to an integration over the longitudinal momentum fraction $x$, one finds that
Fourier-transformations of the amplitudes $\tilde{A}_2(l^2,l\cdot P\eql0)$ 
and $\tilde{A}_7(l^2,l\cdot P\eql0)$, as represented in practice by, e.g.,
the double Gaussian parametrizations, to $k_\perp$-space correspond to the 
lowest $x$-moment of the tmdPDFs $f^{sW}_1$ and $g^{sW}_{1T}$, respectively.
The superscript $sW$ denotes the straight Wilson lines that were 
employed in the definition of the lattice correlators. 
Based on the numerical results for $\text{Re}\tilde{A}_7(l^2,l\cdot P\eql0)$, Fig.~\ref{ReA2},
it turned out that the distribution $g^{sW}_{1T}(k_\perp)$ is positive valued
for up-, and negative valued for down-quarks.

In analogy to the impact parameter space density given in Eq.~\ref{density1},
a transverse momentum density of longitudinally polarized quarks in a transversely polarized nucleon
can be defined using Eqs.~\ref{tmdPDFs1} and \ref{tmdPDFs2}, given by \cite{Diehl:2005jf}
\begin{equation}
	\rho(x,k_\perp;\lambda,S_\perp) = \Phi_{\not n(1+\lambda\gamma^5)/2}(x, k_\perp;P,S_\perp) = 
	\frac{1}{2} \left( f^{}_1(x,k_\perp^2) + \lambda\frac{k_\perp \cdot S_\perp}{m_N} g^{}_{1T}(x,k_\perp^2) \right)\,,
	\label{tmdPDFs4}
\end{equation}
where $\lambda$ denotes the quark helicity. At this point, it is important to note
that the continuum tmdPDFs $f^{}_1(x,k_\perp^2)$ and $g^{}_{1T}(x,k_\perp^2)$ in Eqs.~(\ref{tmdPDFs2},\ref{tmdPDFs4}) 
behave asymptotically like $k_\perp^{-2}$ and $k_\perp^{-4}$ \cite{Bacchetta:2008xw}, respectively, so that the normalization
of $\rho(x,k_\perp;\lambda,S_\perp)$, as given by an integral over $k_\perp$, needs special consideration.
Here, these difficulties have been evaded by the use of a Gaussian ansatz for the $l$-dependence of the amplitudes,
leading to a Gaussian fall-off in momentum space.

\begin{figure}[t]
      \centering
          \includegraphics[angle=0,width=0.65\textwidth,clip=true]{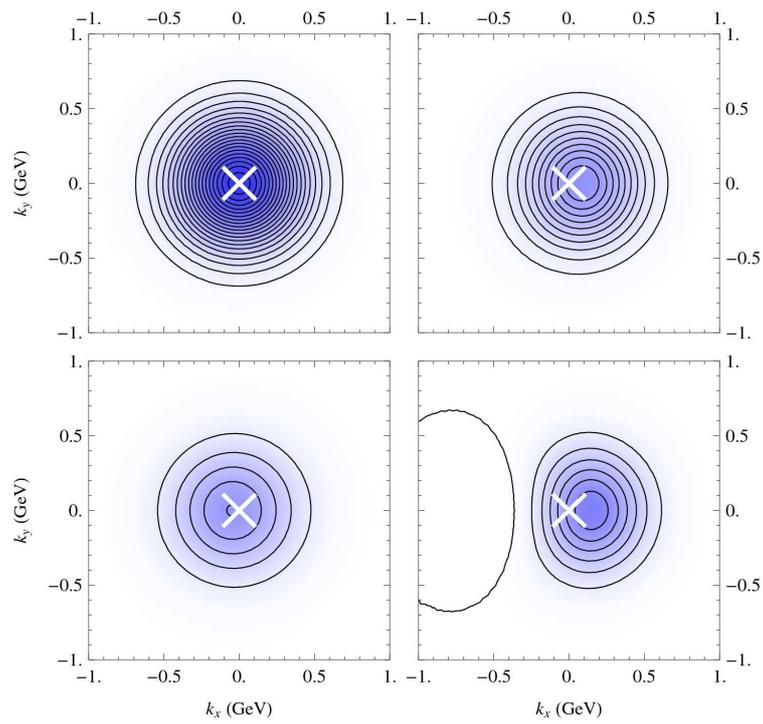}
  \caption{Lowest $x$-moment of the transverse momentum density of
  quarks in the nucleon (from \cite{Musch:2009xy}, 
  see also the proceedings \cite{Musch:2008jd}). Details are given in the text.}
  \label{m020_zZ_densities}
\end{figure}

Figure \ref{m020_zZ_densities} shows the lowest $x$-moment of the density, obtained for
the preliminary lattice results for $f^{sW}_1$ and $g^{sW}_{1T}$, at a pion mass of $m_\pi\approx496\MeV$.
%
%
Compared to the unpolarized density (upper left corner), summed over the
quark helicities $\lambda=\pm1$, the density for up quarks with helicity $\lambda=+1$
in a transversely polarized nucleon with spin $S_\perp=(1,0)$ (upper right corner)
is slightly deformed in $+x$-direction. This is caused by the dipole-like term 
$\propto k_\perp \cdot S_\perp$ in Eq.~\ref{tmdPDFs4} and 
the non-vanishing positive distribution $g^{sW}_{1T}$ for up-quarks.
For down-quarks, $g^{sW}_{1T}$ is negative, so that the corresponding density in the lower left corner
of Fig.~\ref{m020_zZ_densities} is deformed in $-x$-direction.
Clearly, the effects add up in the isovector channel displayed in the lower right corner,
leading to a significant distortion of the density.
We note again that the systematic uncertainties in this exploratory lattice study
of the  tmdPDFs $f^{sW}_1$ and $g^{sW}_{1T}$, in particular due to the renormalization,
are still substantial, so that the results should be regarded as preliminary.
Furthermore, since straight Wilson lines (``$sW$'') were employed in the correlator Eq.~(\ref{tmdPDFs1})
that are significantly different from the Wilson lines of SIDIS and DY-scattering, the lattice results
cannot be directly compared with or used in phenomenological studies.

%% file: Summary.tex
\section{Summary and perspectives}
\label{sec:Summary}
Due to the increasing availability and performance of supercomputers, improved methods and algorithms,
as well as theoretical developments, dynamical lattice hadron structure calculations are by now
routinely carried out at pion masses $\approx350\MeV$ and below, in spatial volumes of $\approx(2\fm)^3$ and larger. 
Significant efforts have been made to get discretization and finite volume effects under control.

In the course of these developments, a large number of increasingly accurate results became available,
some of which have been reviewed in this report:
A summary of lattice studies of form factors and polarizabilities can be found in section \ref{sec:FFsSummary}.
Calculations of moments of PDFs and GPDs as well as distribution amplitudes
are summarized and discussed in sections \ref{sec:PDFsSummary} and \ref{sec:DAs}, respectively.
Finally, in section \ref{sec:InProgress}, we present some promising lattice studies that are beyond mainstream.

Impressive studies of the pion form factor in unquenched lattice QCD have been published by different groups,
in particular in combination with advanced methods and techniques, for example
the ``one-end-trick'' and partially twisted boundary conditions.
In a milestone calculation, $F_\pi(Q^2)$ has been obtained for the first time at a non-zero momentum transfer 
smaller than what could be accessed so far experimentally, and with statistical errors below the
experimental errors, for a comparatively low lattice pion mass of $\approx330\MeV$.
A range of different extrapolations methods, mostly based on 1- and 2-loop chiral
perturbation theory, were used to get estimates of the pion charge radius at the physical point. 
It is noteworthy that a number of low energy constants of chiral effective field theory 
are extracted from such fits with increasing precision.
Although the central values for $\langle r^2\rangle_\pi$ in Fig.~\ref{overview_r2Pi_mPiPhys_v1} 
obtained from different lattice simulations and extrapolation methods 
still scatter significantly, they all nearly overlap within the assigned statistical and systematical uncertainties.
It will be highly interesting to see if the next round of improved lattice calculations of $F_\pi(Q^2)$ will 
finally lead to an agreement with the experimental average from the PDG.

Concerning the nucleon form factors, extensive dynamical lattice studies have been carried out in the recent years
mostly based on conventional methods, reaching down to pion masses of $\approx300\MeV$ in volumes of $\gtrapprox(2\fm)^3$.
The overview plots Fig.~\ref{overview_r1_v2}, \ref{overview_r2_v2} and \ref{overview_kappa_v2} 
for the isovector Dirac and Pauli mean square radii and the anomalous magnetic moment, respectively, 
show a promising overall agreement of the lattice results, and only in some cases discrepancies between individual data points.
Most remarkably, even at the lowest accessible pion masses, the lattice values for $\langle r^2\rangle^{u-d}_{1,2}$ are still
a factor of about two below the respective experimental numbers, and a clear sign of an upwards bending is not yet visible.
However, considering the ChPT predictions in Fig.~\ref{overview_r1_v2}, it is to be expected that very soon at pion masses below
$300\MeV$ chiral dynamics will finally set in and lead to a non-linear increase of the radius as we approach the physical point.
The unambiguous identification of the chiral singularities $\propto\log(m_\pi/\lambda)$ 
in $\langle r^2\rangle^{u-d}_{1}$ and $\propto1/m_\pi$ in $\langle r^2\rangle^{u-d}_{2}$ would indeed represent
a major step in our attempt to calculate these fundamental nucleon structure observables ``ab initio'' on the lattice.

Another benchmark observable, the nucleon isovector axial vector coupling constant $g_A$, received 
a lot of attention in recent years. 
Peculiar to $g_A$ is the importance of the $\Delta(1232)$-resonance, which is largely responsible for the deviation of $g_A$ from $1$, 
furthermore its potentially strong volume dependence, 
and of course the fact that its value is known very precisely from neutron beta decay.
Unquenched lattice results for $g_A$ for a range of pion masses down to $\approx300\MeV$, and
volumes ranging from $\approx(1\fm)^3$ to $\approx(3.5\fm)^3$, displayed in form of a $g_A$-``landscape'' 
in Fig.~\ref{overview_gA}, confirm impressively the expected strong volume dependence at small $m_\pi\times L$.
Previous lattice studies, in one case only including pion masses $\gtrapprox600\MeV$, in the other 
still suffering from limited statistics, observed that the inclusion of explicit $\Delta$-intermediate states 
is indeed crucial for a successful chiral extrapolation, and even found agreement with the experimental number at the physical point. 
More recent lattice data with better statistics and at lower pion masses of $300\ldots350\MeV$, 
however, does not seem to follow the previously predicted $m_\pi$-dependence, but is instead consistently lying 
$\approx10\ldots15\%$ below the experimental value. 
This suggests that the chiral extrapolations of $g_A$, which appeared to be quantitatively successful in the past,
may have to be reconsidered in the future.

With respect to $\Delta$-intermediate states, it is also interesting to note that great progress has been made in lattice calculations of
$\Delta$-baryon form factors, in particular the $\Delta$ magnetic moment, using conventional as well as background field methods.
Together with similar results for nucleon-to-$\Delta$ transition form factors (which we have not discussed in this report),
we may soon have the opportunity to perform simultaneous fits based on ChPT predictions to a large number of lattice data for 
vector and axial-vector nucleon and $\Delta$-baryon (transition) form factors.
Such an approach promises a consistent and improved extraction of the relevant $NN$-, $N\Delta$- and $\Delta\Delta$-coupling constants,
which appear as low energy constants in the ChPT formulae.

Important experience and knowledge has been gained in the last couple of years with respect to
calculations of hadron polarizability using background field methods. 
For some time, apparently contradictory lattice results that differed substantially in magnitude and also in sign 
were presented for the neutron electric polarizability.
By now, these issues seem to be settled, and consistent values for $\alpha^{n}_E$ obtained with different methods 
are available for pion masses $\gtrapprox400\MeV$, as displayed in the overview plot Fig.\ref{overview_alphaN}.
At these pion masses, the lattice results are a factor of $3\ldots5$ below the experimental number.
Possible systematic uncertainties in these calculations certainly have to be studied in greater detail in the future.
However, such low values are not entirely unexpected as ChPT predicts that the neutron electric polarizability diverges 
like $1/m_\pi$ to $+\infty$ in the chiral limit. 
As in the case of the nucleon mean square radii, it is a great challenge to numerically verify the presence 
of such singularities using lattice QCD.

Turning our attention to moments of PDFs and GPDs, we observe that the long-standing issue concerning the 
longitudinal momentum fraction carried by quarks in the nucleon has not yet been resolved:
As displayed in Fig.~\ref{overview_x}, most of the lattice results for $\langle x\rangle_{u-d}$ 
are $\approx50\%$ above the values from global PDF analyses
and are practically flat in $m^2_\pi$ over a wide range of pion masses. 
In addition, there is in some cases a clear disagreement between individual lattice simulations.
Apart from reducing the statistical errors, the central task is therefore to get the apparently substantial systematic
uncertainties, related to possible discretization and finite volume effects, contaminations from excited states
and the operator renormalization, finally under control.
Suitable higher-order results from ChPT may in the end be crucial to a successful extrapolation of the lattice data down to the physical point.

Very valuable information about hadron structure has been obtained from lattice calculations of moments of GPDs.
They provide direct access to many fundamental physics question concerning, e.g., the nucleon spin structure,
the spatial distribution of partons in hadrons, and highly non-trivial correlations between the spin and coordinate degrees of freedom.
Many of the corresponding observables turn out to be rather difficult to access experimentally on a quantitative level.
An important example is the angular momentum carried by quarks in the nucleon, which has
been calculated recently for the first time in dynamical lattice QCD. 
The lattice results, summarized in Table \ref{slj}, are pointing towards a remarkably small total light quark orbital angular momentum, 
$L^{u+d}\approx0$, and negligible total angular momentum of the $d$-quarks, $J^{d}\approx0$.
A directly related and extremely interesting question is if the anomalous gravitomagnetic moment of quarks, $B_{20}^{u+d}(\t0)$,
is in indeed close to zero, as suggested by these first lattice studies. 
Most remarkably, in all the above cases, the small values turn out to be the result of quite precise 
\emph{cancellations} between different non-vanishing flavor, spin and orbital angular momentum contributions.
We stress, however, that these exciting results are currently still subject to significant systematic uncertainties,
and therefore must be regarded as preliminary.
In particular, contributions of so-called disconnected diagrams to singlet observables are not yet routinely included in
lattice hadron structure calculations. 
The computation of disconnected contributions, and also gluonic hadron structure observables, remains a big challenge
due to notoriously low signal-to-noise ratios.
We have seen, however, renewed interest in this area and promising progress recently in particular based on a number of innovative ideas
related to stochastic noise methods and noise reduction techniques.

To the class of observables that are very difficult to access in experiment also belong 
the tensor\footnote{Also denoted by ``quark helicity flip'', ``chiral odd'' or ``transversity''.} 
form factors, PDFs and GPDs, which are related 
to the transverse spin structure of hadrons, as well as hadronic distribution amplitudes.
Dynamical lattice calculations of the former led to fascinating insights into transverse 
spin densities of quarks in the nucleon and the pion, showing in particular 
that the pion has a non-trivial spin structure.

A number of recent dynamical lattice studies provided impressive predictions for 
the lowest moments of 
meson distribution amplitudes.
They are an important ingredient in the QCD-factorization of, e.g.,  
rare $B$-meson decays (that have small branching ratios in the standard model).
Since the latter are sensitive to non-standard model particles through loops, 
high precision lattice results might indeed provide a useful, model-independent 
input in the search for new physics.

Furthermore, moments of the nucleon distribution amplitude have been obtained for the first time in a sophisticated dynamical lattice study.
At the present level of accuracy, the results confirm, for example, the perception that the largest momentum fraction is carried
by up-valence quarks with spin parallel to that of the proton.
It would be very interesting to perform further detailed studies of baryon distribution amplitudes in the future, 
including comparisons with QCD sum rule and model approaches.

In addition to the steady progress in calculations of classical hadron structure observables like form factors and moments of PDFs,
some more unorthodox but promising ideas have been pursued in the recent years.
They include a direct lattice calculation of Wilson coefficients, which may even allow to access the renormalization scale and scheme
independent hadronic \emph{structure functions} rather than the PDFs in the future.
Progress in a new direction has been made very recently in a first lattice study of manifestly non-local quark operators, 
providing access to the intrinsic transverse momentum of quarks in the nucleon.
This novel approach may eventually lead to lattice calculations of the phenomenologically very important 
transverse momentum dependent parton distributions (tmdPDfs).

One of the main limitations repeatedly encountered in lattice QCD calculations is related to very small but non-zero momentum transfers.
They are required for a well-founded extraction of many fundamentally important observables, e.g., 
charge radii, magnetic moments and the anomalous gravitomagnetic moment.
In standard lattice calculations of matrix elements, accessing a small non-zero momentum transfer of about $|Q|\simeq0.1\GeV$ 
would require a very large spatial lattice volume of $V\simeq(12\fm)^3$, corresponding to a box size of 
$L\simeq150$ for a lattice spacing of $a\simeq0.08\fm$.
Lattices of this dimension are clearly out of reach at present, 
and would require an increase of computer power or resources of one to two orders of magnitude.
An alternative to the usual extra- and interpolations in $Q^2$ based on 
certain (model-dependent) ans\"atze is to employ (partially) twisted boundary conditions, which have already
been successfully used in pion form factors studies.
This is particularly promising in combination with results from chiral perturbation theory describing simultaneously
the $m_\pi$- and the $Q^2$-dependence.
Furthermore, background field methods can give access to (otherwise inaccessible) selected observables directly at $Q^2=0$,
as has been demonstrated recently in a dynamical lattice calculation of the $\Delta$-baryon magnetic moment.
All these approaches have their respective merits and limitations, and should be carefully studied and compared in future works.

As a more general remark, we note that
so far, no clear systematic differences can be observed within errors between calculations 
based on $n_f=2$ and $n_f=2+1$ dynamical quarks, for any of the observables discussed in this review.
Notwithstanding this observation, lattice studies are reaching a precision where systematic 
effects and uncertainties, in particular related to
the infinite volume and continuum limits, begin to play a dominant role and have to be carefully considered and estimated.
Concerning the accessible quark masses, we are at the edge of the transition into the chiral regime,
and a clearly visible chiral curvature is expected to show up in many observables very soon.
This will allow hadron structure calculations in lattice QCD and ChPT to reach their full potential:
To give quantitative predictions for a large number of observables and low energy constants from first principles,
and thereby providing unprecedented insight into the low energy dynamics of QCD.

In conclusion, we have seen tremendous advances in lattice hadron structure calculations from the mid-1980's up to now. 
Considering the current status and recent progress of this lively field of research, 
it seems likely that the next major achievements are just a stone's throw away.

\section*{Note added in proof}
It is in the nature of this very active field of research that while this work was being completed, 
a substantial number of new results have been presented and published.
Here we give a short overview of the most relevant recent works that were not discussed in this review.

A calculation and chiral analysis of the pion vector and scalar form factors for $n_f=2$ overlap fermions
was presented in \cite{Aoki:2009qn} by the JLQCD collaboration.
Recent extensive studies of nucleon form factors using $n_f=2+1$ domain wall fermions
by RBC-UKQCD and LHPC can be found in \cite{Yamazaki:2009zq} and \cite{Syritsyn:2009mx}, respectively.
A calculation of the nucleon electromagnetic and axial vector form factors
in twisted mass QCD for two quark flavors has been presented in \cite{Alexandrou:2008rp}.
Interesting results on charge radii, magnetic moments and in particular possible
deformations of decuplet baryons, obtained in the quenched approximation
using FLIC fermions, have been reported in \cite{Boinepalli:2009sq}.
In \cite{Alexandrou:2009hs}, the transverse charge densities of the $\Delta$-baryon have been
studied with $n_f=2$ Wilson fermions, in a mixed action approach using $n_f=2+1$ domain wall valence
and staggered sea quarks, and also in the quenched approximation.
A short review of recent lattice studies of the nucleon spin has been given in \cite{Lin:2009qs}.
Employing background field methods, the electric polarizabilities of the pion and kaon 
have been extracted on an anisotropic lattice for $n_f=2+1$ clover-Wilson fermions \cite{Detmold:2009dx}.
Different ways to extract electric polarizabilities of hadrons in lattice QCD were discussed in \cite{Guerrero:2009yv}.
Refs. \cite{Musch:2009ku,Hagler:2009mb} present a lattice approach to transverse momentum dependent PDFs,
including first results for quark densities in the transverse momentum plane.
An interesting exploratory study of a new quark field smearing approach can be found in \cite{Peardon:2009gh}, and
ref. \cite{Sturm:2009kb} presents and advocates an improved non-perturbative renormalization scheme for quark bilinears.

\section*{Acknowledgments}
This work would not have been possible without the support over the years from numerous colleagues, only
some of which can be mentioned here.
Enduring thanks go to my colleagues from QCDSF(/UKQCD) and LHPC, 
W.~Bietenholz, J.~Bratt, D.~Br\"ommel, R.G.~Edwards, M.~Engelhardt, G.T.~Fleming, 
M.~G\"ockeler, M.~G\"urtler, R.~Horsley, A.~Ali Khan,
H.-W.~Lin, M.F.~Lin, H.B.~Meyer, B.~Musch, Y.~Nakamura, J.W.~Negele,  M.~Ohtani, K.~Orginos,
D.~Pleiter, A.V.~Pochinsky, P.E.L.~Rakow, D.G.~Richards, D.B.~Renner,
A.~Sch\"afer, G.~Schierholz, W.~Schroers,
H.~St\"uben, S.N.~Syritsyn, and J.M.~Zanotti
for years of enjoyable and successful collaboration on many subjects discussed in this review.
It is a pleasure to thank S.J.~Brodsky, M.~Burkardt, M.~Diehl, M.~Dorati, Th.~Feldmann, T.~Gail, Th.R.~Hemmert, N.~Kaiser, P.~Kroll,
M.~Procura, and W.~Weise
for providing interesting results, fruitful collaboration, and for sharing their insights in inspiring discussions. 
Special thanks go to J.M.~Zanotti for helpful remarks on the manuscript.
I would like to thank the cluster of excellence ``Origin and Structure of the Universe'' of the DFG for partial support,
and I am indebted to the DFG for the support in the framework of the Emmy-Noether-Program during the last years.